\documentclass[12pt,oneside,nocenter,noupper,bold,epsfig,psfig]{teseir}
\usepackage[a4paper]{geometry}
\usepackage[dvips]{epsfig}
\usepackage[english]{babel}
\usepackage{graphics}
\usepackage{color}
\usepackage[T1]{fontenc}
\usepackage{setspace}
\usepackage{slashed}
\usepackage{amsmath}
\usepackage{amssymb}
\usepackage{bm}
\usepackage{fancyhdr}
\usepackage{dcolumn}
\usepackage{cite}
\usepackage{dsfont}
\usepackage{caption}
\usepackage{lscape}
\usepackage{multirow}
\usepackage{rotating}

\usepackage{leading}                                        

\def \log {\ln}
\newcommand{\be}{\begin{equation}}       
\newcommand{\ee}{\end{equation}}       
\newcommand{\bear}{\begin{eqnarray}}       
\newcommand{\eear}{\end{eqnarray}}       
\newcommand{\bfg}{\begin{figure}}
\newcommand{\efg}{\end{figure}}
\newcommand{\bi}{\begin{itemize}}
\newcommand{\ei}{\end{itemize}}
\newcommand{\I}{\int^{\infty}_{0}}

\newcommand{\fdd}{\rightarrow}
\newcommand{\tit}{\textit}
\newcommand{\bb}{{\bf b}}
\newcommand{\rr}{{\bf r}}
\newcommand{\qq}{{\bf q}}
\newcommand{\kk}{{\bf k}}
\newcommand{\pp}{{\bf p}}
\newcommand{\jj}{{\bf j}}
\newcommand{\zz}{{\bf z}}
\newcommand{\PP}{{\bf P}}
\newcommand{\bbp}{{\bf b}^{\prime}}

\newcommand{\kkp}{{\bf k}^{\prime}}
\newcommand{\ppp}{{\bf p}^{\prime}}
\newcommand{\rrp}{{\bf r}^{\prime}}
\newcommand{\rrpp}{{\bf r}^{\prime\prime}}

\newcommand{\bh}{\hbar}
\newcommand{\kphi}{\vert\phi\rangle}
\newcommand{\kpsi}{\vert\psi\rangle}

\newcommand{\kpsipm}{\vert\psi^{\pm}\rangle}
\newcommand{\kpsiplus}{\vert\psi^{+}\rangle}
\newcommand{\kpsiminus}{\vert\psi^{-}\rangle}

\newcommand{\krrpp}{\vert\rrpp\rangle}

\newcommand{\brr}{\langle\rr\vert}
\newcommand{\bkk}{\langle\kk\vert}
\newcommand{\brrp}{\langle\rrp\vert}
\newcommand{\bkkp}{\langle\kkp\vert}
\newcommand{\minusbkkp}{\langle-\kkp\vert}

\newcommand{\e}{\epsilon}
\newcommand{\ie}{\pm i\epsilon}
\newcommand{\psirpm}{\brr\psi^{\pm}\rangle}
\newcommand{\psirplus}{\brr\psi^{+}\rangle}
\newcommand{\psirminus}{\brr\psi^{-}\rangle}
\newcommand{\psirpplus}{\brrp\psi^{+}\rangle}
\newcommand{\psirppm}{\brrp\psi^{\pm}\rangle}

\newcommand{\rOr}{\left\langle\rr\left\vert\frac{1}{E\ie-H_{0}}\right\vert\rrp\right\rangle}
\newcommand{\rpVphipm}{\brrp V \kpsipm}
\newcommand{\Gpm}{G_{\pm}(\rr,\rrp)}

\newcommand{\rpArpp}{\brrp \hat{A}\krrpp}
\newcommand{\kpVpsirplus}{\bkkp V\kpsiplus}
\newcommand{\kVpsirplus}{\bkk V\kpsiplus}
\newcommand{\kpVpsirminus}{\minusbkkp V\kpsiminus}
\newcommand{\alp}{\alpha^{\prime}_{\mathds{P}}}
\newcommand{\ap}{\alpha_{\mathds{P}}}
\newcommand{\IP}{\mathds{P}}
\newcommand{\IR}{\mathds{R}}
\newcommand{\aplus}{\alpha_{+}}
\newcommand{\amin}{\alpha_{-}}
\newcommand{\alplus}{\alpha^{\prime}_{+}}
\newcommand{\almin}{\alpha^{\prime}_{-}}

\makeindex
\begin{document}


\pagenumbering{roman}
\onehalfspacing
\cleardoublepage

\begin{titlepage}

\begin{center}
	\includegraphics[width=2cm,clip=true]{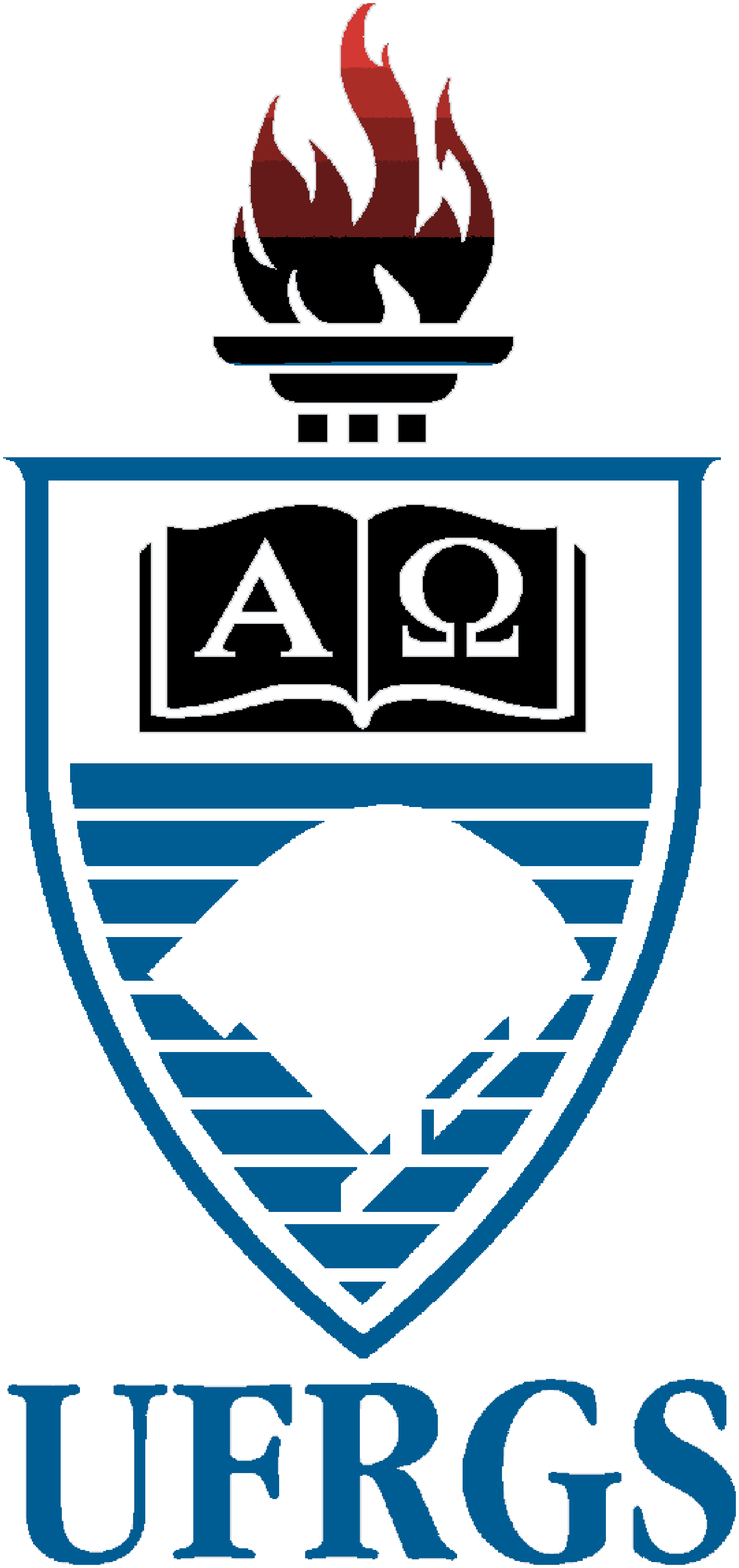}
\end{center}

\begin{center}
\textsc{\large Programa de P\'os-Gradua\c c\~ao em F\'isica}
\end{center}
\begin{center}
\textsc{\Large Tese de Doutoramento}
\end{center}

\setcounter{footnote}{0}
\renewcommand{\thefootnote}{\fnsymbol{footnote}}

\vspace{3 cm}
\begin{center}
\textsc{{\huge\bfseries Din\^amica N\~ao-Perturbativa \\ \vspace{0.4cm} de Colis\~oes Hadr\^onicas}}

\setcounter{footnote}{0}
\renewcommand{\thefootnote}{\arabic{footnote}}

\vspace{2 cm}

\end{center}

\begin{center}
\textsc{{\LARGE Mateus Broilo da Rocha}}
\end{center}

\vspace{7.0 cm} 
\begin{center} 
{Porto Alegre\\ 
2019} 
\end{center}
\newpage
\thispagestyle{plain}
{\huge }
  \begin{quote}
 \vspace*{10 cm}
{\LARGE  } 
 \end{quote}
\vspace*{1cm}
\begin{quote}
\end{quote}





\end{titlepage}

\pagestyle{empty}

\begin{titlepage}

\begin{center}
	\includegraphics[width=2.5cm,clip=true]{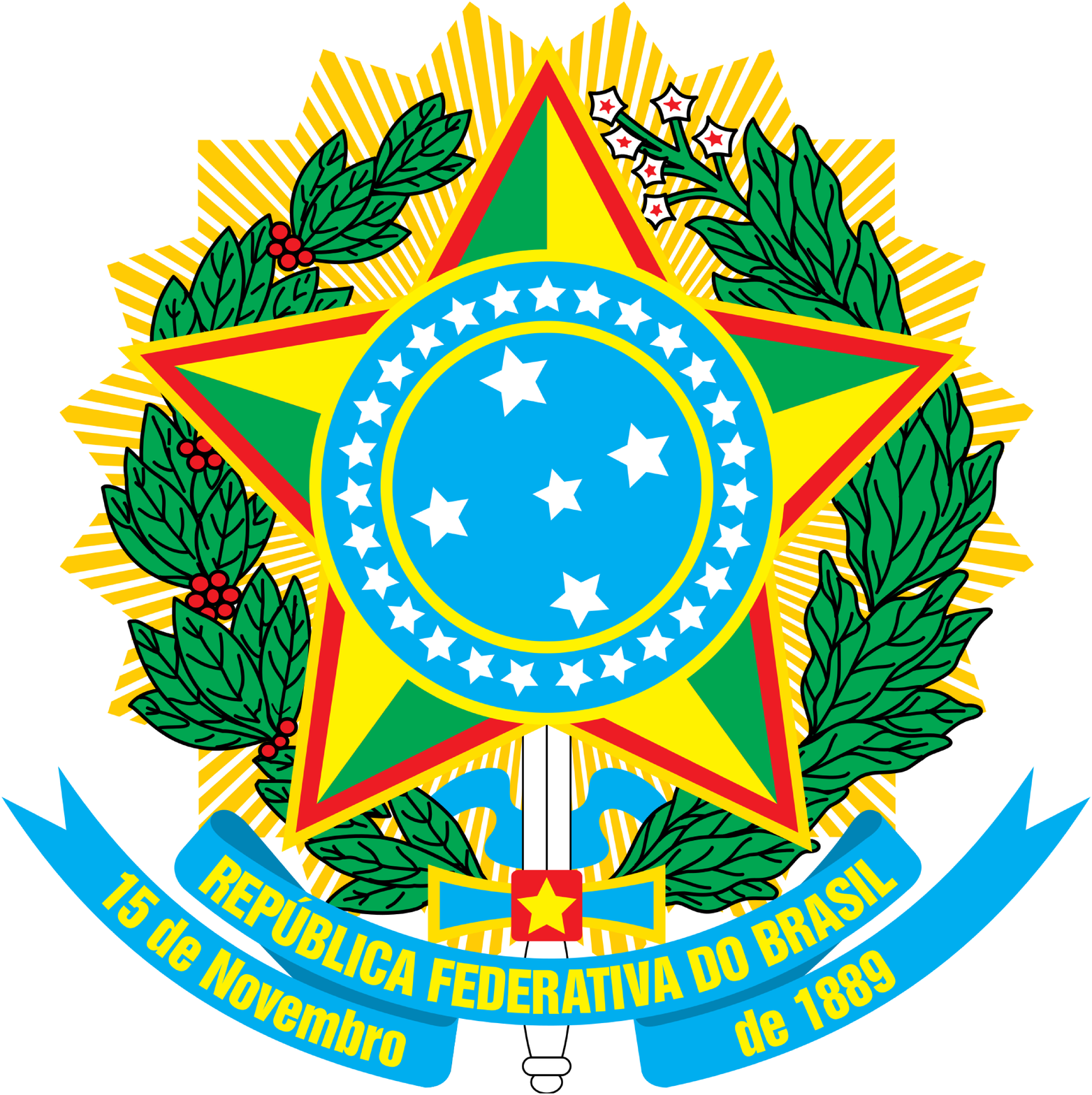}
\end{center}

\begin{center}
\textsc{\Large{Federal University of Rio Grande do Sul}}
\end{center}
\begin{center}
\textsc{\large{Institute of Physics}}
\end{center}
\begin{center}
\textsc{\Large{Doctoral Thesis}}
\end{center}

\setcounter{footnote}{0}
\renewcommand{\thefootnote}{\fnsymbol{footnote}}

\vspace{3.2 cm}
\begin{center}
\textsc{{\huge\bfseries Nonperturbative Dynamics \\ \vspace{0.4cm} of Hadronic Collisions}}
\footnote{Conselho Nacional de Desenvolvimento Cient\'ifico e Tecnol\'ogico (CNPq)}
\end{center}

\setcounter{footnote}{0}
\renewcommand{\thefootnote}{\arabic{footnote}}

\vspace{1 cm}

\begin{center}
\textsc{\Large{ Mateus Broilo da Rocha}}
\end{center}

\vspace{2 cm}
\begin{center}
\textsc{\large{\bf Advisor:} Prof. Dr. Emerson Gustavo de Souza Luna }
\end{center}

\vspace{1.5 cm}
\tit{Doctoral Thesis submitted to the Institute of Physics at Federal University of Rio Grande do Sul in candidacy for the degree of Doctor of Science.}

\vspace{1.5 cm} 
\begin{center} 
{Porto Alegre\\ 
April, 2019} 
\end{center} 
\end{titlepage} 

\thispagestyle{plain} 
\begin{quote}
 \vspace*{15 cm}

\end{quote}
\newpage

\newpage
\thispagestyle{plain}
\vspace*{2 cm}
\textsc{{\huge\bfseries \noindent List of Publications}}

\vspace{0.5cm}
\mbox{}

\section*{\textsc{Articles in Scientific Journals}}

\begin{itemize}

\item[1.] \textsc{Bahia, C. A. S., Broilo, M. and Luna, E. G. S.}, \textit{Nonperturbative QCD effects in forward scattering at the LHC}, {\bf Phys. Rev. D92, 7 (2015), 074039.}
\newline
DOI: 10.1103/PhysRevD.92.074039.

\item[2.] \textsc{Broilo, M., Luna, E. G. S., and Menon, M. J.}, \textit{Soft Pomerons and the Forward LHC Data}, {\bf Phys.Lett. B781 (2018) 616-620.}
\newline
DOI: 10.1016/j.physletb.2018.04.045.

\item[3.] \textsc{Broilo, M., Luna, E. G. S., and Menon, M. J.}, \textit{Forward Elastic Scattering and Pomeron Models}, {\bf Phys. Rev. D98, 7 (2018), 074006.}
\newline
DOI: 10.1103/PhysRevD.98.074006.

\end{itemize}

\section*{\textsc{Conference Proceedings}}

\begin{itemize}

\item[1.] \textsc{Bahia, C. A. S., Broilo, M. and Luna, E. G. S.}, \textit{Energy-dependent dipole form factor in a QCD-inspired model}, {\bf J. Phys. Conf. Ser. 706, 5 (2016), 052006.}
\newline
DOI: 10.1088/1742-6596/706/5/052006.

\item[2.] \textsc{Bahia, C. A. S., Broilo, M. and Luna, E. G. S.}, \textit{Regge phenomenology at LHC energies}, {\bf Int. J. Mod. Phys. Conf. Ser. 45 (2017) 1760064.}
\newline
DOI: 10.1142/S2010194517600643.

\item[3.] \textsc{Broilo, M., Luna, E. G. S., and Menon, M. J.}, \textit{ 	
Leading Pomeron Contributions and the TOTEM Data at 13 TeV}, {\bf Hadron Physics 2018.}
\newline
e-Print: arXiv:1803.06560.

\end{itemize}

\newpage


\thispagestyle{plain} 
\begin{quote}
 \vspace*{15 cm}

\end{quote}
\newpage

\thispagestyle{plain} 
\begin{quote}

 \vspace*{2 cm}
 {\huge\bfseries \noindent }
 \bigskip

\begin{center}
{\bf \tit{To all whom I love, \\ no more nor less.}}
\end{center}

\vspace*{8 cm}

{\begin{flushright}\mbox{} \tit{``Meu lado animal \\ Pratica o Kama Sutra \\ Meu lado humano \\ Vai ao div\~a \\ Meu lado cotidiano \\ Toma tarja preta \\ Meu lado espiritual \\ Reza todas as rezas \\ De vez em quando a gente se encontra.''} \\--\,{\textsc{\large Sobre os Lados}}, \textsc{Michelline Lage.} \\ (Poemas no trem)\end{flushright}} 

{\begin{flushright}\mbox{} \tit{``As for me, I am tormented with an everlasting itch for things remote. \\ I love to sail forbidden seas, and land on barbarous coasts.''} \\-\,{\textsc{\large Moby Dick}}, \textsc{Herman Melville}.\end{flushright}}

\end{quote}
\newpage


\thispagestyle{plain} 
\begin{quote}
 \vspace*{15 cm}

\end{quote}
\newpage


\thispagestyle{plain}
\vspace*{2 cm}
\textsc{{\huge\bfseries \noindent List of Abbreviation}}

\vspace{0.5cm}
\mbox{}

\begin{equation*}
\begin{split}
\tiny
&{\bf ATLAS} \quad \tit{A Toroidal LHC ApparatuS}\\
&{\bf CERN} \quad \tit{Conseil Européen pour la Recherche Nucléaire}\\
&{\bf CL} \quad \tit{Confidence Level}\\
&{\bf CM} \quad \tit{Center of Mass}\\
&{\bf DIS} \quad \tit{Deep Inelastic Scattering}\\
&{\bf DGM} \quad \tit{Dynamical Gluon Mass}\\
&{\bf DGLAP} \quad \tit{Dokshitzer-Gribov-Lipatov-Altarelli-Parisi}\\
&{\bf HERA} \quad \tit{Hadron Electron Ring Accelerator}\\
&{\bf ISR} \quad \tit{Intersecting Storage Ring}\\
&{\bf LHC} \quad \tit{Large Hadron Collider}\\
&{\bf lhs} \quad \tit{left-hand side}\\
&{\bf LO} \quad \tit{Leading Order}\\
&{\bf NLO} \quad \tit{Next to Leading Order}\\
&{\bf OPE} \quad \tit{Operator Product Expansion}\\
&{\bf PDF} \quad \tit{Parton Distribution Function}\\
&{\bf pQCD} \quad \tit{Perturbative Quantum Chromodynamics}\\
&{\bf QCD} \quad \tit{Quantum Chromodynamics}\\
&{\bf QED} \quad \tit{Quantum Electrodynamics}\\
&{\bf QIM} \quad \tit{QCD-Inspired Model}\\
&{\bf rhs} \quad \tit{right-hand side}\\
&{\bf SLAC} \quad \tit{Stanford Linear Accelerator}\\
&{\bf TOTEM} \quad \tit{TOTal Elastic and diffractive cross section Measurement}\\
\end{split}
\end{equation*}

\newpage

\thispagestyle{plain} 
\begin{quote}
 \vspace*{15 cm}

\end{quote}
\newpage


\newpage
\thispagestyle{plain}
\vspace*{2 cm}
\textsc{{\huge\bfseries \noindent Resumo}}

\vspace{1cm}
\hyphenation{cro-mo-di-nâ-mi-ca}
\mbox{\,\,\,\,\,\,\,\,\,}
Nos \'ultimos anos, o LHC tem fornecido medi\c c\~oes precisas de espalhamentos el\'asticos pr\'oton-pr\'oton que t\^em se tornado um guia importante na busca pela sele\c c\^ao de modelos fenomenol\'ogicos e abordagens te\'oricas para se entender, em um n\'ivel mais profundo, a teoria das intera\c c\~oes fortes. Nesta tese, atrav\'es da formula\c c\~ao de dois modelos compat\'iveis com as propriedades de analiticidade e unitaridade, estudamos alguns aspectos relacionados \`a F\'isica por tr\'as das intera\c c\~oes hadr\^onicas. Em especial investigamos o espalhamento el\'astico pr\'oton-pr\'oton e antipr\'oton-pr\'oton em altas energias usando um modelo baseado em teoria de Regge, onde o crescimento da se\c c\~ao de choque total \'e atribu\'ido \`a troca de um estado sem cor possuindo os n\'umeros qu\^anticos do v\'acuo, e um outro modelo baseado no modelo a partons da Cromodin\^amica Qu\^antica, onde o crescimento da se\c c\~ao de choque total \'e associado a espalhamentos semiduros dos p\'artons que comp\~oem os h\'adrons.

\thispagestyle{plain}
\newpage
\thispagestyle{plain}




\newpage

\thispagestyle{plain} 
\begin{quote}
 \vspace*{15 cm}

\end{quote}
\newpage


\thispagestyle{plain}
\vspace*{2 cm}
\textsc{{\huge\bfseries \noindent Abstract}}
\hyphenation{un-res-tric-ted}
\hyphenation{cons-trai-ned}
\vspace{1cm}

\mbox{\,\,\,\,\,\,\,\,\,}
In the last couple of years, the LHC has released precise measurements of elastic proton-proton scattering which has become an important guide in the search for selecting phenomenological models and theoretical approaches to understand, in a deeper level, the theory of strong interactions. In this thesis, through the formulation of two models compatible with analyticity and unitarity constraints, we study some aspects concerning the Physics behind hadronic interactions. In particular, we investigate the proton-proton and the antiproton-proton elastic scattering at high energies using a Regge theory-based model, where the increase of the total cross section is attributed to the exchange of a colorless state having the quantum numbers of the vacuum, and using a model based on the Quantum-Chromodynamics-improved parton model, where the increase of the total cross section is in turn associated with semihard scatterings of partons in the hadrons.

\textsc{
\tableofcontents
\listoffigures
\listoftables
}


\pagenumbering{arabic}
\pagestyle{headings}
\chapter{\textsc{Introduction}}
\hyphenation{in-vol-ving}
\hyphenation{a-naly-sis}
\hyphenation{re-pre-sents}
\hyphenation{con-ti-nu-ed}
\mbox{\,\,\,\,\,\,\,\,\,}
Traditionally, the theory of elementary particle Physics and fields is devoted to study the structure of matter and its interactions at subatomic levels. The whole work and research performed in theoretical particle Physics over the last century has led the scientific community towards the formulation of a well-grounded theory describing three of the four fundamental forces, the so-called Standard Model of elementary particles, or simply known as Standard Model\footnote{Which has just celebrated its $50^{th}$ birthday.} \cite{Weinberg:1967tq}. Speaking of which, these primary fundamental forces are the electromagnetic, weak and strong interactions. More specifically, it is understood that the electromagnetic and weak interactions are different manifestations of the same fundamental force named electroweak interaction, as it was discovered in the early $60$'s by Sheldon Glashow \cite{Glashow:1961tr}. Although the Higgs mechanism was incorporated into the electroweak sector in the late $60$'s by Steven Weinberg and Abdus Salam, only a couple of years ago it was indeed observed the spontaneous symmetry breaking mechanism as the most probable mass generation mechanism for the elementary particles. 

Despite this last breakthrough within the observation of the Higgs boson there are still open problems, specially in the nuclear-strong sector which is described by a Yang-Mills $SU(3)$ theory known as Quantum Chromodynamics. The QCD is considered as the standard theory of strong interactions. It is a local non-Abelian field theory based on the invariant properties of an exact gauge symmetry $SU(3)$ \cite{Ellis:1990pp,Muta:2010xua}. This theory describes the interaction among partons within hadrons and its principal characteristic is the asymptotic freedom of quarks and gluons in the limit of high momentum transfer, which is equivalent to the limit of short distances. However, the non-observation of free physical parton fields in nature implies the hypothesis that confinement sets up a threshold in the limit of low momentum transfer, or the limit of long distances. Therefore, the existence of two different interaction scales, with unique regimes, lead to the central problem in QCD which is the lack of a global method to study strong interactions, since it is physically inconsistent to apply a perturbative approach into confinement region. 

One of the challenges faced by the particle Physics scientific community concerns the nonperturbative aspects of nuclear interactions, which are manifested in processes involving low momenta transfer, defined as soft processes. An example of soft process  is the diffractive hadronic scattering, which can be separated into elastic scattering and single- or double-diffraction dissociation \cite{Barone:2002cv}. The elastic hadronic scattering at high energies represents a rather simple kinematic process, however its complete dynamical description is a fundamental problem in QCD, since the confinement phenomena prohibits a perturbative approach, which is a characteristic of processes with high momenta transfer, also known as hard processes, and a consequence of asymptotic freedom \cite{Politzer:1973fx,Gross:1973id}. The region of low momentum transfer is important not only because confinement is an intrinsically and exclusive characteristic of strong interactions, but rather diffractive processes are dominant in high-energy scatterings. Such process is defined as a reaction in which no quantum numbers are exchanged between the colliding particles where its experimental signature is the presence of large gaps in the rapidity distribution, \tit{i.e.}, the absence of hadronic activity in some regions of the phase space.
 
One approach to the description of elastic hadronic scattering, and independent from QCD, is the Regge theory. This a mathematical framework originally formulated in the picture of nonrelativistic Quantum Mechanics by Tulio Regge \cite{Regge:1959mz,Regge:1960zc} and later developed in the context of an effective field theory \cite{Gribov:2003nw,Collins:1977jy,Baker:1976cv}. Priorly, it was used to study the bound states of an attractive well-behaved spherically symmetric potential in such a way that the partial wave amplitude can be properly analytically continued to complex values of angular momenta. Later on, it was translated into the language of particle Physics by means of the properties of the scattering $S$-matrix \cite{Gribov:1962fx,Gribov:1968fc,Gribov:1968ia}. Regge theory belongs to the class of $t$-channel models and is the theoretical framework used to study diffraction. The bound states, or even sometimes a whole family of resonances, which is known as Regge poles, are related to the description of strong interactions by means of the exchange of Regge trajectories \cite{Collins:1977jy,Gribov:2003nw,Donnachie:2002en,Forshaw:1997dc}.

In the Regge approach, the observed asymptotic increase of the hadronic cross sections with increasing energy is associated with the exchange of a colorless state with the quantum numbers of the vacuum, which is the leading singularity in the $t$-channel, \tit{i.e.} the leading Regge pole, the so-called Pomeron \cite{Foldy:1963zz,Donnachie:1992ny}. It is common to find in the literature that diffraction is synonym of Pomeron physics in the Regge language. However, Regge theory and QCD are two completely different approaches. In the QCD perspective the Pomeron can be described as bound states of reggeized gluons, where in its simplest configuration a color singlet state made up of two reggeized gluons \cite{Donnachie:2002en,Forshaw:1997dc}. Therefore, the former serves as an important handbook in the journey towards a fundamental theory for soft hadronic process. 

Another commonly approach is the QCD-inspired eikonal model. The connection between the elementary dynamics of QCD to physical processes involving hadrons, where the hadronic interactions are described by means of interactions among quarks and gluons, respectively, is made by means of the QCD parton model \cite{Luna:2005nz,Luna:2006qp,Luna:2006sn,Fagundes:2011zx,Bahia:2015gya,Bahia:2015hha,Luna:2006tw,Beggio:2013vfa,Luna:2010tp}. Therefore, the behavior of physical observables is derived using standard QCD cross sections for elementary partonic subprocesses, updated sets of quarks and gluon distribution functions and physically motivated cutoffs that restricts the parton-level processes to semihard ones. In this picture, the increase of the total cross section is associated with parton-parton semihard scatterings, and the high-energy dependence of the cross sections is driven mainly by processes involving gluons. This is quite well understood in the framework of perturbative QCD, but since at high energies there is a close relation between the soft and the semihard components, then clearly the nonperturbative dynamics of the theory is also manifest at the elementary level. However, these elementary processes are plagued by infrared divergences and one natural way to regularize these divergences is by means of a purely nonperturbative effect, the mechanism of gluon mass generation \cite{Brodsky:2001wx,Brodsky:2001ww}. This \tit{ad hoc} mass scale separates the perturbative from the nonperturbative QCD regions, therefore being the natural regulator in our eikonal model.

The usual methodology and schemes adopted to tackle diffractive processes are based on general fundamental principles associated with axiomatic field theory, as for example, the properties of analyticity, unitarity and crossing symmetry of the scattering $S$-matrix. By means of phenomenological models, we search for connections among these fundamental principles and field theories. Our work so far is related to the study of hadronic scatterings, namely the $pp$ and $\bar{p}p$ scatterings, by means of two different phenomenological approaches based on those properties of the scattering amplitudes: \tit{(i)} soft Pomeron models and a Regge-Gribov inspired model, based on Regge dynamics, \tit{(ii)} QCD-inspired eikonal model, based on the parton model of QCD.

At the light of the recent LHC data \cite{0295-5075-96-2-21002,Antchev:2013haa,Antchev:2013iaa,Antchev:2013paa,Antchev:2016vpy,Antchev:2017dia,
Antchev:2017yns,Antchev:2015zza,Antchev:2011zz,Aad:2014dca,Aaboud:2016ijx} which have indicated an unexpected odd decrease in the value of the $\rho$ parameter, the ratio of the real to imaginary part of the forward scattering amplitude, and a $\sigma_{tot}$ value in agreement with previous experimental measurements, the first part of this Thesis is devoted to study the role that soft Pomerons play in strong interactions, which is indisputably of extreme importance. These data at $13$ TeV are not simultaneous described by the predictions of some well-known Pomeron models \cite{Cudell:2002xe} but show agreement with the maximal Odderon dominance hypothesis as it was recently demonstrated \cite{Martynov:2017zjz,Martynov:2018nyb}. Therefore, it will be presented a detailed analysis on the applicability of Pomeron dominance by means of a general class of forward scattering amplitudes. We study the effects of the extrema bounds of the soft Pomeron to forward and nonforward physical observables. Firtly, by means of Born-level amplitudes with single- and double-Pomeron exchange, where the latter is used to restore the unitarity bound since the Pomeron intercept is an effective power valid only over a limited range of energies. Secondly, using eikonalized amplitudes, which is unitarized by construction. More precisely, we consider the possibilities of different combinations of vertices and trajectories for the Pomeron, as for instance we give particular attention to the nearest $t$-channel singularity in the Pomeron trajectory \cite{Anselm:1972ir}, and examine the effects of subleading even and odd Regge contributions.

By means of the revised version of the dynamical gluon mass model, the main focus in the second part of this Thesis is to explore the nonperturbative dynamics of QCD. We bring up the infrared properties of QCD by considering the possibility that this nonperturbative dynamics generates an effective gluon mass scale and the dynamical gluon mass is intrinsically related to an infrared finite strong coupling. Considering the recent elastic data sets in $pp$ and $\bar{p}p$ collisions \cite{0295-5075-96-2-21002,Antchev:2013haa,Antchev:2013iaa,Antchev:2013paa,Antchev:2016vpy,Antchev:2017dia,
Antchev:2017yns,Antchev:2015zza,Antchev:2011zz,Aad:2014dca,Aaboud:2016ijx}, the description of the forward observables follows by considering the eikonal representation, the unitarity condition of the scattering $S$-matrix and the existence of a class of energy-dependent semihard form factors \cite{Bahia:2015gya,Bahia:2015hha}. These form factors represent the overlap distribution of partons within hadrons at a given impact parameter. We are also considering the effects of updated sets of partonic distribution functions in the forward observables, namely CTEQ$6$L \cite{Pumplin:2002vw}, CT$14$ \cite{Dulat:2015mca} and MMHT \cite{Harland-Lang:2014zoa}, respectively.

The outline of this Thesis is organized as follows: Chapter \ref{ch2} introduces some topics related to the formal theory of scattering as well as a brief discussion concerning the kinematics of scattering process, the partial wave expansion and the eikonal formalism. The closing section presents the experimental data compiled and analyzed by the Particle Data Group \cite{Tanabashi:2018oca} and also the recent elastic data sets obtained by the ATLAS and TOTEM Collaboration at the LHC considered in both parts of the Thesis. More precisely, these recent data sets correspond to the highest collider CM energy.

Chapter \ref{chap3} introduces the idea behind Regge theory. It is described the mathematical construction that begins as an analytical continuation of the partial-wave amplitudes to complex values of angular momenta, by bringing up informations of how it effects the convergence domain of the scattering amplitude. Later it is shown how the introduction of a new quantum number, named signature, puts through the problem of analytically continuation in relativistic scattering.

Chapter \ref{ch4} corresponds to the beginning of the first original part of the Thesis. Here for the first time, as far as we are concerned, the asymptotic high-energy behavior of elastic hadron-hadron collisions was studied by taking into account the theoretical uncertainties in a $\chi^{2}$-fitting procedure associated with a general class of parametrization with Pomeron dominance. At the end it is discussed that models with only Pomeron dominance at high-energy region cannot be excluded by the recent experiment results.

Chapter \ref{ch5} presents a quick review of the field theoretical formulation of Regge theory. It is briefly shown that multi-Regge diagrams must be considered in order to tame the asymptotic growth of the cross sections and what sort of diagrams actually leads to the branch-cut singularities. It is also discussed how such diagrams can be summed up by means of Gribov's Reggeon calculus to give the high-energy behavior of the scattering amplitude. 

Chapter \ref{ch6} concludes the first original part of the Thesis. The Regge-Gribov formalism is used to model Born-level amplitudes and eikonalized amplitudes. It is studied the effect of the contribution of double-Pomeron exchange in the Born-level analysis at the limit of high energies and how it is supposed to restore unitarity. Moreover, it is also considered the effects of a nonlinear term corresponding to a two-pion loop in the Pomeron trajectory. At the end it is shown the first results obtained by means of both eikonalized and non-eikonalized amplitudes.

In Chapter \ref{chQCD}, it is presented a brief introduction to Quantum Chromodynamics, the original parton model and the parton model of QCD, where the Bjorken scaling is broken by means of logarithms divergences. It is also shown a brief deduction concerning the leading order and the next-to-leading order expressions for the QCD coupling, a fundamental parameter in the Standard Model.

Chapter \ref{chQIM} corresponds to the second original part of the Thesis, a revised study of the original dynamical gluon mass model. The nonperturbative QCD effects is introduced by means of a QCD-inspired eikonal model whereupon a momentum-dependent gluon mass scale is used to deal with the infrared divergences that usually plague the elementary partonic subprocesses. By means of a class of factorized energy-dependent form factors and also considering updated sets of partonic distribution functions constrained by physically motivated kinematic cutoff, we give predictions to forward observables, namely total cross section and $\rho$-parameter for $pp$ and $\bar{p}p$ interactions.

The conclusions and final remarks of the work so far are included in Chapter \ref{chap9}, as well as some future perspectives and possible secondary research that the present work might branch.

The Appendixes are there just to give a more mathematical meaning to our calculations. And most of all, it is used in the attempt to not overload the text.

\clearpage
\thispagestyle{plain}

\chapter{\textsc{Kinematics of Scattering Processes}}
\label{ch2}
\hyphenation{dif-fe-rent}
\hyphenation{des-crip-tion}
\hyphenation{ob-ser-va-bles}

\mbox{\,\,\,\,\,\,\,\,\,}
This first Chapter is fully dedicated to show in detail the aspects of the cross sections. Within the Quantum Mechanical approach in the high-energy limit, and also by means of the semiclassical approximation, it will be shown that the scattering amplitude, which is the function that defines the whole scattering, turns out to be represented in the impact parameter plane by an eikonal function.

\section{\textsc{Preliminaries}}
\label{sec2.1}
\mbox{\,\,\,\,\,\,\,\,\,}
From Optics to Quantum Mechanics and from nuclear to hadron Physics, diffraction covers a large set of phenomena \cite{Barone:2002cv}. There are many physical experiments concerning different sort of collision processes and particles. At high energies, there is the possibility that the final result shall be a composite system of many new particles. Usually it is said that these collisions lead to scatterings. An easy way to represent it and to understand what is going on is by means of a reaction equation,
\be
1+2\fdd 3+4+5+\text{...}\,,
\label{ch2.1}
\ee
where it represents the collision between particles of types $1$ and $2$ leading to a composite state of new particles $3+4+5+\text{...}$ as the final system, \tit{i.e.} the postscattering result.

In the $60$'s, W.L. Good \& W.D. Walker \cite{Good:1960ba}, were the first authors to give a precise description, and also a modern one, of hadronic diffraction:

\tit{``A phenomenon is predicted in which a high-energy particle beam undergoing diffraction scattering from a nucleus will acquire components corresponding to various products of the virtual dissociations of the incident particles [...] These diffraction-produced systems would have a characteristic extremely narrow distribution in transverse momentum and would have the same quantum numbers of the initial particle.''}

Although, even today this definition still is pretty much the same, it is possible to exploit it by introducing two other different, but equivalent definitions. In the theoretical point of view, a hadronic diffractive process is in general defined as follows \cite{Barone:2002cv}:
\begin{itemize}
	\item[I.] \tit{``A reaction in which no quantum numbers are exchanged between the colliding particles is, at high energies, a diffractive reaction.''}
\end{itemize}
This definition is simple and general enough so that it can cover all the diffractive collision cases at high energies, as depicted in Figure \ref{ch2fig1}:
\begin{itemize}

	\item[i.] Elastic scattering -- $1+2\fdd 1^{\prime}+2^{\prime}$ -- when the incident particles correspond exactly to those ones in the final state.

	\item[ii.] Single-diffraction -- $1+2\fdd 1^{\prime}+X_{2}$ or $1+2\fdd X_{1}+2^{\prime}$ -- when one of the incident particles remains the same after the collision whilst the other one produces a set $X$ of new particles, or a resonance, preserving the quantum numbers of the initial state.

	\item[iii.] Double-diffraction -- $1+2\fdd X_{1}+X_{2}$ -- when both the incident particles give rise to new sets of different particles, or to a resonance, preserving the quantum numbers of the initial state.

\end{itemize}

On the other hand in the experimental point of view, the complete identification of the final state is not always possible to obtain. Hence, in practice it is needed to provide an operational definition of hadronic diffraction which is equivalent to the one given before \cite{Barone:2002cv,Bjorken:1992er}: 
\begin{itemize}
	\item[II.] \tit{``A diffractive reaction is characterized by a large, non-exponentially suppressed, rapidity gap in the final state.''}
\end{itemize}

This definition expresses the fact that the request of a large final state rapidity gap non-exponentially suppressed, $dN/d\Delta\eta\sim constant$, defines the reaction as a diffractive one. Since, contaminations from the distribution of nondiffractive events are of the form $dN/d\Delta\eta\sim e^{-\Delta\eta}$. Therefore, diffractive contributions can be distinguished only asymptotically from nondiffractive ones, after all the latter decrease with energy.

Particularly, the main focus of this Thesis is the diffractive scattering where the two particles remain unaltered after the collision, but in a different kinematic configuration. More specifically, the Thesis kernel is the study of a two-body exclusive scattering special case \tit{viz.} the elastic scattering.
\bfg[hbtp]
  \begin{center}
    \includegraphics[width=12cm,clip=true]{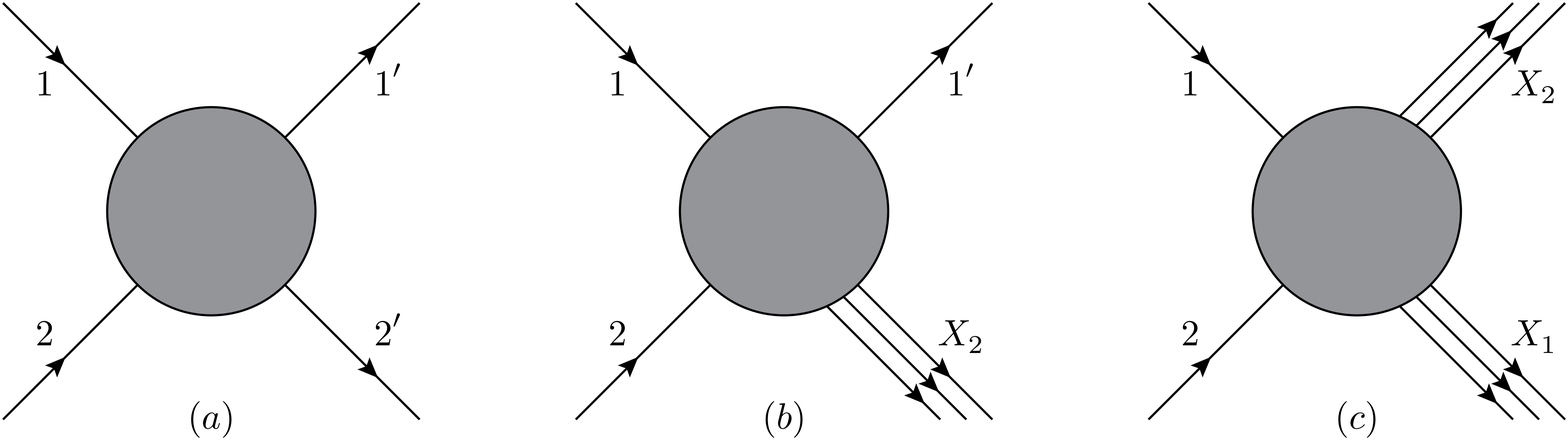}
    \caption{(a) Elastic scattering, (b) single- and (c) double-diffraction.}
    \label{ch2fig1}
  \end{center}
\efg

Despite the fact that it is viewed as the simplest process concerning the kinematics, the theoretical description of elastic scattering is extremely difficult \cite{Barone:2002cv}, because it is linked to processes with low-momentum transfer $q^{2}$, which are usually called soft ones. Actually the description of hadronic scattering with low-$q^{2}$ is one of the most important and challenging concerning the active field of high-energy Physics. 

There is a property called asymptotic freedom of QCD which states that in a well defined energy range the partons can be viewed as free objects, \tit{i.e} as the energy increases the bounds between partons become asymptotically weaker and the corresponding length scale decreases \cite{Politzer:1973fx,Gross:1973id}. This feature is valid for hard processes, meaning high-$q^{2}$. Although part of the process is still of nonperturbative origin the high-$q^{2}$ value allows one to use perturbative QCD. Soft processes are the opposite because confinement, which states that the bounds between partons will not weaken as the energy increases \cite{Muta:2010xua}, prevents a perturbative approach and then new calculation methods are required.

In what follows, it will be shown the basics of two-particles elastic scattering via the Quantum Mechanical formalism by means of the formal theory of scattering.

\section{\textsc{The Lippmann-Schwinger Equation}}
\label{sec2.2}

\mbox{\,\,\,\,\,\,\,\,\,}
There are some basic hypotheses simplifying the scattering, such as one can disregard the coherence effects in the scattered wave function. Therefore, instead of looking at the whole beam one might deal only with two particles interacting, since the scattered flux of $N$ particles equals to $N$ times the flux scattered by one particle. The problem is much simpler then, and the Hamiltonian describing this system is simply
 \be
H=H_{0}+V(\rr),
\label{ch2.3}
\ee
where $V(\rr)$ is the potential and $H_{0}$ stands for the kinetic-energy operator,
\be
H_{0}=\frac{\PP^{2}}{2M},
\label{ch2.4}
\ee
and $\PP=\PP_{1}+\PP_{2}$ is the total momentum, $M=m_{1}m_{2}/(m_{1}+m_{2})$ is the reduced mass and $\rr=\rr_{1}-\rr_{2}$ is the relative distance between particles. 

The absence of a scatter implies that $V(\rr)$ is zero, 
\be
H_{0}\kphi=E\kphi,
\label{ch2.5}
\ee
where $\kphi$ are the energy eigenkets of $H_{0}$. The presence of $V(\rr)$ leads just to a different energy eigenstate from the one usual of a free-particle state. But, when considering elastic scattering processes, one might be interested in obtaining the stationary solution of the full-Hamiltonian Schrödinger equation\footnote{Henceforth, it will be used the natural system of units, however in this section it was explicitly written all the factors $\bh$ to avoid any kind of confusion. Otherwise, when not mentioned, it shall be considered that $\bh=c=1$.},
\be
(H_{0}+V)\kpsi=E\kpsi\,\,\,,\,\,\,E=\frac{\bh^{2}k^{2}}{2M}.
\label{ch2.6}
\ee
\mbox{\,\,\,\,\,\,\,\,\,}
To work with scattering problems always demands a careful look at the boundary conditions so that the evolution description of the wave function is correctly written. Then, the first assumption made is that the initial state should be a state without interaction, \tit{i.e.}, a plane wave at asymptotically $r\fdd-\infty$. The second assumption made is to consider the potential interacting on the initial state and in the range $r\leq r_{0}$ of $V(\rr)$. For this reason, asymptotically when $r\fdd\infty$, the wave function of the whole system is the superposition of the not scattered and the scattered components, \tit{i.e}, the limit $V\fdd 0$ implies $\kpsi\fdd\kphi$.

Within these two assumptions a possible solution could be
\be
\kpsi=\kphi+\frac{1}{E-H_{0}}\,V\,\kpsi,
\label{ch2.7}
\ee
even though there is a singularity in the \tit{rhs}. To avoid further and unnecessary complications one might use a prescription to deal with the singular operator $1/(E-H_{0})$. A nice and easy way to do it is by making the eigenenergy $E$ slightly complex,
\be
\kpsipm=\kphi+\frac{1}{E\ie-H_{0}}\,V\,\kpsipm.
\label{ch2.8}
\ee
\mbox{\,\,\,\,\,\,\,\,\,}
The trick used here will not cause any trouble at all, since the limit that $\e\to 0^{\pm}$ will always be taken once the singularity can be properly contoured by using the residue theorem \cite{Sakurai:2011zz}. Looking at expression (\ref{ch2.8}) it is easy to see that it does not depend on any particular representation, this is known as the Lippmann-Schwinger equation. The physical meaning of $\pm$ will be discussed in the following section, but it is associated with the time $t\gg t_{0}$ and $t\ll t_{0}$, where $t_{0}$ stands for the precise moment when the interaction occurred.

\subsection{\textsc{The Scattering Amplitude}}
\label{sec2.2.1}

\mbox{\,\,\,\,\,\,\,\,\,}
Working in the position basis, expression (\ref{ch2.8}) can be written as an integral equation,
\be
\psirpm=\frac{e^{i\pp\cdot\rr/\bh}}{(2\pi\bh)^{3/2}}+\frac{2M}{\bh^{2}}\,\int d^{3}\rrp\,\Gpm\,\rpVphipm,
\label{ch2.9}
\ee
where $\Gpm$ is defined as
\be
\Gpm\equiv\frac{\bh^{2}}{2M}\,\rOr.
\label{ch2.10}
\ee
\mbox{\,\,\,\,\,\,\,\,\,}
It is straightforward to show that $\Gpm$ can easily be projected into momentum basis,
\be
\Gpm=-\frac{1}{(2\pi)^{3}}\int d^{3}\qq\,\frac{e^{i\qq\cdot\vert\rr-\rrp\vert}}{q^{2}-(k\mp i\e)^{2}}\,\,\,,\,\,\,(k\mp i\e)^{2}\simeq k^{2}\mp i\e,
\label{ch2.11}
\ee
where it was set $\ppp\equiv\bh\qq$. As it was mentioned before, the singularity must be properly contoured by means of the residue theorem, then by doing so and after taking the limit $\e\to0^{\pm}$ one finds that
\be
\Gpm=-\frac{1}{4\pi}\,\frac{e^{\pm ik\vert\rr-\rrp\vert}}{\vert\rr-\rrp\vert},
\label{ch2.12}
\ee
this means that $\Gpm$ is nothing more than the Green function for the Helmholtz equation.

There is a definition which states that a given operator is local if, and only if, it can be written as $\rpArpp=\hat{A}(\rrp)\delta^{(3)}(\rrp-\rrpp)$. In the case where this operator is $V(\rr)$, then a set of local potentials are defined as those ones which are functions of the position operator $\rr$ only, and as a result $\rpVphipm=V(\rrp)\,\psirppm$. Then, the expression for the total wave function (\ref{ch2.9}) is finally  given by
\be
\psirpm=\frac{e^{i\kk\cdot\rr }}{(2\pi\bh)^{3/2}}-\frac{2M}{\bh^{2}}\,\int d^{3}\rrp\,\frac{e^{\pm ik\vert\rr-\rrp\vert}}{4\pi\vert\rr-\rrp\vert}\,V(\rrp)\,\psirppm,
\label{ch2.13}
\ee
where $\kk\equiv\pp/\bh$. It is important to remember that the wave function $\psirpm$ was written in the presence of a scatterer, then the first term in the \tit{rhs} of expression (\ref{ch2.13}) represents the incident wave $\langle\rr\vert\phi\rangle$ whilst the second term represents the scattering effect.

What is really interesting to study in scattering processes is the effect of the scatterer far outside its range. Therefore, in the asymptotic limit where the detector is placed very far away from the scatterer, or similarly $\rr\gg\rrp$, it can be easily shown that 
\be
\vert\rr-\rrp\vert\simeq r-\rrp\cdot\hat{\rr},
\label{ch2.14}
\ee
so, the wave function turns out to be given by
\bear
\psirplus&\underset{r\fdd\infty}{\sim}&\frac{e^{i\kk\cdot\rr}}{(2\pi\bh)^{3/2}}-\frac{1}{4\pi}\,\frac{2M}{\bh^{2}}\,\frac{e^{ikr}}{r}\int d^{3}\rrp\,e^{-i\kkp\cdot\rrp}\,V(\rrp)\psirpplus\nonumber\\
&=&\frac{1}{(2\pi\bh)^{3/2}}\left[e^{i\kk\cdot\rr}+\frac{e^{ikr}}{r}\,f(\kk,\kkp)\right].
\label{ch2.15}
\eear
\mbox{\,\,\,\,\,\,\,\,\,}
In the aforementioned expression, and as it was already mentioned before, it is clear that $\psirplus$ corresponds to the original plane wave in propagation direction $\kk$ plus an outgoing spherical wave with spatial dependence $e^{ikr}/r$ and amplitude $f(\kk,\kkp)$. This is usually referred as the scattering amplitude and it contains all the dynamical information of the collision process. In general it is written as a function of the wave vectors $\kk$ and $\kkp$, or similarly as a function of $k$ and the respectively scattering angles of $\kkp$ relative to $\kk$, 
\be
f(\kk,\kkp)\equiv-\frac{1}{4\pi}\,\frac{2M}{\bh^{2}}\,(2\pi)^{3}\int d^{3}\rrp\,e^{-i\kkp\cdot\rrp}\,V(\rrp)\psirpplus=-\frac{4\pi^{2}M}{\bh^{2}}\,\kpVpsirplus.
\label{ch2.16}
\ee
\mbox{\,\,\,\,\,\,\,\,\,}
From the same point of view $\psirminus$ represents the original plane wave in propagation direction $\kk$ and an incoming spherical wave with spatial  dependence $e^{-ikr}/r$ and amplitude proportional to $\kpVpsirminus$.

\section{\textsc{Cross Sections}}
\label{sec2.3}

\mbox{\,\,\,\,\,\,\,\,\,}
In a simple way, we may say that the differential cross section is viewed as the ratio of the number of particles scattered in an element of solid angle $d\Omega$ per unit of time to the number of incident particles crossing an unit area per unit of time. It is equivalent to say it represents the occurrence of detected events per $d\Omega$,
\be
\frac{d\sigma}{d\Omega}\,d\Omega=\frac{r^{2}\,\vert\jj_{scatt}\vert\,d\Omega}{\vert\jj_{inc}\vert},
\label{ch2.17}
\ee
where $\vert\jj_{scatt}\vert$ and $\vert\jj_{inc}\vert$ account for the scattered and the incident density flux, respectively, and they can be calculated by means of the relation:
\be
\jj=-\frac{i\bh}{2M}\left(\psi^{\ast}\nabla\psi-\psi\nabla\psi^{\ast}\right),
\label{ch2.18}
\ee
which is obtained from Schrödinger's equation.

Within expressions (\ref{ch2.15}) and (\ref{ch2.18}), it is easily seen that\footnote{Considering an incident plane wave in the z direction.}
\be
\jj_{scatt}=\frac{\bh k}{r^{2}M}\,\vert f(\kk,\kkp)\vert^{2}\,\hat{\rr},
\label{ch2.19}
\ee
\be
\jj_{inc}=\frac{\bh k}{M}\,\hat{\zz},
\label{ch2.20}
\ee
and hence, the elastic differential cross section is given by
\be
\frac{d\sigma}{d\Omega}=\vert f(\kk,\kkp)\vert^{2}.
\label{ch2.21}
\ee
\mbox{\,\,\,\,\,\,\,\,\,}
Before pressing on, it should be clear that the elastic scattering processes ought to be only one accessible channel, whereupon the possibility of inelastic events must be taken into account, so that the total cross section contains a probability interpretation of the scattering process. Integrating (\ref{ch2.21}) over the whole solid angle it leads to the elastic cross section, $\sigma_{el}$. Hence,
\be
\sigma_{tot}=\sigma_{el}+\sigma_{in},
\label{ch2.22}
\ee
or in the probabilistic interpretation form,
\be
\frac{\sigma_{el}}{\sigma_{tot}}+\frac{\sigma_{in}}{\sigma_{tot}}=1,
\label{ch2.23}
\ee
which in a sense it tells us that the total cross section measures the overall probability of interaction. This relation represents the unitarity principle, where in the quantum mechanical context it is linked to conservation of probabilities whilst in classical mechanics it is linked to conservation of energy.

There is still an expression, also known as the optical theorem \cite{Barone:2002cv,Sakurai:2011zz}, which relates the total cross section to the imaginary part of the forward scattering amplitude $f(\kk,\kk)=f(\theta=0)$,
\be
\sigma_{tot}=\frac{4\pi}{k}\,\text{Im}\,f(\kk,\kk).
\label{ch2.24}
\ee

\section{\textsc{Asymptotic Theorems}}
\label{sec2.4}

\mbox{\,\,\,\,\,\,\,\,\,}
These theorems were derived by means of the fundamental properties of analyticity, unitarity and crossing symmetry, and usually they are expressed by a set of mathematically rigorous inequalities which the scattering amplitudes must satisfy.

\subsection{\textsc{Optical Theorem}}
\label{sec2.4.1}

\mbox{\,\,\,\,\,\,\,\,\,}
In the beginning of this chapter, it was mentioned that the most important physical observables in diffraction were the cross sections, provided that they are associated with the occurrence probability of one certain scattering process. Starting with Lippmann-Schwinger's equation (\ref{ch2.8}), the imaginary part of the scattering amplitude is
\bear
\text{Im}\,\kVpsirplus & = &-\pi\bkk V \delta(E-H_{0})V\kpsiplus \nonumber\\
&=& -\pi\int d\Omega^{\prime} \, \frac{Mk}{\bh^{2}}\,\vert\bkkp V \kpsiplus\vert^{2},
\label{ch2.25}
\eear
where it was used that
\be
\lim_{\epsilon\to 0^{+}}\left(\frac{1}{E\ie-H_{0}}\right)=\lim_{\epsilon\to 0^{+}}\,\int^{\infty}_{\infty}\,dE^{\prime}\,\frac{\delta(E-E^{\prime})}{E\ie-H_{0}}=i\pi\delta(E-H_{0}).
\label{ch2.26}
\ee
\mbox{\,\,\,\,\,\,\,\,\,}
Thus, by means of the expression for the differential cross section (\ref{ch2.21}) and bearing in mind that $\kkp=\kk$ imposes scattering in the forward direction, \tit{i.e},
\be
\sigma_{tot}=\int\frac{d\sigma}{d\Omega}\,d\Omega,
\label{ch2.27}
\ee 
it is found that
\be
\begin{split}
\text{Im}\,f(\kk,\kk) & =-\frac{\bh^{2}}{4\pi^{2}M}\,\left(-\pi\,\frac{Mk}{\bh^{2}}\int d\Omega^{\prime}\,\vert\bkkp V \kpsiplus\vert^{2}\right)\\
& = \frac{k}{4\pi}\,\sigma_{tot}.
\end{split}
\label{ch.28}
\ee

\subsection{\textsc{The Froissart-Martin-\L ukaszuk Bound}}
\label{sec2.4.2}

\mbox{\,\,\,\,\,\,\,\,\,}
For many years and since the early days of the ISR, it is known that the $pp$ total cross section starts rising after attaining a minimum in the region of $35-40$ mb \cite{Tanabashi:2018oca}. It was found that the $\sigma^{pp}_{tot}$ has a typical-like $\log^{\gamma}s$ growing behavior with $\gamma\sim2$ \cite{Barone:2002cv,Block:1984ru}. As a matter of fact, the Froissart-Martin-\L ukaszuk bound is much more general. By assuming the analyticity and unitarity properties of the scattering $S$-matrix, it states that for any hadronic cross section there is a limit in which it cannot grow faster than $\log^{2}s$ \cite{Froissart:1961ux,Martin:1965jj,Lukaszuk:1967zz},
\be
\sigma_{tot}(s)\le (const) \cdot \log^{2}s, \,\,\,\textnormal{usually}\,\,\,(const)=\frac{\pi}{m^{2}_{\pi}}.
\label{ch2.29}
\ee

By considering first the asymptotic representation of partial wave amplitudes\footnote{The partial wave expansion will be shown in Section \ref{sec2.5}} in the $s$-channel,
\be
A_{\ell}(s)\underset{\ell,s\to\infty}{\sim}f(s)\exp[-\ell\zeta(z_{0})],
\label{ch2.30}
\ee
where $\zeta(z)=\log[z+(z^{2}-1)^{1/2}]$ , $z_{0}=1+2t/s$ and $f(s)$ is a function with a power-like $s$-dependence. And afterwards expanding $\zeta(z)$ as $\log(1+x)$ and taking the limit where $s\to\infty$, therefore expression (\ref{ch2.30}) can be rewritten as
\be
A_{\ell}(s)\sim\exp\left[-\left(\frac{2t}{\sqrt{s}}\right)\ell+\delta\log s\right].
\label{ch2.31}
\ee
\mbox{\,\,\,\,\,\,\,\,\,}
In the limit of high energies one can neglect the partial waves with angular momentum values $\ell\gtrsim c\sqrt{s}\log s$ in such a way that the amplitude series can be truncated as
\be
A(s,t)\underset{s\to\infty}{\simeq}\sum^{c\sqrt{s}\log s}_{\ell=0}(2\ell+1)A_{\ell}(s)P_{\ell}(z)\lesssim 16i\,\pi \sum^{c\sqrt{s}\log s}_{\ell=0}(2\ell+1) \sim i\,Cs\log^{2}s,\,\,\,\text{para}\,\,\,s\to\infty,
\label{ch2.32}
\ee
where it was used the unitarity bound $0\leq Im A_{\ell}(s)/16\pi\leq1$ and also that the associated Legendre polynomial of $\ell$-order are $\vert P_{\ell}(z)\vert\leq1$ for $-1\leq z \leq 1$. The optical theorem finally leads to expression (\ref{ch2.29}) as $s\to\infty$.

\subsection{\textsc{The (revised) Pomeranchuk Theorem for Total Cross Sections}}
\label{sec2.4.3}

\mbox{\,\,\,\,\,\,\,\,\,}
What is usually called today as the original Pomeranchuk theorem states that in general if $ab$ and $a\bar{b}$ cross sections become asymptotically constant and if the $\rho$-parameter, \tit{i.e} the ratio of the real to imaginary part of the forward scattering amplitude, increases slower than $\log s$, hence the two cross sections become asymptotically equal. This definition of the theorem was quite nice when the highest energy available data came from Serpukhov $\sqrt{s}\sim 70$ GeV, where the $\sigma^{pp}_{tot}$ seemed to become constant with increasing energy.

Nowadays this scenario has completely crashed since it is seen that cross sections keep rising as the energy increases. Thus, the original Pomeranchuk theorem must be carefully modified. The scattering processes like $ab$ and $a\bar{b}$ at high energies still obey asymptotically,
\be
\lim_{s\to\infty}\frac{\sigma^{ab}_{tot}(s)}{\sigma^{a\bar{b}}_{tot}(s)}\to 1,
\label{ch2.33}
\ee
if the real part of the scattering amplitude is smaller than its imaginary part\footnote{Pomeranchuk's theorem is a direct manisfestation of the scattering $S$-matrix property of crossing symmetry and can be properly demonstrated by use of dispersion relations for the forward elastic amplitude.}.

The revised version of the theorem \cite{Eden:1965zzb,Grunberg:1973mc} states that if $ab$ and $a\bar{b}$ cross sections grow as $\log^{\gamma}s$, then\footnote{For a better understanding if the cross sections will eventually approach each other, maybe a good thing to do is to look for Fischer's theorem.} the difference between them should be bounded as
\be
\Delta\sigma\leq (const)\cdot\log^{\gamma/2}s.
\label{ch2.34}
\ee

The results shown here have a limited applicability range in the experimental data analysis since they are valid at the asymptotic limit $s\to\infty$. Although asymptotical one can use them to study cross section properties and extrema bounds. Perhaps the optical theorem is in fact one of the most important relation in diffraction.

\section{\textsc{Partial Wave Expansion}}
\label{sec2.5}

\mbox{\,\,\,\,\,\,\,\,\,}
The rotational invariance from a spherically symmetric potential implies an elastic scattering amplitude decomposed as a summation of angular momentum components \cite{Barone:2002cv,Sakurai:2011zz,Block:1984ru}. This decompositon, also known as partial wave expansion, is written as
\be
f(\kk,\kkp)=f(k,\theta)=\sum^{\infty}_{\ell=0}(2\ell+1)\,a_{\ell}(k)P_{\ell}(\cos\theta),
\label{ch2.38}
\ee
where the summation is over all possible values of angular momentum $\ell$, $k$ is the momentum in the CM frame and $a_{\ell}(k)$ is the partial wave amplitude.

To understand the physical meaning of $a_{\ell}(k)$ once again it is useful to analyze the asymptotic behavior of $\psirplus$. Substituting the partial wave expanded scattering amplitude (\ref{ch2.38}) in expression (\ref{ch2.15}), and also using the expansion of a plane wave in terms of spherical waves, one finds that in the limit of large distances \cite{Sakurai:2011zz,Block:1984ru},
\be
\psirplus=\frac{1}{(2\pi\bh)^{3/2}}\,\sum^{\infty}_{\ell=0}(2\ell+1)\,\frac{P_{\ell}(\cos\theta)}{2ik}\left\{\left[1+2ika_{\ell}(k)\right]\,\frac{e^{ikr}}{r}-\frac{e^{-i(kr-\ell\pi)}}{r}\right\}.
\label{ch2.39}
\ee
\mbox{\,\,\,\,\,\,\,\,\,}
Then, the interacting potential acting on the scattering process changes only the coefficient of the outgoing wave, \tit{i.e.} $1\fdd1+2ika_{\ell}(k)$, whilst the incoming wave remains unaltered. By conservation of current the initial incoming flux must equal the outgoing flux, and this must hold for each partial wave because of angular momentum conservation. In principle one could define a quantity $S(k)$ such that \cite{Barone:2002cv}
\be
S_{\ell}(k)\equiv1+2ika_{\ell}(k),
\label{ch2.40}
\ee
this means that by conservation of probability,
\be
\vert S_{\ell}(k)\vert=1,
\label{ch2.41}
\ee
thus, the most that can happen between the wave functions before and after the scattering is a change in the phase of the outgoing wave. Expression (\ref{ch2.41}) is known as the unitarity relation for the $\ell$-th partial wave where $S_{\ell}(k)$ stands for the $\ell$-th diagonal element of the scattering $S$-matrix. Defining this phase shift as $2\delta_{\ell}(k)$ so that
\be
S_{\ell}(k)=e^{2i\delta_{\ell}(k)},
\label{ch2.42}
\ee
hence the amplitudes $a_{\ell}(k)$ can be rewritten as\footnote{And here there are the first hints of the eikonal approximation.}
\be
a_{\ell}(k)=\frac{e^{2i\delta_{\ell}(k)}-1}{2ik}.
\label{ch2.43}
\ee
\mbox{\,\,\,\,\,\,\,\,\,}
By means of expression (\ref{ch2.21}) one finds the elastic cross section,
\bear
\sigma_{el}&=&\int d\Omega\,\frac{d\sigma_{el}}{d\Omega}\,=\int d\Omega\,\vert f(k,\theta)\vert^{2}=\int d\Omega\,\Bigg\vert\sum_{\ell=0}^{\infty}\,(2\ell+1)a_{\ell}(k)P_{\ell}(\cos\theta)\Bigg\vert^{2}\nonumber\\
&=&2\pi\sum_{\ell,\ell^{\prime}=0}^{\infty}\,(2\ell+1)(2\ell^{\prime}+1)\vert a_{\ell}(k)\vert^{2}\int^{+1}_{-1}d(\cos\theta)\, P_{\ell}(\cos\theta)P_{\ell^{\prime}}^{\ast}(\cos\theta)\nonumber\\
&=&4\pi\sum_{\ell=0}^{\infty}(2\ell+1)\vert a_{\ell}(k)\vert^{2},
\label{ch2.44}
\eear
where it was used $P_{\ell}(\cos\theta)=P_{\ell}^{\ast}(\cos\theta)$ and the orthogonality of Lengendre polynomials,
\be
\int_{-1}^{+1}d(\cos\theta)\,P_{\ell}(\cos\theta)P_{\ell^{\prime}}(\cos\theta)=\frac{2}{2\ell+1}\,\delta_{\ell\ell^{\prime}}.
\label{ch2.45}
\ee 
\mbox{\,\,\,\,\,\,\,\,\,}
Finally, the total cross section is found, as usual, by means of the optical theorem (\ref{ch2.24}),
\bear
\sigma_{tot}&=&\frac{4\pi}{k}\,\text{Im}\,f(k,\theta=0)\nonumber\\
&=&\frac{4\pi}{k}\,\sum_{\ell=0}^{\infty}(2\ell+1)\,\text{Im}\,a_{\ell}(k).
\label{ch2.46}
\eear
\mbox{\,\,\,\,\,\,\,\,\,}
Consider the case where only elastic collisions take place, then expression (\ref{ch2.44}) equals expression (\ref{ch2.46}), thus leading to a relation usually called as elastic initarity condition, 
\be
\text{Im}\,a_{\ell}(k)=k\vert a_{\ell}(k)\vert^{2}.
\label{ch2.47}
\ee
Which means that it holds for those cases where the phase shifts $\delta_{\ell}(k)$ are real quantities, \tit{i.e.}, the validity of unitarity relation (\ref{ch2.41}) is ensured only when $\delta_{\ell}(k)$ is real.

In high-energy physics essentially there are two kinds of collisions: elastic ones, where there are conservation of incident particles quantum numbers and the initial and final states are the same; and inelastic ones, where there are changes in the incident particles quantum numbers and the initial and final state are not necessarily the same. In the general case when the inelastic channels are open, the elastic conditions (\ref{ch2.41}) and (\ref{ch2.47}) are no longer true, and somehow absorption effects must be introduced in the region near the potential \cite{Barone:2002cv}. In this case, the unitarity condition implies that
\be
\vert S_{\ell}(k)\vert\leq1,
\label{ch2.48}
\ee
where now the phase shifts $\delta_{\ell}(k)$ are complex quantities. Then, expression (\ref{ch2.42}) must be rewritten as
\be
S_{\ell}(k)=\eta_{\ell}(k)e^{2i\zeta_{\ell}(k)},
\label{ch2.49}
\ee
where $\eta_{\ell}(k)\equiv e^{-2\text{Im}\,\delta_{\ell}(k)}$, with $\text{Im}\,\delta_{\ell}(k)\geq0$ and $\zeta_{\ell}(k)\equiv \text{Re}\,\delta_{\ell}(k)$ are real quantities. Therefore, in this general case the elastic unitarity condition (\ref{ch2.47}) is given by
\be
\text{Im}\,a_{\ell}(k)\geq k\vert a_{\ell}(k)\vert^{2}.
\label{ch2.50}
\ee
\mbox{\,\,\,\,\,\,\,\,\,}
Within these relations, the general unitarity condition satisfied by the partial wave amplitudes reads
\be
\text{Im}\,a_{\ell}(k)-k\vert a_{\ell}(k)\vert^{2}=\frac{1-\eta_{\ell}^{2}(k)}{4k},
\label{ch2.51}
\ee
\bfg[t]
  \begin{center}
    \includegraphics[width=8cm,clip=true]{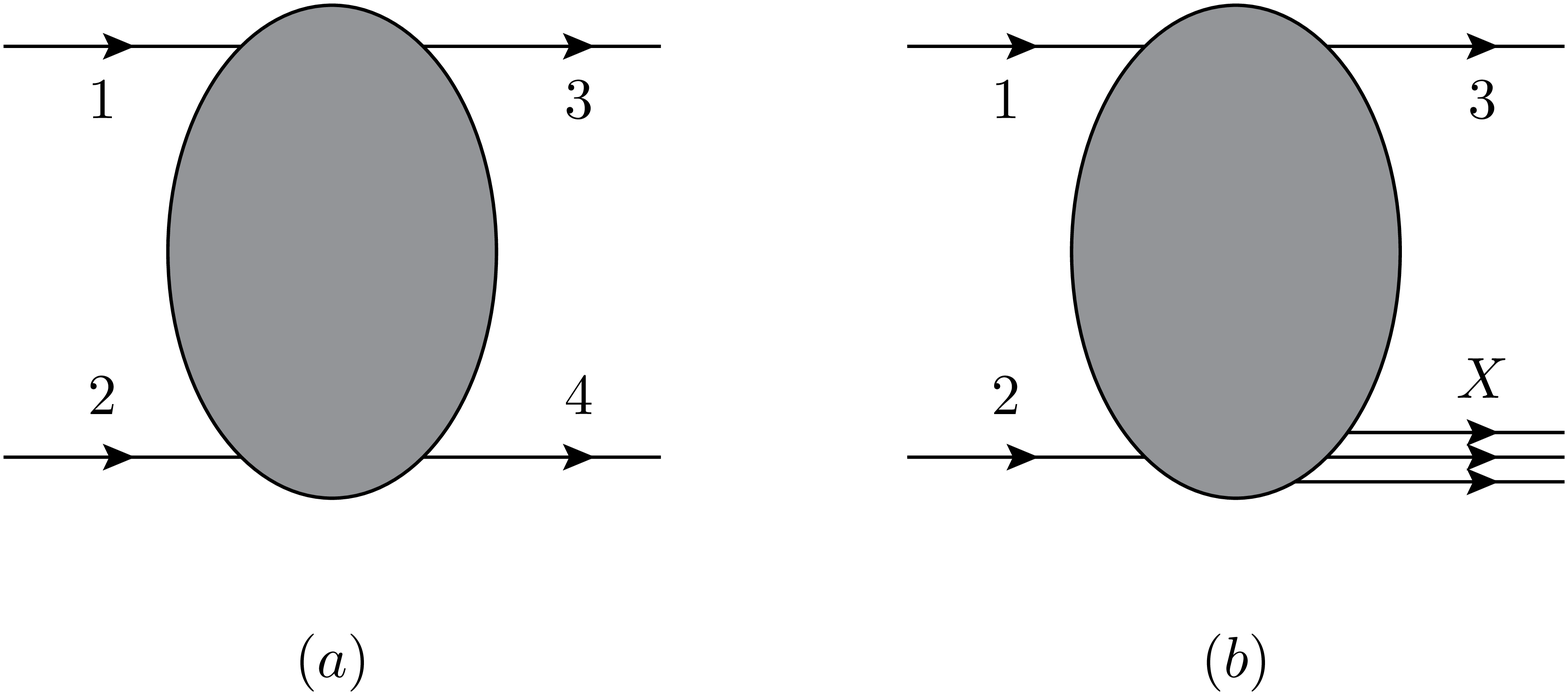}
    \caption{(a) Two-body exclusive and (b) single-particle inclusive scatterings.}
    \label{ch2fig2}
  \end{center}
\efg
where $0\leq\eta_{\ell}(k)\leq1$, are usually called as absorption (or inelastic) coefficients. It is straightforward to see that in the elastic limit, $\eta_{\ell}(k)=1$, its \tit{rhs} reduces to zero and expression (\ref{ch2.47}) is a particular case of (\ref{ch2.51}).

Bearing in mind that $S_{\ell}(k)$ is a diagonal element of the scattering $S$-matrix and it is accordingly written by expression (\ref{ch2.49}) when both elastic and inelastic channels are open, the cross sections can be properly written in terms of the absorption coefficients $\eta_{\ell}(k)$ and the real phase shifts $\zeta_{\ell}(k)$, 
\be
\sigma_{tot}=\sigma_{el}+\sigma_{in}=\frac{2\pi}{k^{2}}\,\sum_{\ell=0}^{\infty}(2\ell+1)[1-\eta_{\ell}(k)\cos2\zeta_{\ell}(k)],
\label{ch2.52}
\ee
\be
\sigma_{el}(k)=\frac{\pi}{k^{2}}\,\sum_{\ell=0}^{\infty}(2\ell+1)[1-2\eta_{\ell}(k)\cos2\zeta_{\ell}(k)+\eta_{\ell}^{2}(k)],
\label{ch2.53}
\ee
\be
\sigma_{in}(k)=\frac{\pi}{k^{2}}\,\sum_{\ell=0}^{\infty}(2\ell+1)[1-\eta_{\ell}^{2}(k)].
\label{ch2.54}
\ee

\section{\textsc{Two-Body Processes and the Mandelstam Invariants}}
\label{sec2.6}

\mbox{\,\,\,\,\,\,\,\,\,}
Each diffractive process has its own experimental signature consisting in an unique final kinematic configuration. However, once it is settled the basic fundamental hypothesis that physical observables are independent from the inertial reference frame, the necessary requirement is that these physical quantities must be grounded on a set of symmetry operations associated with $4-$dimensional space-time which maintains the Physics invariant. This means that symmetry operations can be defined simply as an operation leading a physical system to another, but conditioned that the same properties are preserved and the same equations are satisfied. In summary, it must be invariant under Lorentz transformations \cite{Halzen:1984mc}. Relying on this, it is often used the Mandelstam invariants. 

In a two-body exclusive scattering process, see\footnote{And for completeness (b) was depicted just to show the class where single-diffractive dissociation falls into} Figure \ref{ch2fig2}(a),
\be
1+2\fdd 3+4,\,\,\,\,\,(s-\text{channel}),
\label{ch2.55}
\ee
only two independent kinamatic variables are needed\footnote{In a generic reaction of the type $1+2\to 3+4+...+N$, the number of independent Lorentz invariants is $4(N-1)-N-6=3N-10$.}. As mentioned, these variables are usually chosen between the three Mandelstam invariants defined as \cite{Mandelstam:1958xc}
\bear
\label{ch2.56}
&s=\left(p_{1}+p_{2}\right)^{2}=\left(p_{3}+p_{4}\right)^{2},\\
\label{ch2.57}
&t=(p_{1}-p_{3})^{2}=\left(p_{2}-p_{4}\right)^{2},\\
\label{ch2.58}
&u=\left(p_{1}-p_{4}\right)^{2}=\left(p_{2}-p_{3}\right)^{2},
\eear
and by means of energy-momentum conservation and considering the particles on-shell, then the identity,
\be
s+t+u=\sum_{i=1}^{4}m_{i}^{2},
\label{ch2.59}
\ee
\bfg[hbtp]
  \begin{center}
    \includegraphics[width=12cm,clip=true]{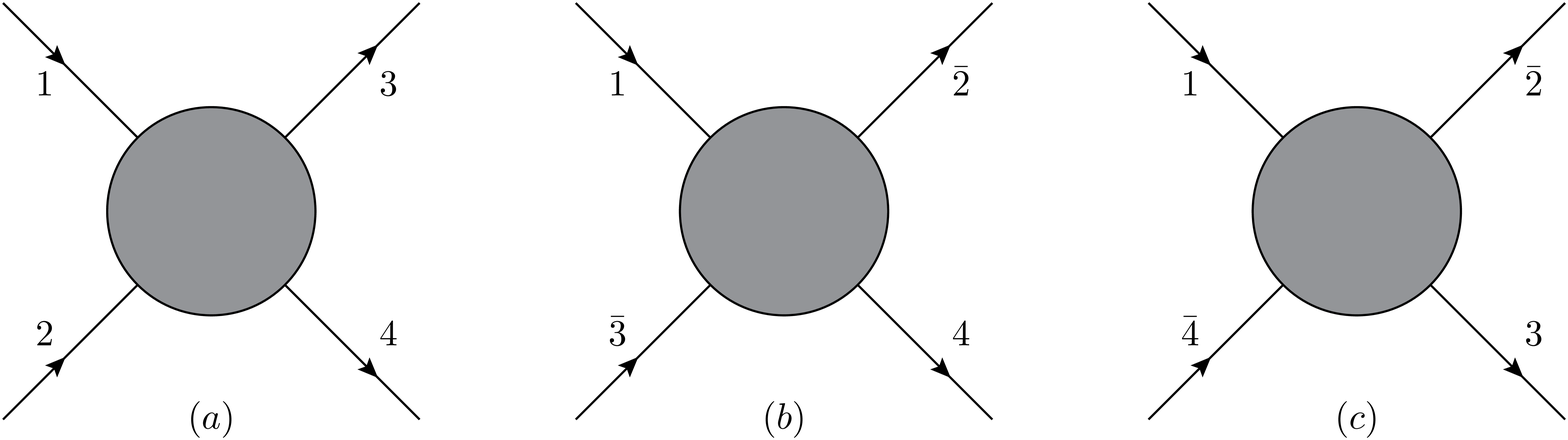}
    \caption{(a) $s$-, (b) $t$- and (c) $u$-channel.}
    \label{ch2fig3}
  \end{center}
\efg
must be respected. Respectively, $p_{i}$, $i=1,...,4$, are the $4-$momenta of particles $1,...,4$. In general, $s$ and $t$ are the chosen independent variables. Where in reaction (\ref{ch2.55}) $s$ is the square of the total CM energy and $t$ is the squared momentum transfer. This reaction is referred to an $s$-channel process. Analogously, for $t$-channel process (and $u$-channel process) means that $t$ $(u)$ defined by expression (\ref{ch2.57}) and (\ref{ch2.58}), is the total squared CM energy, see Figure {\ref{ch2fig3}}. Here, as for example, $\bar{2}$ means that the momentum of particle $2$ has been reversed and all additive quantum numbers have changed sign, \tit{i.e.} $\bar{2}$ is the antiparticle of $2$,
\be
1+\bar{3}\fdd \bar{2}+4,\,\,\,\,\,(t-\text{channel}),
\label{ch2.60}
\ee
\be
1+\bar{4}\fdd \bar{2}+3,\,\,\,\,\,(u-\text{channel}).
\label{ch2.61}
\ee

\subsection{\textsc{The Center-of-Mass Reference Frame}}
\label{sec2.6.1}

\mbox{\,\,\,\,\,\,\,\,\,}
By taking as an example the $s$-channel process (\ref{ch2.55}), the momentum conservation relation in the CM frame, see Figure \ref{ch2fig4}, by definition is
\be
\pp_{1}=-\pp_{2}=\kk,
\label{ch2.62}
\ee 
\be
\pp_{3}=-\pp_{4}=\kk^{\prime},
\label{ch2.63}
\ee
where the $4-$momenta of the particles can be written as
\be
\begin{split}
& p_{1}=(E_{1},\kk) \,\,\,,\,\,\, p_{2}=(E_{2},-\kk),\\
& p_{3}=(E_{3},\kk^{\prime}) \,\,\,,\,\,\, p_{4}=(E_{4},-\kk^{\prime}).\\
\end{split}
\label{ch2.64}
\ee
Moreover, the energies $E_{i}$ can be written in terms of the total squared CM energy $s$, 
\be
E_{1}=\frac{1}{2\sqrt{s}}(s+m^{2}_{1}-m^{2}_{2}),
\label{ch2.65}
\ee
\be
E_{2}=\frac{1}{2\sqrt{s}}(s+m^{2}_{2}-m^{2}_{1}),
\label{ch2.66}
\ee
\be
E_{3}=\frac{1}{2\sqrt{s}}(s+m^{2}_{3}-m^{2}_{4}),
\label{ch2.67}
\ee
\be
E_{4}=\frac{1}{2\sqrt{s}}(s+m^{2}_{4}-m^{2}_{3}).
\label{ch2.68}
\ee

The momenta $\kk$ and $\kk^{\prime}$, and also the momentum transfer $t$, are expressed by means of the mass shell condition $p_{\nu}p^{\nu}=p^{2}=m^{2}$, then
\be
\vert\kk\vert=\frac{1}{2\sqrt{s}}\,\lambda^{1/2}(s,m_{1}^{2},m_{2}^{2}),
\label{ch2.69}
\ee
\be
\vert\kk^{\prime}\vert=\frac{1}{2\sqrt{s}}\,\lambda^{1/2}(s,m_{1}^{2},m_{4}^{2}),
\label{ch2.70}
\ee
\be
t=m^{2}_{1}+m^{2}_{3}-2E_{1}E_{3}+2\vert\kk\vert\vert\kk^{\prime}\vert\cos\theta,
\label{ch2.71}
\ee
and $\cos\theta$ can be written in terms of $s$ and $t$ variables,
\be
\cos\theta=\frac{s^{2}+s(2t-\Sigma_{i}m_{i}^{2})+(m_{1}^{2}-m_{2}^{2})(m_{3}^{2}-m_{4}^{2})}{\lambda^{1/2}(s,m_{1}^{2},m_{2}^{2})\lambda^{1/2}(s,m_{1}^{2},m_{4}^{2})},
\label{ch2.72}
\ee
where $\lambda(x,y,z)$ is a function defined as
\be
\lambda(x,y,z)=x^{2}+y^{2}+z^{2}-2xy-2yz-2xz.
\label{ch.2.73}
\ee
In the high-energy limit, $s\to \infty$, the masses can be neglected and one finds that $E_{i}\simeq \sqrt{s}/2$ and $\vert\kk\vert,\vert\kk^{\prime}\vert\simeq\sqrt{s}/2$.

An important two-body exclusive scattering case is the elastic scattering represented by Figure \ref{ch2fig2}. The relations between the CM variables, $\kk$ and $\theta$ can be much simplified considering the elastic scattering of particles with the same mass,
\be
\vert\kk\vert=\frac{1}{2}\,\sqrt{s-4m^{2}},
\label{ch2.74}
\ee
\be
\cos\theta=1+\frac{2t}{s-4m^{2}}.
\label{ch2.75}
\ee
\mbox{\,\,\,\,\,\,\,\,\,}
It is straightforward to see the following reverse relations,
\be
s=4(k^{2}+m^{2}),
\label{ch2.76}
\ee
\be
t=-4k^{2}\sin^{2}\left(\frac{\theta}{2}\right),
\label{ch2.77}
\ee
\be
u=-2k^{2}(1+\cos\theta),
\label{ch2.78}
\ee
where in the last one it was used the identity (\ref{ch2.59}).
\bfg[hbtp]
  \begin{center}
    \includegraphics[width=5cm,clip=true]{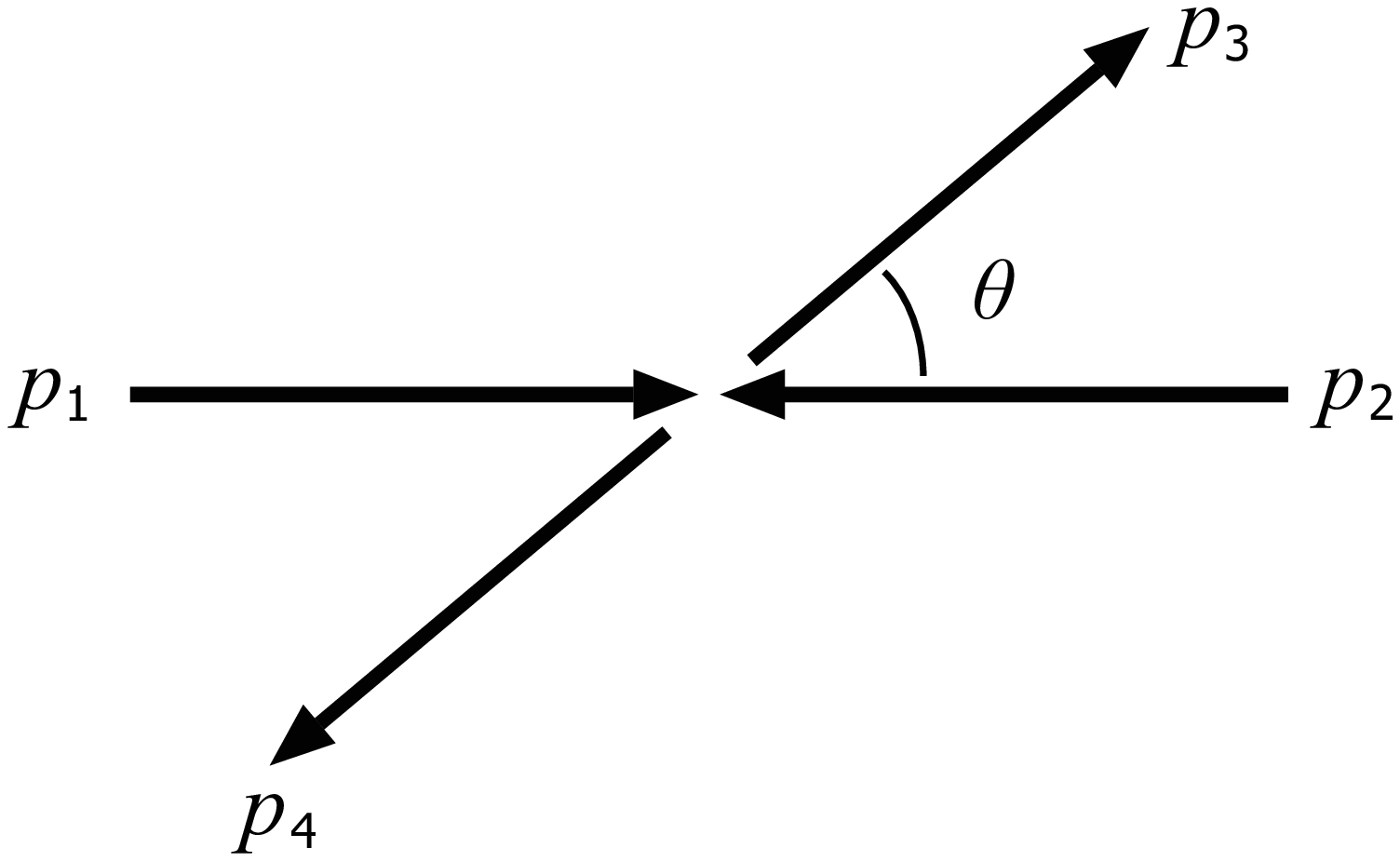}
    \caption{The CM reference frame.}
    \label{ch2fig4}
  \end{center}
\efg

\subsection{\textsc{Physical Domains}}
\label{sec2.6.2}

\mbox{\,\,\,\,\,\,\,\,\,}
Also by considering the case where the collision process is an equal mass scattering, then the physical domain is simply found by means of the kinematic limits of $k$ and $\theta$ in expression (\ref{ch2.76}-\ref{ch2.78}). For $k\geq 0$, $-1\leq\cos\theta\leq 1$ and $0\leq\sin^{2}(\theta/2)\leq 1$, one finds that for the $s$-channel,
\be
s\geq 4m^{2}\,\, , \,\, t\leq 0 \,\, , \,\, u\leq 0.
\label{ch2.79}
\ee
\mbox{\,\,\,\,\,\,\,\,\,}
Following the same path, for a $t$-channel reaction (\ref{ch2.60}), the squared CM energy is now written as $t=(p_{1}+p_{\bar{3}})^{2}=(p_{1}-p_{3})^{2}$ and the momentum transfer $s=(p_{1}-p_{\bar{2}})^{2}=(p_{1}+p_{2})^{2}$,
\be
t=4(k_{t}^{2}+m^{2}),
\label{ch2.80}
\ee
\be
s=-4k_{t}^{2}\sin^{2}\left(\frac{\theta_{t}}{2}\right),
\label{ch2.81}
\ee
where the $t$ subscript means $t$-channel, respectively. The physical domains are then
\be
t\geq 4m^{2}\,\, , \,\, s\leq 0 \,\, , \,\, u\leq 0.
\label{ch2.82}
\ee
\mbox{\,\,\,\,\,\,\,\,\,}
Similarly, for the $u$-channel reaction (\ref{ch2.61}),
\be
u=4(k_{u}^{2}+m^{2}),
\label{ch2.83}
\ee
\be
t=-4k_{u}^{2}\sin^{2}\left(\frac{\theta_{u}}{2}\right),
\label{ch2.84}
\ee
\be
u\geq 4m^{2}\,\, , \,\, s\leq 0 \,\, , \,\, t\leq 0.
\label{ch2.85}
\ee
\bfg[hbtp]
  \begin{center}
    \includegraphics[width=10cm,clip=true]{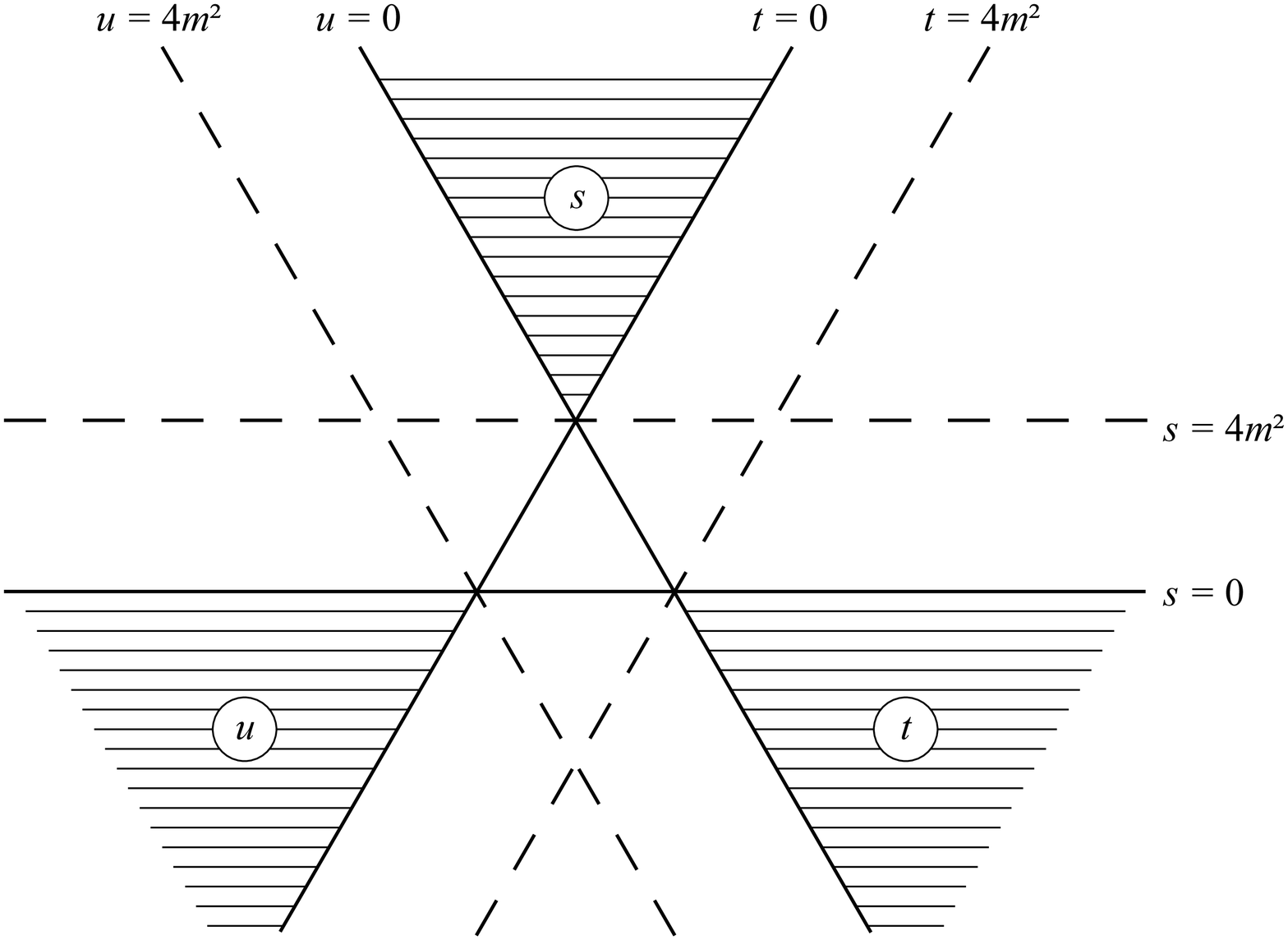}
    \caption{The Mandelstam plot and physical domains.}
    \label{ch2fig5}
  \end{center}
\efg

Although $s,t,u$-channel processes have different and non-overlapping physical domains, see Figure \ref{ch2fig5}, the crossing symmetry of the $S$-matrix ensures they are described by the same scattering amplitude. Hence,
\be
F_{1+2\fdd 3+4}(s,t,u)=F_{1+\bar{3}\fdd\bar{2}+4}(t,s,u),
\label{ch2.86}
\ee
and for the $u$-channel,
\be
F_{1+2\fdd 3+4}(s,t,u)=F_{1+\bar{4}\fdd\bar{2}+3}(u,t,s).
\label{ch2.87}
\ee

Thus, it is easy to see that processes such as those ones given by reactions (\ref{ch2.55}) and (\ref{ch2.60}) are simply related by the change of $s$ and $t$. Usually it is said that (\ref{ch2.60}) is the crossed channel of (\ref{ch2.55}). Our main focus is on $pp$ and $\bar{p}p$ elastic scattering described by
\be
s\text{-channel}:\,\,\,p+p\fdd p+p,
\label{ch2.88}
\ee
\be
t\text{-channel}:\,\,\,\bar{p}+p\fdd \bar{p}+p.
\label{ch2.89}
\ee

As a function of the Mandelstam variables, the elastic differential cross section (\ref{ch2.21}) and the total cross section (\ref{ch2.24}) can be written as
\be
\frac{d\sigma}{dt}(s,t)=\pi\,\vert F(s,t)\vert^{2},
\label{ch2.90}
\ee
\be
\sigma_{tot}(s)=4\pi\,\text{Im}\,F(s,t=0),
\label{ch2.91}
\ee
where it was used $\pi dq^{2}\simeq k^{2}d\Omega$ and $q^{2}=-t$, and also the normalisation $f=F\sqrt{s}/2\simeq kF$ for the scattering amplitude. Beyond these two physical observables, there is another quantity of particular interest in elastic hadronic scattering which defines the phase of the forward amplitude called the $\rho$-parameter and written as the ratio of the real to imaginary part of the forward scattering amplitude,
\be
\rho(s)=\frac{\text{Re}\,F(s,t=0)}{\text{Im}\,F(s,t=0)}.
\label{ch2.92}
\ee
\mbox{\,\,\,\,\,\,\,\,\,}
Strictly and rigorously speaking, this is not a physical quantity in a sense that it is not obtained as a direct measurement, but as a parameter to adjust the elastic differential cross section in the Coulombian-nuclear region. Experimentally, it is known that the nuclear (hadronic) elastic differential cross section can be parametrised in the low-$\vert t\vert$ region, $0.02 < \vert t\vert < 0.5$ GeV$^{2}$, as
\be
\frac{d\sigma_{n}}{dt}(s,t)=\left[\frac{d\sigma_{n}}{dt}\right]_{t=0}\,e^{Bt},
\label{ch2.93}
\ee
where $B(s)$ is the forward slope, \tit{i.e.} a plot $\log(d\sigma_{nucl}/dt)\times t$ in the low-$t$ region would give a straight line of slope $B$. In this case, the slope can be written as
\be
B(s)=\left[\frac{d}{dt}\,\log{\frac{d\sigma_{n}}{dt}(s,t)}\right]_{t=0}=\left[\left(\frac{d\sigma_{n}}{dt}\right)^{-1}\frac{d}{dt}\left(\frac{d\sigma_{n}}{dt}\right)\right]_{t=0}.
\label{ch2.94}
\ee
By means of expressions (\ref{ch2.90}) and (\ref{ch2.91}), the nuclear differential cross section can be written in terms of the $\rho$-parameter as
\be
\frac{d\sigma_{n}}{dt}(s,t)=\pi\,\left\vert (\rho+i)\,\frac{\sigma_{tot}}{4\pi}\,e^{Bt/2}\right\vert^{2}.
\label{ch2.95}
\ee
\mbox{\,\,\,\,\,\,\,\,\,}
The simultaneous presence of Coulombian and nuclear fields does not allow one to disregard interference effects. Instead, the complete differential cross section is given by
\be
\frac{d\sigma}{dt}=\frac{d\sigma_{coulomb}}{dt}+\frac{d\sigma_{cn}}{dt}+\frac{d\sigma_{n}}{dt},
\label{ch2.96}
\ee
where the Coulombian and interference term are respectively given as
\be
\frac{d\sigma_{c}}{dt}=\pi\left\vert (\mp) G^{2}(t)\,\frac{2\alpha}{\vert t \vert} \right\vert^{2},
\label{ch2.coulomb}
\ee
\be
\frac{d\sigma_{cn}}{dt}\approx(\mp)\pi(\rho+\alpha\varphi)\left(\frac{\alpha\sigma_{tot}}{\vert t\vert}\right),
\label{ch2.97}
\ee
the fine-structure constant is $\alpha$, $\varphi$ is the phase factor in the Coulombian region, $G(t)$ is the proton electromagnetic form factor and the upper (lower) sign corresponds to $pp$ $(\bar{p}p)$. The presence of the interference cross section implies direct measurement of the quantity $(\rho+\alpha\varphi)$, thus if one knows $\sigma_{tot}$ then the $\rho$-parameter can be evaluated. What is usually known as the elastic differential cross section is indeed the nuclear differential cross section \cite{Block:1984ru}.

\section{\textsc{Impact Parameter Representation and the Eikonal Formalism}}
\label{sec2.7}
\hyphenation{ge-o-me-tri-cal}
\mbox{\,\,\,\,\,\,\,\,\,}
In the high-energy limit the $\ell$-th partial wave bounds the cross section as a decreasing function of the energy, see expression (\ref{ch2.52}),
\be
\sigma_{tot}\leq\frac{2\pi}{k^{2}}\,(2\ell+1),
\label{ch2.98}
\ee
then a huge increasing number of partial waves must contribute to the high-energy amplitude. There is a subtlety that in this high-energy limit it is possible to construct a geometrical representation based on the impact parameter space, which is the $2$-dimensional space orthogonal to the beam of particles, known as the impact parameter representation. 

The scattering amplitude is written as a summation over all possible values of angular momentum $\ell$, however for finite potentials the energy is bounded by a maximum value $\ell_{max}$, given by $\sqrt{\ell_{max}(\ell_{max}+1)}\simeq kr_{0}$, where $r_{0}$ stands for the potential range. At high energies $kr_{0}\gg 1$, \tit{i.e.} when the energy of the scattering particles is higher than the interacting potential, $E\geq \left\vert V(\rr)\right\vert$, implies the condition $\ell_{max}\gg 1$. Thus, it is sensible to convert the discrete summation over $\ell$ into an integral of element $d\ell$,
$$
\sum_{\ell=0}^{\ell_{max}}\to\I d\ell,
$$
then the scattering amplitude, expression (\ref{ch2.38}), is written as an integral,
\be
f(\kk,\kkp)=\frac{i}{2k}\I d\ell\,(2\ell+1)\left[1-e^{i\chi(k,\ell)}\right]P_{\ell}(\cos\theta),
\label{ch2.99}
\ee
where it was defined $\chi(k,\ell)$ as the continuum form of the phase shifts, also known as the eikonal function\footnote{In the literature sometimes one may also find it by the name of opacity function, because it is related to shadowing, and defined as $\Omega(s,b)=-i\chi(s,b)$.}. For $\ell\gg 1$ and small angles the Legendre polynomials can be written in terms of Bessel's functions of zeroth order,
$$
P_{\ell}(\cos\theta)\approx J_{0}\left[(2\ell+1)\sin(\theta/2)\right]\approx J_{0}(kb\theta).
$$ 
By means of the semiclassical approximation $bk=\ell+1/2$, where the impact parameter $b$ is viewed as the minimal transverse distance between two particles in a collision, then $\int d\ell$ can be properly replaced by $\int db\,k$. Finally, by using expressions (\ref{ch2.76}) and (\ref{ch2.77}) for the energy $s$ and the momentum transfer $q \, (\equiv\sqrt{-t})$, the scattering amplitude can be rewritten,
\be
f(s,q)=ik\I\,db\,b\,J_{0}(qb)\left[1-e^{i\chi(s,b)}\right],
\label{ch2.100}
\ee
and through the normalisation $f=kF$, 
\be
F(s,t)=i\I\,db\,b\,J_{0}(b\sqrt{-t})\,\Gamma(s,b).
\label{ch2.101}
\ee

In the eikonal formalism, the quantity $1-e^{i\chi(s,b)}$ is known as the profile function,
\be
\Gamma(s,b)=1-e^{i\chi(s,b)},
\label{ch2.102}
\ee
which describes the absorption resulting from the opening of inelastic channels by means of a simple optical-geometrical property namely shadowing, a clear bridge towards diffraction \cite{Barone:2002cv}. It is easily seen the aforementioned expression is the Fourier-Bessel transform of the scattering amplitude $F(s,t)$,
\be
\Gamma(s,b)=-i\I\,d\sqrt{-t}\,\sqrt{-t}\,J_{0}(b\sqrt{-t})\,F(s,t).
\label{ch2.103}
\ee
\mbox{\,\,\,\,\,\,\,\,\,}
Expression (\ref{ch2.102}) tells that the profile function is a complex function, and hence in principle there is nothing wrong rewriting it as a combination of a real and an imaginary part,
\be
\Gamma(s,b)=\text{Re}\,\Gamma(s,b)+i\,\text{Im}\,\Gamma(s,b),
\label{ch2.104}
\ee
and equivalently,
\be
\begin{split}
\Gamma(s,b)&=1-e^{i\,\text{Re}\,\chi(s,b)-\text{Im}\,\chi(s,b)} \\
&=\underbrace{\left(1-e^{-\chi_{_{I}}}\cos{\chi_{_{R}}}\right)}_{\text{Re}\,\Gamma(s,b)}+i\underbrace{\left(-e^{-\chi_{_{I}}}\sin{\chi_{_{R}}}\right)}_{\text{Im}\,\Gamma(s,b)},
\label{ch2.105}
\end{split}
\ee
where it was written $\text{Re}\,\chi(s,b)\equiv\chi_{_{R}}$ and $\text{Im}\,\chi(s,b)\equiv\chi_{_{I}}$ just to not to overload the notation. Continuing, the square modulus of $\Gamma(s,b)$,
\be
\begin{split}
\vert\Gamma(s,b)\vert^{2}&=[\text{Re}\,\Gamma(s,b)]^{2}+[\text{Im}\,\Gamma(s,b)]^{2} \\
& =2\underbrace{\left(1-e^{-\chi_{_{I}}}\cos{\chi_{_{R}}}\right)}_{\text{Re}\,\Gamma(s,b)}-\left(1-e^{-2\chi_{_{I}}}\right),
\label{ch2.106}
\end{split}
\ee
so, the real part of the profile function, related to the imaginary part of $F(s,t)$, is associated with contributions from the elastic and inelastic channels via unitarity \cite{Barone:2002cv}, in other words the unitarity of the $S$-matrix requires that the absorptive part of the elastic scattering amplitude receives contributions from both the elastic and the inelastic channels. In impact parameter space this condition may be written as
\be
2\text{Re}\,\Gamma(s,b)=\vert\Gamma(s,b)\vert^{2}+\left(1-e^{-2\chi_{_{I}}}\right).
\label{ch2.107}
\ee
\mbox{\,\,\,\,\,\,\,\,\,}
The cross sections in the impact parameter representation are then written by means of the optical theorem (\ref{ch2.46}), the eikonal representation (\ref{ch2.102}) and the unitarity principle (\ref{ch2.23}),
\bear
\sigma_{el}(s)&=&2\pi\I db\,b\,\vert\Gamma(s,b)\vert^{2}\nonumber\\
&=& 2\pi\I db\,b\,\vert1-e^{-\chi_{_{I}}+i\chi_{_{R}}}\vert^{2}\nonumber\\
&=& 2\pi\I db\,b\,\left[2\left(1-e^{-\chi_{_{I}}}\cos{\chi_{_{R}}}\right)-\left(1-e^{-2\chi_{_{I}}}\right)\right],
\label{ch2.108}
\eear
\bear
\sigma_{in}(s)&=&2\pi\I db\,b\left[2\,\text{Re}\Gamma(s,b)-\vert\Gamma(s,b)\vert^{2}\right]\nonumber\\
&=&2\pi\I db\,b\,\left(1-e^{-2\chi_{_{I}}}\right),
\label{ch2.109}
\eear
\bear
\sigma_{tot}(s)&=&2\pi\I db\,b\,2\,\text{Re}\Gamma(s,b)\nonumber\\
&=&4\pi\I db\,b\,\left(1-e^{-\chi_{_{I}}}\cos{\chi_{_{R}}}\right).
\label{ch2.110}
\eear
\mbox{\,\,\,\,\,\,\,\,\,}
The $\rho$-parameter will be given by
\bear
\rho(s)&=&\frac{\text{Re}\left\{i\I db\,b\,\left(1-e^{i\chi(s,b)}\right)\right\}}{\text{Im}\left\{i\I db\,b\,\left(1-e^{i\chi(s,b)}\right)\right\}}\nonumber\\
&=&\frac{\I db\,b\,e^{-\chi_{_{I}}}\sin\chi_{_{R}}}{\I db\,b\,\left(1-e^{-\chi_{_{I}}}\cos\chi_{_{R}}\right)},
\label{ch2.111}
\eear
the elastic differential cross section,
\be
\begin{split}
& \,\,\,\,\,\,\,\,\,\,\,\,\,\,\,\,\,\,\,\,\,\,\,\,\,\,\,\,\,\,\,\,\,\,\,\,\,\,\,\,\,\,\,\,\,\,\,\,\,\,\,\,\,\,\,\,\frac{d\sigma}{dt}(s,t) =\pi\left\vert \,i\I\,db\,b\,J_{0}(b\sqrt{-t})\,\left[1-e^{i\chi(s,b)}\right]\right\vert^{2} \\
& = \pi\,\left\{\left[\I\,db\,b\,J_{0}(b\sqrt{-t})\,\left(e^{-\chi_{_{I}}}\sin\chi_{_{R}}\right)\right]^{2} + \left[\I\,db\,b\,J_{0}(b\sqrt{-t})\,\left(1-e^{-\chi_{_{I}}}\cos\chi_{_{R}}\right)\right]^{2}\right\},
\end{split}
\label{ch2.112}
\ee
and the $B$-slope,
\be
\begin{split}
B(s)= & \frac{1}{2}\,\Bigg\{\frac{\I db\,b\,e^{-\chi_{_{I}}}\sin\chi_{_{R}}\,\I db\,b^{3}\,e^{-\chi_{_{I}}}\sin\chi_{_{R}}}{[\I db\,b\,e^{-\chi_{_{I}}}\sin\chi_{_{R}}]^{2}+[\I db\,b\,(1-e^{-\chi_{_{I}}}\cos\chi_{_{R}})]^{2}} \, + \\
& +\, \frac{\I db\,b\,(1-e^{-\chi_{_{I}}}\cos\chi_{_{R}})\,\I db\,b^{3}\,(1-e^{-\chi_{_{I}}}\cos\chi_{_{R}})}{[\I db\,b\,e^{-\chi_{_{I}}}\sin\chi_{_{R}}]^{2}+[\I db\,b\,(1-e^{-\chi_{_{I}}}\cos\chi_{_{R}})]^{2}}\Bigg\}.
\end{split}
\label{ch2.113}
\ee
\mbox{\,\,\,\,\,\,\,\,\,}
Notice that, by using the unitarity principle (\ref{ch2.23}) the structure of the cross sections (\ref{ch2.108}-\ref{ch2.110}) could be written in terms of distribution functions, also known as overlap functions defined as
\be
G_{tot}(s,b)=2\,\text{Re}\,\Gamma(s,b),
\label{ch2.114}
\ee
\be
G_{el}(s,b)=\vert\Gamma(s,b)\vert^{2},
\label{ch2.115}
\ee
\be
G_{in}(s,b)=G_{tot}(s,b)-G_{el}(s,b)=2\text{Re}\,\Gamma(s,b)-\vert\Gamma(s,b)\vert^{2}.
\label{ch2.116}
\ee
\mbox{\,\,\,\,\,\,\,\,\,}
In this picture the inelastic overlap function $G_{in}(s,b)$ is the probability occurrence of at least one inelastic event at impact parameter $b$, or in the same way the probability that neither hadron is broken up in a collision at impact parameter $b$ is therefore given by $P(s,b)=e^{-2\chi_{_{I}}(s,b)}$. One direct physical consequence is that no scattering process can be uniquely inelastic, and thus the usual statement that the elastic amplitude results from the shadow scattering from the inelastic channels. Similarly in classical optical theory, according to Babinet's principle, the incidence of plane waves into an obstacle is equivalent to the diffraction by its complementary object. Henceforth, the scattering amplitude tends to be purely imaginary and the elastic scattering purely diffractive with increasing energy.

\section{\textsc{Experimental Data}}
\label{sec2.8}

\mbox{\,\,\,\,\,\,\,\,\,}
The behavior of the hadronic cross sections with increasing CM energy is extremely important to get a good understanding of high-energy diffraction Physics. Through out the whole information we get from regions of low and high momentum transfer one specifically draws attention, the aspects related to the cross sections which play a fundamental role in the study of the smooth transition to perturbative QCD.

The experimental data used in this Thesis are from $pp$ and $\bar{p}p$ scatterings obtained by dedicated experiments in colliders over the past few years. These collisions correspond to events with the highest CM energy that have ever been produced and measured in laboratory.

\section*{\S \,\,\textsc{Forward Physical Quantities}}
\label{sec2.8.1}
\mbox{\,\,\,\,\,\,\,\,\,}
It will be outlined here the experimental data of $\sigma_{tot}^{pp,\bar{p}p}$ and $\rho^{pp,\bar{p}p}$, despite the fact that the complete set for both of these forward quantities is huge, for some reasons it will not be entirely considered the whole amount of data points, see Figure \ref{ch2fig8} where the corresponding collaborations are properly identified. Firstly, because the very low-energy data correspond to the Coulombian-nuclear interference region and we are mainly focused on diffractive processes. Secondly, because the very high-energy data, higher than the LHC present energies, correspond to cosmic ray CM energies\footnote{There are some cosmic ray data points in the middle energy region, but smaller than the typical Tevatron ones, and few points approximately at LHC typically CM energies and a few more farther than the LHC.}. It is worth to be mentioned that cosmic ray predictions are strongly dependent on the Monte Carlo generator for extracting the $\sigma^{pp}_{tot}$ from measurements of proton-air production cross section $\sigma^{p-air}_{prod}$, which causes uncertainties larger than those ones from colliders\footnote{To avoid repeating the same information over and over: concerning the error bars the statistical and systematic uncertainties were combined into quadrature.}, see Figure \ref{ch2fig10}. This concludes the reason why it was decided to use forward, collider only, data in the energy region $\sqrt{s}\geq 5$ GeV.

As mentioned above, although these cosmic ray data are not being considered in the fitting, we extrapolate our results in the energy region far beyond LHC just as a matter of comparison. Respectively we compare our predictions with the AUGER experimental datum at $\sqrt{s}=57$ TeV with $\sigma^{pp}_{tot}=133.0\pm 29.0$ mb \cite{Collaboration:2012wt} and the Telescope Array (TA) datum at $\sqrt{s}=95$ TeV with $\sigma^{pp}_{tot}=170.0 \pm 51.0$ mb \cite{Abbasi:2015fdr}.

In the case of $\bar{p}p$ collisions, there are results for $\sigma^{\bar{p}p}_{tot}$ and $\rho^{\bar{p}p}$ only at energies $\sqrt{s}\leq 1.8$ TeV, which represents the CM energy of Tevatron at Fermilab. As for the case of $pp$ collisions, the $\sigma^{pp}_{tot}$ and $\rho^{pp}$ results at energies $\sqrt{s}\leq62.8$ were obtained at CERN-ISR, and the recent runs at the LHC at the energy interval $\sqrt{s}=7-13$ TeV obtained by the TOTEM Collaboration as well as the measurements of $\sigma^{pp}_{tot}$ at $\sqrt{s}=7-8$ TeV obtained by ATLAS. It is shown in Table \ref{ch2tab1} the compiled set of the highest collider energy data for $\sigma_{tot}^{pp}$ and $\rho^{pp}$ obtained very recently at the LHC \cite{0295-5075-96-2-21002,Antchev:2013haa,Antchev:2013iaa,Antchev:2013paa,Antchev:2016vpy,Aad:2014dca,Aaboud:2016ijx} by means of three specific methods:

\subsubsection{\textsc{I. Elastic Scattering Extrapolating to the Optical Point $(t=0)$}}
\hyphenation{i-ma-gi-na-ry}
\mbox{\,\,\,\,\,\,\,\,\,}
According to expression (\ref{ch2.90}) the elastic differential cross section can be written in terms of the scattering amplitude. Hence a relation between the total cross section and the imaginary part of the forward scattering amplitude can be found by means of the optical theorem (\ref{ch2.91}),
\be
\sigma_{tot}(s)=4\pi\,\text{Im}\,F(s,t=0)=\frac{4\sqrt{\pi}}{\sqrt{1+\rho^{2}}}\,\left[\frac{d\sigma}{dt}\Bigg\vert_{t=0}\right]^{1/2},
\label{ch2.117}
\ee
where it was used $F(s,0)=(\rho+i)\,\text{Im}\,F(s,0)$ and $\rho$ stands for the ratio of the real to imaginary part of the scattering amplitude. 

Exponentially extrapolating the elastic differential cross section to $t=0$, see expression (\ref{ch2.93}), and using the prediction of $\rho\simeq 0.141\pm 0.007$ at $\sqrt{s}=7$ TeV from the COMPETE  Group \cite{Cudell:2002xe}, the TOTEM Collaboration arrived at $\sigma_{tot}^{pp}=98.30\pm 2.80$ mb \cite{0295-5075-96-2-21002} and $\sigma_{tot}^{pp}=98.58\pm 2.23$ mb \cite{Antchev:2013haa}, both using the luminosity provided by the CMS. 

Similarly at $\sqrt{s}=8$ TeV and by using the COMPETE prediction  of $\rho\simeq 0.140\pm 0.007$ \cite{Cudell:2002xe} the TOTEM Collaboration arrived at $\sigma_{tot}^{pp}=101.5\pm 2.1$ mb and $\sigma_{tot}^{pp}=101.9\pm 2.1$ mb \cite{Antchev:2015zza}. Later for the first time at the LHC the $\rho$-parameter was thouroughly extracted via the Coulombian-nuclear interference region, \tit{viz.} $\rho^{pp}=0.120\pm 0.030$ at $\sqrt{s}=8$ TeV where $\sigma_{tot}^{pp}=102.90\pm 2.3$ mb and $\sigma_{tot}^{pp}=103.0\pm 2.3$ mb, respectively \cite{Antchev:2016vpy}.

The first measurements of the $\rho$-parameter at $\sqrt{s}=13$ TeV was obtained by the TOTEM Collaboration through the differential cross section due to the effects of the Coulombian-nuclear interference region. It was observed an unexpected and quite odd decrease which yielded $\rho^{pp}=0.09\pm 0.01$ and $\rho^{pp}=0.10\pm 0.01$ \cite{Antchev:2017yns}, respectively, therefore excluding all the models classified and published by the COMPETE Group \cite{Cudell:2002xe}. These result obtained by TOTEM have enhanced an old discussion from alternative phenomenological models, based on Regge poles and in the QCD perturbative framework, and are compatible with a colourless $3$-gluon bound state exchange in the $t$-channel of the $pp$ elastic scattering. However, it must be kept in mind that nothing has been proved ideed, at least up so forth.

Last, but not least, the ATLAS Collaboration at LHC has also made some data analyses at energies of $\sqrt{s}= 7$ and $8$ TeV. For the former energy the corresponding result for $pp$ total cross section reads $\sigma^{pp}_{tot}=95.4 \pm 1.4$ mb \cite{Aad:2014dca} considering $\rho=0.140 \pm 0.008$ \cite{Cudell:2002xe} and for the latter the ATLAS arrived at $\sigma^{pp}_{tot}=96.07 \pm 0.92$ mb \cite{Aaboud:2016ijx} using $\rho=0.1362 \pm 0.0034$, respectively.

Despite the lack of agreement between the measurements obtained by the TOTEM and ATLAS Collaboration at the LHC, it seems interesting and very meaningful to study what sort of effects these elusive differences could actually result in the predictions for the considered physical quantities. By considering either ensembles, at high-energies, with ATLAS or TOTEM only data and ATLAS $+$ TOTEM as an unique data set. 

\subsubsection{\textsc{II. $\rho$-Independent Determination}}
\mbox{\,\,\,\,\,\,\,\,\,}
The measured quantity in a scattering experiment is a counting rate and not a cross section \tit{per se}. Thus, for an elastic scattering, the quantity measured is the (number of counts)/seconds/(interval of time),
\be
\Delta N(t)={\cal L}_{int}\,\frac{d\sigma}{dt},
\label{ch2.118}
\ee
where ${\cal L}_{int}$ is the integrated luminosity with units $(area)^{-1}\times(time)^{-1}$. The cross section can be obtained by summing directly the elastic and inelastic cross section,
\be
\sigma_{tot}=\frac{N_{el}+N_{in}}{{\cal L}^{CMS}_{int}}=\sigma_{el}+\sigma_{in},
\label{ch2.119}
\ee
where $N_{el}$ and $N_{in}$ represents, respectively, the elastic and inelastic rates integrated over a given data taking period. 

Using this technique the TOTEM Collaboration arrived at $\sigma_{tot}^{pp}=99.10\pm 4.30$ mb at $\sqrt{s}=7$ TeV \cite{Antchev:2013iaa}. It is included in the data set the first estimate for the $\rho$-parameter made by the TOTEM Collaboration in their $\rho$-independent measurement. This estimate was obtained by combining the elastic and inelastic measurements in order to determine $\rho^{2}$,
\be
\rho^{2}=16\pi\,(\bh c)^{2}\, {\cal L}^{CMS}_{int}\,\frac{dN_{el}/dt\vert_{t=0}}{(N_{el}+N_{in})^{2}}-1,
\label{ch2.120}
\ee
which yielded $\rho^{2}=0.009\pm 0.056$. Taking a uniform $\vert \rho\vert$ distribution, then at $\sqrt{s}=7$ TeV, $\rho^{pp}=0.145\pm 0.091$ \cite{Antchev:2013iaa}.

\subsubsection{\textsc{III. Luminosity-Independent Technique}}
\mbox{\,\,\,\,\,\,\,\,\,}
Using the optical theorem (\ref{ch2.91}), the elastic and inelastic measurements can be combined in such a way that total cross section can be written without the knowledge of the luminosity,
\be
\sigma_{tot}=\frac{16\pi(\bh c)^{2}}{1+\rho^{2}}\,\frac{dN_{el}/dt\vert_{t=0}}{N_{el}+N_{in}}.
\label{ch2.121}
\ee

Taking $\rho\simeq 0.141\pm 0.007$ from the COMPETE \cite{Cudell:2002xe} extrapolation at $\sqrt{s}=7$ TeV yields the luminosity-independent total cross section $\sigma_{tot}^{pp}=98.0\pm 2.5$ mb \cite{Antchev:2013iaa}. 

Similarly at the energy $\sqrt{s}=8$ TeV and taking $\rho\simeq 0.140\pm 0.007$ from COMPETE \cite{Cudell:2002xe} preferred model extrapolation, the TOTEM Collaboration arrived at $\sigma_{tot}^{pp}=101.7\pm 2.9$ \cite{Antchev:2013paa}.

The first measurement of $pp$ total cross section at $\sqrt{s}=13$ TeV was obtained by the TOTEM which yields $\sigma_{tot}^{pp}=110.6\pm 3.4$ mb \cite{Antchev:2017dia} where it was assumed a value of $\rho = 0.1$ \cite{Antchev:2017yns}.

\section*{\S \,\,\textsc{The Elastic Differential Cross Section}}
\mbox{\,\,\,\,\,\,\,\,\,}
As it was mentioned above, the fundamental quantity measured in an elastic scattering experiment is the counting rate in a fixed energy. By means of expression (\ref{ch2.118}) the differential cross section can be determined as
\be
\frac{d\sigma}{dt}=\frac{1}{{\cal L}_{int}}\,\frac{dN_{el}}{dt}.
\label{ch2.121}
\ee 
\mbox{\,\,\,\,\,\,\,\,\,}
In general, the elastic differential cross section can be divided in three well behaved and specific contributions, see expression (\ref{ch2.96}), and written as
\be
\frac{d\sigma}{dt}=\pi\left\vert F_{c}\,e^{i\alpha\varphi(t)}+F_{n}\right\vert^{2},
\label{ch2.122}
\ee
where $F_{c}$ and $F_{n}$ stands for the scattering amplitudes. In the case of $pp$ scattering \cite{Block:1984ru}:
\begin{itemize}

\item[i.] $d\sigma_{c}/dt\sim4\pi^{2}(\alpha/\vert t \vert)^{2}$, which is the purely Coulombian component and dominates in the region $\vert t \vert <10^{-3}$ GeV$^{2}$.

\item[ii.] $d\sigma_{cn}/dt\sim-\pi(\rho+\alpha\varphi)(\alpha\sigma_{tot}/\vert t \vert)$, which is the Coulombian-nuclear interference component and dominates in the region $\vert t \vert \simeq10^{-3}$ GeV$^{2}$. The term $\alpha\varphi$ leads to the distortion in the purely hadronic component induced by the presence of the Coulombian field.

\item[iii.] $d\sigma_{n}/dt=(1+\rho^{2})\,\sigma^{2}_{tot}\,e^{-B\vert t\vert}/16$, which is the purely hadronic component parametrised in the region of the diffractive peak and dominates at $\vert t \vert >10^{-2}$ GeV$^{2}$.

\end{itemize}

Thus in the region where the momentum transfer is typically higher than $10^{-2}$ GeV$^{2}$, the elastic differential cross section is completely determined only by the purely hadronic component. The consequence is that the diffraction pattern observed in Figure \ref{ch2fig11} is exclusively due to the strong interaction dynamical evolution as a function of momentum transfer in the collision.

The experimental data set for elastic differential cross section used in this Thesis so far corresponds to the recent measurements made by the TOTEM Collaboration at CM energies $\sqrt{s}=7$, $8$ and $13$ TeV \cite{Antchev:2013haa,Antchev:2011zz,Antchev:2016vpy,Antchev:2017yns} for $\vert t \vert \leq 0.1$ GeV$^{2}$, see Figure \ref{ch2fig12}.

\begin{table}[h]
\centering
\scalebox{0.9}{
\begin{tabular}{c@{\quad}|c@{\quad}|c@{\quad}|c@{\quad}}
\hline \hline
& & & \\[-0.2cm]
\,\,\,\,\,\,\,Collaboration \,\,\,\,\,\,\,  &  \,\,\,\,\,\,\,   Reference  \,\,\,\,\,\,\,    &   \,\,\,\,\,\,\,   $\sqrt{s}$ (TeV)  \,\,\,\,\,\,\,   &  \,\,\,\,\,\,\,   $\sigma_{tot}$ (mb)  \,\,\,\,\,\,\, \\ [0.7ex] \hline
& & & \\[-0.3cm]
\multirow{13}{*}{TOTEM} &  CERN-EP-2017-321     &            $2.76$         &           $84.7 \pm 3.3$  \\ [0.7ex]

		       &  EPL96 (2011) 21002    &            $7$            &           $98.3 \pm 2.8$  \\ [0.7ex]

                       &  EPL101 (2013) 21003   &            $7$            &           $98.6 \pm 2.2$  \\ [0.7ex]

                       &  EPL101 (2013) 21004   &            $7$            &           $98.0 \pm 2.5$  \\ [0.7ex]

                       &  EPL101 (2013) 21004   &            $7$            &           $99.1 \pm 4.3$  \\ [0.7ex]

                       &  PRL111 (2013) 012001  &            $8$            &           $101.7 \pm 2.9$ \\ [0.7ex]
                    
                       &  NPB 899 (2015) 527    &            $8$            &           $101.5 \pm 2.1$ \\ [0.7ex]

                       &  NPB 899 (2015) 527    &            $8$            &           $101.9 \pm 2.1$ \\ [0.7ex]

                       &  EPJ C76 (2016) 661    &            $8$            &           $102.9 \pm 2.3$ \\ [0.7ex]

                       &  EPJ C76 (2016) 661    &            $8$            &           $103.0 \pm 2.3$ \\ [0.7ex]
                       
                       &  CERN-EP-2017-321      &            $13$           &           $110.6 \pm 3.4$ \\ [0.7ex]
                       
                       &  CERN-EP-2017-335      &            $13$           &           $110.3 \pm 3.5$ \\ [0.7ex]\hline
& & & \\[-0.2cm]                       
\multirow{2}{*}{ATLAS} &  NPB 899 (2014) 486    &            $7$            &           $95.4 \pm 1.4$  \\ [0.7ex]

                       &  PLB 761 (2016) 158    &            $8$            &           $96.07 \pm 0.92$ \\ [0.7ex]\hline \hline

& & & \\[-0.2cm]
Collaboration          &     Reference          &      $\sqrt{s}$ (TeV)     &           $\rho$            \\ [0.7ex] \hline
& & & \\[-0.3cm]
\multirow{4}{*}{TOTEM} &  EPL101 (2013) 21004   &            $7$            &           $0.145 \pm 0.091$ \\ [0.7ex]
           
                       &  EPJ C76 (2016) 661    &            $8$            &           $0.120 \pm 0.030$ \\ [0.7ex]
           
                       &  CERN-EP-2017-335      &            $13$           &           $0.090 \pm 0.010$ \\ [0.7ex]

                       &  CERN-EP-2017-335      &            $13$           &           $0.100 \pm 0.010$ \\\hline \hline
\end{tabular}}
\caption{Total cross section and the ratio of the real to imaginary part of the forward scattering amplitude data obtained at the LHC.}
\label{ch2tab1}
\end{table}

\bfg[hbtp]
  \begin{center}
    \includegraphics[width=15cm,height=20cm,clip=true]{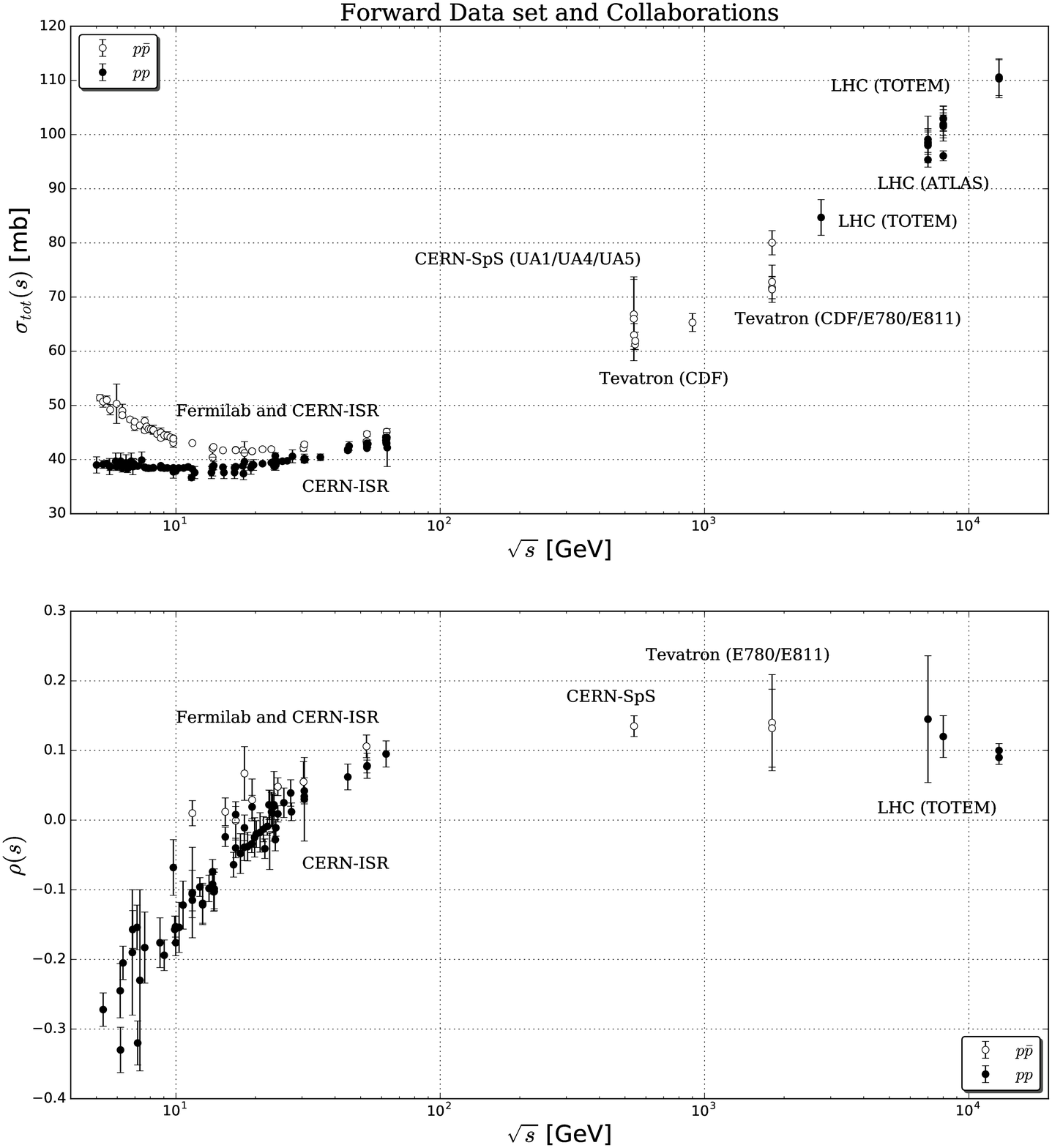}
    \caption{Forward data set at $\sqrt{s}\geq 5$ GeV.}
    \label{ch2fig8}
  \end{center}
\efg

\bfg[hbtp]
  \begin{center}
    \includegraphics[width=12cm,height=8cm,clip=true]{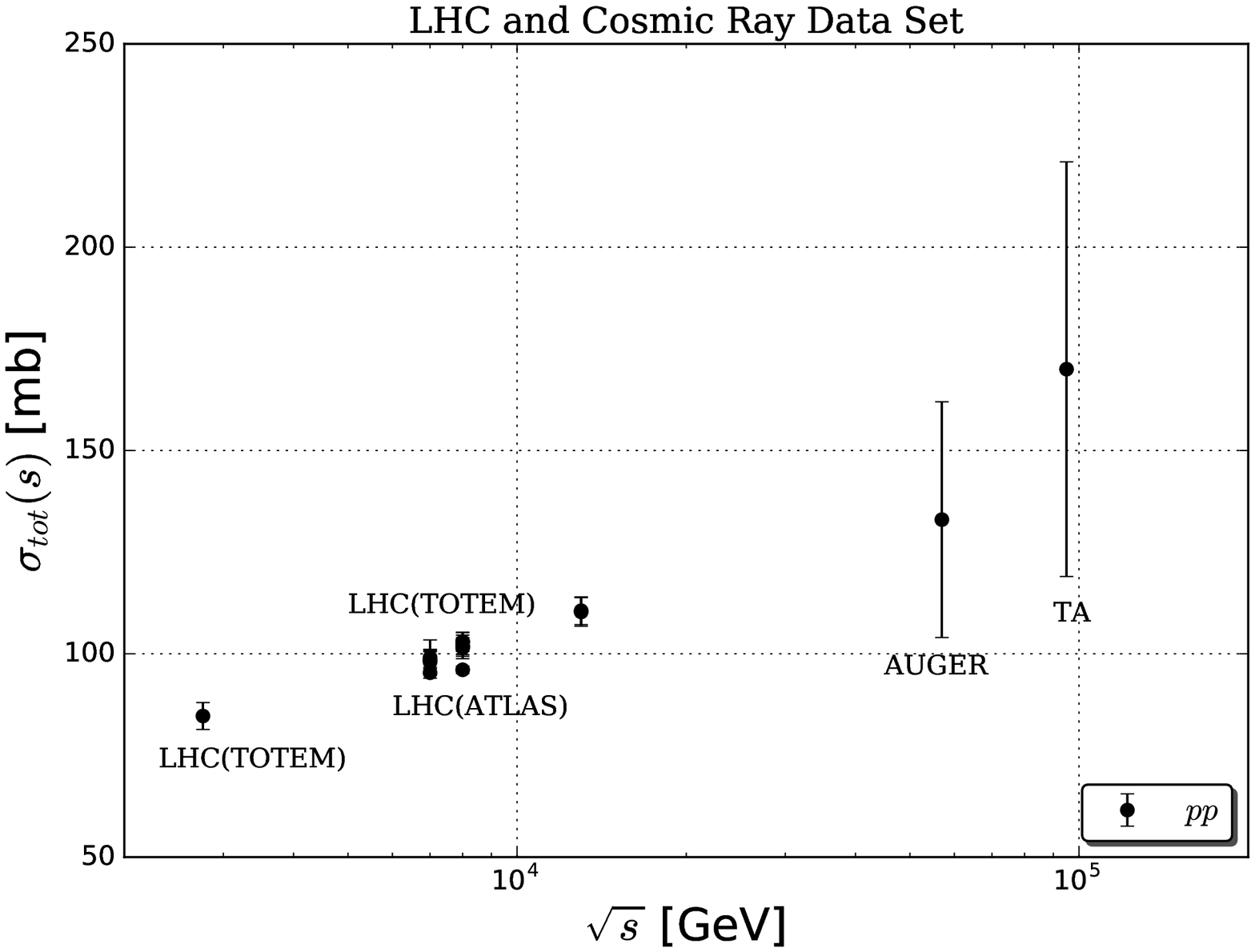}
    \caption{Experimental data points at cosmic ray energies.}
    \label{ch2fig10}
  \end{center}
\efg

\bfg[hbtp]
  \begin{center}
    \includegraphics[width=15cm,height=8cm,clip=true]{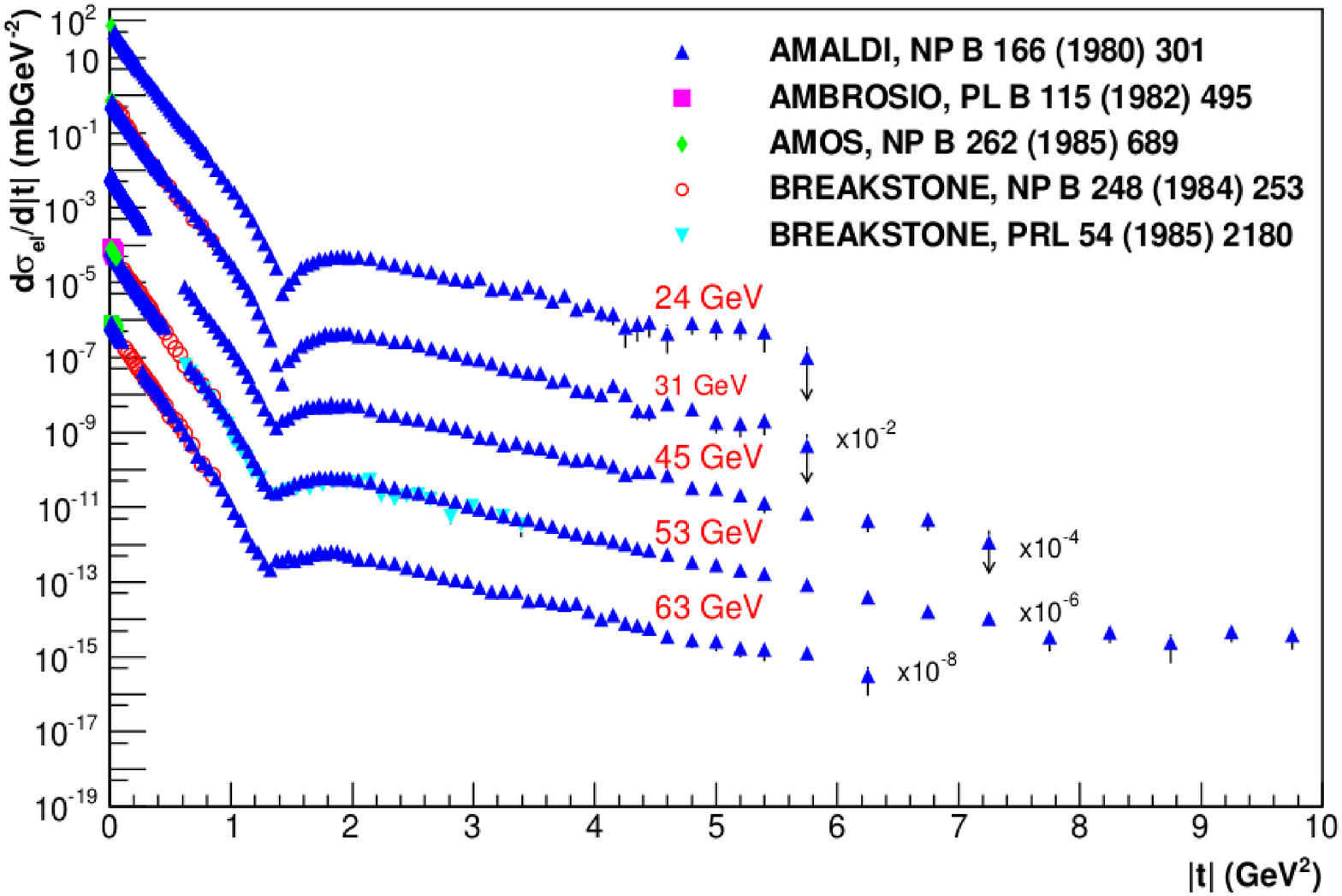}
    \caption{Elastic differential cross section data set at $\sqrt{s}=24-63$ GeV, obtained by ISR experiment at CERN. This figure was taken from Reference \cite{Fagundes:2013aja}.}
    \label{ch2fig11}
  \end{center}
\efg

\bfg[hbtp]
  \begin{center}
    \includegraphics[width=15cm,height=20cm,clip=true]{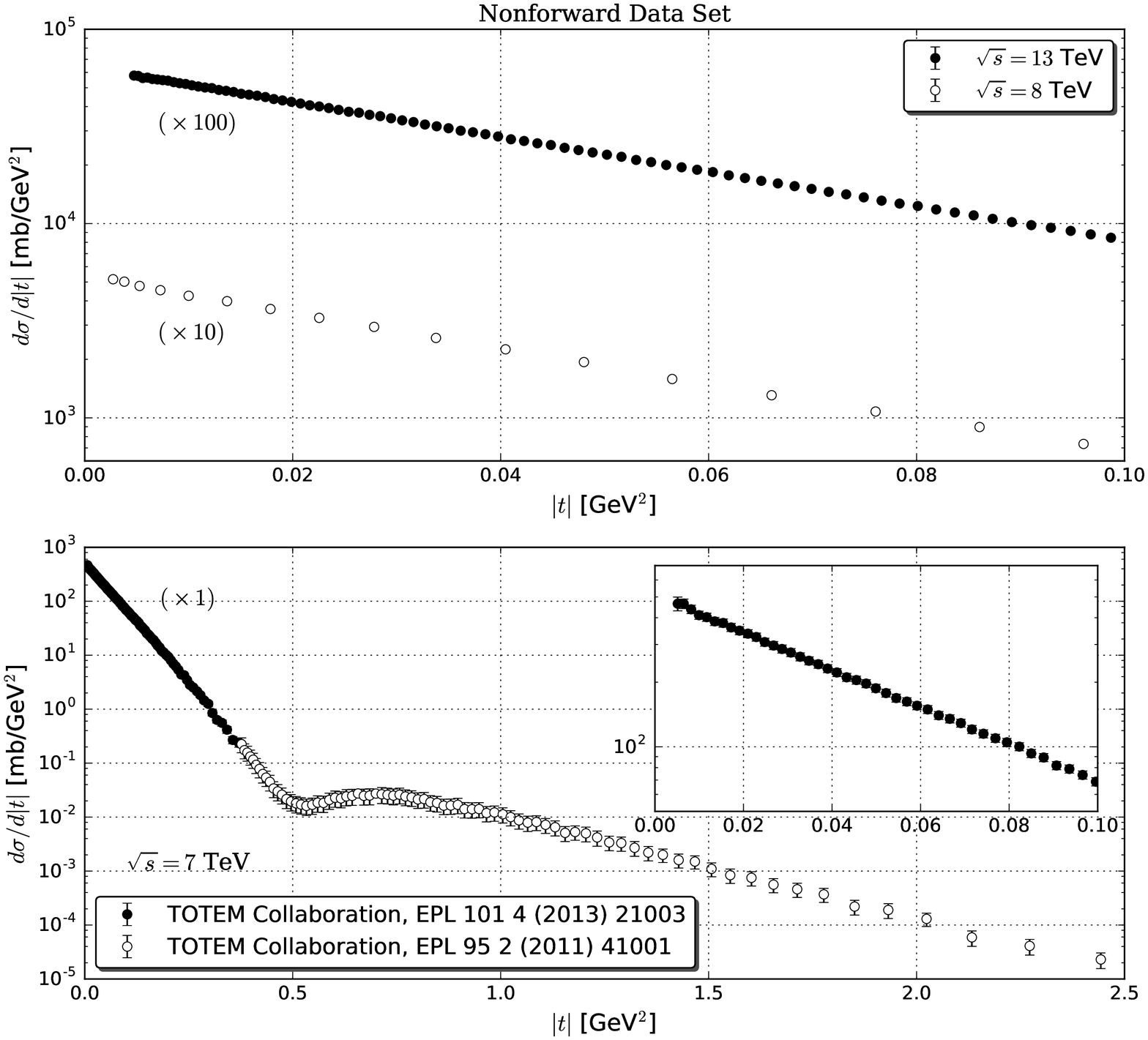}
    \caption{Data set for $pp$ elastic differential cross section at $\sqrt{s}=7$, $8$ and $13$ TeV.}
    \label{ch2fig12}
  \end{center}
\efg

\clearpage
\thispagestyle{plain}

\chapter{\textsc{Regge Theory}}
\label{chap3}
\mbox{\,\,\,\,\,\,\,\,\,}
Traditionally, Regge theory is the theoretical framework used to study diffraction, and it belongs to the class of $t$-channel models, where the description of strong interaction is performed by means of the exchange of something which in fact is not viewed as particles with a definite spin, but rather by a class of objects known as Regge trajectories \cite{Barone:2002cv,Gribov:2003nw,Donnachie:2002en,Collins:1977jy}. In the particle physics terminology, Regge trajectories are often called Reggeons and the so-called Pomeron stands for the Reggeon with the vacuum quantum numbers which asymptotically dominates at increasing energy. Thus, in the Regge language diffractive reactions are those ones described by Pomeron exchange between particles $1$ and $2$ whereas no quantum numbers are exchanged in the reaction.

Through out the scientific community, many consider the Pomeron as a misleading,  an ill-defined or even in some cases a meaningless concept. However, one cannot dispute the phenomenological success of Regge theory in describing a wide class of reactions in which no other framework was capable of. Once it was said by A. Donnachie and P.V. Landshoff \cite{Donnachie:1992ny} that although this less optimistic point of view ``\tit{Regge theory remains one of the great truths of particle Physics''}, moreover because a large class of processes are very well described using just simple predictions. On the other hand, it is worth mentioning that when extended to high energies, it relies on a series of assumptions, such as the introduction of a new quantum number called signature. Therefore, its merging with a fundamental theory as QCD would be definitely a huge step towards a full description and understanding of strong interactions. This remains as one of the open problems in the physics of $20th$ century, and with a little bit of luck and huge effort by the community perhaps it might be solved in the $21th$ one. In simple words it could be briefly stated that: is Regge theory the limit of QCD at $q\to 0$?

\section{\textsc{Regge Poles}}
\label{sec3.1}
\mbox{\,\,\,\,\,\,\,\,\,}
Despite the fact that the main ideas of Regge theory have been used in the theoretical field of particle physics, more specifically in the study of hadronic phenomena, it was originally formulated in the context of nonrelativistic Quantum Mechanics \cite{Regge:1959mz,Regge:1960zc}. The basic idea of T. Regge was to study the bound states for an attractive spherically symmetric potential which appears as poles of the partial wave amplitude $a_{\ell}(k)$ for integer values of $\ell$, and then analytically to continue these values to complex ones, therefore obtaining an interpolated function $a(\ell,k)$, which reduces to $a_{\ell}(k)$ for $\ell=0,1,2,...$, respectively. In the case of well behaved potential, \tit{i.e.} $V(\rr)\to0$ at $r\to\infty$, Yukawa-type potentials, and in order to constrain the functional form of the scattering amplitude when analytically continued to complex $\ell$ values the general mathematical properties of the $S$-matrix, namely unitarity, analyticity, and crossing symmetry, are essential to define the singularities of $a(\ell,k)$. More specifically, they will be given by simple removable poles, which is known as Regge poles, and located at values defined by
\be
\ell=\alpha(k),
\label{ch3.1}
\ee
where $\alpha(k)$ is a function of the energy named Regge trajectory. Each bound state or resonance corresponds to a single Regge trajectory as given by expression (\ref{ch3.1}). By giving integer values to the angular momentum $\ell$, \tit{i.e.} by attributing a physical value, then it can be properly obtained the energy of each state.

In the picture of relativistic scattering, and by means of the general properties of the $S$-matrix, it can be demonstrated that the relativistic scattering amplitude $A_{\ell}(t)$ can be analytically continued to complex $\ell$ values, thus obtaining an interpolation function $A(\ell,t)$, still with simple poles defined by
\be
\ell=\alpha(t).
\label{ch3.2}
\ee 
The contribution of each singularity to the scattering amplitude is given by
\be
\lim_{s\to\infty} A(s,t)\sim s^{\alpha(t)},
\label{ch3.3}
\ee
where the singularity with the largest real part, or the leading singularity in the $t$-channel, determines the asymptotic behavior of the scattering amplitude in the $s$-channel.

\section{\textsc{Partial Wave Expansion and the Complex Angular Momenta}}
\label{sec3.2}
\mbox{\,\,\,\,\,\,\,\,\,}
Although the idea of complex angular momenta is rather old and originally due to Poincaré, and latter used by Sommerfeld to study the propagation of electromagnetic waves, this technique is central in Regge theory. The continuation to complex angular momenta naturally emerges when studying the convergence domain of the scattering amplitude, in a sense that it allows a correct analytic continuation to arbitrarily large energies. To find out the convergence domain, the starting point is the partial wave expansion of the scattering amplitude in the $t$-channel \cite{Barone:2002cv},
\be
A(s,t)=\sum_{\ell=0}^{\infty}(2\ell+1)\,A_{\ell}(t)P_{\ell}(z),
\label{ch3.4}
\ee
where the partial wave amplitude is given by
\be
A_\ell(t)=\frac{1}{2}\int^{+1}_{-1} dz\,P_{\ell}(z)\,A(s(z,t),t),
\label{ch3.5}
\ee
with
\be
z\equiv\cos\vartheta_t=1+\frac{2s}{t-4m^{2}}.
\label{ch3.6}
\ee
\mbox{\,\,\,\,\,\,\,\,\,}
Although expression (\ref{ch3.4}) is a correct representation of scattering in the physical $t$-channel domain $t\geq 4m^{2}$ and $-1\leq z \leq 1$, this partial wave series cannot be readly used to represent the $t$-channel exchange crossing symmetric amplitude for high-energy $s$-channel scattering. This is due to the singularities of $A(s,t)$ appearing in the partial wave amplitude $A_{\ell}(t)$ and the $s$ dependence embodied in the Legendre polynomials, which are entire functions of $z$. More specifically, as $s\to\infty$, $z$ becomes proportional to $s$ and the series diverges. Thus, the partial wave series does not converge in a domain of the complex $s$, $t$ and $u$ variables larger than the physical $t$-channel domain.

\subsection{\textsc{Convergence Domain}}
\label{ch3.2.1}
\mbox{\,\,\,\,\,\,\,\,\,}
The asymptotic behavior of $P_{\ell}(z)$ for real $\ell$ is given by \cite{Barone:2002cv}
\be
\lim_{\ell\to\infty}P_{\ell}(\cos\vartheta)={\cal O}(e^{\ell\vert\textnormal{Im}\,\vartheta\vert}),
\label{ch3.7}
\ee
thus the partial wave series in expression (\ref{ch3.4}) at $\ell\to\infty$ converges only if $A_{\ell}(t)\,e^{\ell\vert\textnormal{Im}\,\vartheta\vert}\leq 1$. By using these relation and also writing the partial wave amplitude as
\be
\lim_{\ell\to\infty}A_{\ell}(t)\sim e^{-\ell\eta(t)},
\label{ch3.8}
\ee
one finds that the corresponding convergence region of $A(s,t)$ in the complex plane is $\vert\textnormal{Im}\,\vartheta\vert\leq\eta(t)$, which represents a symmetric horizontal strip in the imaginary $\vartheta$-axis of width $\eta(t)$. By rewriting $z=\cos\vartheta=x+i\,y$, and also $\vartheta=\textnormal{Re}\,\vartheta+i\,\textnormal{Im}\,\vartheta$, then,
\be
z=\cos(\textnormal{Re}\,\vartheta)\cosh(\textnormal{Im}\vartheta)-i\,\sin(\textnormal{Re}\,\vartheta)\sinh(\textnormal{Im}\,\vartheta),
\label{ch3.9}
\ee
\bfg[hbtp]
  \begin{center}
    \includegraphics[width=15cm,clip=true]{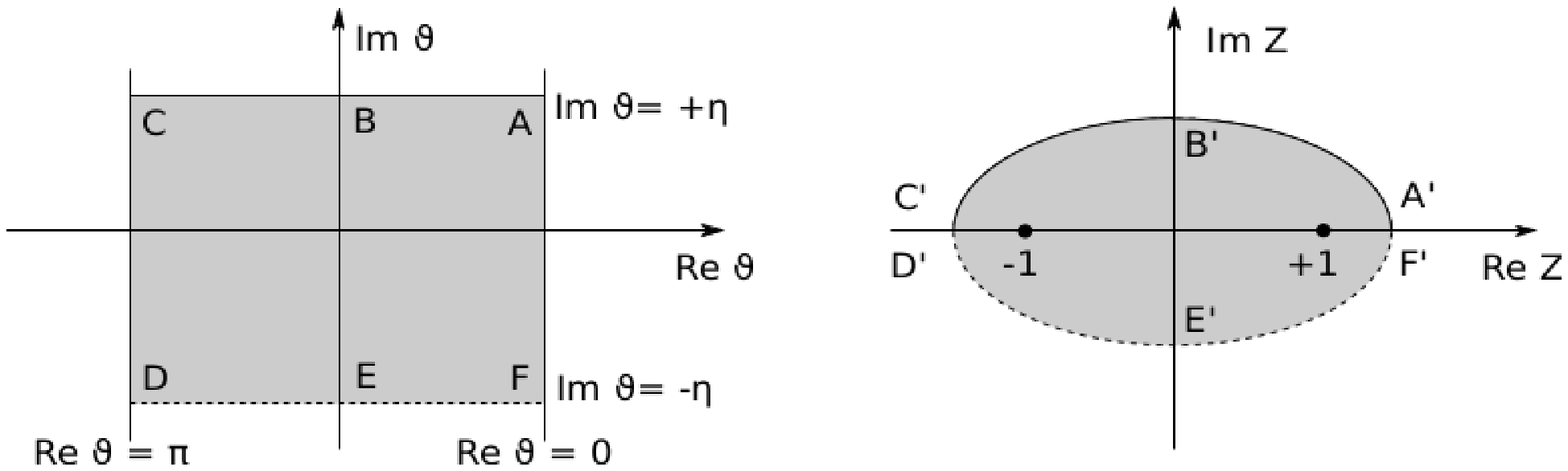}
    \caption{The convergence domain of the scattering amplitude for $\vert\textnormal{Im}\,\vartheta\vert\leq\eta(t)$.}
    \label{ch3fig1}
  \end{center}
\efg
\newline
a straightforward calculation shows that the convergence domain of the partial wave expansion in expression (\ref{ch3.4}) in the complex plane is given by an ellipsis with foci $z=\pm 1$, see Figure \ref{ch3fig1},
\be
\frac{x^{2}}{\chi^{2}}+\frac{y^{2}}{\chi^{2}-1}=1,
\label{ch3.10}
\ee
where this quantity $\chi=\cosh\eta(t)$ was defined. This result implies that the partial wave expansion converges in a domain slightly larger than the elementary physical domain $-1\leq z \leq 1$, but never to arbitrarily large values of $z$. This means that expression (\ref{ch3.4}) cannot be continued to regions where $s$ and $u$ becomes arbitrarily large.

In the opposite case where the partial wave series is obtained using purely imaginary values of $\ell$ the asymptotic behavior of the Legendre polynomials would be given by \cite{Barone:2002cv}
\be
\lim_{\ell\to i\infty}P_{\ell}(\cos\vartheta)={\cal O}(e^{\vert\ell\vert\vert\textnormal{Re}\,\vartheta\vert}).
\label{ch3.11}
\ee
In this case, and provided that the partial wave amplitude behaves as
\be
\lim_{\ell\to i\infty}A_{\vert\ell\vert}(t)\sim e^{-\vert\ell\vert\delta(t)},
\label{ch3.12}
\ee
the convergence condition for $A(s,t)$ in the complex plane would be ensured in a symmetric vertical line in the real $\vartheta$-axis of width $\delta(t)$, \tit{i.e} in $\vert\textnormal{Re}\,\vartheta\vert\leq\delta(t)$. Following the same procedure as before, a straightforward calculation would show that the convergence domain is given by an hyperbola with foci $z=\pm1$, where convergence is ensured outside its halves, see Figure \ref{ch3fig2},
\be
\frac{x^{2}}{\xi^{2}}+\frac{y^{2}}{1-\xi^{2}}=1,
\label{ch3.13}
\ee
\bfg[hbtp]
  \begin{center}
    \includegraphics[width=15cm,clip=true]{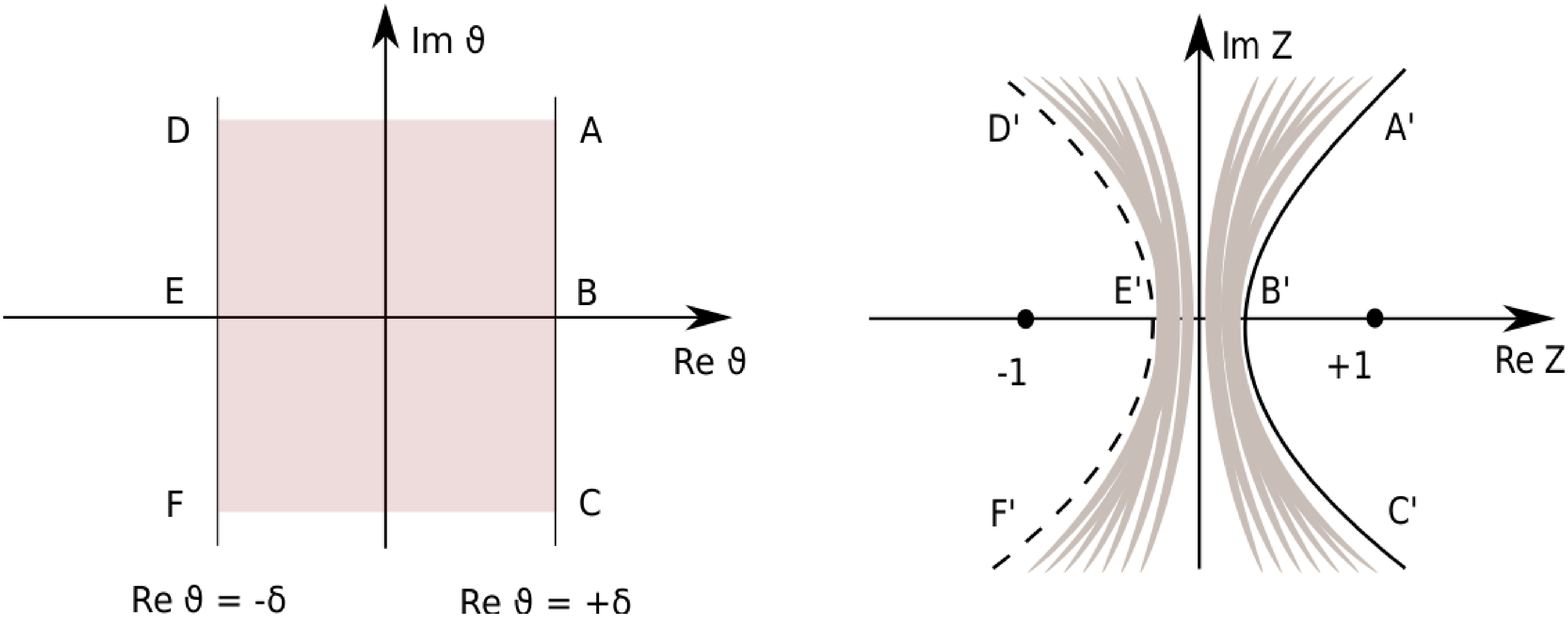}
    \caption{The convergence domain of the scattering amplitude for $\vert\textnormal{Re}\,\vartheta\vert\leq\delta(t)$.}
    \label{ch3fig2}
  \end{center}
\efg
where it was defined $\xi=\cos\delta(t)$. The difference with respect to the previous case is that the hyperbola has an open domain and overlaps the ellipsis. The new expansion will represent the same scattering function in a domain where $s$ or $u$ can become arbitrarily large and guarantees the analytic continuation of the partial wave series over complex values of $\ell$.

\subsection{\textsc{Introduction of Complex Angular Momenta}}
\label{ch3.2.2}
\mbox{\,\,\,\,\,\,\,\,\,}
The introduction of complex angular momenta rely on a series of assumptions, such as assuming that it is possible to continue the partial wave amplitude $A_{\ell}(t)$ to complex values of $\ell$ and construct an interpolating function $A(\ell,t)$ which reduces to $A_{\ell}(t)$ for real integer values of $\ell$ and has the following properties \cite{Barone:2002cv}:
\bi

\item[I.] $A(\ell,t)$ has only simple singularities in the complex $\ell$-plane.

\item[II.] $A(\ell,t)$ is analytic for $\textnormal{Re}\,\ell\geq L$.

\item[III.] $\lim\limits_{\vert\ell\vert\to\infty}A(\ell,t)=0$ for $\textnormal{Re}\,\ell> 0$.

\ei

It is possible to show that if $A(\ell,t)$ exists and satisfies properties II and III, then is uniquely determined by the values it takes for integer $\ell$ \cite{titch}. By using I and II the partial wave expansion (\ref{ch3.4}) can be rewritten as
\be
A(s,t)=\sum^{N-1}_{\ell=0}(2\ell+1)A_{\ell}(t)P_{\ell}(z)+\sum_{\ell=\textnormal N}^{\infty}(2\ell+1)A_{\ell}(t)P_{\ell}(z),
\label{ch3.14}
\ee
where $N$ is the first integer value of momenta greater than L, and the second summation represents the contribution of the amplitude which carries the singularity in the real axis. Using the relation for the Legendre polynomials $P_{\ell}(-z)=(-1)^{\ell}P_{\ell}(z)$, and also that \cite{Gribov:2003nw},
\be
\sum_{\ell}(2\ell+1)A_{\ell}(t)P_{\ell}(z)=\frac{i}{2}\int_{C} d\ell\,\frac{(-1)^{\ell}(2\ell+1)A(\ell,t)P_{\ell}(z)}{\sin\pi \ell},
\label{ch3.15}
\ee
one finds that expression (\ref{ch3.14}) is written in terms of an integral over a contour $C$ which avoids all singularities of $A(\ell,t)$.
In order to involve all the singularities of $A(\ell,t)$ in the complex plane, the contour $C$ is deformed into a line parallel to the imaginary $\ell$-axis, namely $C^{\prime}=(a-i\infty,a+i\infty)$. Thus, this new contour $C^{\prime}$ is located to the right of all singularities. The property III and the asymptotic behavior of the Legendre polynomials in the limit at $\ell\to\infty$, $\vert P_{\ell}(-z)/\sin\pi\ell\vert<{\cal O}(\ell^{-1/2})$, ensures that the semicircle integral which closes $C^{\prime}$ at infinity vanishes \cite{Barone:2002cv},
\be
A(s,t)=\sum^{N-1}_{\ell=0}(2\ell+1)A_{\ell}(t)P_{\ell}(z)+\frac{i}{2}\int^{a+i\infty}_{a-i\infty}d\ell\,(2\ell+1)A(\ell,t)\,\frac{P_{\ell}(-z)}{\sin\pi \ell}.
\label{ch3.17}
\ee

If the singularities of the partial wave series are simple poles, and if the contour is moved towards to even smaller values, then the contributions from the residues of the poles together with the residues from the poles of $\sin^{-1}\pi\ell$ cancel some of the terms of the truncated series in the above expression. Thus, by displacing the contour $C^{\prime}$ to the left in the interval $-1/2\leq\textnormal{Re}\,\ell<0$, then,
%
\bear
\!\!\!\!\!\!A(s,t)&=&\sum^{\infty}_{\ell=N}(2\ell+1)A_{\ell}(t)P_{\ell}(z)+\frac{i}{2}\int^{a+i\infty}_{a-i\infty}d\ell\,(2\ell+1)A(\ell,t)\,\frac{P_{\ell}(-z)}{\sin\pi \ell}\nonumber\\
&=&\frac{i}{2}\int_{C}d\ell\,(2\ell+1)A(\ell,t)\,\frac{P_{\ell}(-z)}{\sin\pi \ell}+\frac{i}{2}\int^{a+i\infty}_{a-i\infty}d\ell\,(2\ell+1)A(\ell,t)\,\frac{P_{\ell}(-z)}{\sin\pi \ell}.
\label{ch3.18}
\eear
By means of the residue theorem, where it states that the line integral of a function is related to the sum of the residues of the function at the poles,
\be
\int_{C}d\ell\,f(\ell)=2\pi i\sum_{i} \textnormal{Res}\,f(\ell)\vert_{\ell\to\alpha_{i}},
\label{ch3.19}
\ee
where $\alpha_{i}$ stands for the $i$-th pole of $A(\ell,t)$, then the residue can be calculated by means of the following relation \cite{butkov1973mathematical},
\be
\textnormal{Res}\,f(\ell)\vert_{\ell\to\alpha_{i}}=\frac{1}{(k-1)!}\,\lim_{\ell\to\alpha_{i}}\,\frac{d^{k-1}}{d\ell^{k-1}}\,\left[(\ell-\alpha_{i})^{k}f(\ell)\right],
\label{ch3.20}
\ee
where for a simple pole $\alpha_{i}$, \tit{i.e} $k=1$. Therefore, the residue of the first integral in the \tit{rhs} of expression (\ref{ch3.18}) is given by
\be
\textnormal{Res}\,f(\ell)\vert_{\ell\to\alpha_{i}}=\gamma_{i}(t)\,\frac{(2\alpha_{i}+1)P_{\alpha_{i}}(-z)}{\sin\pi\alpha_{i}},
\label{ch3.21}
\ee
and the residue function at that pole is defined by
\be
\gamma_{i}(t)=\lim_{\ell\to\alpha_{i}}(\ell-\alpha_{i})A(\ell,t).
\label{ch3.22}
\ee
By substituting the last result (\ref{ch3.21}) into expression (\ref{ch3.18}) one finds \cite{Barone:2002cv,Collins:1977jy},
\be
A(s,t)=-\pi\sum_{i}\gamma_{i}(t)\,\frac{(2\alpha_{i}(t)+1)P_{\alpha_{i}}(-z)}{\sin\pi\alpha_{i}(t)}+\frac{i}{2}\int^{a+i\infty}_{a-i\infty}d\ell\,(2\ell+1)A(\ell,t)\,\frac{P_{\ell}(-z)}{\sin\pi \ell},
\label{ch3.23}
\ee
which is known as the Watson-Sommerfeld transform of the scattering amplitude $A(s,t)$, where $\alpha_{i}$ represents the location in the complex $\ell$-plane of the $i$-th pole of $A(\ell,t)$, called Regge pole. As previously shown in the case of purely imaginary angular momenta, the convergence domain of expression (\ref{ch3.23}) is valid outside the halves of the hyperbola in the $z$-plane, see expression (\ref{ch3.13}).

The large-$s$ behavior of the Watson-Sommerfeld transformed amplitude for fixed $t$, which is exactly the same as large-$z$, is given by \cite{california1954tables}
\be
\lim_{\vert z \vert\to\infty}P_{\ell}(z){\sim}z^{\ell},\,\,\,\,\textnormal{Re}\,\ell\geq-\frac{1}{2}.
\label{ch3.24}
\ee
Therefore, in the large $\vert z\vert$-limit the integral in the \tit{rhs} of expression (\ref{ch3.23}) behaves asymptotically as $\vert z\vert^{-1/2}$ and gives a negligible contribution to the scattering amplitude $A(s,t)$, where only the pole contribution survives,
\be
\lim_{\vert z \vert\to\infty}A(s,t)\simeq-\pi\sum_{i}\gamma_{i}(t)(2\alpha_{i}(t)+1)\,\frac{(-z)^{\alpha_{i}(t)}}{\sin\pi\alpha_{i}(t)}.
\label{ch3.25}
\ee
\mbox{\,\,\,\,\,\,\,\,\,}
The dominant term will be the one with the largest $\textnormal{Re}\,\alpha_{i}$. Thus, in the case of a $t$-channel exchange, and by taking into account only the right-most Regge trajectory $\alpha(t)$, the asymptotic behavior of the scattering amplitude for fixed $t$ will be written as \cite{Barone:2002cv}
\be
A(s,t)\underset{s\to\infty}{\simeq}-\gamma(t)\,\frac{s^{\alpha(t)}}{\sin\pi\alpha(t)},
\label{ch3.26}
\ee
where some constants and $t$-dependent factors were absorbed into $\gamma(t)$. Since the convergence domain relies outside of the halves of the hyperbola, the asymptotic amplitude in the crossed $s$-channel exchange for asymptotic large-$t$ behavior can be properly obtained by switching $t\leftrightarrow s$, \tit{i.e.} by invoking the $S$-matrix crossing symmetry property,
\be
A(s,t)\underset{t\to\infty}{\simeq}-\gamma(s)\,\frac{t^{\alpha(s)}}{\sin\pi\alpha(s)}.
\label{ch3.27}
\ee
Hence, expression (\ref{ch3.26}) represents the prediction of the Regge theory for the large-$s$ behavior of the scattering amplitude, and also that the leading singularity governs the asymptotic behavior of $A(s,t)$ as $s\to\infty$.

\section{\textsc{Regge Poles in Relativistic Scattering}}
\label{ch3.3}
\mbox{\,\,\,\,\,\,\,\,\,}
If the scattering amplitude admits the existence of $N$-times subtracted dispersion relation, the partial wave scattering amplitude can be written as an integral representation for $A(s,t)$ valid for $\ell\geq N$, known as the Froissart-Gribov projection \cite{Froissart:1961ux,Gribov:1968fc},
\be
A_{\ell}(t)=\frac{1}{\pi}\int^{+\infty}_{z_{0}}dz_{t}\,D_{s}(s(z_{t},t),t)\,Q_{\ell}(z_{t})+\frac{1}{\pi}\int^{-\infty}_{-z_{0}}dz_{t}\,D_{u}(u(z_{t},t),t)\,Q_{\ell}(z_{t}),
\label{ch3.28}
\ee
where $z_{t}$ is the scattering angle in the $t$-channel,
\be
z_{t}\equiv\cos\vartheta_{t}=1+\frac{2s}{t-4m^{2}},
\label{ch3.29}
\ee
$Q_{\ell}(z^{\prime}_{t})$ stands for the Legendre function of the second kind,
\be
Q_{\ell}(z^{\prime}_{t})=\frac{1}{2}\int^{1}_{-1}dz_{t}\,P_{\ell}(z_{t})\,\frac{z^{N}_{t}}{z^{\prime N}_{t}(z^{\prime N}_{t}-z_{t})},
\label{ch3.30}
\ee
and finally, $D_{s}(s,t)$ and $D_{u}(u,t)$ are the discontinuity functions in the $s$ and $u$ channels. Respectively, they are given by \cite{Barone:2002cv}
\be
D_{s}(s,t)=\frac{1}{2i}\lim_{\e\to0^{+}}\left[A(s+i\e,t)-A(s-i\e,t)\right],
\label{ch3.31}
\ee
\be
D_{u}(u,t)=\frac{1}{2i}\lim_{\e\to0^{+}}\left[A(4m^{2}-u-t-i\e,t)-A(4m^{2}-u-t+i\e,t)\right],
\label{ch3.32}
\ee
and it coincides with the imaginary part of the scattering amplitude, $D_{s}(s,t)=\textnormal{Im}\,A(s,t)$. By rewriting the second integral in expression (\ref{ch3.28}) with the change of variables $z_{t}\to-z_{t}$, and using $u(-z_{t},t)=s(-z_{t},t)$, one finds that 
\be
Q_{\ell}(-z_{t})=-e^{-i\pi\ell}Q_{\ell}(z_{t}),
\label{ch3.33}
\ee
and the partial wave scattering amplitude can be written as \cite{Barone:2002cv}
\be
A_{\ell}(t)=\frac{1}{\pi}\int^{+\infty}_{z_{0}}dz_{t}\,\left[D_{s}(s(z_{t},t),t)+e^{-i\pi \ell}D_{u}(u(z_{t},t),t)\right]\,Q_{\ell}(z_{t}),
\label{ch3.34}
\ee
where $e^{-i\pi \ell}=(-1)^{\ell}$ for integer values of $\ell$. The first term in the \tit{rhs} of the above expression vanishes exponentially, because of the asymptotic behavior of $Q_{\ell}(z)$ at $\ell\to\infty$ \cite{california1954tables},
\be
\lim_{\ell\to \infty}Q_{\ell}(z_{t})\sim \ell^{-1/2}\,\exp\left\{-\left(\ell+\frac{1}{2}\right)\,\ln\left[z_{t}+(z^{2}_{t}-1)^{1/2}\right]\right\}.
\label{ch3.35}
\ee
The amplitude cannot be continued to complex values of $\ell$, because the second term in expression (\ref{ch3.34}) diverges when $\textnormal{Im}\,\ell$ becomes asymptotically large at $\ell\to\infty$,
\be
\lim_{\ell\to \infty}A_{\ell}(t)=\lim_{\ell\to \infty}\frac{1}{\pi}\int^{+\infty}_{z_{0}}dz_{t}\,e^{-i\pi\left(\textnormal{Re}\,\ell+i\textnormal{Im}\,\ell\right)}\,D_{u}(u(z_{t},t),t)\,Q_{\ell}(z_{t}).
\label{ch3.36}
\ee
\mbox{\,\,\,\,\,\,\,\,\,}
The way to overcome this problem can be properly done by introducing a new quantum number, the signature $\xi=\pm1$ \cite{Barone:2002cv,Collins:1977jy,Gribov:2003nw}, in such a way that the amplitude is written by means of crossing-even and crossing-odd terms,
\be
A^{\xi}_{\ell}(t)=\frac{1}{\pi}\int^{\infty}_{z_{0}}dz_{t}\,D^{\xi}_{s}(s,t)\,Q_{\ell}(z_{t}),
\label{ch3.37}
\ee
where the signature discontinuity function $D^{\xi}_{s}(s,t)$ is defined as
\be
D^{\xi}_{s}(s,t)=D_{s}(s,t)+\xi D_{u}(s,t).
\label{ch3.38}
\ee
Thus, expression (\ref{ch3.37}) coincides with expression (\ref{ch3.34}) for even and odd integer $\ell$-values,
\be
A^{+}_{\ell}(t)=A_{\ell}(t),\,\,\,\,\textnormal{for}\,\,\,\,\ell \,\,\textnormal{even},
\label{ch3.39}
\ee
\be
A^{-}_{\ell}(t)=A_{\ell}(t),\,\,\,\,\textnormal{for}\,\,\,\,\ell \,\,\textnormal{odd},
\label{ch3.40}
\ee
where it is usually said that the even-amplitude has positive signature, $\xi=+1$, whereas the odd-amplitude has negative signature, $\xi=-1$. By means of expressions (\ref{ch3.37}) and (\ref{ch3.38}), and notice that $D_{s}+\xi D_{u}=D_{s}+(-1)^{\ell} D_{u}$, then,
\bear
\sum_{\xi=\pm1}\left(1+\xi e^{-i\pi\ell}\right)A^{\xi}_{\ell}(t)&=&\frac{1}{\pi}\int^{\infty}_{z_{0}}dz_{t}\sum_{\xi=\pm1}\left(1+\xi e^{-i\pi\ell}\right)\left[D_{s}(s,t)+(-1)^{\ell} D_{u}(s,t)\right]Q_{\ell}(z_{t})\nonumber\\
&=&\frac{2}{\pi}\int^{\infty}_{z_{0}}dz_{t}\left[D_{s}(s,t)+(-1)^{\ell} D_{u}(s,t)\right]Q_{\ell}(z_{t}),
\label{ch3.41}
\eear
where the last line is simply the partial wave amplitude as given by expression (\ref{ch3.34}). Therefore, we obtain
\be
A_{\ell}(t)=\frac{1}{2}\sum_{\xi=\pm1}\left(1+\xi e^{-i\pi\ell}\right)A^{\xi}_{\ell}(t).
\label{ch3.42}
\ee
\mbox{\,\,\,\,\,\,\,\,\,}
It can be defined a function called definite-signature scattering amplitude which is written by the partial wave expansion,
\be
A^{\xi}(t)=\sum^{\infty}_{\ell=0}(2\ell+1)A^{\xi}_{\ell}(t)P_{\ell}(z_{t}),
\label{ch3.43}
\ee
where it is a well-behaved function at $\ell\to\infty$, and hence can be analytically continued to the complex $\ell$-plane by means of the Watson-Sommerfeld transform,
\be
A^{\xi}(z_{t},t)=\frac{i}{2}\int_{C}d\ell\,(2\ell+1)A^{\xi}(\ell,t)\,\frac{P_{\ell}(-z_{t})}{\sin\pi\ell},
\label{ch3.44}
\ee
and following the same procedure as before \cite{Barone:2002cv},
\be
A^{\xi}(z_{t},t)=\frac{i}{2}\int_{C}d\ell\,(2\ell+1)A^{\xi}(\ell,t)\,\frac{P_{\ell}(-z_{t})}{\sin\pi\ell}+\frac{i}{2}\int^{c+i\infty}_{c-i\infty}d\ell\,(2\ell+1)A^{\xi}(\ell,t)\,\frac{P_{\ell}(-z_{t})}{\sin\pi\ell},
\label{ch3.45}
\ee
where the first integral is calculated by means of the theorem of residues at $\ell\to\alpha_{i_{\xi}}$, and $\alpha_{i_{\xi}}$ defines the location of the $i$-th pole of the definite-signature amplitude $A^{\xi}(z_{t},t)$,
\be
A^{\xi}(z_{t},t)=-\pi\sum_{i_{\xi}}\gamma_{i_{\xi}}(t)\,(2\alpha_{i_{\xi}}(t)+1)\,\frac{P_{\alpha_{i_{\xi}}}(-z_{t})}{\sin\pi\alpha_{i_{\xi}}}+\frac{i}{2}\int^{c+i\infty}_{c-i\infty}d\ell\,(2\ell+1)A(\ell,t)\,\frac{P_{\ell}(-z_{t})}{\sin\pi\ell}.
\label{ch3.46}
\ee
with the summation taken over definite-signature Regge poles. It is easy to show that the full amplitude can be obtained by applying $(2\ell+1)P_{\ell}(z_{t})$ in expression (\ref{ch3.42}),
\be
A(z_{t},t)=\sum^{\infty}_{\ell=0}(2\ell+1)A_{\ell}(t)P_{\ell}(z_{t})=\frac{1}{2}\sum_{\xi=\pm1}\sum^{\infty}_{\ell=0}\left(1+\xi e^{-i\pi\ell}\right)(2\ell+1)A_{\ell}(t)P_{\ell}(z_{t}),
\label{ch3.47}
\ee
and now, the full amplitude is given by
\bear
A(z_{t},t)=&-&\pi\sum_{\xi=\pm1}\sum_{i_{\xi}}\frac{1+\xi e^{-i\pi\ell}}{2}\,\gamma_{i_{\xi}}(t)\,(2\alpha_{i_{\xi}}(t)+1)\,\frac{P_{\alpha_{i_{\xi}}}(-z_{t})}{\sin\pi\alpha_{i_{\xi}}(t)}+\nonumber\\
&+&\frac{i}{2}\sum_{\xi=\pm1}\int^{c+i\infty}_{c-i\infty}d\ell\,\frac{1+\xi e^{-i\pi\ell}}{2}(2\ell+1)A(\ell,t)\frac{P_{\ell}(-z_{t})}{\sin\pi\ell}d\ell.
\label{ch3.48}
\eear
\mbox{\,\,\,\,\,\,\,\,\,}
There are two asymptotic limits to analyze in the above expression. The first one is the behavior of the full amplitude at large-$\vert z_{t}\vert$, which is given by the Legendre polynomials,
\bear
\lim_{z_{t}\to\infty}A(z_{t},t)\simeq&-&\sum_{\xi=\pm1}\sum_{i_{\xi}}\gamma_{i_{\xi}}(t)\,\frac{1+\xi e^{-i\pi\alpha_{i_{\xi}}(t)}}{\sin\pi\alpha_{i_{\xi}}(t)}\,(-z_{t})^{\alpha_{i_{\xi}}(t)}\nonumber\\
&+&\frac{i}{2}\sum_{\xi=\pm1}\int^{c+i\infty}_{c-i\infty}d\ell\,\frac{1+\xi e^{-i\pi\ell}}{2}(2\ell+1)A(\ell,t)\frac{(-z_{t})^{\ell}}{\sin\pi\ell},
\label{ch3.49}
\eear
where some factors have been absorbed into the residue function $\gamma_{i_{\xi}}(t)$. The argument of the integral in the above expression vanishes exponentially in the limit at $\ell\to\infty$, therefore the pole series gives the dominant contribution,
\be
A(z_{t},t)\underset{z_{t}\to\infty}{\simeq}-\sum_{\xi=\pm1}\sum_{i_{\xi}}\gamma_{i_{\xi}}(t)\,\frac{1+\xi e^{-i\pi\alpha_{i_{\xi}}(t)}}{\sin\pi\alpha_{i_{\xi}}(t)}\,(-z_{t})^{\alpha_{i_{\xi}}(t)},
\label{ch3.50}
\ee
by keeping only the leading pole, respectively it is the pole with the largest $\textnormal{Re}\,\alpha_{i_{\xi}}$, and defining $\alpha(t)$ as its trajectory and $\gamma(t)$ the residue. Thus, in the case of a $t$-channel exchange the asymptotic scattering amplitude at large-$s$ will be given by \cite{Barone:2002cv}
\be
A(s,t)\underset{s\to\infty}{\simeq}-\gamma_{i}(t)\,\frac{1+\xi e^{-i\pi\alpha(t)}}{\sin\pi\alpha(t)}\,s^{\alpha(t)}.
\label{ch3.51}
\ee
As before, the amplitude in the crossed $s$-channel for asymptotic large-$t$ can be obtained by the exchange of $t\leftrightarrow s$,
\be
A(s,t)\underset{t\to\infty}{\simeq}-\gamma_{i}(s)\,\frac{1+\xi e^{-i\pi\alpha(s)}}{\sin\pi\alpha(s)}\,t^{\alpha(s)}.
\label{ch3.52}
\ee
\mbox{\,\,\,\,\,\,\,\,\,}
These expressions shows a fundamental result of Regge theory which states that the leading complex angular momentum singularity of the partial wave amplitude in a given channel, that is precisely the Regge pole, determines the asymptotic behavior of the scattering amplitude in the crossed channel. 

\section{\textsc{Regge Trajectories}}
\label{ch3.4}
\mbox{\,\,\,\,\,\,\,\,\,}
The partial wave amplitude, which asymptotically is given by $A(\ell,t)\propto s^{\alpha}$ for $s\to\infty$, in the presence of a Regge pole at $\ell\to\alpha(t)$ behaves as \cite{Barone:2002cv,Donnachie:2002en}
\be
A(\ell,t)\underset{\ell\to\alpha(t)}{\sim}\,\frac{\gamma(t)}{\ell-\alpha(t)},
\label{ch3.53}
\ee
where for real and positive values of $t$ the Regge poles represent resonances or bound states of increasing angular momentum, spin. In relativistic scattering theory the Reggeons are associated with the exchange of families of particles, thus the Regge trajectory interpolates such bound states. Values of $t$ such that $\alpha(t)$ in a non negative integer correspond to the squared mass of a bound state or resonance with that spin. 

The denominator in the scattering amplitude in expression (\ref{ch3.51}) vanishes whenever $\alpha(t)$ crosses an integer. But, as a consequence, to avoid that the same thing happens to the numerator at every other integer value of $\ell$, it is imposed that a trajectory with positive signature, $\xi=+1$, interpolates between resonances with non negative even integer angular momentum, whereas trajectories with negative signatures, $\xi=-1$, interpolates positive odd integer angular momentum resonances. Summarizing, one needs to identify the poles with the exchange of physical particles of spin $J_{i}^{\pm}$, where it represents positive integer values, and mass $m_{i}$, and $\alpha(m^{2}_{i})=J_{i}$. Therefore, the $s$-channel asymptotic behavior of the amplitude is determined by the exchange of a whole family of resonances in the crossed $t$-channel.

A simple way to visualize the Regge trajectories, is to expand $\alpha(t)$ in power series around $t=0$,
\be
\alpha(t)=\alpha(0)+\alpha^{\prime}t,
\label{ch3.54}
\ee
where $\alpha(0)$ and $\alpha^{\prime}$ are known as the intercept and the slope of the trajectory, respectively. Although the above linear expansion for the Regge trajectory was supposed to be justified only for low-$t$, it can be seen, by means of Chew \& Frautschi plots of the spins of low lying mesons against square mass, that by interpolating resonances with the same quantum numbers one finds that they lie in a straight line, and hence expression (\ref{ch3.54}) holds for rather larger values of $t$ \cite{Chew:1961ev,Chew:1962eu}.

By means of the optical theorem, and considering the simplest version of $t$-channel models where nuclear forces are usually attributed to the exchange of mesons, the total cross section would be given by $\sigma_{tot}\sim s^{2J-2}$ at large-$s$. However if the exchanged meson have spin greater than $1$ it would require the cross section to increase, at least, quadratically with $s$. It is clear that this simple model leads to violation of the Froissart-Martin-\L ukaszuk bound, \tit{i.e.} violation of unitarity. 

In the Regge language, the cross section is given asymptotically by $\sigma_{tot}\sim s^{\alpha(0)-1}$. Following the Donnachie-Landshoff model \cite{Donnachie:1992ny}, all the leading mesonic trajectories are degenerated and lie on an exchange-degenerate linear trajectory with $\alpha(0)\simeq 0.45$ and $\alpha^{\prime}\simeq 0.93$ GeV$^{-2}$. By considering only the leading mesonic exchange, it leads to total cross sections decreasing with energy and experimentally none of this is observed. Total cross section do not vanish asymptotically, but rather rise slowly as $s$ increases, an indication that something else must contribute to the observed rise of hadronic total cross sections. Moreover, by retaining the picture of particle exchange then all the resonance contributions in the $t$-channel must act collectively and combine in a way, or another, to give the observed energy dependence. This is Regge theory, the mathematical framework for adding resonances together.

\section{\textsc{The Pomeron}}
\label{ch3.5}
\mbox{\,\,\,\,\,\,\,\,\,}
There was a time where the total cross section seemed to become constant with increasing energy when the highest CM energy data came from Serpukhov. This is the basis of the Pomeranchuk theorem which states that the cross section for $ab$ and $a\bar{b}$ should become asymptotically equal. In $1961$, in order to account for the asymptotically expected constant behavior, Chew, Frautschi and Gribov introduced a Regge trajectory with intercept exactly equal to $1$ and named Pomeron, after its inventor I. Ya. Pomeranchuk, a soviet theoretical physicist who idealized its concept in $1958$.

The Pomeron trajectory does not correspond to any known particle, and from the fits to elastic hadronic scattering data, it was found that its trajectory is much flatter than the other Reggeons, $\alp\simeq 0.25$ GeV$^{-2}$. As a consequence of the constancy required of the total cross section, it should correspond to the pole with the largest intercept, originally $\ap(0)=1$. However in the mid-$60$'s, it was already known that hadronic total cross sections were almost flat at the energy interval $\sqrt{s} \sim 10$ -- $20$ GeV and should slowly increase with increasing energy. Latter, Foldy \& Peierls \cite{Foldy:1963zz} noticed that if for a particular scattering the cross section does not fall as the energy increases then that process must be dominated by the exchange of vacuum quantum numbers. Thus, by attributing this observed rise to the exchange of a single Regge pole and in order to describe the increase of all total hadronic cross section at high energies it follows that the Pomeron should have an intercept greater than $1$, \tit{i.e.} an effective intercept such that $\ap(0)=1+\e$ with $\e > 0$. Despite the fact that general unitarity arguments constrain the physical Pomeron intercept by $\ap(0)\leq 1$, since an intercerpt larger than $1$ would violate the Froissart-Marin-\L ukaszuk bound, there are some claims that we are far from asymptopia and therefore $\ap(0)> 1$ shall be no problem at presently collider energies. However, the supercritical intercept value can be arrived at by taking into account, in addition to Regge poles, multi-Pomeron cuts in the complex angular momentum plane. Reggeon field theory, sometimes called as Reggeon calculus, enables one to systematically analyze the exchange of poles and branch points in high-energy scattering processes, then allowing that the bare Pomeron intercept may be greater than one and these contribute to tame the growth of cross sections.

The physical particles which would provide the resonances for integer values of the Pomeron trajectory have not been yet identified. Particles with the quantum numbers of the vacuum are difficult to detect, but such particles can exist in perturbative QCD as bound states of gluons, in the simplest case by a color singlet of two reggeized gluons.

\section{\textsc{Regge Phenomenology}}
\label{secRGb.1}
\mbox{\,\,\,\,\,\,\,\,\,}
Diffraction processes account for a substantial fraction of hadron-hadron total cross section at high energies. The elastic scattering events are very difficult to be detected at very high energies since the particles scatter through small angles. On the other hand, the Pomeron has a place into several different processes which include, besides elastic scattering, single- or double-diffractive dissociation and are characterized by the presence of one or more large rapidity gaps. In the simplest scenario the diffractive processes are driven by an isolated pole at $J=\alpha(t)$, resulting in an elastic amplitude $A(s,t)$ written in terms of the Regge pole trajectory $\alpha(t)$, where $A(s,t)\propto s^{\alpha(t)}$. However, if more than one pole contributes, the elastic scattering amplitude is expressed in the $s$-channel as a descending asymptotic series of powers of $s$,
\be
A(s,t)=\sum_{i}\gamma_{i}(t)\eta_{i}(t)s^{\alpha_{i}(t)},
\label{chRGb.1}
\ee
where $\eta_{i}(t)$ is the signature factor, which completely determines the phase of the scattering amplitude because the Regge trajectories and the residue functions are expected to be real below the threshold,
\be
\eta_{i}(t)=-\frac{1+\xi e^{-i\pi\alpha_{i}(t)}}{\sin\left[\pi\alpha_{i}(t)\right]}.
\label{chRGb.2}
\ee
Straightforward calculation shows that the signature factor for even- and odd-trajectory can be written respectively as
\be
\begin{split}
&\eta(t)=-\frac{e^{-i\pi\alpha_{i}(t)/2}}{\sin\left[\frac{\pi}{2}\,\alpha_{i}(t)\right]}\underset{t\ll 1}{\sim}-e^{-i\pi\alpha_{i}(t)/2}, \,\,\,\,\,\,\,\textnormal{for}\,\,\,\,\xi=+1,\\
&\eta(t)=-i\,\frac{e^{-i\pi\alpha_{i}(t)/2}}{\cos\left[\frac{\pi}{2}\,\alpha_{i}(t)\right]}\underset{t\ll 1}{\sim}ie^{-i\pi\alpha_{i}(t)/2}, \,\,\,\,\textnormal{for}\,\,\,\,\xi=-1.
\end{split}
\label{chRGb.3}
\ee
\mbox{\,\,\,\,\,\,\,\,\,}
The ratio of the real to imaginary part of $A(s,t)$, by considering a single Regge pole exchange, is written as
\be
\frac{\textnormal{Re}\,A(s,t)}{\textnormal{Im}\,A(s,t)}=-\frac{1+\xi\cos\pi\alpha(t)}{\sin\pi\alpha(t)},
\label{chRGb.4}
\ee
thus the ratio is determined by the intercept of the exchanged trajectory. In the forward direction, \tit{i.e.} $t=0$, this expression defines the $\rho$-parameter.

Each Regge trajectory has its own respectively associated quantum numbers. As for example the Pomeron is defined by
\be
\IP:\,\,\, P=+1,\,\,\,C=+1,\,\,\,G=+1,\,\,\,I=0,\,\,\,\xi=+1,
\label{chRGb.5}
\ee
whereas for the other leading Reggeons,
\bear
f_{2}&:&\,\,\, P=+1,\,\,\,C=+1,\,\,\,G=+1,\,\,\,I=0,\,\,\,\xi=+1,\nonumber\\
\rho&:&\,\,\, P=-1,\,\,\,C=-1,\,\,\,G=+1,\,\,\,I=1,\,\,\,\xi=-1,\nonumber\\
\omega&:&\,\,\, P=-1,\,\,\,C=-1,\,\,\,G=-1,\,\,\,I=0,\,\,\,\xi=-1,\nonumber\\
a_{2}&:&\,\,\, P=+1,\,\,\,C=+1,\,\,\,G=-1,\,\,\,I=1,\,\,\,\xi=+1.\nonumber
\eear

Since the Pomeron has a positive-signature trajectory, then in the limit $\ap(0)\to 1$, for an asymptotically constant cross sections, the signature factor behaves as
\be
\lim_{\ap(0)\to1}\eta_{\IP}(0)=-\frac{1+e^{-i\pi \ap(0)}}{\sin\pi \ap(0)}=i.
\label{chRGb.6}
\ee
Thus the leading $\IP$ scattering amplitude has $\eta(0)=i$, and for large-$s$ at $t=0$ behaves as a purely imaginary amplitude,
\be
A_{\IP}(s,t=0)\underset{s\to\infty}{\sim}i\gamma_{\IP}(0)s^{\ap(0)}.
\label{chRGb.7}
\ee

Regge poles are only part of the possible singularities allowed by Regge theory. There are other types of singularities in the complex $\ell$-plane which correspond to the exchange of two or more Reggeon, as for example multiple-Pomeron exchange also known as Regge cuts. In the case of $t$-channel double-Pomeron exchange, the Regge cut structure leads to scattering amplitudes which behaves asymptotically at $s\to\infty$ as $A_{\IP\IP}(s,t)\sim -s^{\alpha_{\IP\IP}(t)}/\log s$, where $\alpha_{\IP\IP}(t)=2\ap(0)-1+\alp t/2$. Another possibility to take into account unitarity restoration is to construct an eikonalized amplitude, which is unitarized by definition. In this picture the eikonal expansion could be related somehow to the exchanged series. 

If, however, the high-energy behavior of the total cross section is, indeed, a result of the superposition of many exchanged Reggeons, and since the intercepts are universal, then one should therefore expect that they are able to contribute in the description of other hadronic processes,
\bear
\pi^{-}p&\sim&\IP+f_{2}+\rho,\nonumber\\
\pi^{+}p&\sim&\IP+f_{2}-\rho,\nonumber\\
K^{-}p&\sim&\IP+f_{2}+\rho+a_{2}+\omega,\nonumber\\
K^{+}p&\sim&\IP+f_{2}-\rho+a_{2}-\omega,\nonumber\\
pp&\sim&\IP+f_{2}-\rho+a_{2}-\omega,\nonumber\\
\bar{p}p&\sim&\IP+f_{2}+\rho+a_{2}+\omega,\nonumber\\
pn&\sim&\IP+f_{2}+\rho-a_{2}-\omega.\nonumber
\eear
In the particle-particle and antiparticle-particle total cross section differences the Pomeron contribution cancels out, 
\bear
\sigma(\pi^{-}p)-\sigma(\pi^{+}p)&\sim&2\rho,\nonumber\\
\sigma(K^{-}p)-\sigma(K^{+}p)&\sim&2(\omega+\rho),\nonumber\\
\sigma(\bar{p}p)-\sigma(pp)&\sim&2(\omega+\rho),\nonumber\\
\sigma(pn)-\sigma(pp)&\sim&2(\rho-a_{2}),\nonumber
\eear
since these total cross section differences are determined by sub-leading Regge trajectories which vanish at the asymptotic limit $s\to\infty$, in agreement with the revised Pomeranchuk theorem.

\subsection{\textsc{Total and Elastic Cross Section}}
\label{secRGb.1.2}
\mbox{\,\,\,\,\,\,\,\,\,}
Each term of the series in expression (\ref{chRGb.1}) represents a specific exchange in the $t$-channel. From the optical theorem, the total cross section reads
\be
\sigma_{tot}(s)=4\pi\sum_{i}g_{i}s^{\alpha_{i}(0)-1},
\label{chRGb.8}
\ee
with $g_{i}\equiv\gamma_{i}(0)\textnormal{Im}\,\eta_{i}(0)$. The Pomeron as it emerges from fits to forward observables is called soft Pomeron. The magnitude of its intercept $\alpha_{\mathds{P}}$ plays a central role in Regge theory, since the Pomeron is the Reggeon with the largest intercept. However, as it was mentioned in the last section, in order to describe the observed increase with $s$ of all hadronic cross sections at high-energy, see (upper) Figure \ref{ch2fig8}, the Pomeron should have an effective intercept such that $\alpha_{\mathds{P}}=1+\epsilon$, with $\epsilon>0$, where it can be viewed as an \tit{ad hoc} inserted phenomenological parameter or can be arrived at by means of Reggeon calculus.

The elastic differential cross section in Regge theory is written as
\be
\frac{d\sigma}{d\vert t \vert}(s,t)=F(t)\,s^{2\alpha(t)-2},
\label{chRGb.9}
\ee
where $F(t)$ is a function of $t$ and it has absorbed the residue function, the signature factor and other constants. In general, if many poles contribute, then interference terms will appear. By considering only one single Reggeon with linear trajectory,
\be
\frac{d\sigma}{d\vert t \vert}(s,t)=F(t)\,s^{2\alpha(0)-2}\,e^{-2\alpha^{\prime}\vert t \vert\log s},
\label{chRGb.10}
\ee
where $s^{\alpha^{\prime 2}t}=e^{2\alpha^{\prime}t\log s}$. If the colliding particles are alike and by assuming a simple exponential parametrization for the residue function, such as $\gamma(t)=\gamma(0)e^{B_{0}t}$, thus
\be
\frac{d\sigma}{d\vert t \vert}(s,t)\sim s^{2\alpha(0)-2}\,e^{-B\vert t \vert}\,,
\label{chRGb.11}
\ee
where $B=B_{0}+2\alpha^{\prime}\log s$ is the $t$-slope of the scattering amplitude. The width of the forward peak, $\Delta\vert t \vert=(B_{0}+2\alpha^{\prime}\log s)^{-1}$, decreases with increasing energy. This phenomena is known as the shrinkage of the forward diffraction peak, and it can be interpreted as an increase of the interaction radius $R_{int}\sim\sqrt{\alpha^{\prime}\log s}$ at $s\to\infty$. This shrinkage is indeed observed, however it is not suggested by any optical analogy.

\clearpage
\thispagestyle{plain}

\chapter{\textsc{Soft Pomeron and Analytic Models}}
\label{ch4}
\mbox{\,\,\,\,\,\,\,\,\,}

In elastic hadron-hadron collisions, the forward scattering is characterized by two physical observables, the total cross section and the $\rho$-parameter. In terms of the scattering amplitude and its Mandelstam variables, the former is given by the optical theorem and the later associated with the phase of the amplitude.


Since the real and imaginary parts of the amplitude can be formally correlated by means of dispersion relations, $\sigma_{tot}(s)$ and $\rho(s)$
provide a fundamental physical connection between the phase of the amplitude and the total probability of the hadronic interaction as a function of the energy. However, despite their rather simple analytic forms, the investigation of these two quantities in terms of the energy, constitute a long standing challenge in the study of the hadronic interactions \cite{Pancheri:2016yel}. 

In the experimental context, to access the forward and near forward region demands complex and sophisticated instrumentation and data analyses. In addition, the difficulties grow progressively as the energy increases. For example, the $\rho$-parameter is determined in the region of interference between the Coulomb and nuclear interactions, which are of the same magnitude at values of the momentum transfer proportional to the inverse of the total cross section \cite{Block:1984ru}. As a consequence of the rise of $\sigma_{tot}$ at the highest energies, it becomes extremely difficult to reach this region as the energy increases.

In the theoretical QCD context, this deep infrared region ($t \rightarrow 0$) is not expected to be accessed by perturbative techniques. A crucial point concerns the absence of a nonperturbative approach able to predict the energy dependence of $\sigma_{tot}$ and $\rho$ from first principles and without model assumptions.

In the phenomenological context, beyond classes of models including other physical quantities, $\sigma_{tot}(s)$ and $\rho(s)$ are usually investigated by means of amplitude analyses, an approach based on Regge theory and analytic $S$-matrix concepts. In this formalism \cite{Collins:1977jy,Forshaw:1997dc,Donnachie:2002en}, the singularities in the complex angular momentum $J$-plane ($t$-channel) are associated with the asymptotic behavior of the elastic scattering amplitude in terms of the energy ($s$-channel), as it was discussed in the last chapter.

\section{\textsc{Single, Double and Triple Poles}}
\label{sdt}
\mbox{\,\,\,\,\,\,\,\,\,}
In the general case, associated with a pole of order $N$, the contribution to the imaginary part of the forward amplitude in the $s$-channel is $s^{\alpha_0} \ln^{N-1}s$, where $\alpha_{0}$ is the intercept of the trajectory. Therefore, for the total cross section,
\be
\sigma_{tot}(s)  \propto s^{\alpha_0 - 1}\ln^{N-1} s.
\ee

There may be the possibility that higher order poles in the complex $J$-plane could be generated by means of derivatives of the simple pole \cite{Donnachie:2002en,Fagundes:2017iwb},
\be
\frac{d^{n}}{d\alpha^{n}}\left[\frac{1}{J-\alpha}\right]=\frac{n!}{\left(J-\alpha\right)^{n+1}}\,\,\,\,\textnormal{for}\,\,\,\,n=1,2,...,
\ee
where $N=n+1$ stands for the order of the pole. Hence, applying this differentiation to the power-law form of the scattering amplitude, one finds
\be
\frac{d^{n}}{d\alpha^{n}}\,s^{\alpha}=s^{\alpha}\ln^{n}s^{\alpha},
\ee
and the contribution of the scattering amplitude in the $s$-channel associated with a $t$-channel pole of order $N$ is $s^{\alpha}\ln^{N-1}s$. In the case of the Pomeron contribution, considering a pole at $J=\alpha=1$, the total cross section is given by
\be
\sigma^{\mathds{P}}\propto \ln^{N-1}s.
\ee

It was repeatedly mentioned that in order to preserve the unitarity property of the scattering $S$-matrix the Froissart-Martin-\L ukaszuk bound constrains the rise of the total cross section. Bearing in mind that this constraint is written in the form of $\sigma\leq (cte)\ln^{2}s$, for this reason the allowed extra leading contributions, besides the simple pole, are double and triple poles, respectively. Therefore, there are the following possibilities connecting the singularities and the asymptotic behavior:
\begin{itemize}

\item[i. ]
simple pole ($N=1$) at $J = \alpha_0$, with $\alpha_0 = 1$\ \ $\Rightarrow$\ \ $\sigma$ \ constant;

\item[ii. ]
simple pole ($N=1$) at $J=\alpha_0$\ \ $\Rightarrow$\ \ $\sigma \propto s^{\alpha_0-1}$;

\item[iii. ]
double pole ($N=2$) at $J=\alpha_0$, with $\alpha_0=1$\ \ $\Rightarrow$\ \ $\sigma \propto \ln(s)$;

\item[iv. ]
triple pole ($N=3$) at $J=\alpha_0$, with $\alpha_0=1$\ \ $\Rightarrow$\ \ $\sigma \propto \ln^2(s)$.

\end{itemize}

For an elastic particle-particle and antiparticle-particle scattering, given the above inputs for $\sigma_{tot}(s)$,  the even and odd contributions associated with $\textnormal{Im}\,A(s,t=0)/s$ through crossing symmetry are defined and the corresponding real parts are obtained through analyticity by means of dispersion relations, leading to $\rho(s)$ defined as the ratio of the real to imaginary part of the forward scattering amplitude.

\section{\textsc{Motivation and the Forward LHC$13$ Data}}
\mbox{\,\,\,\,\,\,\,\,\,}
Historically, the leading contribution to $\sigma_{tot}$ at the highest energies has been associated with an even-under-crossing object named Pomeron. Typical Pomeron models consider contributions associated with either a simple pole at $J=\alpha_0$ (for example, Donnachie and Landshoff \cite{Donnachie:1979yu,Donnachie:2013xia} and some QCD-inspired models \cite{Beggio:2013vfa,Fagundes:2011zx,Luna:2006qp,Luna:2005nz,Block:1998hu}) or a triple pole at $J=1$ (as selected in the detailed analysis by the COMPETE Collaboration \cite{Cudell:2001pn,Cudell:2002xe} and used in the successive editions of the Review of Particle Physics, by the Particle Data Group (PDG) \cite{Patrignani:2016xqp}).

However, recently, new experimental information on $\sigma_{tot}$ and $\rho$ at $\sqrt{s}=13$ TeV were reported by the TOTEM Collaboration at the LHC \cite{Antchev:2017dia,Antchev:2017yns}, namely,
\begin{eqnarray}
\sigma^{pp}_{tot} &=& 110.6 \pm 3.4\ \textnormal{mb}, \nonumber \\
\rho^{pp} &=& 0.10 \pm 0.01\,\,\, \textnormal{and}\,\,\, 0.09 \pm 0.01,
\label{totem13}
\end{eqnarray}
indicating an unexpected decrease in the value of the $\rho^{pp}$ and $\sigma^{pp}_{tot}$ in agreement with the trend of previous measurements at $7$ and $8$ TeV. Indeed, recent investigation concerning bounds on the rise of $\sigma_{tot}(s)$, including all TOTEM data at 7 and 8 TeV and the L$\gamma$ parametrization \cite{Fagundes:2017cmp}, has predicated at $13$ TeV the value $\sigma^{pp}_{tot} = 110.7\pm 1.2$ mb, which is in full agreement with the above measurement. However, for $\rho^{pp}$ at $13$ TeV the extrapolation yielded $0.1417\pm 0.0047$, indicating complete disagreement with the data and far above the experimental result \cite{Fagundes:2017cmp}. Moreover, the results (\ref{totem13}) are not simultaneously described by all the predictions of the Pomeron models from the detailed analysis by the COMPETE Collaboration in 2002 \cite{Cudell:2002xe}. 

Remarkably, the odd-under-crossing asymptotic contribution, introduced by \L ukaszuk and Nicolescu \cite{Lukaszuk:1973nt} and named Odderon\footnote{From a QCD viewpoint, a color singlet made up of three reggeized gluons in the simplest configuration \cite{Ewerz:2003xi,Ewerz:2005rg,Braun:1998fs}.} \cite{Joynson:1975az}, provides quite good descriptions of the experimental data, as predicted by the AGN model \cite{Avila:2006wy} and demonstrated recently in the analyses by Martynov and Nicolescu \cite{Martynov:2017zjz,Martynov:2018nyb}.

On the other hand, also recently, the above data at $13$ TeV have been analyzed by the Durham Group in the context of a QCD-based multichannel eikonal model with Pomeron dominance, tuned in 2013 with data up to 7 TeV \cite{Khoze:2013jsa,Khoze:2014aca}. The analysis indicates that the data at 13 TeV are reasonably described without an odd-signature term \cite{Khoze:2017swe}. Moreover, the authors also argument that the  maximal Odderon is inconsistent with the black disk limit \cite{Khoze:2018bus}. Also very recently, subsequent articles have also discussed possible effects related to Odderon contributions in different contexts \cite{Khoze:2018kna,Broniowski:2018xbg,Troshin:2018ihb,Csorgo:2018uyp,Gotsman:2018buo}.

In view of all these recent informations and the fact that forward amplitude analyses have favored the Pomeron dominance, at least, up to $8$ TeV, it seems important to develop detailed tests on the applicability of the Pomeron models by means of a general class of forward scattering amplitudes. With that in mind, we have already reported two forward analyses with Pomeron dominance and including, for the first time, the TOTEM data at $13$ TeV. In the first work, two models have been tested, without taking into account the uncertainty regions in the data reductions \cite{Broilo:2018brv} (as in the Martynov and Nicolescu analyses \cite{Martynov:2017zjz,Martynov:2018nyb}). We concluded that the models are not able to satisfactorily describe the $\sigma_{tot}$ and $\rho$ data at $13$ TeV. In the second analysis, we have considered one Pomeron model with $6$ free parameters and have evaluated the uncertainty regions with CL of $90$\% which, in the case of normal errors, the $\chi^{2}$ fitting procedure correspond to the projection of the $\chi^{2}$ hypersurface in the interval $\chi^{2}-\chi^{2}_{min}=10.65$. We have concluded that the model seems not to be excluded by the bulk of experimental data presently available \cite{Broilo:2018els}.

This work so far shall extend our investigation in several important aspects. The main point is to consider classes of even leading contributions by incorporating different components of several models and investigating the effect of several combinations, with focus on the uncertainties involved in the data reductions. To this end, we shall treat a general parametrization for $\sigma_{tot}(s)$ consisting of constant, power, logarithmic and logarithmic squared functions of the energy, together with even and odd Reggeons, namely $a_2/f_2$ and $\rho/\omega$ trajectories, respectively for the low energy region. The analytic connection with  $\rho(s)$ is obtained by means of even and odd singly subtracted dispersion relations. 


However, there is an intrinsic difficulty with this kind of analysis deserving attention from the beginning. In what concerns the $\sigma_{tot}$ and $\rho$ data, despite the great expectations with the LHC, the experimental information presently available are characterized by discrepancies between the measurements of $\sigma_{tot}$ by the TOTEM Collaboration and by the ATLAS Collaboration at $7$ TeV and mainly at $8$ TeV. Some consequences of these discrepancies have already been discussed in References \cite{Fagundes:2017cmp,Broilo:2018els}. In the first analysis made by Martynov and Nicolescu, it was presented arguments for not including the ATLAS data \cite{Martynov:2017zjz}, which, however, have been included in their second analysis \cite{Martynov:2018nyb}. 

It should be also noted that the uncertainties in the TOTEM measurements of $\sigma_{tot}$ are essentially systematic (uniform distribution) and not statistical  (Gaussian distribution). This fact puts limitations in a strict interpretation of the $\chi^{2}$ test of goodness-of-fit.


Here, to address the above question and as in previous analyses \cite{Fagundes:2017cmp,Fagundes:2017iwb,Broilo:2018brv,Broilo:2018els}, we shall adopt two variants for defining our data set: all the experimental data below $7$ TeV (above $5$ GeV) and two independent ensembles by adding either only the TOTEM data at $7$, $8$ and $13$ TeV (ensemble T) or by including also the ATLAS data at $7$ and $8$ TeV (ensemble T+A). In addition, in order to investigate and stress the importance of the uncertainty regions in the fit results, we shall consider data reductions with two different CL, associated with both one and two standard deviations, namely $68.27$\% and $95.45$\% CL, respectively.


\section{\textsc{The Odderon}}
\mbox{\,\,\,\,\,\,\,\,\,}
Since it was briefly mentioned the possible effects of the Odderon contribution at the LHC13 in the last section, it seems wise to give a short statement regarding the nature of this odd fellow. The Odderon is a hypothetical Regge trajectory, first introduced by \L ukaszuk and Nicolescu in $1973$ \cite{Lukaszuk:1973nt} and developed over the years in the context of pertubative QCD, which could play some important contribution in high-energy hadronic collisions. It is conceived in the terms of Regge language as the $C=P=-1$ partner of the Pomeron and related to a singularity of the crossing-odd scattering amplitude in the complex angular momentum plane. Since the leading soft Pomeron was assumed to saturate unitarity, the Odderon contribution was supposed to rise less rapidly as the energy increases. More specifically, it was associated with a double pole near $J=1$, giving for large $s$,
\be
\sigma^{ab}(s)-\sigma^{\bar{a}b}(s)\sim\ln s.
\ee
In the literature this form is known by the name of Maximal Odderon contribution. Later, it was also considered the presence of simple poles with intercept $\alpha_{\mathds{O}}(t=0)\simeq 1$ and named Minimal Odderon contribution.

The existence of the Odderon clearly predicts differences in the asymptotic behavior of scaterring amplitutes for $pp$ and $\bar{p}p$ interactions. For example, its presence would therefore entail a distinct contribution at high energies for $\sigma^{pp}$ and $\sigma^{\bar{p}p}$ and the same goes for the case of the $\rho$-parameter. As a matter of fact, $pp$ and $\bar{p}p$ differential cross sections indeed differ in the dip region and, as mentioned in Chapter \ref{ch2} and displayed in Figure \ref{ch2fig11}, the dip seems to recede towards zero as the energy increases where its $\vert t \vert$ value are roughly estimated as $1/\sigma_{tot}$. This subtlety difference is that on the one hand whilst $pp$ scattering shows a pronounced dip, on the other hand $\bar{p}p$ scattering shows no dip at all but only a shoulder. This sort of behavior can be seen in Regge theory as a phenomenological support to the existence of the Odderon. 

At the high-$\vert t \vert$ region, the realm of perturbative QCD, it is known that $pp$ scaterring is dominated by three gluon exchange with $C=P=-1$ configuration and this is expected to correspond to the ``hard'' Odderon. The natural path would be the search for an Odderon exchange at low-$\vert t \vert$, for which there is no presently and plesantly evidence from the available data so far. Therefore the ``soft'' Odderon has not been yet observed at least at $t=0$.

\section{\textsc{Analytic Models}}
\mbox{\,\,\,\,\,\,\,\,\,}
As commented in section \ref{sdt}, in the Regge theory, simple, double and triple poles in the complex angular momentum plane are associated with power, logarithmic and logarithmic squared functions of the energy for the total cross section. In this context, for $pp$ and $\bar{p}p$ scattering,  we consider a general parametrization for $\sigma_{tot}(s)$ consisting of two Reggeons (even and odd under crossing) and four (even) Pomeron contributions,
\be
\sigma_{tot}(s) =  a_1 \left[\frac{s}{s_0}\right]^{-b_1} \!\!\!\!\!\!+ \tau a_2 \left[\frac{s}{s_0}\right]^{-b_2} \!\!\!\!\!\!+ A + B \left[\frac{s}{s_0}\right]^{\epsilon}+ C \ln\left(\frac{s}{s_0}\right) + D \ln^2\left(\frac{s}{s_0}\right),
\label{stg}
\ee
where $\tau = -1\,(+1)$ for $pp\,(\bar{p}p)$, while $a_1$, $b_1$, $a_2$, $b_2$, are free fit parameters associated with the secondary Reggeons, and $A$, $B$, $\epsilon$, $C$, $D$ are the free parameters associated with Pomeron components. The analytic results for $\rho(s)$ have been obtained by means of singly subtracted derivative dispersion relations \cite{Avila:2003cu}, taking into account an effective subtraction constant $K$,
\be
\begin{split}
\rho(s) & =  \frac{1}{\sigma_{tot}(s)}\Bigg\{ \frac{K}{s}- a_1\,\tan \left( \frac{\pi\, b_1}{2}\right) \left[\frac{s}{s_0}\right]^{-b_1} \!\!\!\!\!\!+\tau \, a_2\, \cot \left(\frac{\pi\, b_2}{2}\right) \left[\frac{s}{s_0}\right]^{-b_2}\\
& +  B\,\tan \left( \frac{\pi\, \epsilon}{2}\right) \left[\frac{s}{s_0}\right]^{\epsilon} + \frac{\pi C}{2} + \pi D \ln\left(\frac{s}{s_0}\right)\Bigg\},
\label{rhog}
\end{split}
\ee
where the presence of $K$ avoids the full high-energy approximation in dispersion relation approaches \cite{Fagundes:2017iwb}. By following \cite{Fagundes:2017cmp,Fagundes:2017iwb,Broilo:2018els,Broilo:2018brv}, we consider the energy scale $s_{0}$ fixed at the physical threshold for scattering states \cite{Menon:2013vka}, 
\be
s_0 = 4m_p^2 \sim  3.521 \mathrm{GeV}^2,
\nonumber
\ee
where $m_p$ is the proton mass.


Strictly speaking, expressions (\ref{stg}) and (\ref{rhog}) bring enclosed analytic structures similar to those appearing in some well known models in the literature, as for example, Donnachie and Landshoff (DL) \cite{Donnachie:1979yu,Donnachie:2013xia}, Block and Halzen (BH) \cite{Block:2012ym,Block:2012nj}, COMPETE and PDG parametrizations (COMPETE) \cite{Cudell:2001pn,Cudell:2002xe,Patrignani:2016xqp}. We shall consider four particular cases, distinguished by the corresponding Pomeron contributions, defined and denoted as follows:

\begin{itemize}

\item[] Model I:\
$A=C=D=0$ \quad $\Rightarrow$ \quad $\sigma_{I}^\mathds{P} = B \left[\frac{s}{s_0}\right]^{\epsilon}$ \quad (DL-type)

\item[] Model II:\ 
$B=C=0$, $\epsilon=0$ \quad $\Rightarrow$ \quad $\sigma_{II}^\mathds{P} = A + D \ln^{2}\left(\frac{s}{s_0}\right)$ \quad  (COMPETE-type)

\item[] Model III:\
$A=B=0$, $\epsilon=0$ \quad $\Rightarrow$ \quad $\sigma_{III}^\mathds{P} = C \ln \left(\frac{s}{s_0}\right) + 
D \ln^{2}\left(\frac{s}{s_0}\right)$ \quad  (BH-type)

\item[] Model IV:\
$A=D=0$ \quad $\Rightarrow$ \quad $\sigma_{IV}^\mathds{P} = B \left[\frac{s}{s_0}\right]^{\epsilon}
+ C \ln \left(\frac{s}{s_0}\right)$
\quad (hybrid power-log)

\end{itemize}

We note that models II and III are analytically similar. The difference concerns the phenomenological interpretation of the singularities as single and double poles. As far as we know, model IV was never considered in the literature. Its use here is related to tests on the power law (single pole) in attempts to describe simultaneously the $\sigma_{tot}$ and the $\rho$ data at 13 TeV.
 
In the General Model,  expressions (\ref{stg}) and (\ref{rhog}), we have $10$ free parameters, $a_1$, $b_1$, $a_2$, $b_2$, $A$, $B$, $\epsilon$, $C$, $D$ and $K$, which are determined through fits to the experimental data on $\sigma_{tot}$ and $\rho$ from $pp$ and $\bar{p}p$ elastic scattering in the interval $\sqrt{s}=5$ GeV - $13$ TeV.

\section{\textsc{Ensembles and Data Reductions}}
\mbox{\,\,\,\,\,\,\,\,\,}
The data above $5$ GeV and below $7$ TeV have been collected from the PDG database \cite{Patrignani:2016xqp}, without any kind of data selection or sieve procedure. We have used all the published data by the experimental collaborations. The data at $7$ and $8$ TeV by the TOTEM and ATLAS Collaborations can be found in Table \ref{ch2tab1}, together with further information and complete list of references in section \ref{sec2.8}, respectively. The TOTEM data at $13$ TeV are those in (\ref{totem13}) \cite{Antchev:2017dia,Antchev:2017yns}.

Given the tension between the TOTEM and ATLAS measurements on $\sigma_{tot}$ at $7$ TeV and mainly $8$ TeV, we shall consider two ensembles of $pp$ and $\bar{p}p$ data above $5$ GeV, both comprising the same data set in the region below $7$ TeV. We then construct:

\begin{itemize}

\item[i.] Ensemble TOTEM (T) by adding only the TOTEM data at $7$, $8$ and $13$ TeV;

\item[ii.] Ensemble TOTEM + ATLAS (T + A)  by adding to ensemble T the ATLAS data at $7$ and $8$ TeV.

\end{itemize}

The fits were performed with the objects of the TMinuit package and using the default MINUIT error analysis \cite{minuit,James:2004xla}. We have carried out global fits using a $\chi^{2}$ fitting procedure, where the value of $\chi^{2}_{min}$ is distributed as a $\chi^{2}$ distribution with  $\nu$ degrees of freedom. The global fits to $\sigma_{tot}$ and $\rho$ data were performed adopting an interval $\chi^{2}-\chi^{2}_{min}$ corresponding, in the case of normal errors, to the projection of the $\chi^{2}$ hypersurface containing first $\sim 68$\% of probability  and in a second step, $\sim 95$\% of probability, namely 1 $\sigma$ and 2 $\sigma$.

As a convergence criteria we consider only minimization results which imply positive-definite covariance matrices, since theoretically the covariance matrix for a physically motivated function must be positive-definite at the minimum. As tests of goodness-of-fit we shall adopt the chi-square per degree of freedom $\chi^2/\nu$ and the integrated probability $P(\chi^2)$ \cite{bev}.

\section{\textsc{Results}}
\mbox{\,\,\,\,\,\,\,\,\,}
The data reductions with the general model given by expressions (\ref{stg}) and (\ref{rhog}) did not comply with the above convergence requirements and thus can not be regarded as a possible solution. This may be due to an excessive number of free parameters. On the other hand,  in the particular cases given by Models I, II, III and IV, the convergence criteria were reached. 

In each case, the values of the free fit parameters with uncertainty of 1 $\sigma$, together with the corresponding statistical information, are displayed in Table \ref{ch4t1} in case of ensemble T and Table \ref{ch4t2} within ensemble T+A. 

Through  error propagation from the fit parameters, we determine the uncertainty regions for the theoretical results (curves), within 1 $\sigma$ and 2 $\sigma$. The results for $\sigma_{tot}(s)$ and $\rho(s)$ with Models I, II, III and IV (ensembles T and T+A) are compared with the experimental data in Figures \ref{ch4fig1}, \ref{ch4fig2}, \ref{ch4fig3} and \ref{ch4fig4}, respectively. In each Figure, the insets highlight the LHC energy region.

\subsection{\textsc{Theoretical Uncertainty Propagation}}
\mbox{\,\,\,\,\,\,\,\,\,}
Before pressing on, it might be interesting to briefly discuss how we calculated the error propagation in each case considered. The uncertainty region is calculated by means of the covariance error matrix, which measures the level of correlation between the parameters, and is obtained at the end of the minimization process for each model and ensemble \cite{minuit,James:2004xla}. The theoretical error propagation in Models I, II, III and IV can be derived by means of a general formula which takes into account a high-level of correlation. For example, by considering some given quantity $w$ that is calculated as a function of other quantities $x,\,y,\,z,...$, one finds \cite{vuolo}
\be
\begin{split}
& \sigma^{2}_{w}(x,y,z,...) = \left(\frac{\partial w}{\partial x}\right)^{2}\sigma^{2}_{x} + \left(\frac{\partial w}{\partial y}\right)^{2}\sigma^{2}_{y} +\left(\frac{\partial w}{\partial z}\right)^{2}\sigma^{2}_{z} + ... \\
& + 2 \left(\frac{\partial w}{\partial x}\right)\left(\frac{\partial w}{\partial y}\right)\sigma^{2}_{xy} + 2 \left(\frac{\partial w}{\partial x}\right)\left(\frac{\partial w}{\partial z}\right)\sigma^{2}_{xz} + 2 \left(\frac{\partial w}{\partial y}\right)\left(\frac{\partial w}{\partial z}\right)\sigma^{2}_{yz} + ...
\end{split}
\label{gf}
\ee
The uncertainties are positive by definition, then the error associated will always be considered as the positive square root. The factor $2$ multiplying the off-diagonal elements comes from the fact that the covariance error matrix is symmetric.

For our case, we are studying the theoretical uncertainty propagation in models I - IV by considering the covariance error matrix associated to the forward scattering observables, total cross section and $\rho$-parameter. It should be stressed that although our interest remains on the high-energy behavior of the aforementioned quantities, the parameters involved present high correlation. For this reason, in treating this sort of plight we cannot disregard the low-energy parameters associated to secondary Reggeons. This makes the calculation tedious, but since all the four models are at Born-level it turns out that error calculus will be much easier than from an eikonalized model.

Returning to the task in hand, by means of the general formula (\ref{gf}) and the expressions for $\sigma_{tot}(s)$ and $\rho(s)$ in the general model, it is easy to find a recurrence expression for the associated errors. For the total cross section one finds
\be
\sigma^{2}_{\sigma_{_{tot}}}(s;i,j)=\sum_{i,j}
\left(\frac{\partial\sigma_{tot}(s)}{\partial i}\right)\left(\frac{\partial\sigma_{tot}(s)}{\partial j}\right)\sigma^{2}_{ij},
\ee
as for the case of the $\rho$-parameter,
\be
\sigma^{2}_{\rho}(s;i,j)=\frac{1}{\sigma^{2}_{tot}(s)}\sum_{i,j}
\left(\frac{\partial A^{R}(s;i,j)}{\partial i}-\rho(s)\,\frac{\partial\sigma_{tot}(s)}{\partial i}\right)
\left(\frac{\partial A^{R}(s;i,j)}{\partial j}-\rho(s)\,\frac{\partial\sigma_{tot}(s)}{\partial j}\right)
\sigma^{2}_{ij},
\ee
where $i,j$ stands for the free parameter and $A^{R}$ is the real part of the forward scattering amplitude, \tit{i.e.} the curly brackets in expression (\ref{rhog}). The corresponding partial differentiations are the following:
\begin{itemize}
\item[i.] with respect to the total cross section
\end{itemize}

\be
\frac{\partial\sigma_{tot}(s)}{\partial a_{1}} = \left(\frac{s}{s_{0}}\right)^{-b_{1}} 
\ee

\be
\frac{\partial\sigma_{tot}(s)}{\partial b_{1}} = -a_{1}\left(\frac{s}{s_{0}}\right)^{-b_{1}}\ln\left(\frac{s}{s_{0}}\right) 
\ee

\be
\frac{\partial\sigma_{tot}(s)}{\partial a_{2}} = \tau \left(\frac{s}{s_{0}}\right)^{-b_{2}} 
\ee

\be
\frac{\partial\sigma_{tot}(s)}{\partial b_{2}} = -\tau a_{2}\left(\frac{s}{s_{0}}\right)^{-b_{2}}\ln\left(\frac{s}{s_{0}}\right) 
\ee

\be
\frac{\partial\sigma_{tot}(s)}{\partial A} = 1 
\ee

\be
\frac{\partial\sigma_{tot}(s)}{\partial B} = \left(\frac{s}{s_{0}}\right)^{\e} 
\ee

\be
\frac{\partial\sigma_{tot}(s)}{\partial \e} = B\left(\frac{s}{s_{0}}\right)^{\e}\ln\left(\frac{s}{s_{0}}\right) 
\ee

\be
\frac{\partial\sigma_{tot}(s)}{\partial C} =  \ln\left(\frac{s}{s_{0}}\right) 
\ee

\be
\frac{\partial\sigma_{tot}(s)}{\partial D} = \ln^{2}\left(\frac{s}{s_{0}}\right) 
\ee

\be
\frac{\partial\sigma_{tot}(s)}{\partial K} = 0 
\ee

\begin{itemize}
\item[ii.] with respect to the real component of the forward scattering amplitude
\end{itemize}

\be
\frac{\partial A^{R}(s)}{\partial a_{1}} = -\tan\left(\frac{\pi\,b_{1}}{2}\right)\left(\frac{s}{s_{0}}\right)^{-b_{1}} 
\ee

\be
\frac{\partial A^{R}(s)}{\partial b_{1}} = -\frac{a_{1}\,\pi}{2}\,\sec^{2}\left(\frac{\pi\,b_{1}}{2}\right)\left(\frac{s}{s_{0}}\right)^{-b_{1}} +
 a_{1}\tan\left(\frac{\pi\,b_{1}}{2}\right) \ln\left(\frac{s}{s_{0}}\right) \left(\frac{s}{s_{0}}\right)^{-b_{1}}  
\ee

\be
\frac{\partial A^{R}(s)}{\partial a_{2}} = \tau \, \textnormal{ctan}\left(\frac{\pi\,b_{2}}{2}\right)\left(\frac{s}{s_{0}}\right)^{-b_{2}} 
\ee

\be
\frac{\partial A^{R}(s)}{\partial b_{2}} = - \tau \, \frac{a_{2}\,\pi}{2}\,\textnormal{csec}^{2}\left(\frac{\pi\,b_{2}}{2}\right)\left(\frac{s}{s_{0}}\right)^{-b_{2}} -
 \tau\, a_{2}\,\textnormal{ctan}\left(\frac{\pi\,b_{2}}{2}\right) \ln\left(\frac{s}{s_{0}}\right) \left(\frac{s}{s_{0}}\right)^{-b_{2}} 
\ee

\be
\frac{\partial A^{R}(s)}{\partial A} = 0 
\ee

\be
\frac{\partial A^{R}(s)}{\partial B} = \tan\left(\frac{\pi\,\e}{2}\right)\left(\frac{s}{s_{0}}\right)^{\e} 
\ee

\be
\frac{\partial A^{R}(s)}{\partial \e} = \frac{B\,\pi}{2}\,\left(\frac{s}{s_{0}}\right)^{\e} \sec^{2}\left(\frac{\pi\,\e}{2}\right) + 
 B \left(\frac{s}{s_{0}}\right)^{\e} \ln\left(\frac{s}{s_{0}}\right) \tan\left(\frac{\pi\,\e}{2}\right) 
\ee

\be
\frac{\partial A^{R}(s)}{\partial C} =  \frac{\pi}{2} 
\ee

\be
\frac{\partial A^{R}(s)}{\partial D} = \frac{\pi}{2}\,\ln\left(\frac{s}{s_{0}}\right) 
\ee

\be
\frac{\partial A^{R}(s)}{\partial K} = \frac{1}{s} 
\ee

Each uncertainty region will be calculated by considering the set of parameters associated with the respective model.

\subsection{\textsc{Ensembles T and T+A}}
\mbox{\,\,\,\,\,\,\,\,\,}
Ensemble T+A encompasses all the experimental data presently available on forward $pp$ and $\bar{p}p$ scattering at high energies. However, as commented in the first impressions of this Chapter and discussed in \cite{Fagundes:2017cmp,Broilo:2018els}, the TOTEM and ATLAS data at $7$ and $8$ TeV present discrepant values. In special, at $8$ TeV, the ATLAS measurement of $\sigma_{tot}$ differs from the latest TOTEM result at this energy by $3$ standard deviation,
\be
\frac{\sigma_{tot}^{\mathrm{TOTEM}} - \sigma_{tot}^{\mathrm{ATLAS}}}{\Delta \sigma_{tot}^{\mathrm{TOTEM}}} = 
\frac{103 - 96.07}{2.3} = 3.0.
\nonumber
\ee 

On the one hand, TOTEM published $4$ measurements at $7$ TeV and $5$ at $8$ TeV (all consistent among them) and ATLAS only one point at each energy. On the other hand, the ATLAS uncertainties in these results are much smaller than the TOTEM uncertainties. For example, at $8$ TeV, if the ATLAS uncertainty is considered, the aforementioned ratio results $7.5$ standard deviations. Besides the TOTEM results for $\sigma_{tot}$ being larger than the ATLAS values at $7$ and $8$ TeV, the TOTEM data indicate a rise of the total cross section faster than the ATLAS data \cite {Fagundes:2017cmp}.

Obviously, these facts make any amplitude analyses more difficult and put serious limitations in  secure interpretations of the results and  unquestionable conclusions that may be reached. It is expected that these discrepancies might be resolved through further re-analyses or new data, but it can also happen that these systematic differences may persist. We recall the discrepancies characterizing the experimental information at the highest energy reached in $\bar{p}p$ scattering, namely $1.8$ TeV. The CDF and E$710$ results differ by $2.3$ standard deviation (respect the E$710$ uncertainty) and predictions from most phenomenological models lies between these points.

Anyway, presently we understand that ensemble T+A is the effective representative of the experimental information available, so that an efficient model should be able to access all points within the corresponding uncertainties. More precisely, the predicted uncertainty region must present agreement with the error bars of the experimental points, by reaching all of them, even if in a barely way, but never excluding one or another data, namely TOTEM or ATLAS results.

On the basis of these comments and before discussing the efficiency of each model in the fit results, three characteristics of ensembles T and T+A in our data reductions deserve to be highlighted. 
\begin{itemize}

\item[i. ]
From the Figures, for all models the main visual difference in the results within ensembles T and T+A concerns $\sigma_{tot}$ at the highest energies but not $\rho$ at these energies. Indeed, for example, with model I (Figure \ref{ch4fig1}) the uncertainty region in the fit result for $\sigma_{tot}$ at $13$ TeV within ensemble T goes through the lower error bar, the central value and half of the upper error bar, but within ensemble  T + A, goes through only the lower error bar; on the other hand, for $\rho$ at $13$ TeV the uncertainty regions within T and T + A are essentially the same, lying far above the experimental data and error bars. Analogous behaviors can be seen in Figures \ref{ch4fig2}, \ref{ch4fig3} and \ref{ch4fig4}. This is a consequence of the large number of experimental data on $\sigma_{tot}$ at the highest energies (mainly LHC region) as compared with those respect to $\rho$.

\item[ii. ]
From Tables \ref{ch4t1} and \ref{ch4t2}, in all cases, independently of the ensemble or model, for $\nu \sim 250$, the $\chi^2/\nu$ lies in the region $\sim$ $1.2$ - $1.3$ and the integrated probability  $P(\chi^2) \sim 10^{-2} - 10^{-3}$. Taking into account the discrepant values between TOTEM and ATLAS data, the fits can be considered as reasonably accurate.

\item[iii. ]
For models I, II and III the integrated probability $P(\chi^2)$ is one order of magnitude smaller within ensemble T + A than within T and for model IV nearly $1/2$. This is a consequence of the aforementioned tension between the TOTEM and ATLAS data at $7$ and mainly $8$ TeV.

\end{itemize}

\subsection{\textsc{Models}}
\mbox{\,\,\,\,\,\,\,\,\,}
First, notice that from the figures and within the uncertainties, all models present quite good descriptions of the experimental data up to $7$ TeV, as expected. Therefore, let us focus the discussion in the region $8$ - $13$ TeV (mainly $13$ TeV) and in the goodness of the fits.

\begin{itemize}

\item[i. ] Model I (DL-type)

The fit result in Figure \ref{ch4fig1} is in plenty agreement with the $\sigma_{tot}$ datum at $13$ TeV within ensemble T and the uncertainty region crosses the lower error bar in case of ensemble T + A. However, for $\rho$ the curves do not decrease in the region $10^3 - 10^4$ GeV (see insets) and even with $2$ $\sigma$ the results at $13$ TeV lie far above the upper error bars. Within both ensembles the integrated probability is the smallest among the models ($10^{-3} - 10^{-4}$). We conclude that this model is not  in agreement with the TOTEM data at $13$ TeV.

\item[ii. ] Model II (COMPETE-type)

From Figure \ref{ch4fig2} and ensemble T, the fit result (uncertainty region) for $\sigma_{tot}$ at $13$ TeV crosses the central value and the lower error bar and reaches half this bar within ensemble T + A. For $\rho$, the curves decrease in the region $10^3 - 10^4$ GeV, but as in the previous case, the uncertainty regions lie far above the upper error bars (insets). We conclude this model does not present a satisfactory description of the new data at $13$ TeV.

\item[iii. ] Model III (BH-type)

From the tables, the integrated probability is one of the highest among the models. From Figure \ref{ch4fig3} for $\sigma_{tot}$ and ensemble T + A, the uncertainty region with $1$ $\sigma$ reaches the upper error bar of the ATLAS datum at 8 TeV and the lower bar  of the TOTEM datum at $13$ TeV (similar with $2$ $\sigma$ in case of ensemble T). For $\rho$ the curves present the faster decrease among the models in the region $10^3 - 10^4$ GeV (insets) and at $13$ TeV, with $2$ $\sigma$, the uncertainty region reaches the upper extremum of the error bar with ensemble T + A (barely reach this point with ensemble T). We understand that this model is not excluded by the bulk of experimental data presently available.

\item[iv. ] Model IV (Hybrid power-log)

Based on the disagreement of Model I with the TOTEM data at $13$ TeV and given the efficiency of the power law (simple pole Pomeron) below $13$ TeV, we have tested hybrid contributions by adding either a double pole or triple pole contributions. In the latter case the fits did not converge and in the former case the fit results are presented in Tables I and II and Figure \ref{ch4fig4}. In this case we have one more parameter (as compared with $7$ parameters in the other $3$ models), resulting in lager uncertainty regions. For $\sigma_{tot}$ the uncertainty regions with $1$ $\sigma$ encompass all the experimental data at the LHC energy region. However, from the tables the integrated probabilities are the smallest among the models  and although the results for $\rho$ (Figure \ref{ch4fig4}) present a small decrease in the region $10^3 - 10^4$ GeV, the uncertainty regions lie far above the TOTEM data. We conclude that the model does not present agreement with the TOTEM data at $13$ TeV. 

\end{itemize}

Based on the above discussion, we understand that models I, II and IV are not able to describe simultaneously the TOTEM data on $\sigma_{tot}$ and $\rho$ at 13 TeV. On the other hand, taking into account the bulk of experimental data presently available (ensemble T + A) and the uncertainties in both theoretical and experimental results, Model III seems not to be excluded.

\subsection{\textsc{Further Tests}}
\mbox{\,\,\,\,\,\,\,\,\,}
Looking for possible improvements in the efficiency of Model III, we have also developed further investigations. Here, in all fits we have considered the energy cutoff at $\sqrt{s}_{min} = 5$ GeV and the subtraction constant as a free fit parameter. In order to investigate the effect of the energy cutoff and the role of the subtraction constant  we have also carried out fits without this parameter, namely by fixing $K = 0$ and rising the energy cutoff to $7.5$ and $10$ GeV. 

Firstly, still with the subtraction constant as a free fit parameter, we develop fits with energy cutoff at $7.5$ and $10$ GeV. The results are displayed in Table \ref{ch4t3} and Figures \ref{ch4fig6} and \ref{ch4fig7}. In a second step the subtraction constant is fixed at zero and the fits are developed with energy cutoff at $5$, $7.5$ and $10$ GeV. The results are shown in Table \ref{ch4t3} and Figures \ref{ch4fig8}, \ref{ch4fig9} and \ref{ch4fig10}. As before, in all the cases we employ ensembles T and T + A and CL with one and two standard deviations.

For K as a free fit parameter, comparison of Table \ref{ch4t1} (cutoff at $5$ GeV) with Table \ref{ch4t3} (cutoffs at $7.5$ and $10$ GeV), shows that for both ensembles, rising the cutoff results in a slightly increase in $P(\chi^2)$ and from Figures \ref{ch4fig3}, \ref{ch4fig6} and \ref{ch4fig7}, the uncertainty regions become larger, mainly at lower energies. The same effect is observed by fixing $K = 0$  (Table \ref{ch4t4} and Figures \ref{ch4fig8}, \ref{ch4fig9} and \ref{ch4fig10}). The rise of the cutoff does not led to an improvement in the fit results, within the uncertainty region, at $13$ TeV.

For cutoff at $5$ GeV, the results with $K$ free (Table \ref{ch4t1}, Figure \ref{ch4fig3}) and $K=0$ fixed (Table \ref{ch4t4}, Figure \ref{ch4fig8}) show the following features:

\begin{itemize}

\item[i. ]
within both ensembles, the integrated probability is slightly larger for $K$ free;

\item[ii. ]
for $\rho$ at 13 TeV and ensemble T, the distance between the minimum of the uncertainty region and the extreme of the upper error bar is smaller with $K$ free than with $K=0$;

\item[iii. ]
for $\rho$ at $13$ TeV, ensemble T + A and $K=0$, the uncertainty region lies slightly above the extreme of the upper error bar (Figure \ref{ch4fig8}) and for $K$ free the uncertainty region reaches this point (Figure \ref{ch4fig3}).

\end{itemize}

In conclusion, the rising of the cutoff does not lead to improvements in the fit results, neither fixing $K=0$. The results with $K$ free and cutoff at $5$ GeV present best agreement with the TOTEM data at $13$ TeV.

Taking into account the energy region analyzed, $5$ GeV - $13$ TeV and the $pp$ and $\bar{p}p$ scattering, we did not find remarkable or considerable improvements. Indeed, with the cutoff at $5$ GeV, the results with and without the subtraction constant are similar, with integrated probability slightly greater in case of $K$ free and the uncertainty region reaching the extreme of the upper error bar of $\rho$ at $13$ TeV (ensemble T + A).

Therefore, we select as our best result those obtained here with model III, cutoff at $5$ GeV and the subtraction constant as a free fit parameter (Figure \ref{ch4fig3} and Tables \ref{ch4t1} and \ref{ch4t2}). For this case we present in Figure \ref{ch4fig5} a detail of the predictions for  $\sigma_{tot}$ and $\rho$ at $13$ TeV and the experimental data; the numerical values are given in Table \ref{ch4t5}, together with the corresponding predictions at $14$ TeV and uncertainties associated with $1$ $\sigma$ and also $2$ $\sigma$.

\section{\textsc{Conclusions on Pomeron Models}}
\mbox{\,\,\,\,\,\,\,\,\,}
We have presented a forward amplitude analysis on the experimental data presently available from $pp$ and $\bar{p}p$ scattering in the energy region $5$ GeV - $13$ TeV. The analysis consists of tests with different analytic parameterizations for $\sigma_{tot}(s)$ and $\rho(s)$, all of them characterized by Pomeron leading contributions (even-under-crossing). The data reductions show that most models present no simultaneous agreement with the recent $\sigma_{tot}$ and $\rho$ measurements at $13$ TeV by the TOTEM Collaboration. Different models and variants have been tested and among them, Model III (two simple poles Reggeons, one double pole and one triple pole Pomerons), with only seven free fit parameters, led to the best results. 

Two aspects have been stressed along this present work. The first concerns the TOTEM results at $13$ TeV, indicating an expected rise of the total cross section but an unexpected decrease in the value of the $\rho$ parameter. The extrapolation from the recent analysis with data up to $8$ TeV shows clearly the plenty agreement with the $\sigma_{tot}$ result and the overestimation of the $\rho$ data \cite{Fagundes:2017cmp}. Notice also that the value here obtained for the Pomeron intercept with Model I and Ensemble T, $\epsilon = 0.0914 \pm 0.0039$ (Table \ref{ch4t1}) is consistent with results of fits up to $8$ TeV, for example those obtained in Reference \cite{Menon:2013vka}: $\epsilon = 0.0926 \pm 0.0016$. However, the Model I result for $\rho$ at $13$ TeV is in complete disagreement with the TOTEM data (Figure \ref{ch4fig1}). 

The second aspect concerns the tension between the TOTEM and ATLAS data at $7$ TeV and mainly at $8$ TeV, discussed in certain detail in the previous sections. That led us to consider separately the two ensembles denoted T (excluding the ATLAS data) and T + A (including the ATLAS data). We have shown that these discrepancies play an important role in the interpretations of the fit results.

Another aspect deserves attention when interpreting the data reductions. As discussed in Reference \cite{Fagundes:2017iwb}, the TOTEM uncertainties are essentially systematics (uniform distribution) and not statistical (Gaussian distribution). Therefore, a model result crossing the central value of an experimental result may have a limited significance on statistical grounds.

The unexpected decrease in the $\rho$ value has been well described in the recent analyses by Martynov and Nicolescu. The first paper treated only the TOTEM data \cite{Martynov:2017zjz} and in the second one the ATLAS data have been included \cite{Martynov:2018nyb}. The $\chi^2/\nu$ are similar in both cases, namely $1.075$ without ATLAS and $1.100$ including ATLAS, corresponding to an increase of $2.3$\%. For $\rho$ at $13$ TeV, in both cases the curves seems to cross the central value of the experimental points. However, for $\sigma_{tot}$ with ATLAS excluded the curve crosses the lower error bar at $13$ TeV, but lies above the error bars of the ATLAS data at $7$ TeV and mainly $8$ TeV. With ATLAS included, the curve crosses the ATLAS data, but lies below the lower error bar of the TOTEM data at $8$ TeV and mainly $13$ TeV. Summarizing, the curve does not reach the upper error bars of the ATLAS data on $\sigma_{tot}$ at $7$ and $8$ TeV in the former case and does not reach the lower error bar of the TOTEM datum at $13$ TeV in the latter case.

In what concerns our results with Model III, the $\chi^2/\nu$ are also similar in both cases: $1.210$ (T) and $1.234$ (T + A), corresponding to an increase slightly small, $2.0$\%. The uncertainty regions of the fit results do not cross the central values of the $\rho$ data at $13$ TeV, but barely reach the upper error bar. However, the same is true for the ATLAS datum on $\sigma_{tot}$ at $8$ TeV. Therefore, we conclude that the agreement between the phenomenological model and the experimental points is reasonably compatible within the uncertainties. In other words, in case of fits to ensembles T or T + A (all the experimental data presently available) and within the uncertainties, the Pomeron Model III, with $7$ free fit parameters, seems not to be excluded by the experimental data presently available on forward $pp$ and $\bar{p}p$ elastic scattering.

The Odderon is a well-founded concept in perturbative QCD \cite{Braun:1998fs,Ewerz:2003xi,Ewerz:2005rg}. However, it is not clear if this formulation can be directly extended to the extreme infrared region where $\sigma_{tot}$ and $\rho$ are defined, namely $t \rightarrow 0$, or distinct and independent objects, such as soft and hard Odderons, should be considered. In this infrared extreme, despite the consistent description of the unexpected decrease of the $\rho$ parameter at $13$ TeV, the Odderon model predicts a crossing in the $pp$ and $\bar{p}p$ total cross sections at high energies. Although in agreement with high-energy theorems \cite{Grunberg:1973mc}, it seems still lacking a pure (model independent) nonperturbative QCD explanation (from first principles) for an asymptotic rise of the total cross section faster for hadron-hadron than for antihadron-hadron collisions.

Finally, we understand that further re-analysis and new experimental data at $13$ TeV and $14$ TeV, by the TOTEM and ATLAS collaborations, shall be crucial for confronting, in a conclusive way, the possible dominance of Odderon or Pomeron in forward elastic hadron scattering at high energies.

\begin{table}[h]
\centering
\scalebox{0.9}{
\begin{tabular}{c@{\quad}c@{\quad}c@{\quad}c@{\quad}c@{\quad}}
\hline \hline
& & &  \\[-0.4cm]
Model:     & I  & II  & III  & IV    \\[0.05ex]
\hline
& & & &\\[-0.4cm]
$a_{1}$ [mb]        & 41.4\,$\pm$\,1.8       & 32.2\,$\pm$\,1.8       & 58.8\,$\pm$\,1.5       & 51.5\,$\pm$\,7.1      \\[0.05ex]
$b_{1}$             & 0.378\,$\pm$\,0.028    & 0.392\,$\pm$\,0.049    & 0.229\,$\pm$\,0.017    & 0.296\,$\pm$\,0.037   \\[0.05ex]
$a_{2}$ [mb]        & 17.0\,$\pm$\,2.0       & 17.0\,$\pm$\,2.1       & 16.9\,$\pm$\,2.0       & 17.0\,$\pm$\,2.1      \\[0.05ex]
$b_{2}$             & 0.545\,$\pm$\,0.037    & 0.545\,$\pm$\,0.037    & 0.543\,$\pm$\,0.036    & 0.544\,$\pm$\,0.037   \\[0.05ex]
$A$ [mb]            & -                      & 29.6\,$\pm$\,1.2       & -                      & -                     \\[0.05ex]
$B$ [mb]            & 21.62\,$\pm$\,0.73     & -                      & -                      & 9.6\,$\pm$\,7.5       \\[0.05ex]
$\epsilon$          & 0.0914\,$\pm$\,0.0030  & -                      & -                      & 0.108\,$\pm$\,0.019   \\[0.05ex]
$C$ [mb]            & -                      & -                      & 3.67\,$\pm$\,0.34      & 2.4\,$\pm$\,1.6       \\[0.05ex]
$D$ [mb]            & -                      & 0.251\,$\pm$\,0.010    & 0.132\,$\pm$\,0.024    & - \\[0.05ex]
$K$ [mbGeV$^{2}$]   & 69\,$\pm$\,47          & 55\,$\pm$\,50          & 20\,$\pm$\,44          & 45\,$\pm$\,47         \\[0.05ex]
\hline
& & & & \\[-0.4cm]
$\nu$               & 248                    & 248                    & 248                    & 247                   \\[0.05ex]
$\chi^2/\nu$        & 1.273                  & 1.193                  & 1.210                  & 1.249                 \\[0.05ex]
$P(\chi^2)$         & 2.3 $\times$ 10$^{-3}$ & 2.0 $\times$ 10$^{-2}$ & 1.4 $\times$ 10$^{-2}$ & 4.8 $\times$ 10$^{-3}$\\[0.05ex]
\hline
& & & & \\[-0.4cm] 
Figure:            & \ref{ch4fig1}          & \ref{ch4fig2}          & \ref{ch4fig3}           & \ref{ch4fig4}         \\[0.05ex]
\hline \hline 
\end{tabular}
}
\caption{Values of the fitted parameters from ensemble T through models I - IV, by considering one standard deviation, energy cutoff at $5$ GeV and $K$ as a free fit parameter.}
\label{ch4t1}
\end{table}

\begin{table}[h]
\centering
\scalebox{0.9}{
\begin{tabular}{c@{\quad}c@{\quad}c@{\quad}c@{\quad}c@{\quad}}
\hline \hline
& & &  \\[-0.4cm]
Model:     & I  & II  & III  & IV    \\[0.05ex]
\hline
& & & &\\[-0.4cm]
$a_{1}$ [mb]        & 41.4\,$\pm$\,1.8       & 32.3\,$\pm$\,2.0       & 59.1\,$\pm$\,1.5        & 53.1\,$\pm$\,9.6     \\[0.05ex]
$b_{1}$             & 0.386\,$\pm$\,0.028    & 0.412\,$\pm$\,0.045    & 0.234\,$\pm$\,0.016     & 0.291\,$\pm$\,0.044  \\[0.05ex]
$a_{2}$ [mb]        & 17.0\,$\pm$\,2.1       & 17.0\,$\pm$\,2.0       & 16.9\,$\pm$\,2.0        & 17.0\,$\pm$\,2.1     \\[0.05ex]
$b_{2}$             & 0.545\,$\pm$\,0.037    & 0.545\,$\pm$\,0.036    & 0.543\,$\pm$\,0.036     & 0.544\,$\pm$\,0.038  \\[0.05ex]
$A$ [mb]            & -                      & 30.20\,$\pm$\,0.90     & -                       & -                    \\[0.05ex]
$B$ [mb]            & 22.01\,$\pm$\,0.64     & -                      & -                       & 8.0\,$\pm$\,10       \\[0.05ex]
$\epsilon$          & 0.0895\,$\pm$\,0.0024  & -                      & -                       & 0.110\,$\pm$\,0.033  \\[0.05ex]
$C$ [mb]            & -                      & -                      & 3.81\,$\pm$\,0.30       & 2.8\,$\pm$\,2.1      \\[0.05ex]
$D$ [mb]            & -                      & 0.244\,$\pm$\,0.077    & 0.119\,$\pm$\,0.020     & -                    \\[0.05ex]
$K$ [mbGeV$^{2}$]   & 73\,$\pm$\,48          & 64\,$\pm$\,50          & 23\,$\pm$\,43           & 46\,$\pm$\,48        \\[0.05ex]
\hline
& & & & \\[-0.4cm]
$\nu$               & 250                    & 250                    & 250                     & 249\\[0.05ex]
$\chi^2/\nu$        & 1.307                  & 1.227                  & 1.234                   & 1.273\\[0.05ex]
$P(\chi^2)$         & 7.9 $\times$ 10$^{-4}$ & 8.2 $\times$ 10$^{-3}$ & 6.9 $\times$ 10$^{-3}$  & 2.3 $\times$ 10$^{-3}$\\[0.05ex]
\hline
& & & & \\[-0.4cm]
Figure:             & \ref{ch4fig1}          & \ref{ch4fig2}          & \ref{ch4fig3}           & \ref{ch4fig4}         \\[0.05ex]
\hline \hline 
\end{tabular}
}
\caption{Values of the fitted parameters from ensemble T + A through models I - IV, by considering one standard deviation, energy cutoff at $5$ GeV and $K$ as a free fit parameter.}
\label{ch4t2}
\end{table}

\newpage

\bfg[hbtp]
  \begin{center}
    \includegraphics[width=8.0cm,height=8.0cm]{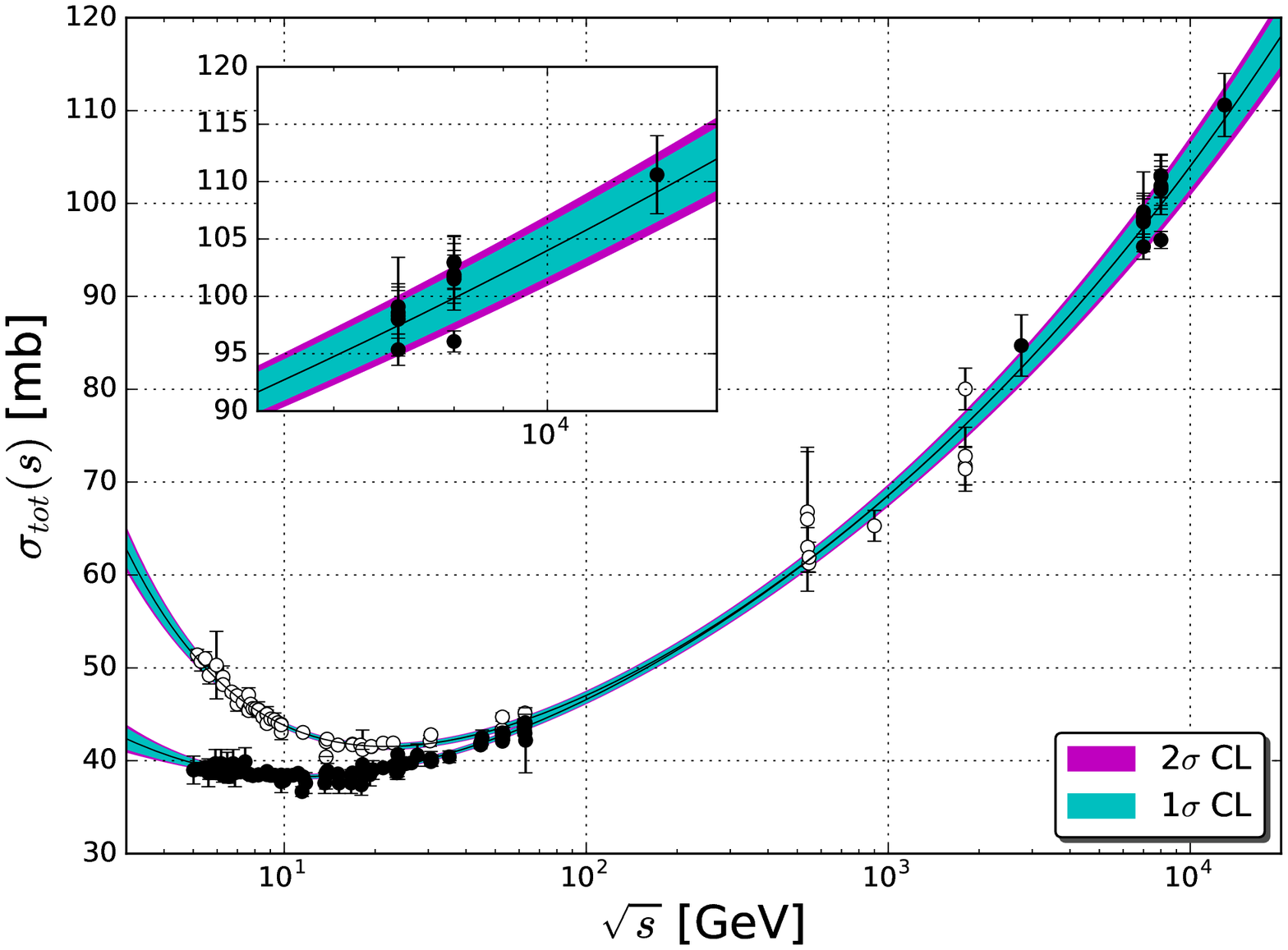}
    \includegraphics[width=8.0cm,height=8.0cm]{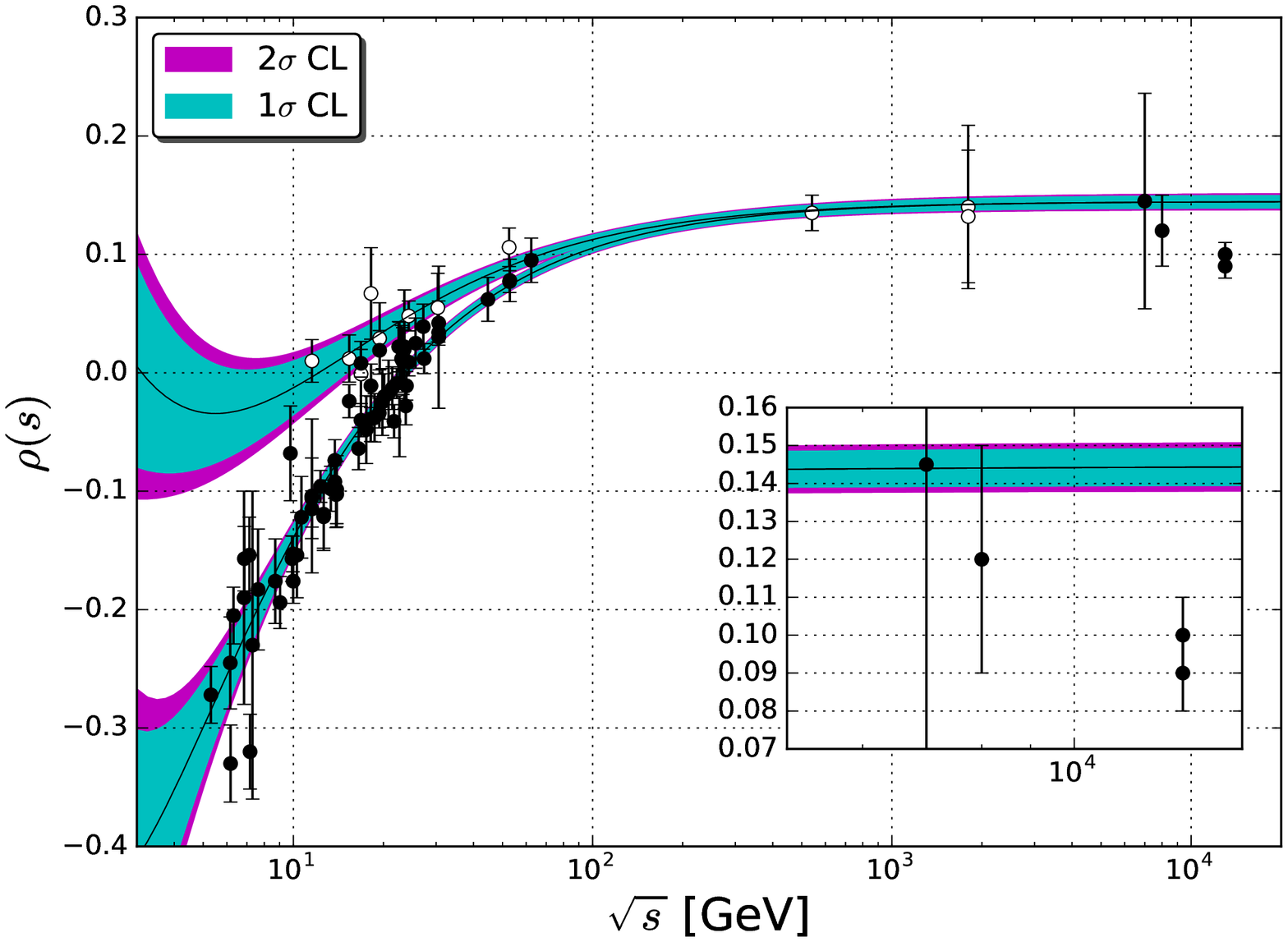}
    \includegraphics[width=8.0cm,height=8.0cm]{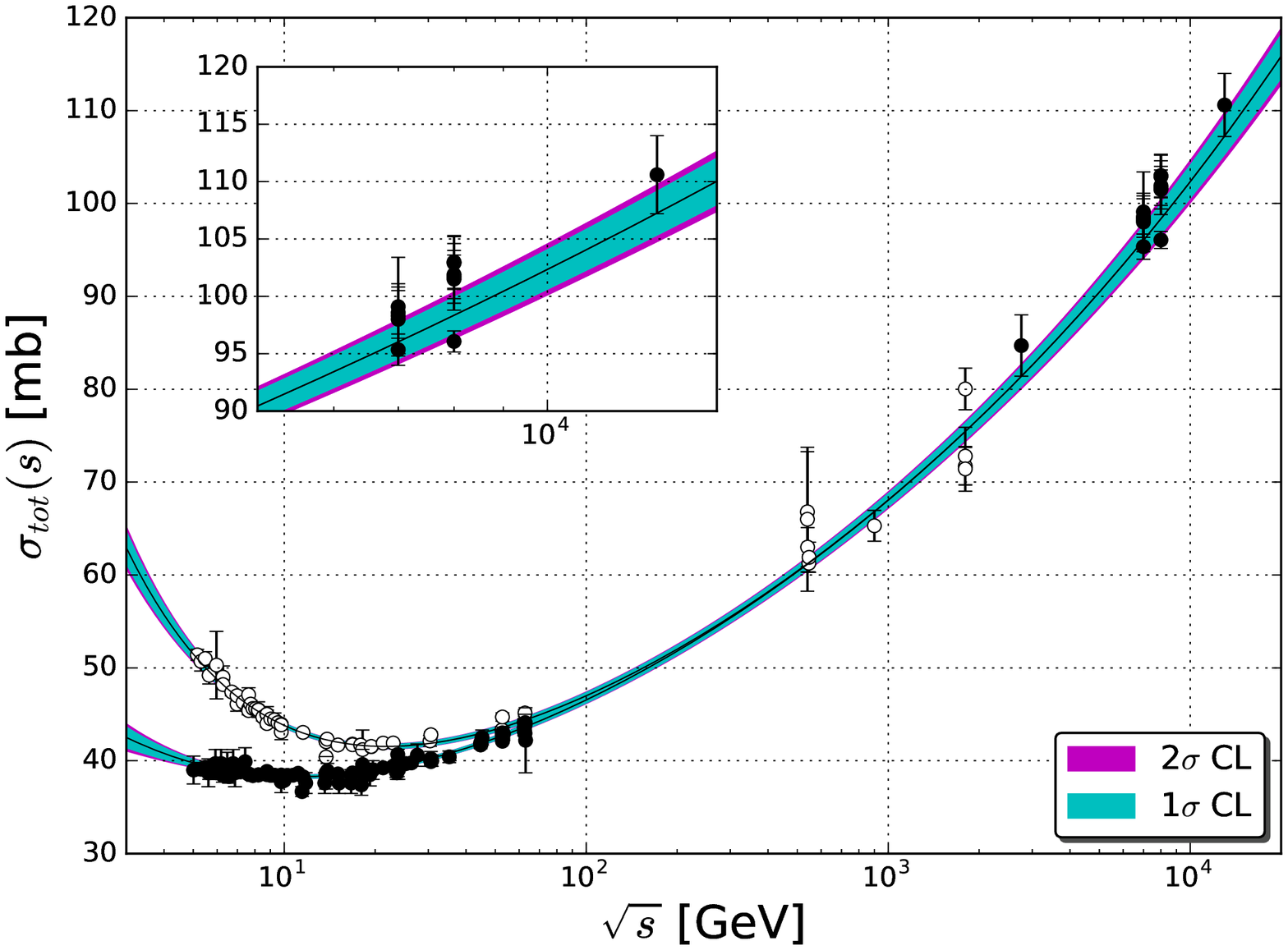}
    \includegraphics[width=8.0cm,height=8.0cm]{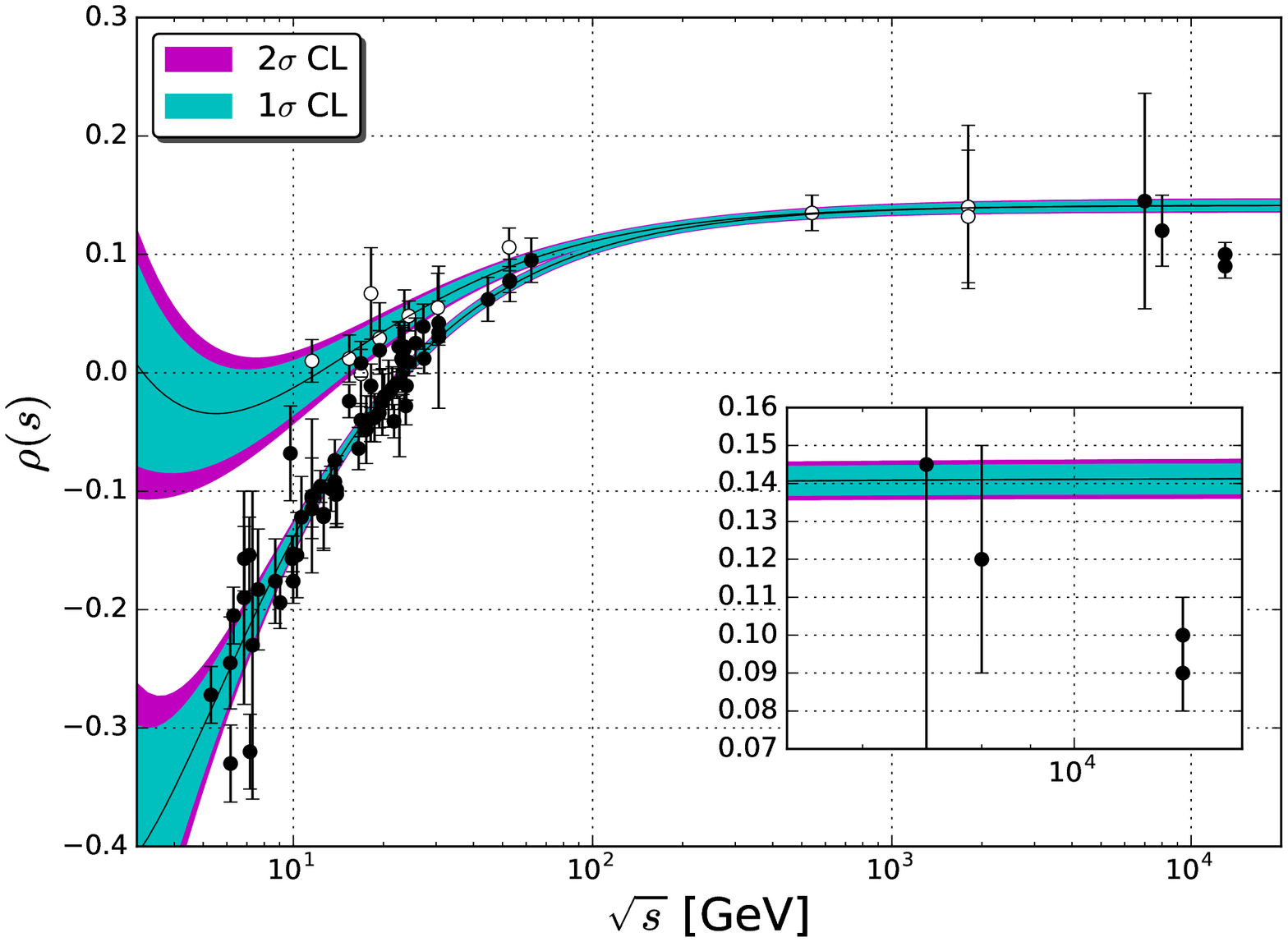}
    \caption{Fit results with Model I to ensembles T (above) and T+A (below).}
    \label{ch4fig1}
  \end{center}
\efg

\bfg[hbtp]
  \begin{center}
    \includegraphics[width=8.0cm,height=8.0cm]{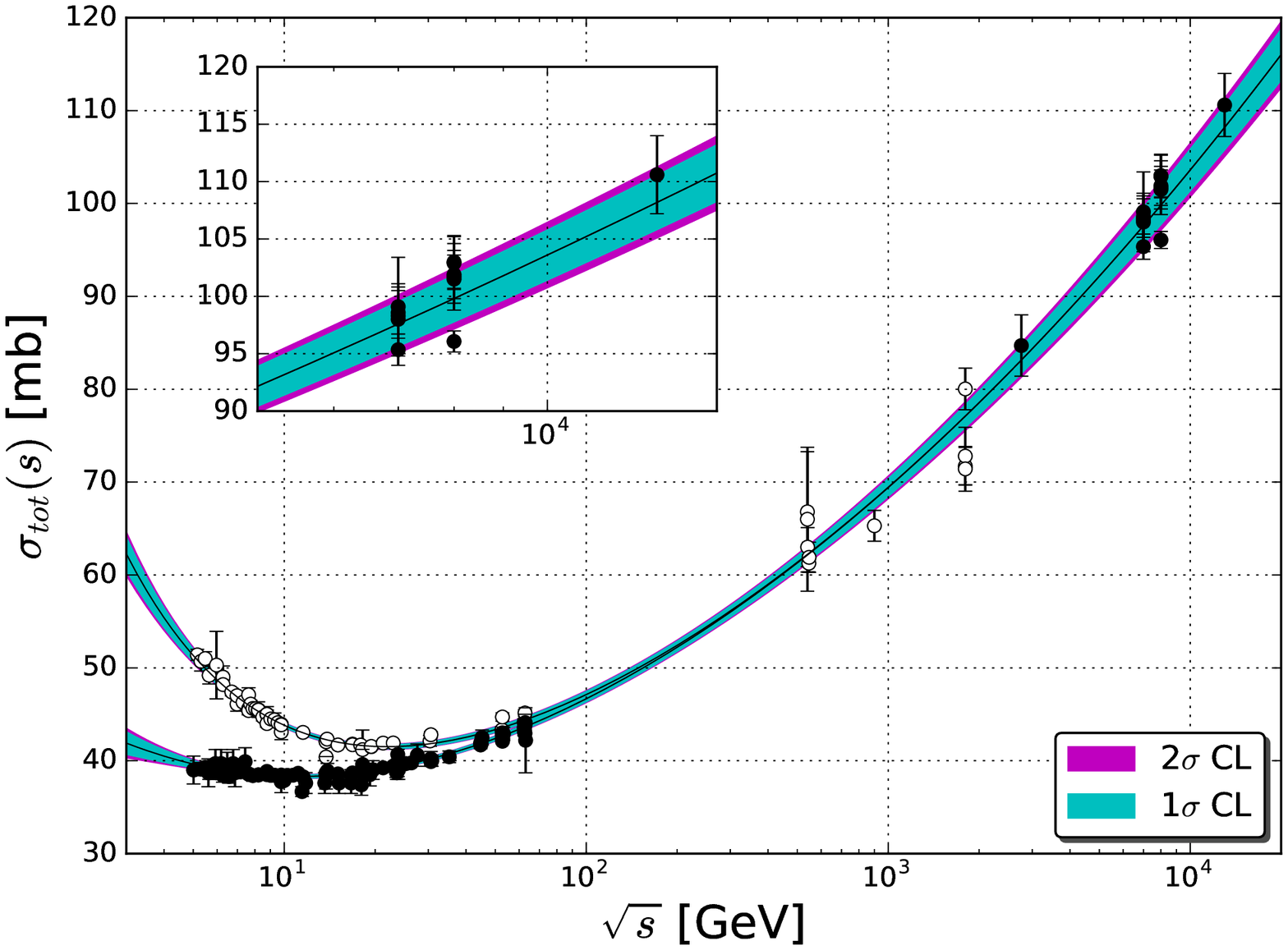}
    \includegraphics[width=8.0cm,height=8.0cm]{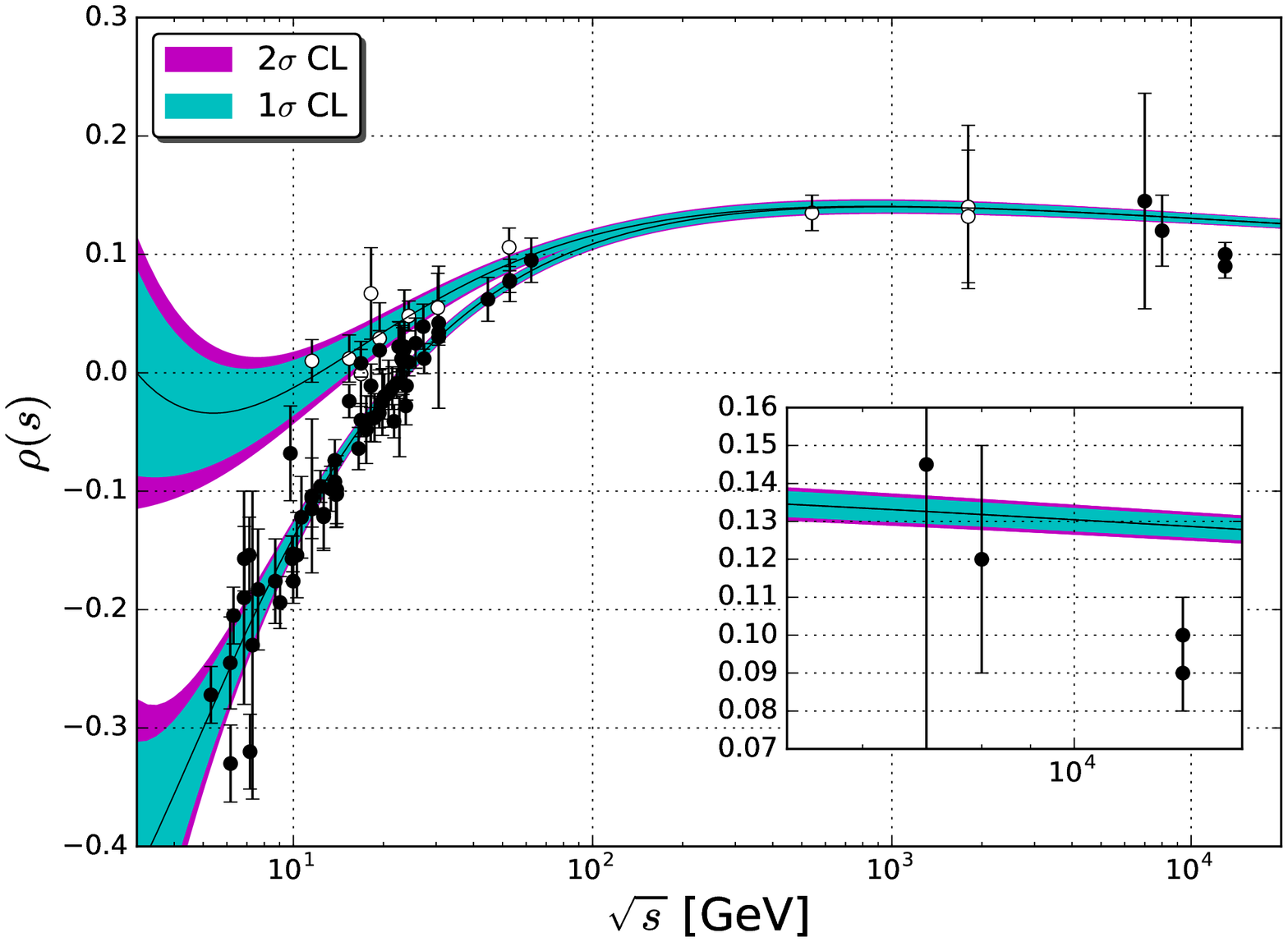}
    \includegraphics[width=8.0cm,height=8.0cm]{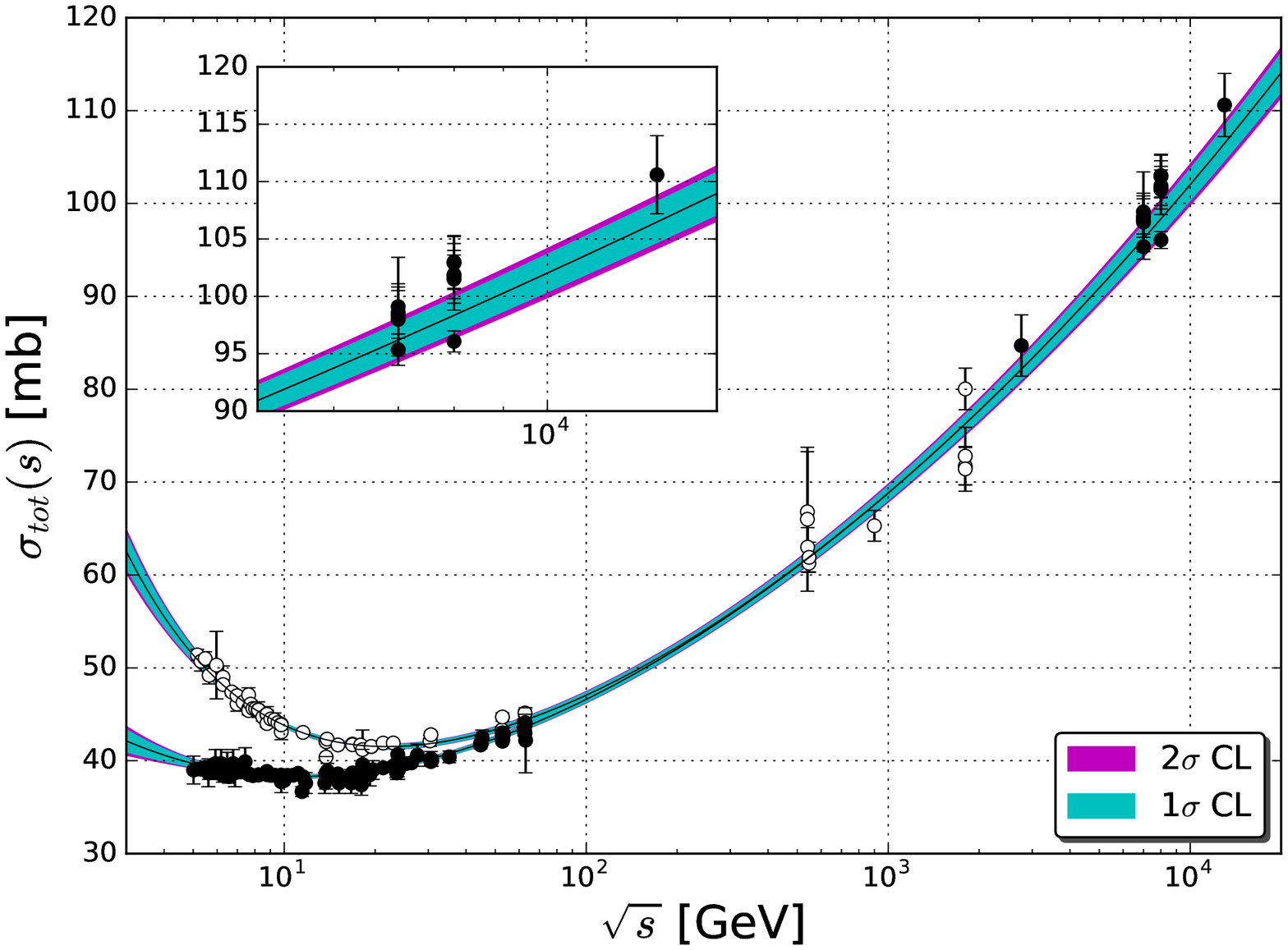}
    \includegraphics[width=8.0cm,height=8.0cm]{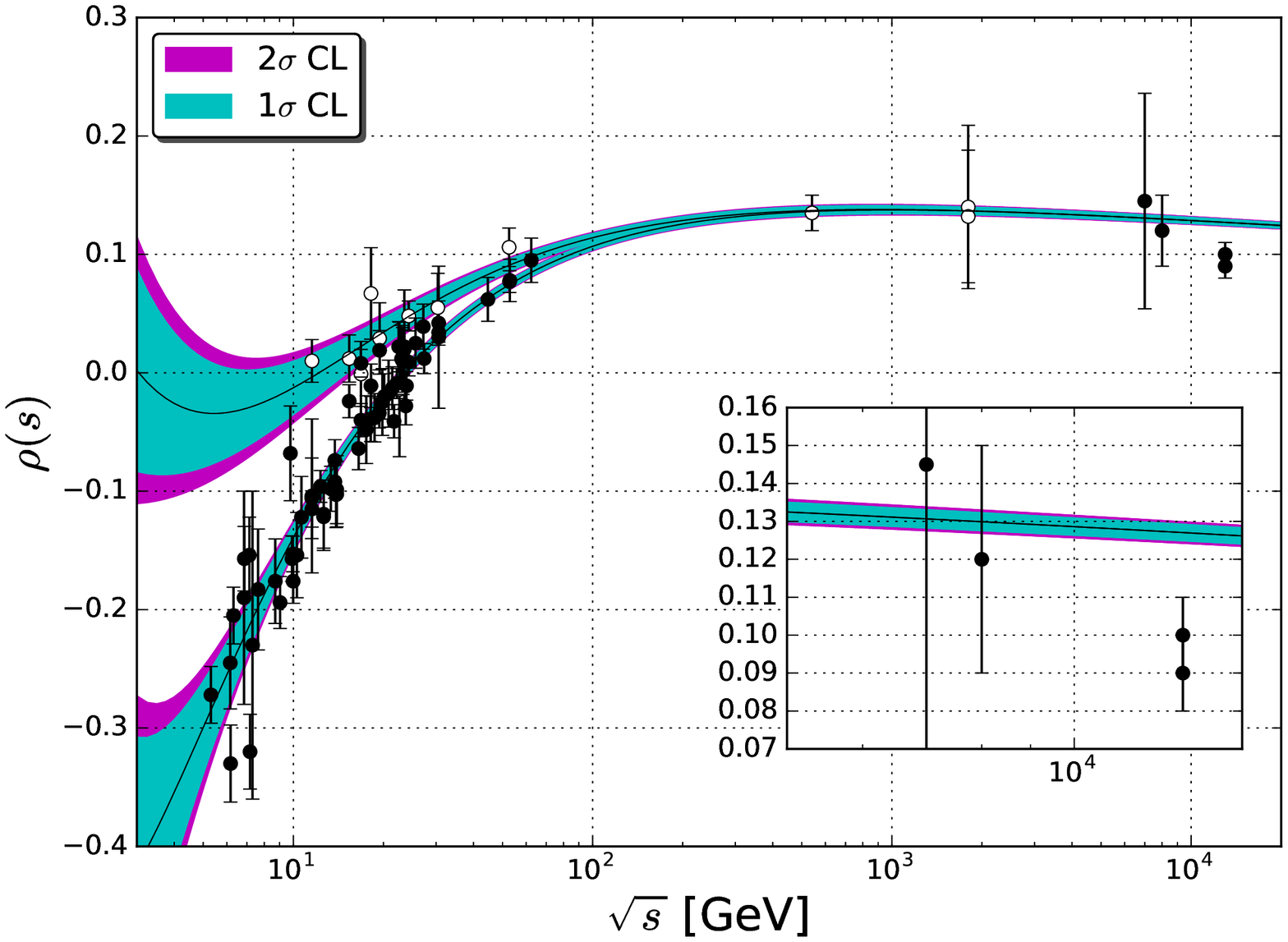}
    \caption{Fit results with Model II to ensembles T (above) and T + A (below).}
    \label{ch4fig2}
  \end{center}
\efg

\bfg[hbtp]
  \begin{center}
    \includegraphics[width=8.0cm,height=8.0cm]{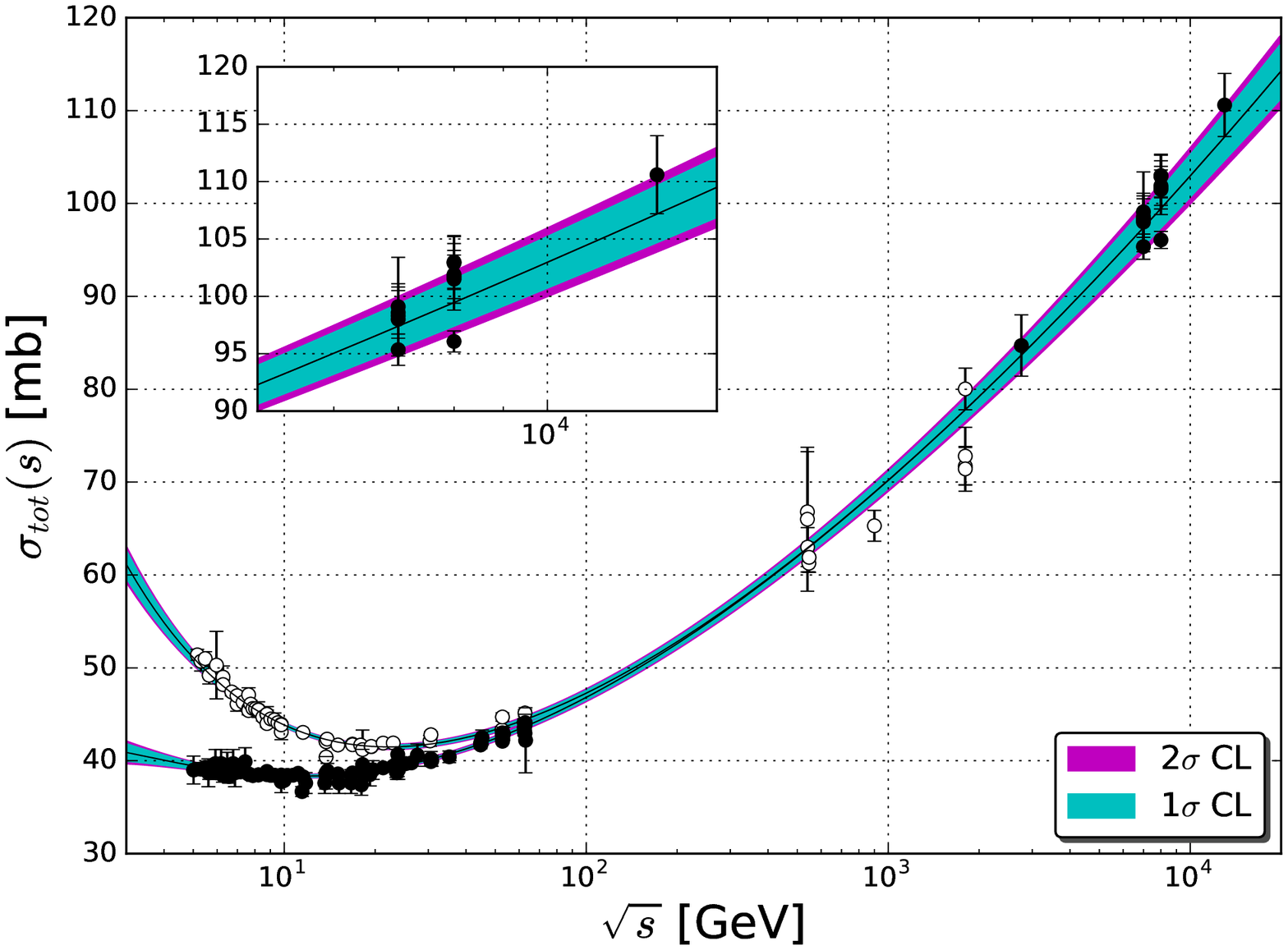}
    \includegraphics[width=8.0cm,height=8.0cm]{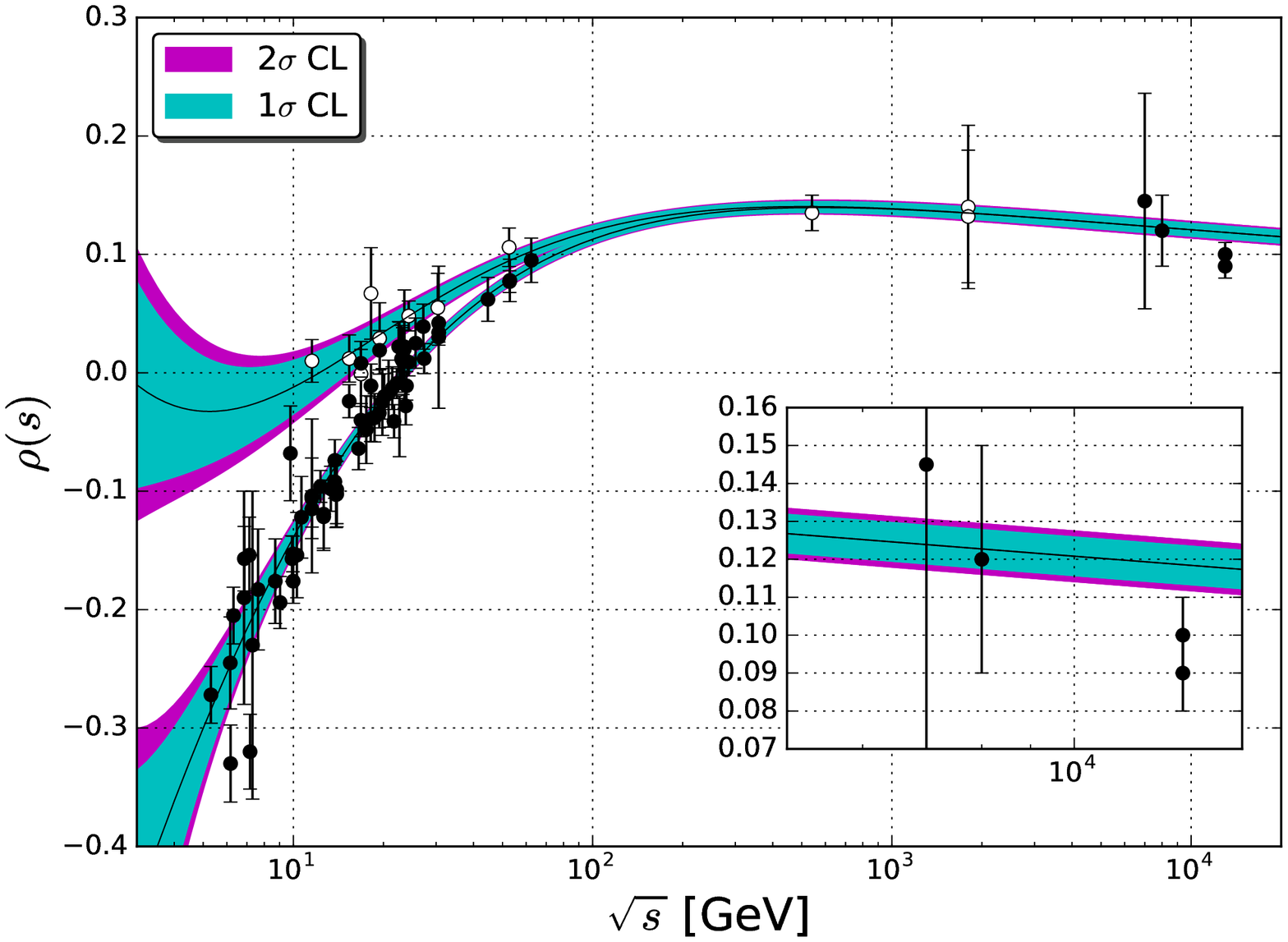}
    \includegraphics[width=8.0cm,height=8.0cm]{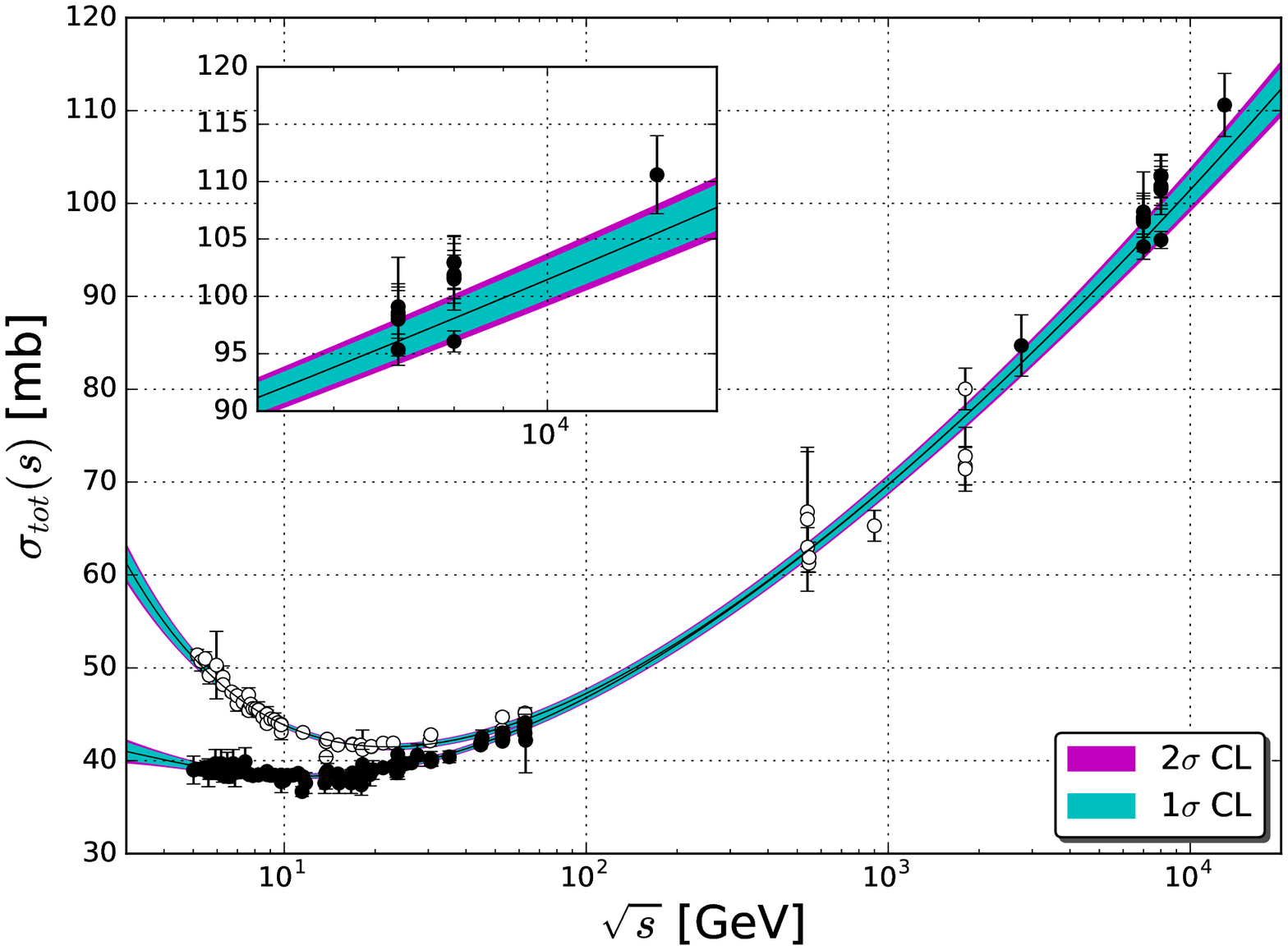}
    \includegraphics[width=8.0cm,height=8.0cm]{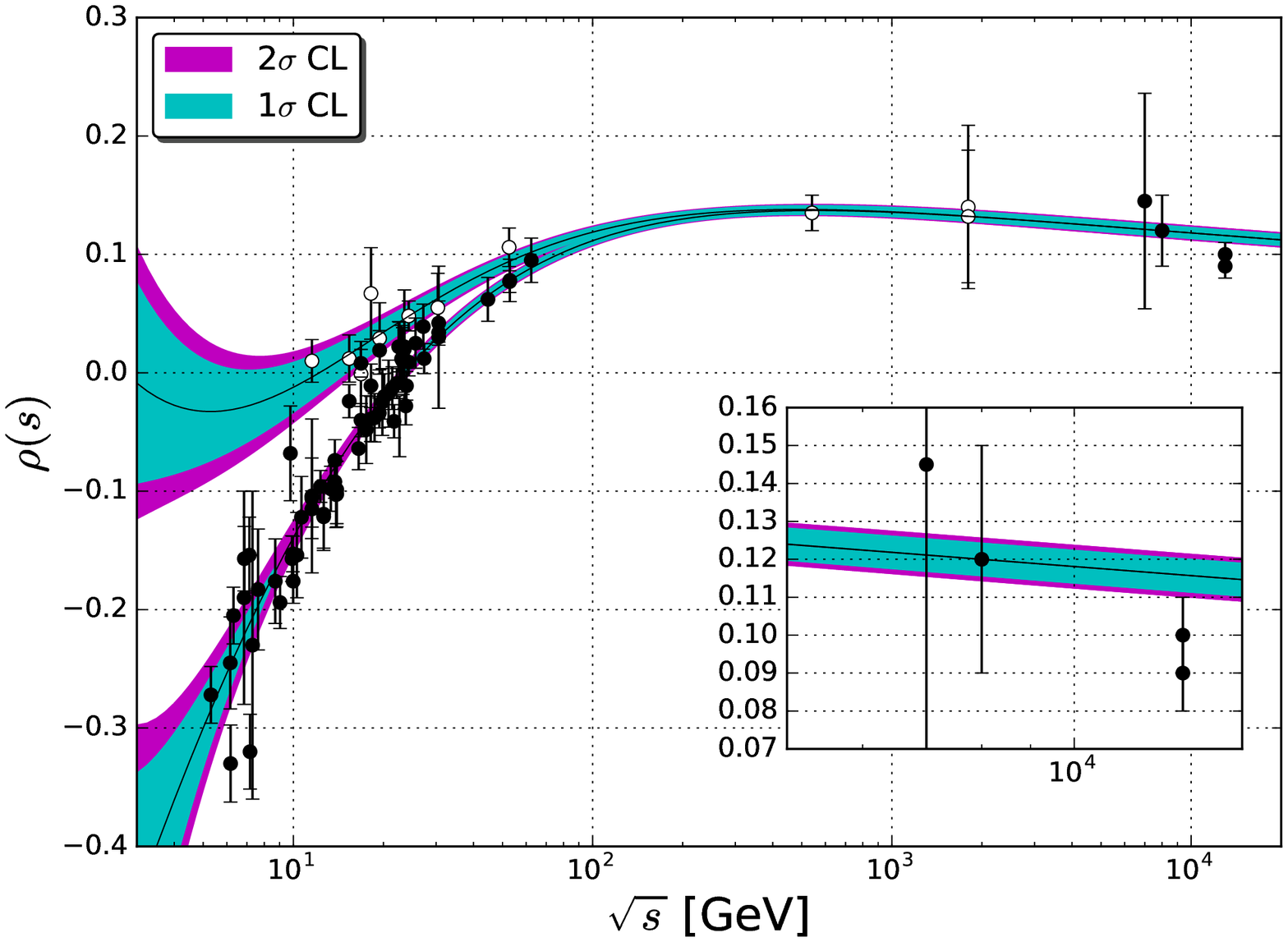}
    \caption{Fit results with Model III to ensembles T (above) and T + A (below).}
    \label{ch4fig3}
  \end{center}
\efg

\bfg[hbtp]
  \begin{center}
    \includegraphics[width=8.0cm,height=8.0cm]{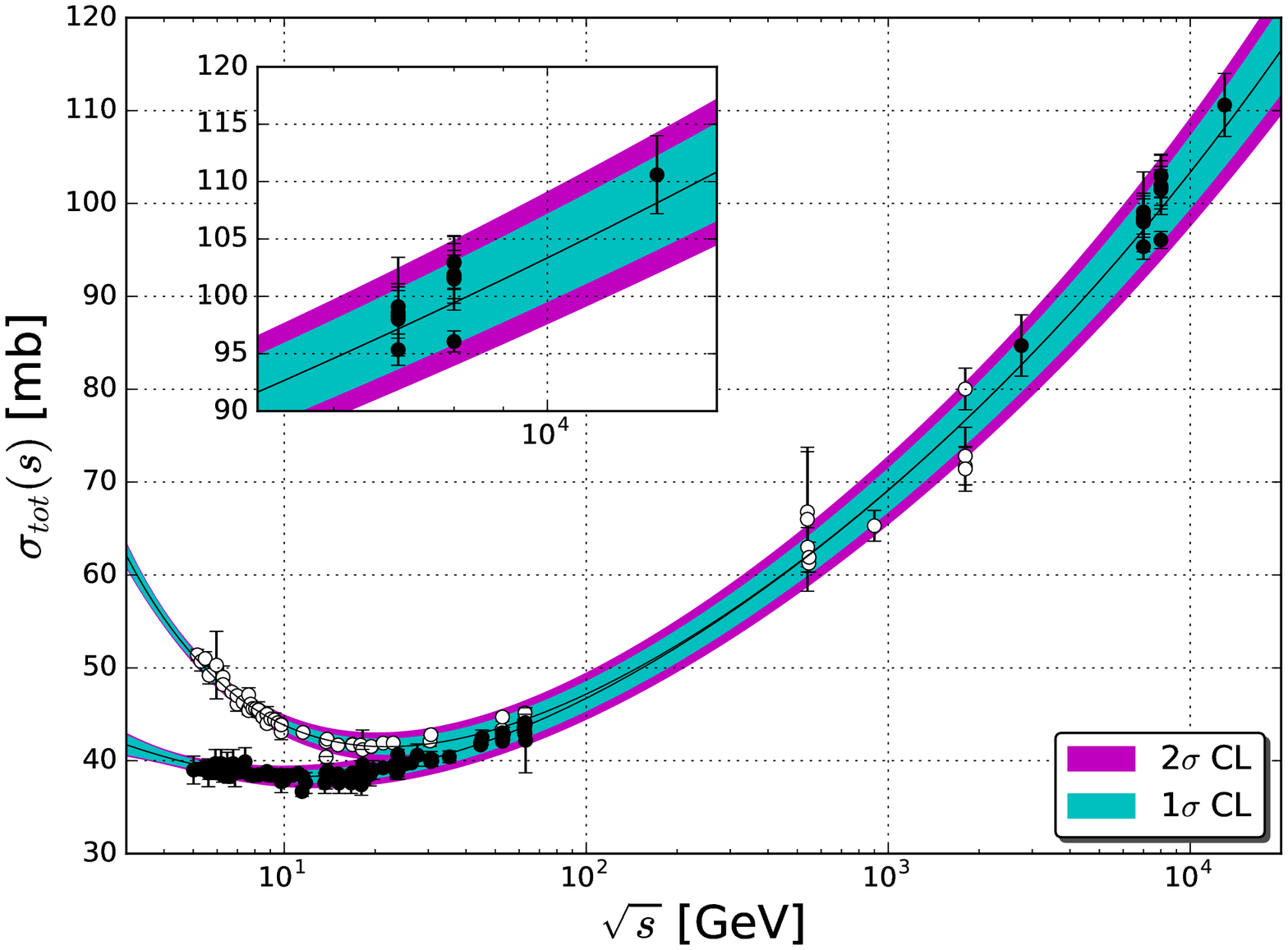}
    \includegraphics[width=8.0cm,height=8.0cm]{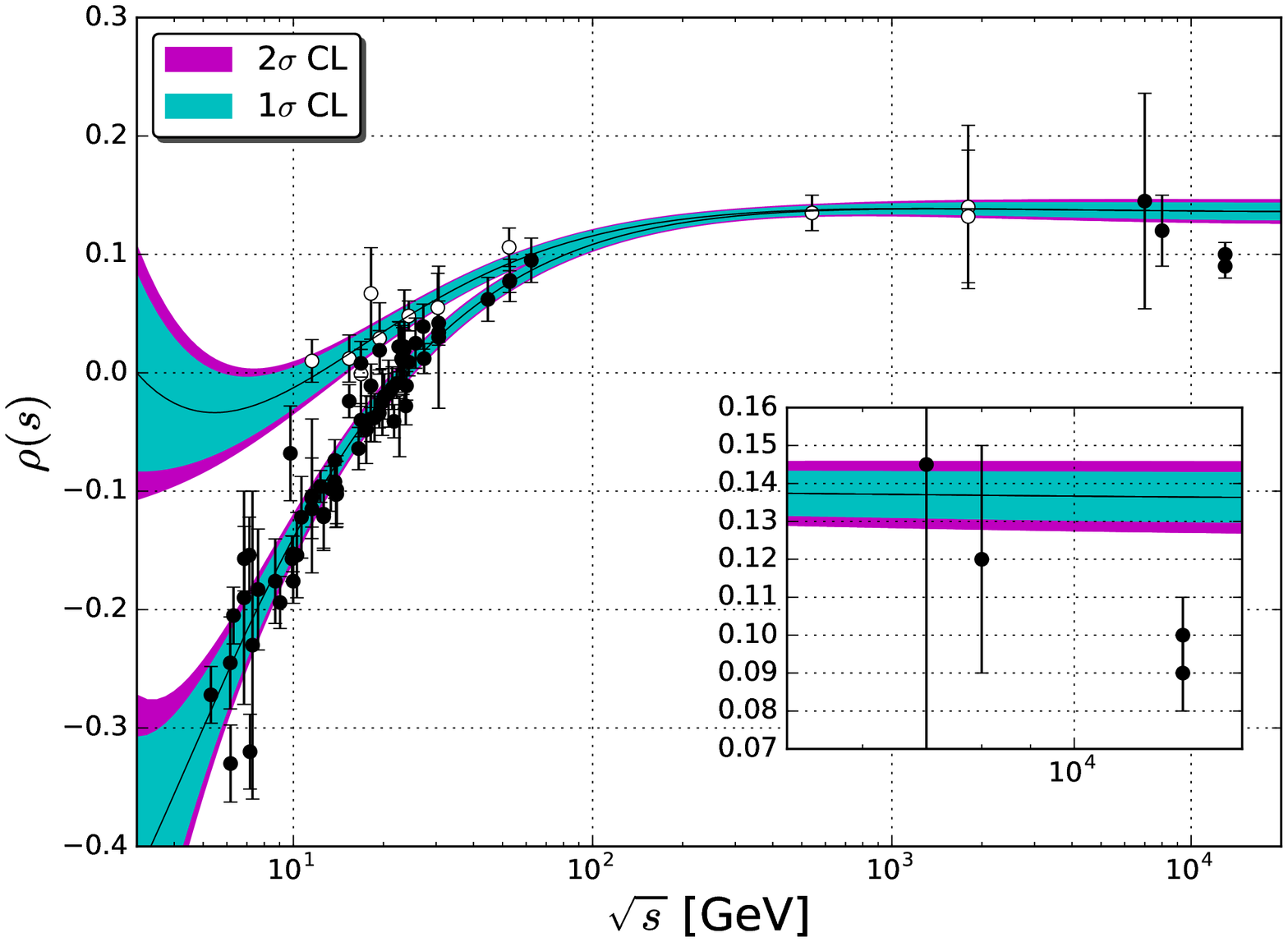}
    \includegraphics[width=8.0cm,height=8.0cm]{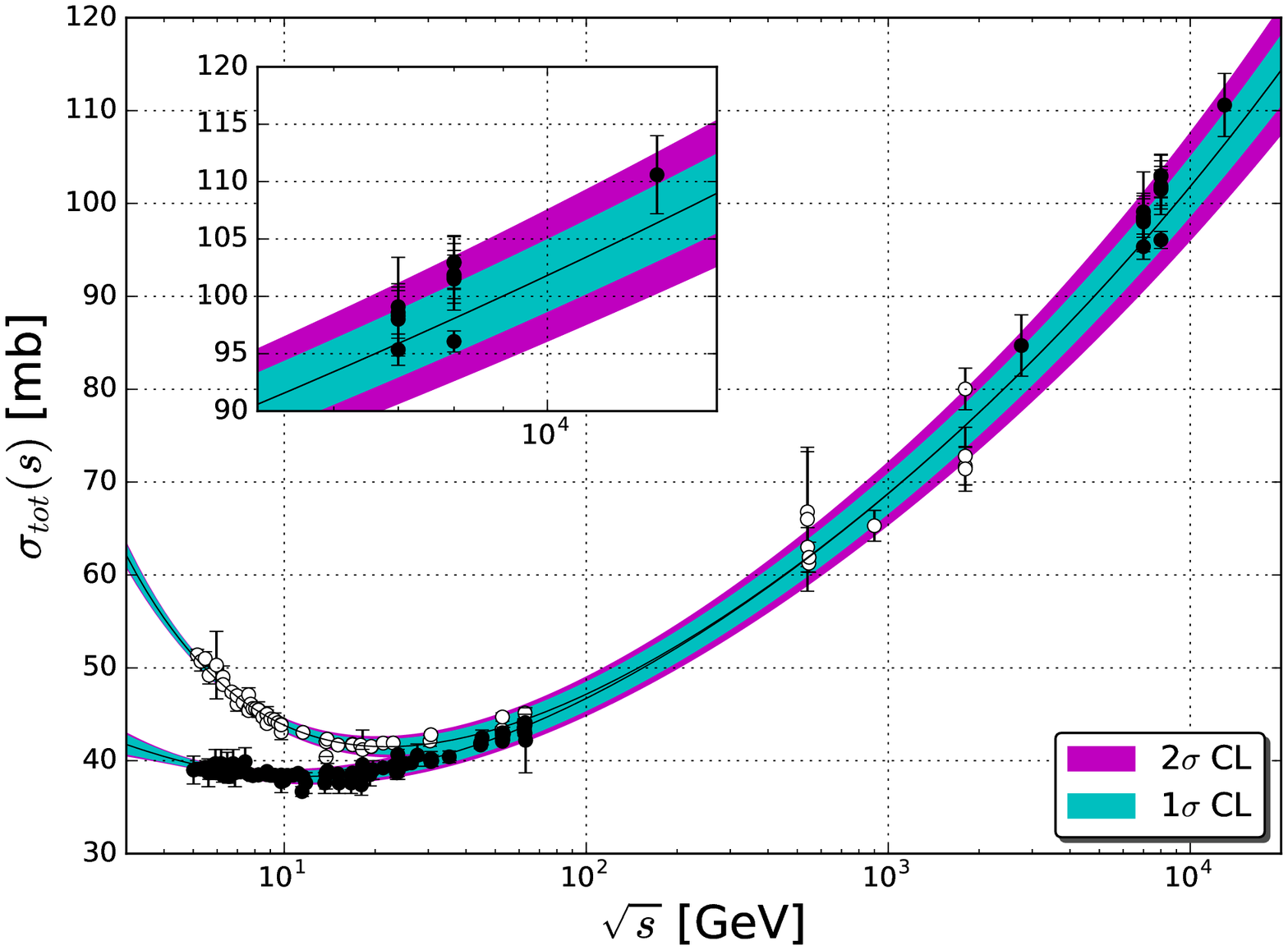}
    \includegraphics[width=8.0cm,height=8.0cm]{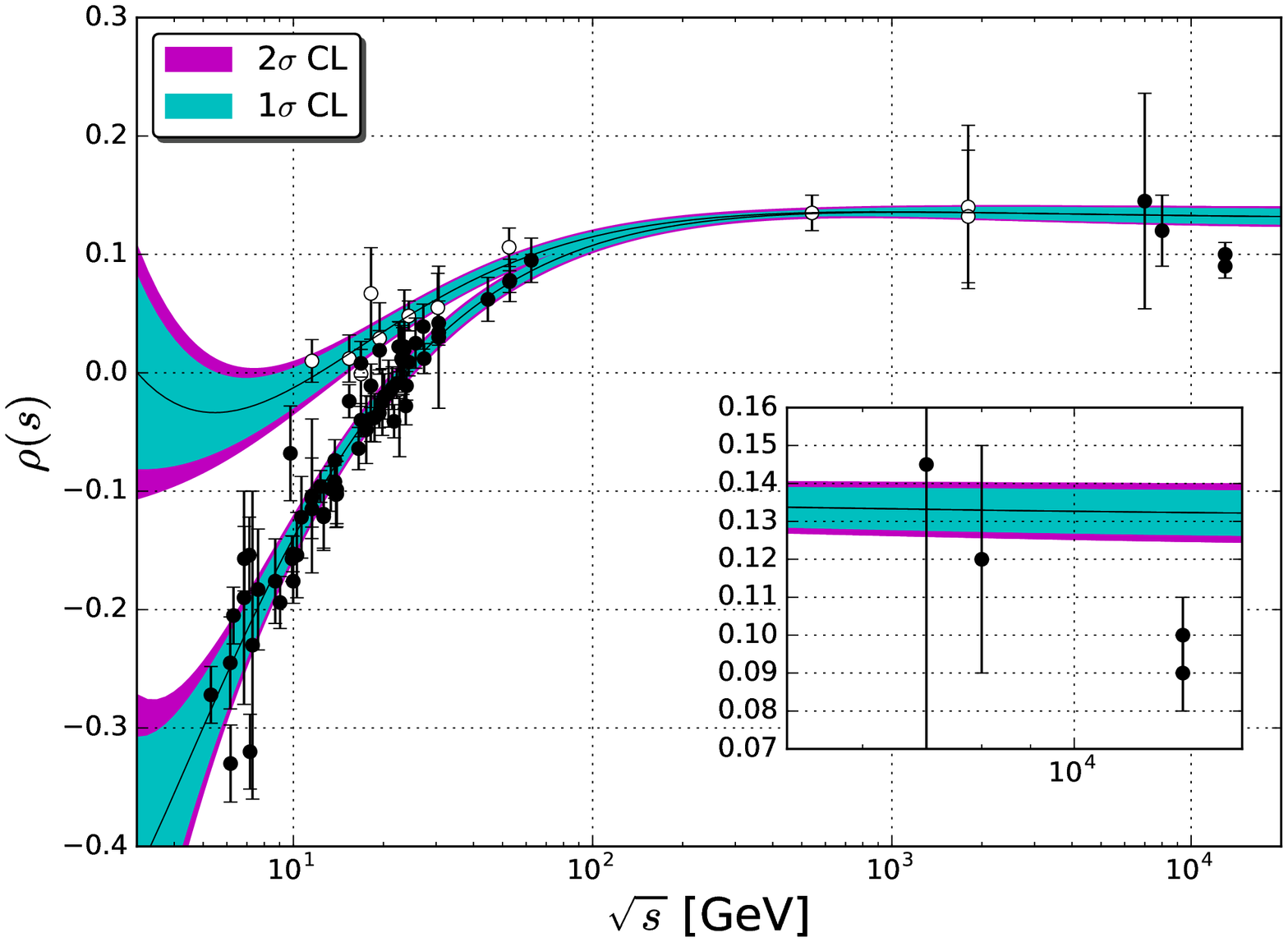}
    \caption{Fit results with Model IV to ensembles T (above) and T + A (below).}
    \label{ch4fig4}
  \end{center}
\efg

\bfg[hbtp]
  \begin{center}
    \includegraphics[width=14.0cm,height=14cm]{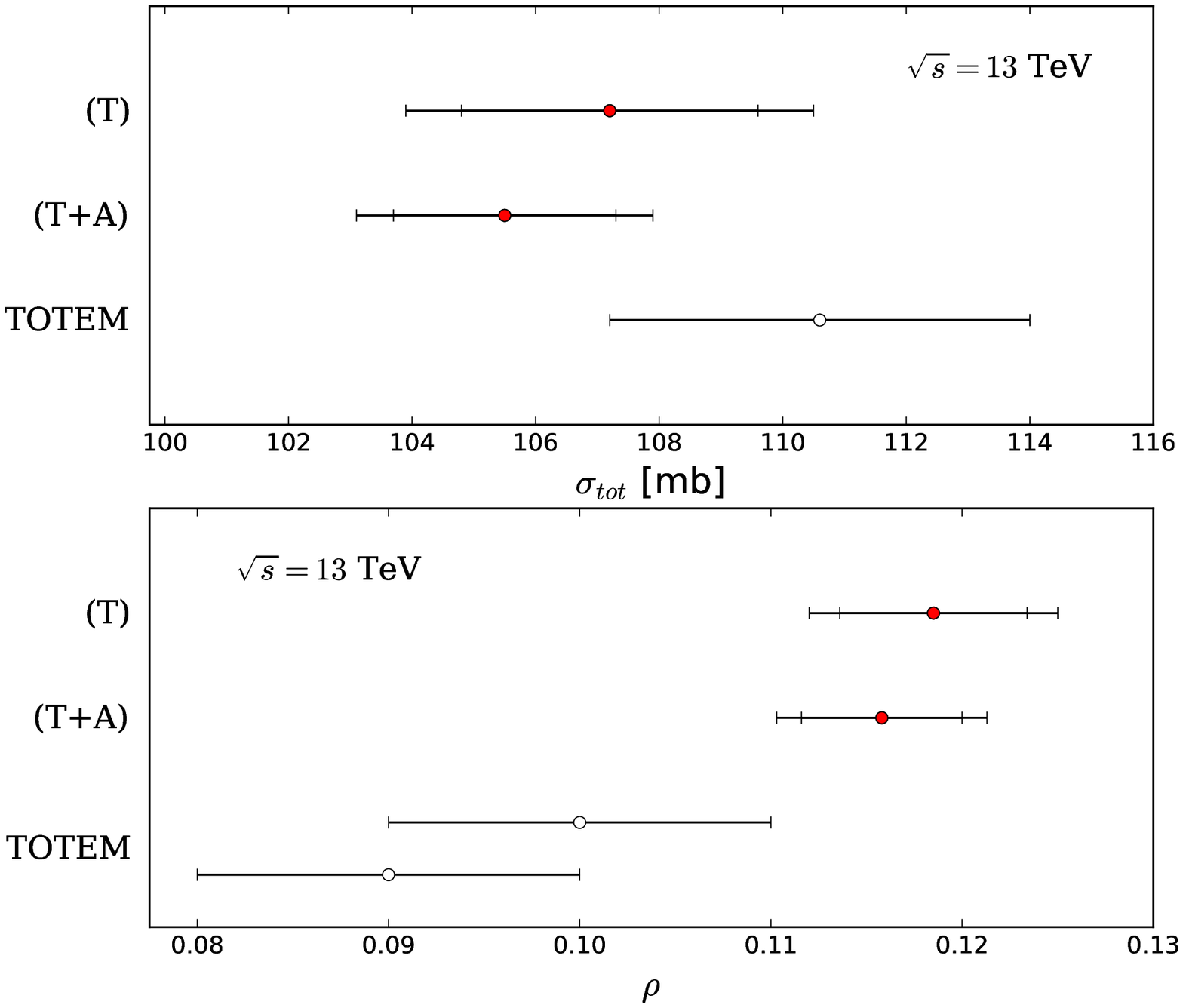}
    \caption{Predictions of Model III for $\sigma_{tot}$ and $\rho$ at $13$ TeV with $1$ and $2$ standard deviations from fits to ensemble T and T + A (filled circles) together with the TOTEM measurements (\ref{totem13}) (empty circles).}
    \label{ch4fig5}
  \end{center}
\efg

\begin{table}[h]
\centering
\scalebox{0.9}{
\begin{tabular}{c@{\quad}c@{\quad}c@{\quad}c@{\quad}c@{\quad}c@{\quad}c@{\quad}c@{\quad}}
\hline \hline
& & & &    \\[-0.4cm]
Ensemble: & \multicolumn{2}{c} {T} & & \multicolumn{2}{c} {T+A} \\
\cline{2-3} \cline{5-6}
& & & & & &  \\[-0.4cm]
$\sqrt{s_{min}}$ [GeV]    & 7.5 & 10 &  & 7.5 & 10 \\[0.05ex]
\hline
& & & & & & \\[-0.4cm] 
$a_{1}$ [mb]       & 57.5\,$\pm$\,2.1    & 55.8\,$\pm$\,4.0    &  & 57.9\,$\pm$\,2.1    & 56.5\,$\pm$\,4.1                \\[0.05ex]
$b_{1}$            & 0.217\,$\pm$\,0.023 & 0.202\,$\pm$\,0.037 &  & 0.224\,$\pm$\,0.021 & 0.212\,$\pm$\,0.037             \\[0.05ex]
$a_{2}$ [mb]       & 16.8\,$\pm$\,2.7    & 15.1\,$\pm$\,4.6    &  & 16.8\,$\pm$\,2.7    & 15.1\,$\pm$\,4.8                \\[0.05ex]
$b_{2}$            & 0.542\,$\pm$\,0.046 & 0.520\,$\pm$\,0.070 &  & 0.542\,$\pm$\,0.046 & 0.520\,$\pm$\,0.072             \\[0.05ex]
$C$ [mb]           & 3.48\,$\pm$\,0.44   & 3.25\,$\pm$\,0.66   &  & 3.66\,$\pm$\,0.38   & 3.48\,$\pm$\,0.59               \\[0.05ex]
$D$ [mb]           & 0.143\,$\pm$\,0.030 & 0.156\,$\pm$\,0.040 &  & 0.128\,$\pm$\,0.024 & 0.138\,$\pm$\,0.035             \\[0.05ex]
$K$ [mbGeV$^{2}$]  & -15\,$\pm$\,74      & 4.17\,$\pm$\,116    &  & -9.5\,$\pm$\,73     & 14.3\,$\pm$\,117                \\[0.05ex]
\hline
& & & & & & \\[-0.4cm]
$\nu$              & 205                    & 164                    &  & 207                    & 166                    \\[0.05ex]
$\chi^2/\nu$       & 1.217                  & 1.213                  &  & 1.253                  & 1.263                  \\[0.05ex]
$P(\chi^2)$        & 1.8 $\times$ 10$^{-2}$ & 3.3 $\times$ 10$^{-2}$ &  & 7.8 $\times$ 10$^{-3}$ & 1.2 $\times$ 10$^{-2}$ \\[0.05ex]
\hline
& & & & & & \\[-0.4cm] 
Figure:            & \ref{ch4fig6}          & \ref{ch4fig7}          &  & \ref{ch4fig6}          & \ref{ch4fig7}          \\[0.05ex]
\hline \hline 
\end{tabular}
}
\caption{Fitted parameters with model III to ensembles T and T + A by considering one standard deviation, energy cutoffs 
at $7.5$ and $10$ GeV and $K$ as a free fit parameter.}
\label{ch4t3}
\end{table}

\begin{table}[h]
\centering
\scalebox{0.8}{
\begin{tabular}{c@{\quad}c@{\quad}c@{\quad}c@{\quad}c@{\quad}c@{\quad}c@{\quad}c@{\quad}}
\hline \hline
& & & & & & \\[-0.4cm]
Ensemble: & \multicolumn{3}{c} {T} & & \multicolumn{3}{c} {T+A} \\
\cline{2-4} \cline{6-8}
& & & & & &  \\[-0.3cm]
$\sqrt{s_{min}}$ [GeV]    & 5 & 7.5 & 10 & & 5 & 7.5 & 10 \\[0.05ex]
\hline
& & & & & & \\[-0.4cm]
$a_{1}$ [mb]      & 58.6\,$\pm$\,1.3    & 57.7\,$\pm$\,1.8    & 55.8\,$\pm$\,3.1    & & 58.8\,$\pm$\,1.3    & 58.0\,$\pm$\,1.8    & 56.2\,$\pm$\,3.1 \\[0.05ex]
$b_{1}$           & 0.226\,$\pm$\,0.015 & 0.219\,$\pm$\,0.019 & 0.202\,$\pm$\,0.029 & & 0.231\,$\pm$\,0.014 & 0.225\,$\pm$\,0.018 & 0.209\,$\pm$\,0.028 \\[0.05ex]
$a_{2}$ [mb]      & 17.0\,$\pm$\,1.8    & 16.6\,$\pm$\,2.3    & 15.2\,$\pm$\,4.2    & & 17.1\,$\pm$\,1.8    & 16.6\,$\pm$\,2.3    & 15.3\,$\pm$\,4.3 \\[0.05ex]
$b_{2}$           & 0.547\,$\pm$\,0.032 & 0.538\,$\pm$\,0.038 & 0.521\,$\pm$\,0.064 & & 0.548\,$\pm$\,0.032 & 0.539\,$\pm$\,0.038 & 0.522\,$\pm$\,0.063 \\[0.05ex]
$C$ [mb]          & 3.62\,$\pm$\,0.30   & 3.51\,$\pm$\,0.38   & 3.24\,$\pm$\,0.53   & & 3.76\,$\pm$\,0.26   & 3.67\,$\pm$\,0.33   & 3.44\,$\pm$\,0.47 \\[0.05ex]
$D$ [mb]          & 0.135\,$\pm$\,0.022 & 0.141\,$\pm$\,0.026 & 0.157\,$\pm$\,0.033 & & 0.122\,$\pm$\,0.018 & 0.127\,$\pm$\,0.021 & 0.140\,$\pm$\,0.029 \\[0.05ex]
\hline
& & & & & & \\[-0.4cm]
$\nu$             & 249       & 206       & 165                 & & 251                 & 208                 & 167 \\[0.05ex]
$\chi^2/\nu$      & 1.210     & 1.213     & 1.206               & & 1.238               & 1.248               & 1.256 \\[0.05ex]
$P(\chi^2)$       & 1.3 $\times$ 10$^{-2}$ & 2.0 $\times$ 10$^{-2}$ & 3.7 $\times$ 10$^{-2}$ & & 6.1 $\times$ 10$^{-3}$ & 8.8 $\times$ 10$^{-3}$ & 1.4 $\times$ 10$^{-2}$ \\[0.05ex]
\hline
& & & & & & \\[-0.4cm]
Figure:            &  \ref{ch4fig8} & \ref{ch4fig9} &  \ref{ch4fig10}  & &  \ref{ch4fig8} & \ref{ch4fig9}   &  \ref{ch4fig10}    \\[0.05ex]
\hline \hline 
\end{tabular}
}
\caption{Fitted parameters with model III to ensembles T and T + A by considering one standard deviation, energy cutoffs at $5$, $7.5$ and $10$ GeV  and the subtraction constant $K=0$ (fixed).}
\label{ch4t4}
\end{table}

\bfg[hbtp]
  \begin{center}
 \includegraphics[width=8.0cm,height=8.0cm]{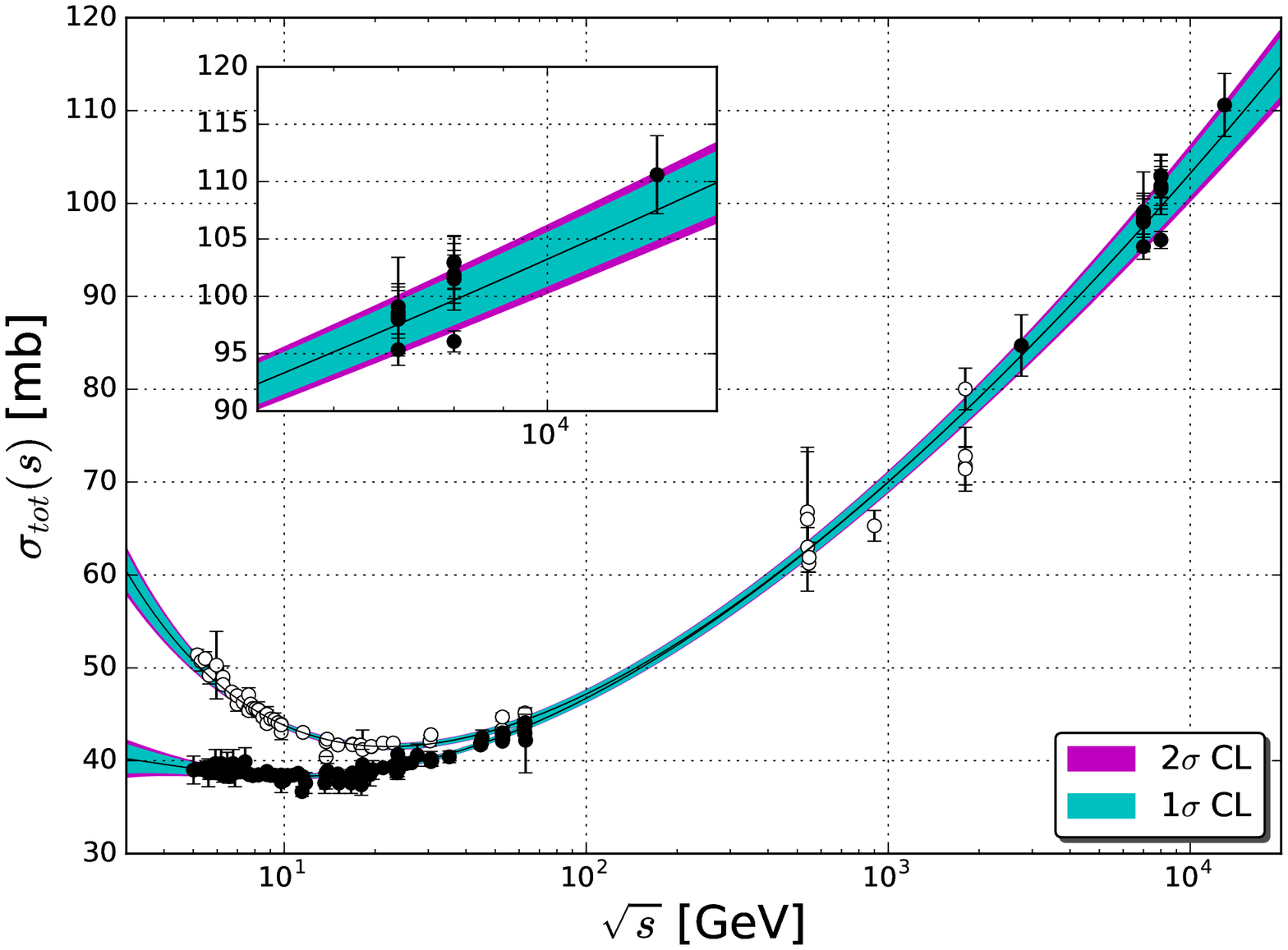}
 \includegraphics[width=8.0cm,height=8.0cm]{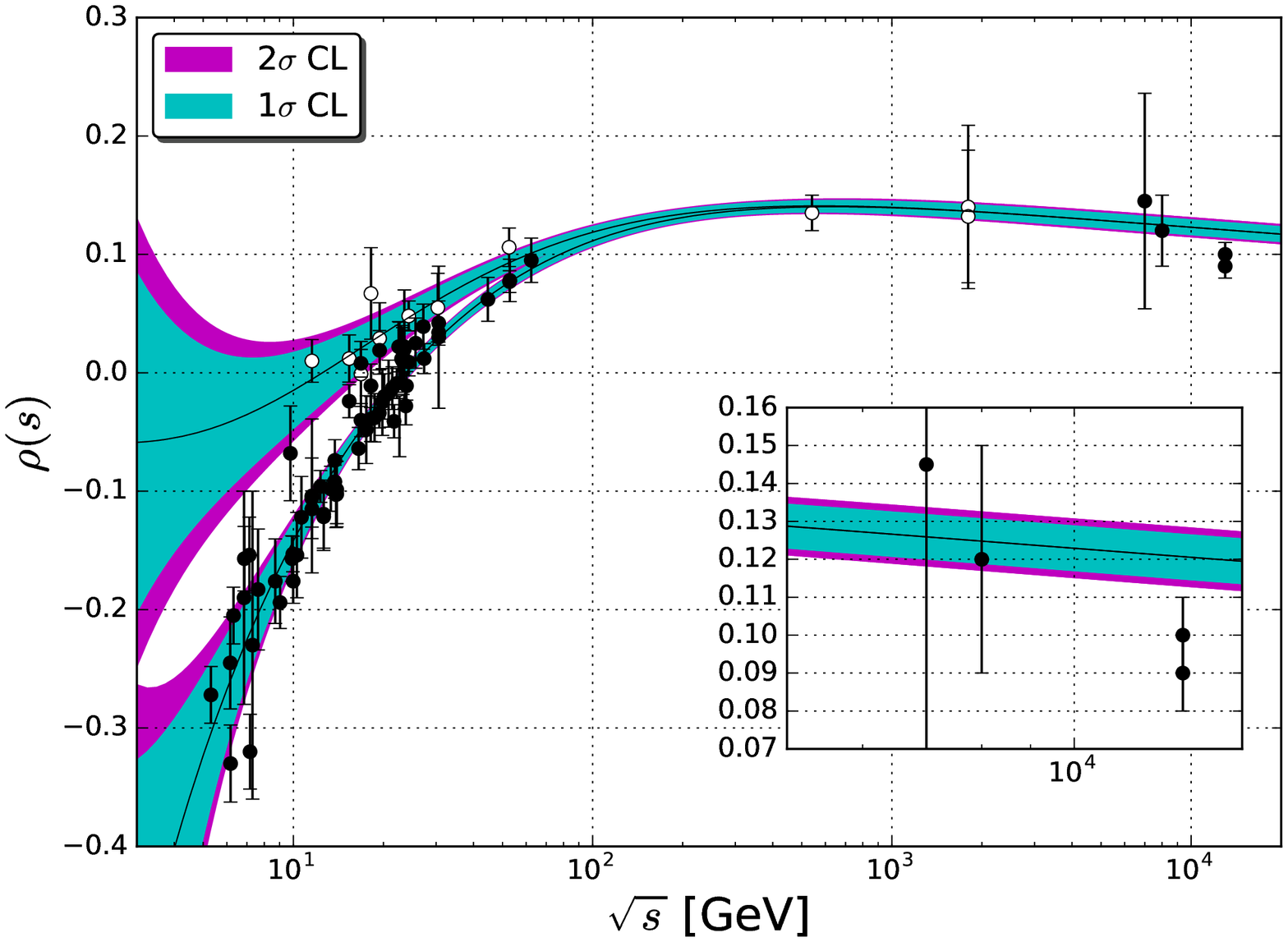}
 \includegraphics[width=8.0cm,height=8.0cm]{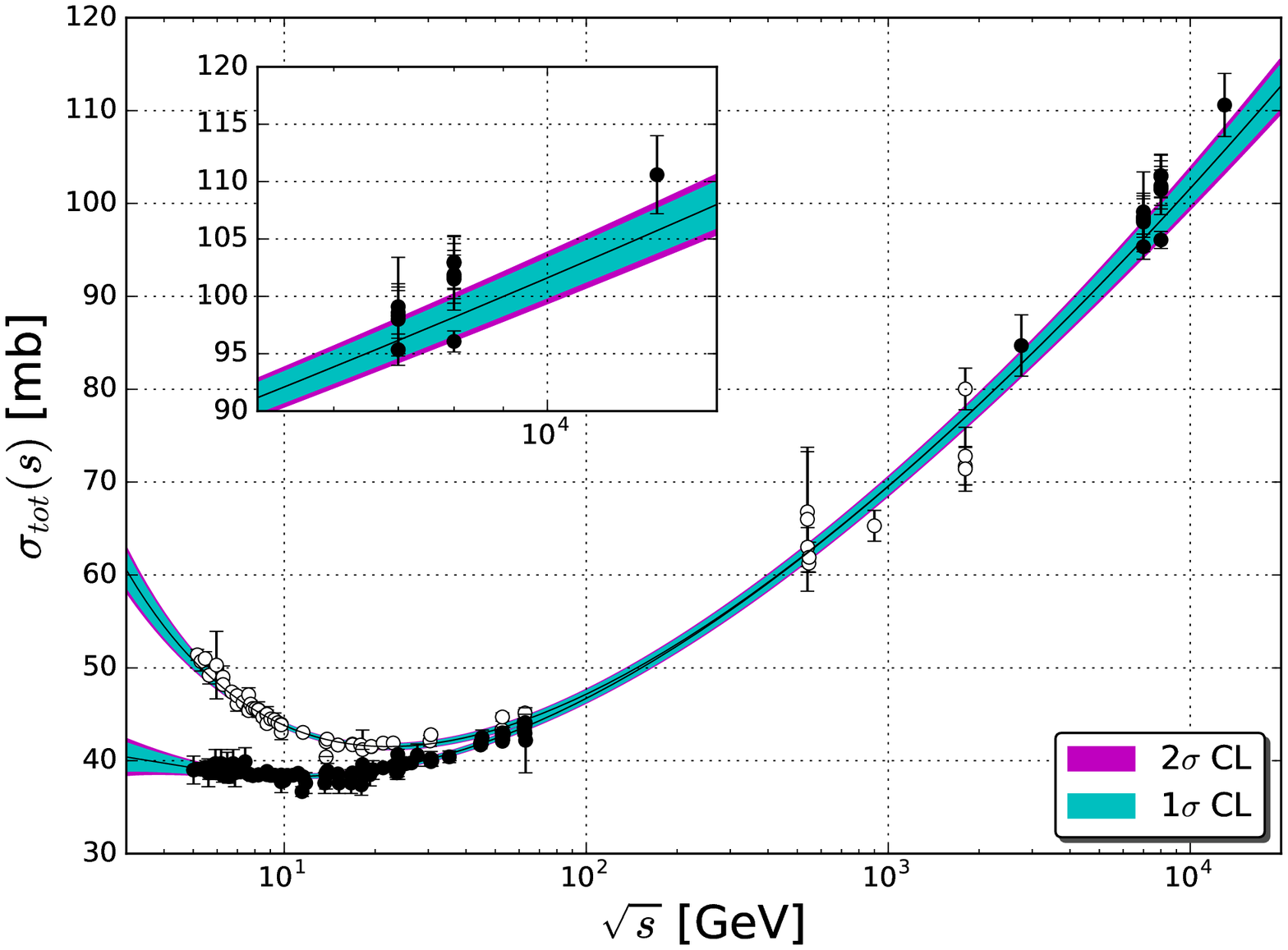}
 \includegraphics[width=8.0cm,height=8.0cm]{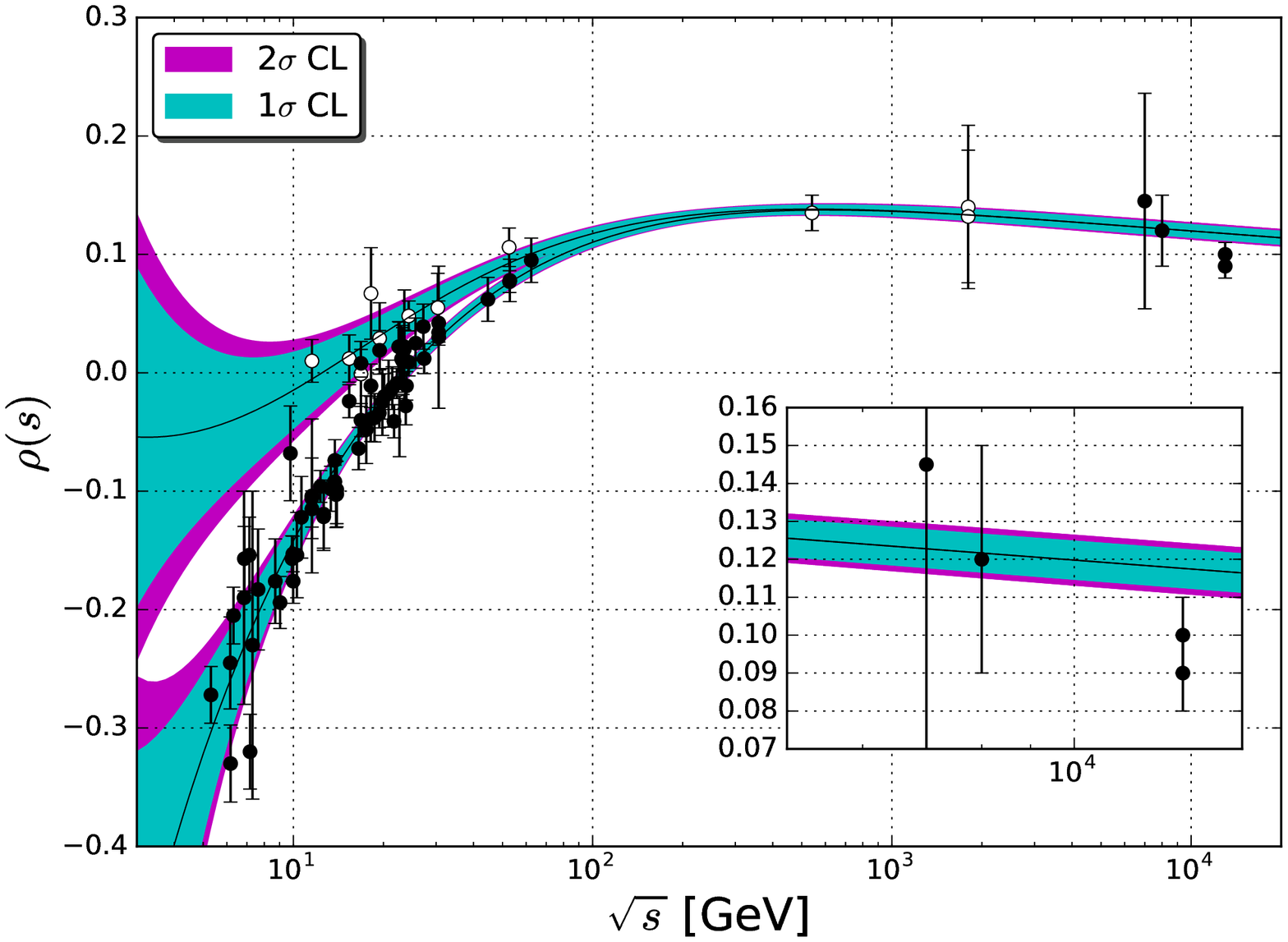}
 \caption{Fit results with Model III to ensembles T (above) and T + A (below) by considering the 
energy cutoff at $\sqrt{s}=7.5$ GeV and $K$ as a free fit parameter.}
    \label{ch4fig6}
  \end{center}
\efg

\bfg[hbtp]
  \begin{center}
 \includegraphics[width=8.0cm,height=8.0cm]{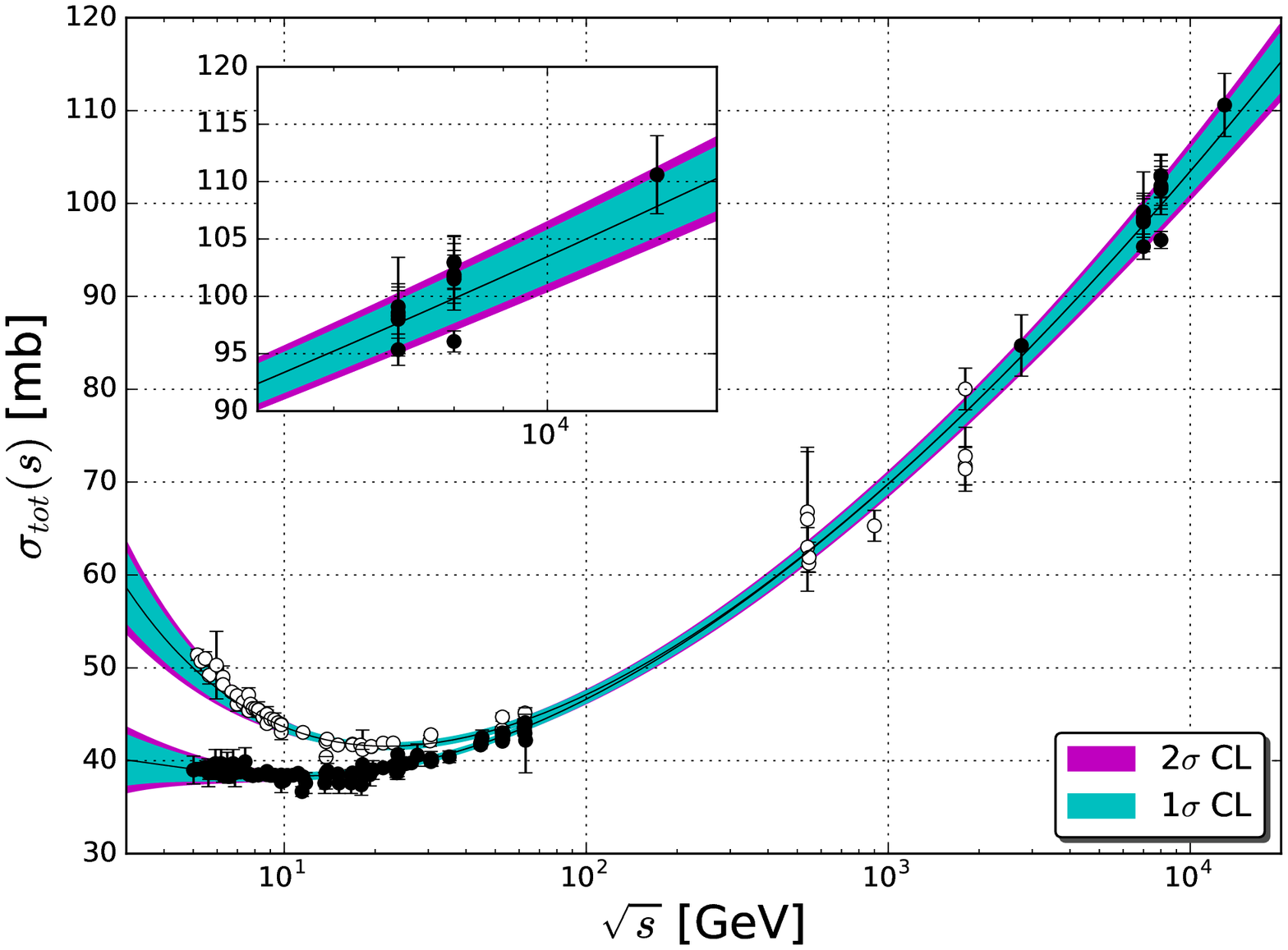}
 \includegraphics[width=8.0cm,height=8.0cm]{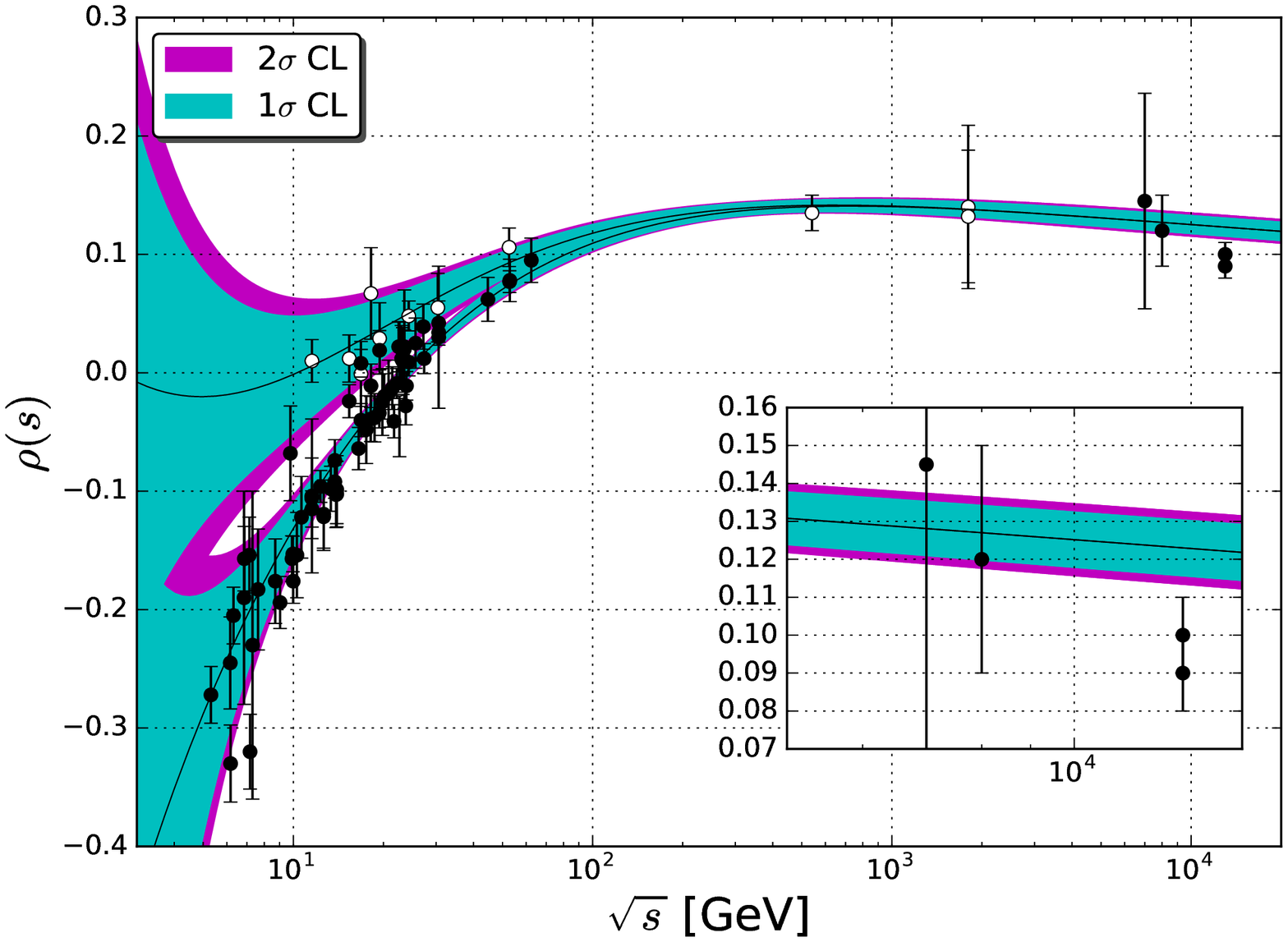}
 \includegraphics[width=8.0cm,height=8.0cm]{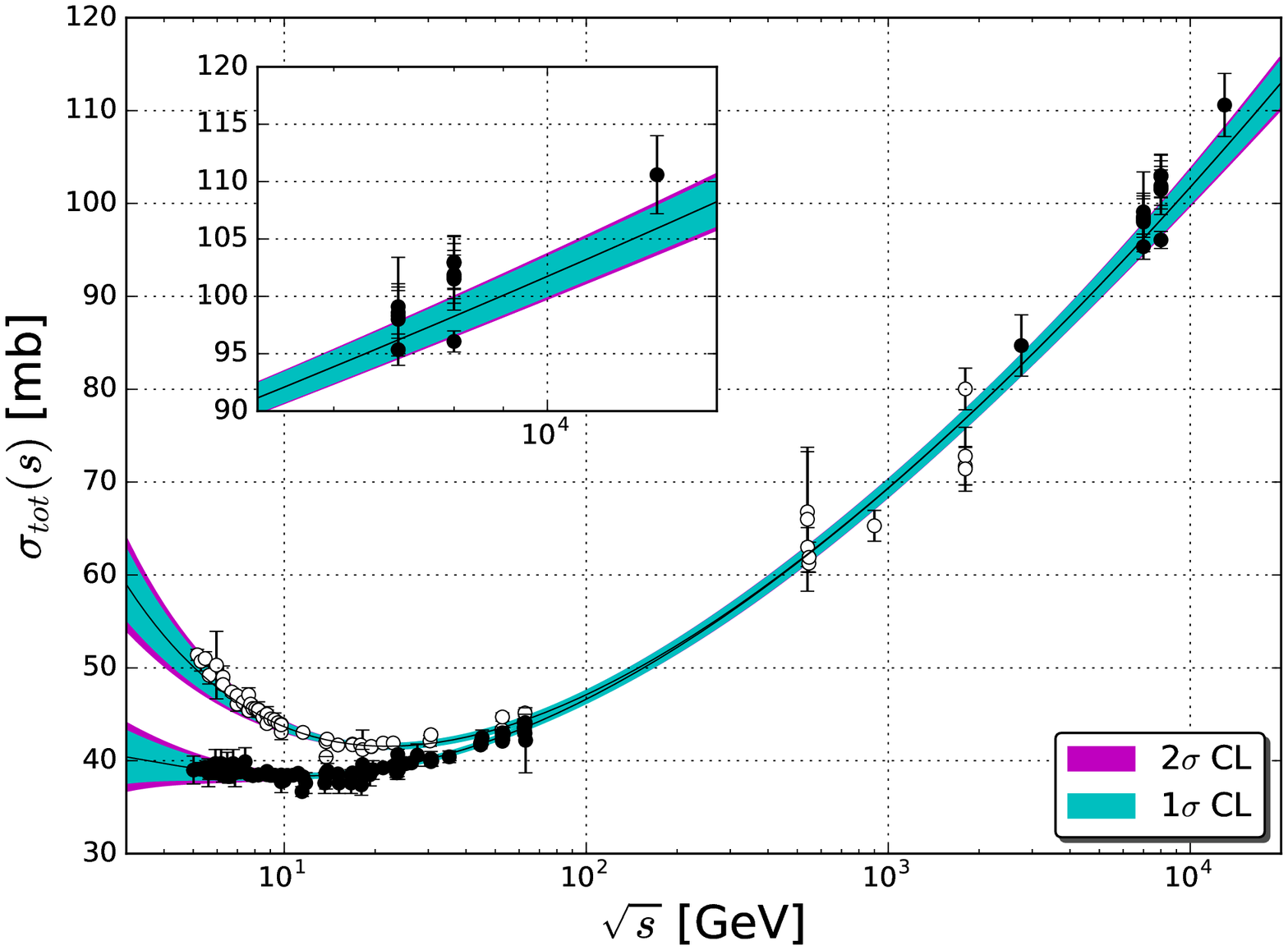}
 \includegraphics[width=8.0cm,height=8.0cm]{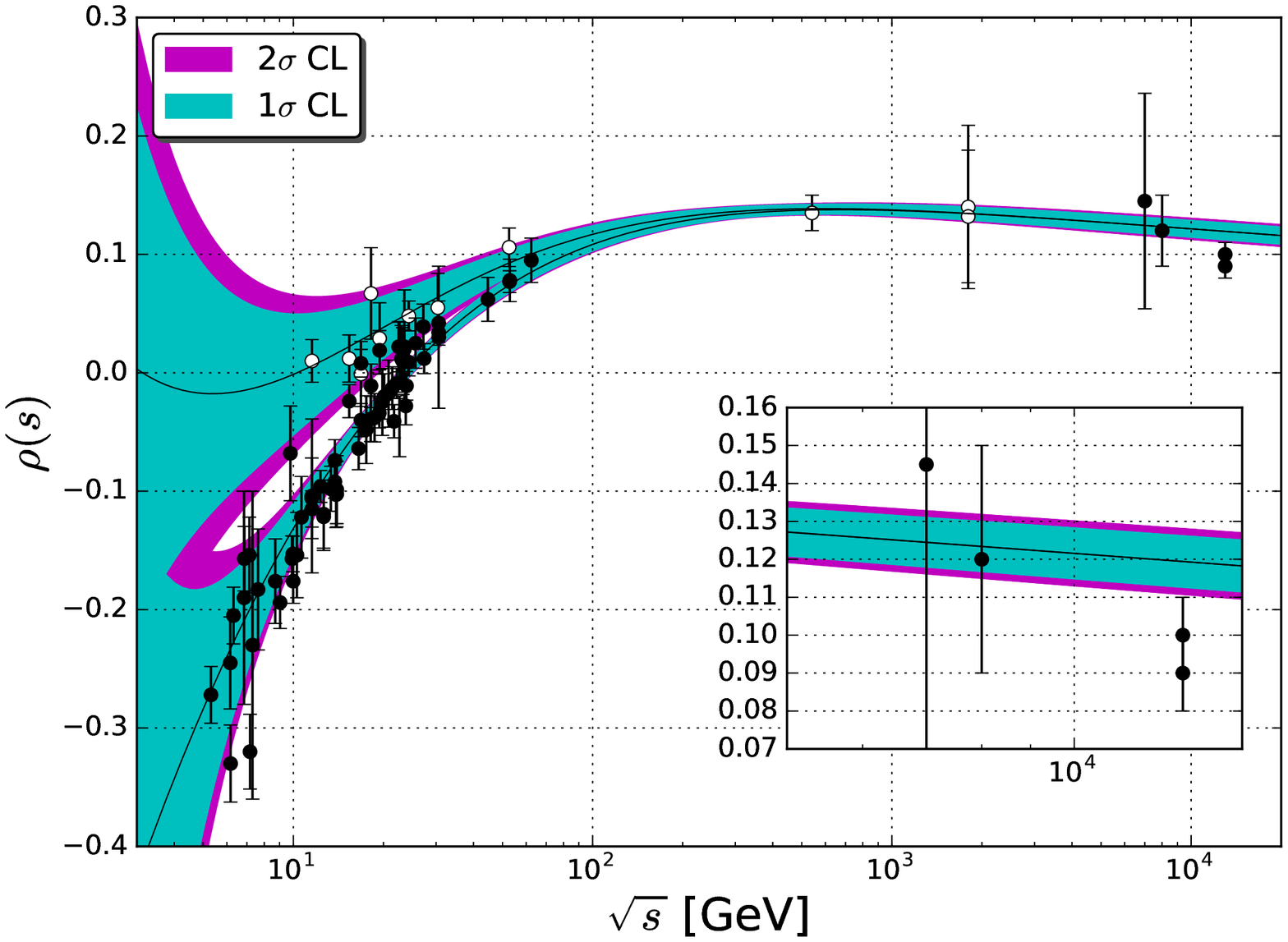}
 \caption{Fit results with Model III to ensembles T (above) and T + A (below) by considering 
the energy cutoff at $\sqrt{s}=10$ GeV and $K$ as a free fit parameter.}
    \label{ch4fig7}
  \end{center}
\efg

\bfg[hbtp]
  \begin{center}
 \includegraphics[width=8.0cm,height=8.0cm]{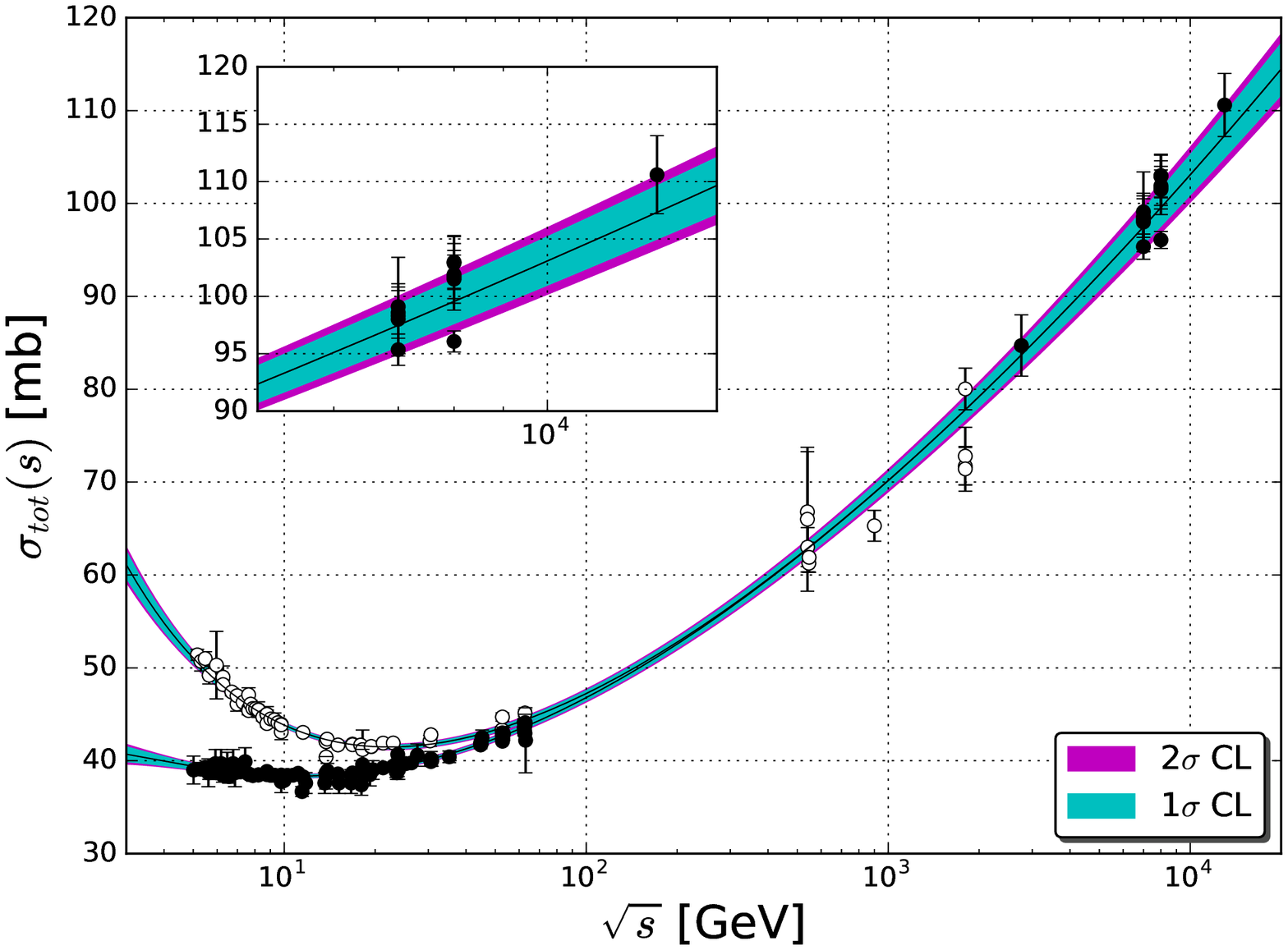}
 \includegraphics[width=8.0cm,height=8.0cm]{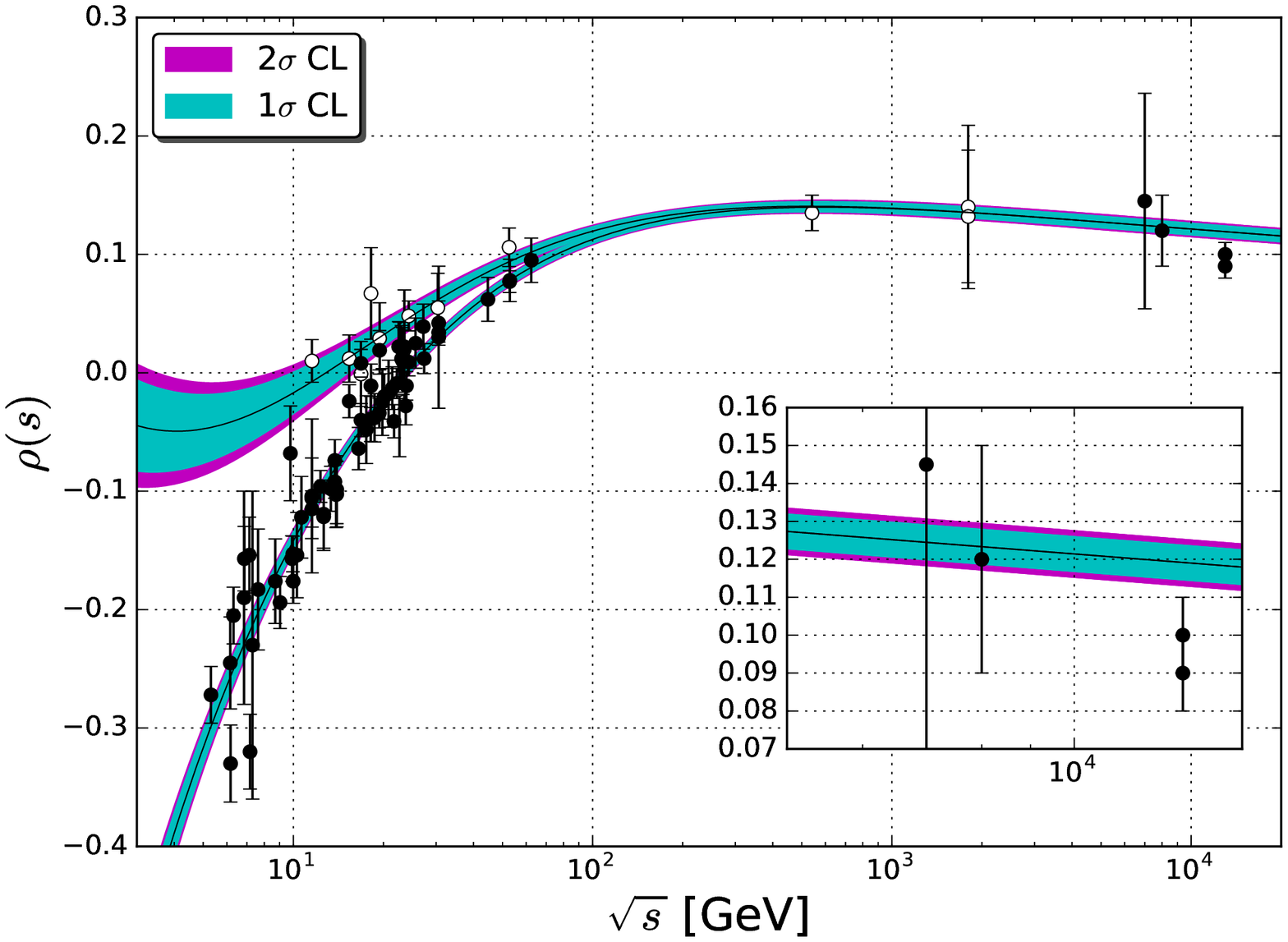}
 \includegraphics[width=8.0cm,height=8.0cm]{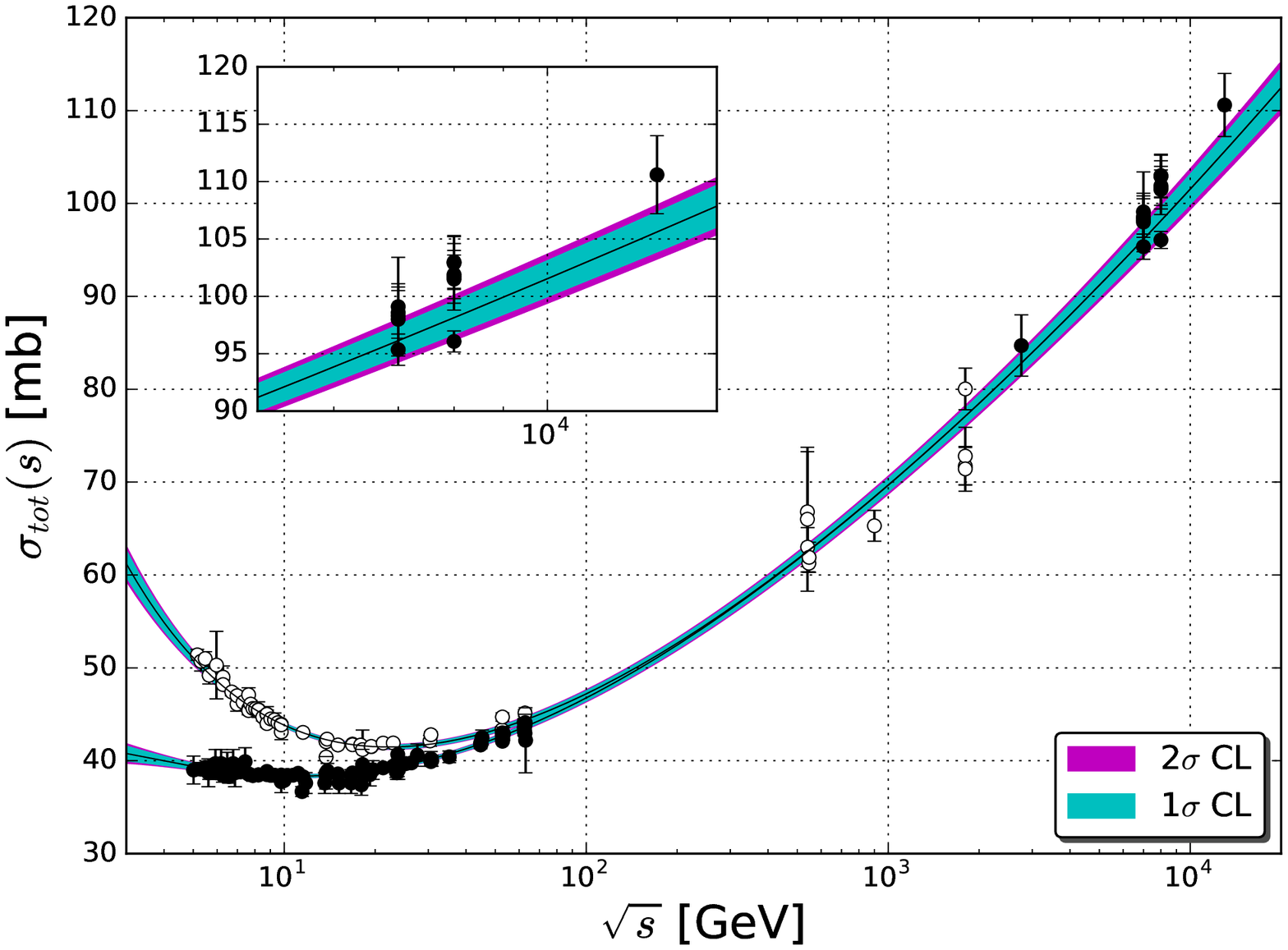}
 \includegraphics[width=8.0cm,height=8.0cm]{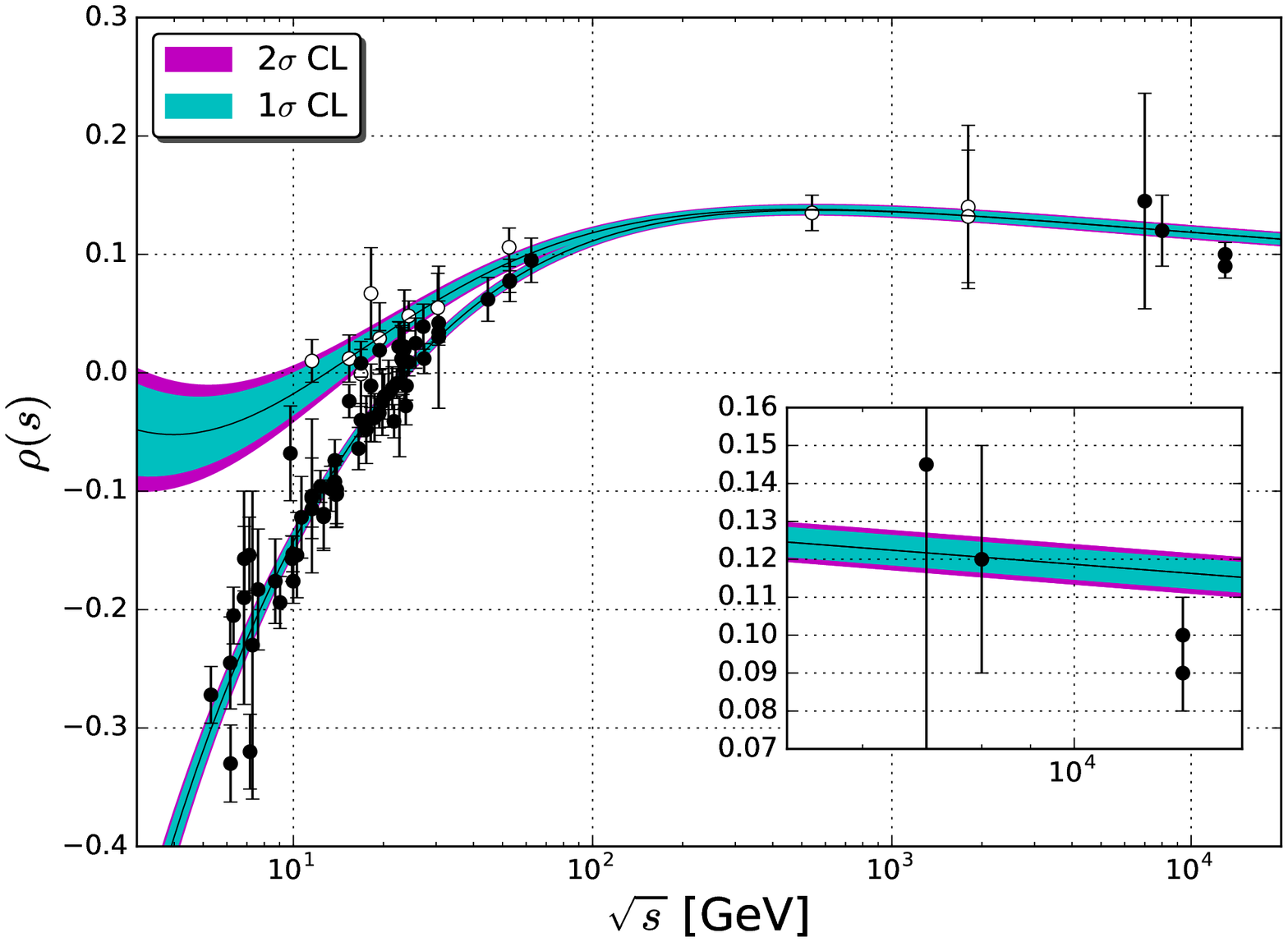}
 \caption{Fit results with Model III to ensembles T (above) and T + A (below) by considering 
the energy cutoff at $\sqrt{s}=5$ GeV and $K=0$ (fixed).}
    \label{ch4fig8}
  \end{center}
\efg

\bfg[hbtp]
  \begin{center}
 \includegraphics[width=8.0cm,height=8.0cm]{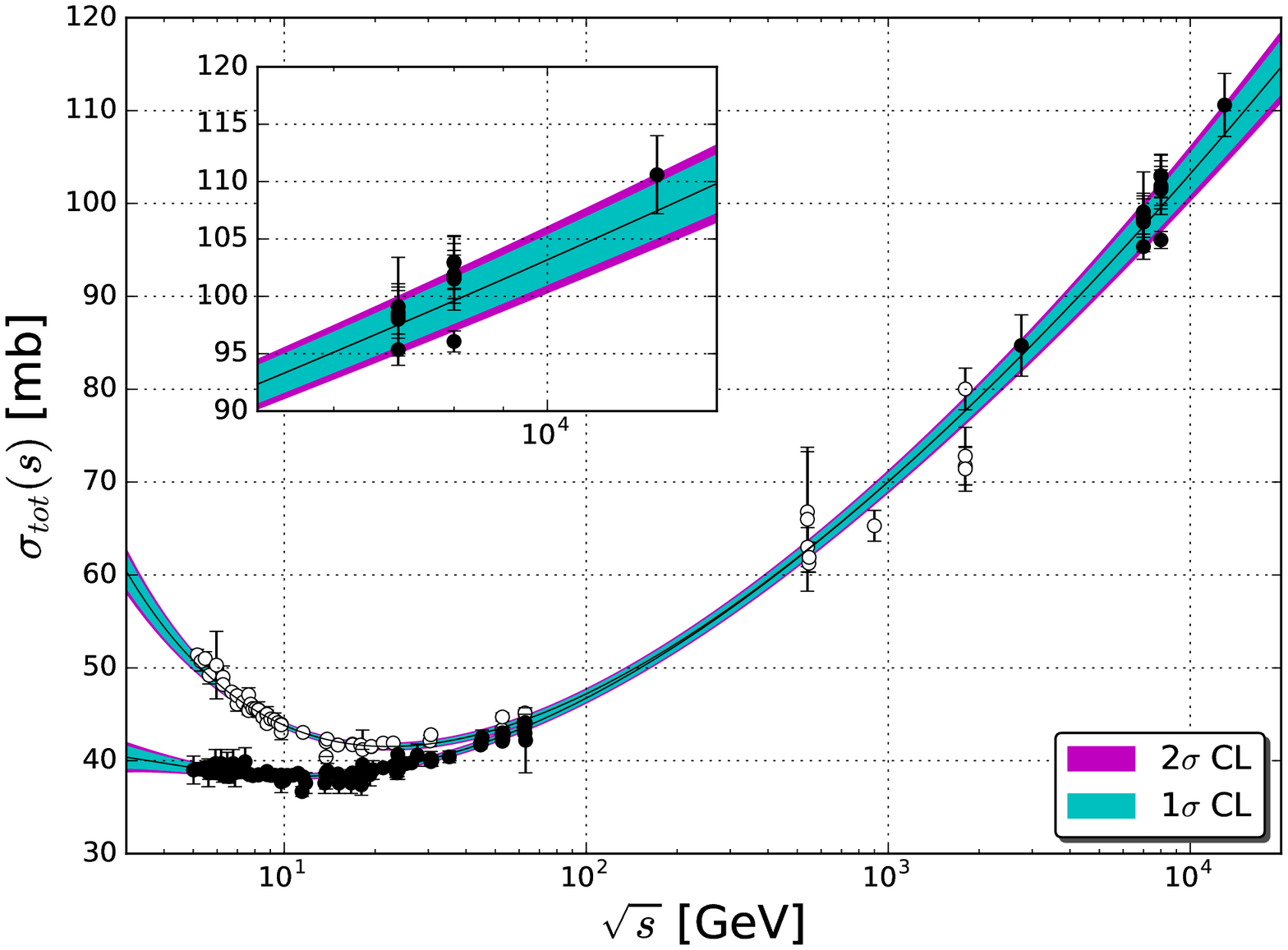}
 \includegraphics[width=8.0cm,height=8.0cm]{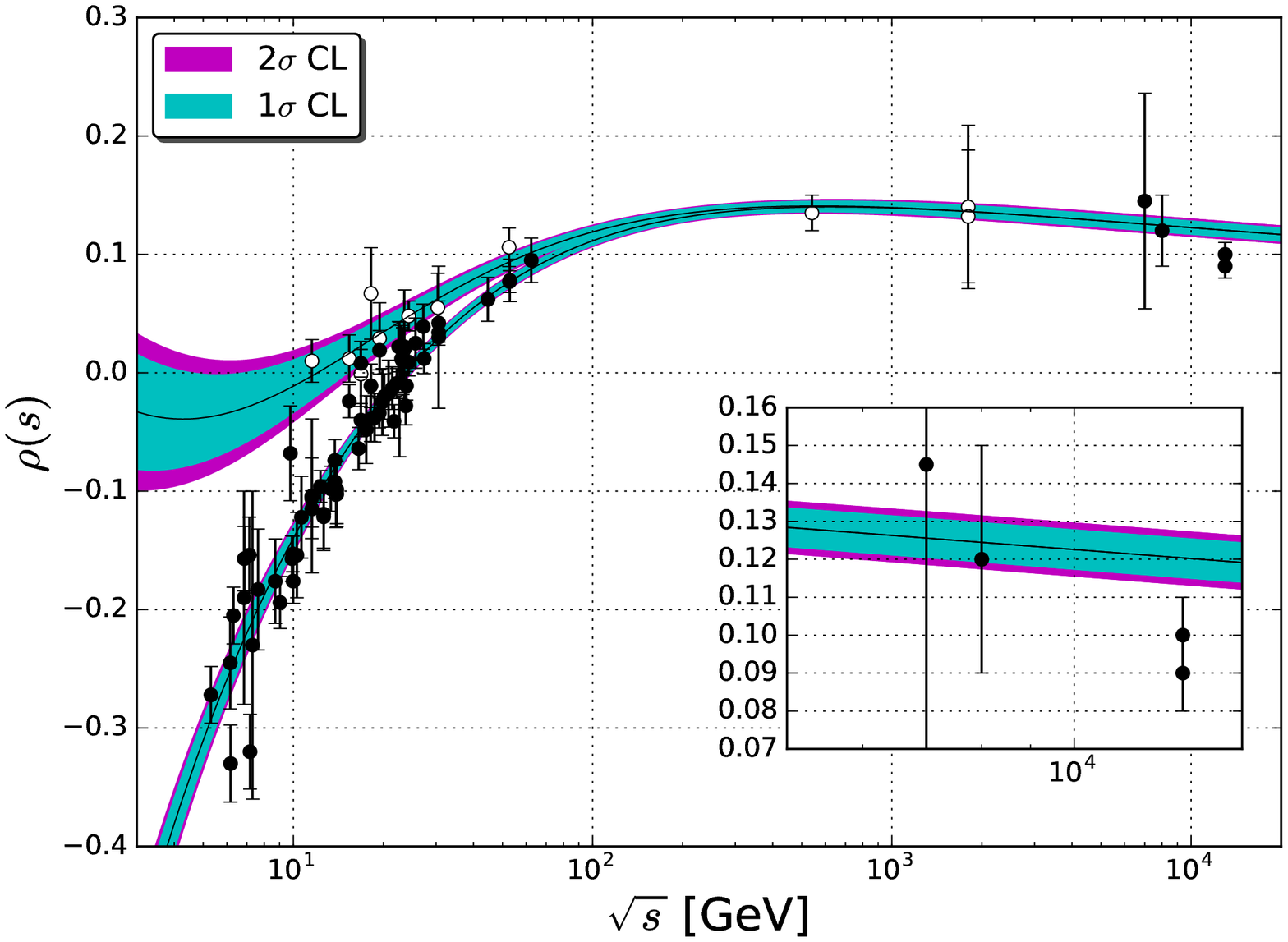}
 \includegraphics[width=8.0cm,height=8.0cm]{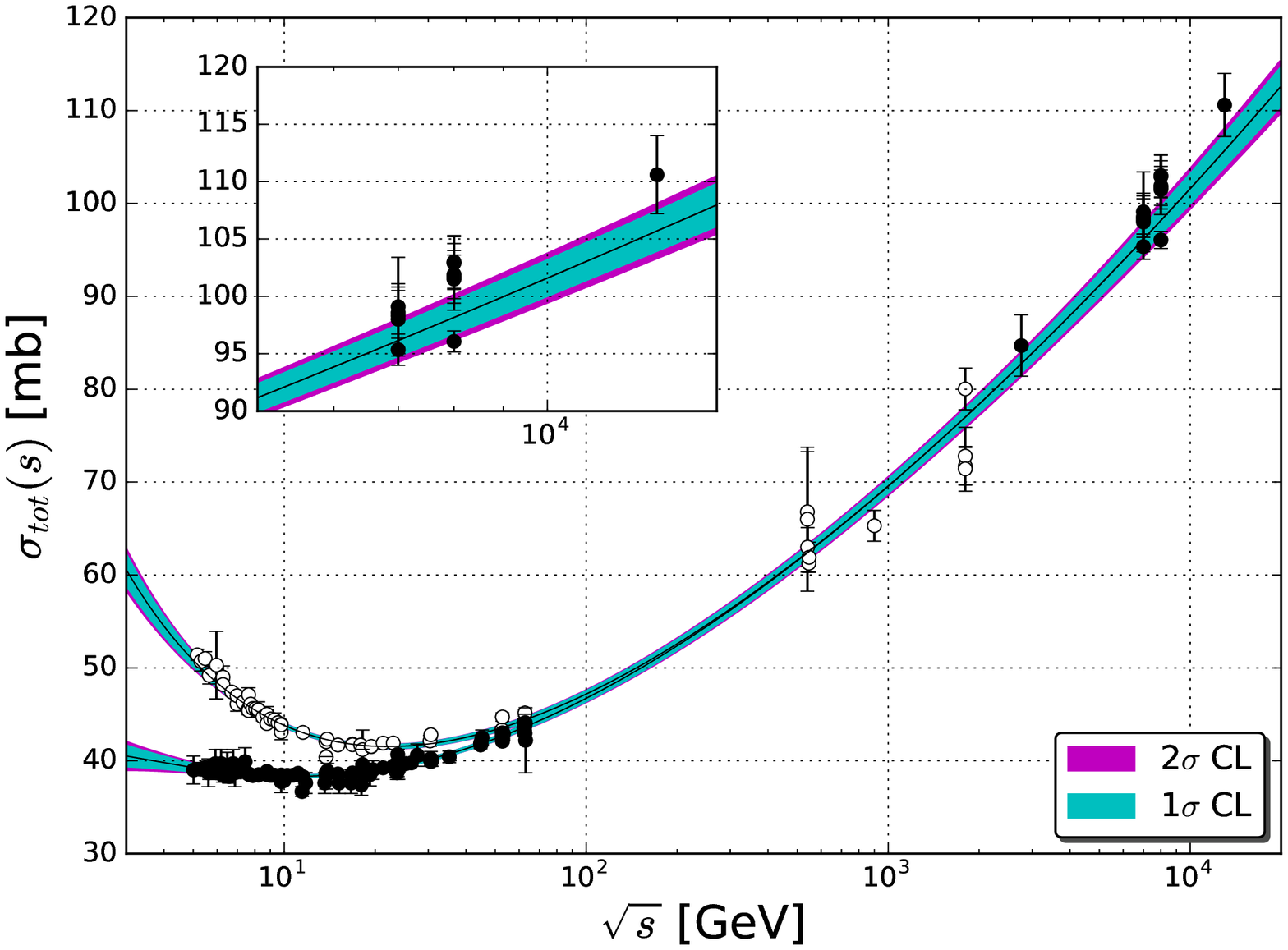}
 \includegraphics[width=8.0cm,height=8.0cm]{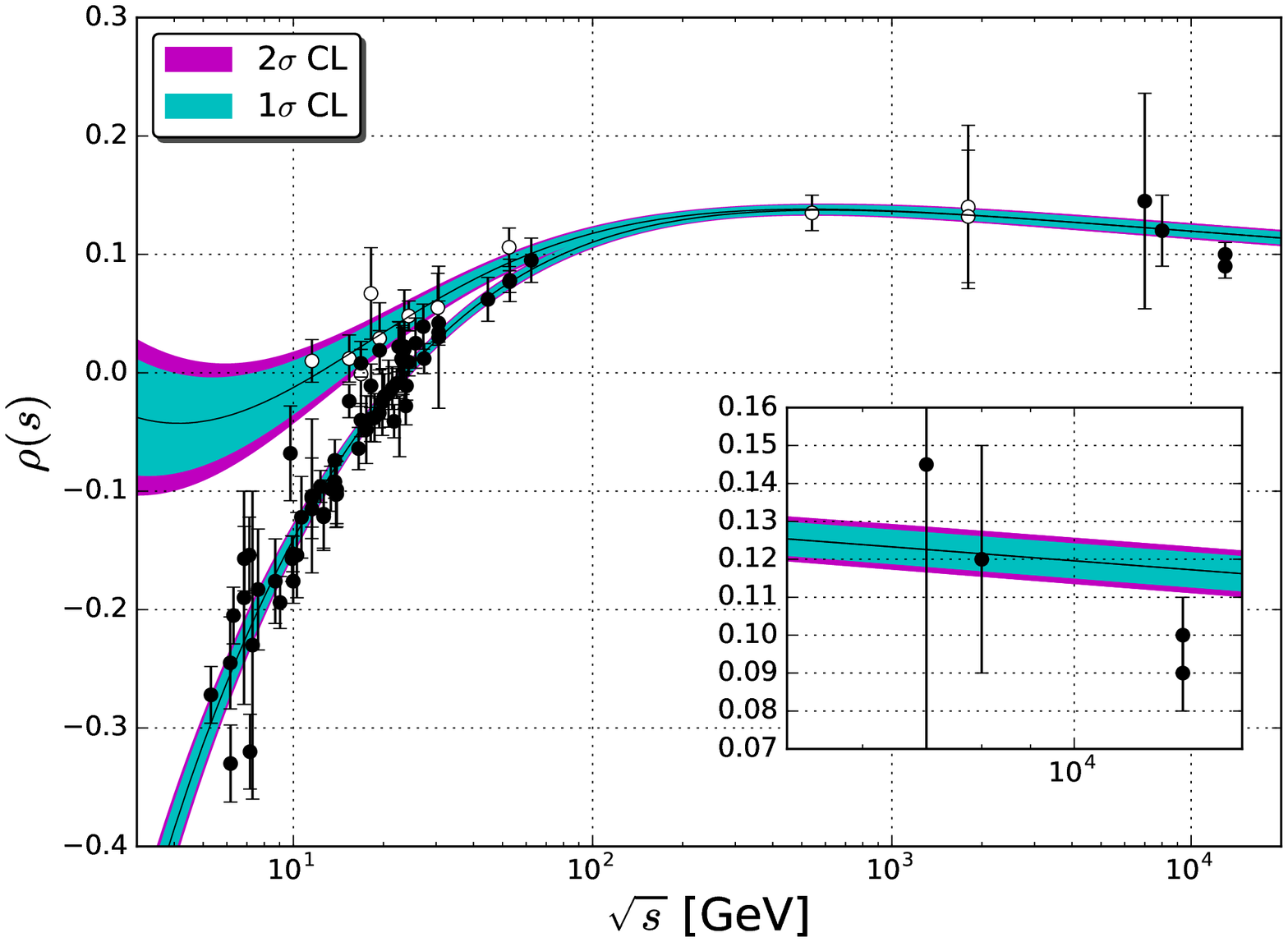}
 \caption{Fit results with Model III to ensembles T (above) and T + A (below) by considering 
the energy cutoff at $\sqrt{s}=7.5$ GeV and $K=0$ (fixed).}
    \label{ch4fig9}
  \end{center}
\efg

\bfg[hbtp]
  \begin{center}
 \includegraphics[width=8.0cm,height=8.0cm]{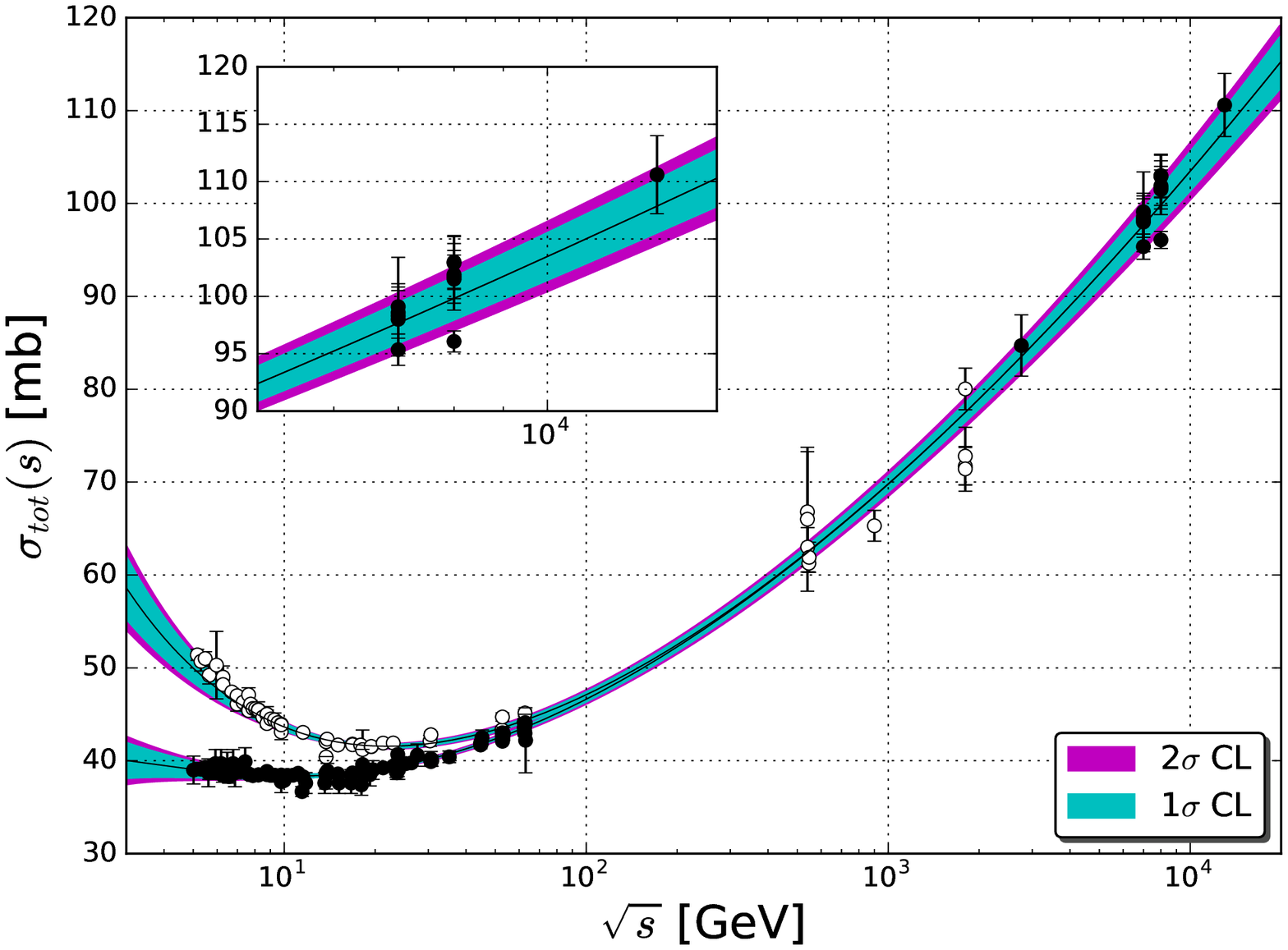}
 \includegraphics[width=8.0cm,height=8.0cm]{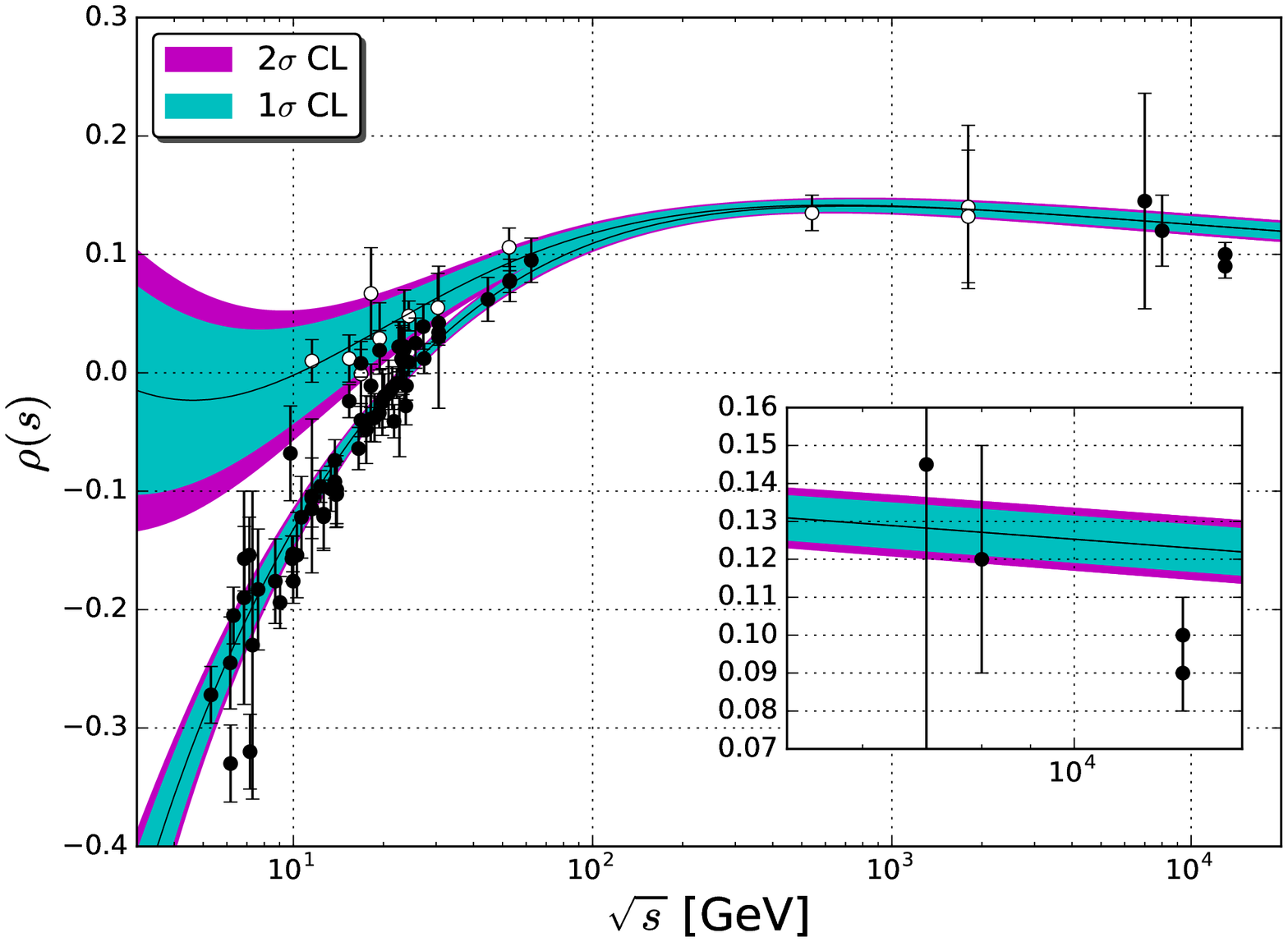}
 \includegraphics[width=8.0cm,height=8.0cm]{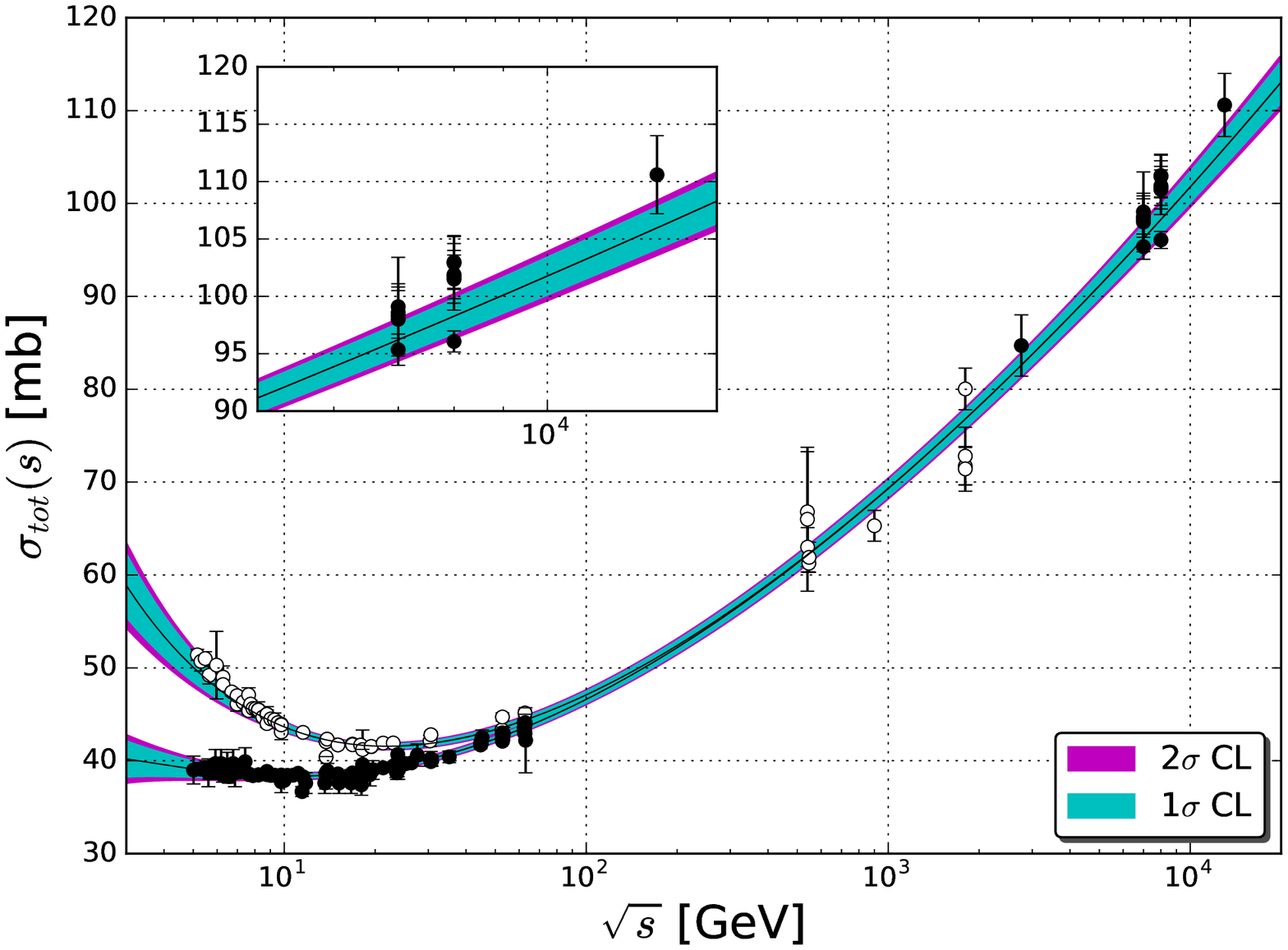}
 \includegraphics[width=8.0cm,height=8.0cm]{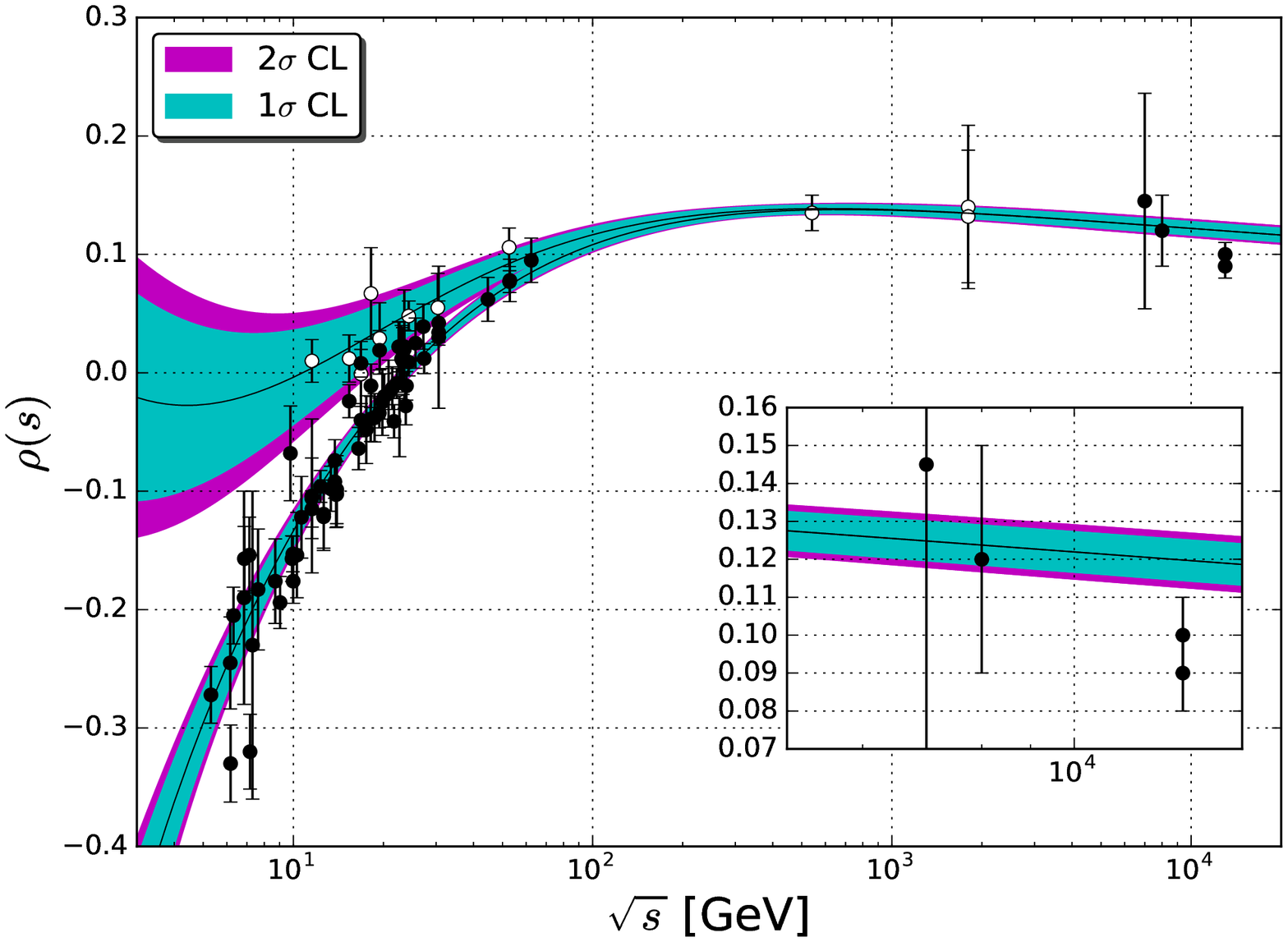}
 \caption{Fit results with Model III to ensembles T (above) and T + A (below) by considering the 
energy cutoff at $\sqrt{s}=10$ GeV and $K=0$ (fixed).}
    \label{ch4fig10}
  \end{center}
\efg

\begin{table}[h]
\centering
\scalebox{0.9}{
\begin{tabular}{c@{\quad}c@{\quad}c@{\quad}c@{\quad}c@{\quad}c@{\quad}c@{\quad}c@{\quad}c@{\quad}}
\hline \hline
& & & & & & & &  \\[-0.4cm]
& & \multicolumn{3}{c} {$\sigma_{tot}$ (mb)} & & \multicolumn{3}{c}
{$\rho$} \\
\cline{3-5} \cline{7-9}
& & & & & & & &  \\[-0.4cm]
$\sqrt{s}$ [TeV] & Ensemble    & Central & $1\sigma$ & $2\sigma$ & & Central &
$1\sigma$ & $2\sigma$ \\[0.05ex]
\hline
& & & & & & & & \\[-0.4cm]
\multirow{2}{*}{13} & T     & 107.2 &  $\pm$\,2.4 & $\pm$\,3.3    & & 0.1185 &
$\pm$\,0.0049   & $\pm$\,0.0065 \\[0.05ex]
                    & T+A   & 105.5 &  $\pm$\,1.8 & $\pm$\,2.4    & & 0.1158 &
$\pm$\,0.0042   & $\pm$\,0.0055 \\[0.05ex]
\hline 
& & & & & & & & \\[-0.3cm]
\multirow{2}{*}{14} & T     & 108.4 & $\pm$\,2.5 & $\pm$\,3.3     & & 0.1179 &
$\pm$\,0.0049   & $\pm$\,0.0065 \\[0.05ex]
                    & T+A   & 106.7 & $\pm$\,1.8 & $\pm$\,2.5     & & 0.1152 &
$\pm$\,0.0042   & $\pm$\,0.0055 \\[0.05ex]
\hline \hline 
\end{tabular}
}
\caption{Predictions of Model III for $\sigma_{tot}$ and $\rho$ at 13 TeV and 14 TeV for $pp$ and $\bar{p}p$ scattering: central values and uncertainties with $1$ $\sigma$ and $2$ $\sigma$ (Tables \ref{ch4t1} and \ref{ch4t2}).}
\label{ch4t5}
\end{table}

\clearpage
\thispagestyle{plain}

\chapter{\textsc{Reggeon Calculus in a Nutshell}}
\label{ch5}
\mbox{\,\,\,\,\,\,\,\,\,}
Over the last few decades Regge pole phenomenology had been consolidated as a great success in hadronic diffraction. Despite the hopes and glories, there are some essential features in the experimental point of view that poles alone cannot explain. These are mainly due to: failures of factorization, since Regge factorization breaks down when there is the presence of more than one Regge trajectory, and the Pomeron supercritical intercept, in order to describe the observed increase of all hadronic total cross sections.

At high energies Reggeon diagrams have several singularities closed to each other. In this scenario, it is necessary to take into account their mutual influence, as for example the interaction between Pomerons begins to play an important role to high-energy collisions \cite{Gribov:2003nw,Barone:2002cv}, see Figure \ref{RGCfig1}. For this reason, Regge poles, which is viewed as corresponding to a single scattering, are not the only singularities of the amplitude. There are also the presence of branch points corresponding to the case of the exchange of several Reggeons, and cut singularities, contrarily to poles' situation, it can be interpreted as multiple scatterings of hadrons' constituents, \tit{i.e.} rescattering effects.

The Regge pole hypothesis is not sufficient to describe the complete asymptotic behavior of the scattering amplitude. Moreover, the interaction vertices, \tit{e.g.} multi-Pomeron and Pomeron-hadron are not known theoretically. For this reason Regge theory becomes unsafe. However, the modern viewpoint is that with a Reggeon field theory, which is first motivated by a consideration of hybrid Feynman graphs, it turns out to be possible to deduce and analyze the exchange of poles and branch points in high-energy scatterings.
\bfg[hbtp]
  \begin{center}
    \includegraphics[width=13cm,height=4cm]{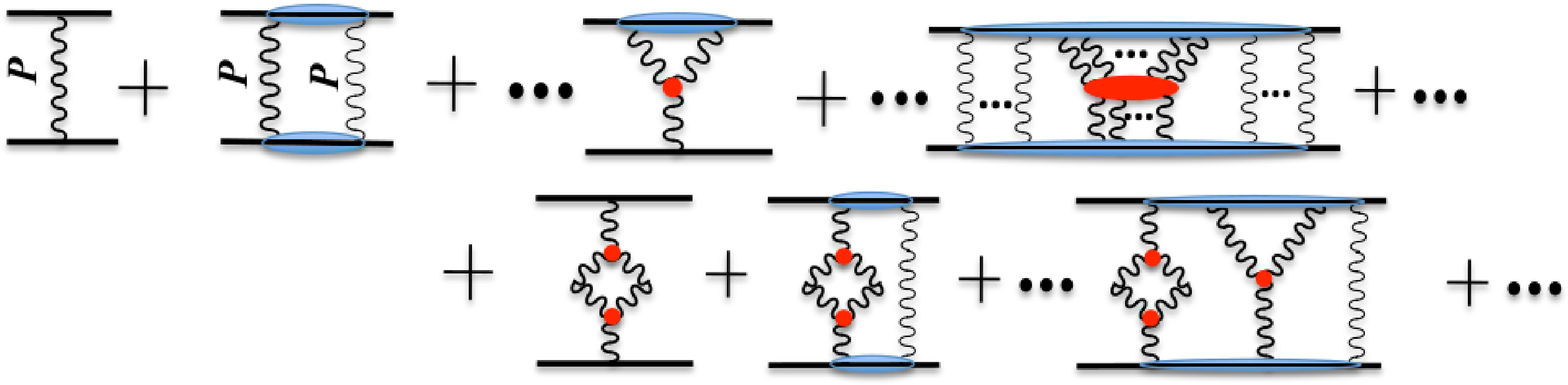}
    \caption{Pole plus the Branch-cut diagrams of multiple Pomeron influence. This figure was taken from Reference \cite{Poghosyan}.}
    \label{RGCfig1}
  \end{center}
\efg
\section{\textsc{Regge Poles in Field Theory}}
\label{ch5sec1}
\mbox{\,\,\,\,\,\,\,\,\,}
The interacting Lagrangian ${\cal L}_{int}=g\phi^{3}$ corresponds to a massive $\phi^{3}$ theory, which is viewed as the simplest approach of a field theoretical model of Regge poles, where the factor $g$ stands for the coupling constant. By considering the Feynman rules \cite{Eden:1966dnq}, the amplitude for an arbitrary diagram can be written respectively as \cite{Barone:2002cv,Collins:1977jy}

\be
A\propto \int\frac{d^{4}k_{1}...d^{4}k_{l}}{\displaystyle\prod^{n}_{i=1}(q^{2}_{i}-m^{2}_{i}+i\epsilon)},
\label{eq5.1}
\ee
where the $k_{l}$ are the independent loop momenta and the $q's$ are constrained at each vertex by $\delta$-functions. The aforementioned expression can be rewritten by using the Feynman relation:
\be
\frac{1}{u_{1}...u_{n}}=(n-1)!\int^{1}_{0}d\alpha_{1}...d\alpha_{n}\,\frac{\delta\left(1-\displaystyle\sum^{n}_{i=1}\alpha_{i}\right)}{\left[ \displaystyle\sum^{n}_{i=1}\alpha_{i}u_{i} \right]^{n}},
\label{ch5.2}
\ee
one finds,
\be
A \propto \int^{1}_{0}d\alpha_{1}...d\alpha_{n} \int d^{4}k_{1}...d^{4}k_{l} \,\frac{\delta\left(1-\displaystyle\sum^{n}_{i=1}\alpha_{i}\right)}{\left[\displaystyle\sum^{n}_{i=1}\alpha_{i}(q^{2}_{i}-m^{2}_{i})+i\epsilon \right]^{n}},
\label{ch5.3}
\ee
where the singularities of the integrand at $q^{2}-m^{2}_{i}$ result in singularities of the scattering amplitude if either: $q^{2}_{i}=m^{2}_{i}$ or $\alpha_{i}=0$ $\forall$ $i=1,...,n$ and $\frac{\partial}{\partial k_{j}}\displaystyle\sum^{n}_{i=1}\alpha_{i}(q^{2}_{i}-m^{2}_{i})=0$ for $j=1,...,l$ \cite{Collins:1977jy}.
\bfg[hbtp]
  \begin{center}
    \includegraphics[width=4cm,height=4cm]{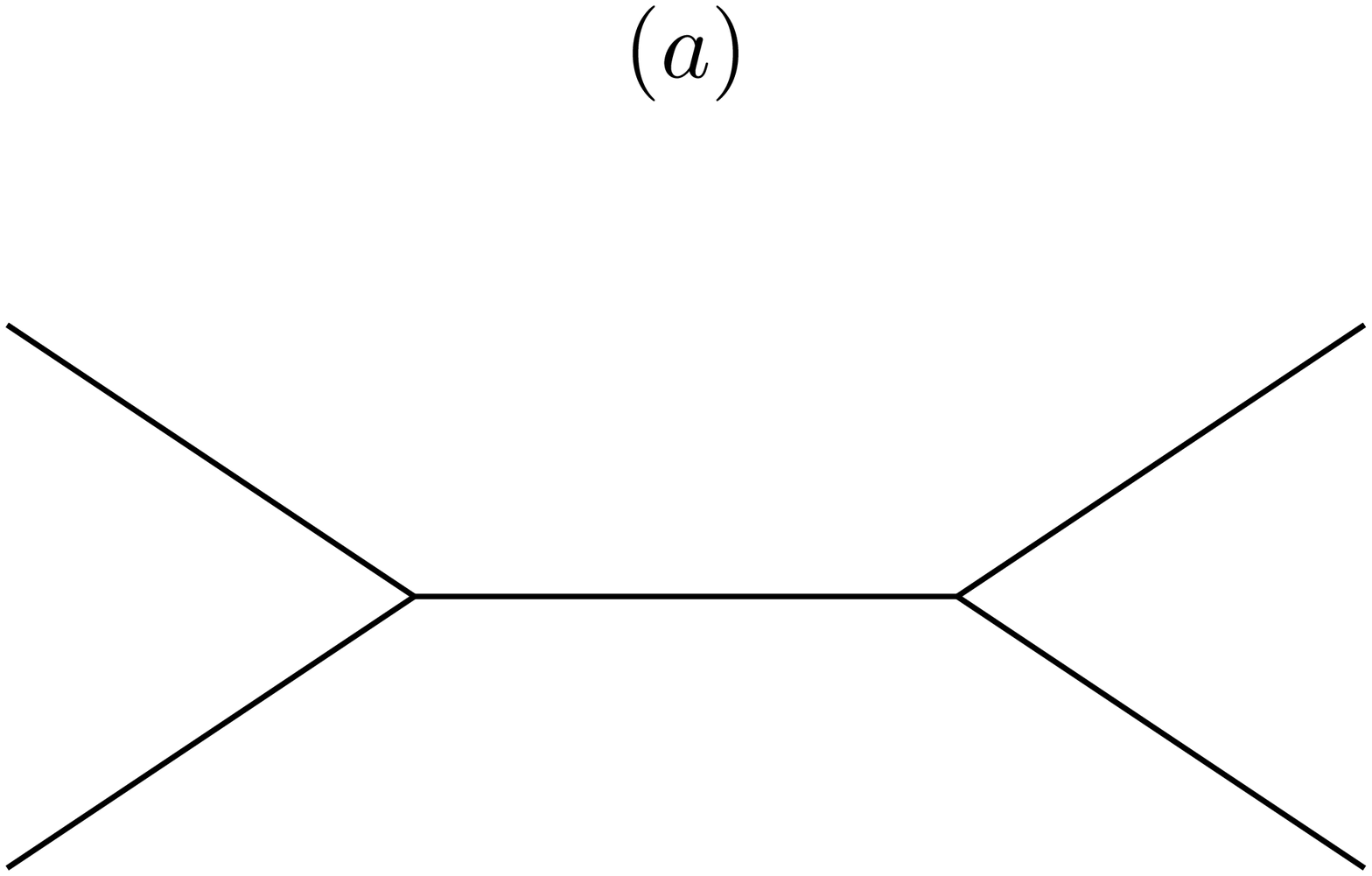}
    \qquad
    \includegraphics[width=4cm,height=4cm]{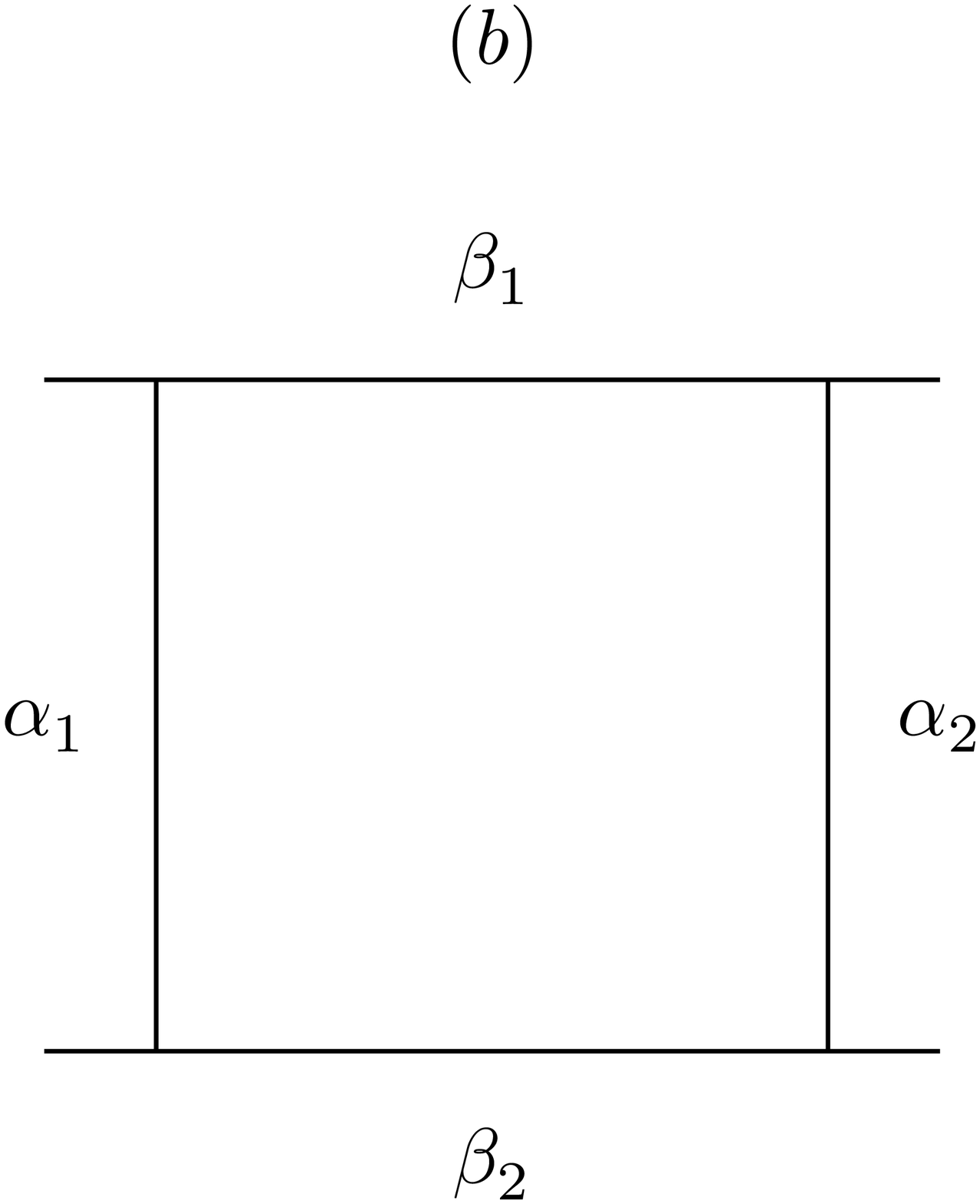}
    \qquad
    \includegraphics[width=4cm,height=4cm]{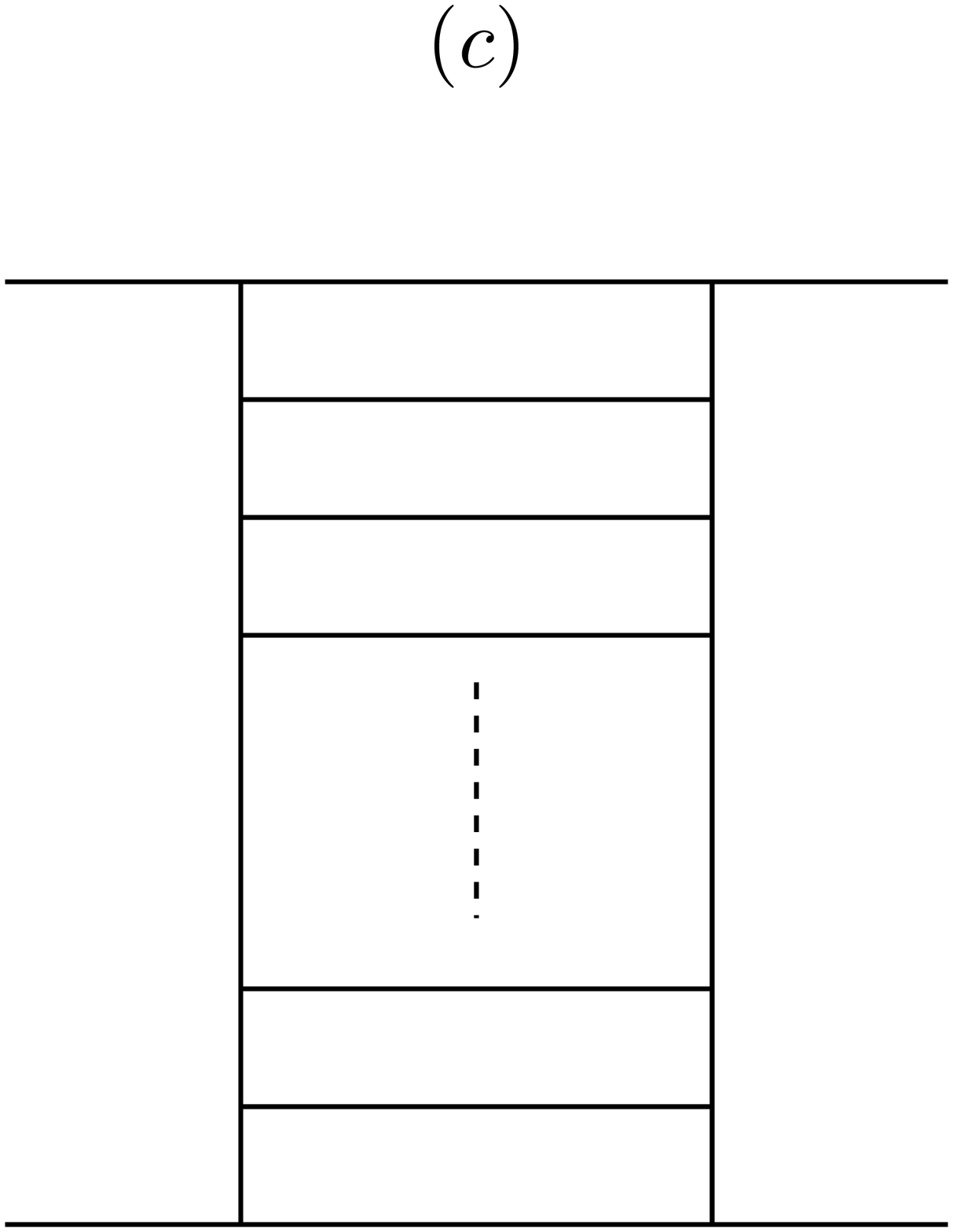}
    \caption{Ladder Feynman diagrams in $\phi^{3}$ field theory. (a) One, (b) two and (c) $n$ rungs.}
    \label{RGCfig2}
  \end{center}
\efg

For a general Feynman integral like the above one, the amplitude for the $2\to 2$ scattering with two independent invariants $s$ and $t$ can be expressed as 
\be
A=\frac{\displaystyle\int^{1}_{0}\displaystyle\prod^{n}_{i=1}d\alpha_{i}\,\delta\left(1-\displaystyle\sum^{n}_{i=1}\alpha_{i}\right)\left[c(\alpha)\right]^{n-2l-2}}{\left[g(\alpha)s+d(t,\alpha)\right]^{n-2l}},
\label{ch5.4}
\ee
where $c$, $d$, and $g$, are some functions \cite{Collins:1977jy}.

The single particle exchange in a two-body scattering process is depicted by the pole diagram in Figure \ref{RGCfig2}(a),
\be
A^{(1)}=\frac{g^{2}}{m^{2}-s}\underset{s\to\infty}{\sim} \frac{g}{s},
\label{ch5.5}
\ee
which is the Born approximation for the $t$-channel scattering amplitude. The next order is represented by the box diagram in Figure \ref{RGCfig2}(b), whose amplitude is given by
\be
A^{(2)}=g^{2}\left(-\frac{g^{2}}{16\pi^{2}}\right)\frac{\displaystyle\int^{1}_{0}\displaystyle\prod^{2}_{i=1}d\alpha_{1}d\beta_{1}\,\delta\left(1-\displaystyle\sum^{2}_{i=1}\alpha_{i}-\displaystyle\sum^{2}_{i=1}\beta_{i}\right)}{\left[\alpha_{1}\alpha_{2}s+d_{2}(\alpha,\beta,t)\right]^{2}}\underset{s\to\infty}{\sim} g^{2}K(t)\,\frac{\log s}{s},
\label{ch5.6}
\ee
where $K(t)$ is written as
\be
K(t) \sim g^{2} \int \frac{d^{2}\kk_{\perp}}{(\kk_{\perp}+m^{2})\left[(\kk_{\perp}+q)^{2}+m^{2}\right]},\, \,\,\,t=-q^{2}\,\,\, \textnormal{and}\,\,\,q\simeq\qq_{\perp}.
\label{ch5.7}
\ee

And for the $n$-rung ladder diagram, see Figure \ref{RGCfig2}(c), 
\be
A^{(n)}=g^{2}\left(-\frac{g^{2}}{16\pi^{2}}\right)^{n-1}(n-1)! \,\frac{\displaystyle\int^{1}_{0}\displaystyle\prod^{2}_{i=1}d\alpha_{i}d\beta_{i}\,\delta\left(1-\displaystyle\sum^{2}_{i=1}\alpha_{i}-\displaystyle\sum^{2}_{i=1}\beta_{i}\right)c^{n-2}(\alpha,\beta)}{\left[\alpha_{1}...\alpha_{n}s+d_{n}(\alpha,\beta,t)\right]^{n}},
\label{ch5.8}
\ee
hence, in the $s\to \infty$, the ladder diagram amplitude can be expressed by
\be
A^{(n)} \sim \frac{g^{2}}{s}\,\frac{\left[K(t)\,\log s\right]^{n-1}}{(n-1)!}.
\label{ch5.9}
\ee

A quick inspection in expressions (\ref{ch5.5}), (\ref{ch5.6}) and (\ref{ch5.8}) show that all the diagrams in Figure \ref{RGCfig2} have a power behavior like $s^{-1}$. This is an indication that just a single-particle propagator is necessary to get across the diagram. However, the power of the $\log s$ terms depends on the number of propagators \cite{Collins:1977jy}. Finally, taking into account the asymptotic behavior of the sum of an infinite series of ladder diagrams with any number of rungs, and also by considering that the asymptotic behavior of the sum is the sum of the asymptotic behaviors, then the resulting amplitude will be given by \cite{Collins:1977jy,Barone:2002cv}
\be
\begin{split}
A(s,t) & = \sum_{n} A^{(n)} \sim \sum^{\infty}_{n=1} \frac{g^{2}}{s}\,\frac{\left[K(t)\log s \right]^{n-1}}{(n-1)!} \\
& \sim \frac{g^{2}}{s}\,e^{K(t)\log s}\sim g^{2}s^{\alpha(t)},
\end{split}
\label{ch5.10}
\ee
where
\be
\alpha(t)=-1+K(t).
\label{ch5.11}
\ee

Therefore, the power behavior of $s$ in expression (\ref{ch5.10}) clearly may be identified with the leading $t$-channel exchange Regge trajectory. Hence, one is able to see that by summing up and infinite number of diagrams, where each one of them contribute with a power of $\log s$ from the successive interactions of the two particle scattering in the $t$-channel, will build the Regge behavior in the large-$s$ limit \cite{Collins:1977jy,Barone:2002cv}. As it was briefly mentioned in the opening remarks of this section, even though this is a simple model of Regge poles in the field theoretical point of view, the central idea that the Regge behavior results as the summation of leading $\log s$ terms is essentially correct \cite{Barone:2002cv}. This idea is carried out in QCD, where the reggeization process arises as the sum of a series of gluon ladders in the high-energy limit.

\section{\textsc{Regge Cuts}}
\label{ch5sec1}
\mbox{\,\,\,\,\,\,\,\,\,}
A single Regge pole exchange can be interpreted as a set of ladder diagrams where it is performed the summation over all number of available rungs. However, in the $\ell$-plane there are other types of singularity known as Regge cuts which arise from the exchange of two or more Reggeons \cite{Collins:1977jy}, as depicted in Figure \ref{RGCfig3}. In terms of Regge theory the cut structure is important to understand $s$-channel unitarity \cite{Barone:2002cv}.
\bfg[hbtp]
  \begin{center}
    \includegraphics[width=5cm,height=4cm]{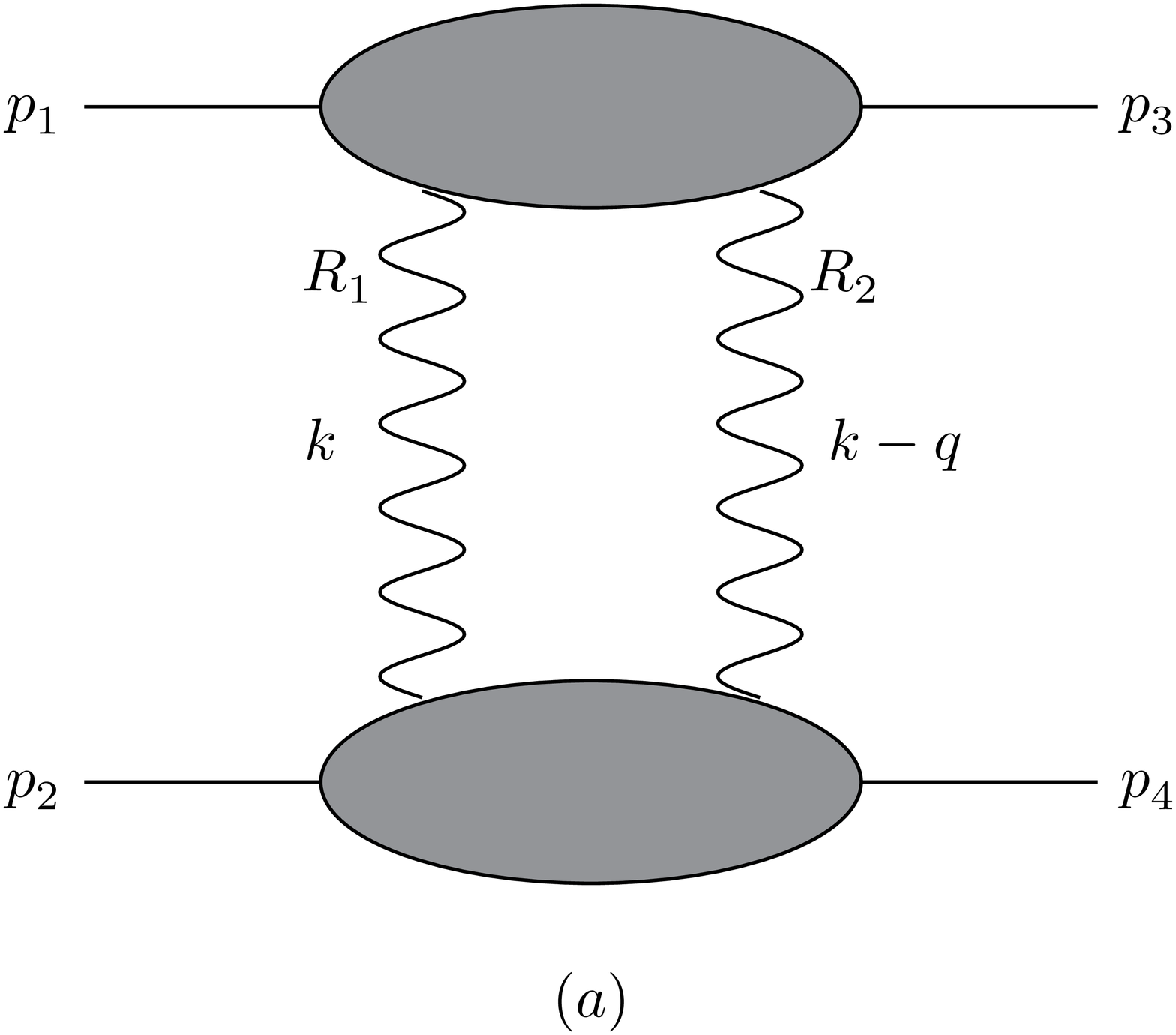}
    \qquad
    \includegraphics[width=5cm,height=4cm]{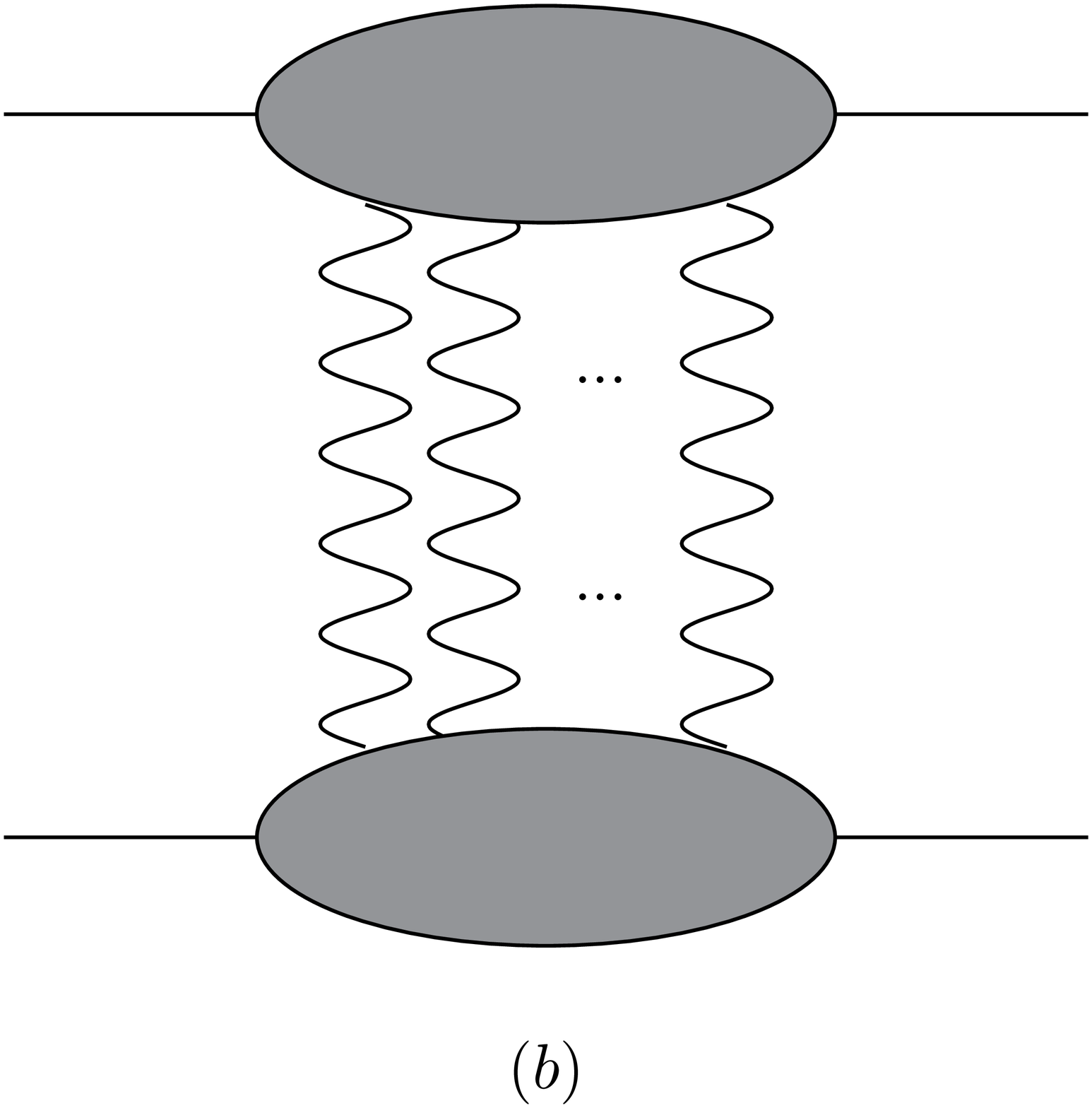}
    \caption{(a) Exchange of two and (b) $n$ Reggeons.}
    \label{RGCfig3}
  \end{center}
\efg

\bfg[hbtp]
  \begin{center}
    \includegraphics[width=5cm,height=5cm]{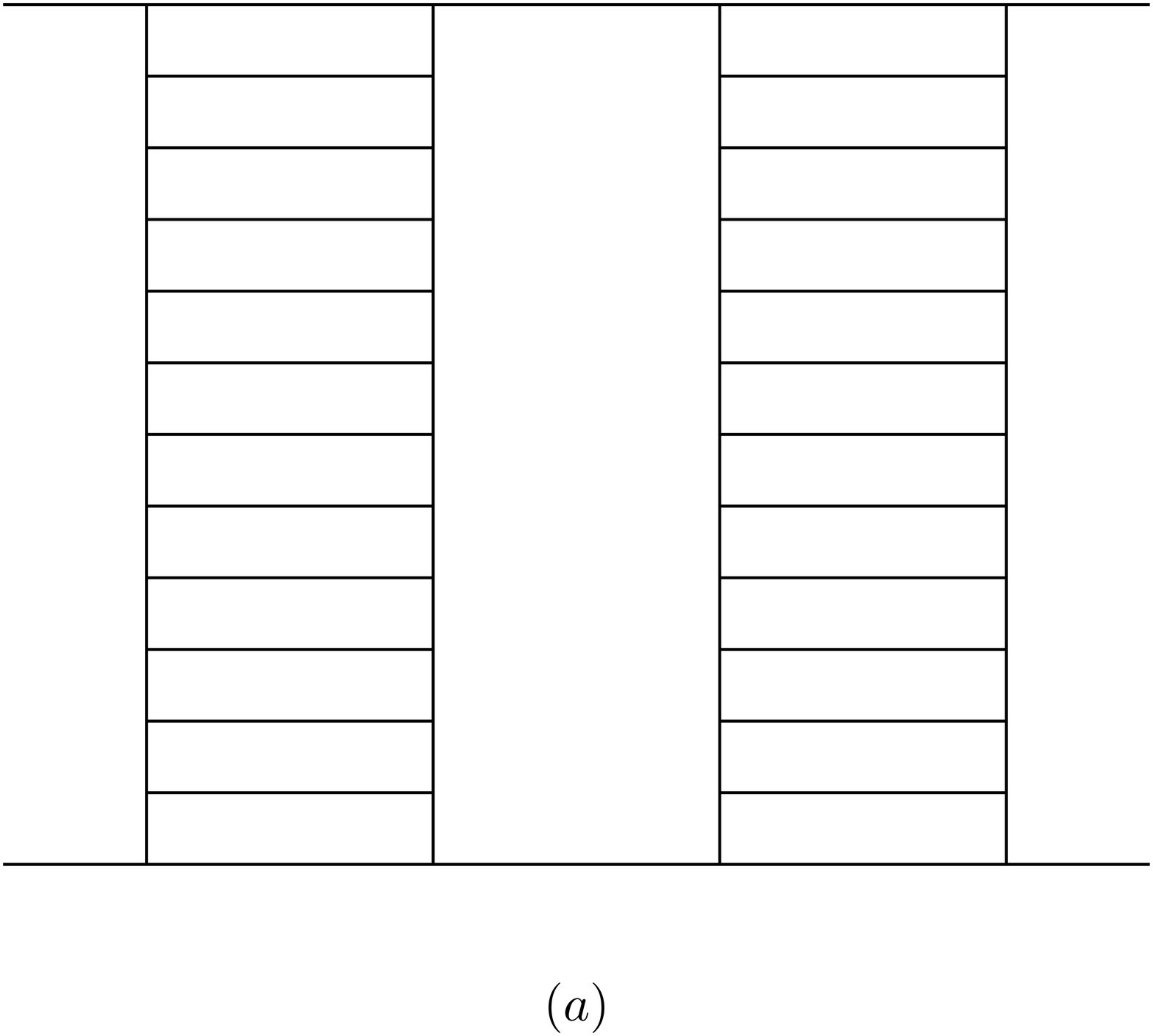}
    \qquad
    \includegraphics[width=5cm,height=5cm]{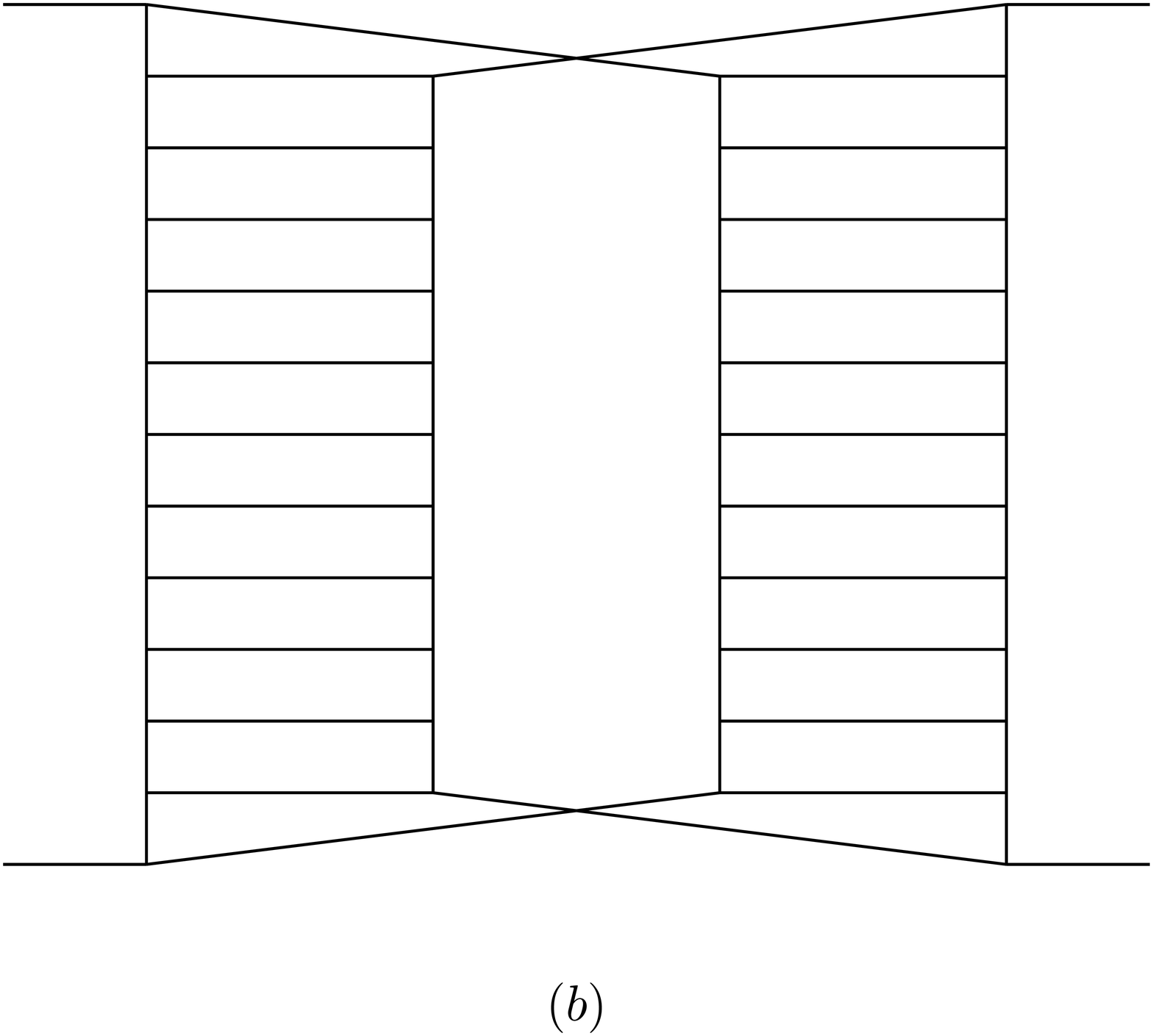}
    \caption{(a) Two-ladder and (b) double-cross diagram.}
    \label{RGCfig4}
  \end{center}
\efg

The simplest case where a Regge cut might be produced corresponds to the exchange of two Reggeons, as shown in Figure \ref{RGCfig3}(a), and also in Figure \ref{RGCfig4} where it can properly be represented by a double ladder diagram. However, it turns out that in all sorts of two-ladder diagrams, like the one in Figure \ref{RGCfig4}(a), the amplitude behaves as \cite{Barone:2002cv}
\be
A(s,t) \underset{s\to\infty}{\sim} s^{-3}\log s,
\label{ch5.12}
\ee
independent of the number of rungs in the two ladders. Moreover, the summation of all such diagrams are expected to have this same asymptotic behavior \cite{Collins:1977jy,Eden:1966dnq}. The Regge behavior comes from the sum over all powers of $\log s$, for this reason by summing this kind of diagrams, one would find that planar diagrams contribute only to the Regge pole. Therefore, the two-ladder diagram in Figure \ref{RGCfig4}(a) does not result in a cut-structure \cite{Collins:1977jy}.

In order to obtain a branch point singularity it is necessary to take into account a more complicated case which is to consider the influence of nonplanar diagrams. The simplest form is the double-cross graph, see Figure \ref{RGCfig4}(b), where its leading behavior for large-$s$ is given by \cite{Donnachie:2002en,Barone:2002cv}
\be
A(s,t) \sim \frac{s^{\alpha_{c}(t)}}{\log s},
\label{ch5.13}
\ee
where the exchange of two Reggeons, and considering that each one of them has a linear trajectory, will yeld a cut in the $\ell$-plane with trajectory given by \cite{Donnachie:2002en}
\be
\alpha_{c}(t)=\alpha_{c}(0)+\alpha^{\prime}_{c}t,
\label{ch5.14}
\ee
with
\bear
\alpha_{c}(0) & = & \alpha_{1}(0)+\alpha_{2}(0) - 1,\\
\label{ch5.15}
\alpha^{\prime}_{c} & = & \frac{\alpha^{\prime}_{1}\alpha^{\prime}_{2}}{\alpha^{\prime}_{1}+\alpha^{\prime}_{2}}.
\label{ch5.16}
\eear

Notice that, the cuts will have lower intercepts than the poles, unless one of the intercepts of Reggeons $1$ or $2$ is greater than or equal to one. At sufficiently large-$\vert t\vert$ values, the cut contribution will dominate over the poles' one in the asymptotic limit of high energies for the amplitude, because the slope $\alpha^{\prime}_{c}$ is smaller than either slopes from both Reggeons \cite{Donnachie:2002en}.

In addition, one might consider how to obtain the contribution from multi-Reggeons diagrams. By means of the above discussion, the intecerpt of the $n$-Reggeon cut is given by
\be
\alpha_{12...n}(0)=\alpha_{1}(0)+\alpha_{2}(0)+...+\alpha_{n}(0)-n+1,
\label{ch5.17}
\ee
and the slope is calculated by following expression (\ref{ch5.14}). Whereupon in a three-Reggeon exchange, for example, the slope would respectively be written as
\be
\alpha^{\prime}_{123}=\frac{\alpha^{\prime}_{1}\alpha^{\prime}_{2}\alpha^{\prime}_{3}}{\alpha^{\prime}_{1}\alpha^{\prime}_{2}+\alpha^{\prime}_{1}\alpha^{\prime}_{3}+\alpha^{\prime}_{2}\alpha^{\prime}_{3}}.
\label{ch5.18}
\ee

An ingenious theoretical tool to calculate multi-Reggeon diagrams was conceived in the $60$'s by Gribov \cite{Gribov:1968fc}. This method consists in inserting Regge poles into Feynman diagrams so that the discontinuities across the angular momentum plane can be deduced. Although this technique is very useful to analyze the analytic structure of Regge amplitudes and to obtain insights of their singularities, it was not able to provide a clear comprehension in how large the two-Reggeon cut contribution really is. Up so foth, no way has been found to sum up all the Regge diagrams and to solve Reggeon field theory.

\section{\textsc{Gribov's Reggeon Calculus}}
\label{ch5sec3}
\mbox{\,\,\,\,\,\,\,\,\,}
As mentioned above, Gribov found a clever way of summing diagrams by using Feynman integrals for the Reggeons' coupling and directly substituting the ladders by Regge lines with propagators $\eta(k^{2})s^{\alpha(k^{2})}$. 

\subsection{\textsc{Two-Reggeon Exchange}}
\mbox{\,\,\,\,\,\,\,\,\,}
For example, by considering the simplest case of multiple-Reggeons exchange as shown in Figure \ref{RGCfig3}(a), the two-Reggeon exchange amplitude is written as \cite{Barone:2002cv}
\be
A^{(2)}(s,t)= -\frac{i}{2!}\int\frac{d^{4}k}{(2\pi)^{4}}\,\eta(k^{2})\,\eta((q-k)^{2})\,s^{\alpha(k^{2})+\alpha((q-k)^{2})}\,T(p_{1},k,q)T(p_{2},k,q),
\label{ch5.19}
\ee
where $T's$ stands for the Reggeon-particle scattering amplitude, and each amplitude represents the sum of many different vertex contributions. The $\eta$ terms are related to the signature factor of the exchanged Reggeons. 

In the asymptotic large-$s$ limit the on-shell condition implies that the exchanged momenta of the intermediate particles are predominantly transverse, \tit{i.e.} $k^{2}\simeq-\kk^{2}_{\perp}$ and $(q-k)^{2}\simeq -(\qq_{\perp}-\kk_{\perp})^{2}$. Therefore, the two-Reggeons exchange amplitude, see Figure \ref{RGCfig5}(a), can be put in the following form \cite{Barone:2002cv}
\be
\begin{split}
A^{(2)}(s,t)& = \frac{i}{2s2!}\int\frac{d^{2}\kk_{\perp}}{(2\pi)^{2}}\,\gamma(\kk^{2}_{\perp})\,\eta(\kk^{2}_{\perp})\,s^{\alpha(\kk^{2}_{\perp})}\\
& \times  \gamma((\qq_{\perp}-\kk_{\perp})^{2})\,\eta((\qq_{\perp}-\kk_{\perp})^{2})\,s^{\alpha((\qq_{\perp}-\kk_{\perp})^{2})}.
\end{split}
\label{ch5.20}
\ee
Notice that the above expression can be rewritten as
\be
A^{(2)}(s,t) = \frac{i}{2s2!}\int\frac{d^{2}\kk_{\perp}}{(2\pi)^{2}}\,A^{(1)}(s,\kk^{2}_{\perp})\,A^{(1)}(s,(\qq_{\perp}-\kk_{\perp})^{2}),
\label{ch5.21}
\ee
where $A^{(1)}$ represents the Regge pole scattering amplitude and $\gamma$ is the residue function. By considering the case where the trajectories are approximately linear at low-$t$ values, the Regge pole amplitude is given by
\be
A^{(1)}(s,\kk^{2}_{\perp})=\gamma(0)\,\eta(0)\,s^{\alpha(0)}\,e^{-\Lambda(s)\kk^{2}_{\perp}},
\label{ch5.22}
\ee
where $\Lambda(s)=B_{0}/2+\alpha^{\prime}(\log s - i\pi/2)$. Hence, integrating expression (\ref{ch5.21}) over the transverse momentum \cite{Barone:2002cv,Collins:1977jy},
\be
A^{(2)}(s,t)=\frac{i}{16\pi}\,\gamma^{2}(0)\,\eta^{2}(0)\,\frac{s^{2\alpha(0)-1}}{2\Lambda(s)}\,e^{\Lambda(s)t/2} \underset{s\to\infty}{\sim} \frac{s^{\alpha_{c}(t)}}{\log s},
\label{ch5.23}
\ee
which are related with a cut singularity located at
\be
\alpha_{c}(t)=2\alpha(0)-1+\frac{\alpha^{\prime}}{2}\,t.
\label{ch5.24}
\ee
\bfg[hbtp]
  \begin{center}
    \includegraphics[width=5cm,height=4cm]{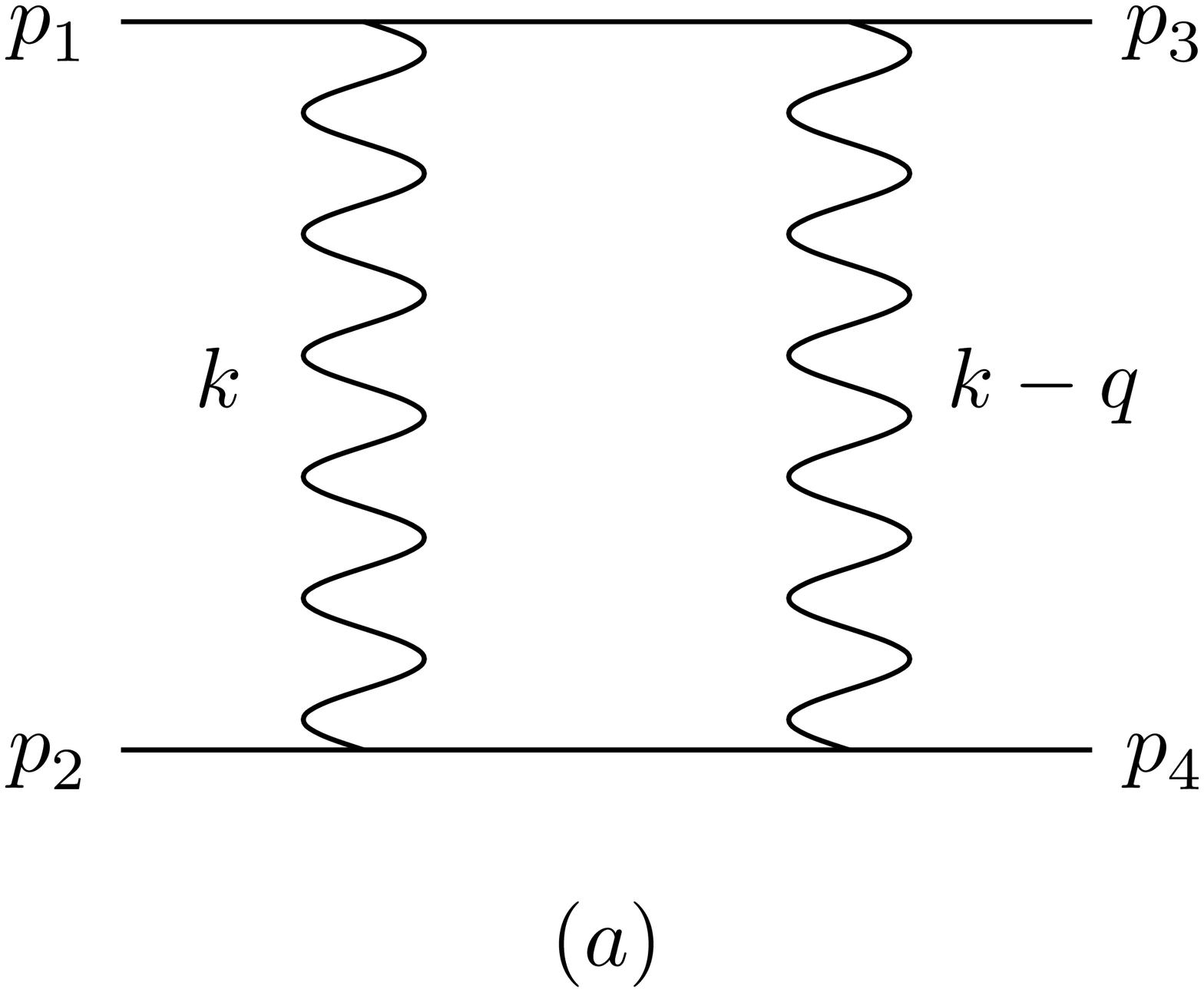}
    \qquad
    \includegraphics[width=6cm,height=4cm]{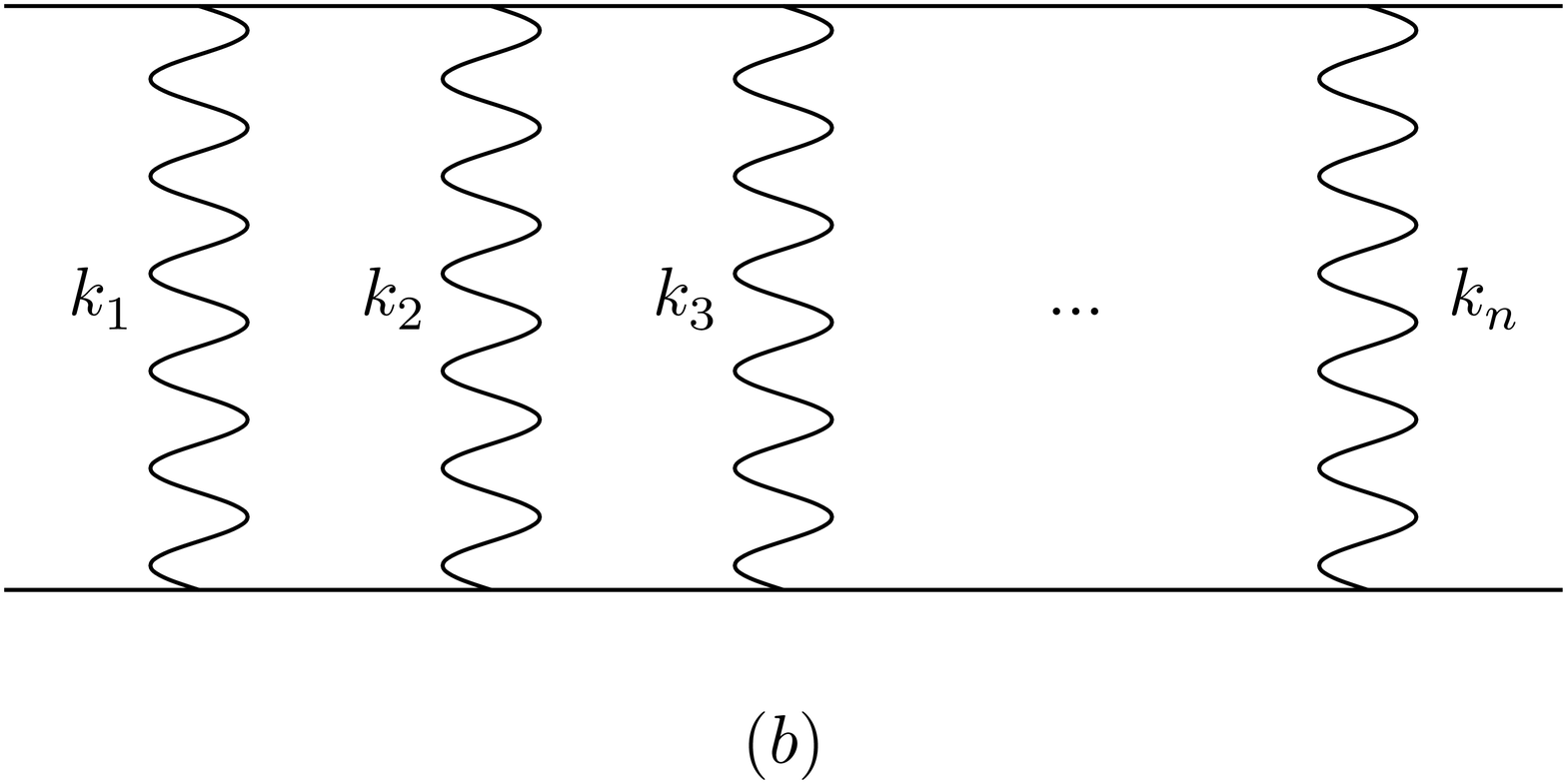}
    \caption{(a) Two-Reggeon and (b) multi-Reggeon exchange.}
    \label{RGCfig5}
  \end{center}
\efg

It is instructive to consider the case of the double-Pomeron exchange. By using the optical theorem and taking into account that $\eta_{\IP}(0)=i$, one finds
\be
\sigma_{tot} \sim A\,s^{\alpha_{\IP}(0)-1} - B\,\frac{s^{2(\alpha_{\IP}(0)-1)}}{\log s},
\label{ch5.25}
\ee
which means that the contribution of $\IP\otimes\IP$ cut to the total cross section is negative, where $A$ and $B$ are positive values. Moreover, because of the $\IP$ signature, the cut amplitude has an opposite phase with respect to the Pomeron pole's one.

\subsection{\textsc{Multi-Reggeon Exchange}}
\mbox{\,\,\,\,\,\,\,\,\,}
It still remains to be checked the scenario where $n$ identical Reggeons are exchanged, see Figure \ref{RGCfig5}(b). Similar to the case of the two-Reggeon, one finds that the $n^{th}$-Reggeon exchange amplitude is written as \cite{Barone:2002cv}
\be
A^{(n)}(s,t)=\frac{i^{n-1}}{n!}\frac{1}{(2s)^{n-1}}\,(2\pi)^{2}\displaystyle\int\displaystyle\prod^{n}_{i=1}\left[\frac{d^{2}\kk_{i\perp}}{(2\pi)^{2}}\,A^{(1)}(s,\kk_{i\perp})\right]\delta\left(\qq_{\perp}-\sum^{n}_{i=1}\kk_{i\perp}\right),
\label{ch5.26}
\ee
again assuming low-$\vert t\vert$ values and linear trajectories, and integrating over the transverse momentum,
\be
A^{(n)}(s,t) \sim i^{n-1}\gamma^{n}(0)\eta^{n}(0)\,\frac{s^{n(\alpha(0)-1)+1)}}{\Lambda^{n-1}(s)}\,e^{\Lambda(s)t/n},
\label{ch5.27}
\ee
where the signature of the cut is the product of the signatures of the poles.

At the asymptotic limit of high energies, this amplitude behaves as
\be
A^{(n)}(s,t)\underset{s\to\infty}{\sim} \frac{s^{\alpha_{c}(t)}}{\log^{n-1} s},
\label{ch5.28}
\ee
respectively with the cut located at
\be
\alpha_{c}(t)=n(\alpha(0)-1)+1+\frac{\alpha^{\prime}}{n}\,t.
\label{ch5.29}
\ee

More generally, for nonlinear trajectories, the position of the cut is \cite{Collins:1977jy}
\be
\alpha_{c}(t)=n\alpha\left(\frac{t}{n^{2}}\right)-n+1.
\label{ch5.30}
\ee

The cuts generated by the exchange of $n$ Pomerons, whose $\alpha_{\IP}(t)=\alpha^{0}_{\IP}+\alpha^{\prime}_{\IP}t$ , gives a branch point respectively at
\be
\alpha_{c}(t)=\alpha^{0}_{\IP}+\frac{\alpha^{\prime}_{\IP}}{n}\,t,
\label{ch5.31}
\ee
which means that the trajectory associated with the cut becomes flatter as $n$ increases. Hence higher order cuts are even flatter and they will lie above lower order cuts for $\vert t\vert<0$. 

It is interesting to notice that whilst the Pomeron pole dominates at low-$\vert t\vert$ values, the double-Pomeron cut contribution becomes important at high-$\vert t \vert$ region, and the destructive interference between the two terms, which have opposite phases because of the Pomeron signature, implies a reduction of the cross section. Moreover, this effect is associated with the dip region observed in $pp$ elastic cross section \cite{Barone:2002cv}. As the number of exchanged Reggeons grows, the trajectories get flatter and flatter, therefore at asymptotic hight-$\vert t\vert$ the scattering process will be dominated by the exchange of many Reggeons. On the one hand this means that a complete description based on Regge theory becomes extremely difficult. On the other hand the region of asymptotically higher values of $\vert t \vert$ belongs to perturbative QCD, thus different theoretical frameworks stars to play an important role.

Returning to the task in hand, all these cuts coincide at $t=0$ approximately at $1$ since $\alpha_{\IP}(0)\approx 1$. Similarly for an $\IR\otimes\IP$ cut will be located at
\be
\alpha_{c}(t)=\alpha^{0}_{\IR}+\left(\frac{\alpha^{\prime}_{\IR}\alpha^{\prime}_{\IP}}{\alpha^{\prime}_{\IR}+\alpha^{\prime}_{\IP}}\right)t,
\label{ch5.32}
\ee
and an $\IR\otimes(\IP)^{(n)}$ cut,
\be
\alpha_{c}(t)=\alpha^{0}_{\IR}+\left(\frac{\alpha^{\prime}_{\IR}(\alpha^{\prime}_{\IP})^{n}}{\alpha^{\prime}_{\IR}+n\alpha^{\prime}_{\IP}}\right)t,
\label{ch5.33}
\ee
so all the cuts coincide with $\alpha_{\IR}(t)$ at $t=0$ and will lie above it at $\vert t\vert<0$ region \cite{Collins:1977jy}.

\section{\textsc{The Eikonal Model}}
\label{ch5sec4}
\mbox{\,\,\,\,\,\,\,\,\,}
As it was previously mentioned, although Reggeon calculus, in fact, can provide a good deal to analyze the properties of Regge cuts, there is no well formulated theory that allows one to calculate the actual strength of the cuts relative to the poles. But there are models, and the most popular is the eikonal model.

The amplitude relative to a typical $n$-rung ladder diagram, where eventually some of the rungs cross each other, is given by\footnote{Firstly, by considering the exchange of scalar particle rather than Reggeons itsel.}
\bear
& A^{(n)}(s,t) = g^{2n}\displaystyle\prod^{n}_{i=1}\displaystyle\int\frac{d^{4}k_{i}}{(2\pi)^{4}}\,\frac{1}{k^{2}_{i}-m^{2}}\,(2\pi)^{4}\delta\left(p_{1}-p^{\prime}_{1}-\displaystyle\sum^{n}_{i=1} k_{i}\right)\nonumber\\
& \times \{\left[(p_{1}-k_{1})^{2}-m^{2}\right]
\left[(p_{1}-k_{1}-k_{2})^{2}-m^{2}\right]
...
\left[(p_{1}-k_{1}-...-k_{n})^{2}-m^{2}\right]\nonumber\\
& \times \left[(p_{2}+k_{1})^{2}-m^{2}\right]
...
\left[(p_{2}+k_{1}+...+k_{n})^{2}-m^{2}\right]
\}^{-1},
\label{ch5.34}
\eear
where $(p_{1}\pm k)^{2}-m^{2}\approx \pm 2p_{1}\cdot k $ by means of the high-energy small angle scattering approximation. Summing over all possible permutations of the ordering of the rungs, it is possible to show that through out some calculations and introducing the impact parameter $b$, one arrives respectively at \cite{Collins:1977jy}
\be
A^{(n)}(s,t)=\frac{ s}{n!}\int^{\infty}_{0} db\,b\,J_{0}(b\sqrt{-t})\,\chi(s,b)\left(i\chi(s,b)\right)^{n-1},
\label{ch5.35}
\ee
and by summing over all possible numbers of rungs, 
\be
A(s,t)=\sum^{\infty}_{n=1}A^{(n)}(s,t)= is\int^{\infty}_{0}db\,b\,J_{0}(b\sqrt{-t})\left[1-e^{i\chi(s,b)}\right].
\label{ch5.36}
\ee

Notice that the first term of the series is associated with the single-particle exchange Born approximation,
\be
A_{B}(s,t)= s\int^{\infty}_{0}db\,b\,J_{0}(b\sqrt{-t})\,\chi(s,b),
\label{ch5.37}
\ee
and inverting, one is able to find what is known as the eikonal function,
\be
\chi(s,b)=\frac{1}{ s}\int^{0}_{-\infty}dt\,J_{0}(b\sqrt{-t})\,A_{B}(s,t).
\label{ch5.38}
\ee

The second term of the amplitude is the summation of all the two-particle exchange diagrams,
\be
A^{(2)}(s,t)= s \int^{\infty}_{0} db\,b\,J_{0}(b\sqrt{-t})\,\frac{i\chi^{2}(s,b)}{2},
\label{ch5.39}
\ee
which inserting expression (\ref{ch5.38}) gives
\bear
& A^{(2)}(s,t)= is \displaystyle\int^{\infty}_{0}db\,b\,J_{0}(b\sqrt{-t})\,\frac{1}{ s^{2}}\nonumber\\
& \times \displaystyle\int^{0}_{-\infty}dt_{1}\,J_{0}(b\sqrt{-t_{1}})\,A_{B}(s,t_{1}) \displaystyle\int^{0}_{-\infty}dt_{2}\,J_{0}(b\sqrt{-t_{2}})\,A_{B}(s,t_{2}).
\label{ch5.40}
\eear

In fact it can be shown that if the particle exchange are replaced by ladders, therefore the eikonal series can be interpreted as the summation over all the Regge cuts due to any given number of Reggeons exchanged \cite{Collins:1977jy,Donnachie:2002en}.

\section{\textsc{Regge Cut Phenomenology}}
\label{ch5sec4}
\mbox{\,\,\,\,\,\,\,\,\,}
Once it is known the Regge pole trajectories and the respective position of each of the branch points in the complex angular momentum plane, one will be able to discover if the multi-Reggeon, more specifically multi-Pomeron, exchange will significantly contribute to the asymptotic large-$s$ behavior of the hadronic scattering amplitude.

As it was repeatedly mentioned in early Chapter \ref{chap3} on Regge theory, the Pomeron represents the Reggeon responsible for the asymptotic behavior of the scattering amplitude at high energies. However, the high-energy limit of hadronic collisions back in the $60$'s, and today (2019) is called low or intermediate energy region, gave experimental indications in favor of not growing cross sections. For this reason, the Pomeron was interpreted as the pole with the largest intercept $\alpha_{\IP}(0)=1$ and thus initiated the study of soft hadronic interactions. This Pomeron intercept, as it was first introduced, saturates the unitarity bound, and the fact is that an intercept slightly larger than one will eventually lead to a violation of the Froissart-Martin-\L ukaszuk bound. 

The asymptotic rise of total hadronic cross sections was already forseen by Grivov. However, in the eve $70$'s scientific community gradually understood that the asymptotic high energy at that time was indeed not that high to be really consider as an asymptotic regime and the true Regge dynamics of scattering amplitude could be accessed only at energies higher by order values \cite{book:832857}. In the late $80$'s the observed rise of all the total hadronic cross sections was directly implying a supercritical Pomeron $\alpha_{\IP}(0)>1$. As shown by Donnachie \& Landshoff \cite{Donnachie:1992ny}, accounting for secondary Reggeons, total cross sections could be described by a supercritical Pomeron intercept $\alpha_{\IP}(0)-1=\epsilon\simeq 0.08$.

On the one hand, the influence of Regge cuts such as: $\IP\IP$, $\IP\IP\IP$ and so on, indeed indicate the preasymptotic growth of total cross section within $\alpha_{\IP}(0)=1$. On the other hand, this growth is not enough to be related with the actual observed behavior. Therefore the concept of a supercritical Pomeron pole became widely accepted. The physics beyond the exchange of single Pomeron poles should ensure the preservation of unitarity, and it is known that multiple Pomeron exchanges are, in fact, able to tame the asymptotic rise of the total cross section, even though the break of unitarity only occurs in astonishing high energies. Another point of view to way out this dilemma is to say that an intercept $\alpha_{\IP}(0)=1+\epsilon>1$ is just an effective intercept which is the result of not just entirely single Pomeron exchange, but has contributions from the Regge cuts \cite{Forshaw:1997dc}.

In fact, the present scenario is unquestionable. The experimental data available on hadronic collisions so far suggest a Pomeron intercept larger than one. But on the theoretical side,  the study of the singularity of the Pomeron exact Green's function by means of Reggeon perturbation theory implies that the bare Pomeron intercept may, indeed, be slightly greater than one \cite{Baker:1976cv}, and this can only be compatible with the Froissart-Martin-\L ukaszuk bound if unitarization produces strong canceling cuts \cite{Collins:1977jy}.

\subsection{\textsc{Total Cross Section}}
\mbox{\,\,\,\,\,\,\,\,\,}
As it was mentioned in the previous section, the Regge cuts are responsible to restore the $s$-channel unitarity. However, it is interesting to notice that the eikonal model may provide a way to achieve the Froissart-Martin-\L ukaszuk bound even if there is a Reggeon contribution with an intercept larger than one. Considering this Reggeon as the usual Pomeron pole with $\alpha_{\IP}> 1$ and approximately with a linear trajectory, more specifically taking into account only the Pomeron contribution since it is responsible for the large-$s$ dynamics, the scattering amplitude shall be given as\footnote{Consider an implicit division by $s_{0}$.} \cite{Collins:1977jy,Barone:2002cv}
\be
A_{\IP}(s,t)=\gamma(t)\left[e^{-i\pi\alpha_{\IP}(t)/2}\right]s^{\alpha_{\IP}(t)},
\label{ch5.41}
\ee
where $\gamma(t)$ stands for the residue function and here it represents the Pomeron-hadron vertex and will be parametrized as $\gamma(t)=\beta_{\IP}e^{at}$ and the term in square brackets is the low-$t$ Pomeron signature.

The first term in the eikonal expansion stands for the Born-level amplitude. Therefore, inverting the Fourier-Bessel transform and by considering the Born-level as the Pomeron pole amplitude, the eikonal function will be given by
\be
\chi(s,b)=\frac{1}{s}\int^{0}_{-\infty}dt\,J_{0}(b\sqrt{-t})\,\beta_{\IP}e^{at}\left[e^{-i\pi\alpha_{\IP}(t)/2}\right]s^{\alpha_{\IP}(t)},
\label{ch5.42}
\ee
and using the relation $\log(-is)=\log(s)-i\pi/2$, one easily finds
\be
\chi(s,b)=\beta_{\IP}e^{-i\pi\alpha^{0}_{\IP}/2}s^{\alpha^{0}_{\IP}-1}\int^{0}_{-\infty}dt\,J_{0}(b\sqrt{-t})\,e^{\left[a+\alpha^{\prime}_{\IP}\log(-is)\right]t}.
\label{ch5.43}
\ee
By means of the following relation \cite{Collins:1977jy}
\be
\int^{0}_{-\infty}dt\,J_{n}(b\sqrt{-t})e^{ct}(-t)^{n/2+m}=\left(\frac{b}{2}\right)^{n}\left(-\frac{\partial}{\partial c}\right)^{m}\left[\frac{e^{-b^{2}/4c}}{c}\right],
\label{ch5.44}
\ee
one is able to arrive at
\be
\chi(s,b)=\frac{\beta_{\IP}e^{-i\pi\alpha^{0}_{\IP}/2}e^{(b^{2}_{0}-b^{2})/4\alpha^{\prime}_{\IP}\log(-is)}}{\alpha^{\prime}_{\IP}\log(-is)},
\label{ch5.45}
\ee
where for simplicity it was assumed $a=0$ and defined $b^{2}_{0}=4(\alpha^{0}_{\IP}-1)\alpha^{\prime}_{\IP}\log^{2}s$. When $s\to \infty$ one sees that $\textnormal{Re}\chi\to \infty$, unless $b^{2}<b^{2}_{0}$. Hence in the limit and by applying the optical theorem in the forward eikonalized amplitude, see expression (\ref{ch5.36}), one arrives at the asymptotic form of the total cross section,
\be
\sigma_{tot}(s)=\frac{4\pi}{s}\textnormal{Im}\,A_{eik}(s,t=0)=4\pi\int^{b_{0}}_{0}db\,b=8\pi(\alpha^{0}_{\IP}-1)\alpha^{\prime}_{\IP}\log^{2}s.
\label{ch5.46}
\ee
Therefore, the Froissart-Martin-\L ukaszuk bound is satisfied provided that $8\alpha^{\prime}_{\IP}m^{2}_{\pi}\epsilon<1$.

\clearpage
\thispagestyle{plain}

\chapter{\textsc{Regge-Gribov Based Model}}
\label{ch6}
\mbox{\,\,\,\,\,\,\,\,\,}
It is well known that good descriptions of forward data up to the Tevatron energy have been obtained by using a linear Pomeron trajectory \cite{Donnachie:1992ny,Cudell:2002xe,Cudell:1996sh,Covolan:1996uy,Luna:2003kw,Luna:2004gr,Luna:2004zp,Avila:2002tk}, namely $\ap(t)=1+\epsilon+\alp t$. The energy dependence of the total and diffractive cross sections is driven by $\e$ while $\alp$ determines the energy dependence of the forward slopes. However, for example, ZEUS and H$1$ low-$t$ data for exclusive $\rho$ and $\phi$ photoproduction call forth a rather nonlinear Pomeron trajectory \cite{Derrick:1992kk,Ahmed:1992sp,Breitweg:1997aa}. Most recently, the TOTEM experiment at the CERN LHC has released new data on total cross sections and diffractive processes \cite{Antchev:2017dia,Antchev:2017yns} that have enhanced the interest in high-energy hadron physics and become a pivotal source of information for selecting theoretical methods and models. 

Currently, these measurements provide an unique constraint on the Pomeron parameters and allow us to study its behavior more thoroughly since the contribution of the Pomeron component to the $\chi^{2}$ is absolutely dominant in the LHC regime. Hence, the TOTEM data allow us to address more effectively the question of linearity versus nonlinearity of the Pomeron trajectory. Moreover, the small value of $\alp$ usually obtained from screened Regge models indicates that the soft Pomeron may be treated perturbatively since in the Gribov Reggeon calculus the mean transverse momentum of the partons is given by $\langle p_{T}\rangle=1/\sqrt{\alp}$ \cite{Gribov:1968fc,Gribov:1962fx,Gribov:1968ia,Baker:1976cv}. This perturbative approach sets up the stage for building a fundamental theory for soft processes based upon QCD. The screening effects can be calculated in terms of a two-channel eikonal model and again the TOTEM data are instrumental in determining the effects of the eikonalization on the Pomeron parameters in both the one- and two-channel models.

Given the central role that the soft Pomeron plays in strong processes, its close scrutiny continues to be a core task in hadron physics. This Chapter is devoted to a detailed study of the soft Pomeron at the light of these recent LHC data \cite{Antchev:2017dia,Antchev:2017yns}. More precisely, it will be evaluated the relative plausibilities of different combinations of vertices and trajectories of the soft Pomeron.

\section{\textsc{Born-Level Analysis}}
\label{secRGb.2}
\mbox{\,\,\,\,\,\,\,\,\,}
The forward Born-level Regge amplitude introduced some time ago by Donnachie and 
Landshoff has two contributions \cite{Donnachie:1992ny}, one representing an effective single Pomeron and the other representing the exchange of the highest-spin meson trajectories, namely $a_{2}$, $f_{2}$, $\omega$ and $\rho$. However, more recent analysis have indicated that the assumption of degeneracy of the mesons trajectories is not supported by the forward data \cite{Cudell:1996sh,Covolan:1996uy,Luna:2003kw,Luna:2004gr,Luna:2004zp}. The best result are obtained with a Born-level amplitude decomposed into three contributions,
\be
A_{B}(s,t)=A_{\IP}(s,t)+A_{+}(s,t)+\tau\,A_{-}(s,t).
\label{chRGb.12}
\ee
The term $A_{\IP}(s,t)$ represents the exchange of the Pomeron, $A_{+}(s,t)$ the exchange of the Reggeons with $C=+1$, namely $a_{2}$ and $f_{2}$, and $A_{-}(s,t)$ that of the Reggeons with $C=-1$, namely $\omega$ and $\rho$. Specifically, the amplitude for single exchange is
\be
A_{i}(s,t)=\beta^{2}_{i}(t)\eta_{i}(t)\left(\frac{s}{s_{0}}\right)^{\alpha_{i}(t)},
\label{chRGb.13}
\ee
where $\beta_{i}$ is the elastic proton-Reggeon vertex, $\eta_{i}(t)$ is the signature factor and $\alpha_{i}(t)$ is the Regge pole trajectory, with $i=\IP,+,-$. Here $s_{0}$ is a mass scale usually chosen to be of the order of $1$ GeV$^{2}$. By comparing expressions (\ref{chRGb.1}) and (\ref{chRGb.13}) it can be seen that the residue function factorizes as $\gamma_{i}(t)=\beta^{2}_{i}(t)$. The signature factor is given by expression (\ref{chRGb.2}), where $\xi=+1$ for the Pomeron and the Reggeons $a_{2}$ and $f_{2}$, and $\xi=-1$ for the Reggeons $\omega$ and $\rho$. Thus, the $pp$ and $\bar{p}p$ scatterings are described in terms of Pomeron, positive- and negative-signature Regge exchange. 

However, in order to simplify the numerical calculations involved in the forthcoming eikonal analysis, it was adopted the asymptotic form of the signatures at the very low-$t$ region, namely $\eta_{i}(t)=-e^{-i\pi\alpha_{i}(t)/2}$ for even-signature trajectories and $\eta_{i}(t)=ie^{-i\pi\alpha_{i}(t)/2}$ for odd-signature ones \cite{Covolan:1996uy}. The choice of these simplified signatures do not affect the Pomeron parameters $\e$ and $\alp$, but simply introduces the vertex transformations,
\be
\begin{split}
&\beta^{2}_{\IP}(t)\to\sin\left[\frac{\pi}{2}\,\ap(t)\right]\beta^{2}_{\IP}(t),\\
&\beta^{2}_{+}(t)\to\sin\left[\frac{\pi}{2}\,\aplus(t)\right]\beta^{2}_{+}(t),\\
&\beta^{2}_{-}(t)\to-\cos\left[\frac{\pi}{2}\,\amin(t)\right]\beta^{2}_{-}(t).
\end{split}
\label{chRGb.14}
\ee
\mbox{\,\,\,\,\,\,\,\,\,}
Therefore, using these simplified form of the Reggeon signatures, each term in the Born-level amplitude, see expression (\ref{chRGb.3}), can be separated into its real and imaginary parts, respectively. For the Pomeron contribution,
\be
A_{\IP}(s,t)=-\beta^{2}_{\IP}(t)\cos\left[\frac{\pi}{2}\,\ap(t)\right]\left(\frac{s}{s_{0}}\right)^{\ap(t)}\!\!\!+i\beta^{2}_{\IP}(t)\sin\left[\frac{\pi}{2}\,\ap(t)\right]\left(\frac{s}{s_{0}}\right)^{\ap(t)},
\label{chRGb.15}
\ee
for Reggeons with $C=+1$,
\be
A_{+}(s,t)=-\beta^{2}_{+}(t)\cos\left[\frac{\pi}{2}\,\aplus(t)\right]\left(\frac{s}{s_{0}}\right)^{\aplus(t)}\!\!\!+i\beta^{2}_{+}(t)\sin\left[\frac{\pi}{2}\,\aplus(t)\right]\left(\frac{s}{s_{0}}\right)^{\aplus(t)},
\label{chRGb.16}
\ee
and the term corresponding to Reggeons with $C=-1$,
\be
A_{-}(s,t)=\beta^{2}_{-}(t)\sin\left[\frac{\pi}{2}\,\amin(t)\right]\left(\frac{s}{s_{0}}\right)^{\amin(t)}\!\!\!+i\beta^{2}_{-}(t)\cos\left[\frac{\pi}{2}\,\amin(t)\right]\left(\frac{s}{s_{0}}\right)^{\amin(t)}.
\label{chRGb.17}
\ee
Thus the complete real and imaginary Born-level amplitude is written by
\be
\textnormal{Re}\,A_{B}(s,t)=\textnormal{Re}\,A_{\IP}(s,t)+\textnormal{Re}\,A_{+}(s,t)+\tau\,\textnormal{Re}\,A_{-}(s,t),
\label{chRGb.18}
\ee
\be
\textnormal{Im}\,A_{B}(s,t)=\textnormal{Im}\,A_{\IP}(s,t)+\textnormal{Im}\,A_{+}(s,t)+\tau\,\textnormal{Im}\,A_{-}(s,t),
\label{chRGb.19}
\ee
where $\tau$ flips sign when going from $pp\,(\tau=-1)$ to $\bar{p}p\,(\tau=+1)$.

The positive-signature secondary Reggeons, namely $a_{2}$ and $f_{2}$, are taken to have an exponential form for the proton-Reggeon vertex,
\be
\beta_{+}(t)=\beta_{+}(0)e^{b_{+}t/2},
\label{chRGb.20}
\ee
and to lie on an exchange-degenerate linear trajectory of form,
\be
\alpha_{+}(t)=1-\eta_{+}+\alplus t,
\label{chRGb.21}
\ee
where $\alpha_{+}(0)=1-\eta_{+}$. Similarly, the exchange-degenerate negative-signature secondary Reggeons, namely $\omega$ and $\rho$, are described by the parameters $\beta_{-}(0)$, $b_{-}$, $\eta_{-}$ and $\almin$.

For Pomeron exchange it will be investigated two different types of proton-Pomeron vertex and two different types of trajectory, one of which being nonlinear. The methodology is, using the standard statistical $\chi^{2}$ test, to evaluate the relative plausibilities of these vertices and trajectories in the light of the LHC data, \tit{i.e.} to consider different combinations of $\beta_{\IP}$ and $\ap(t)$, and the effectiveness of these combinations at describing the high-energy forward data. 

In the first combination, referred to as BI model, it will be adopted an exponential form for the proton-Pomeron vertex,
\be
\beta_{\IP}(t)=\beta_{\IP}(0)e^{b_{\IP}t/2},
\label{chRGb.22}
\ee
and a linear Pomeron trajectory,
\be
\alpha_{\IP}(t)=\ap(0)+\alp t,
\label{chRGb.23}
\ee
where henceforth is defined $\ap(0)\equiv 1+\e$. 

In the second model, named BII, it will be adopted an exponential vertex, see expression (\ref{chRGb.15}) and the nonlinear Pomeron trajectory \cite{Anselm:1972ir,Khoze:2000wk,Luna:2008pp,Luna:2010ch},
\be
\ap(t)=\ap(0)+\alp t-\frac{\beta^{2}_{\pi}m^{2}_{\pi}}{32\pi^{3}}\,h\left(\frac{4m^{2}_{\pi}}{\vert t\vert}\right),
\label{chRGb.24}
\ee
where
\be
h(x)=\frac{4}{x}\,F^{2}_{\pi}(t)\left[2x-(1+x)^{3/2}\log\left(\frac{\sqrt{1+x}+1}{\sqrt{1+x}-1}\right)+\log\left(\frac{m^{2}}{m^{2}_{\pi}}\right)\right],
\label{chRGb.25}
\ee
with $x=4m^{2}_{\pi}/\vert t\vert$, $m_{\pi}=139.6$ MeV and $m=1$ GeV. The nonlinear term in the Pomeron trajectory comes from the nearest $t$-channel singularity, a two-pion loop \cite{Anselm:1972ir}. In the above expression $F_{\pi}(t)$ is the form factor of the pion-Pomeron vertex, for which it will be taken the standard pole expression $F_{\pi}(t)=\beta_{\pi}/(1-t/a_{1})$. The coefficient $\beta_{\pi}$ specifies the value of the pion-Pomeron coupling and for this it will be adopted the additive quark model relation $\beta_{\pi}/\beta_{\IP}=2/3$. 

In the third combination, BIII model, we adopt the nonlinear Pomeron trajectory, see expression (\ref{chRGb.24}) and the power-like form for the proton-Pomeron vertex \cite{Anselm:1972ir,Khoze:2000wk,Luna:2008pp,Luna:2010ch,Khoze:2014nia},
\be
\beta_{\IP}(t)=\frac{\beta_{\IP}(0)}{(1-t/a_{1})(1-t/a_{2})},
\label{chRGb.26}
\ee
where the free parameter $a_{1}$ is the same as the one in the form factor of the pion-Pomeron vertex $F_{\pi}(t)$. 

The total cross section, the elastic differential cross section and the $\rho$-parameter are expressed in terms of the amplitude in expression (\ref{chRGb.3}),
\be
\begin{split}
\sigma_{tot}(s)&=\frac{4\pi}{s}\,\textnormal{Im}\,A(s,t=0)\\
&=Xs^{\e}+Y_{+}s^{-\eta_{+}}+\tau\,Y_{-}s^{-\eta_{-}},
\end{split}
\label{chRGb.27}
\ee
\be
\frac{d\sigma}{d\vert t \vert}(s,t)=\frac{\pi}{s^{2}}\,\left\vert \textnormal{Im}\,A(s,t)\right\vert^{2},
\label{chRGb.28}
\ee
\be
\rho(s)=\frac{\textnormal{Re}\,A(s,t=0)}{\textnormal{Im}\,A(s,t=0)},
\label{chRGb.29}
\ee
where in the above expressions the scattering amplitude is $A(s,t)=A_{B}(s,t)$. The parameters $X$ and $Y_{\pm}$ represents the imaginary parts of the forward scattering amplitude, and are respectively given by
\be
X\equiv 4\pi\beta^{2}_{\IP}(0)\sin\left[\frac{\pi}{2}\,\ap(0)\right]s^{-(1+\e)}_{0},
\ee
\be
Y_{+}\equiv 4\pi\beta^{2}_{+}(0)\sin\left[\frac{\pi}{2}\,\aplus(0)\right]s^{-(1+\eta_{+})}_{0},
\ee
\be
Y_{-}\equiv 4\pi\beta^{2}_{-}(0)\cos\left[\frac{\pi}{2}\,\amin(0)\right]s^{-(1+\eta_{-})}_{0}.
\ee

\subsection{\textsc{Double-Pomeron Exchange}}
\label{secRGb2.1}
\mbox{\,\,\,\,\,\,\,\,\,}
The lack of a footprint of unitarization breaking up to LHC energies can be confirmed by investigating the role of multiple Pomeron exchanges on the scattering amplitude. Unfortunately, despite the advances in theoretical understanding of the Pomeron in the last four decades, we still do not know how to do it. However, there is a consensus that the contribution of the double-Pomeron exchange $(\IP\IP)$ is negative and has energy dependence $s^{\alpha_{\IP\IP}(t)}$ divided by some function of $\log s$ \cite{Donnachie:2013xia,Collins:1977jy}, where
\be
\alpha_{\IP\IP}(t)=1+2\e+\frac{1}{2}\,\alp t.
\label{chRGb.30}
\ee

Thus, the $\IP\IP$ contribution is flatter in $t$ than the single $\IP$ exchange, becoming more important for higher values of $t$. In order to estimate an upper bound on the ratio $R\equiv\beta^{2}_{\IP\IP}(0)/\beta^{2}_{\IP}(0)$, we add the phenomenological term to the scattering amplitude in expression (\ref{chRGb.3}),
\be
A_{\IP\IP}(s,t)=-\beta^{2}_{\IP\IP}(t)\,\eta_{\IP\IP}(t)\left(\frac{s}{s_{0}}\right)^{\alpha_{\IP\IP}(t)}\left[\log\left(\frac{s}{s_{0}}\right)\right]^{-1},
\label{chRGb.31}
\ee
where $\eta_{\IP\IP}(t)=-e^{-i\pi\alpha_{\IP\IP}(t)/2}$ and $\beta_{\IP\IP}(t)=\beta_{\IP\IP}(0)e^{b_{\IP}t/4}$. We include this double-Pomeron exchange term in the model BI. This combination is henceforth called BI+$\IP\IP$ or either simply referred as BIV model.

\section{\textsc{Eikonal Analysis}}
\label{secRGb.3}
\mbox{\,\,\,\,\,\,\,\,\,}
It is well known that expression (\ref{chRGb.13}) leads to total cross sections which violate the Froissart-Martin bound. The usual justification for using such Born-level models is that it is viewed as an effective scattering amplitude where the unitarity violation only occurs at extremely high energies far beyond LHC. 

As it was mentioned in the previous section, in the case of Born-level amplitudes the breakdown of unitarity can be avoided by introducing the exchange series $\IP+\IP\IP+\IP\IP\IP+...$ in such a way that the amplitude in expression (\ref{chRGb.13}) is some kind of ``Born approximation'', that is why it is usually called as Born-level amplitude. Although some general analytic properties of these multiple-exchange terms are known, it is less clear how to carry out a full computation of them. On the other hand, it is well established that eikonalization is an effective procedure to take into account some properties of high-energy $s$-channel unitarity. 

The Froissart-Martin bound imposes a strict restriction on the rate of growth of any total cross section. It is worth mentioning that while the Froissart-Martin bound holds for all eikonalized amplitudes studied here, it is not necessarily synonymous with total unitarization: it was shown some time ago that any model for input Pomeron with intercept $\alpha_{\IP}(0) > 1$, but with linear trajectory, is affected by small asymptotic violations of unitarity \cite{Giffon:1996gm}. We also notice that eikonal unitarization corresponds to one of the two solutions of the unitarity equation
\be
\Gamma(s,b) = \frac{1}{2} \left[ 1 \pm \sqrt{1-4G_{inel}(s,b)} \right],
\label{unitar008}
\ee
the one with minus sign. Choosing the plus sign in (\ref{unitar008}) we get the alternative solution \cite{Troshin:2017zmg}
\be
\Gamma(s,b) = \frac{\textnormal{Im}\,\tilde{\chi}(s,b)}{1-i\tilde{\chi}(s,b)} ,
\label{unitar009}
\ee
where $\tilde{\chi}(s,b)$ is the analogue of the eikonal $\chi(s,b)$. Thus, we see that different unitarization procedures are possible in impact parameter representation. 

Thus in the sense to restore the Froissart-Martin bound, the Born-level amplitude is eikonalized using the impact parameter representation, see expression (\ref{ch2.100}),
\be
A_{eik}(s,t)=is\I db\,b\,J_{0}(b\sqrt{-t})\left[1-e^{i\chi(s,b)}\right],
\label{chRGb.37}
\ee 
where the eikonalized amplitude is written in terms of a complex eikonal function $\chi(s,b)=\chi_{_{R}}(s,b)+i\chi_{_{I}}(s,b)$. As long as the Pomeron term in the Born-level amplitude represents a single-Pomeron exchange, one in principle could expand $1-e^{i\chi(s,b)}$ and connect each term to the exchange series,
\be
1-\sum^{\infty}_{n=0}\frac{(i\chi)^{n}}{n!}=-i\chi+\frac{\chi^{2}}{2!}+\frac{i\chi^{3}}{3!}+...\leftrightarrow \IP+\IP\IP+\IP\IP\IP+...
\label{chRGb.38}
\ee
\mbox{\,\,\,\,\,\,\,\,\,}
This is not an absolute truth, because the analyticity properties of poles and cuts in multiple-Pomeron exchange is much more complicated than a simple exponential expansion, and therefore does not tell us the whole story. Despite the criticism, this is just a phenomenological way to give some meaning to eikonal unitarization, which as mentioned above is just one possible solution of the unitarity equation \cite{Desgrolard:1999pr,Giffon:1996gm,Troshin:2017zmg}. The procedure consists by expanding the exponential up to the first order, \tit{i.e.} retaining only the first term which is linear in $\chi$, then the first-approximation amplitude, identified as the Born-level amplitude, is given by
\be
A_{B}(s,t)=s\I db\,b\,J_{0}(b\sqrt{-t})\,\chi(s,b),
\label{chRGb.39}
\ee
hence, the eikonal function is related to the Born-level amplitude by the Fourier-Bessel transform,
\be
\chi(s,b)=\frac{1}{s}\I d\sqrt{-t}\,\sqrt{-t}\,J_{0}(b\sqrt{-t})\,A_{B}(s,t),
\label{chRGb.40}
\ee
and then it is inserted back into expression (\ref{chRGb.30}) to provide the ``full eikonalized'' amplitude. The total cross section, the elastic differential cross section and the $\rho$-parameter are calculated using expressions (\ref{chRGb.20}-\ref{chRGb.22}) with $A(s,t)=A_{eik}(s,t)$, see expressions (\ref{ch2.110}-\ref{ch2.112}).

\subsection{\textsc{Two-Channel Eikonal Model}}
\label{secRGb3.1}
\mbox{\,\,\,\,\,\,\,\,\,}
An effective Pomeron intercept $\alpha_{\IP}(0) > 1$ is obtained taking into account multi-Pomeron cuts (moving branch points) in the $j$-plane. These singularities are required in order to assure $s$-channel unitarity. In the models considered in the preceding sections we have not accounted for the possibility of diffractive proton excitation in intermediate states, such as $p\to N^{\ast}$. However, it is possible to incorporate the $s$-channel unitarity with elastic and a low-mass intermediate state $N^{\ast}$ by using a two-channel eikonal approach. 

The Good-Walker formalism \cite{Good:1960ba} provides an elegant and convenient form to incorporate $p\to N^{\ast}$ diffractive dissociation. In this approach we introduce diffractive eigenstates $\vert\phi_{i}\rangle$ that diagonalise the interaction $T$-matrix, where $S=\mathds{1}+iT$. As a result the incoming hadron wave functions $\vert h \rangle$ (in our case the `beam'
and `target' proton wave functions) can be written as superpositions of these diffractive eigenstates, namely 
\be
\vert h \rangle_{beam} = \sum_{i} a_{ki}\vert\phi_{i}\rangle, \,\,\,\,\,\textnormal{and}\,\,\,\,\, \vert h \rangle_{target} = \sum_{k} a_{ik}|\phi_{k}\rangle.
\ee
Since we need at least two diffractive eigenstates, in a two-channel eikonal model $i,k=1,2$. The extension to $n$-channel eikonal models is straightforward, however, it is well known that a two-channel model is sufficient to capture the single- or double-diffractive dissociation behavior very accurately \cite{Khoze:2000wk,Luna:2008pp,Luna:2010ch,Khoze:2014nia,Khoze:2000cy,Gotsman:1999ri,Gotsman:1993vd}. In this Thesis we adopt a two-channel eikonal model in which the Pomeron couplings to the two diffractive eigenstates $k$ are
\be
\beta_{\IP,k}(t)=(1\pm \gamma)\beta_{\IP}(t),
\ee
\tit{i.e.}, the eigenvalues of the two-channel vertex are $1\pm \gamma$, where $\gamma \simeq 0.55$ \cite{Luna:2008pp,Luna:2010ch}. This value is in accordance with $p \to N^{\ast}$ dissociation observed at CERN-ISR energies, more specifically, it is the value required in order to obtain the experimental value of the cross section for low-mass diffraction, namely $\sigma_{SD}^{lowM} \simeq 2$ mb, measured at $\sqrt{s}=31$ GeV. 

Since each amplitude has two vertices, the total cross section is controlled by an elastic scattering amplitude with three different exponents,
\be
A_{eik}(s,t) = is \int^{\infty}_{0} b\, db\, J_{0}(bq) \left[ 1 -\frac{1}{4}\, e^{i(1+\gamma)^{2}\chi (s,b)} -\frac{1}{2}\, e^{i(1+\gamma^{2})\chi (s,b)} - \frac{1}{4}\, e^{i(1-\gamma)^{2}\chi (s,b)} \right].
\ee

In the calculation of the eikonal function (\ref{chRGb.40}) the input amplitudes $(A_{Born} (s,t))$ are simply the ones related to BI,
BII and BIII models. These single-channel eikonal models are referred to, respectively, as OI, OII and OIII models, and as for the case of the two-channel ones, TI, TII and TIII models, respectively.

\section{\textsc{Results and Conclusions}}

\mbox{\,\,\,\,\,\,\,\,\,}
In our analyses we carry out global fits to forward $pp$ and $\bar{p}p$ scattering data above $\sqrt{s}=10$ GeV and to the
elastic differential scattering cross section for $pp$ at LHC energy. Specifically, we fit to the total cross section,
$\sigma_{tot}^{pp,\bar{p}p}$, the ratio of the real to imaginary part of the forward scattering amplitude, $\rho^{pp,\bar{p}p}$ and to the elastic differential scattering cross section, $d\sigma^{pp}/dt$, at $\sqrt{s}=7$, $8$ and $13$ TeV with $|t|\leq 0.1$ GeV$^{2}$ (this range for $|t|$ is enough for an appropriate evaluation of $\alpha^{\prime}_{\IP}$). We use data sets compiled and analyzed by the Particle Data Groups\cite{Tanabashi:2018oca} as well as the recent data at LHC, specifically from the TOTEM Collaboration \cite{0295-5075-96-2-21002,Antchev:2013haa,Antchev:2013iaa,Antchev:2013paa,Antchev:2016vpy,Antchev:2017dia,
Antchev:2017yns,Antchev:2015zza,Antchev:2011zz} with the statistic and systematic errors added in quadrature.

The TOTEM data set includes the first and second measurements of the total $pp$ cross section at $\sqrt{s}=7$ TeV, namely $\sigma^{pp}_{tot}=98.3\pm2.8$ mb \cite{0295-5075-96-2-21002} and $\sigma^{pp}_{tot}=98.6\pm 2.22$ mb \cite{Antchev:2013haa}, both using the optical theorem together with the luminosity provided by the CMS, the luminosity-independent measurement at $\sqrt{s}=7$ TeV, namely $\sigma^{pp}_{tot}=98.0\pm 2.5$ mb \cite{Antchev:2013iaa}, the $\rho$-independent measurements at $\sqrt{s}=7$ TeV of $\sigma_{pp}$ and $\rho$-parameter, namely $\sigma^{pp}_{tot}=98.0\pm2.5$ mb and $\rho^{pp}=0.145\pm0.091$ \cite{Antchev:2013iaa}, the luminosity-independent measurement at $\sqrt{s}=8$ TeV, namely $\sigma^{pp}_{tot}=101.7\pm 2.9$ mb \cite{Antchev:2013paa}, and the measurement in the Coulombian-nuclear interference region at $\sqrt{s}=8$ TeV of $\sigma_{pp}$ and $\rho$-parameter, namely $\sigma^{pp}_{tot}=102.0\pm 2.3$ mb and $\sigma^{pp}_{tot}=103.0\pm 2.3$ mb, for central and peripheral phase formulations, respectively, and $\rho^{pp}=0.120\pm 0.030$ \cite{Antchev:2016vpy}. The TOTEM Collaboration has also measured elastic differential cross section at $\sqrt{s}=7$ TeV and $4$-momentum transfers squared $\vert t\vert$ in the intervals $0.00515\leq\vert t\vert\leq 0.235$ GeV$^{2}$ \cite{Antchev:2013haa} and $0.377\leq\vert t\vert\leq 2.443$ GeV$^{2}$ \cite{Antchev:2011zz}, and at $\sqrt{s}=8$ TeV in the intervals $6\times10^{-4}\leq\vert t\vert\leq0.2$ GeV$^{2}$ \cite{Antchev:2016vpy}. More recently, the TOTEM Collaboration has provided the first measurements of total cross section and $\rho$-parameter at $\sqrt{s}=13$ TeV, respectively $\sigma^{pp}_{tot}=110.6\pm 3.4$ mb \cite{Antchev:2017dia} where it was assumed a value of $\rho = 0.1$ \cite{Antchev:2017yns} in the luminosity-independent technique, and $\rho^{pp}=0.09\pm 0.01$ and $\rho^{pp}=0.10\pm 0.01$ \cite{Antchev:2017yns} obtained through the differential cross section due to the effects of the Coulombian-nuclear interference region. Also at this CM energy squared, $\sqrt{s}=13$ TeV, the TOTEM has also given precise measurements of the elastic differential cross section in the range of $8\times 10^{-4}\leq\vert t\vert\leq 0.201$ GeV$^{2}$ \cite{Antchev:2017yns}.

The LHC results reveal some tension between the TOTEM and ATLAS measurements. If we compare the ATLAS result for $\sigma_{tot}^{pp}$ at $\sqrt{s}=7$ TeV, $\sigma_{tot}^{pp}=95.4\pm 1.4$ mb, with the most precise value measured by TOTEM at the same energy, $\sigma_{tot}^{pp}=98.6\pm 2.2$ mb, the difference between the values corresponds to
\be
\frac{\sigma_{\text{TOTEM}}-\sigma_{\text{ATLAS}}}{\Delta\sigma^{\text{TOTEM}}_{tot}}=\frac{98.6-95.4}{2.2}=1.5,\nonumber
\ee
\tit{i.e} differs from almost $1.5$ standard deviations (assuming that the uncertainties are uncorrelated). Even if we compare the ATLAS result for $\sigma_{tot}^{pp}$ at $\sqrt{s}=8$ TeV, $\sigma_{tot}^{pp}=96.07 \pm 0.92$ mb, with the lowest value obtained by TOTEM at the same $8$ TeV, $\sigma_{tot}^{pp}=101.7 \pm 2.9$ mb, 
\be
\frac{\sigma_{\text{TOTEM}}-\sigma_{\text{ATLAS}}}{\Delta\sigma^{\text{TOTEM}}_{tot}}=\frac{101.5-96.07}{2.1}=2.6,\nonumber
\ee
we see that the difference is equivalent to $2.6\sigma$. In comparison with the latest TOTEM result at $8$ TeV, $\sigma_{tot}^{pp}=103.0 \pm 2.3$ mb, the deviation is yet more significant,
\be
\frac{\sigma_{\text{TOTEM}}-\sigma_{\text{ATLAS}}}{\Delta\sigma^{\text{TOTEM}}_{tot}}=\frac{103-96.07}{2.3}=3.\nonumber
\ee 

This strong disagreement clearly indicates the possibility of different scenarios for the rise of the total cross section and consequently for the parameters of the soft Pomeron. Presently we understand that an ensemble T+A, \tit{i.e.} and ensemble containing both TOTEM and ATLAS forward data, represents the effective set of the experimental information available \cite{Broilo:2018els,Broilo:2018qqs}. On the other hand, notice that our data set includes the elastic differential cross section data at $\sqrt{s}=7$, $8$, and $13$ TeV, respectively, where all of them were obtained by the TOTEM Collaboration. Although we consider that the forward ensemble must be T+A like, our ensemble is build within nonforward data. For this reason, we are not taking into account the ATLAS $pp$ total cross section data points. There is a relation which says that the optical point, \tit{i.e} the elastic differential cross section extrapolated to $t=0$, is proportional to $\sigma^{2}_{tot}(1+\rho^{2})$, see expression (\ref{ch2.117}). Therefore, the inclusion of ATLAS may cause some type of bias in our best-$\chi^{2}$ fitting procedure.

In all the fits performed was used, as test of goodness-of-fit, a $\chi^{2}$ fitting procedure, where the value of $\chi^{2}_{min}$ is distributed as a $\chi^{2}$ distribution with $N$ degrees of freedom. The fits to the experimental data sets were performed adopting an interval $\chi^{2}-\chi^{2}_{min}$ corresponding, in the case of normal errors, to the projection of $\chi^{2}$ hypersurface containing $68.27\%$ of probability, representing one standard deviation. This corresponds to $\chi^{2}-\chi^{2}_{min}=8.18$ and $9.30$ in the case of $7$ and $8$ free parameters, respectively.

Following the methodology of using the minimum number of free parameters, in the following analyses the slopes of the secondary-Reggeon linear trajectories, namely $\alplus$ and $\almin$, are fixed at $0.9$ GeV$^{-1}$. These values are in agreement with those usually obtained in Chew-Frautschi plots. Also, the slopes associated with the form factors are fixed at $b_{+}=0.5$ GeV$^{-2}$ and $b_{-}=3.1$ GeV$^{-2}$. These parameters have very little statistical correlation with the Pomeron parameters and their fixed values are consistent with those obtained in previous studies \cite{Covolan:1996uy,Luna:2008pp,Luna:2010ch} and do not have enough statistical weight to constrain the Pomeron parameters. We also fix the scale of the pion-Pomeron vertex at $a_{1}=m^{2}_{\rho}=(0.776$ GeV)$^{2}$ \cite{Khoze:2014nia}.

In the case of Born-level amplitudes, the values of the Regge parameters determined by global fits to $pp$ and $\bar{p}p$ data are
listed in Table \ref{chRGbtab1}. The descriptions of the data are displayed in Figure \ref{chRGbfig1}. Notice that in the case of BI and BII models we fixed the parameter $b_{\IP}$ at $5.5$ GeV$^{-2}$ since, as discussed in Reference \cite{Khoze:2000wk}, it is the natural choice for the computation of double-diffractive central Higgs production via $WW$-fusion (since the $W$ boson is radiated from a quark, like the photon). Moreover, our analyses show that at this $b_{\IP}$ value, which corresponds to the slope of the electromagnetic proton form factor, the Pomeron is described by trajectories with $\alpha^{\prime}_{\IP}\simeq 0.25$ GeV$^{-2}$. Interestingly enough, this values for $\alpha^{\prime}_{\IP}$ are consistent with the ones recently obtained from holographic QCD models \cite{Capossoli:2013kb,Ballon-Bayona:2015wra,Capossoli:2016kcr,Capossoli:2015ywa,Rodrigues:2016cdb}. However, if we perform the global fit at another value of $b_{\IP}$, say $4.0$ GeV$^{-2}$ (which is not atypical \cite{Khoze:2000wk}), we obtain the values $\alpha^{\prime}_{\IP}=0.336\pm 0.011$ GeV$^{-2}$ and $0.334\pm 0.011$ GeV$^{-2}$ in the case of BI and BII models, respectively, whilst the remaining free parameters remain with the same values. 

It is important to notice that the Pomeron intercept $\ap(0)=1+\e$ is an effective power, valid over a limited range of energies, otherwise the forward amplitude $A(s,t=0)$ would grow so large that unitarity bound would be violated. Thus, the parameter $\e$ represents not only the exchange of single Pomeron, but also $n$-Pomeron exchange processes, $n\geq2$. These multiple exchanges must tame the rise of the total cross section so that the breakdown of unitarity is avoided and, as a consequence, the value of $\e$ should decrease slowly with increasing energy. However, the results already seen in Figure \ref{chRGbfig1} clearly indicate that a very good description of forward data up to LHC energy is obtained  using a constant value of $\e$.

The values of the Regge parameters arising from this double-Pomeron analysis are shown in the last column of Table \ref{chRGbtab1}. The description of $\sigma_{tot}$ and $\rho$ is displayed in Figure \ref{chRGbfig1}. Our results show that the simple introduction of the nearest-$t$ channel singularity in the Pomeron trajectory, \tit{i.e}, just by considering a nonlinear trajectory for single-Pomeron exchange, is not sufficient to play a substantial statistical weight when compared to the fitted results from model BI. The same happens with the inclusion of double-Pomeron exchange in model BI$+\IP \IP$. Although each one of the four Born-level amplitudes are built in a unique form, the curves obtained in models BI, BII and BI$+\IP\IP$ are very close to each other. In special, the curves obtained in BI are indistinguishable from BII.

Therefore, these pictorical smooth differences among these Born-level models depicted in Figure \ref{chRGbfig1} must be attributed mainly to the presence of the power-like proton-Pomeron vertex plus the introduction of a nonlinear Pomeron trajectory. Bearing in mind that Figure \ref{chRGbfig1} corresponds to forward observables, we must take into account the significant $t$ dependence in the Born-level amplitude, more precisely in the Pomeron contribution since it gives the asymptotic high-energy behavior. The fits to the $d\sigma/dt$ are responsible for this small difference. More specifically, the small deviations in the central values of the Pomeron intercept and in the coupling of the $p-\IP$ vertex at $t=0$, $\beta_{\IP}(0)$, as listed in Table \ref{chRGbtab1}.

The preceding results have demonstrated that it is possible a very good description of forward data up to LHC energy by using a constant value of $\epsilon$. In fact, from the Table \ref{chRGbtab1}, we see that the value of $R$ is very close to zero,
\be
R=\frac{\beta^{2}_{\IP\IP}(0)}{\beta^{2}_{\IP}(0)}=\frac{\vert \IP\IP_{coupling}\vert}{\IP_{coupling}}<1.8\times 10^{-4}.
\label{chRGb.32}
\ee

The Born-level fit results for the total cross section are summarized below, where it was used only the central values listed in Table \ref{chRGbtab1}. In addition, $s$ is in GeV$^{2}$ and the total cross section in mb, where $s$ is divided by $s_{0}$.
\be
\textnormal{(BI)}\,\,\,\sigma^{pp,\bar{p}p}_{tot}=18.635\,s^{0.0938}+59.601\,s^{-0.344}\mp 30.639\,s^{-0.530},
\label{chRGb.33}
\ee
\be
\textnormal{(BII)}\,\,\,\sigma^{pp,\bar{p}p}_{tot}=18.630\,s^{0.0938}+59.592\,s^{-0.343}\mp 30.639\,s^{-0.530},
\label{chRGb.34}
\ee
\be
\textnormal{(BIII)}\,\,\,\sigma^{pp,\bar{p}p}_{tot}=18.508\,s^{0.0944}+59.364\,s^{-0.341}\mp 30.606\,s^{-0.530},
\label{chRGb.35}
\ee
\be
\textnormal{(BIV)}\,\,\,\sigma^{pp,\bar{p}p}_{tot}=18.547\,s^{0.0941}-\,0.124\,\frac{s^{1.188}}{\log s}+57.863\,s^{-0.335}\mp 32.725\,s^{-0.543}.
\label{chRGb.36}
\ee

The values of the Regge parameters obtained using the one- and two-channel eikonalized amplitude are listed in Table \ref{chRGbtab2}.
Similarly to the case of Born-level analysis, the Regge parameters for models with the same exponential form for the proton-Pomeron vertex, but different Pomeron trajectories (linear and nonlinear) are very close to each other. The one- and two-channel description of the data is displayed in Figure \ref{chRGbfig2}. 

In what concerns the one-channel analysis, our results reveal that practically there is no difference between OI and OII models, and once again the difference is properly constrained by the functional form of the third one, OIII model. As for the case of the two-channel analysis, wee that the $\sigma_{tot}$ result for models TI and TIII are very similar, however the $\rho$ result in the high-energy region suggests that there is practically no difference in the curves obtained with models TII and TIII.

Despite the fact that in the eikonal formalism there is a ``running'' in $t$, see expression (\ref{chRGb.40}), and only after was set $t=0$, at least for the description of the forward physical observables, it is not strong enough, \tit{per se}, to distinguish between OI and OII results. Even though, the eikonalized amplitude is build up from the Born-level amplitude, however, we could not say that such results were then expected. Again, even after the eikonalization, the results are very sensitive to the choice of the proton-Pomeron vertex and rather insensitive to the choice of Pomeron trajectory. We obtained higher values for the $\chi^{2}/\nu$ in the two-channel analysis case than the one-channel, but this test of goodness-of-fit was not able to distinguish the best result obtained from both eikonalized analysis. Once more, the role of a power-like proton-Pomeron vertex within the nearest-$t$ channel singularity in the Pomeron trajectory accounts for the dynamical differences. 

\begin{table}[hbtp]
\centering
\scalebox{0.9}{
\begin{tabular}{c@{\quad}c@{\quad}c@{\quad}c@{\quad}c@{\quad}}
\hline \hline
& & & & \\[-0.3cm]
& \multicolumn{4}{c}{Born-level amplitudes}  \\
\cline{2-5}
& & & & \\[-0.3cm]
 & BI & BII & BIII & BIV\\
& & & &  \\[-0.4cm]
\hline
 & & & & \\[-0.3cm]
$\epsilon$ 		             & 0.0938\,$\pm$\,0.0023 & 0.0938\,$\pm$\,0.0023      & 0.0944\,$\pm$\,0.0025      & 0.0941\,$\pm$0.0028  \\[1.2ex]
$\alpha^{\prime}_{\IP}$ [GeV$^{-2}$] & 0.253\,$\pm$\,0.011   & 0.252\,$\pm$\,0.011        & 0.27\,$\pm$\,0.16          & 0.254\,$\pm$\,0.011 \\[1.2ex]
$\beta_{\IP}(0)$ [GeV$^{-1}$] 	     & 1.962\,$\pm$\,0.038   & 1.962\,$\pm$\,0.038        & 1.956\,$\pm$\,0.041        & 1.958\,$\pm$\,0.047   \\[1.2ex]
$b_{\IP}$ [GeV$^{-2}$] 		     & {\bf 5.5 [fixed]} & {\bf 5.5 [fixed]}      &  -                     & {\bf 5.5 [fixed]} \\[1.2ex]
$\eta_{+}$ 		             & 0.344\,$\pm$\,0.041   & 0.343\,$\pm$\,0.041        & 0.341\,$\pm$\,0.043        & 0.335\,$\pm$\,0.045   \\[1.2ex]
$\alpha^{\prime}_{+}$ [GeV$^{-2}$]   & {\bf 0.9 [fixed]} & {\bf 0.9 [fixed]}      & {\bf 0.9 [fixed]}      & {\bf 0.9 [fixed]} \\[1.2ex]
$\beta_{+}(0)$ [GeV$^{-1}$] 	     & 3.77\,$\pm$\,0.33     & 3.77\,$\pm$\,0.33          & 3.76\,$\pm$\,0.35          & 3.70\,$\pm$\,0.36     \\[1.2ex]
$b_{+}$ [GeV$^{-2}$] 		     & {\bf 0.5 [fixed]} & {\bf 0.5 [fixed]}      & {\bf 0.5 [fixed]}      & {\bf 0.5 [fixed]} \\[1.2ex]
$\eta_{-}$ 			     & 0.530\,$\pm$\,0.070   & 0.530\,$\pm$\,0.070        & 0.530\,$\pm$\,0.075        & 0.543\,$\pm$\,0.075   \\[1.2ex]
$\alpha^{\prime}_{-}$ [GeV$^{-2}$]   & {\bf 0.9 [fixed]} & {\bf 0.9 [fixed]}      & {\bf 0.9 [fixed]}      & {\bf 0.9 [fixed]} \\[1.2ex]
$\beta_{-}(0)$ [GeV$^{-1}$] 	     & 2.91\,$\pm$\,0.43     & 2.91\,$\pm$\,0.43          & 2.91\,$\pm$\,0.46          & 2.98\,$\pm$\,0.47     \\[1.2ex]
$b_{-}$ [GeV$^{-2}$] 		     & {\bf 3.1 [fixed]} & {\bf 3.1 [fixed]}      & {\bf 3.1 [fixed]}      & {\bf 3.1 [fixed]} \\[1.2ex]
$a_{1}$ [GeV$^{2}$] 		     & -                 & {\bf $m_{\rho}^{2}$ [fixed]} & {\bf $m_{\rho}^{2}$ [fixed]} & -                 \\[1.2ex]
$a_{2}$ [GeV$^{2}$] 		     & -                 & -                      & 0.90\,$\pm$\,0.48            & -                 \\[1.2ex]
$\beta_{\IP\IP}(0)$ [GeV$^{-1}$]     & -                 & -                      & -                      & 0.026\,$\pm$\,0.017   \\[1.2ex]
\hline
& & & & \\[-0.4cm]
$\nu$                                & 305               & 305                    & 304                    & 304               \\[1.2ex]
$\chi^{2}/\nu$ 			     & 0.823             & 0.823                  & 0.831                  & 0.820             \\[1.2ex]
\hline \hline 
\end{tabular}}
\caption{The values of the Pomeron and secondary Reggeon parameters obtained in global fits to the $\sigma_{tot}^{pp,\bar{p}p}$,
$\rho^{pp,\bar{p}p}$ and $d\sigma^{pp}/dt$ data within Born-level amplitudes by considering one standard deviation.}
\label{chRGbtab1}
\end{table}

\bfg[htbp]
  \begin{center}
    \includegraphics[width=8.0cm,height=8.0cm]{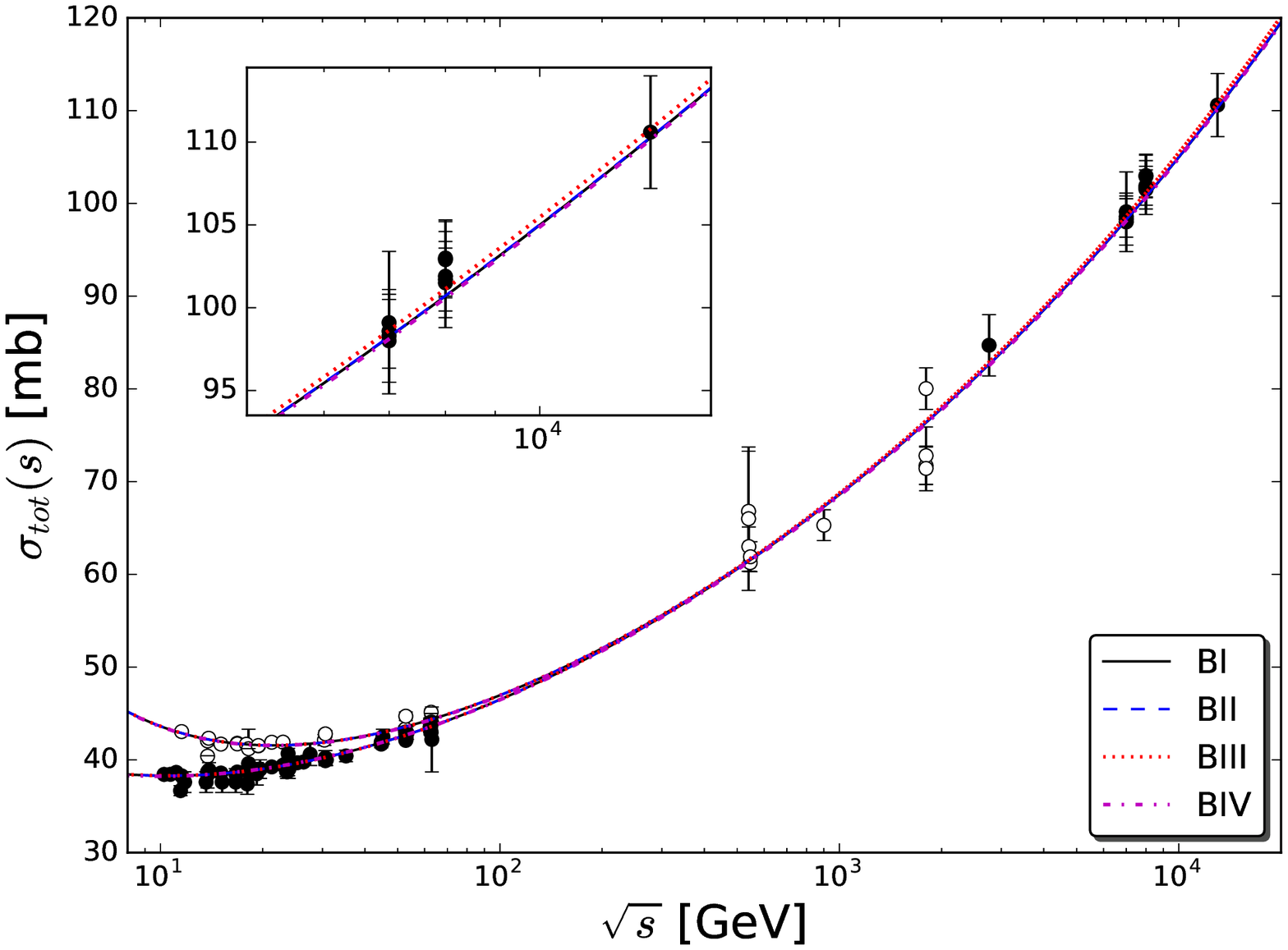}
    \includegraphics[width=8.0cm,height=8.0cm]{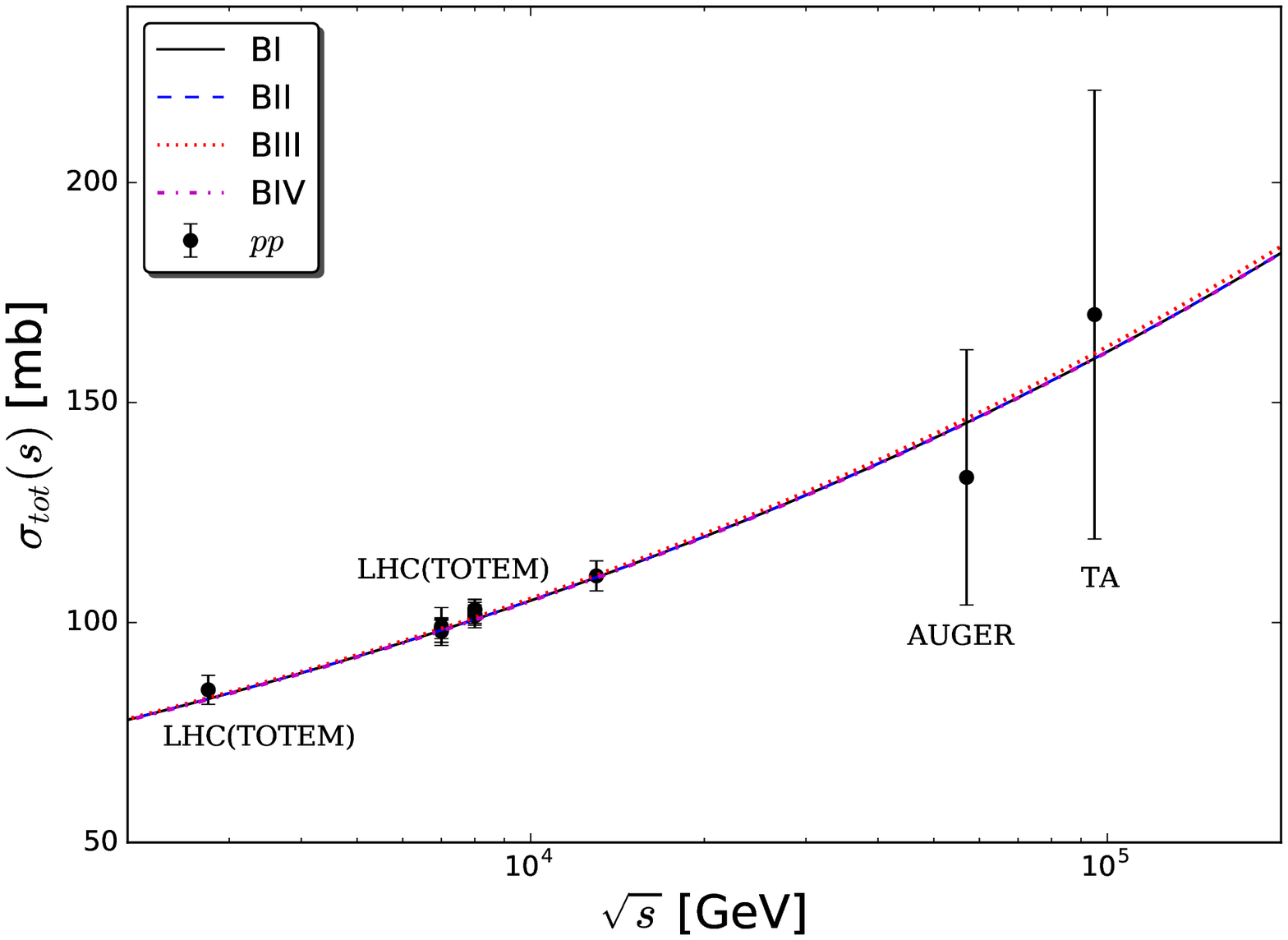}
    \includegraphics[width=8.0cm,height=8.0cm]{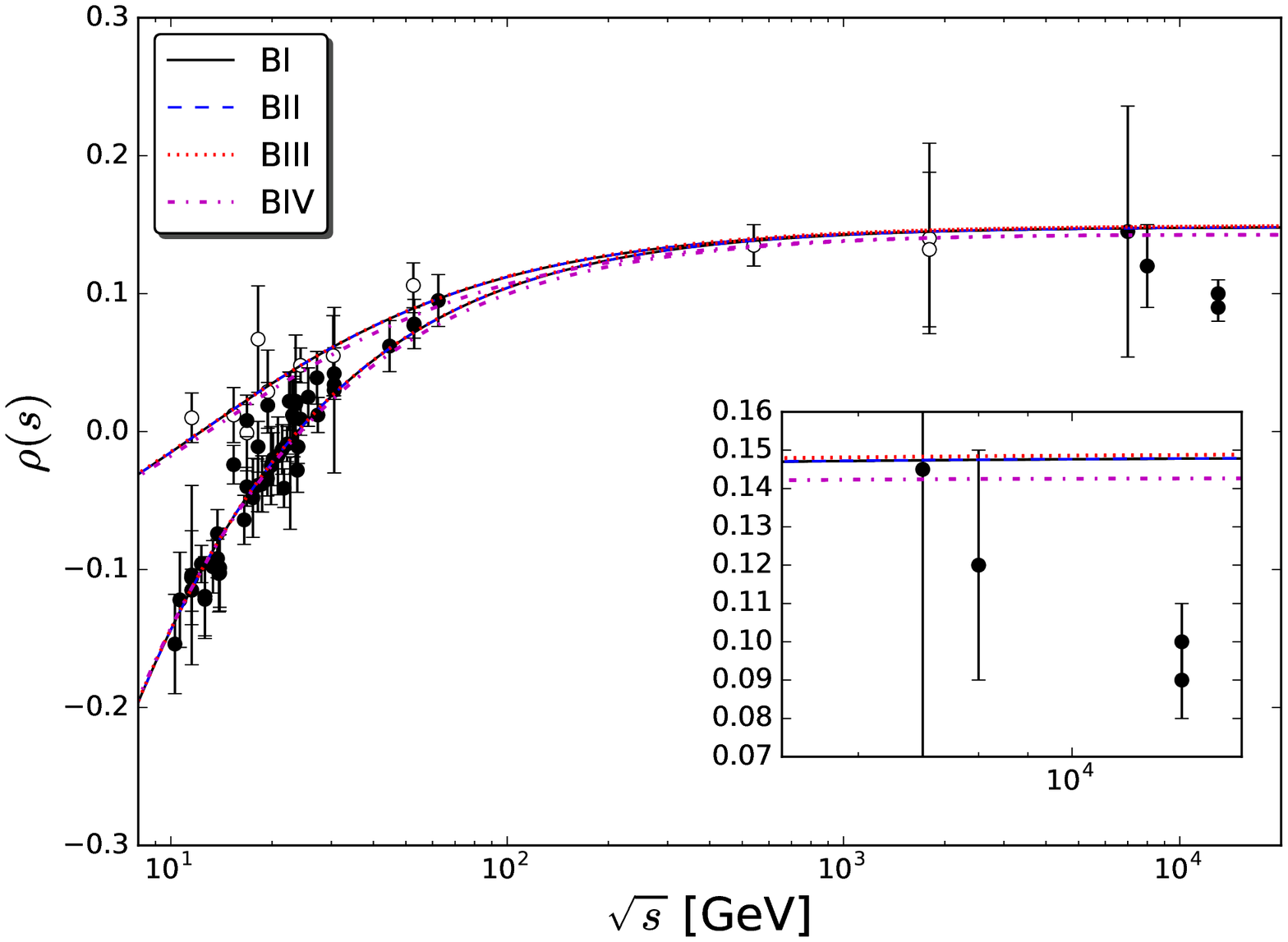}
    \includegraphics[width=8.0cm,height=8.0cm]{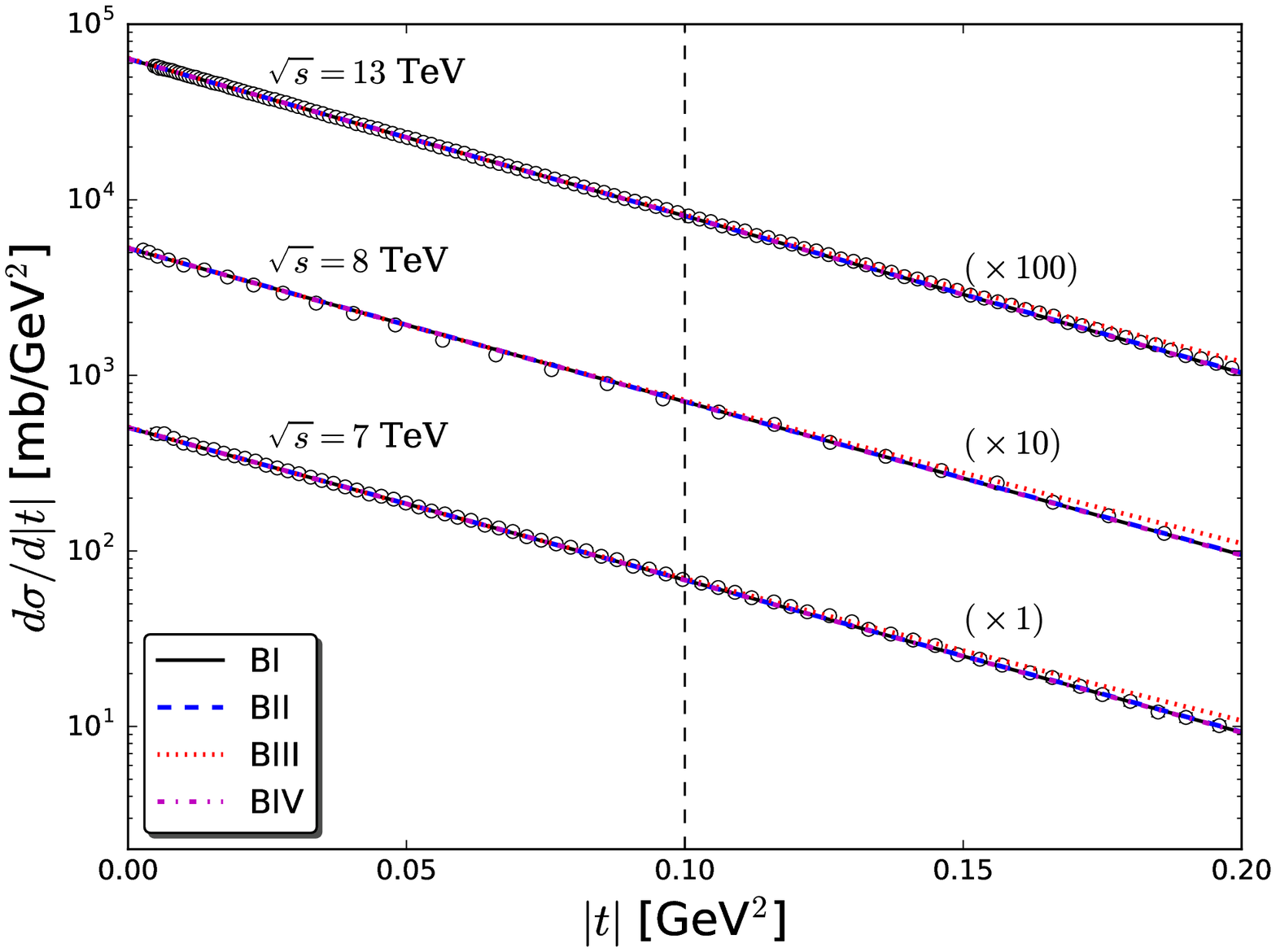}
    \caption{Predictions for the forward and nonforward observables in the Born-level analysis.}
    \label{chRGbfig1}
  \end{center}
\efg

\begin{table}[hbtp]
\centering
\scalebox{0.8}{
\begin{tabular}{c@{\quad}c@{\quad}c@{\quad}c@{\quad}c@{\quad}}
\hline \hline
& & &  \\[-0.2cm]
\,\,\,\,\,\,\,\,\,\,\,\,\,\,\,\,\,\,\,\,& \multicolumn{3}{c}{One-channel eikonalized amplitudes}  \\
\cline{2-4}
& & &  \\[-0.3cm]
 & OI & OII & OIII  \\
& & &  \\[-0.4cm]
\hline
& & &  \\[-0.3cm]
$\epsilon$                              & 0.1246$\pm$0.0014   & 0.124620$\pm$0.000090  & 0.1250$\pm$0.0033     \\[1.2ex]
$\alpha^{\prime}_{\IP}$ [GeV$^{-2}$]    & 0.03005$\pm$0.00030 & 0.0300$\pm$0.0087      & 0.09018$\pm$0.00033   \\[1.2ex]
$\beta_{\IP}(0)$ [GeV$^{-1}$] 	        & 1.827$\pm$0.010     & 1.826$\pm$0.019        & 1.858$\pm$0.053       \\[1.2ex]
$b_{\IP}$ [GeV$^{-2}$] 		        & 7.2690$\pm$0.0011   & 7.24$\pm$0.48          &  -                    \\[1.2ex]
$\eta_{+}$ 				& 0.2992$\pm$0.0067   & 0.299$\pm$0.016        & 0.307$\pm$0.038       \\[1.2ex]
$\alpha^{\prime}_{+}$ [GeV$^{-2}$]      & {\bf 0.9 [fixed]}   & {\bf 0.9 [fixed]}      & {\bf 0.9 [fixed]}     \\[1.2ex]
$\beta_{+}(0)$ [GeV$^{-1}$] 		& 4.288$\pm$0.010     & 4.286$\pm$0.034        & 4..42$\pm$0.34        \\[1.2ex]
$b_{+}$ [GeV$^{-2}$] 		        & {\bf 0.5 [fixed]}   & {\bf 0.5 [fixed]}      & {\bf 0.5 [fixed]}     \\[1.2ex]
$\eta_{-}$ 				& 0.5398$\pm$0.0041   & 0.539$\pm$0.018        & 0.541$\pm$0.072       \\[1.2ex]
$\alpha^{\prime}_{-}$ [GeV$^{-2}$]      & {\bf 0.9 [fixed]}   & {\bf 0.9 [fixed]}      & {\bf 0.9 [fixed]}     \\[1.2ex]
$\beta_{-}(0)$ [GeV$^{-1}$]	        & 3.557$\pm$0.030     & 3.55$\pm$0.13          & 3.60$\pm$0.53         \\[1.2ex]
$b_{-}$ [GeV$^{-2}$] 		        & {\bf 3.1 [fixed]}   & {\bf 3.1 [fixed]}      & {\bf 3.1 [fixed]}     \\[1.2ex]
$a_{1}$ [GeV$^{2}$] 			& -                   & $m_{\rho}^{2}$ [fixed] & $m_{\rho}^{2}$ [fixed]\\[1.2ex]
$a_{2}$ [GeV$^{2}$] 			& - 		      & -                      & 0.1250 $\pm$0.0033    \\[1.2ex]
\hline
& & &  \\[-0.2cm]
$\chi^{2}/\nu$ 				& 0.807 	      & 0.807                  & 0.823                  \\[1.2ex]
\hline\hline
& & &  \\[-0.2cm]
\,\,\,\,\,\,\,\,\,\,\,\,\,\,\,\,\,\,\,\,& \multicolumn{3}{c}{Two-channel eikonalized amplitudes}  \\
\cline{2-4}
& & &  \\[-0.3cm]
 & \,\,\,\,\,\,\,\,\,\,\,\,\,\,\,\,\,\,\,\, TI \,\,\,\,\,\,\,\,\,\,\,\,\,\,\,\,\,\,\,\, & \,\,\,\,\,\,\,\,\,\,\,\,\,\,\,\,\,\,\,\, TII  \,\,\,\,\,\,\,\,\,\,\,\,\,\,\,\,\,\,\,\, & \,\,\,\,\,\,\,\,\,\,\,\,\,\,\,\,\,\,\,\, TIII \,\,\,\,\,\,\,\,\,\,\,\,\,\,\,\,\,\,\,\, \\
& & &  \\[-0.4cm]
\hline
& & &  \\[-0.3cm]
$\epsilon$                              & 0.1505$\pm$0.0048   & 0.1508$\pm$0.0048      & 0.1535$\pm$0.0016     \\[1.2ex]
$\alpha^{\prime}_{\IP}$ [GeV$^{-2}$]    & 0.0100$\pm$0.0022   & 0.0208$\pm$0.0066      & 0.019$\pm$0.016       \\[1.2ex]
$\beta_{\IP}(0)$ [GeV$^{-1}$]           & 1.737$\pm$0.075     & 1.734$\pm$0.074        & 1.776$\pm$0.030       \\[1.2ex]
$b_{\IP}$ [GeV$^{-2}$]                  & 6.61$\pm$0.31       & 6.58$\pm$0.18          &  -                    \\[1.2ex]
$\eta_{+}$                              & 0.289$\pm$0.039     & 0.2875$\pm$0.0039      & 0.2983$\pm$0.0039     \\[1.2ex]
$\alpha^{\prime}_{+}$ [GeV$^{-2}$]      & {\bf 0.9 [fixed]}   & {\bf 0.9 [fixed]}      & {\bf 0.9 [fixed]}     \\[1.2ex]
$\beta_{+}(0)$ [GeV$^{-1}$]             & 4.86$\pm$0.36       & 4.85$\pm$0.36          & 5.034$\pm$0.013       \\[1.2ex]
$b_{+}$ [GeV$^{-2}$] 		        & {\bf 0.5 [fixed]}   & {\bf 0.5 [fixed]}      & {\bf 0.5 [fixed]}     \\[1.2ex]
$\eta_{-}$                              & 0.543$\pm$0.069     & 0.543$\pm$0.070        & 0.545$\pm$0.057       \\[1.2ex]
$\alpha^{\prime}_{-}$ [GeV$^{-2}$]      & {\bf 0.9 [fixed]}   & {\bf 0.9 [fixed]}      & {\bf 0.9 [fixed]}     \\[1.2ex]
$\beta_{-}(0)$ [GeV$^{-1}$]             & 4.03$\pm$0.58       & 4.03$\pm$0.58          & 4.08$\pm$0.48         \\[1.2ex]
$b_{-}$ [GeV$^{-2}$] 		        & {\bf 3.1 [fixed]}   & {\bf 3.1 [fixed]}      & {\bf 3.1 [fixed]}     \\[1.2ex]
$a_{1}$ [GeV$^{2}$]                     & -                   & $m_{\rho}^{2}$ [fixed] & $m_{\rho}^{2}$ [fixed]\\[1.2ex]
$a_{2}$ [GeV$^{2}$]                     & -                   & -                      & 0.80$\pm$0.19         \\[1.2ex]
\hline
& & &  \\[-0.2cm]
$\chi^{2}/\nu$                          & 0.896               & 0.895                   & 0.892                \\[1.2ex]
\hline \hline 
\end{tabular}
}
\caption{The values of the Pomeron and secondary Reggeon parameters obtained in global fits to the $\sigma_{tot}^{pp,\bar{p}p}$,
$\rho^{pp,\bar{p}p}$ and $d\sigma^{pp,\bar{p}p}/dt$ data within one- and two-channel eikonalized amplitudes by considering one standard deviation.}
\label{chRGbtab2}
\end{table}

\bfg[htbp]
  \begin{center}
    \includegraphics[width=8.0cm,height=8.0cm]{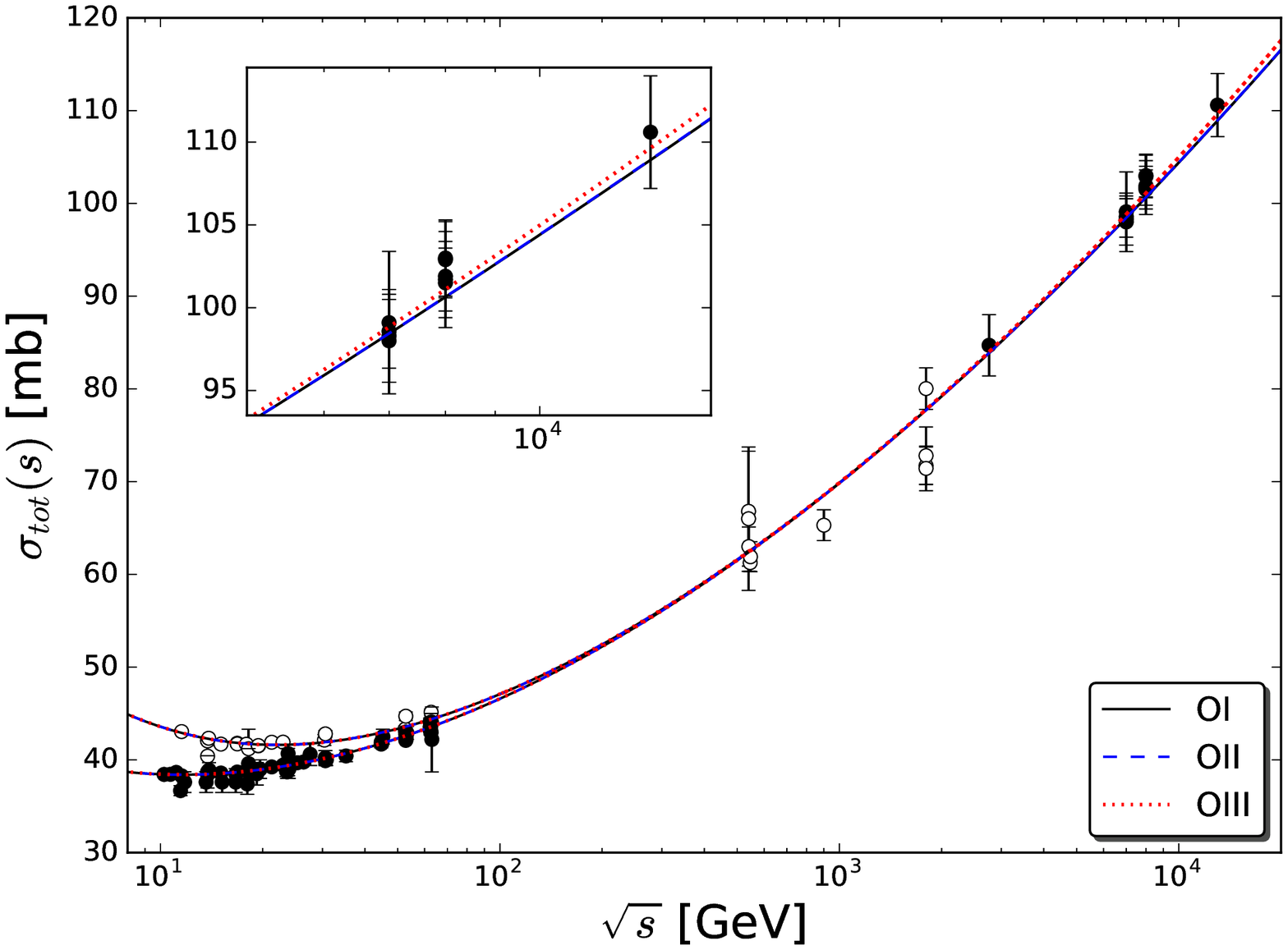}
    \includegraphics[width=8.0cm,height=8.0cm]{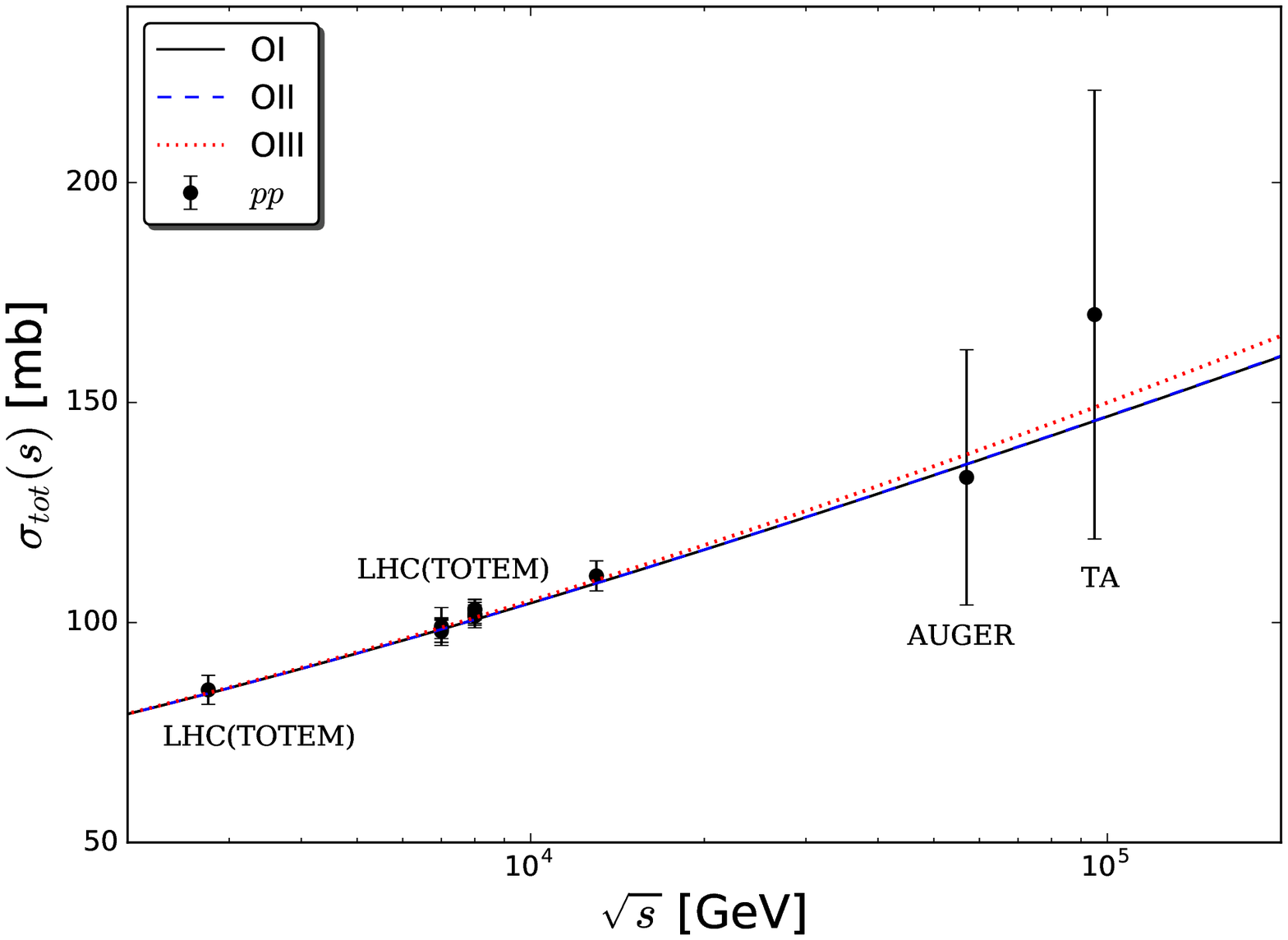}
    \includegraphics[width=8.0cm,height=8.0cm]{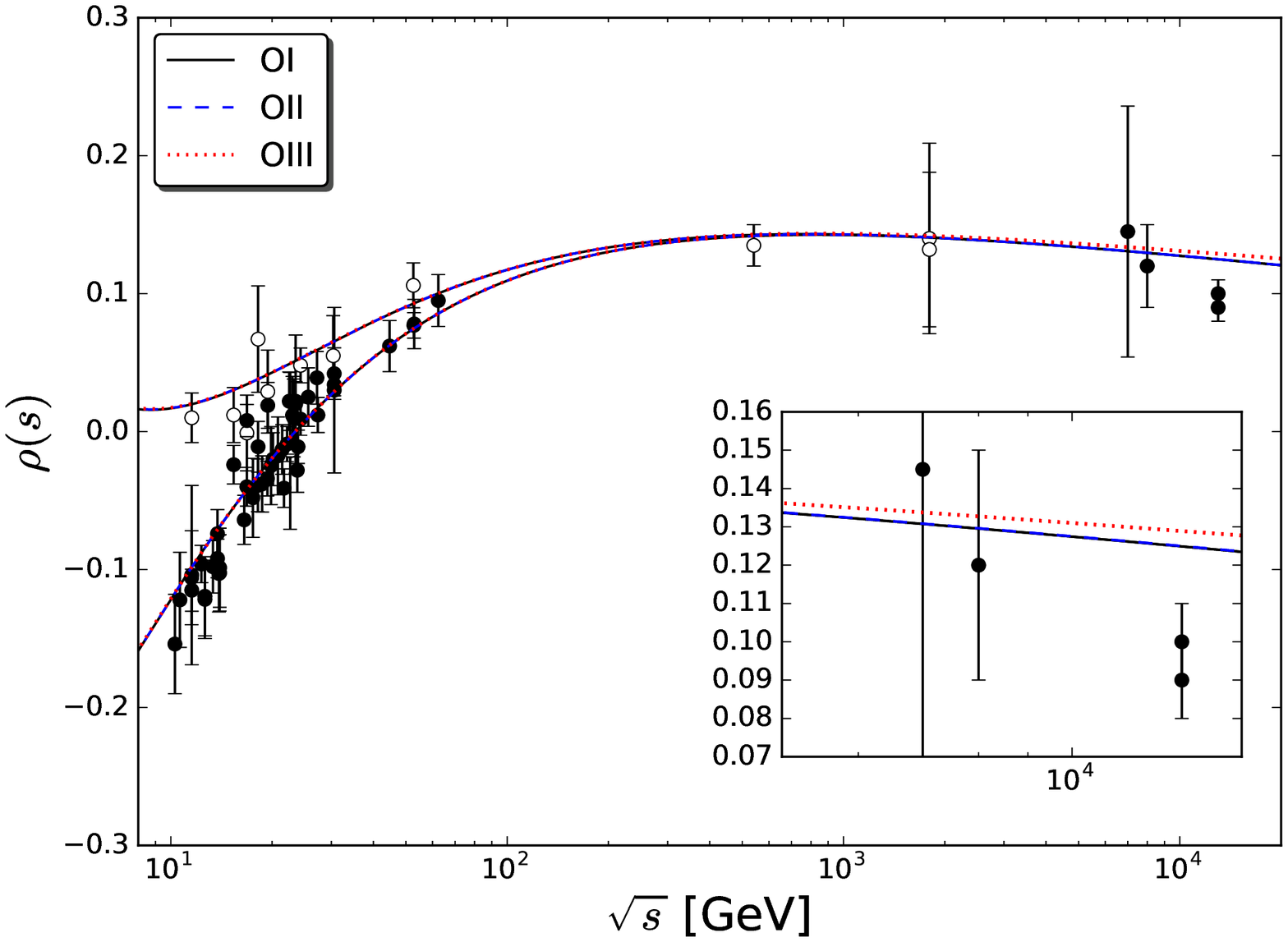}
    \includegraphics[width=8.0cm,height=8.0cm]{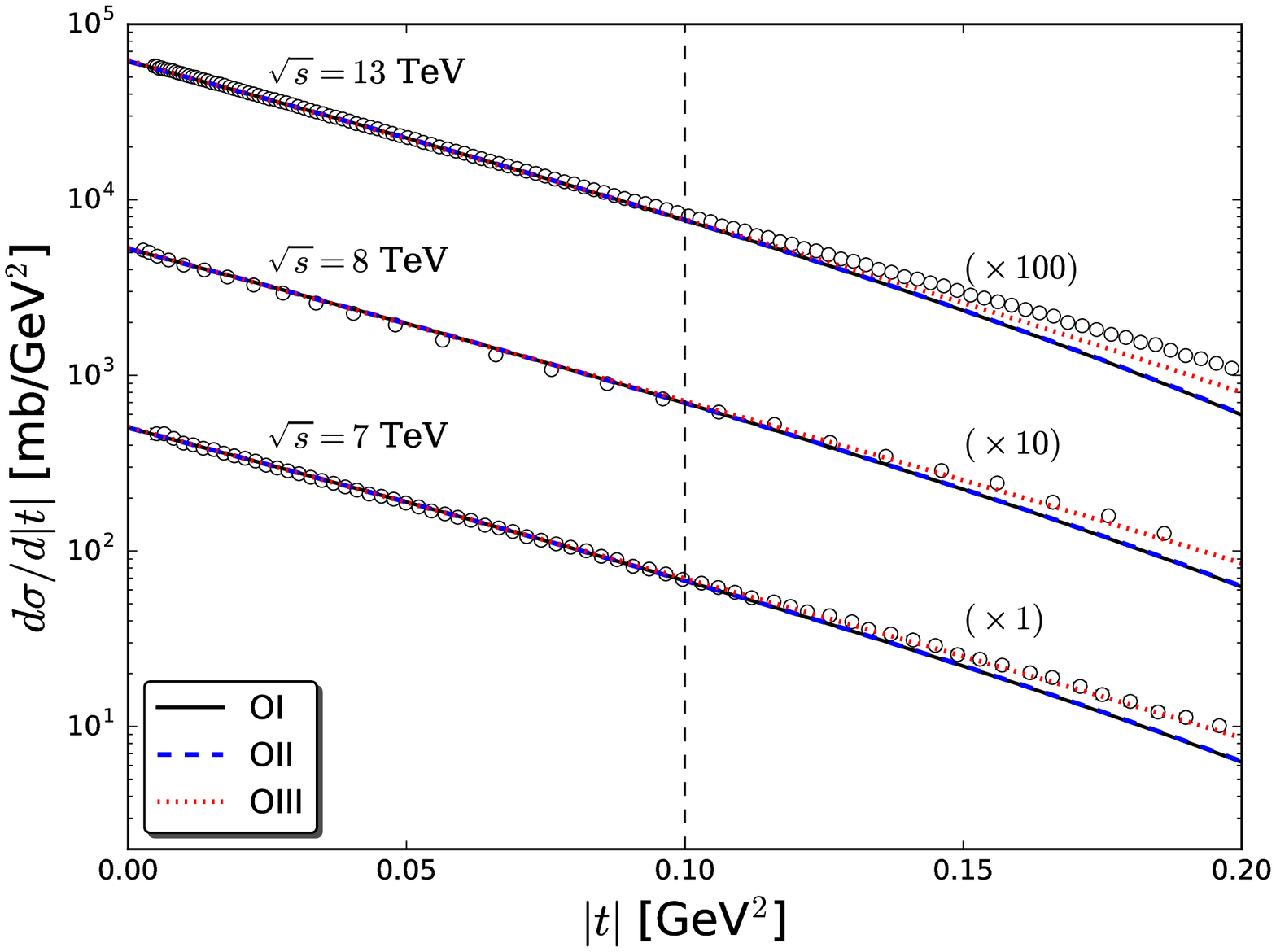}
    \caption{Predictions for the forward and nonforward observables in the one-channel eikonal analysis.}
    \label{chRGbfig2}
  \end{center}
\efg

\bfg[htbp]
  \begin{center}
    \includegraphics[width=8.0cm,height=8.0cm]{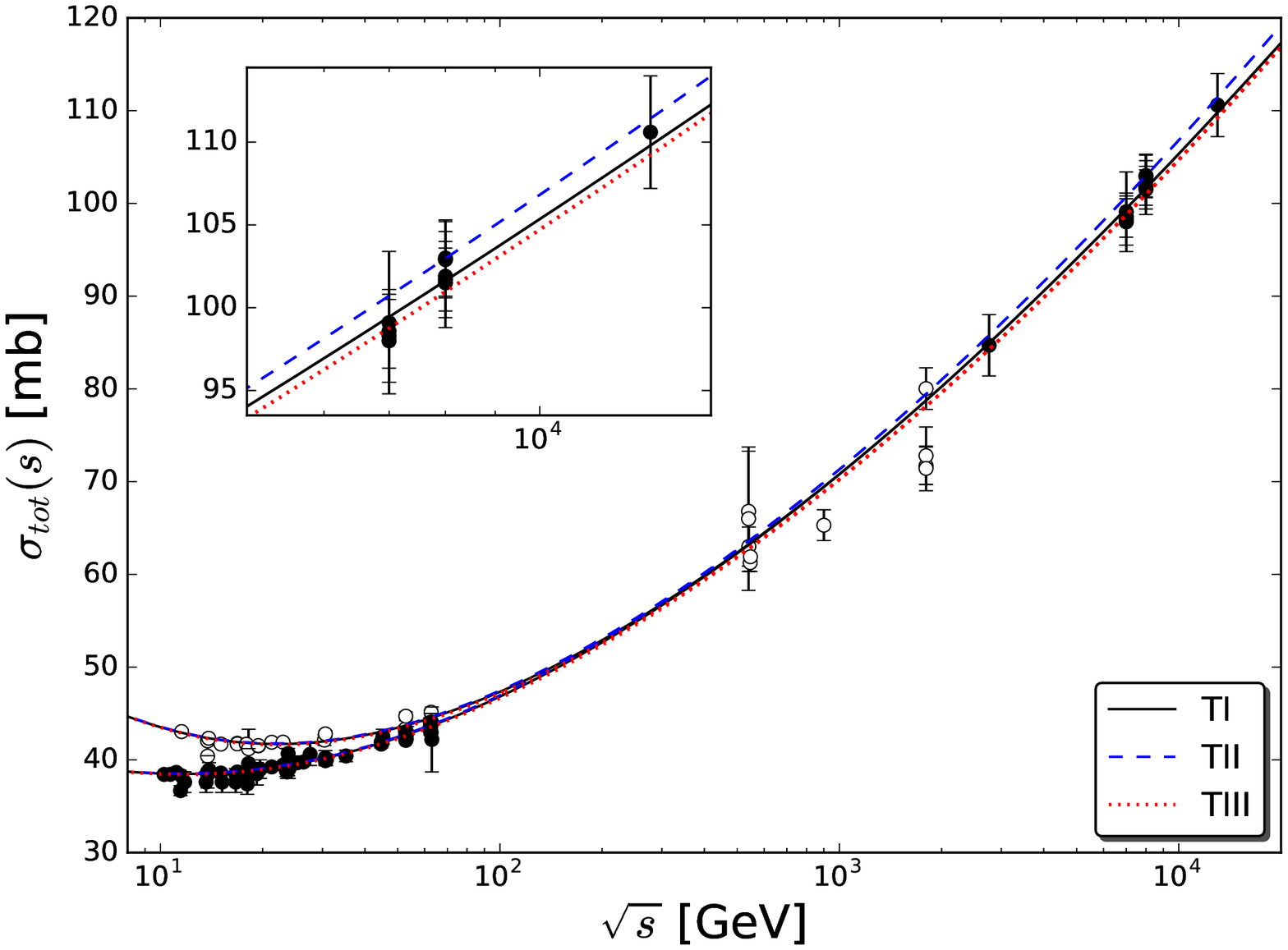}
    \includegraphics[width=8.0cm,height=8.0cm]{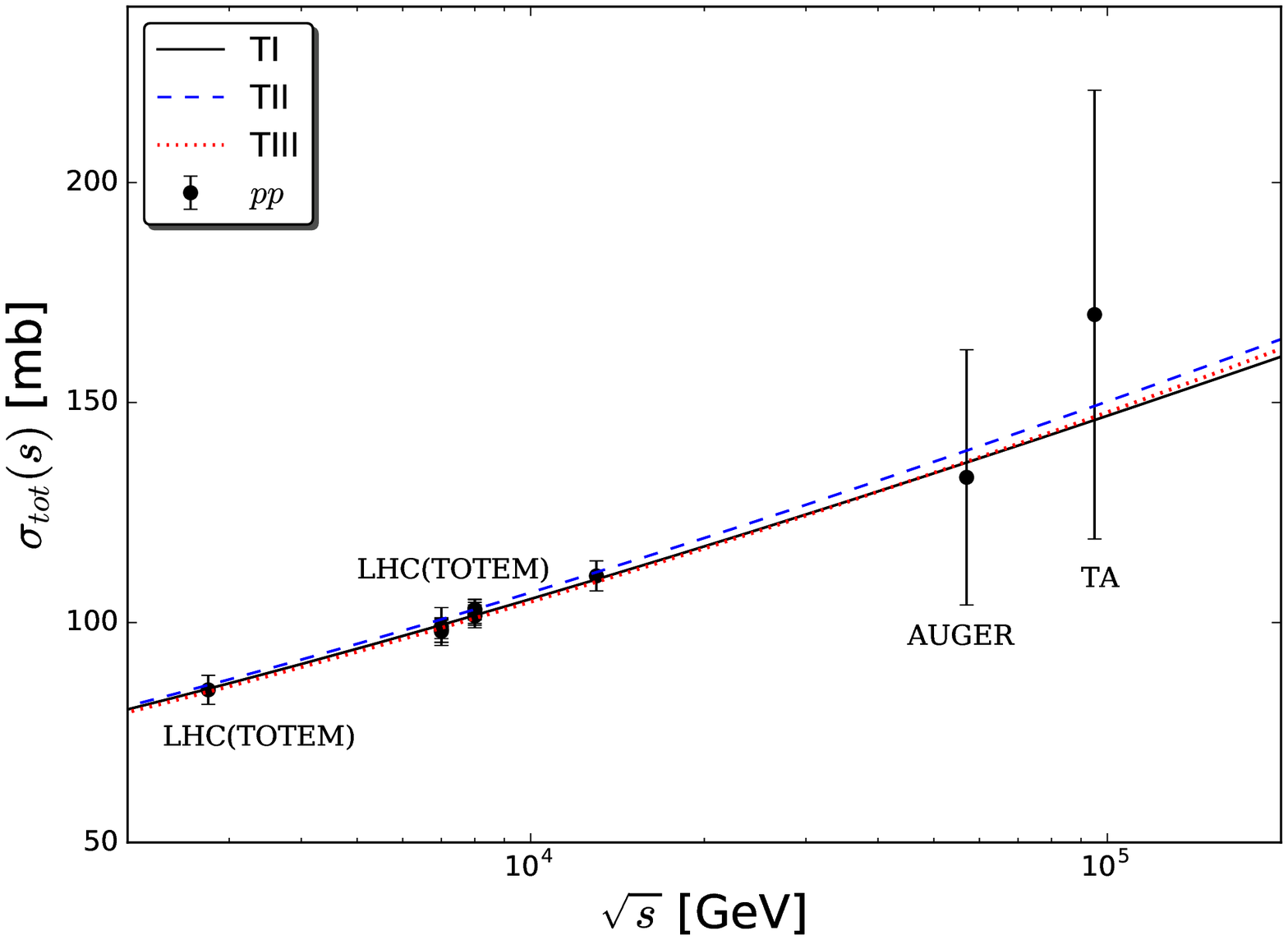}
    \includegraphics[width=8.0cm,height=8.0cm]{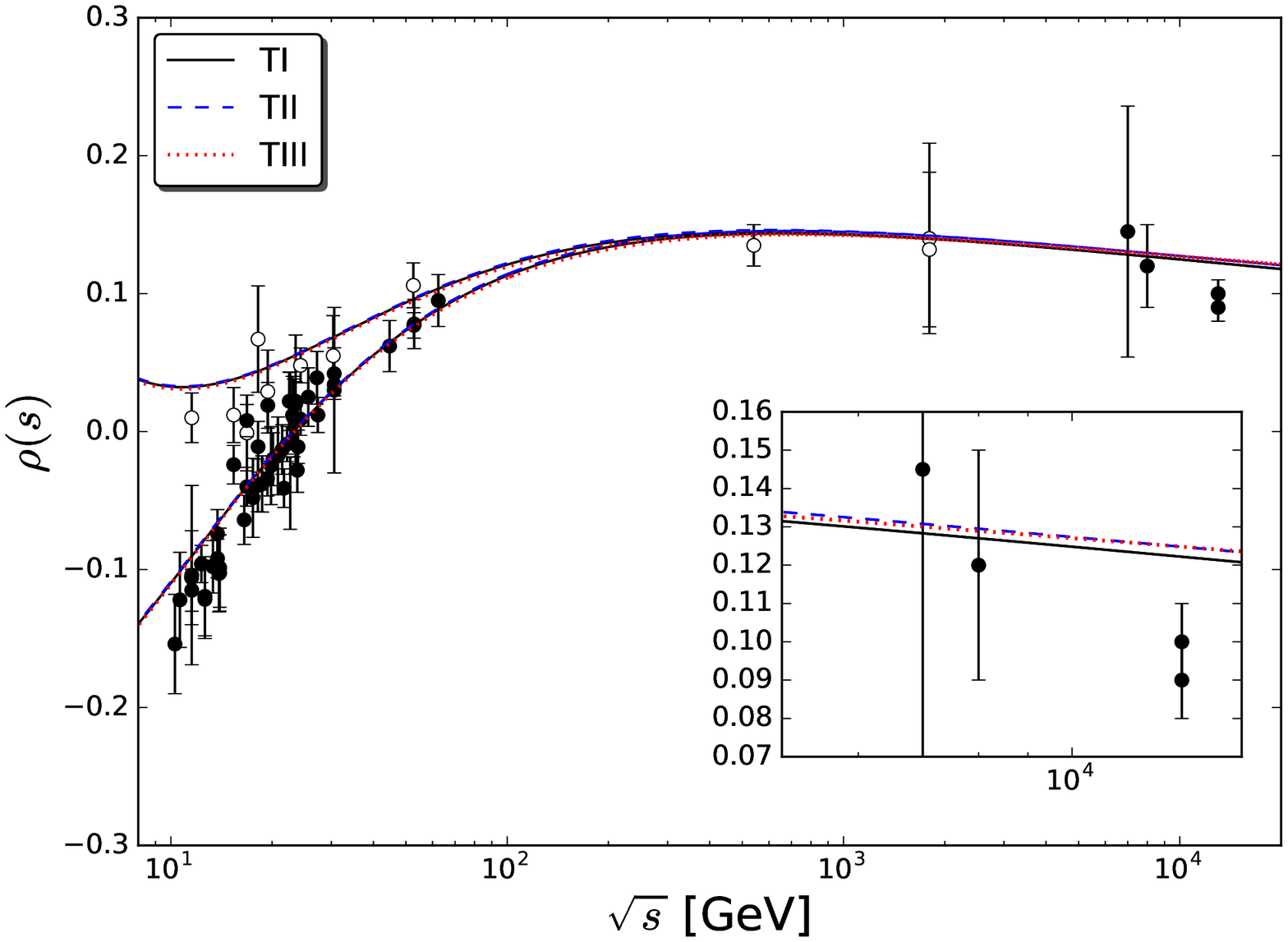}
    \includegraphics[width=8.0cm,height=8.0cm]{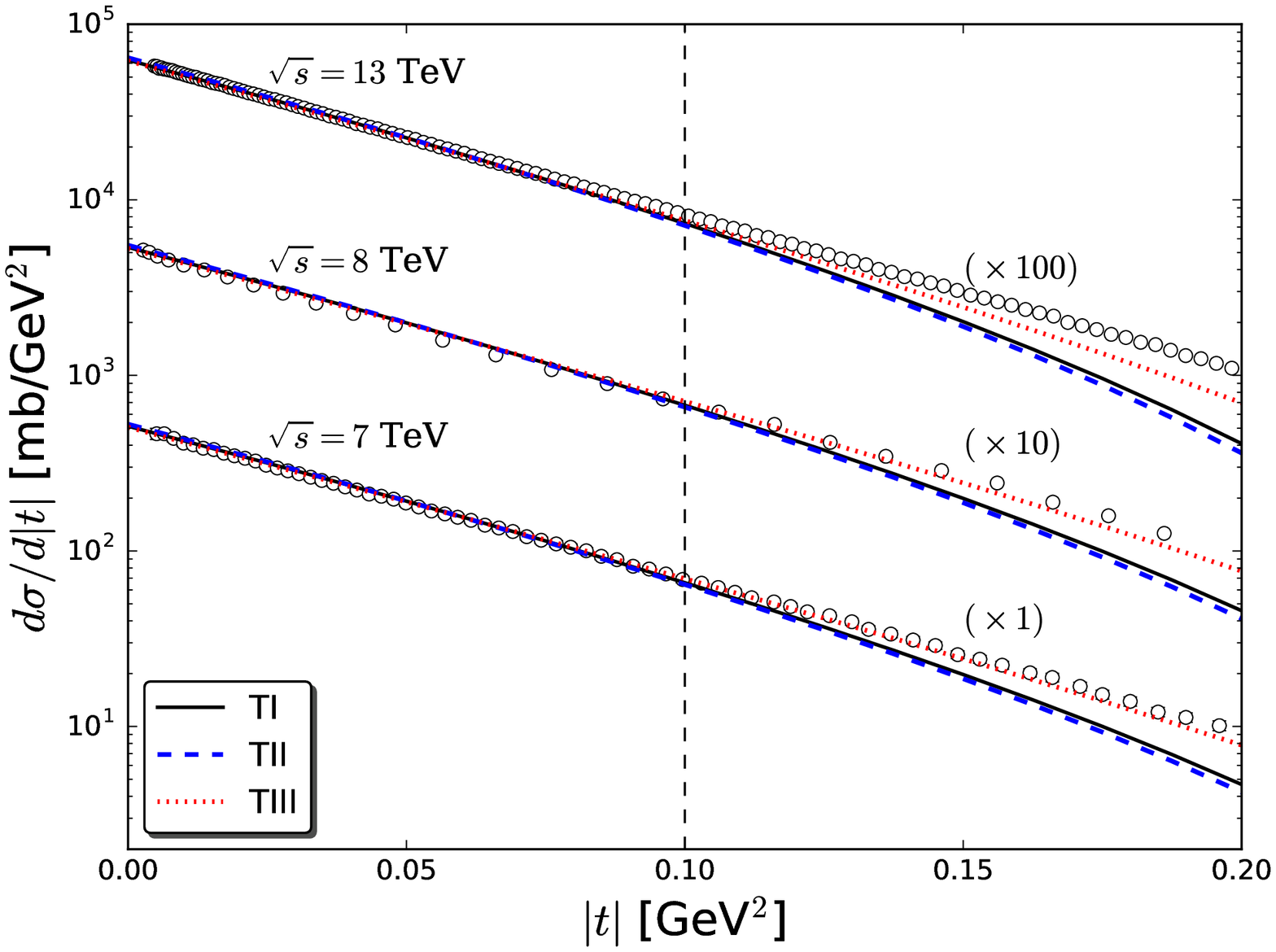}
    \caption{Predictions for the forward and nonforward observables in the two-channel eikonal analysis.}
    \label{chRGbfig3}
  \end{center}
\efg

\clearpage
\thispagestyle{plain}

\chapter{\textsc{QCD Parton Model}}
\label{chQCD}
\mbox{\,\,\,\,\,\,\,\,\,}
Quantum Chromodynamics is the non-Abelian gauge theory that describes the strong interaction among quarks and gluons, which collectively are called partons, and it is one of the pillars of the Standard Model of elementary particles \cite{Muta:2010xua,Ellis:1990pp}. The QCD is described by means of the invariant properties of the symmetry group $SU(N_{c})$, where $N_{c}=3$ defines the dimension of the group and introduces a new quantum number, the color charge. Historically, the color degrees of freedom appeared as a solution to the problem of the barion $\Delta^{++}$ wave function. This barion is compounded by three $u$ quarks, and therefore has spin $3/2$ since it is well known that quarks has spin $1/2$. The Fermi-Dirac statistics properly implies a antisymmetric wave function, moreover the Pauli exclusion principle assures that each one of the three quarks must exhibit distinct quantum numbers. And in fact they do, once it is considered that the wave function is totally antisymmetric with respect to the color degrees of freedom.

The quark fields transform following the $SU(3)_{c}$ fundamental representation and given by the spinors $\psi_{q}$, where each quark flavor has one in three possible color charges. The gauge fields are represented by the $T^{A}$ generator matrices of $SU(3)_{c}$, and given by eight generators, or also by eight vector gluons, identified by $A=1,...,8$. However, quarks were never measured as free particles, but as hadronic states of color singlet linear combinations of quarks, antiquarks and gluons. Then, the interaction force among quarks should increase with increasing distances, and this behavior can be explained by means of a mechanism called confinement. The non-Abelian nature of the gauge group allows the existence of colored gluons and also that they can interact with each other, therefore leading to a property known as asymptotic freedom, where the strong coupling, calculated by means of perturbation theory and renormalization group, monotonically decreases in short distances or equivalently at high values of momentum transfer.

\section{\textsc{Quantum Chromodynamics}}
\label{secQCD.1}
\hyphenation{res-pec-ti-ve-ly}
\mbox{\,\,\,\,\,\,\,\,\,}
There are two distinct categories in QCD: the ultraviolet regime, also known as perturbative, and the infrared regime, called nonperturbative. The former is characterized by the asymptotic free behavior of the theory, \tit{i.e.} it is identified by processes where the mass scale is higher than $\Lambda_{QCD}\sim200$ MeV. The latter one, instead of what usually happens in short distance regime, the quarks and gluons are not observed as free states of particles, therefore hadrons are not considered as color states. Moreover, the gauge symmetry implies that gluons are massless, but recent studies indicate that the gluons may develop a dynamical momentum-dependent mass in the infrared sector \cite{Luna:2005nz,Fagundes:2011zx,Bahia:2015gya,Bahia:2015hha} by means of Dyson-Schwinger equations and its existence is strongly supported by QCD lattice simmulations \cite{Bowman:2004jm,Sternbeck:2005vs,Boucaud:2006if,Bowman:2007du,Bogolubsky:2009dc,Oliveira:2009nn,Cucchieri:2009kk,Cucchieri:2009zt,Dudal:2010tf,Cucchieri:2011ig,Cucchieri:2014via}.

\subsection{\textsc{The Lagrangian of QCD}}
\label{secQCD.1.1}
\mbox{\,\,\,\,\,\,\,\,\,}
The Feynman rules required for a perturbative analysis of QCD can be derived from a Lagrangian density, which is given by
\be
{\cal L}_{QCD}={\cal L}_{classical}+{\cal L}_{gauge-fixing}+{\cal L}_{ghost}.
\label{chQCD.1}
\ee
\mbox{\,\,\,\,\,\,\,\,\,}
The classical Lagrangian term corresponding to the dynamics of massive quarks and massless gluons is
\be
{\cal L}_{cl}(x)=-\frac{1}{4}F_{\mu \nu}^{A}(x)F^{\mu \nu}_{A}(x)+ \sum_{q}\bar{\psi}^{r}_{q}(x) (i\slashed{D}-m)_{rs}\psi^{s}_{q}(x),
\label{chQCD.2}
\ee
where $r,s=1,2,3$ are color index, $q=u,d,s,c,b,t$ is the flavor and $\slashed{D}=\gamma^{\mu}D_{\mu}$. The Dirac matrices satisfy the Clifford algebra anti-commutation relation $\{\gamma^{\mu},\gamma^{\nu}\}=2g^{\mu\nu}$, where the metric is $g^{\mu\nu}=\textnormal{diag}(1,-1,-1,-1)$. The quark fields $\psi^{r}_{q}$ are in the triplet representation of the color group. Then, the covariant derivative acting on triplet fields takes the form
\be
(D_{\mu})_{rs}=\partial_{\mu}\delta_{rs}+ig_{s}(t^{C}G_{\mu}^{C}(x))_{rs},
\label{chQCD.3}
\ee
and similarly acting on octet fields,
\be
(D_{\mu})_{AB}=\partial_{\mu}\delta_{AB}+ig_{s}(T^{C}G_{\mu}^{C}(x))_{AB},
\label{chQCD.4}
\ee
where $t$ and $T$ are matrices in the fundamental and adjoint representations of $SU(3)_{c}$ respectively,
\be
[\lambda^{A},\lambda^{B}]=if^{ABC}\lambda^{C},
\label{chQCD.5}
\ee
and a representation for the generators $t^{A}$ is provided by the eight Gell-Mann matrices, which are hermitian and traceless \cite{Ellis:1990pp},
\be
t^{A}=\frac{1}{2}\,\lambda^{A}.
\label{chQCD.6}
\ee

By convention, it is chosen that the $t^{A}$ generators of $SU(N_{c})$ are normalized as
\be
Tr(t^{A}t^{B})=\frac{1}{2}\delta^{AB},
\label{chQCD.7}
\ee
and obey the following relations:
\be
t^{A}_{ab}t^{A}_{bc}=C_{F}\delta_{ac},
\label{chQCD.8}
\ee
\be
Tr(T^{C}T^{D})=f^{ABC}f^{ABD}=C_{A}\delta^{CD},
\label{chQCD.9}
\ee
where
\be
C_{F}=\frac{N^{2}_{c}-1}{2N_{c}},\,\,\,\,C_{A}=N_{c}.
\label{chQCD.10}
\ee

The field strength tensor is derived from the gluon field $G^{A}_{\mu}$,
\be
F_{\mu \nu}^{A}(x)=\partial_{\mu}G_{\nu}^{A}(x)-\partial_{\nu}G_{\mu}^{A}(x)-g_{s}f^{ABC}G_{\mu}^{B}G_{\nu}^{C},
\label{chQCD.11}
\ee
where $g_{s}$ is the strong coupling, which determines the strength of the interaction between colored fields, and $f^{ABC}$, with $A,B,C=1,...8$, are the structure functions of QCD which satisfy the Jacobi identity,
\be
f_{ABE}f_{ECD}+f_{CBE}f_{AED}+f_{DBE}f_{ACE}=0.
\label{chQCD.12}
\ee

The Feynman rules for $qq$, $qg$ and $gg$ interactions can then be obtained by expanding expression (\ref{chQCD.2}) term by term,
\be
\begin{split}
{\cal L}_{cl}&=\underbrace{-\frac{1}{4}\,\left(\partial_{\mu}G^{A}_{\nu}(x)-\partial_{\nu}G^{A}_{\mu}(x)\right)\left(\partial^{\mu}G_{A}^{\nu}(x)-\partial^{\nu}G_{A}^{\mu}(x)\right)}_{\textnormal{kinetic gluon term}}+\\
& \underbrace{\sum_{q}\bar{\psi}^{r}_{q}(x)\left(i\gamma^{\mu}\partial_{\mu}-m_{q}\right)_{rs}\psi^{s}_{q}(x)}_{\textnormal{kinetic and massive quark term}}-\underbrace{g_{s}G^{A}_{\mu}(x)\sum_{q}\bar{\psi}^{r}_{q}(x)\gamma^{\mu}\,\left(\frac{\lambda^{A}}{2}\,\right)_{rs}\psi^{s}_{q}(x)}_{\textnormal{quark-gluon interaction term}}+\\
&+\underbrace{\frac{g_{s}}{2}\,f^{ABC}\left(\partial_{\mu}G^{A}_{\nu}(x)-\partial_{\nu}G^{A}_{\mu}(x)\right)G^{B}_{\mu}(x)G^{C}_{\nu}(x)}_{\textnormal{3-gluon interaction term}}-\underbrace{\frac{g_{s}}{4}\,f^{ABC}f^{ADE}G^{B}_{\mu}(x)G^{C}_{\nu}(x)G^{D}_{\mu}(x)G^{E}_{\nu}(x)}_{\textnormal{4-gluon interaction term}}.
\label{chQCD.13}
\end{split}
\ee

QCD is a local gauge theory, therefore one can perform a redefinition of the fields independently at every point in space and time without changing the physical content of the theory. Thus, these redefinitions of the quark and gluon fields lead to transformations like the following ones:
\be
\begin{split}
&\psi_{q}(x) \fdd \psi_{q}^{\prime}(x)=U(x)\left[\psi_{q}(x)\right]=\left[e^{ig_{s}t^{A}\theta^{A}(x)}\right]\psi_{q}(x),\\
&G^{A}_{\mu} \fdd G^{\prime A}_{\mu}(x)=G^{A}_{\mu}+\partial_{\mu}\theta^{A}(x)+g_{s}f^{ABC}\theta^{B}(x)G^{C}_{\mu}(x),
\end{split}
\label{chQCD.14}
\ee
where $\theta^{A}(x)$ stands for the set of arbitrarily infinitesimal functions known as gauge angle, from which the physical observables are independent. The covariant derivative transforms in the same way as the field itself,
\be
D_{\mu}\psi_{q}(x)\to D^{\prime}_{\mu}\psi^{\prime}_{q}(x)=\left(\partial_{\mu}+ig_{s}t^{A}G^{\prime A}_{\mu}\right)\psi^{\prime}_{q}(x)=U(x)\left[D_{\mu}\right]\psi_{q}(x),
\label{chQCD.15}
\ee
then the transformation property for the gluon field is given by
\be
t^{A}G'^{A}_{\mu}=U(x)\,t^{A}G_{\mu}^{A}U^{-1}(x)+\frac{i}{g_{s}}\left[\partial_{\mu}U(x)\right]U^{-1}(x).
\label{chQCD.16}
\ee
A straightforward calculation shows that the transformation property of the non-Abelian field strength tensor is
\be
T^{A}F'^{A}_{\mu \nu}=U(x)T^{A}F_{\mu \nu}^{A}U^{-1}(x).
\label{chQCD.17}
\ee

It is impossible to define a gluon propagator without fixing the gauge first. The Feynman rules can only be properly deduced after a previous insertion of a gauge-fixing term ${\cal L}_{gf}$. One possible choice is the Lorenz gauge $\partial_{\mu}G^{\mu}_{A}=0$,
\be
{\cal L}_{fc}=-\frac{1}{2\lambda}(\partial_{\mu}G^{\mu}_{A})^{2},
\label{chQCD.18}
\ee  
where $\lambda$ is a gauge parameter. The Lorenz gauge defines the class of the so-called covariant gauges \cite{Ryder:1985wq}. By means of expression (\ref{chQCD.16}), the gluon propagator is written as
\be
i\Delta^{\mu\nu}(k)=-\frac{i}{k^{2}}\left(g^{\mu\nu}+(\lambda-1)\frac{k^{\mu}k^{\nu}}{k^{2}}\right),
\label{chQCD.19}
\ee
where $k^{\mu}$ stands for the particles $4-$momenta. It is usually defined that $\lambda=1$ $(\lambda=0)$ represents the Feynman (Landau) gauge. However, in a non-Abelian theory, such as QCD, the gauge-fixing term must be accompanied by terms involving complex scalar fields $\eta^{A}$,
\be
{\cal L}_{gt}=\partial_{\mu}\eta^{A\dagger}(D^{\mu}_{AB}\eta^{B}),
\label{chQCD.20}
\ee
and usually called as ghost fields or Faddeev-Popov fields \cite{Faddeev:1967fc,Faddeev:1980be}. In fact, it is required the presence of these fields because they properly cancel the nonphysical degrees of freedom that one usually encounters in working with covariant gauges\footnote{This brief introduction on QCD does not aim to discuss Gribov's copies, therefore, it will not be considered here.}. 

Another possible choice for the gauge-fixing term is the axial gauge,
\be
{\cal L}_{fc}=\partial_{\mu}(n^{\mu}G_{\mu}^{A})^{2}.
\label{chQCD.21}
\ee
This kind of choice has the advantage of avoiding the presence of nonphysical degrees of freedom, but the disadvantage is that the gluon propagator is more complicated,
\be
i\Delta^{\mu\nu}(k)=\frac{i}{k^{2}}\left(-g^{\mu\nu}+\frac{n^{\mu}k^{\nu}+n^{\nu}k^{\mu}}{n\cdot k}-\frac{(n^{2}+\lambda k^{2})k^{\mu}k^{\nu}}{(n\cdot k)^{2}}\right).
\label{chQCD.22}
\ee
Assuming that $n^{2}=0$ and $\lambda=0$, then it defines the light-cone gauge also called the physical gauge, because at $k^{2}\to 0$ there are only two polarizations,
\be
k\cdot\epsilon^{(i)}(k)=0\,\,\,\,\textnormal{and}\,\,\,\,n\cdot\epsilon^{(i)}(k)=0,
\label{chQCD.23}
\ee  
where $\epsilon(k)$ stands for the polarization states.

\subsection{\textsc{Renormalization and the Effective Coupling}}
\label{secQCD.1.2}
\mbox{\,\,\,\,\,\,\,\,\,}
The calculation of Feynman diagrams with loops leads to divergences in the $4-$momentum integrals, which must be properly regularized. These divergences can be treated by means of regularization processes, for instance: adding extra parameters into the theory (infrared mass scale $m_{g}$), imposing an \tit{ad hoc} upper integration limit (regularization with a momentum cutoff in the ultraviolet region) or even working in a space with non-integer dimensions (dimensional regularization $D=4-2\epsilon$). 

An alternative process is the possibility to perform a field rescaling in such a way that the divergences can be absorbed into redefined physical quantities, such as mass, coupling constant and field strength, by means of a renormalization process of the theory\footnote{Formally, the renormalization is performed by adding connected Feynman diagrams, \tit{i.e.}, irreducible $2$-point Green functions. In a theory such as QCD, it is possible to define a renormalized propagator using products of $\gamma_{\Gamma}(g_{s})$ known as anomalous dimension. Therefore, one is able to find the class of equations which defines the renormalization group, the Callan-Szymanski equations.}. Although the physical quantities must be independent of the renormalization scheme, the choice of the perturbative expansion parameter is not unique. Even the renormalization scales (subtraction points that remove the ultraviolet divergences) are not necessarily the same whenever applied into different processes. The renormalization process introduces one, or more than one, scale parameter with mass dimension, but as mentioned the physical observables must be not scale-dependent, and this invariance can be studied by means of the renormalization group equations. Therefore, the subtraction of ultraviolet divergences leads to the emergence of a renormalization scale $\mu$, thus all the renormalized quantities turn out to be scale-dependent.

By considering a dimensionless physical observable ${\cal R}$, which depends on a single energy scale $Q$, calculated as a perturbation series in the effective coupling $\alpha_{s}\equiv\alpha_{s}(\mu^{2})=g^{2}_{s}/4\pi$, and supposing either that $Q$ is much bigger than all other dimensionful parameters. Since this renormalization scale introduces a second mass scale $\mu$, then in the general case ${\cal R}$ depends only on the ratio $Q^{2}/\mu^{2}$ and the renormalized coupling $\alpha_{s}$, but not $\mu$ itself,
\be
{\cal R}\equiv{\cal R}\left(\frac{Q^{2}}{\mu^{2}},\alpha_{s}(\mu^{2})\right),
\label{chQCD.24}
\ee
mathematically, the $\mu$-independence is given by
\be
\mu^{2}\,\frac{d}{d\mu^{2}}{\cal R}(Q^{2}/\mu^{2},\alpha_{s})=\left(\mu^{2}\,\frac{\partial}{\partial\mu^{2}}+\mu^{2}\,\frac{\partial\alpha_{s}}{\partial\mu^{2}}\frac{\partial}{\partial\alpha_{s}}\right)\,{\cal R}(Q^{2}/\mu^{2},\alpha_{s})=0.
\label{chQCD.25}
\ee
This expression can be rewritten in a more compact form,
\be
\frac{d}{d\mu^{2}}{\cal{R}}(e^{\tau},\alpha_{s})=\left(\frac{\partial e^{\tau}}{\partial\mu^{2}}\frac{\partial}{\partial\alpha}+ \frac{\partial\alpha}{\partial\mu^{2}}\frac{\partial}{\partial\alpha}\right){\cal{R}}(e^{\tau},\alpha_{s})=0,
\label{chQCD.26}
\ee
where it was defined
\be
\tau\equiv\log\left(\frac{Q^{2}}{\mu^{2}}\right),\,\,\,\,\textnormal{and}\,\,\,\,\beta(\alpha_{s})\equiv\mu^{2}\,\frac{\partial\alpha_{s}}{\partial\mu^{2}}=\frac{\partial\alpha_{s}}{\partial\log\mu^{2}}.
\label{chQCD.27}
\ee
where $\tau$ is just a parameter and $\beta(\alpha_{s})$ is the well known beta function of the renormalization group, and it gives the asymptotic behavior of the theory in the ultraviolet regime. Since $\partial e^{\tau}/\partial \mu^{2}=-Q^{2}/\mu^{2}$, then expression (\ref{chQCD.26}) can be written as
\be
\left(-\frac{Q^{2}}{\mu^{2}}\frac{\partial}{\partial\tau}+\mu^{2} \frac{\partial\alpha_{s}}{\partial\mu^{2}}\frac{\partial}{\partial\alpha_{s}}\right){\cal{R}}(e^{\tau}, \alpha_s)=0,
\label{chQCD.28}
\ee
hence, taking the limit at $\mu^{2}\to Q^{2}$,
\be
\left(-\frac{\partial}{\partial\tau}+\beta(\alpha_{s})\,\frac{\partial}{\partial\alpha_{s}}\right){\cal R}(e^{\tau},\alpha_{s})=0,
\label{chQCD.29}
\ee
which is known as the renormalization equation of the QCD group. It is possible to show that ${\cal R}(e^{\tau},\alpha_{s})={\cal R}(1,\alpha_{s}(\tau))=\alpha_{s}(\tau)$ is solution of the above expression (\ref{chQCD.29}) with boundary condition $\alpha_{s}(\tau=0)=\alpha_{s}(\mu^{2})=\alpha_{s}$. By implicitly defining a new function, the running coupling $\alpha_{s}(\tau)$, such that
\be
\tau=\int^{\alpha_{s}(\tau)}_{\alpha_{s}(0)}\frac{d\alpha^{\prime}}{\beta(\alpha^{\prime})},
\label{chQCD.30}
\ee
then all the scale dependence in ${\cal R}$ enters through the running of the coupling constant. Hence, by taking ${\cal R}\equiv\alpha(\tau)$ in expression (\ref{chQCD.29}) and differentiating (\ref{chQCD.30}),
\be
\frac{d\alpha_{s}(\tau)}{d\tau}=\beta(\alpha_{s}(\tau)),\,\,\,\,\textnormal{and}\,\,\,\,\frac{d\alpha_{s}(\tau)}{d\alpha_{s}}=\frac{\beta(\alpha_{s}(\tau))}{\beta(\alpha_{s})}.
\label{chQCD.31}
\ee
Once it is known the behavior of the running coupling then one can predict the variation of ${\cal R}$ with any given $Q$.

The running of the coupling constant is determined by the renormalization group equation once it is known the $\beta$ function. In QCD, the $\beta$ function has the perturbative expansion \cite{Ellis:1990pp,Politzer:1973fx,Gross:1973id,Caswell:1974gg}
\be
-\beta({\alpha_{s}(\tau)})=b_{0}\alpha_{s}^{2}(\tau)+b_{1}\alpha_{s}^{3}(\tau)+b_{2}\alpha_{s}^{4}(\tau)+...,
\label{chQCD.32}
\ee
where
\be
b_{0}=\frac{\beta_{0}}{4\pi}=\frac{1}{4\pi}\left(11-\frac{2}{3}N_{f}\right),
\label{chQCD.33}
\ee
\be
b_{1}=\frac{\beta_{1}}{16\pi^{2}}=\frac{1}{16\pi^{2}}\left(102-\frac{38}{3}N_{f}\right),
\label{chQCD.34}
\ee
and $N_{f}$ stands for the number of active light flavors, $m_{q}\ll Q$ and terms $b_{\geq2}$ are dependent on the renormalization scheme. 

By means of the perturbative expansion of the $\beta$ function, see expression (\ref{chQCD.32}), and rewriting expression (\ref{chQCD.31}) as
\be
\frac{d\alpha_{s}(\tau)}{d\tau}\to\frac{d\alpha_{s}(Q^{2})}{dQ^{2}/Q^{2}},
\label{chQCD.35}
\ee
then the running of the coupling with the energy scale $Q$ is given by
\be
\frac{d\alpha_{s}(\tau)}{d\tau}=\beta(\alpha_{s}(\tau))=Q^{2}\,\frac{d\alpha_{s}(Q^{2})}{dQ^{2}}=-b_{0}\alpha_{s}^{2}(Q^{2})\left(1+\frac{b_{1}}{b_{0}}\alpha_{s}(Q^{2})+\frac{b_{2}}{b_{0}}\alpha_{s}^{2}(Q^{2})+...\right).
\label{chQCD.36}
\ee

The perturbative expansion of the $\beta$ function can be truncated so that only the terms $b_{0}$ and $b_{1}$ survive. As a matter of fact, it is possible to do such thing because physical quantities are usually known up to NLO in perturbative calculations. Therefore, in phenomenology only LO and NLO terms are considered. The corresponding behavior of $\alpha_{s}(Q^{2})$ can be calculated rewriting expression (\ref{chQCD.36}) as a geometric series,
\be
Q^{2}\,\frac{d\alpha_{s}(Q^{2})}{dQ^{2}}=-b_{0}\,\alpha_{s}^{2}(Q^{2})\sum^{\infty}_{n=0}\left(\frac{b_{1}}{b_{0}}\,\alpha_{s}(Q^{2})\right)^{n}=-\,\frac{b_{0}\,\alpha_{s}^{2}(Q^{2})}{1-\frac{b_{1}}{b_{0}}\,\alpha_{s}(Q^{2})},
\label{chQCD.37}
\ee
where it was considered only the first order of perturbation, and the $b_{n}$ terms in expression (\ref{chQCD.32}) are written as \cite{Ellis:1990pp,Muta:2010xua}
\be
b_{n}=b_{0}\left(\frac{b_{1}}{b_{0}}\right)^{n}.
\label{chQCD.38}
\ee

Since expressions (\ref{chQCD.36}) and (\ref{chQCD.37}) are equivalent up to ${\cal O}(\alpha^{3}_{s})$, then the latter one can be used to calculate $\alpha_{s}^{LO}(Q^{2})$ and $\alpha_{s}^{N\!LO}(Q^{2})$. Hence, the integration of expression (\ref{chQCD.31}) where the lower limit is the cutoff $\mu^{2}$, and by means of the definition of the parameter $\tau$, see expression (\ref{chQCD.27}), takes the form
\be
\int^{\alpha_{s}(Q^{2})}_{\alpha_{s}(\mu^{2})}\frac{d\alpha_{s}(Q^{2})}{\beta(\alpha_{s})}=\tau=\log\left(\frac{Q^{2}}{\mu^{2}}\right),
\label{chQCD.39}
\ee
and by expression (\ref{chQCD.37}),
\be
\int^{\alpha_{s}(Q^{2})}_{\alpha_{s}(\mu^{2})}d\alpha_{s}(Q^{2})\left(-\,\frac{1-\frac{b_{1}}{b_{0}}\alpha_{s}(Q^{2})}{b_{0}\alpha^{2}_{s}(Q^{2})}\right)=\log\left(\frac{Q^{2}}{\mu^{2}}\right),
\label{chQCD.40}
\ee
where the solution can be obtained by solving the integral. Thus, one gets
\be
\frac{1}{b_{0}\alpha_{s}(Q^{2})}-\,\frac{1}{b_{0}\alpha_{s}(\mu^{2})}+\frac{b_{1}}{b_{0}^{2}}\,\log\left(\frac{\alpha_{s}(Q^{2})}{\alpha_{s}(\mu^{2})}\right)=\log\left(\frac{Q^{2}}{\mu^{2}}\right).
\label{chQCD.41}
\ee
Collecting terms with explicit dependence on the energy scale $Q$ and on the subtraction point $\mu$, then one is able to find
\be
-\,\frac{1}{b_{0}\alpha_{s}(Q^{2})}-\,\frac{b_{1}}{b_{0}^{2}}\,\log\alpha_{s}(Q^{2})+\log Q^{2}=-\,\frac{1}{b_{0}\alpha_{s}(\mu^{2})}-\,\frac{b_{1}}{b_{0}^{2}}\,\log\alpha_{s}(\mu^{2})+\log \mu^{2}=C,
\label{chQCD.42}
\ee
and this expression is nothing more than a $c$-number. So in principle there ought not to be a problem at all to equal the above expression to a constant $C$.

The perturbative QCD tells us how the coupling behaves with the renormalization scale, but it says nothing on the scale itself. Therefore, this scale must be chosen as a fundamental parameter in the theory so that the coupling is set at some reference scale defined by convention. One common choice is to use the neutral vector $Z$ boson mass as reference,  $\mu=M_{Z}\simeq 92$, GeV, which is larger than the typical QCD scale, and thus it is sufficient to ensure that the calculations lie in the perturbative region. Hence, the QCD fundamental parameter is given by the experimental value of $\alpha_{s}(M_{Z}^{2})$.

A clever way to deal with that arbitrarily $C$ constant is to introduce a dimensional parameter $\Lambda$ into the definition of the effective coupling $\alpha_{s}(Q^{2})$. This parameter sets the scale at $\alpha_{s}\to\infty$. Then expression (\ref{chQCD.30}) can be rewritten as
\be
\log\left(\frac{Q^{2}}{\Lambda^{2}}\right)=\int^{\alpha_{s}(Q^{2})}_{\infty}\frac{d\alpha_{s}(Q^{2})}{\beta(\alpha_{s}(Q^{2}))}=\int^{\infty}_{\alpha_{s}(Q^{2})}\frac{d\alpha_{s}(Q^{2})}{b_{0}\alpha_{s}^{2}(\tau)+b_{1}\alpha_{s}^{3}(\tau)+b_{2}\alpha_{s}^{4}(\tau)+...}\,\,,
\label{chQCD.43}
\ee
and using the perturbative expansion of the $\beta$ function, see expression (\ref{chQCD.32}), and integrating up to ${\cal O}(\alpha^{3}_{s})$, one arrives at the precise functional forms at LO and NLO for the running coupling.

Alternatively, it is possible to obtain the same result for the running coupling at LO level, and an approximate one at NLO \cite{Ellis:1990pp}, by defining the constant $C$ as
\be
C=\log\Lambda^{2}+\frac{b_{1}}{b^{2}_{0}}\,\log b_{0},
\label{chQCD.44}
\ee
so that expression (\ref{chQCD.42}) are given by,
\be
\alpha_{s}(Q^{2})=\frac{1}{b_{0}\left[\log\left(\frac{Q^{2}}{\Lambda^{2}}\right)-\,\frac{b_{1}}{b^{2}_{0}}\,\log(b_{0}\,\alpha_{s}(Q^{2}))\right]}=\frac{1}{b_{0}\log\left(\frac{Q^{2}}{\Lambda^{2}}\right)\left[1-\frac{b_{1}}{b^{2}_{0}} \frac{\log(b_{0}\,\alpha_{s}(Q^{2}))}{\log\left(Q^{2}/\Lambda^{2}\right)}\right]}.
\label{chQCD.45}
\ee
Expanding into inverse powers of $\log\left(Q^{2}/\Lambda^{2}\right)$ up to $1$-order, and using that $1/(1-x)=1+x+x^{2}+...$, one arrives at the following result:
\be
\alpha_{s}(Q^{2})=\frac{1}{b_{0}\log\left(\frac{Q^{2}}{\Lambda^{2}}\right)}\left[1+\frac{\frac{b_{1}}{b^{2}_{0}}\,\log(b_{0}\,\alpha_{s}(Q^{2}))}{\log\left(\frac{Q^{2}}{\Lambda^{2}}\right)}\right]=\underbrace{\frac{1}{b_{0}\log\left(\frac{Q^{2}}{\Lambda^{2}}\right)}}_{\textnormal{LO term}}+\underbrace{\frac{\frac{b_{1}}{b^{2}_{0}}\,\log(b_{0}\,\alpha_{s}(Q^{2}))}{b_{0}\log\left(\frac{Q^{2}}{\Lambda^{2}}\right)\log\left(\frac{Q^{2}}{\Lambda^{2}}\right)}}_{\textnormal{NLO term}}.
\label{chQCD.46}
\ee
Hence, the running coupling at LO is obtained by considering only the first term in the \tit{rhs},
\be
\alpha_{s}^{LO}(Q^{2})=\frac{4\pi}{\beta_{0}\log\left(\frac{Q^{2}}{\Lambda^{2}}\right)},
\label{chQCD.47}
\ee
whereas for the case of the running coupling at NLO, both terms must be taken into account and a straightforward calculation leads to
\be
\alpha_{s}^{N\!LO}(Q^{2})=\frac{4\pi}{\beta_{0}\log\left(\frac{Q^{2}}{\Lambda^{2}}\right)}\left[1-\,\frac{\beta_{1}}{\beta_{0}^{2}}\,\frac{\log\log\left(\frac{Q^{2}}{\Lambda^{2}}\right)}{\log\left(\frac{Q^{2}}{\Lambda^{2}}\right)}\right].
\label{chQCD.49}
\ee
\mbox{\,\,\,\,\,\,\,\,\,}
It is usually said that expressions (\ref{chQCD.47}) and (\ref{chQCD.49}) are the $1$-loop and $2$-loop approximations in perturbation theory.

\section{\textsc{The Parton Model}}
\label{secQCD.2}
\mbox{\,\,\,\,\,\,\,\,\,}
Confinement prohibits the formation of free states of quarks an gluons, therefore the colorless hadrons observed in laboratory must be somehow linked to colored quarks and gluons described by QCD. The mechanism, which establishes this link, is known as the parton model and it was built under the hypothesis that hadrons are compounded by punctual particles collectively called partons, thus hadron-hadron interactions results from interactions among these partons. This hypothesis was verified in the late $60$'s by means of deep inelastic experiments at SLAC \cite{Taylor:1991ew}, and since then many other experiments have performed deep inelastic lepton-hadron scattering at increasingly higher energies and provided some of the most precise tests of perturbative QCD. It also represents the most direct way to probe the internal structure of hadrons

\subsection{\textsc{Deep Inelastic Scattering}}
\label{secQCD.2.1}
\mbox{\,\,\,\,\,\,\,\,\,}
It is described by the interaction of a high-energy charged lepton off a nucleon target. This interaction occurs by means of gauge bosons exchange ($\gamma^{\ast}$, $Z^{0}$ for neutral current and $W^{\pm}$ for charged current), where the process with virtual photon exchange has significant contributions for the cross section compared to those ones with $Z^{0}$ and $W^{\pm}$, because massive vector bosons are suppressed due to their mass appearing as a pole in the propagator $1/(Q^{2}+M^{2})$ \cite{Barone:2002cv,Halzen:1984mc,Ellis:1990pp}. In the final state, it is measured a lepton $\ell$ and a hadronic state $X$ formed by the nucleon fragmentation,
\be
\ell+N\to \ell^{\prime}+X.
\label{chQCD.50}
\ee
There are different names for different kinds of processes. For instance, processes where it is measured only the lepton beyond the final hadronic state $X$, are usually called as inclusive, see Figure \ref{ch2fig2}. There are also the class of semi-inclusive processes where it is selected a given final $X$ state or even the exclusive one described by a particular configuration where the nucleon is not completely dissociated.
\bfg[hbtp]
  \begin{center}
    \includegraphics[width=8.5cm,clip=true]{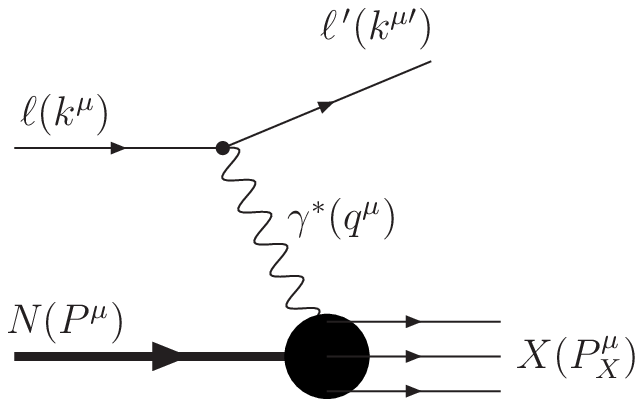}
    \caption{LO perturbative Feynman diagram for DIS $ep\to eX$.}
    \label{chQCDfig1}
  \end{center}
\efg

The inclusive process in reaction (\ref{chQCD.50}) is represented by the LO perturbative diagram in Figure \ref{chQCDfig1}, where the incoming and outgoing lepton $4-$momenta (assumed hereafter to be an electron) are labeled by $k^{\mu}$ and $k^{\prime \mu}$, respectively, the momentum of the target (assumed hereafter to be a proton) by $P^{\mu}$ and the momentum transfer by $q^{\mu}=k^{\mu}-k^{\prime\mu}$. Then the standard deep inelastic variables are described by three kinematic invariants, where one of them is the incoming lepton energy $E$, or alternatively the Mandelstam invariant that represents the CM energy \cite{Barone:2002cv,Halzen:1984mc},
\be
s=(k+P)^{2},
\label{chQCD.51}
\ee
and the other two are usually chosen among the following:
\be
Q^{2}=-q^{2}=(k-k^{\prime})^{2}>0,
\label{chQCD.52}
\ee
\be
W^{2}=(P+q)^{2}
\label{chQCD.53}
\ee
\be
\nu=\frac{P\cdot q}{m_{p}}=\frac{W^{2}+Q^{2}-m_{p}^{2}}{2m_{p}},
\label{chQCD.54}
\ee
\be
x=\frac{Q^{2}}{2P\cdot q}=\frac{Q^{2}}{2m_{p}\nu}=\frac{Q^{2}}{W^{2}+Q^{2}-m_{p}^{2}},
\label{chQCD.55}
\ee
\be
y=\frac{\nu}{E}=\frac{P\cdot q}{P\cdot k}=\frac{W^{2}+Q^{2}-m_{p}^{2}}{s-m_{p}^{2}},
\label{chQCD.56}
\ee
where the energy variables refer to the target rest frame, $m_{p}$ is the proton mass, $W^{2}$ is the CM energy squared of $\gamma^{\ast}p$ system, that is the invariant mass squared of the hadronic system $X$, $\nu=E-E^{\prime}$ is the transferred energy.

DIS is described by the dimensionless Bjorken variable $x$ defined by expression (\ref{chQCD.55}). Since $W^{2}\geq m_{p}^{2}$ and by means of expression (\ref{chQCD.55}), one finds that $W^{2}=m_{p}^{2}+2p\cdot q (1-x)$ and hence the Bjorken $x$ variable takes values $0\leq x \leq 1$. The variable $y=\nu/E$, sometimes called inelasticity, is the fraction of the incoming lepton energy carried by the exchanged photon and also takes values $0\leq y \leq 1$. A useful relation connecting $x$, $y$ and $Q^{2}$ is
\be
xy=\frac{Q^{2}}{s-m^{2}_{p}}\simeq\frac{Q^{2}}{s},\,\,\,\,s\gg m_{p}.
\label{chQCD.57}
\ee
The term deep inelastic is related to the kinematic regime where both $m_{p}\nu\gg m_{p}^{2}$ and $Q^{2}\gg m_{p}^{p}$, with $x$ fixed and finite. Thus one can safely neglect the proton mass with respect to other large energy scales in the process \cite{Barone:2002cv}.

In the lowest order the inclusive QED cross section for the $ep\to eX$ scattering can be written in terms of the leptonic $L_{\mu\nu}$ and hadronic $W_{\mu\nu}$ tensors,
\be
\frac{d^{2}\sigma}{dE^{\prime}d\Omega}=\frac{\alpha^{2}_{em}}{2m_{p}Q^{4}}\,\frac{E^{\prime}}{E}\,L^{\mu \nu}W_{\mu \nu},
\label{chQCD.58}
\ee
where $\alpha_{em}=e^{2}/4\pi\sim 1/137$ is the fine structure constant, also called the electromagnetic coupling constant, and $\Omega \equiv (\theta, \phi)$ is the scattering solid angle. The leptonic tensor is calculated by means of the QED Feynman rules \cite{Halzen:1984mc},
\be
L_{\mu\nu}=\frac{1}{2}\,\text{Tr}\left[\slashed{k}\gamma_{\mu}\slashed{k}^{\prime}\gamma_{\nu}\right]=2(k_{\mu}k_{\nu}^{\prime}+k_{\nu}k_{\mu}^{\prime}-g_{\mu\nu}k\cdot k^{\prime}),
\label{chQCD.59}
\ee
and the hadronic tensor is given by \cite{Barone:2002cv}
\be
W_{\mu\nu}=\frac{1}{2\pi}\int d^{4}z\,e^{iq\cdot z}\,\langle P\vert J_{\mu}(z)J_{\nu}(0)\vert P\rangle,
\label{chQCD.60}
\ee
where $J_{\mu}$ stands for the current density operator and represents the probability of transition between the initial to the final state, and $\vert P\rangle$ is the proton eigenstate. The leptonic tensor can be explicitly calculated in any type of vertex. However, since it is not fully understood the confinement mechanism, the hadronic tensor must be written as a parametrization combining all the possibilities between its $4-$momenta $P$ and $q$ in the $\gamma^{\ast}p$ vertex \cite{Halzen:1984mc},
\be
W_{\mu\nu}=W_{1}g_{\mu\nu}+\frac{W_{2}}{m^{2}_{p}}P_{\mu} P_{\nu}+\frac{W_{3}}{m^{2}_{p}}q_{\mu} q_{\nu}+\frac{W_{4}}{m^{2}_{p}}\left(q_{\mu} P_{\nu}+q_{\nu} P_{\mu}\right),
\label{chQCD.61}
\ee
where $W_{i}$ are scalar functions of $P$ and $q$. By means of some constraints, such that: the hadroninc tensor is symmetric, $W_{\mu\nu}=W_{\nu\mu}$; the hermitiancy of electromagnetic density current, $W_{\mu\nu}=W^{\ast}_{\mu\nu}$; and the conservation of electromagnetic density current, $q^{\mu} W_{\mu\nu}=q^{\nu} W_{\mu\nu}=0$; it is found that one possible parametrization is \cite{Barone:2002cv,Halzen:1984mc,Barger:1987nn}
\be
\begin{split}
\frac{1}{2m_{p}}\,W_{\mu\nu}&=\left(-g_{\mu\nu}+\frac{q_{\mu}q_{\nu}}{q^{2}}\right)W_{1}(P\cdot q,q^{2})+\\
&+\left[\left(P_{\mu}-\,\frac{P\cdot q}{q^{2}}\,q_{\mu}\right)\,\left(P_{\nu}-\,\frac{P\cdot q}{q^{2}}\,q_{\nu}\right)\right]\frac{W_{2}(P\cdot q,q^{2})}{m^{2}_{p}}.
\end{split}
\label{chQCD.62}
\ee
Therefore, the cross section is written in terms of the two structure functions $W_{1}$ and $W_{2}$,
\be
\frac{d^{2}\sigma}{dE^{\prime}d\Omega}=\frac{4\alpha_{em}^{2}E^{\prime 2}}{Q^{4}}\left(2W_{1}\sin^{2}\frac{\theta}{2}+W_{2}\cos^{2}\frac{\theta}{2}\right),
\label{chQCD.63}
\ee
where $\theta$ is the electron scattering angle. It is customary to introduce the dimensionless structure functions:
\be
F_{1}(x,Q^{2})\equiv m_{p}W_{1}(\nu.Q^{2}),
\label{chQCD.64}
\ee
\be
F_{2}(x,Q^{2})\equiv\nu W_{2}(\nu.Q^{2}).
\label{chQCD.65}
\ee
In terms of $F_{1}$ and $F_{2}$, the hadronic tensor reads
\be
W_{\mu\nu}=2\left(-g_{\mu\nu}+\frac{q_{\mu}q_{\nu}}{q^{2}}\right)F_{1}(x,Q^{2})
+\frac{2}{P\cdot q}\left[\left(P_{\mu}-\,\frac{P\cdot q}{q^{2}}\,q_{\mu}\right)\,\left(P_{\nu}-\,\frac{P\cdot q}{q^{2}}\,q_{\nu}\right)\right]F_{2}(x,Q^{2}).
\label{chQCD.66}
\ee
An usual form of the DIS inclusive $ep\to eX$ cross section, which is typically found in the literature \cite{Barone:2002cv,Halzen:1984mc,Barger:1987nn}, is given in terms of the dimensionless structure function and also $x$ and $y$ variables. But first, one might be able to rewrite the cross section into cylindric symmetry coordinates,
\be
\frac{d^{2}\sigma}{dE^{\prime}d\Omega}=\frac{d^{2}\sigma}{dE^{\prime}2\pi \sin\theta d\theta},
\label{chQCD.67}
\ee
and by means of the Jacobi determinant,
\bear
\frac{d^{2}\sigma}{dE^{\prime}d\Omega}&=&\frac{1}{2\pi \sin\theta}\left|\frac{\partial(x,y)}{\partial(E^{\prime},\theta)}\right|\frac{d^{2}\sigma}{dxdy} \nonumber \\
&=&\frac{1}{2\pi \sin\theta}\begin{vmatrix}
\frac{\partial x}{\partial E^{\prime}}&\frac{\partial x}{\partial \theta}\\
\frac{\partial y}{\partial E^{\prime}}&\frac{\partial y}{\partial \theta}\\
\end{vmatrix}\frac{d^{2}\sigma}{dxdy} \nonumber \\
&=&\frac{1}{2\pi \sin\theta}\begin{vmatrix} 
\frac{2E\sin^{2}\frac{\theta}{2}}{M}\left[\frac{E^{\prime}+(E-E^{\prime})}{(E-E^{\prime})^{2}}\right]&\frac{2EE^{\prime}}{(E-E^{\prime})M}\sin \frac{\theta}{2}\cos \frac{\theta}{2}\\
-\frac{1}{E}&0\\
\end{vmatrix}\frac{d^{2}\sigma}{dxdy} \nonumber \\
&=&\frac{1}{2\pi \sin \theta}\frac{2E^{\prime}}{(E-E^{\prime})M}\frac{\sin\theta}{2}\frac{d^{2}\sigma}{dxdy} \nonumber \\
&=&\frac{E^{\prime}}{2\pi MEy}\frac{d^{2}\sigma}{dxdy}.
\label{chQCD.68}
\eear 
Finally, the inclusive cross section in terms of $x$ and $y$ is given by
\be
\frac{d^{2}\sigma}{dxdy}=\frac{4\pi\alpha^{2}_{em}s}{Q^{4}}\left[xy^{2}F_{1}(x,Q^{2})+\left(1-y-\frac{xym_{p}^{2}}{s}\right)F_{2}(x,Q^{2})\right].
\label{chQCD.69}
\ee

An alternative derivation of these results is given in terms of the Mandelstam invariants:
\be
s\simeq 2k\cdot P=2m_{p}E,
\label{chQCD.70}
\ee
\be
t\simeq -Q^{2}=-4EE^{\prime}\sin^{2}\frac{\theta}{2},
\label{chQCD.71}
\ee
\be
u\simeq -2k^{\prime}\cdot P=-2m_{p}E^{\prime}.
\label{chQCD.72}
\ee
and a straightforward calculation gives
\be
\frac{d^{2}\sigma}{dtdu}=\frac{4\pi\alpha^{2}_{em}}{t^{2}s^{2}}\,\frac{1}{s+u}\,\left[(s+u^{2})xF_{1}(x,t)-usF_{2}(x,t)\right],
\label{chQCD.73}
\ee
where there is a relation among the Bjorken variable $x$ and the invariants of Mandelstam,
\be
x=\frac{Q^2}{2P\cdot q}=\frac{Q^2}{2k\cdot p-2k^\prime \cdot p}=\frac{-t}{s+u}.
\label{chQCD.74}
\ee

In principle, measurements of cross section (\ref{chQCD.69}) at different values of $x$ and $y$, or either the cross section as given by expression (\ref{chQCD.73}), allows one to determine the structure functions $F_{1}(x,Q^{2})$ and $F_{2}(x,Q^{2})$. However, since the contribution of $F_{1}(x,Q^{2})$ to the cross section is relatively small, the measurements are usually performed only to determine $F_{2}(x,Q^{2})$, thus $F_{1}(x,Q^{2})$ is theoretically estimated. In Figure \ref{chQCDfig2} is depicted different measurements of $F_{2}(x,Q^{2})$ at different values of $x$ and $Q^{2}$, obtained by ZEUS, H1, NMS, BDCMS and E665 experiments \cite{Chekanov:2002pv}. The values of $F_{2}$ at $x\sim0.25$ are in the region where it was originally observed the scaling invariance at SLAC.

\subsection{\textsc{Bjorken Scaling}}
\label{secQCD.3.1}
\hyphenation{ap-pro-xi-ma-te-ly}
\mbox{\,\,\,\,\,\,\,\,\,}
In DIS, the signal that there are actually structureless particles inside the proton is that a short wavelength photon beam resolves the quarks within the proton provided that $\lambda(\simeq 1/\sqrt{Q^{2}})\ll 1$ fm. Thus, the $ep\to eX$ cross section described by expression (\ref{chQCD.64}), or similarly by expressions (\ref{chQCD.69}) and (\ref{chQCD.73}), suddenly starts behaving like that one for free Dirac particle, and expression (\ref{chQCD.64}) turns into the cross section for a generic electron-lepton elastic scattering process, as in the case of $e\mu\to e\mu$ given by the Mott formula \cite{Halzen:1984mc},
\be
\frac{d^{2}\sigma}{dE^{\prime}d\Omega}=\frac{4\alpha^{2} E^{\prime 2}}{q^{2}}\left(\cos^{2}\frac{\theta}{2}-\frac{q^{2}}{2m^{2}}\sin^{2}\frac{\theta}{2}\right) \delta \left(\nu+\frac{q^{2}}{2m^{2}}\right).
\label{chQCD.75}
\ee
Thus, the proton structure functions become
\be
\begin{split}
& 2m_{p}W^{point}_{1}(\nu,Q^{2})=2F^{point}_{1}(\nu,Q^{2})=\frac{Q^{2}}{2m^{2}_{p}\nu}\,\delta\left(1-\frac{Q^{2}}{2m_{p}\nu}\right)\\
& \nu W^{point}_{1}(\nu,Q^{2})=F^{point}_{2}(\nu,Q^{2})=\delta\left(1-\frac{Q^{2}}{2m_{p}\nu}\right).
\end{split}
\label{chQCD.76}
\ee

These point-like structure functions depend only on the ratio $Q^{2}/2m_{p}\nu$, instead of $Q^{2}$ and $\nu$, independently. However, this is not the case in elastic $ep$ scattering, because protons actually have structure and it can be properly implemented by introducing a form factor,
\be
\begin{split}
& 2m_{p}W^{elastic}_{1}(\nu,Q^{2})=2F^{elastic}_{1}(\nu,Q^{2})=\frac{Q^{2}}{2m^{2}_{p}\nu}\,G^{2}(Q^{2})\delta\left(1-\frac{Q^{2}}{2m_{p}\nu}\right),\\
& \nu W^{elatic}_{1}(\nu,Q^{2})=F^{elastic}_{2}(\nu,Q^{2})=G^{2}(Q^{2})\delta\left(1-\frac{Q^{2}}{2m_{p}\nu}\right),
\end{split}
\label{chQCD.77}
\ee
thus the structure functions cannot be rearranged to be functions of a single dimensionless variable. When $Q^{2}$ increases above $0.71$ GeV$^{2}$, as in a dipole-like picture, this value reflects the inverse size of the proton. The form factor weaken the chance of elastic scattering to occur, so the proton is more likely to break up. Hence, at the high-energy limit, when asymptotically one has
\be
\nu,Q^{2}\fdd \infty,\,\,\,\,\textnormal{but the ratio}\,\,\,\,x=\frac{Q^{2}}{2m_{p}\nu}\to \textnormal{fixed},
\label{chQCD.78}
\ee
now called the Bjorken limit, structure functions $F_{1}$ and $F_{2}$ should approximately scale, \tit{i.e.} depend only on $x$ \cite{Bjorken:1968dy,Bjorken:1969ja},
\be
F_{i}(x,Q^{2})\to F_{i}(x).
\label{chQCD.79}
\ee
\mbox{\,\,\,\,\,\,\,\,\,}
One might be able to conclude that in a process such as DIS, the virtual photon probes the proton as a bag of spin $1/2$ particles. 

\bfg[hbtp]
  \begin{center}
    \includegraphics[width=15cm,clip=true]{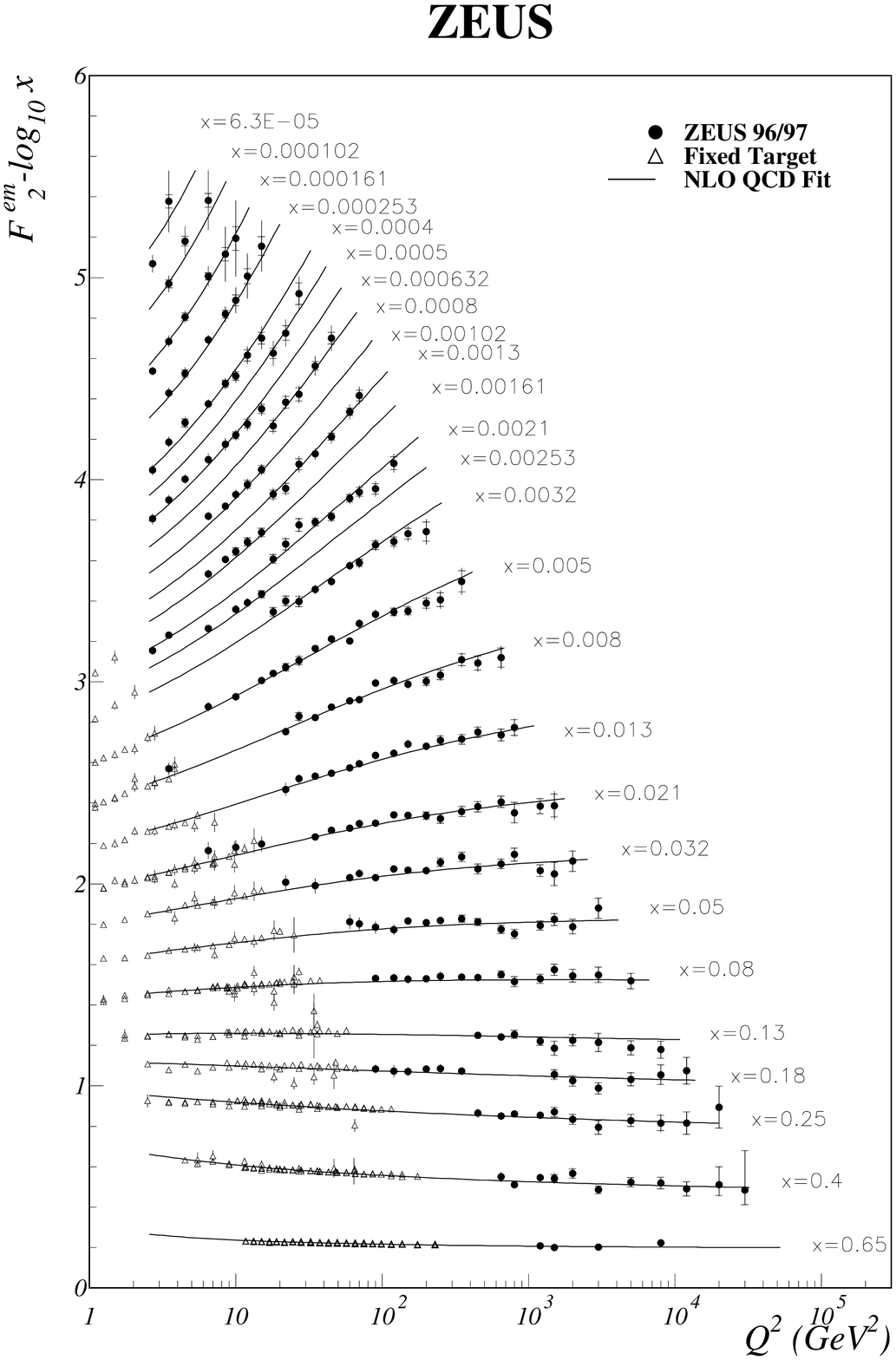}
    \caption{Measurements of the structure function $F_{2}(x,Q^{2})$ for several $x$-values. Respectively, measurements at at low-$Q^{2}$ were obtained by the NMS, BDCMS and E$665$ fixed target experiments, and those ones at high-$Q^{2}$ were obtained by the ZEUS and H$1$ experiments \cite{Chekanov:2002pv}.}
    \label{chQCDfig2}
  \end{center}
\efg

\section{\textsc{The Original Parton Model}}
\label{secQCD.4}
\mbox{\,\,\,\,\,\,\,\,\,}
In the late $60$'s, the experimental verification that in the Bjorken limit the structure functions are only $x$-dependent gave birth to the parton model. It is based on the assumption that the virtual photon incoherently scatters off the constituents of the nucleon, which in the limit at $Q^{2}\to \infty$ are treated as free particles, see Figure \ref{chQCDfig3}. The parton model is most easily formulated in the infinite momentum frame in which the proton is moving very fast, $p^{\mu}=(P,0,0,P)$ with $P\gg m_{p}$. In this picture, the quarks are essentially free during the interaction time interval with the virtual photon, since there are effects of Lorentz contraction (the longitudinal size of the proton is contracted by a factor of $m_{p}/P$ with respect to its original size in the rest frame), therefore the $\gamma^{\ast}p$ interaction time scales are much smaller than the interaction time among quarks itself\cite{Ellis:1990pp,Miller:1971qb}. 

Various types of point partons make up the proton, and each one of them carry a small fraction $x$ of their host proton's momentum and energy. Following this line of thought, the $ep$ interaction can be properly written as an incoherent summation of scattering probabilities among the electrons and free quarks,
\be
\left(\frac{d^{2}\sigma}{dxdy}\right)_{ep\to eX}=\sum_{q}\int^{1}_{0}d\xi\,f_{q}(\xi)\,\left(\frac{d^{2}\sigma}{dxdy}\right)_{eq_{i}\to eq_{i}},
\label{chQCD.80}
\ee
where $f_{q}(\xi)$ are the quark distribution functions and $d\xi\,f_{q}(\xi)$ are the probability to find a quark $q$ carrying a fraction of the proton's momentum between $\xi$ e $\xi+d\xi$, with $0 \leq \xi \leq1$. The actual number of partons within the proton is given by \cite{Halzen:1984mc}
\be
N_{q}=\int^{1}_{0}d\xi\,f_{q}(\xi),
\label{chQCD.81}
\ee
and by conservation of momentum, the sum over all fractions carried by each partons must equal the proton's momentum,
\be
\sum_{q}\int^{1}_{0}d\xi\,\xi\,f_{q}(\xi)=1.
\label{chQCD.82}
\ee

To obtain the structure function $F_{2}(x,Q^{2})$, first it is necessary to calculate the elementary $eq_{i}\to eq_{i}$ cross section \cite{Barger:1987nn,Ellis:1990pp},
\be
\left(\frac{d^{2}\sigma}{dxdy}\right)_{eq}=\frac{2\pi\alpha_{em}^{2}e_{q}^{2}}{Q^{4}}\left[1+(1-y)^{2}\right]\delta(x-\xi),
\label{chQCD.83}
\ee
\bfg[hbtp]
  \begin{center}
    \includegraphics[width=7cm,clip=true]{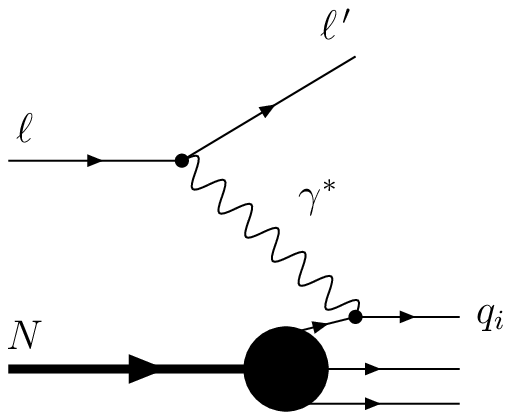}
    \caption{Parton model DIS diagram. The virtual photon interacts with a parton within the proton.}
    \label{chQCDfig3}
  \end{center}
\efg
where $e_{q}$ stands for the quark electric charge. The mass-shell condition\footnote{A real particle is in the mass-shell when $p^{\mu}=p=(E,\vec{p})$ and satisfies $p^{2}\equiv p\cdot p=E^{2}-\vec{p}^{2}=m^{2}=m^{2}$. In the case of a virtual particle, $p^{2}\neq m^{2}$. More specifically, for a virtual photon $q^{2}\neq 0$.} for the outgoing quark with $Q^{2}\gg m^{2}_{q}$,
\be
p^{\prime}_{q}=(p_{q}+q)^{2}=q^{2}+2p_{q}\cdot q=-2P\cdot q(x-\xi)=0,
\label{chQCD.84}
\ee
implies $x=\xi$, and justifies $\delta(x-\xi)$ in expression (\ref{chQCD.84}). Within the elementary $eq_{i}\to eq_{i}$ cross section, see expression (\ref{chQCD.83}), the differential cross section for the quark scattering is written as
\be
\frac{d^{2}\sigma}{dxdy}=\sum_{q}\int^{1}_{0}d\xi\,f_{q}(\xi)\,\frac{2\pi\alpha_{em}^{2}e_{q}^{2}}{Q^{4}}\left[1+(1-y)^{2}\right]\delta(x-\xi).
\label{chQCD.85}
\ee
\mbox{\,\,\,\,\,\,\,\,\,}
By comparing this result with expression (\ref{chQCD.69}), and neglecting the $m_{p}$, one finds the parton model prediction for the structure functions,
\be
F_{2}(x)=2xF_{1}(x)=x\sum_{q}e^{2}_{q}\int^{1}_{0}d\xi\,f_{q}(\xi)\,\delta(x-\xi)=\sum_{q}e^{2}_{q}\,x\,f_{q}(x).
\label{chQCD.86}
\ee
This result is known as the Callan-Gross relation \cite{Callan:1969uq} and suggests that the structure function $F_{2}(x)$ probes a quark with momentum fraction $x=\xi$. Therefore, the measured structure function is in fact a distribution in $x$ rather than a delta function, such that the quarks carry a range of momentum fractions, see Figure \ref{chQCDfig2}. The Callan-Gross relation is a direct consequence of the fermionic nature of the quarks \cite{Ellis:1990pp,Barone:2002cv}.

The measurements of the structure function $F_{2}$ can be used to reveal the internal structure of hadrons. In the case of $ep\to eX$ process, neglecting the presence of $c$ quarks and also heavier quarks, expression (\ref{chQCD.86}) can be written as \cite{Ellis:1990pp,Halzen:1984mc}
\be
\frac{1}{x}F^{ep}_{2}(x)=\left(\frac{2}{3}\right)^{2}\left[u^{p}(x)+\bar{u}^{p}(x)\right]+\left(\frac{1}{3}\right)^{2}\left[d^{p}(x)+\bar{d}^{p}(x)\right]+\left(\frac{1}{3}\right)^{2}\left[s^{p}(x)+\bar{s}^{p}(x)\right].
\label{chQCD.87}
\ee
where it was used the following notation for the quark distribution functions:
\be
\begin{split}
&f_{u}(x)=u(x)=u_{V}(x)+u_{S}(x)\\
&f_{\bar{u}}(x)=\bar{u}(x)=u_{S}(x)\\
&f_{d}(x)=d(x)=d_{V}(x)+d_{S}(x)\\
&f_{\bar{d}}(x)=\bar{d}(x)=\bar{d}_{S}(x)\\
&f_{s}(x)=s(x)=s_{S}(x)=\bar{s}(x).\\
\end{split}
\label{chQCD.88}
\ee
\mbox{\,\,\,\,\,\,\,\,\,}
Hadrons are formed by two different kinds of quarks: the valence ones $(V)$, which define each type of known hadron, and the sea quarks $(S)$, which come in pairs of virtual $q\bar{q}$ created by the vacuum polarization of the color field. Since protons and neutrons form an isospin doublet (same spin-$1/2$, $m_{p}\sim m_{n}$, and interacts by means of the strong force) it is said that protons and neutrons are different quantum states of the same entity called nucleon. Therefore, its structure functions are correlated. Since there are many $u$ quarks within the proton as $d$ quarks within the neutron, then the distribution functions of these quarks will be the same, $u^{p}(x)=d^{n}(x)$, and in the same way $d^{p}(x)=u^{n}(x)$ e $s^{p}(x)=s^{n}(x)$. So, the structure function of neutrons is given by \cite{Halzen:1984mc}
\be
\frac{1}{x}F^{en}_{2}(x)=\left(\frac{2}{3}\right)^{2}\left[u^{n}(x)+\bar{u}^{n}(x)\right]+\left(\frac{1}{3}\right)^{2}\left[d^{n}(x)+\bar{d}^{n}(x)\right]+\left(\frac{1}{3}\right)^{2}\left[s^{n}(x)+\bar{s}^{n}(x)\right].
\label{chQCD.89}
\ee
\mbox{\,\,\,\,\,\,\,\,\,}
In first approximation, one can consider that sea quark constituents, which are lighter than the valence ones, occur with the same frequency and same momentum distribution,
\be
u_{M}(x)=\bar{u}_{S}(x)=d_{S}(x)=\bar{d}_{S}(x)=s_{S}(x)=\bar{s}_{S}(x)=S(x),
\label{chQCD.90}
\ee
where $S(x)$ stands for the momentum distribution of the sea quarks. Moreover, since protons and neutrons do not have valence $\bar{u},\bar{d}$ and $\bar{s}$ quarks, hence
\be
\begin{split}
&u-\bar{u}=u-\bar{u}_{M}=u-u_{M}=u_{V},\\
&d-\bar{d}=d-\bar{d}_{M}=d-d_{M}=d_{V},\\
&s-\bar{s}=s_{M}-\bar{s}_{M}=0.
\end{split}
\label{chQCD.91}
\ee
By integrating these relations, one obtains the number of valence quarks in a proton (neutron),
\be
\int^{1}_{0} dx\left[u(x)-\bar{u}(x)\right]=2\,(1)
\label{chQCD.92}
\ee
\be
\int^{1}_{0} dx\left[d(x)-\bar{d}(x)\right]=1\,(2)
\label{chQCD.93}
\ee
\be
\int^{1}_{0} dx\left[s(x)-\bar{s}(x)\right]=0.
\label{chQCD.94}
\ee
\mbox{\,\,\,\,\,\,\,\,\,}
Respectively, the proton structure function, see expression (\ref{chQCD.87}), can be written in the form
\be
\begin{split}
\frac{1}{x}F^{ep}_{2}&=\left(\frac{2}{3}\right)^{2}\left[u_{V}+u_{S}+\bar{u}_{V}+\bar{u}_{S}\right]+\left(\frac{1}{3}\right)^{2}\left[d_{V}+d_{S}+\bar{d}_{V}+\bar{d}_{S}\right]+\left(\frac{1}{3}\right)^{2}\left[S+S\right] \\
& =\left(\frac{2}{3}\right)^{2}\left[u_{V}+S+S\right]+\left(\frac{1}{3}\right)^{2}\left[d_{V}+S+S\right]+\left(\frac{1}{3}\right)^{2}\left[S+S\right] \\
& =\frac{1}{9}\left[4u_{V}+d_{V}\right]+\frac{4}{9}S,
\end{split}
\label{chQCD.95}
\ee
and for the neutron structure function, see (\ref{chQCD.89}),
\be
\frac{1}{x}F^{en}_{2}(x)=\frac{1}{9}\left[u_{V}+4d_{V}\right]+\frac{4}{9}S.
\label{chQCD.96}
\ee
\mbox{\,\,\,\,\,\,\,\,\,}
It is expected that $S(x)$ has a \tit{Bremsstrahlung} spectrum in the region of small-$x$, since gluons create $q\bar{q}$ sea pairs. The number of sea quarks grows logarithmically at $x\to 0$. Then, in the region $x\sim 0$, the momentum fraction of the valence quarks are much smaller than from the sea quark pairs,
\be
\lim_{x\fdd 0}\frac{F^{en}_{2}(x)}{F^{ep}_{2}(x)}\fdd 1,
\label{chQCD.97}
\ee
as for the case at $x\to 1$, the valence quarks $u_{V}$ and $d_{V}$ have most of the hadron's momentum fraction leaving a small amount for the sea, hence \cite{Halzen:1984mc}, 
\be
\lim_{x\fdd 1}\frac{F^{en}_{2}(x)}{F^{ep}_{2}(x)}\fdd \frac{u_{V}+4d_{V}}{4u_{V}+d_{V}}.
\label{chQCD.98}
\ee
\mbox{\,\,\,\,\,\,\,\,\,}
Although the success of the parton model with respect to the experimental verification of Bjorken scaling, there was still some kind of anxiety by the scientific community. The hypothesis that quarks behave like free particles in small distances seems inconsistent with the lack of direct measurements of quarks, whose confinement suggested a different behavior. Despite the theoretical issues, the DIS measurements have shown that the sum over all fraction of momentum carried by the quarks within the proton did not equal to $1$, see expression (\ref{chQCD.82}), but only $50\%$ of the total proton's momentum. By summing up all the parton's momenta, one should arrive at the total proton's momentum,
\be
\int^{1}_{0}dx\,xp\left[u+\bar{u}+d+\bar{d}+s+\bar{s}\right]=p-p_{g},
\label{chQCD.99}
\ee
therefore this difference must be related to the momentum of neutral partons. However, they are not detected in direct measurements of DIS. By multiplying expression (\ref{chQCD.99}) and defining the neutral parton's momentum fraction $\epsilon_{g}\equiv p_{g}/p$,
\be
\int^{1}_{0}dx\,x\left[u+\bar{u}+d+\bar{d}+s+\bar{s}\right]=1-\epsilon_{g}.
\label{chQCD.100}
\ee
\mbox{\,\,\,\,\,\,\,\,\,}
By means of expressions (\ref{chQCD.95}) and (\ref{chQCD.96}), and neglecting the $s$ quark's momentum since their contribution are rather small, then
\be
\begin{split}
&\int^{1}_{0}dx\,xF^{ep}_{2}(x)=\frac{4}{9}\int^{1}_{0}dx\,x\left[u+\bar{u}\right]+\frac{1}{9}\int^{1}_{0}dx\,x\left[d+\bar{d}\, \right]=\frac{4}{9}\epsilon_{u}+\frac{1}{9}\epsilon_{d}=0.18, \\
&\int^{1}_{0}dx\,xF^{en}_{2}(x)=\frac{1}{9}\int^{1}_{0}dx\,x\left[u+\bar{u}\right]+\frac{4}{9}\int^{1}_{0}dx\,x\left[d+\bar{d}\, \right]=\frac{1}{9}\epsilon_{u}+\frac{4}{9}\epsilon_{d}=0.12,
\end{split}
\label{chQCD.101}
\ee
\bfg[hbtp]
  \begin{center}
    \includegraphics[width=10cm,clip=true]{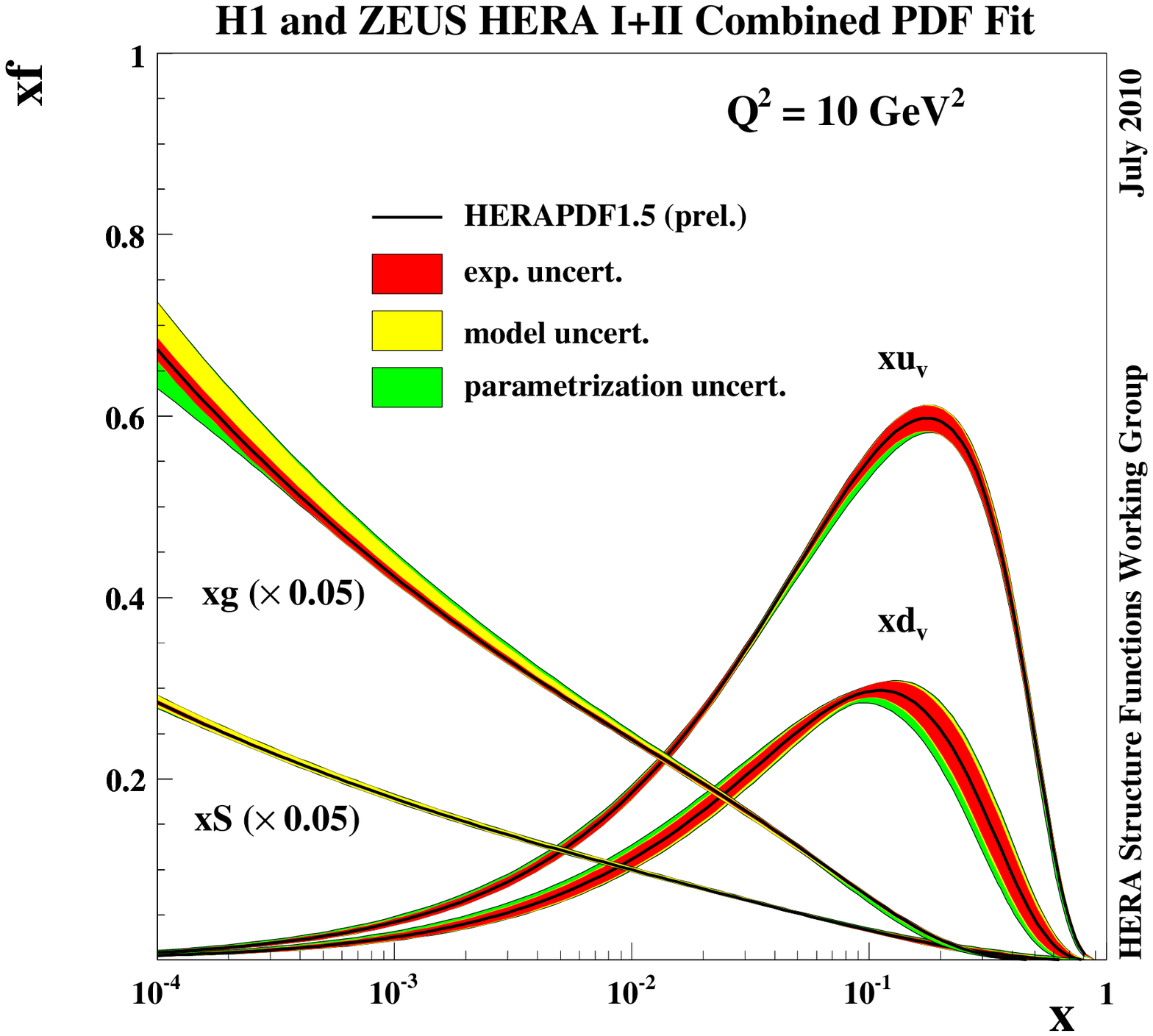}
    \caption{Partons distribution within a proton \cite{Aaron:2009aa}.}
    \label{chQCDfig4}
  \end{center}
\efg
\newline
where $\epsilon_{u}$ and $\epsilon_{d}$ stand for the momentum fractions carried by $(u+\bar{u})$ and $(d+\bar{d})$, and $\epsilon_{g}\sim 1-\epsilon_{u}-\epsilon_{d}$, thus one arrives at $\epsilon_{u}=0.36$, $\epsilon_{d}=0.18$ and $\epsilon_{g}=0.48$, respectively.

Therefore, half of the total amount of the proton's momentum is carried by these neutral partons, which in fact were not detected in DIS $ep$ experiments. Only with QCD and the asymptotic behavior of the effective coupling at $\alpha_{s}(Q^{2}\to\infty)\to 0$ justifies the theoretical hypothesis that quarks are asymptotic free in small distances of order $Q^{-1}$. In addition, the presence of gluons, which evince that they carry the other half of the proton's momentum, and also they are the gauge bosons of the strong force where the coupling is therefore given by the color quantum number. These results, plus the experimental verification that in some kinematic regions the Bjorken scaling is broken has led to the formulation of the QCD parton model. In Figure \ref{chQCDfig4} are depicted the partons distributions, respectively \cite{Aaron:2009aa}.

\section{\textsc{The Parton Model and QCD}}
\label{secQCD.5}
\mbox{\,\,\,\,\,\,\,\,\,}
\bfg[hbtp]
  \begin{center}
    \includegraphics[width=15cm,clip=true]{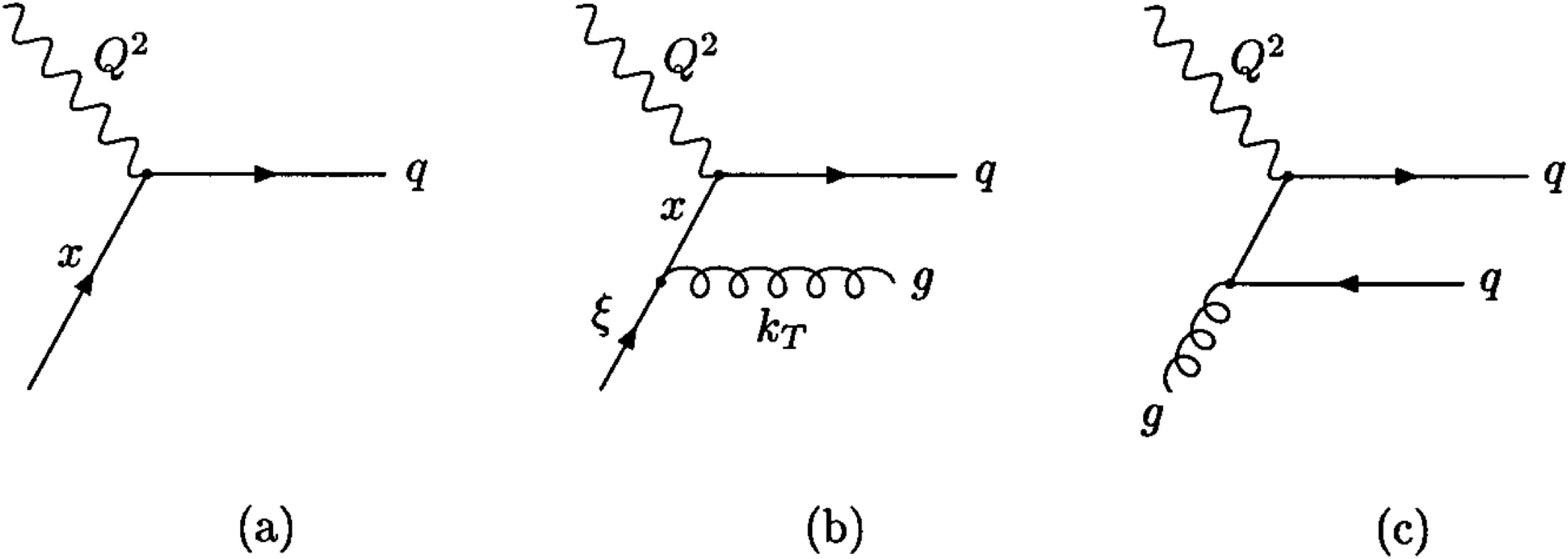}
    \caption{DIS partonic process diagrams: (a) zeroth-order diagram related to the original parton model, (b,c) QCD ${\cal O}(\alpha_{s})$ diagrams  related to the QCD parton model.}
    \label{chQCDfig5}
  \end{center}
\efg
The original parton model, see Figure \ref{chQCDfig3}, is only the zeroth-order approximation of the $\gamma^{\ast}N$ scattering. After all, the partonic constituents within hadrons are not free objects, they are in fact described by QCD. It was shown in the last section that at the asymptotic Bjorken limit the structure functions scale, \tit{i.e.} $F(x,Q^{2})\to F(x)$. However, one observes that this scaling is broken in QCD and structure functions appear to depend on logarithms of $Q^{2}$. According to perturbative QCD, new subprocess can contribute to the DIS cross sections, \tit{e.g.}, a quark can emit a gluon and acquire large transverse momentum $k_{T}$, see Figure \ref{chQCDfig5}(b), thus the integrals extend up to the kinematic limit at $k^{2}_{T}\sim Q^{2}$ and leads to contributions proportional to $\alpha_{s}\log Q^{2}$, which break scaling.

In the case of collinear singularity, \tit{i.e.} in the limit at $t\to0$ the gluon is emitted parallel to the quark, the contribution of diagram $(b)$ from Figure \ref{chQCDfig5} to the structure function $F_{2}(x,Q^{2})$ leads to divergences at $k^{2}_{T}=0$.  To calculate this emission, one must take into account all the possible values of $\xi$ and $k_{T}$ that describe the collinear gluon. More specifically, the logarithms of $Q^{2}$ appear through out the integration over the gluon momentum spectrum emission \cite{Altarelli:2002wg},
\be
\frac{\alpha_{s}}{2\pi}\,\int^{Q^{2}}_{\kappa^{2}}\frac{dk_{T}^{2}}{k_{T}^{2}}=\frac{\alpha_{s}}{2\pi}\,\log\left(\frac{Q^{2}}{\kappa^{2}}\right),
\label{chQCD.102}
\ee
where the upper limit of integration is defined by the photon virtuality $Q^{2}$ interacting within the quarks, and the $\kappa$ in the lower limit stands for an infrared arbitrary parameter responsible to regularize the divergence at $k^{2}_{T}=0$. Hence, once it is known the single gluon vertex contribution\footnote{The contribution of diagram with $n$-gluon emission is proportional to $[\alpha_{s}\log(Q^{2}/\kappa^{2})]^{n}$.}, see expression (\ref{chQCD.102}), it can be shown that the structure function is given by \cite{Barone:2002cv}
\be
F_{2}(x,Q^{2})=\sum_{q}e^{2}_{q}\,x\,\int^{1}_{x}\frac{d\xi}{\xi}\,f_{q}(\xi)\left\{\delta\left(1-\frac{x}{\xi}\right)+\frac{\alpha_{s}}{2\pi}\left[P_{qq}\left(\frac{x}{\xi}\right)\log\left(\frac{Q^{2}}{\kappa^{2}}\right)+h\left(\frac{x}{\xi}\right)\right]+...\right\},
\label{chQCD.103}
\ee
where $P_{qq}$ are the splitting functions and $h(x/\xi)$ is a finite function, respectively. The zeroth-order term in expression (\ref{chQCD.103}) reflects the original parton model\footnote{Delta function property, $$\delta[f(x)]=\sum_{i}\frac{\delta(x-x_{i})}{\vert f^{(\prime)}(x=x_{i})\vert},\,\,\,\,\,\, \text{then}\,\,\,\,\,\, \delta\left(1-\frac{x}{\xi}\right)=\xi\delta(x-\xi).$$}, see expression (\ref{chQCD.86}), with $\xi=x$. After the gluon emission, the momentum fraction $\xi$ carried by the quark reduces to $x$, then the integration limit is now given at $x\leq\xi<1$.

The presence of logarithms of $Q^{2}$ and $\kappa$ implies the nonconvergence of the perturbative expansion in expression (\ref{chQCD.103}), since $\log Q^{2}$ terms can be rather large. Following the same path as the renormalization of the ultraviolet divergences \cite{Barone:2002cv}, the collinear divergences can be properly absorbed by defining a factorization scale $\mu_{f}$, which plays a similar role to the renormalization scale,
\be
\log\left(\frac{Q^{2}}{\kappa^{2}}\right)=\log\left(\frac{Q^{2}}{\mu_{f}^{2}}\right)+\log\left(\frac{\mu_{f}^{2}}{\kappa^{2}}\right),
\label{chQCD.104}
\ee 
and by convenience the $h(x/\xi)$ will be rewritten as
\be
h(z)=h^{(1)}(z)+h^{(2)}(z),
\label{chQCD.105}
\ee
where the singularity $\log\left(\mu^{2}/\kappa\right)$ and the term $h^{(2)}(z)$ will be absorbed into a redefinition of the quark distribution function,
\be
f_{q}(x,\mu_{f}^{2})=f_{q}(x)+\frac{\alpha_{s}}{2\pi}\int^{1}_{x}\frac{d\xi}{\xi}\,f_{q}(\xi)\left[P_{qq}\left(\frac{x}{\xi}\right)\,\log\left(\frac{\mu_{f}^{2}}{\kappa^{2}}\right)+h^{(2)}\left(\frac{x}{\xi}\right)\right]+...
\label{chQCD.106}
\ee
The separation in expression (\ref{chQCD.105}) defines the factorization scheme, and in this case, the structure function  $F_{2}$ can be written in terms of this renormalized quark distribution,
\be
F_{2}(x,Q^{2})=\sum_{q}e_{q}^{2}\,x\int_{x}^{1}\frac{d\xi}{\xi}\,f_{q}(x,\mu_{f}^{2})\,C\left(\frac{x}{\xi},Q^{2},\mu_{f}^{2}\right),
\label{chQCD.107}
\ee
where this result is known as the collinear factorization theorem \cite{Barone:2002cv,Ellis:1990pp}. The $C\left(z,Q^{2},\mu_{f}^{2}\right)$, called coefficient function, is given by the renormalized partonic structure function,
\be
C\left(z,Q^{2},\mu_{f}^{2}\right)=\delta(1-z)+\frac{\alpha_{s}}{2\pi}\left[P_{qq}(z)\,\log\left(\frac{Q^{2}}{\mu_{f}^{2}}\right)+h^{(1)}(z)\right]+...
\label{chQCD.108}
\ee

\subsection{\textsc{Scaling Violation}}
\label{secQCD.5.1}
\mbox{\,\,\,\,\,\,\,\,\,}
Since $F_{2}(x,Q^{2})$ is a physical observable, and thus cannot depend on the unphysical quantity $\mu^{2}$, then expression (\ref{chQCD.107}) implies that $f_{q}(x,\mu_{f}^{2})$ must be finite and analytic. The $\kappa$ parameter can be removed by differentiating expression (\ref{chQCD.106}) with respect to $\log Q^{2}$ at the limit at $\mu_{f}\to Q^{2}$,
\be
\frac{\partial f_{q}(x,Q^{2})}{\partial\log Q^{2}}=\frac{\alpha_{s}}{2\pi}\int_{x}^{1}\frac{d\xi}{\xi}\,f_{q}(\xi,Q^{2})P_{qq}\left(\frac{x}{\xi}\right).
\label{chQCD.109}
\ee
This integral-differential equation is a LO example of a class of evolution equations, namely DGLAP equation \cite{Gribov:1972ri,Lipatov:1974qm,Altarelli:1977zs,Dokshitzer:1977sg}, and it describes how $f_{q}(x,Q^{2})$ evolves with $Q^{2}$ by means of an initial scale $Q_{0}$, once it is known $f_{q}(x,Q^{2}_{0})$. The DGLAP equations describe the perturbative QCD correction to the partonic distribution functions by effectively resuming contributions of the type $[\alpha_{s}\log Q^{2}]^{n}$.

At the limit where $Q^{2}$ is bigger than the quark masses, the gluon distribution affects the quark and antiquark distributions $Q^{2}$-dependence. As a matter of fact the gluon distribution receives contributions from quark and antiquark distributions. Taking into account the diagram $(c)$ from Figure \ref{chQCDfig5}, where the gluon produces a $q\bar{q}$ pair, the complete evolution equation for $f_{q}$ is given by
\be
\frac{\partial f_{q}(x,Q^{2})}{\partial\log Q^{2}}=\frac{\alpha_{s}}{2\pi}\int_{x}^{1}\frac{d\xi}{\xi}\left[f_{q}(\xi,Q^{2})P_{qq}\left(\frac{x}{\xi}\right)+2N_{f}\,f_{g}(\xi,Q^{2})P_{qg}\left(\frac{x}{\xi}\right)\right]+{\cal O}(\alpha_{s}^{2}),
\label{chQCD.110}
\ee
where $f_{g}(x,Q^{2})$ stands for the renormalized gluon distribution function,
\be
\frac{\partial f_{g}(x,Q^{2})}{\partial\log Q^{2}}=\frac{\alpha_{s}}{2\pi}\int_{x}^{1}\frac{d\xi}{\xi}\left[f_{q}(\xi,Q^{2})P_{gq}\left(\frac{x}{\xi}\right)+f_{g}(\xi,Q^{2})P_{gg}\left(\frac{x}{\xi}\right)\right]+{\cal O}(\alpha_{s}^{2}).
\label{chQCD.111}
\ee
\mbox{\,\,\,\,\,\,\,\,\,}
At leading order the splitting functions are written as
\be
P_{qq}(z)=\frac{4}{3}\,\left(\frac{1+z^{2}}{1-z}\right)_{+},
\label{chQCD.112}
\ee
\be
P_{qg}(z)=\frac{1}{2}[z^{2}+(1-z)^{2}],
\label{chQCD.113}
\ee
\be
P_{gq}(z)=\frac{4}{3}\,\frac{[1+(1-z)^{2}]}{z},
\label{chQCD.114}
\ee
\be
P_{gg}=6\left[\frac{1-z}{z}+z(1-z)+\frac{z}{(1-z)_{+}}+\left(\frac{11}{12}-\frac{N_{f}}{3}\right)\delta(1-z)\right],
\label{chQCD.115}
\ee
where the ``plus'' distributions are defined as
\be
\int^{1}_{0}dx\,\frac{f(x)}{(1-x)_{+}}=\int^{1}_{0}dx\,\frac{f(x)-f(1)}{1-x}, \,\,\,\, \text{where}\,\,\,\, \frac{1}{(1-x)_{+}}=\frac{1}{1-x},\,\,\,\,\text{para}\,\,\,\, 0\leq x<1.
\label{chQCD.116}
\ee
\mbox{\,\,\,\,\,\,\,\,\,}
It is possible to properly summarize this result by defining:
\be
{\cal U}(x,Q^{2})=\begin{pmatrix}
f_{q}(x,Q^{2})\\
f_{g}(x,Q^{2})
\end{pmatrix},
\label{chQCD.117}
\ee 
thus, expressions (\ref{chQCD.110}) and (\ref{chQCD.111}) will be rewritten as
\be
\frac{\partial \,{\cal U}(x,Q^{2})}{\partial \log Q^{2}}=\frac{\alpha_{s}}{2\pi}\int^{1}_{x} \frac{d\xi}{\xi}\,{\cal P}\left(\frac{x}{\xi},Q^{2}\right)\,{\cal U}(\xi,Q^{2}),
\label{chQCD.118}
\ee
where,
\be
{\cal P}(z,Q^{2})=\begin{pmatrix}
P_{qq}(z,Q^{2})&2N_{f}P_{qg}(z,Q^{2})\\
P_{gq}(z,Q^{2})&P_{gg}(z,Q^{2})
\end{pmatrix}.
\label{chQCD.119}
\ee
The above relation is called splitting matrix, because $P_{ij}$ physically represents the probability that each elementary vertex ``j emits i'' and has its momentum fraction reduced by a factor $z$.

The predictions with respect to hard and semihard scatterings of hadrons are related to the precise knowledge of the partonic distribution functions. These are universal functions, since they describe the partonic hadronic content at any given scattering. The PDF's, $f_{q}(x,Q^{2})$ and $f_{g}(x,Q^{2})$, used in our QCD-inspired eikonal model are specific for $pp$ and $\bar{p}p$ collisions, where they were obtained by means of a global analysis of all data involving deep inelastic process and some other process involving protons. The better the precision in which the data are obtained and the wider the kinematic region, will lead the scientific community to a better understanding of the PDF's. The parametrization of the PDF reflects its $x$ dependence in an initial $Q_{0}$ scale, where its value must be high enough to be in the perturbative region. Some authors adopt $Q_{0}=1$ GeV and parametrizations such that \cite{Martin:1987vw}
\be
xf_{i}(x,Q^{2}_{0})=A_{i}x^{-\lambda_{i}}(1-x)^{\eta_{i}}(1+\epsilon_{i}\sqrt{x}+\gamma_{i}x),
\label{chQCD.120}
\ee
with $i=u_{V},d_{V},S$, where $S$ stands for the total sea quark distribution, and $A_{i}, \lambda_{i}, \epsilon_{i},\eta_{i}$ e $\gamma_{i}$ are determined by the best fit to the data. By means of this parametrization, it is used the DGLAP equations to evolve $f_{i}(x,Q^{2})$ in all values of $x$ and $Q^{2}$ in which there are available experimental data. 

It is most noteworthy to state that several authors use different sets of parametrizations, and different choices to deal with uncertainties and correlated systematic errors between new and previous experimental data.

\clearpage
\thispagestyle{plain}

\chapter{\textsc{QCD-Inspired Eikonal Model}}
\label{chQIM}
\hyphenation{mo-dels}
\mbox{\,\,\,\,\,\,\,\,\,}
It is well known that processes with low-$q^{2}$ contribute to most of the total cross section and also that diffractive reactions cannot be treated perturbatively and calculated in a reliable way within QCD. A wide variety of models in high-energy particle scattering belong to the so-called class of QCD-inspired models \cite{Halzen:1993xy}. These type of models represent an attempt to create a solid background towards a future theoretical description fully based on QCD. Moreover, they aim to describe some hadronic processes linked to the transition region between the perturbative and nonperturbative domains by means of the QCD parton model \cite{Matthiae:1994uw}.

The unitarity condition of the scattering $S$-matrix demands that the absorbed part of the elastic scattering amplitude receives contributions from both the elastic and inelastic channel. Following the impact-parameter picture, this condition can be seen by expressions (\ref{ch2.107}) and (\ref{ch2.116}). In the QIM the description of the elastic scattering appears as the shadow of the inelastic processes and the diffracted waves will be added up coherently in the forward direction giving rise to a sharp peak at the optical point. Therefore, it implies that the scattering amplitude can be properly addressed in the eikonal approximation, see expression (\ref{ch2.100}). A common feature in this kind of models is to consider that part of the growth of the total cross section at high energies is associated with the rapidly increase of the PDF's, mainly gluons, at small-$x$.

An analysis made by Amaldi and Schubert \cite{Amaldi:1979kd} at the ISR energy region showed that those models which were constructed through out factorized eikonals in energy and impact-parameter, \tit{e.g.} $\chi(s,b)=f(s)w(b)$, were incompatible with the experimental results. Henceforth, the QIM narrows the choice of physically motivated eikonals with some kind of hybrid properties, or either sometimes called as semi-factorized, in energy and impact-parameter by means of the QCD parton-model.

Various models fall into the class of QIM \cite{Luna:2005nz,Fagundes:2011zx,Luna:2006sn,Luna:2006qp,Bahia:2015hha,Bahia:2015gya,LHeureux:1985qwr,Durand:1987yv,Durand:1988cr,Durand:1988ax}. Albeit they often lack of mathematical rigor, they claim to reformulate old concepts in a modern approach based on QCD. Over the past few decades other attempts have been made to provide a picture of soft diffraction based upon QCD, for instance considering that the growth of the total cross section with energy is intrinsically driven by the rise of the gluon contribution, since it gives the dominant contribution at small-$x$. The break in this kind of approach lies in the well known fact that perturbative QCD is inadequate to deal with low-$\vert t\vert$ processes and hence in order to obtain quantitative results each model is strongly dependent on assumptions.

\section{\textsc{Nonperturbative QCD Effects}}
\label{secDGM.1}
\mbox{\,\,\,\,\,\,\,\,\,}
The recent measurements of $pp$ elastic, inelastic and total cross sections at the LHC by the TOTEM Collaboration \cite{0295-5075-96-2-21002,Antchev:2013haa,Antchev:2013iaa,Antchev:2013paa,Antchev:2016vpy,Antchev:2011zz} have enhanced the interest in the theoretical and experimental study of hadron-hadron interactions. Furthermore, it also has become a pivotal source of information for selecting models and theoretical methods. Presently, the LHC provides us with the highest available collider CM energy, $\sqrt{s}=13$ TeV. One of the main theoretical approaches for the description of the observed increase of hadron-hadron total cross sections, which was predicted many years ago \cite{Cheng:1970bi} and accurately verified by experiment \cite{Tanabashi:2018oca}, is the QCD-inspired formalism, as well as the class of Regge-pole models. 

The latter attributes the increase of the total cross section to the exchange of a colorless state having the quantum numbers of the vacuum, \tit{viz.} the Pomeron \cite{Barone:2002cv,Collins:1977jy,Gribov:2003nw,Donnachie:2002en}. In the QCD framework a possible interpretation for the Pomeron is that it can be understood as the exchange of at least two gluons in a color-singlet state \cite{Low:1975sv,Nussinov:1975mw}. An interesting model for the Pomeron was evaluated in reference \cite{Landshoff:1986yj}, where it was pointed out the importance of the QCD nonperturbative vacuum. One important aspect of this nonperturbative-type physics appears as an infrared gluon mass scale which regulates the divergent behavior of the Pomeron exchange.

In the QIM approach the energy dependence of the total cross sections is obtained from the QCD using an eikonal formulation compatible with analyticity and unitarity constraints, as it was already shown before. More precisely, the behavior of the forward observables is derived from the QCD parton model using standard QCD cross sections for elementary parton-parton processes, updated sets of PDF's and physically motivated cutoffs that restrict the parton-level processes to semihard ones.

These semihard processes arise from hard scatterings of partons carrying very small fraction of their host hadron's momenta, leading to the appearance of jets with transverse energy $E_{T}$ much smaller than the total CM energy available in the hadronic collision. In this picture the scattering of hadrons is an incoherent summation over all possible constituent scattering and the increase of the total cross sections is directly associated with parton-parton semihard scatterings. As it was repeatedly mentioned before, the high-energy dependence of the cross sections is driven mainly by processes involving the gluon contribution, since it gives the dominant contribution at small-$x$. However, despite this scenario being quantitatively understood in the framework of perturbative QCD, the nonperturbative character of the theory is also manifest at the elementary level since at high energies the soft and the semihard components of the scattering amplitude are closely related \cite{Gribov:1984tu,Levin:1990gg}. Thus, in considering the forward scattering amplitude, it becomes important to distinguish between semihard gluons, which participate in hard parton-parton scattering, and soft gluons, emitted in any given parton-parton QCD radiation process.

There is no easy way out on the task of describing forward observables in hadron-hadron collision bringing up information on the infrared properties of QCD, but fortunately it can be properly addressed by considering the possibility that the nonperturbative dynamics of QCD generate an effective gluon mass. This dynamical gluon mass is intrinsically related to an infrared finite strong coupling constant, and its existence is strongly supported by recent QCD lattice simulations \cite{Bowman:2004jm,Sternbeck:2005vs,Boucaud:2006if,Bowman:2007du,Bogolubsky:2009dc,Oliveira:2009nn,Cucchieri:2009kk,Cucchieri:2009zt,Dudal:2010tf,
Cucchieri:2011ig,Cucchieri:2014via,Aguilar:2009nf} as well as by phenomenological results \cite{Luna:2005nz,Fagundes:2011zx,Luna:2006sn,Luna:2006qp,Bahia:2015hha,Bahia:2015gya,Luna:2006tw,Beggio:2013vfa,Luna:2010tp,LHeureux:1985qwr,
Durand:1987yv,Durand:1988cr,Durand:1988ax,AbouElNaga:1991ib,Cheung:2011gp,Sauli:2011xr,Jia:2012wf,Lipari:2013kta,Giannini:2013jla,Sidorov:2013aza,
Gomez:2014mpa,Ayala:2014pha,Cvetic:2013gta,Allendes:2014fua}. More specifically, a global description of $\sigma^{pp,\bar{p}p}_{tot}(s)$ and $\rho^{pp,\bar{p}p}(s)$ can succeed in a consistent way by introducing a nonperturbative QCD effective charge in the calculation of the parton-level processes involving gluons, which dominate at high-energy and determine the asymptotic behavior of hadron-hadron cross sections.

\subsection{\textsc{Infrared Mass Scale}}
\label{secDGM.1.1}

\mbox{\,\,\,\,\,\,\,\,\,}
Recently, it has been discussed in the literature some possible ways of merging the nonperturbative aspects of QCD with the perturbative expansion. Somehow, the freezing of the QCD running coupling constant at low-energy scales suggests that, in principle, it could be possible to capture nonperturbative effects in a reliable way \cite{Brodsky:2001wx,Brodsky:2001ww}. The existence of a dynaminal gluon mass is intimately related with the freezing of the running coupling constant \cite{Aguilar:2002tc}. Therefore, it should systematically be present in the perturbative expansion.

One attempt to understand the effects of dynamically massive gluons was performed by Forshaw \tit{et al.} \cite{Forshaw:1998tf}. They have introduced bare massive gluons and study the amplitude for some tree and one-loop diagrams that could be relevant for diffractive scattering. Despite to be instructive, the calculation does not recover the massless QCD result, \tit{i.e.} it does not reproduce the high-energy limit of massless gluons with $2$ degrees of freedom. By the way, this is a very good question because in principle for a gluon with a momentum-dependent dynamical mass, even though it is not a physical mass, in the literal and strictly precise meaning of mass, it should still have $3$ degrees of freedom. Actually the dynamical masses go to zero at large momenta and it should be expected to recover the elementary cross sections of perturbative QCD in the high-energy limit. So, how exactly happens this transition $3\to 2$ degrees of freedom?

The elementary processes are plagued by infrared divergences, which have to be regularized by means of some cutoff procedure. One natural regulator for these infrared divergences was introduced some time ago \cite{Cornwall:1980zw,Cornwall:1981zr} and has become an important ingredient to the class of dynamical mass eikonal models \cite{Luna:2005nz,Fagundes:2011zx,Luna:2006sn,Luna:2006qp,Bahia:2015hha,Bahia:2015gya,Luna:2006tw,Beggio:2013vfa,Luna:2010tp}. It is based on the increasing evidence that the QCD develops an effective, momentum-dependent mass for the gluons, while preserving the local $SU(3)_{c}$ invariance of the theory. This dynamical gluon mass $M_{g}(Q^{2})$ introduces a natural scale that, in principle, sets up a threshold for gluons to pop up from the vacuum \cite{Cornwall:1982zn,Cornwall:1983zb}. Moreover, it is intrinsically linked to an infrared-finite QCD effective charge $\bar{\alpha}_{s}(Q^{2})$, therefore being the natural infrared regulator in this new class of DGM eikonal model.

Since the gluon mass generation is purely a dynamical effect, the formal tool to tackle this nonperturbative phenomenon, in the continuum, is provided by the Dyson-Schwinger equations \cite{Dyson:1949ha,Schwinger:1951ex}. These equations constitute an infinite set of coupled nonlinear integral equation governing the dynamics of QCD Green's functions. The functional forms of $M_{g}$ and $\bar{\alpha}_{s}$, obtained by J.M. Cornwall through the use of the pinch technique in order to derive a gauge invariant Dyson-Schwinger equation for the gluon propagator and the triple gluon vertex, are given by \cite{Cornwall:1980zw,Cornwall:1981zr}
\be
\bar{\alpha}_{s}(Q^{2})=\frac{4\pi}{\beta_{0}\log\left[\left(Q^{2}+4M^{2}_{g}(Q^{2})\right)/\Lambda^{2}\right]},
\label{chDGM.1}
\ee
\be
M^{2}_{g}(Q^{2})=m^{2}_{g}\left[\frac{\log\left[\left(Q^{2}+4M^{2}_{g}(Q^{2}\right)/\Lambda^{2}\right]}{\log\left(4m^{2}_{g}/\Lambda^{2}\right)}\right]^{-12/11},
\label{chDGM.2}
\ee
where $\Lambda(\equiv \Lambda_{QCD})$ is the QCD scale parameter, $\beta_{0}=11-2n_{f}/3$, $n_{f}$ stands for the number of flavors and $m_{g}$ is an infrared mass scale to be adjusted in order to provide reliable results concerning calculations of strongly interacting processes. As mentioned in the previous section, the existence of the gluon mass scale $m_{g}$ is strongly supported by QCD lattice simulations and phenomenological results, and its value is typically found to be of the order  $m_{g}=500\pm 200$ MeV. Notice that in the limit $Q^{2}\gg \Lambda^{2}$ the dynamical mass $M^{2}_{g}(Q^{2})$ vanishes and the effective charge matches with the one-loop perturbative QCD coupling $\alpha^{pQCD}_{s}(Q^{2})$. It means that the asymptotic ultraviolet behavior of the LO running coupling, obtained from the renormalization group equation in perturbation theory, is reproduced in solutions of Dyson-Schwinger equations,
\be
\bar{\alpha}_{s}(Q^{2}\gg \Lambda^{2})\sim \frac{4\pi}{\beta_{0}\log(Q^{2}/\Lambda^{2})}=\alpha^{pQCD}_{s}(Q^{2}),
\label{chDGM.3}
\ee
provided only that the truncation method employed in the analysis preserves the multiplicative renormalizability \cite{Luna:2010tp}. There is also a different functional expression for the dynamical gluon mass \cite{Aguilar:2001zy,Aguilar:2004td} given by
\be
M^{2}_{g}(Q^{2})=\frac{m^{4}_{g}}{Q^{2}+m^{2}_{g}},
\label{chDGM.4}
\ee
which is consistent with the asymptotic behavior of $M^{2}_{g}(Q^{2})$ in the presence of the gluon condensates \cite{Aguilar:2004kt}. Even though the calculation of the hadronic cross section does not depend strongly on the specific form of $M_{g}(Q^{2})$, but more on its infrared mass-scale value $m_{g}$ \cite{Luna:2005nz,Fagundes:2011zx,Bahia:2015gya,Bahia:2015hha,Aguilar:2004kt}.

However, in the infrared region, the coupling $\alpha^{pQCD}_{s}(Q^{2})$ has Landau singularities on the spacelike semiaxis $0\leq Q^{2}\leq \Lambda^{2}$, \tit{i.e.,} it has a nonholomorphic (singular) behavior at low $Q^{2}$ \cite{Stefanis:2009kv}. This problem has been faced in the past years with analytic versions of QCD whose coupling $\alpha_{s}(Q^{2})$ is holomorphic (analytic) in the entire complex plane except the timelike axis $(Q^{2}< 0)$ \cite{Shirkov:1997wi,Webber:1998um,Nesterenko:1999np,Nesterenko:2004tg,Alekseev:2005he,Cvetic:2006mk,Cvetic:2006gc,Cvetic:2008bn,
Cvetic:2010di,Cvetic:2012pw,
Ayala:2012yw,Contreras:2014aka}. The effective charge $\bar{\alpha}_{s}(Q^{2})$, on the other hand, shows the existence of an infrared fixed point as $Q^{2}\to0$, \tit{i.e.,} the dynamical mass term tames the Landau pole and $\bar{\alpha_{s}}$ freezes at a finite value in the infrared limit. Thus, providing that the gluon mass scale is set larger than half of the QCD scale parameter, namely $m_{g}/\Lambda > 1/2$, the analyticity of $\bar{\alpha}_{s}(Q^{2})$ is preserved. This ratio is also phenomenologically determined \cite{Luna:2005nz,Fagundes:2011zx,Luna:2006sn,Luna:2006qp,Luna:2006tw,Beggio:2013vfa,Luna:2010tp,LHeureux:1985qwr,Durand:1987yv,Durand:1988cr,
Durand:1988ax,
AbouElNaga:1991ib,Cheung:2011gp,Sauli:2011xr,Jia:2012wf,Lipari:2013kta,Giannini:2013jla,Sidorov:2013aza,Gomez:2014mpa,Ayala:2014pha,
Cvetic:2013gta,Allendes:2014fua} and typically lies in the interval $m_{g}/\Lambda\,\in\,[1.1;2]$. More over, as recently pointed out by G. Cveti\v{c} \cite{Cvetic:2013gza}, the evaluation of renormalization scale-invariant spacelike quantities at low-$Q^{2}$, in terms of infrared freezing couplings, can be done as a truncated series in derivatives of the coupling with respect to the logarithm of $Q^{2}$, which in turn exhibit significantly better convergence properties.

\section{\textsc{The Revised Dynamical Gluon Mass Model}}
\label{secDGM.2}

\mbox{\,\,\,\,\,\,\,\,\,}
In the QCD-based (or ``minijet'') models the increase of the total cross sections is associated with semihard scatterings of partons in the hadrons. These models incorporate soft and semihard processes in the treatment of high-energy hadron-hadron interactions and a consistent calculation must take into account a formulation compatible with analyticity and unitarity constraints.

Following the L. Durand \& H. Pi model prescription \cite{Durand:1987yv,Durand:1988cr,Durand:1988ax}, \tit{i.e.} $\Gamma(s,b)=1-e^{-\chi(s,b)}$, in the eikonal representation, the cross sections, $\rho$-parameter and $B$-slope still will be written according to expressions (\ref{ch2.108}-\ref{ch2.113}), but changing $\text{Re}\,\chi\fdd-\text{Im}\,\chi$ and $\text{Im}\,\chi\fdd\text{Re}\,\chi$, thus,
\be
\sigma_{tot}(s)=4\pi\I db\,b\,\left[1-e^{-\chi_{_{R}}(s,b)}\cos \chi_{_{I}}(s,b)\right],
\label{chDGM.5}
\ee
\be
\begin{split}
\sigma_{inel}(s)&=\sigma_{tot}(s)-\sigma_{el}(s)\\
&= 2\pi\I db\,b\,\left[1-e^{-2\chi_{_{R}}(s,b)}\right],
\end{split}
\label{chDGM.6}
\ee
as well as the $\rho$-parameter,
\be
\rho(s)=-\frac{\I db\,b\, e^{-\chi_{_{R}}(s,b)}\sin \chi_{_{I}}(s,b)}{\I db\,b\,\left[1-e^{-\chi_{_{R}}(s,b)}\cos \chi_{_{I}}(s,b)\right]} ,
\label{chDGM.7}
\ee
respectively, where as usual $s$ is the square of the total CM energy, $b$ is the impact parameter, the quantity in square brackets in the \tit{rhs} of expression (\ref{chDGM.6}) stands for the inelastic overlap function $G_{in}(s,b)$, and the complex eikonal function is written as
\be
\chi(s,b)=\text{Re}\,\chi(s,b)+i\,\text{Im}\,\chi(s,b)\equiv\chi_{_{R}}(s,b)+i\,\chi_{_{I}}(s,b).
\label{chDGM.8}
\ee
In this picture the probability that neither hadron is broken up in a collision at impact parameter $b$ is therefore given by $P(s,b)=e^{-2\chi_{_{R}}}(s,b)$.

In this revised version of the DGM model, hereinafter referred to as DGM$15$ \cite{Bahia:2015hha,Bahia:2015gya}, it was assumed that the eikonal function for $pp$ and $\bar{p}p$ scattering is additive with respect to the soft and semihard parton interactions in the hadron-hadron collision,
\be
\chi(s,b) = \chi_{_{soft}}(s,b) + \chi_{_{SH}}(s,b).
\label{chDGM.9}
\ee

In the semihard limit of strong interactions hadron-hadron collisions can be treated as an incoherent sum of the interactions among quarks and gluons. More specifically, the QCD cross section $\sigma_{QCD}$ is obtained by convoluting the cross section $\hat{\sigma}$ for the QCD subprocesses with their associated parton distributions. It follows from the QCD parton model that the eikonal $\chi_{_{SH}}(s,b)$ can be factorized as \cite{Durand:1987yv,Durand:1988cr,Durand:1988ax}
\be
\textnormal{Re}\,\chi_{_{SH}}(s,b) = \frac{1}{2}\, W_{_{SH}}(b)\,\sigma_{QCD}(s),
\label{chDGM.10}
\ee
where $W_{_{SH}}(b)$ is an overlap density for the partons at impact parameter space $b$,
\be
W_{_{SH}}(b) = \int d^{2}b^{\prime}\, \rho_{A}(\vert \bb -\bbp\vert)\, \rho_{B}(b^{\prime}),
\label{chDGM.11}
\ee
and $\sigma_{QCD}(s)$ is the usual QCD cross section,
\be
\begin{split}
\sigma_{QCD}(s)&=\sum_{ij}\,\frac{1}{1+\delta_{ij}}\,\int^{1}_{0} dx_{1}\int^{1}_{0} dx_{2}\int^{\infty}_{Q^{2}_{min}}\!\!\!\!d\vert\hat{t}\vert\,\frac{d\hat{\sigma}_{ij}}{d\vert\hat{t}\vert}(\hat{s},\hat{t})\\
& \times f_{i/A}(x_{1},\vert\hat{t}\vert,\vert)\,f_{j/B}(x_{2},\vert\hat{t}\vert)\,\Theta\left(\frac{\hat{s}}{2}-\vert \hat{t} \vert\right),
\end{split}
\label{chDGM.12}
\ee
with $\vert \hat{t}\vert\equiv Q^{2}$ and $i,j=q,\bar{q},g$. In the above expression the integration limits satisfy $x_{1}x_{2}s>2\vert \hat{t}\vert>2Q^{2}_{min}$, where $Q^{2}_{min}$ is a minimum momentum transfer in the semihard scattering, $\hat{s}$ and $\hat{t}$ are the Mandelstam variables of the parton-parton subsystem, and $x_{1}$ and $x_{2}$ are the fractions of the momenta of the host hadrons $A$ and $B$ carried by the partons $i$ and $j$. The term $d\hat{\sigma}_{ij}/d\vert \hat{t}\vert $ is the differential cross section for $ij$ scattering, and $f_{i/A}(x_{1},\vert\hat{t}\vert)$ ($f_{j/B}(x_{2},\vert\hat{t}\vert)$) is the usual parton $i$ ($j$) distribution in the hadron $A$ ($B$).

The eikonal function is written in terms of even and odd eikonal parts connected by crossing symmetry, similarly to the case of the Durand \& Pi model \cite{Durand:1987yv,Durand:1988cr,Durand:1988ax}. By considering $pp$ and $\bar{p}p$ scatterings, this combination reads
\be
\chi_{pp}^{\bar{p}p}(s,b) = \chi^{+} (s,b) \pm \chi^{-} (s,b),
\label{chDGM.13}
\ee
with the even part written in term of soft and semihard even eikonals
\be
\chi^{+}(s,b) = \chi^{+}_{_{soft}}(s,b) + \chi^{+}_{_{SH}}(s,b),
\label{chDGM.14}
\ee
and similarly the odd eikonal,
\be
\chi^{-}(s,b) = \chi^{-}_{_{soft}}(s,b) + \chi^{-}_{_{SH}}(s,b).
\label{chDGM.15}
\ee

In the QCD parton model $\chi^{-}_{_{SH}}(s,b)$ decreases rapidly with increasing $s$, since the difference between $pp$ and $\bar{p}p$ cross sections is due only to the different weighting of the quark-antiquark (valence) annihilation cross sections in the two channels. Hence the crossing-odd eikonal $\chi^{-}(s,b)$ receives no contribution from semihard processes at high energies. As a result, it is sufficient to take $\chi_{_{SH}}=\chi^{+}_{_{SH}}$ and, consequently, $\chi^{-}=\chi^{-}_{_{soft}}$, since the main interest relies on the high-energy scattering region. The connection between the real and imaginary parts of $\chi^{+}(s,b)$ and $\chi^{-}(s,b)$ was obtained by means of dispersion relations. It deserves a careful reading, thus it will be discussed in a separated section elsewhere in the text. Henceforth, further details on integral and derivative dispersion relations, respectively, can be found in Appendix \ref{APX3}.

\subsection{\textsc{Energy-Dependent Form Factors}}
\label{secDGM.2.1}

\mbox{\,\,\,\,\,\,\,\,\,}
For the overlap densities, the simplest hypothesis is to assume $W_{_{SH}}(b)$ is the same as $W_{soft}(b)$. This prescription is not however true in the QCD parton model, since soft interactions are mainly related to interactions among valence quarks, whilst semihard interactions are dominated by gluons. Moreover, a scenario where quarks and gluons exhibit a somewhat different spatial distribution seems plausible \cite{Durand:1988ax}, since gluons are expected to be distributed around the quarks. Furthermore, in contrast with gluons, quarks have electric charges, and the (matter) distribution of the valence quarks can be associated in a reasonable way with the proton's charge distribution. As a consequence, a commonly used choice for the soft overlap densities $W^{-}_{soft}(b)$ and $W^{+}_{soft}(b)$ comes from the charge dipole approximation to the form factors $G_{A}(k_{\perp})$ and $G_{B}(k_{\perp})$ of the colliding hadrons $A$ and $B$, see Appendix \ref{secAPX2.1},
\bear
A(b) &=& \int d^{2}b^{\prime}\, \rho_{A}(\vert\bb-\bbp\vert)\, \rho_{B}(b^{\prime})  \nonumber \\
 &=& 2\pi\,\int_{0}^{\infty}dk_{\perp}\, k_{\perp}\, J_{0}(k_{\perp}b)\,G_{A}(k_{\perp})\,G_{B}(k_{\perp}),
\label{chDGM.16}
\eear
and
\be
G_{A}(k_{\perp})=G_{B}(k_{\perp})\equiv G_{dip}(k_{\perp};\mu)=\left( \frac{\mu^{2}}{k_{\perp}^{2}+\mu^{2}} \right)^{2}.
\label{chDGM.17}
\ee
Here, $\rho(b)$ is the parton density, which gives the probability density for finding a parton in the area $d^{2}b$ at impact parameter $b$. In terms of the form factor it is simply written as
\be
\rho(b)=\frac{1}{2\pi}\int d^{2}k_{\perp}\, G(k_{\perp})e^{i{\bf k}_{\perp}\cdot {\bf b}}.
\label{chDGM.18}
\ee
Thus, using the dipole form factor $G_{dip}(k_{\perp};\mu)$, see Appendix \ref{secAPX2.2.2},
\bear
W^{+}_{_{soft}}(b;\mu^{+}_{_{soft}}) &=& \frac{1}{2\pi}\int_{0}^{\infty}dk_{\perp}\, k_{\perp}\, J_{0}(k_{\perp}b)\,G_{dip}^{2}(k_{\perp};\mu^{+}_{_{soft}}) \nonumber \\
 &=& \frac{(\mu^{+}_{_{soft}})^{2}}{96\pi} (\mu^{+}_{_{soft}} b)^{3} K_{3}(\mu^{+}_{_{soft}} b),
\label{chDGM.19}
\eear
where $\mu^{+}_{_{soft}}$ is a free adjustable parameter that accounts for the matter (valence quark) distribution inside the hadron. The $W(b;\mu)$ function is normalized so that $\int d^{2}\,W(b;\mu)=1$. In the same way, the odd soft density is written as
\be
W^{-}_{_{soft}}(b;\mu^{-}_{_{soft}}) = \frac{(\mu^{-}_{_{soft}})^{2}}{96\pi}
(\mu^{-}_{_{soft}} b)^{3} K_{3}(\mu^{-}_{_{soft}} b),
\label{chDGM.20}
\ee
where $\mu^{-}_{_{soft}}\equiv0.5$ GeV, its value is fixed since the odd eikonal just accounts for the difference between $pp$ and $\bar{p}p$ channels at low energies.

In the case of semihard gluons, which dominate at high-energy, it was considered the possibility of a ``broadening'' of the spatial distribution. Our assumption suggests an increase of the average gluon radius when $\sqrt{s}$ increases. The way for introducing this effect can be paved by looking at previous approaches, particularly in geometrical ones, in which the role of phenomenological energy-dependent form factors is central \cite{Carreras:1972fu,White:1973fr,Menon:1991ma,Menon:1996es,Beggio:1999gt,Lipari:2009rm,Fagundes:2013aja}. The assumption considered in the DGM$15$ model, based on the QCD parton model, can be properly implemented using two \tit{Ansätze} for the energy-dependent form factors, namely a monopole,
\be
G^{(m)}_{_{SH}}(s,k_{\perp};\nu_{_{SH}})=\frac{\nu_{_{SH}}^{2}}{k_{\perp}^{2}+\nu_{_{SH}}^{2}},
\label{chDGM.21}
\ee
and a dipole,
\be
G^{(d)}_{_{SH}}(s,k_{\perp};\nu_{_{SH}})=\left( \frac{\nu_{_{SH}}^{2}}{k_{\perp}^{2}+\nu_{_{SH}}^{2}} \right)^{2},
\label{chDGM.22}
\ee
where $\nu_{_{SH}}= \nu_{1}-\nu_{2}\log (s/s_{0})$, with $\sqrt{s_{0}}\equiv5$ GeV. Here, $\nu_{1}$ and $\nu_{2}$ are constants to be fitted. In the case of the monopole the overlap density is, see Appendix \ref{secAPX2.2.1},
\bear
W^{(m)}_{_{SH}}(s,b;\nu_{\!\!_{SH}}) &=& \frac{1}{2\pi}\int_{0}^{\infty}dk_{\perp}\, k_{\perp}\, J_{0}(k_{\perp}b)\,
[G^{(m)}_{_{SH}}(s,k_{\perp};\nu_{_{SH}})]^{2} \nonumber \\
 &=& \frac{\nu^{2}_{_{SH}}}{4\pi} (\nu_{_{SH}} b) K_{1}(\nu_{_{SH}} b).
\label{chDGM.23}
\eear
In analogy with expression (\ref{chDGM.19}), in the case of the dipole one is led to
\be
W^{(d)}_{_{SH}}(s,b;\nu_{\!\!_{SH}}) = \frac{\nu^{2}_{_{SH}}}{96\pi} (\nu_{_{SH}} b)^{3} K_{3}(\nu_{_{SH}} b).
\label{chDGM.24}
\ee
\mbox{\,\,\,\,\,\,\,\,\,}
Notice that, as mentioned earlier, semihard interactions dominate at high energies. Thus, it was considered an energy-dependence behavior for the spatial distribution exclusively in the case of $W_{_{SH}}(s,b)$. In this way, the soft overlap densities $W^{+}_{_{soft}}(b)$ and $W^{-}_{_{soft}}(b)$ will merge only from the ``static'' dipole form factor, \tit{i.e.}, from expressions (\ref{chDGM.19}) and (\ref{chDGM.20}), whereas the semihard overlap density $W_{_{SH}}(s,b)$ will be directly associated with expressions (\ref{chDGM.23}) and (\ref{chDGM.24}). Moreover, in the semihard sector there is another form in which the eikonal can be factorized into the QCD parton model, since now $\text{Re}\,\chi_{_{SH}}(s,b)=\frac{1}{2}\,W_{_{SH}}(s,b)\,\sigma_{QCD}(s)$.

\subsection{\textsc{Integral Dispersion Relations and High-Energy Eikonal Construction}}
\label{secDGM.2.2}
\hyphenation{phe-no-me-no-lo-gy}
\mbox{\,\,\,\,\,\,\,\,\,}
The analyticity of the scattering amplitude $f(s,t)$ leads to dispersion relations with crossing symmetry condition. In the case of elastic processes in the forward direction, the crossing variable is the energy $E$ of the incident particle in the laboratory frame \cite{Block:1984ru}. For an even amplitude, the real and the imaginary parts of $f^{+}(E)$ are connected by the dispersion relation,
\be
\text{Re}\,f^{+}(E)=\frac{2}{\pi}\,{\cal P}\int^{\infty}_{m}dE^{\prime}\left[\frac{E^{\prime}}{E^{\prime 2}-E^{2}}\right]\,\text{Im}\,f^{+}(E).
\label{chDGM.25}
\ee
\mbox{\,\,\,\,\,\,\,\,\,}
The eikonals are written in terms of even and odd eikonal parts connected by crossing symmetry, namely $\chi_{pp}^{\bar{p}p} = \chi^{+} \pm \chi^{-} $, where $\chi^{+} $ and $\chi^{-} $ are therefore real analytic functions of $E$, \tit{i.e.} they take real values on a real-axis segment, with the same cut structure as $f^{+}$ and $f^{-}$, respectively. Hence, taking the limit $E\gg m$ and changing the variables from $E\to s$, one finds that the even eikonal also satisfies the reverse dispersion relation,
\be
\text{Im}\,\chi^{+}(s,b)=-\frac{2s}{\pi}\,{\cal P}\I ds^{\prime}\,\frac{\text{Re}\,\chi^{+}(s,b)}{s^{\prime 2}-s^{2}},\,\,\,\,\,\,\text{in the limit at}\,\,\,\,s\gg m.
\label{chDGM.26}
\ee
Thus, integrating by parts,
\begin{eqnarray}
\textnormal{Im}\,\chi^{+}(s,b) &=& \lim_{\substack{\epsilon \to 0 \\ s^{\prime\prime}\to \infty}} -\frac{2s}{\pi} \left[
\int_{0}^{s-\epsilon}ds^{\prime}\,\frac{\textnormal{Re}\,\chi^{+}(s^{\prime},b)}{s^{\prime 2}-s^{2}}    +
\int_{s+\epsilon}^{s^{\prime\prime}}ds^{\prime}\,\frac{\textnormal{Re}\,\chi^{+}(s^{\prime},b)}{s^{\prime 2}-s^{2}}
\right] \nonumber \\
 &=& \lim_{s^{\prime\prime}\to \infty} \frac{1}{\pi} \left[ \textnormal{Re}\,\chi^{+}(s^{\prime\prime},b)\ln \left( \frac{s^{\prime\prime}+s}{s^{\prime\prime}-s} \right) -
\int_{0}^{\infty}ds^{\prime}\,\ln \left( \frac{s^{\prime}+s}{\vert s^{\prime}-s\vert } \right) \frac{d\,\textnormal{Re}\,\chi^{+}(s^{\prime},b)}{ds^{\prime}}
\right] \nonumber \\
 &=& -\frac{1}{\pi} \int_{0}^{\infty}ds^{\prime}\,\ln \left( \frac{s^{\prime}+s}{\vert s^{\prime}-s\vert } \right) \frac{d\,
\textnormal{Re}\,\chi^{+}(s^{\prime},b)}{ds^{\prime}} ,
\label{disrel001}
\end{eqnarray}
where in the last step we have observed that the first term vanishes in the limit $s''\to \infty$, and it was used,
\be
\frac{s}{s^{\prime 2}-s^{2}}=\frac{1}{2(s^{\prime}+s)}-\frac{1}{2(s^{\prime}-s)}.
\label{apx6.9}
\ee
\mbox{\,\,\,\,\,\,\,\,\,}
Applying this dispersion relation to $\textnormal{Re}\,\chi_{_{SH}}(s,b)=\textnormal{Re}\,\chi^{+}_{_{SH}}(s,b)=\frac{1}{2}\,W_{\!\!_{SH}}(s,b)\,\sigma_{_{QCD}}(s)$, one finds,
\be
\begin{split}
\textnormal{Im}\,\chi_{_{SH}}(s,b) =&-\frac{1}{2\pi}\, \int_{0}^{\infty} ds^{\prime}\,\ln \left( \frac{s^{\prime}+s}{\vert s^{\prime}-s\vert} \right) \left[ \sigma_{_{QCD}}(s^{\prime})\, \frac{dW_{_{SH}}(s^{\prime},b)}{ds^{\prime}} \right] \\
&-\frac{1}{2\pi}\, \int_{0}^{\infty} ds^{\prime}\,\ln \left( \frac{s^{\prime}+s}{\vert s^{\prime}-s\vert} \right) \left[ W_{_{SH}}(s^{\prime},b)\, \frac{d\sigma_{QCD}(s^{\prime})}{ds^{\prime}} \right].
\label{chDGM.27}
\end{split}
\ee
\mbox{\,\,\,\,\,\,\,\,\,}
The second integral on the \tit{rhs} involves the derivatives of the QCD cross section $\sigma_{QCD}(s^{\prime})$. One should at this point notice that the $s^{\prime}$ dependence in $d\hat{\sigma}_{ij}/d\vert\hat{t}\vert$ terms can be ignored, since their derivatives are of order $1/s^{\prime 2}$. In this way, the only energy dependence appears in the Heaviside function $\Theta(x-y)$, in which
\begin{eqnarray}
\frac{d}{ds^{\prime}}\, \Theta \! \left( \frac{\hat{s}^{\prime}}{2}-\vert\hat{t}\vert \right) = 
\frac{d}{ds^{\prime}}\, \Theta \! \left( s^{\prime} -\frac{2\vert\hat{t}\vert}{x_{1}x_{2}} \right) =
\delta \! \left( s^{\prime} -\frac{2\vert\hat{t}\vert}{x_{1}x_{2}} \right),
\label{delta001}
\end{eqnarray}
where $\hat{s}=x_{1}x_{2}s$ e $\hat{s}^{\prime}=x_{1}x_{2}s^{\prime}$. The $\delta$-function removes the integration over $ds^{\prime}$, thus, the second integral can be expressed as
\bear
I_{2}(s,b) &=& -\frac{1}{2\pi}\, \int_{0}^{\infty} ds^{\prime}\, \ln\left( \frac{s^{\prime}+s}{\vert s^{\prime}-s\vert}\right) W_{_{SH}}(s^{\prime},b)\, \frac{d\sigma_{QCD}(s^{\prime})}{ds^{\prime}} \nonumber \\
&=& -\frac{1}{2\pi}\sum_{ij} \frac{1}{1+\delta_{ij}} \, W_{_{SH}}\!\left(\frac{2\vert \hat{t}\vert}{x_{1}x_{2}},b\right) \int_{0}^{1}\!\!dx_{1}\int_{0}^{1}\!\!dx_{2} \int_{Q^{2}_{min}}^{\infty}\!\!\!\!d\vert \hat{t}\vert\,\frac{d\hat{\sigma}_{ij}}{d\vert \hat{t}\vert}(\hat{s},\hat{t}) \nonumber \\ 
&\times& f_{i/A}(x_{1},\vert \hat{t}\vert)f_{j/B}(x_{2},\vert \hat{t}\vert) \ln \left( \frac{\hat{s}/2+\vert \hat{t}\vert}{\hat{s}/2-\vert \hat{t}\vert}\right).
\label{chDGM.28}
\eear
\mbox{\,\,\,\,\,\,\,\,\,}
The energy-dependent form factor $W_{_{SH}}(s,b)$ can have a monopole or a dipole form, namely $W^{(m)}_{_{SH}}(s,b;\nu_{_{SH}})$ or $W^{(d)}_{_{SH}}(s,b;\nu_{_{SH}})$, see expressions (\ref{chDGM.23}) and (\ref{chDGM.24}). In the case of a monopole-like form, the first integral on \tit{rhs} of (\ref{chDGM.27}) can be rewritten as
\bear
\!\!\!\!\!\!\!\!\!\!\!\!\!\!\!\!\!I^{(m)}_{1}(s,b) \!\!\!\!&=&\!\!\!\! -\frac{1}{2\pi}\, \int_{0}^{\infty} ds^{\prime}\, \ln\left( \frac{s^{\prime}+s}{\vert s^{\prime}-s\vert } \right) 
\sigma_{QCD}(s^{\prime})\,\frac{dW^{(m)}_{_{SH}}(s^{\prime},b;\nu_{_{SH}})}{ds^{\prime}} \nonumber \\
\!\!\!\!&=&\!\!\!\! -\frac{b}{8\pi^{2}}\sum_{ij} \frac{1}{1+\delta_{ij}} \int_{0}^{\infty} \frac{ds^{\prime}}{s^{\prime}}\, 
\ln\left( \frac{s^{\prime}+s}{\vert s^{\prime}-s\vert } \right)\int_{0}^{1}\!\!dx_{1} \int_{0}^{1}\!\!dx_{2} \int_{Q^{2}_{min}}^{\infty}\!\!\!\!d\vert \hat{t}\vert \, \frac{d\hat{\sigma}_{ij}}{d\vert \hat{t}\vert }(\hat{s}^{\prime},\hat{t}) \nonumber \\ 
\!\!\!\!&\times& \!\!\!\!f_{i/A}(x_{1},\vert \hat{t}\vert )f_{j/B}(x_{2},\vert \hat{t}\vert )\left[ b\nu_{2}\nu_{_{SH}}^{3}K_{0}(\nu_{_{SH}} b)- 2\nu_{2}\nu_{_{SH}}^{2}K_{1}(\nu_{_{SH}} b)  \right]\! \Theta\! \left( \frac{\hat{s}^{\prime}}{2} - \vert \hat{t}\vert  \right),
\label{chDGM.29}
\eear
and in the case of a dipole-like form factor one finds,
\bear
\!\!\!\!\!\!\!\!\!\!\!\!\!\!\!\!\!I^{(d)}_{1}(s,b) \!\!\!\!&=&\!\!\!\! -\frac{1}{2\pi}\, \int_{0}^{\infty} ds^{\prime}\, \ln\left( \frac{s^{\prime}+s}{\vert s^{\prime}-s\vert } \right) 
\sigma_{_{QCD}}(s^{\prime})\,\frac{dW^{(d)}_{_{SH}}(s^{\prime},b;\nu_{_{SH}})}{ds^{\prime}} \nonumber \\
\!\!\!\!&=&\!\!\!\! -\frac{b^{3}}{192\pi^{2}}\sum_{ij} \frac{1}{1+\delta_{ij}} \int_{0}^{\infty} \frac{ds^{\prime}}{s^{\prime}}\, \ln\left(
\frac{s^{\prime}+s}{\vert s^{\prime}-s\vert } \right) 
\int_{0}^{1}\!\!dx_{1} \int_{0}^{1}\!\!dx_{2} \int_{Q^{2}_{min}}^{\infty}\!\!\!\!d\vert \hat{t}\vert \,
\frac{d\hat{\sigma}_{ij}}{d\vert \hat{t}\vert }(\hat{s}^{\prime},\hat{t}) \nonumber \\ 
\!\!\!\!&\times&\!\!\!\! f_{i/A}(x_{1},\vert \hat{t}\vert )f_{j/B}(x_{2},\vert \hat{t}\vert )\left[ b\nu_{2}\nu_{_{SH}}^{5}K_{2}(\nu_{_{SH}} b)-
2\nu_{2}\nu_{_{SH}}^{4}K_{3}(\nu_{_{SH}} b)  \right] \Theta \left( \frac{\hat{s}^{\prime}}{2} - \vert \hat{t}\vert  \right).
\label{chDGM.30}
\eear
\mbox{\,\,\,\,\,\,\,\,\,}
The soft eikonal is needed only to describe the lower-energy forward data, since the main contribution to the asymptotic behavior of the hadron-hadron total cross section comes from parton-parton semihard collisions. Therefore, it is enough to build an instrumental parametrization for the soft eikonal with terms dictated by the Regge phenomenology \cite{Luna:2003kw,Luna:2004gr,Luna:2004zp,Luna:2008pp,Luna:2010ch}. For the even part of the soft eikonal it was taken,
\be
\chi^{+}_{_{soft}}(s,b) = \frac{1}{2}\, W^{+}_{soft}(b;\mu^{+}_{_{soft}})\, \left\{ A +\frac{B}{(s/s_{0})^{\gamma}}\, e^{i\pi\gamma/2}+\,C\left[ \ln\left(\frac{s}{s_{0}}\right) -i\,\frac{\pi}{2} \right] \right\} ,
\label{chDGM.31}
\ee
where $\sqrt{s_{0}}\equiv 5$ GeV and $A$, $B$, $C$, $\gamma$ and $\mu^{+}_{_{soft}}$ are fitting parameters. The phase factor $e^{i\pi\gamma/2}$, which ensures the correct analyticity properties of the amplitude, is a result of the integral dispersion relation (\ref{chDGM.25}).

The odd eikonal $\chi^{-}(s,b)$ which accounts for the difference between $pp$ and $\bar{p}p$ channels and vanishes at high-energy, is given by
\be
\chi^{-}(s,b) = \frac{1}{2}\, W^{-}_{_{soft}}(b;\mu^{-}_{_{soft}})\,D\, \frac{e^{-i\pi/4}}{\sqrt{s/s_{0}}},
\label{chDGM.32}
\ee
where $D$, the strength of the odd term, is also a fitting parameter. The expression (\ref{chDGM.32}) was written with its correct analyticity property, since the phase factor $e^{-i\pi/4}$ is a result of the dispersion relation,
\be
\textnormal{Im}\,\chi^{-}(s,b) = -\frac{2s^{2}}{\pi}\, {\cal P}\int_{0}^{\infty}ds^{\prime}\, \frac{\textnormal{Re}\,\chi^{-}(s^{\prime},b)}{s^{\prime}(s^{\prime 2}-s^{2})},\,\,\,\,\,\,\text{valid at}\,\,\,\,s\gg m.
\label{chDGM.33}
\ee

\subsection{\textsc{The Role of Gluons}}
\label{secDGM.2.3}

\mbox{\,\,\,\,\,\,\,\,\,}
The calculation of the QCD cross section implies the sum over all possible parton types, but it is sufficiently accurate for our purpose in the DGM$15$ model to fix the number of flavors $n_{f}=4$ and keep only the gluon $g$ and the quarks $u,d,s$ and $c$. As a matter of fact $\text{Re}\,\chi_{_{SH}}(s,b)$ and $\text{Im}\,\chi_{_{SH}}(s,b)$ have to be determined taking into account all the heavy quarks, where each heavy quarks $h=c,b,t$ with mass $M_{h}$ is effectively decoupled from physical cross section at momenta scales below the threshold $Q_{h}=M_{h}$, $n_{f}$ being an increasing function of $Q_{h}$.

However, our numerical results show that the contributions of the quarks $b$ and $t$ to $\chi_{_{SH}}$ are very small indeed. In fact even the charm contribution is tiny, and was included only for high-precision purposes. Hence, there is no fundamental role for heavy quarks $(m_{q}\approx M_{h},\,\,h=c,b,t)$ in our analyses, and this result can be understood as follows: heavy quarks are produced, perturbatively, from the splitting of gluons into $\bar{h}h$ pairs at energies above the threshold $Q_{h}=M_{h}$. At sufficiently small-$x$, the ratio of the heavy-quarks parton distribution function, $h(x,Q^{2})$, to the gluon one, $g(x,Q^{2})$, is \cite{Barnett:1987jw,Olness:1987ep,Aivazis:1993pi,Stelzer:1997ns}
\be
\frac{h(x,Q^{2})}{g(x,Q^{2})}\sim\frac{\alpha_{s}(Q^{2})}{2\pi}\,\log\left(\frac{Q^{2}}{M^{2}_{h}}\right),
\label{chDGM.34}
\ee
where $h(x,Q^{2})=0$ at $Q=M_{h}$. However, the angular dependencies of the dominant subprocesses in (\ref{chDGM.12}) are very similar and all dominated by the $t$-channel angular distribution. As a consequence, the parton-parton differential cross sections vary essentially as $d\hat{\sigma}_{ij}/d\vert \hat{t}\vert\sim 1/Q^{4}$. Hence, the effects of distribution functions as well as current masses of heavy quarks on $\sigma_{QCD}(s)$ are absolutely negligible.

In order to obtain $\chi_{_{SH}}(s,b)$ it was selected parton-parton scattering processes containing at least one gluon in the initial state. The reason for this choice comes from the behavior of the partonic splitting dictated by DGLAP evolution equations at leading order \cite{Gribov:1972ri,Lipatov:1974qm,Altarelli:1977zs,Dokshitzer:1977sg}, in which the gluon splitting functions $P_{gq}\to 4/(3z)$ and $P_{gg}\to 6/z$ are singular as $z\to 0$. As a result, the gluon distribution becomes very large as $x\to 0$ (in the convolution integrals $z<x$), and its role in the evolution of parton distributions becomes central. Thus, taking into account the mechanism of dynamical mass generation in QCD, it was selected the following required parton-parton processes for calculating $\sigma_{QCD}(s)$:
\bi

\item[i.] gluon-gluon elastic scattering,
\be
\frac{d\hat{\sigma}}{d\hat{t}}(gg\to gg)=\frac{9\pi\bar{\alpha}^{2}_{s}}{2\hat{s}^{2}}\left(3 -\frac{\hat{t}\hat{u}}{\hat{s}^{2}}-
\frac{\hat{s}\hat{u}}{\hat{t}^{2}}-\frac{\hat{t}\hat{s}}{\hat{u}^{2}} \right) ,
\label{chDGM.35}
\ee

\item[ii.] quark-gluon elastic scattering,
\be
\frac{d\hat{\sigma}}{d\hat{t}}(qg\to qg)=\frac{\pi\bar{\alpha}^{2}_{s}}{\hat{s}^{2}}\, (\hat{s}^{2}+\hat{u}^{2}) \left(
\frac{1}{\hat{t}^{2}}-\frac{4}{9\hat{s}\hat{u}} \right) ,
\label{chDGM.36}
\ee

\item[iii.] gluon fusion into a quark pair,
\be
\frac{d\hat{\sigma}}{d\hat{t}}(gg\to \bar{q}q)=\frac{3\pi\bar{\alpha}^{2}_{s}}{8\hat{s}^{2}}\, (\hat{t}^{2}+\hat{u}^{2}) \left(
\frac{4}{9\hat{t}\hat{u}}-\frac{1}{\hat{s}^{2}} \right),
\label{chDGM.37}
\ee

\ei
where $\hat{s}=x_{1}x_{2}s$ and $\hat{t}=-Q^{2}$. The gluon-gluon and quark-gluon scattering processes in fact dominate at high energies. For example, at $\sqrt{s}=7$ TeV and with $Q_{min}=1.3$ GeV, their relative contribution to the cross section $\sigma_{QCD}(s)$ is around $98.84\%\,\,(98.66\%)$ for the CTEQ$6$L (MSTW) set of parton distributions. The relative contribution of the process $gg\to\bar{q}q$ is rather tiny, nevertheless, it was included just for completeness.

In the limit of large enough $Q^{2}$, the expressions (\ref{chDGM.35}-\ref{chDGM.37}) reproduce their pQCD counterparts. In these expressions the kinematic constraints under consideration are given by $\hat{s}+\hat{t}+\hat{u}=4M^{2}_{g}(Q^{2})$ in the case of gluon-gluon scattering, and $\hat{s}+\hat{t}+\hat{u}=2M^{2}_{g}(Q^{2})+2M^{2}_{q}(Q^{2})$ in the case of quark-gluon scattering and gluon fusion into a quark pair. Here, $M^{2}_{q}(Q^{2})$ is the dynamical quark mass,
\be
M_{q}(Q^{2})=\frac{m^{3}_{q}}{Q^{2}+m^{2}_{q}},
\label{chDGM.38}
\ee
which assumes a nonzero infrared mass scale $m_{q}$, to be phenomenologically adjusted. Notice that the effective mass for quarks is a sum of the dynamical mass and the running one. However, as discussed, only the contributions of lighter quarks are relevant in calculating $\sigma_{QCD}(s)$ and as a result the effective mass behavior is dominated by the dynamical part. The expression (\ref{chDGM.38}), which rapidly decreases with increasing $Q$, is the simplest \tit{Ansatz} for a dynamical quark mass in agreement with OPE \cite{Politzer:1976tv,Pascual:1981jr,Shifman:1978zq,Vainshtein:1978wd,Shifman:1978bx,Lavelle:1991ve,Dudal:2008sp}. According to the OPE the dynamical mass is a function of the quark condensate $\langle \bar{\psi}\psi\rangle$. More specifically, $M_{q}(P^{2})\propto\langle \bar{\psi}\psi\rangle/P^{2}$, where $P^{2}=-p^{2}$ is the momentum in Euclidean space. The quark mass scale $m_{q}$ can be related to the quark condensate ($\langle \bar{\psi}\psi\rangle\propto m^{3}_{q}$ by dimensional considerations) and general constraints are satisfied for $m_{q}\in[100;250]$ MeV. The simple power-law behavior of $M_{q}(Q^{2})$ is finally obtained by introducing the factor $m^{2}_{q}$ in the denominator in order to get the right infrared limit $M^{2}_{q}(Q^{2}\to 0)=m^{2}_{q}.$

\section{\textsc{The Parton Distribution Functions}}
\label{secDGM.3}
\hyphenation{mo-dern}\mbox{\,\,\,\,\,\,\,\,\,}
Although the QCD-inspired eikonal models have passed through many changes and in the phenomenological point of view evolved in some other aspects in the last couple of decades, the gluon distribution function usually adopted is still based on (naive) parametrizations of the form $g(x)\propto(1-x)^{5}/x^{J}$, where the parameter $J=\alpha_{_{\mathds{P}}}(0)=1+\epsilon$, with $\epsilon>0$, controls the evolution of the gluon distribution at small-$x$. The quark-quark and quark-gluon contributions are written by means of parametrizations based on Regge phenomenology. In the Regge language the quantity $J$ controls the asymptotically behavior of the total cross section and is the so-called intercept of the Pomeron. Hence, the total cross section should behave asymptotic as a Pomeron power-law $s^{J-1}$, and a consistent value of $J$ should be determined by fitting forward quantities data by means of a Regge pole model.

However, the validity of the functional form of $g(x)$ is approximately correct only in the limits at $x\to 0$ and $x\to 1$, whereas in intermediate $x$ regions it does not reproduce the behavior of any other distribution function $g(x,Q^{2})$ running with $Q^{2}$, whatever the values of the momentum scale $Q^{2}$ and the parameter $J$. The coupling $\alpha_{s}(Q^{2})$ is one of the basic parameter in QCD, since its dependence with $Q$ reflects the property of asymptotic freedom of QCD. Thus, by fixing the coupling $\alpha_{s}(Q^{2})$, as some models used to do, it represents a very unsatisfactory approximation for the partonic distribution. This resembles to a loss of bond to the QCD parton model, is such a way that it turned out to be contested the terminology ``QCD-inpired model''. The partonic distributions must run with the momentum scale $Q$ according to the DGLAP equations \cite{Gribov:1972ri,Lipatov:1974qm,Altarelli:1977zs,Dokshitzer:1977sg}, thus allowing to determine the distribution in $x$ and $Q$ by means of an initial scale $Q_{0}$. It is extremely important that PDF's should exhibit a dependence with the momentum scale. 

In the literature there are a great variety of different kinds of PDF's, where some authors usually named the ``first generation'' the distributions set formed by EHLQ \cite{Duke:1983gd} and DO \cite{Duke:1983gd}, respectively. But this first generation set of PDF's are now obsolete when compared to the rigorous calculations necessary to describe hadronic processes, since both theoretical and experimental developments have had breakthroughs in the last few years. The modern generation is compounded by distribution sets obtained by means of the most recent DIS structure functions data and other related processes, where in this set of PDF's one finds for instance the following ones: ABM \cite{Alekhin:1996za,Alekhin:2000ch,Alekhin:2002fv}, CTEQ/CT14 \cite{Botts:1992yi,Lai:1994bb,Lai:1996mg,Lai:1999wy,Pumplin:2002vw,Pumplin:2002vw,Stump:2003yu,Hou:2016nqm,Dulat:2015mca}, GRV/GJR \cite{Gluck:1994uf,Gluck:1998xa,Gluck:2007ck}, MRS/MRST/MSTW/MMHT \cite{Martin:1994kn,Martin:1998sq,Martin:1999ww,Martin:2009iq,Martin:2002dr,Harland-Lang:2014zoa}, NNPDF \cite{Ball:2008by}. These distributions differ from each other basically by the experimental data used, the types of parametrizations initially adopted, the different choices of the initial momentum scale $Q_{0}$ and the statistical treatment of the systematic errors.

In what follows, it will be discussed in general lines two different sets of PDF's, namely CTEQ6 and MSTW. The main objective is to give a hint on it so that one can understand its basics features, to show its functional forms, to discuss the limits of applicability as well as some general aspects.

\subsection{\textsc{The Partonic Distribution CTEQ6}}
\label{secDGM.3.1}
\mbox{\,\,\,\,\,\,\,\,\,}
The development of PDF's through global analysis of hard scattering processes is extremely important to the search for a theoretical description involving the QCD parton model phenomenology, and also the search for new Physics in lepton-hadron and hadron-hadron colliders. Over the past few years there have been efforts beyond the conventional analysis used by the most popular PDF's \cite{Botts:1992yi,Lai:1994bb,Lai:1996mg,Lai:1999wy,Gluck:1994uf,Gluck:1998xa,Martin:1994kn,Martin:1998sq}. One recent distribution was developed by the CTEQ Collaboration and it extends the previous generations, mainly due to the treatment between previous and new experimental data, and the treatment of the systematic uncertainties associated with partonic distributions and its physical predictions. In this set \cite{Pumplin:2002vw,Pumplin:2002vw,Stump:2003yu}, the conventional strategy methodology of producing the best global fit to the data was largely improved introducing new statistical tools which allow to characterize the partonic distribution space parameter around the global minimum. Within this new methodology \cite{Pumplin:2000vx,Stump:2001gu,Pumplin:2001ct} it turned out to be possible to explore the systematical uncertainties in the partonic distributions and its physical predictions, resulting in a better understanding of the hadronic content, mostly the gluon distribution.

The functional form of the partonic parametrizations used in CTEQ6 with initial momentum scale fixed in $Q_{0}=1.3$ GeV is given by \cite{Pumplin:2002vw,Pumplin:2002vw,Stump:2003yu}
\be
xf(x,Q_{0})=A_{0}\,x^{A_{1}}(1-x)^{A_{2}}\,e^{A_{3}x}\,(1+e^{A_{4}}\,x)^{A_{5}},
\label{chDGM.39}
\ee
with independent parameters for the parton combinations $u_{V}\equiv u-\bar{u}$, $d_{V}\equiv d-\bar{d}$, $g$ e $\bar{u}+\bar{d}$. The functional behavior of expression (\ref{chDGM.39}) at $x=0$ and $x=1$ represents the singularity associated with Regge phenomenology at small-$x$ and to the quark counting rule at bigger values of $x$, respectively.

Since the previous distribution set, CTEQ5 \cite{Lai:1999wy}, a great quantity of new experimental data have contributed to the statistical analysis of an updated PDF. More specifically, it was very important in this new analysis the recent measurements of structure function in DIS with neutral currents by H1 \cite{Adloff:1999ah,Adloff:2000qj} and ZEUS \cite{Chekanov:2001qu} experiments, the measurements of inclusive jets cross section at D\O{} \cite{Abbott:2000ew,Abbott:2000kp}, the measurements of Drell-Yan deuteron/proton ratio at FNAL E866/NuSea \cite{Towell:2001nh} and the reanalyzed measurements of $F_{2}$ at CCFR \cite{Yang:2000ju}.

In the CTEQ6 distribution, the QCD coupling constant $\alpha_{s}(Q^{2})$ is written in its LO and NLO forms. Respectively, for the case where one uses CTEQ6L1,
\be
\alpha^{LO}_{s}(Q^{2})=\frac{4\pi}{\beta_{0}\log\left(Q^{2}/\Lambda^{2}\right)},
\label{chDGM.40}
\ee
and for the case of CTEQ6L,
\be
\alpha^{NLO}_{s}(Q^{2})=\frac{4\pi}{\beta_{0}\log\left(Q^{2}/\Lambda^{2}\right)}\,\left[1-\frac{\beta_{1}}{\beta^{2}_{0}}\,\frac{\log\log\left(Q^{2}/\Lambda^{2}\right)}{\log\left(Q^{2}/\Lambda^{2}\right)}\right].
\label{chDGM.41}
\ee
\mbox{\,\,\,\,\,\,\,\,\,}
An effective number of quarks $n_{f}$ could be fixed if they were massless, in this case expressions (\ref{chDGM.40}) and (\ref{chDGM.41}) would be determined by only one $\Lambda$. However, the decoupling theorem \cite{Appelquist:1974tg} says that each heavy quark with mass $m_{i}$ decouples from the physical cross sections at energy scales $\mu<m_{i}$. Thus, the effective number of quark flavors depends on the renormalization scale $\mu$ implying that the resolution of the strong coupling $\alpha_{s}$ and the determination of $\Lambda$ are not unique in the presence of massive quarks, \tit{i.e.} it depends on the renormalization scheme adopted. One natural choice is based on the requirement that $\alpha_{s}(\mu)$ is a continuous function of $\mu$, but respectively the values of $\Lambda$ are discontinuous at $\mu=m_{i}$.

The $u$, $d$ and $s$ quarks are considered massless in the CTEQ6 distribution\footnote{$m_{u}\approx 1.7-3.1$ MeV, $m_{d}\approx 4.1-5.7$ MeV and $m_{s}\approx 80-130$ MeV.}, and the mass scales are defined by the masses of $c$ and $b$ quarks, with $m_{c}=1.3$ GeV and $m_{b}=4.2$ GeV. The leading order coupling is determined by the $M_{Z}$ scale where $\alpha^{LO}_{s}(M^{2}_{Z})=0.130$ and $\Lambda$'s are defined as $\Lambda_{4}=215$ MeV and $\Lambda_{5}=165$ MeV. In the next-to-leading order case, $\alpha^{NLO}_{s}(M^{2}_{Z})=0.118$ with $\Lambda_{4}=326$ MeV and $\Lambda_{5}=226$ MeV.

\subsection{\textsc{The Partonic Distribution MSTW}}
\label{secDGM.3.2}

\mbox{\,\,\,\,\,\,\,\,\,}
The year of $2008$ was the $20^{\textnormal{th}}$ anniversary of the first publication of MRS distribution, which contained the first NLO global analysis of the partonic distributions \cite{Martin:1987vw}. It is natural that new experimental data and new theoretical methods to treat these data imply in the development of more sophisticated PDF's. The MRST98 \cite{Martin:1998sq} was the first updated set of MRS to use the new measurements of structure function obatained at HERA. Moreover, the MRST98 is also known for being the first one to apply the study of heavy quarks in the partonic analysis. 

One of the recent distribution sets, originally based on MRS, extends the previous versions such as MRST2001 LO \cite{Martin:2002dr}, MRST2004 NLO \cite{Martin:2004ir}, MRST2006 NNLO \cite{Martin:2007bv}. In the recent set, namely MSTW \cite{Martin:2009iq}, the technique used to obtain the best fit to the data presents some improvements because of the advances in the study of error propagations, hence implying in a better understanding of the partonic distribution uncertainties.

The functional form of the partonic parametrizations used in MSTW with initial momentum scale fixed in $Q_{0}=1.0$ GeV is given by the following expressions \cite{Martin:2009iq}
\be
xu_{V}(x,Q^{2}_{0})=A_{u}x^{\eta_{1}}(1-x)^{\eta_{2}}(1+\epsilon_{u}\sqrt{x}+\gamma_{u}x),
\label{chDGM.42}
\ee
\be
xd_{V}(x,Q^{2}_{0})=A_{d}x^{\eta_{3}}(1-x)^{\eta_{4}}(1+\epsilon_{d}\sqrt{x}+\gamma_{d}x),
\label{chDGM.43}
\ee
\be
xS(x,Q^{2}_{0})=A_{S}x^{\delta_{S}}(1-x)^{\eta_{S}}(1+\epsilon_{S}\sqrt{x}+\gamma_{S}x),
\label{chDGM.44}
\ee
\be
x\Delta(x,Q^{2}_{0})=A_{\Delta}x^{\eta_{\Delta}}(1-x)^{\eta_{S}+2}(1+\gamma_{\Delta}x+\delta_{\Delta}x^{2}),
\label{chDGM.45}
\ee
\be
xg(x,Q^{2}_{0})=A_{g}x^{\delta_{g}}(1-x)^{\eta_{g}}(1+\epsilon_{g}\sqrt{x}+\gamma_{g}x)+A_{g^{\prime}}x^{\delta_{g^{\prime}}}(1-x)^{\eta_{g^{\prime}}},
\label{chDGM.46}
\ee
\be
x(s+\bar{s})(x,Q^{2}_{0})=A_{+}x^{\delta_{S}}(1-x)^{\eta_{+}}(1+\epsilon_{S}\sqrt{x}+\gamma_{S}x),
\label{chDGM.47}
\ee
\be
x(s-\bar{s})(x,Q^{2}_{0})=A_{-}x^{\delta_{-}}(1-x)^{\eta_{-}}(1-x/x_{0}),
\label{chDGM.48}
\ee
where $\Delta=\bar{d}-\bar{u}$, $q_{V}=q-\bar{q}$ and the light sea quarks are defined as $S\equiv2(\bar{u}+\bar{d})+s+\bar{s}$. The above expressions are constrained by four normalizations, the counting rules
\be
\int^{1}_{0}dx\,u_{V}(x,Q_{0}^{2})=2,
\label{chDGM.49}
\ee
\be
\int^{1}_{0}dx\,d_{V}(x,Q_{0}^{2})=1,
\label{chDGM.50}
\ee
\be
\int^{1}_{0}dx\,s_{V}(x,Q_{0}^{2})=0,
\label{chDGM.51}
\ee
and the conservation of momentum
\be
\int^{1}_{0}dx\,x\,\left[u_{V}(x,Q_{0}^{2})+d_{V}(x,Q_{0}^{2})+S(x,Q_{0}^{2})+g(x,Q_{0}^{2})\right]=1.
\label{chDGM.52}
\ee
\mbox{\,\,\,\,\,\,\,\,\,}
In the MSTW the $u$, $d$ and $s$ quarks are considered massless and the mass scales are defined by the mass of $c$ and $b$ quarks. As stressed out by the authors \cite{Martin:2009iq}, to obtain the best fit at LO, then the second term in the \tit{rhs} of expression (\ref{chDGM.46}) should not be considered.

\section{\textsc{Results of DGM15}}
\label{secDGM.4}

\mbox{\,\,\,\,\,\,\,\,\,}
First, in order to determine the model parameters, we fix  $n_{f}=4$ and set the values of the gluon and quark mass scales to $m_{g}=400$ MeV and $m_{q}=250$ MeV, respectively. These choices for the mass scales are not only consistent with our LO procedures, but are also the ones usually obtained in other calculations of strongly interacting processes \cite{Luna:2005nz,Luna:2006qp,Luna:2006sn,Fagundes:2011zx,Luna:2006tw,Beggio:2013vfa,Luna:2010tp,Doff:2013dda,Luna:2014kza}. Next, a global fit to high-energy forward $pp$ and $\bar{p}p$ scattering data was carried out above $\sqrt{s}=10$ GeV, namely, the total cross section $\sigma_{tot}^{pp,\bar{p}p}$ and the ratio of the real to imaginary part of the forward scattering amplitude $\rho^{pp,\bar{p}p}$.

It was used data sets compiled and analyzed  by the Particle Data Group \cite{Tanabashi:2018oca} as well as the LHC data from the TOTEM Collaboration, with the statistical and systematical errors added in quadrature. The TOTEM data set includes the first and the second measurements of the total $pp$ cross section at $\sqrt{s}=7$ TeV, $\sigma_{tot}^{pp}=98.30\pm2.80$ mb \cite{0295-5075-96-2-21002} and $\sigma_{tot}^{pp}=98.58\pm2.23$ mb \cite{Antchev:2013haa}, both using the optical theorem together with the luminosity provided by the CMS, the luminosity-independent measurements at $\sqrt{s}=7$ TeV, $\sigma_{tot}^{pp}=98.0\pm2.50$ mb \cite{Antchev:2013iaa}, the $\rho$-independent measurements at $\sqrt{s}=7$ TeV, $\sigma_{tot}^{pp}=99.10\pm4.30$ mb \cite{Antchev:2013iaa}, and the luminosity-independent measurement at $\sqrt{8}$ TeV, $\sigma_{tot}^{pp}=101.70\pm2.90$ \cite{Antchev:2013paa}. The data set includes the first estimate for the $\rho$-parameter made by the TOTEM Collaboration in their $\rho$-independent measurement at $\sqrt{s}=7$ TeV, namely $\rho^{pp}=0145\pm0.091$ \cite{Antchev:2013iaa}. 

Unfortunately, the DGM$15$ model \cite{Bahia:2015gya,Bahia:2015hha} does not include the corresponding total cross section at $\sqrt{s}=8$ TeV, namely $\sigma_{tot}^{pp}=101.5 \pm 2.1$ \cite{Antchev:2015zza}, $\sigma_{tot}^{pp}=101.9 \pm 2.1$ \cite{Antchev:2015zza}, $\sigma_{tot}^{pp}=102.9 \pm 2.3$ \cite{Antchev:2016vpy} and $\sigma_{tot}^{pp}=103.0 \pm 2.3$ \cite{Antchev:2016vpy}, and the corresponding $\rho$-parameter at $\sqrt{s}=8$ TeV, namely $\rho^{pp}=0.120 \pm 0.030$ \cite{Antchev:2016vpy}. Neither the most recent measurements obtained by TOTEM at $\sqrt{s}=13$ TeV, $\sigma_{tot}^{pp}=110.6 \pm 3.4$ \cite{Antchev:2017dia}, $\rho^{pp}=0.090 \pm 0.010$ \cite{Antchev:2017yns} and $\rho^{pp}=0.100 \pm 0.010$ \cite{Antchev:2017yns}. The reason why can be traced back when the paper was published, because at that time these data were not available yet. Despite the fact that presently we understand that the correct (forward) data set corresponds to the one within both ATLAS and TOTEM results \cite{Broilo:2018qqs,Broilo:2018els}, it is worth to be mentioned that we also did not consider such T+A ensemble. So, it was decided just for completeness to maintain the results as given in Reference \cite{Bahia:2015hha}, where it as used TOTEM only data.

In all the fits performed was used a $\chi^{2}$ fitting procedure, assuming an interval $\chi^{2}-\chi_{min}^{2}$ corresponding, in the case of normal errors, to the projection of the $\chi^{2}$ hypersurface containing $90$\% of probability. In this DGM$15$ version (8 fitting parameters) this corresponds to the interval $\chi^{2}-\chi_{min}^{2}=13.36$.

In our analysis we have investigated the effects of some sets of PDF's on the high-energy cross sections. In performing the fits one uses tree-level formulas for the parton-parton cross sections. In this way we have to choose PDF's evolved with LO splitting functions, as in case of LO sets CTEQ6L, CTEQ6L1 and MSTW. For the coupling $\alpha_{s}(Q^{2})$ it is usual to use either the LO formula for formal consistency or even the NLO one. In the specific case of CTEQ distributions \cite{Pumplin:2002vw,Abbott:2000ew}, the CTEQ6L1 uses LO formula for $\alpha_{s}(Q^{2})$ with $\Lambda^{(4flavor)}_{CTEQ6L1}=215$ MeV, whereas CTEQ6L uses NLO formula for $\alpha_{s}(Q^{2})$ with $\alpha_{s}(M^{2}_{Z})=0.118$, consistent with the value $\Lambda^{(4flavor)}_{CTEQ6L}=326$ MeV. Since the dynamical mass $M_{g}(Q^{2})$ practically vanishes at scales where four flavors are active, we choose these same values of $\Lambda^{(4flavor)}$ in our effective charges $\bar{\alpha}_{s}^{LO}(Q^{2})$ and $\bar{\alpha}_{s}^{N\!LO}(Q^{2})$, where $\bar{\alpha}_{s}^{LO}$ is given by the expression (\ref{chDGM.1}) whereas $\bar{\alpha}_{s}^{NLO}(Q^{2})$ is given by \cite{Luna:2010tp}
\be
\bar{\alpha}^{N\!LO}_{s}(Q^{2}) = \frac{4\pi}{\beta_{0}\ln\left[\left(Q^{2} +
4M^{2}_{g}(Q^{2})\right)/\Lambda^{2}\right]}\left[1-\frac{\beta_{1}}{\beta_{0}^{2}}\frac{\ln\ln\left[\left(Q^{2}+4M^{2}_{g}(Q^{2})\right)/\Lambda^{2}\right]}{\ln\left[\left(Q^{2} + 4M^{2}_{g}(Q^{2})\right)/\Lambda^{2}\right]} \right],
\label{chDGM.53}
\ee
where $\beta_{1} =102 -38n_{f}/3$ and $\Lambda = \Lambda^{(4flavor)}_{CTEQ6L}$. This NLO nonperturbative coupling is built by saturating the two-loop perturbative strong coupling $\alpha_{s}^{N\!LO}$, that is, by introducing the replacement $\alpha_{s}^{N\!LO}(Q^{2}) \to \bar{\alpha}_{s}^{N\!LO}(Q^{2}) = \alpha_{s}^{N\!LO}(Q^{2} + 4M^{2}_{g}(Q^{2}))$ into the perturbative result. Notice that we are using the same dynamical mass $M^{2}_{g}(Q^{2})$ expression for both LO and NLO couplings, since the results from Reference \cite{Luna:2010tp} give support to the statement that the dynamical mass scale $m_{g}$ is not strongly dependent on the perturbation order. 

The MSTW set uses an alternative definition of $\alpha_{s}$, where the renormalization group equation for $\alpha_{s}$ is truncated at the appropriate order and solved starting from an initial value $\alpha_{s}(Q_{0}^{2})$. This input value is one of their fit parameters and replaces the $\Lambda$ parameter \cite{Martin:2009iq}, respectively $\alpha(M^{2}_{z})=0.140$ corresponds to $\Lambda^{(5flavor)}=0.255$ GeV. In the usual matching-prescription scheme \cite{Owens:1992hd}, see Figure \ref{alfa_LO_MSTW}, the behavior of $\alpha_{s_{MSTW}}(Q^{2})$ can be properly reproduced from the choice $\Lambda^{(4flavor)}_{MSTW}\sim 319$ MeV. 

\bfg[hbtp]
  \begin{center}
    \includegraphics[width=10cm,height=8cm]{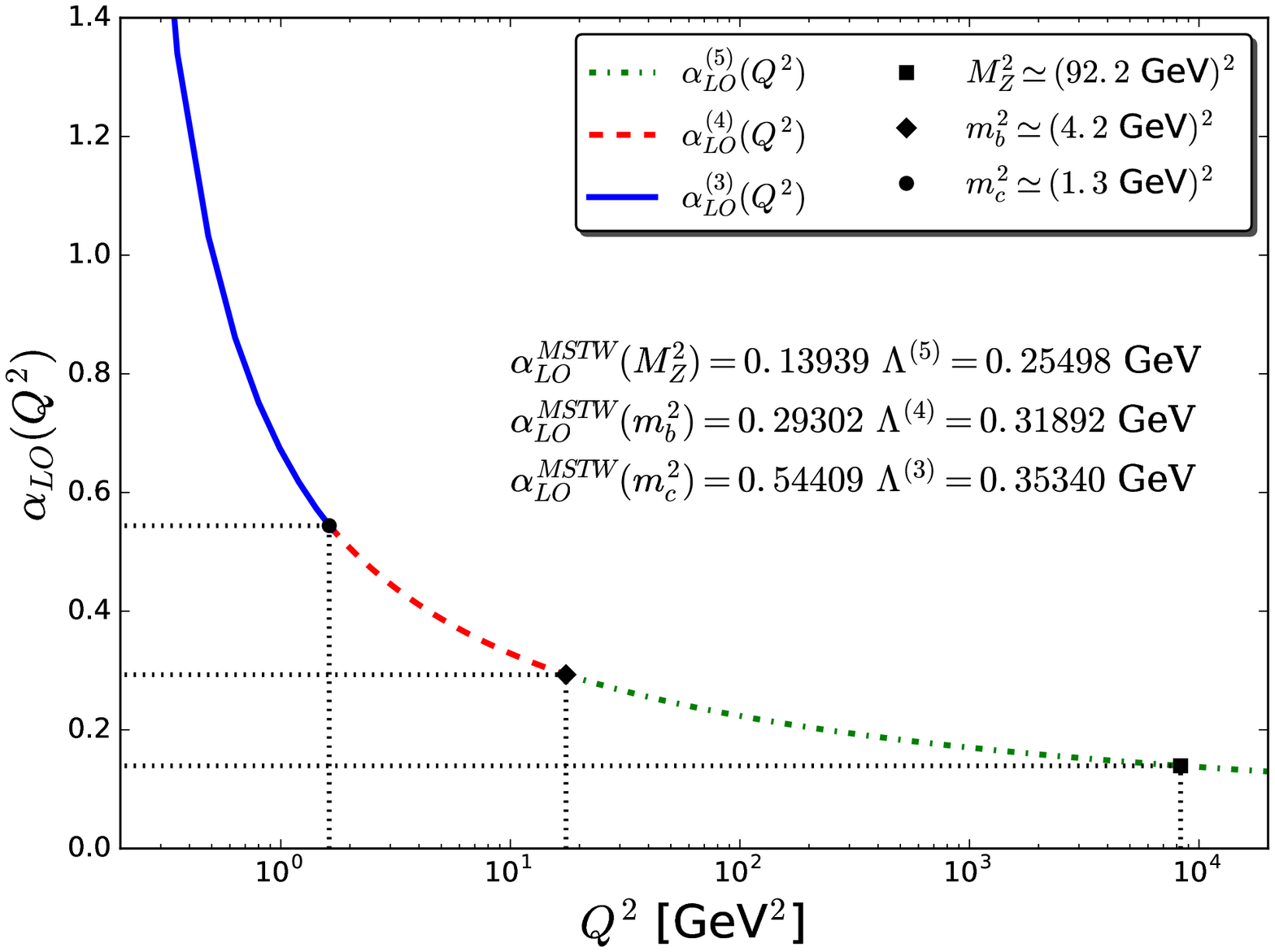}
    \caption{The representation of the decoupling theorem: the coupling function is associated with different values of $\Lambda_{QCD}$, which also depends on the number of effective quark flavors.}
    \label{alfa_LO_MSTW}
  \end{center}
\efg

The values of the fitted parameters are given in Table \ref{chDGMtab1}. It is shown the values of the parameters in the case of a monopole and dipole form factor in the semihard sector. The $\chi^{2}/\nu$ for all fits was obtained for $154$ degrees of freedom. The sensitivity of the $\chi^{2}/\nu$ to the cutoff $Q_{min}$ is shown in Figure \ref{chDGMfig1}. We observe that the $\chi^{2}/\nu$ is not very sensitive to $Q_{min}$ in the interval $[1.0, 1.5]$ GeV for all PDF's we have considered. The results of the fits to $\sigma_{tot}$ and $\rho$ for both $pp$ and $\bar{p}p$ channels are displayed in Figure \ref{chDGMfig2}, together with the theoretical predictions for the $pp$ cross sections at cosmic-ray energies. The comparison of the curves with the AUGER experimental datum at $\sqrt{s} = 57$ TeV \cite{Collaboration:2012wt} and the Telescope Array datum at $\sqrt{s} = 95$ TeV \cite{Abbasi:2015fdr} shows good agreement. The curves depicted in Figures \ref{chDGMfig2} were all calculated using the cutoff $Q_{min}=1.3$ GeV, the value of the CTEQ6 fixed initial scale $Q_{0}$. In the case of MSTW set the slightly lower value $Q_{0}\equiv 1$ GeV is adopted, and the condition $Q_{min}\ge Q_{0}$ is always satisfied in our analysis. In the case of fits using the CTEQ6 set, calculations in the region $Q_{min}< Q \leq Q_{0}$ were carried out with PDF's fixed at the scale $Q=Q_{0}=1.3$ GeV. In Table \ref{chDGMtab2} we show the theoretical predictions for the forward scattering quantities $\sigma_{tot}^{pp,\bar{p}p}$ and $\rho^{pp,\bar{p}p}$ using different sets of parton distributions. 

\section{\textsc{Conclusions on DGM15}}
\label{secDGM.5}

\mbox{\,\,\,\,\,\,\,\,\,}
In the calculation of $\sigma_{tot}^{pp,\bar{p}p}$ and $\rho^{pp,\bar{p}p}$ we have investigated the behavior of the forward amplitude for a range of different cutoffs and parton distribution functions, namely CTEQ6L, CTEQ6L1 and MSTW, and considered the phenomenological implications of a class of energy-dependent form factors for semihard partons. We introduce integral dispersion relations specially tailored to connect the real and imaginary parts of eikonals with energy-dependent form factors. In our analysis we have included the LHC data from the TOTEM Collaboration available at that time \cite{Antchev:2016vpy}. 

We have paid attention to the sensitivity of the $\chi^{2}/\nu$ to the cutoff $Q_{min}$, which restricts the parton-parton processes to semihard interactions. Our results show that very good descriptions of $\sigma_{tot}^{pp,\bar{p}p}$ and $\rho^{pp,\bar{p}p}$ data are obtained by constraining the cutoff to the interval $1.0 \le Q_{min}\lesssim 1.5$ GeV. The $\chi^{2}/\nu$ for the best global fits lie in the range $[1.05;1.06]$ for $154$ degrees of freedom. This good statistical result shows that our eikonal model, where nonperturbative effects are naturally included via a QCD effective charge, is well suited for detailed predictions of the forward quantities to be measured at higher energies. In fact our predictions for $pp$ total cross section are statistically compatible with the AUGER result at $\sqrt{s}=57$ TeV, namely $\sigma^{pp}_{tot}=[133\pm 13\,(\textnormal{stat})^{+17}_{-20}\,(\textnormal{syst})\pm 16\,(\textnormal{Glauber})]$ mb \cite{Collaboration:2012wt}, as well as with the Telescope Array result at $\sqrt{s}=95$ TeV, namely $\sigma^{pp}_{tot}=[170^{+48}_{-44}\,(\textnormal{stat})^{+17}_{-19}\,(\textnormal{syst})]$ mb \cite{Abbasi:2015fdr}. However it is worth noting that both results are model dependent, since they come from the conversion of the proton-air production cross section via a Glauber calculation. Moreover, as stressed by AUGER group, the total uncertainty of converting the proton-air to $pp$ cross section may be larger than the published. Clearly new results from AUGER and Telescope Array at higher energies would be extremely informative.

The uncertainty in our theoretical predictions for the forward observables at $\sqrt{s}=8,\,13,\,14,\,57$ and $95$ TeV, see Table \ref{chDGMtab2}, have been estimated by varying the gluon mass scale within a typical uncertainty $\delta m_{g}$ while keeping all other model parameters constant, and by exploring the uncertainties of parton distributions on production cross sections. This procedure does not determines the formal theoretical uncertainty in $\sigma_{tot}$ and $\rho$, since the variance-covariance matrix method, necessary for proving this quantity, was not employed. However, at high energies the forward observables are dominated by semihard interactions represented by the eikonal term $\chi_{_{SH}}(s,b)$, which depends only on $3$ parameters, namely $\nu_{1}$, $\nu_{2}$ and $m_{g}$.

In all $\chi^{2}$ analyses we have observed that the correlation coefficients of these parameters are very small. Moreover, the values of $\sigma_{tot}$ and $\rho$ are actually more sensitive to the gluon mass scale than to variations of others parameters of the model. A reliable estimate of $\delta m_{g}$, namely around $7.1$\% of the value of $m_{g}$, was obtained from the analysis of the proton structure function $F_{2}(x,Q^{2})$ at small-$x$ \cite{Luna:2010tp}. Hence in our case, where $m_{g}$ was set at $400$ MeV, the gluon mass uncertainty is $\delta m_{g}\sim 28$ GeV. In order to estimate the uncertainty of parton distributions on the forward predictions we simply adopt the conservative stance that the PDF's uncertainties on the total cross sections are of the same order of magnitude as the uncertainties on the production cross sections of the $W$ and $Z$ bosons at the LHC. The uncertainties on the production cross sections are estimated to be $\pm5\%$ by the CTEQ group \cite{Pumplin:2002vw,Pumplin:2002vw,Stump:2003yu,Pumplin:2000vx,Stump:2001gu,Pumplin:2001ct,Lai:1999wy}. Summarizing, the total uncertainty of our theoretical predictions is obtained from the quadrature sum of the uncertainties coming from the gluon mass uncertainty $\delta m_{g}$ and the parton distributions.

In the semihard sector we have considered a new class of form factors in which the average gluon radius increases with $\sqrt{s}$. With this assumption we have obtained another form in which the eikonal can be factored into the QCD parton model, more specifically $\textnormal{Re}\,\chi_{_{SH}}(s,b) = \frac{1}{2}\, W_{_{SH}}(s,b)\,\sigma_{QCD}(s)$. The imaginary part of this semi-factorizable eikonal was obtained by means of appropriate integral dispersion relations which take into account eikonals with energy-dependent form factors. Although these dispersion relations are quite accurate at high energies, detailed studies using derivative dispersion relations \cite{Avila:2002tk,Avila:2001qz,Avila:2003cu} would be needed to quantify the effect of dispersion-relation subtractions on the imaginary part of the eikonal.

\begin{table}[hbtp]
\centering
\scalebox{0.9}{
\begin{tabular}{c@{\quad}c@{\quad}c@{\quad}c@{\quad}}
\multicolumn{4}{c}{Monopole form factor}\\
\hline\hline
& & & \\[-0.2cm]
& CTEQ6L & CTEQ6L1 & MSTW \\[0.7ex] 
\hline
& & & \\[-0.2cm]
$\nu_{1}$ [GeV]        & 1.71\,$\pm$\,0.54                    & 1.98\,$\pm$\,0.75                  & 1.52\,$\pm$\,0.767                  \\[0.7ex]
$\nu_{2}$ [GeV]        & 0.0034\,$\pm$\,0.0013                & 0.0052\,$\pm$\,0.0016              & 0.00095\,$\pm$\,0.00087             \\[0.7ex]
$A$ [GeV$^{-2}$]       & 125.3\,$\pm$\,14.7                   & 107.3\,$\pm$\,9.0                  & 107.2\,$\pm$\,13.6                  \\[0.7ex]
$B$ [GeV$^{-2}$]       & 43.0\,$\pm$\,24.9                    & 28.7\,$\pm$\,14.7                  & 30.5\,$\pm$\,16.2                   \\[0.7ex]
$C$ [GeV$^{-2}$]       & 1.98\,$\pm$\,0.68                    & 1.22\,$\pm$\,0.40                  & 1.19\,$\pm$\,0.47                   \\[0.7ex]
$\gamma$               & 0.76\,$\pm$\,0.19                    & 0.70\,$\pm$\,0.21                  & 0.64\,$\pm$\,0.25                   \\[0.7ex]
$\mu^{+}_{soft}$ [GeV] & 0.78\,$\pm$\,0.18                    & 0.41\,$\pm$\,0.27                  & 0.48\,$\pm$\,0.30                   \\[0.7ex]
$D$ [GeV$^{-2}$]       & 23.8\,$\pm$\,2.0                     & 21.4\,$\pm$\,2.7                   & 21.9\,$\pm$\,2.8                    \\[0.7ex]
$\mu^{-}_{soft}$ [GeV] & {\bf 0.5 [fixed]}                    & {\bf 0.5 [fixed]}                  & {\bf 0.5 [fixed]}                   \\[0.7ex]
\hline
& & & \\[-0.2cm]
$\chi^{2}/154$         & 1.060                           & 1.063                              & 1.049                     \\[0.7ex]
\hline\hline\\[0.4cm]

\multicolumn{4}{c}{Dipole form factor}\\
\hline\hline
& & & \\[-0.2cm]
& CTEQ6L & CTEQ6L1 & MSTW \\[0.7ex] 
\hline
& & & \\[-0.2cm]
$\nu_{1}$ [GeV]            & 2.36\,$\pm$\,0.62              & 2.77\,$\pm$\,0.87                   & 2.27\,$\pm$\,0.85                   \\[0.7ex]
$\nu_{2}$ [GeV]            & 0.0051\,$\pm$\,0.0042          & 0.0079\,$\pm$\,0.0054               & 0.0031\,$\pm$\,0.0029               \\[0.7ex]
$A$ [GeV$^{-2}$]           & 128.9\,$\pm$\,13.9             & 108.9\,$\pm$\,8.6                   & 108.5\,$\pm$\,11.5                  \\[0.7ex]
$B$ [GeV$^{-2}$]           & 46.7\,$\pm$\,26.1              & 30.2\,$\pm$\,5.8                    & 31.6\,$\pm$\,16.2                   \\[0.7ex]
$C$ [GeV$^{-2}$]           & 2.10\,$\pm$\, 0.67             & 1.26\,$\pm$\,0.44                   & 1.23\,$\pm$\,0.47                   \\[0.7ex]
$\gamma$                   & 0.78\,$\pm$\,0.17              & 0.72\,$\pm$\,0.20                   & 0.66\,$\pm$\,0.23                   \\[0.7ex]
$\mu^{+}_{soft}$ [GeV]     & 0.82\,$\pm$\,0.15              & 0.46\,$\pm$\,0.21                   & 0.51\,$\pm$\,0.24                   \\[0.7ex]
$D$ [GeV$^{-2}$]           & 24.0\,$\pm$\,1.9               & 21.7\,$\pm$\,2.3                    & 22.1\,$\pm$\,2.4                    \\[0.7ex]
$\mu^{-}_{soft}$ [GeV]     & {\bf 0.5 [fixed]}              & {\bf 0.5 [fixed]}                   & {\bf 0.5 [fixed]}                   \\[0.7ex]
\hline
& & & \\[-0.2cm]
$\chi^{2}/154$             & 1.064                            & 1.062                            & 1.047                      \\[0.7ex]
\hline \hline
\end{tabular}
}
\caption{Values of the DGM$15$ model parameters from the global fit to the scattering $pp$ and $\bar{p}p$ forward data.}
\label{chDGMtab1}
\end{table}

\begin{table}[hbtp]
\centering
\scalebox{0.9}{
\begin{tabular}{c@{\quad}c@{\quad}c@{\quad}c@{\quad}c@{\quad}c@{\quad}c@{\quad}}
\hline \hline
& & & & & &\\[-0.2cm]
        & \,\,\,\,\,  \,\,\,\,\, &  \multicolumn{2}{c}{$\sigma_{tot}$ [mb]}  & &  \multicolumn{2}{c}{$\rho$}  \\
\cline{3-4} \cline{6-7}
& & & & & &\\[-0.2cm]
        &   $\sqrt{s}$ [TeV]     & \,\,\,\,\,  monopole      \,\,\,\,\,        &\,\,\,\,\,  dipole \,\,\,\,\,              &  & \,\,\,\,\, monopole       \,\,\,\,\,           &\,\,\,\,\, dipole    \,\,\,\,\,                 \\
& & & & & &\\[-0.4cm]
\hline
 & & & & & &\\[-0.2cm]
        & $8.0$  &  $100.9^{+8.6}_{-7.3}$    &  $101.0^{+8.6}_{-7.3}$   & & $0.115^{+0.009}_{-0.008}$ & $0.106^{+0.009}_{-0.007}$\\[1.2ex]
        & $13.0$ &  $111.5^{+9.7}_{-8.4}$    &  $111.7^{+9.7}_{-8.4}$   & & $0.110^{+0.010}_{-0.008}$ & $0.101^{+0.009}_{-0.008}$\\[1.2ex]
 CTEQ6L	& $14.0$ &  $113.2^{+9.9}_{-8.6}$    &  $113.5^{+9.9}_{-8.6}$   & & $0.110^{+0.010}_{-0.008}$ & $0.100^{+0.009}_{-0.008}$\\[1.2ex]
 	& $57.0$ &  $152.5^{+15.4}_{-14.7}$  &  $154.1^{+15.6}_{-14.9}$ & & $0.097^{+0.010}_{-0.010}$ & $0.088^{+0.009}_{-0.009}$\\[1.2ex]
 	& $95.0$ &  $170.3^{+17.2}_{-16.5}$  &  $172.9^{+17.5}_{-16.8}$ & & $0.092^{+0.010}_{-0.010}$ & $0.083^{+0.009}_{-0.009}$\\[1.2ex] \hline
 & & & & & &\\[-0.2cm]
        & $8.0$  &  $101.1^{+8.6}_{-7.3}$    &  $101.2^{+8.6}_{-7.3}$   & & $0.134^{+0.012}_{-0.009}$ & $0.124^{+0.011}_{-0.009}$\\[1.2ex]
 	& $13.0$ &  $112.4^{+9.8}_{-8.5}$    &  $112.9^{+9.8}_{-8.5}$   & & $0.131^{+0.012}_{-0.010}$ & $0.120^{+0.011}_{-0.009}$\\[1.2ex]
CTEQ6L1	& $14.0$ &  $114.2^{+10.0}_{-8.7}$   &  $114.9^{+10.0}_{-8.7}$  & & $0.130^{+0.012}_{-0.010}$ & $0.119^{+0.011}_{-0.009}$\\[1.2ex]
 	& $57.0$ &  $159.3^{+16.1}_{-15.4}$  &  $163.7^{+16.5}_{-15.8}$ & & $0.117^{+0.012}_{-0.012}$ & $0.106^{+0.011}_{-0.011}$\\[1.2ex]
	& $95.0$ &  $181.5^{+18.3}_{-17.6}$  &  $188.9^{+19.0}_{-18.4}$ & & $0.112^{+0.012}_{-0.012}$ & $0.101^{+0.011}_{-0.011}$\\[1.2ex]\hline
 & & & & & &\\[-0.2cm]
        & $8.0$  &  $101.3^{+8.6}_{-7.3}$    &  $101.3^{+8.7}_{-7.3}$   & & $0.142^{+0.013}_{-0.010}$ & $0.131^{+0.012}_{-0.009}$\\[1.2ex]
 	& $13.0$ &  $113.3^{+9.9}_{-8.5}$    &  $113.6^{+9.9}_{-8.5}$   & & $0.139^{+0.012}_{-0.011}$ & $0.128^{+0.011}_{-0.010}$\\[1.2ex]
 MSTW	& $14.0$ &  $115.4^{+10.1}_{-8.7}$   &  $115.7^{+10.1}_{-8.8}$  & & $0.139^{+0.013}_{-0.011}$ & $0.128^{+0.012}_{-0.010}$\\[1.2ex]
	& $57.0$ &  $162.1^{+16.4}_{-15.6}$  &  $164.7^{+16.6}_{-15.9}$ & & $0.127^{+0.013}_{-0.013}$ & $0.116^{+0.012}_{-0.011}$\\[1.2ex]
 	& $95.0$ &  $183.0^{+18.5}_{-17.8}$  &  $187.3^{+18.9}_{-18.2}$ & & $0.123^{+0.013}_{-0.013}$ & $0.112^{+0.012}_{-0.012}$\\[1.2ex]
\hline \hline
\end{tabular}}
\caption{Predictions for the forward scattering quantities $\sigma_{tot}^{pp,\bar{p}p}$ and $\rho^{pp,\bar{p}p}$ using different sets of parton distributions.}
\label{chDGMtab2}
\end{table}

\bfg[hbtp]
  \begin{center}
    \includegraphics[width=8.0cm,height=9.0cm]{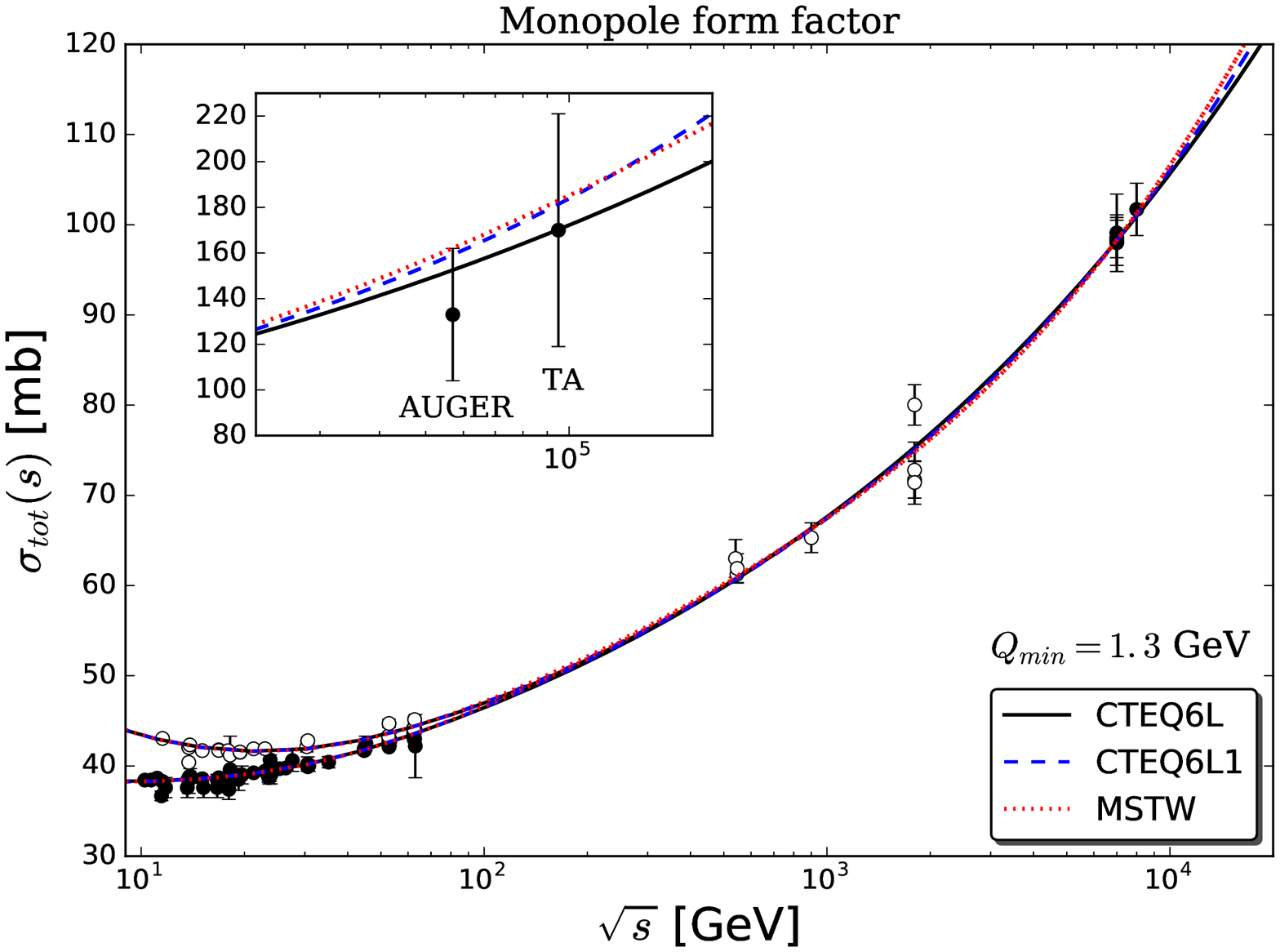}
    \includegraphics[width=8.0cm,height=9.0cm]{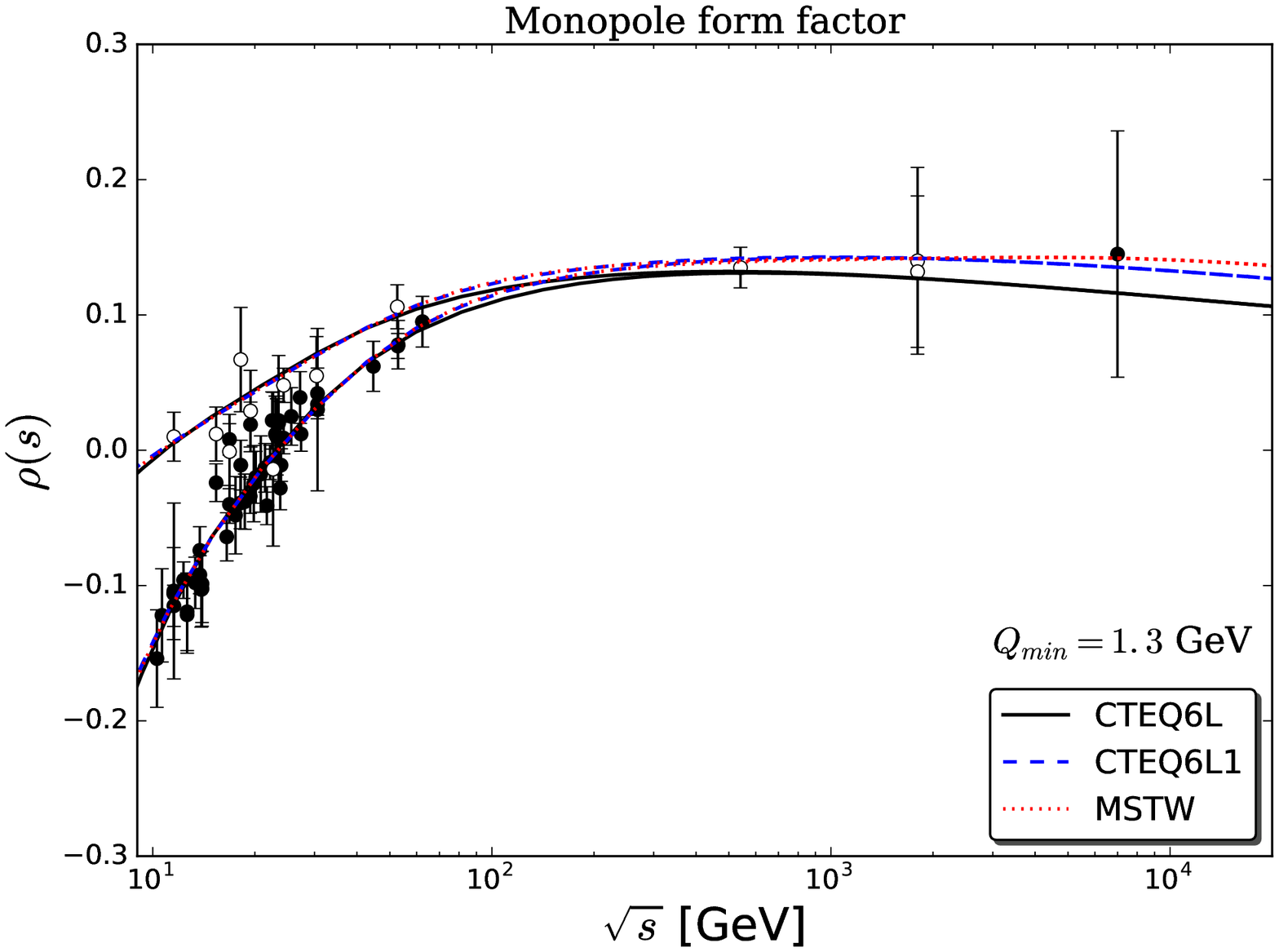}
    \includegraphics[width=8.0cm,height=9.0cm]{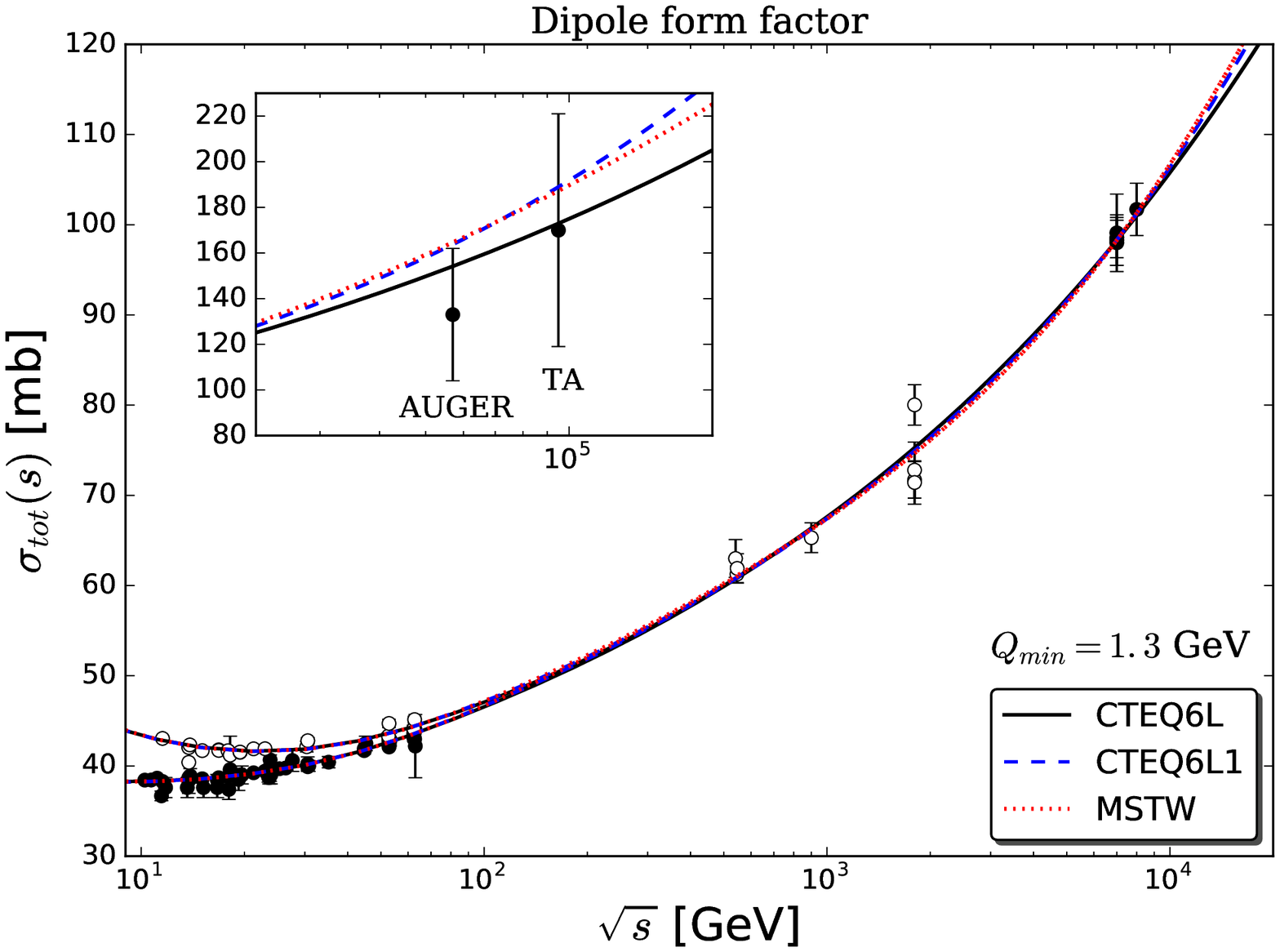}
    \includegraphics[width=8.0cm,height=9.0cm]{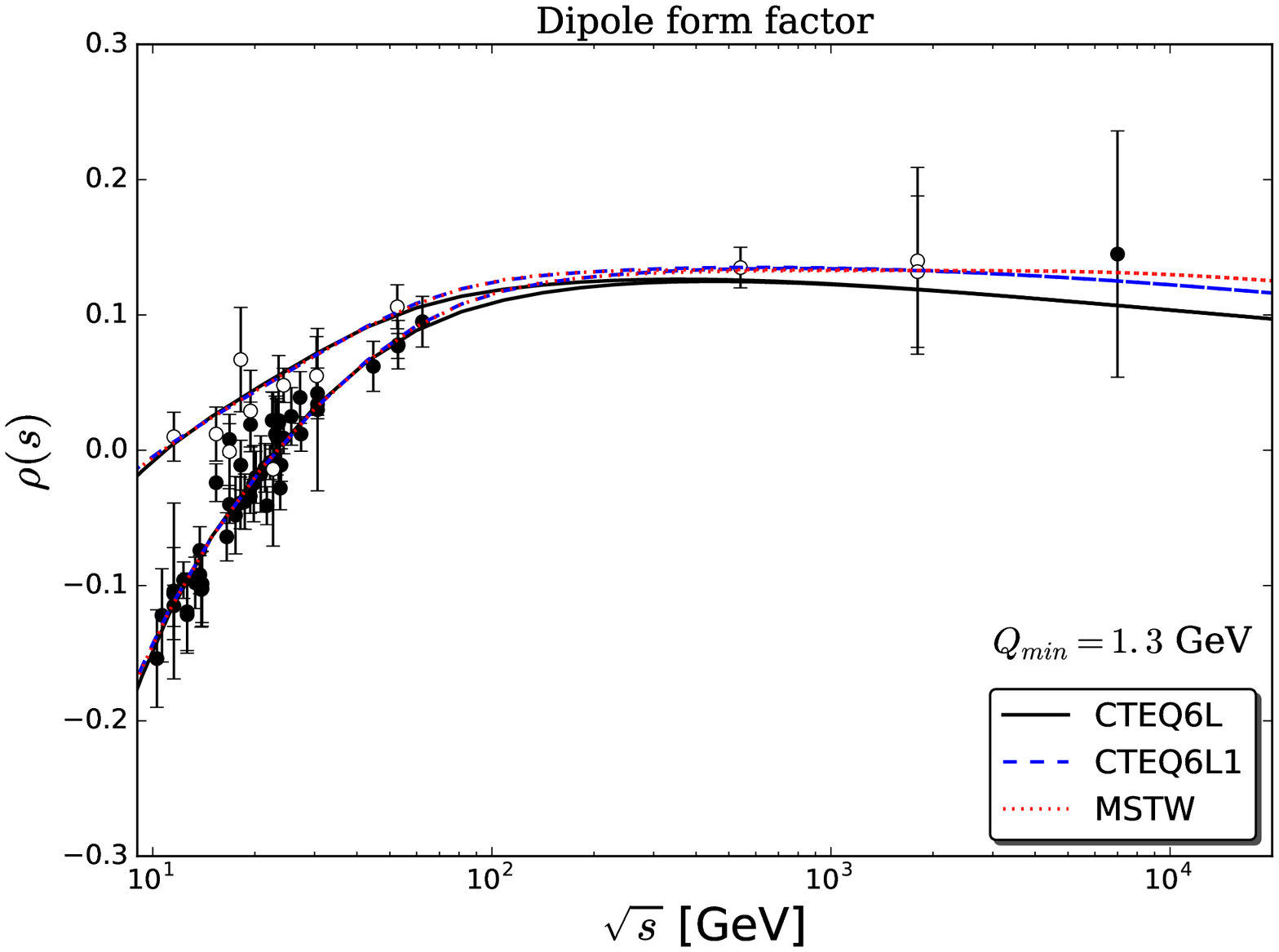}
    \caption{Total cross section and $\rho$-parameter for $pp$ ($\bullet$) and $\bar{p}p$ ($\circ$).}
    \label{chDGMfig2}
  \end{center}
\efg

\bfg[hbtp]
  \begin{center}
    \includegraphics[width=10cm,height=10cm]{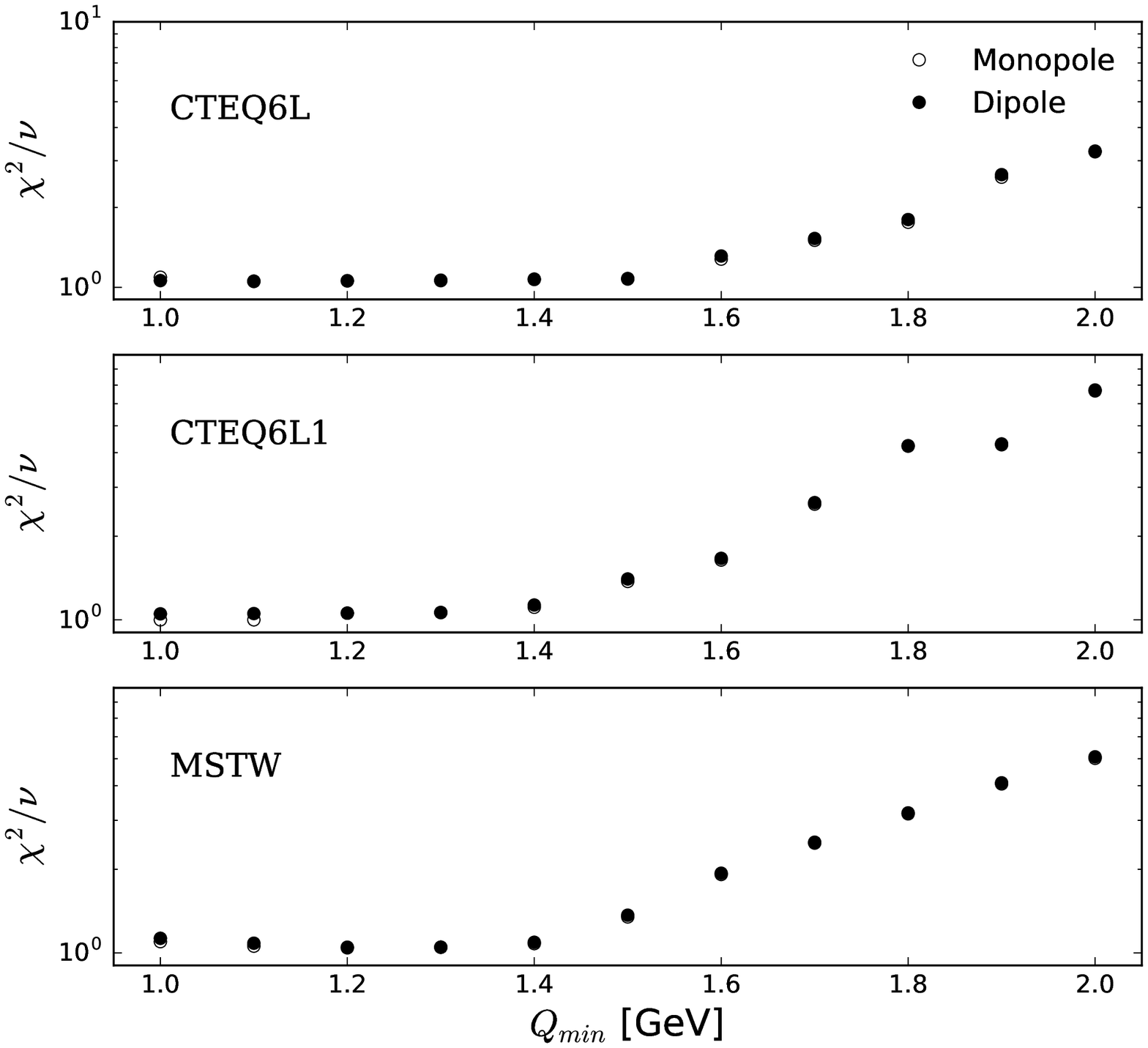}
    \caption{The $\chi^{2}/\nu$ as a function of the cutoff $Q_{min}$.}
    \label{chDGMfig1}
  \end{center}
\efg


\section{\textsc{The 2019 Dynamical Gluon Mass Model}}
\label{secDGM19.1}

\mbox{\,\,\,\,\,\,\,\,\,}
Recently, the TOTEM Collaboration has provided new experimental measurements on $\sigma_{tot}$ and $\rho$ from LHC13, the highest energy reached in accelerators. In a first paper \cite{Antchev:2017dia}, by using as input $\rho = 0.10$, the measurement of the total cross section yielded
\be
 \sigma_{tot}=110.6 \pm 3.4 \, \text{mb}.
\label{datast1}
\ee
In a subsequent work \cite{Antchev:2017yns}, an independent measurement of the total cross section was reported,
\be
\sigma_{tot}=110.3 \pm 3.5 \, \text{mb},
\label{datast2}
\ee
together with the first measurements of the $\rho$ parameter:
\be
\rho = 0.10 \pm 0.01 \,\, \text{and} \,\,\, \rho = 0.09 \pm 0.01. 
\label{datarho}
\ee

Although the values of $\sigma_{tot}$ are in consensus with the increase of previous measurements by TOTEM, the $\rho$ values indicate a rather unexpected decrease, as compared with measurements at lower energies and predictions from the wide majority of phenomenological models. This new information has originated a series of recent papers and  debates on possible phenomenological explanations for the rather small $\rho$-value. The main theoretical ingredients involve either the concept of an Odderon (a crossing odd color-singlet with at least three gluons) \cite{Lukaszuk:1973nt,Ewerz:2003xi,Ewerz:2005rg} or of a Pomeron (a crossing even color-singlet with at least three gluons) \cite{Forshaw:1997dc,Donnachie:2002en}.

The variety of recent phenomenological analyses treats different aspects involved, pointing to some distinct scenarios, and might be grouped in some classes according to their main characteristics:

\begin{itemize}

\item[i.] Maximal Odderon (\tit{e.g.}, Martynov, Nicolescu \cite{Martynov:2017zjz,Martynov:2018nyb,Martynov:2018sga,Martynov:2018pye}) and Odderon effects in elastic hadron scattering (\tit{e.g.}, Cs\"org\"o, Pasechnik, Ster \cite{Csorgo:2018uyp,Csorgo:2019rsr}, Gonçalves, Silva \cite{Goncalves:2018nsp});

\item[ii.] discussions on Odderon effects in other reactions (\tit{e.g.},  Harland-Lang, Khose, Martin, Ryskin
\cite{Harland-Lang:2018ytk}, Gonçalves \cite{Goncalves:2018pbr});

\item[iii.] Pomeron dominance with small Odderon contribution (\tit{e.g.}, Khoze, Martin, Ryskin \cite{Khoze:2018kna}, Gotsman, Levin, Potashnikova \cite{Gotsman:2018buo,Gotsman:2017ncs}, Lebiedowicz, Nachtmann, Szczurek \cite{Lebiedowicz:2018plp}, Bence, Jenkovszky, Szanyi \cite{Bence:2018ain});

\item[iv.] leading Pomeron without Odderon contribution in elastic scattering (\tit{e.g.}, Shabelski, Shuvaev \cite{Shabelski:2018jfq}, Broilo, Luna, Menon \cite{Broilo:2018els,Broilo:2018brv,Broilo:2018qqs}, Durand and Ha \cite{Durand:2018irx}) and in other reactions (\tit{e.g.}, Lebiedowicz, Nachtmann, Szczurek \cite{Lebiedowicz:2018eui});

\item[v.] reanalyzes of the differential cross section data from TOTEM \cite{Antchev:2017yns}, indicating results for $\sigma_{tot}$ and $\rho$ at 13 TeV different from the aforequoted values (\tit{e.g.}, Pacetti, Srivastava, Pancheri \cite{Pancheri:2018yhd}, Kohara, Ferreira, Rangel \cite{Kohara:2018wng}, Cudell, Selyugin \cite{Cudell:2019mbe}).

\end{itemize}

In this rather ``complex'' scenario, we present here a phenomenological description of the forward $pp$ and $\bar{p}p$ elastic scattering data in the region $10$ GeV - $13$ TeV. The formalism has well founded  bases on QCD (perturbative and nonperturbative aspects), ingredients from Regge-Gribov phenomenology, and is characterized by leading Pomeron component, without Odderon contribution. 

As mentioned before, one of the main phenomenological methods presently available and successfully able to explain the observed 
characteristics of the elastic scattering, including the rise of hadron-hadron total cross section, is based upon QCD. At the presently available collider energy region, the behavior of the forward quantities $\sigma_{tot}(s)$ and $\rho(s)$ are expected to be asymptotically dominated by the semihard interactions. The essential concepts and motivations behind this approach can be summarized by taking into account the same eikonal approach as in the DGM$15$ model \cite{Bahia:2015gya,Bahia:2015hha}, but the high-energy eikonal will be properly constructed by considering different dispersion relations. 

In this newest version of the model, hereinafter referred to as DGM$19$, we selected the best result\footnote{Based on the best prediction at $\sqrt{s}=13$ TeV, see Table \ref{chDGMtab2}.} of the previous DGM$15$ model, namely the corresponding fit obtained by means of CTEQ$6$L with a dipole form factor. As a matter of comparison with the previous results, we also study the effects concerning different updated sets of parton distributions in our forward analysis. More specifically, CT$14$ \cite{Hou:2016nqm,Dulat:2015mca} and MMHT \cite{Harland-Lang:2014zoa}.

\subsection{\textsc{Derivative Dispersion Relations and High-Energy Eikonal Construction}}
\label{secDGM19.2}

\mbox{\,\,\,\,\,\,\,\,\,}
Two components are considered in our eikonal representation, one associated with the semihard interactions and calculated from QCD and a second one associated with soft contributions and based on the Regge phenomenology. Except for an odd under crossing Reggeon contribution, necessary to distinguish between $pp$ and $\bar{p}p$ scattering at low energies, all the dominant components at high energies (soft and semihard) are associated with even under crossing contributions, namely we have Pomeron dominance and absence of Odderon.

The fundamental basis of models inspired upon QCD is that the semihard scatterings of partons in hadrons are responsible for the observed increase of the total cross section. Here by assuming assuming a Pomeron dominance, represented by a crossing even contribution, namely we consider that the semihard odd component does not contribute with the scattering process. Hence, as before $\chi^{-}_{_{SH}} = 0$, \tit{i.e.} $\chi_{_{SH}}=\chi^{+}_{_{SH}}$ and  $\chi^{-}=\chi^{-}_{_{soft}}$.

In respect the even contribution, analyticity allows to connect the real and imaginary parts by means of derivative dispersion relation, see Appendix \ref{secAPX3.1.2},
\be
\text{Im}\,\chi^{+}_{_{SH}}(s,b) = -\tan\left[\frac{\pi}{2}\,\frac{d}{d\ln s}\right]
\text{Re}\,\chi^{+}_{_{SH}}(s,b),
\label{eikish}
\ee
so that our main input here is the real part of the crossing even contribution. It follows from the QCD improved parton model that, at leading order, this semihard eikonal can be fatorized \cite{Durand:1987yv,Durand:1988cr,Durand:1988ax} as in expression (\ref{chDGM.10}). 

Taking into account the expansion of the tangent operator only up to the first order, one arrives at the following structure for the imaginary part of the semihard eikonal
\be
\text{Im}\,\chi_{_{SH}}(s,b) = -\frac{\pi}{4}\,s\,\left[W_{_{SH}}(s,b)\,\frac{d\sigma_{QCD}(s)}{ds}+\sigma_{QCD}(s)\,\frac{d W_{_{SH}}(s,b)}{ds}\right],
\label{eikrdd}
\ee
where the overlap density distribution of hard parton scattering, $W_{_{SH}}(s,b)$, will be given by the dipole \tit{Ansatz}, see expression (\ref{chDGM.24}). Hence, one can easily find:
\begin{eqnarray}
\text{Im}\,\chi_{SH}(s,b)&=& \sigma_{QCD}(s)\,\frac{\nu_{2}b^{3}}{384}\left[2\nu^{4}_{SH}K_{3}(\nu_{_{SH}}b)-b\nu^{5}_{SH}K_{2}(\nu_{_{SH}}b)\right] \nonumber\\
&-&\frac{\pi}{4}\,s\,W_{_{SH}}(s,b)\frac{d\sigma_{QCD}(s)}{ds}.
\label{imeiksh}
\end{eqnarray}

However, it is widely known that the energy change of prescription $s\to-is$ is equivalent to an even inverse derivative dispersion relation like (\ref{eikish}) \cite{Martin:1973qm,Henzi:1984aw}, see expression (\ref{apxeq8}). The use of this prescription would simplify a lot the numerical calculus of DGM$19$. However, as far as we are concern the modified Bessel function, $K_{n}(x)$ does not seem to be well-defined for complex arguments. Therefore, the mathematical point of view tells us that perhaps $W_{_{SH}}(s,b)$ shall not be considered in the derivative dispersion relation process. From the physical point of view, and currently it is what we understand and accept, there is no meaning to apply the dispersion relation into the form factor because by definition it is the Fourier transform of the parton density at impact parameter $b$. More precisely, it gives the partons' distribution within hadrons and thus there is no actual reason to consider this contribution, as in the second term in \tit{rhs} of expression (\ref{eikrdd}). Finally, the semihard eikonal would be simply given by
\be
\text{Im}\,\chi_{SH}(s,b)=\frac{1}{2}\,W_{_{SH}}(s,b)\,\sigma_{QCD}(s\to-is)\simeq -\frac{\pi}{4}\,s\,W_{_{SH}}(s,b)\,\frac{d\sigma_{QCD}(s)}{ds},
\label{dgm19eq63}
\ee
where $\sigma_{QCD}(s)$ is the usual QCD cross section given by expression (\ref{chDGM.12}).

\subsection{\textsc{Soft Contribution}}
\label{secDGM19.3}

\mbox{\,\,\,\,\,\,\,\,\,}
The full complex even and odd soft contributions are based on the Regge formalism and are constructed in accordance with Asymptotic Uniqueness (Phragm\'en-Lindel\"off theorems). Assuming also leading even component, they are parametrized by
\be
\chi^{+}_{_{soft}}(s,b) = \frac{1}{2}\,W^{+}_{_{soft}}(b;\mu^{+}_{soft})\left\{A+\frac{B}{\sqrt{s/s_{0}}}e^{i\pi/4}+C\left[\ln\left(\frac{s}{s_{0}}\right)-i\frac{\pi}{2}\right]^{2}\right\},
\label{dgm19eq64}
\ee
where in analogy with the QCD cross section, 
\be
\sigma_{soft}(s)=A+\frac{B}{\sqrt{s/s_{0}}}e^{i\pi/4}+C\ln^{2}\left(\frac{s}{s_{0}}\right),
\label{dgm19eq65} 
\ee
and the odd eikonal,
\be
\chi_{_{soft}}^{-}(s,b) = \frac{1}{2}\, W^{-}_{_{soft}}(b;\mu^{-}_{_{soft}})\,D\, \frac{e^{-i\pi/4}}{\sqrt{s/s_{0}}},
\label{dgm19eq66}
\ee
where $A$, $B$, $C$ and $D$ are free fit parameters and from the corresponding dipole form factors, see expressions (\ref{chDGM.19}) and (\ref{chDGM.20}).

We notice that in the Regge context, the soft even contribution consists of a Regge pole with intercept $1/2$, a critical Pomeron and a triple pole Pomeron, both with intercept $1$. The odd contribution is associated with only a Regge pole, with interpept $1/2$. 

Summarizing the model has $7$ free fit parameters, $5$ associated with the soft contribution, $A, B, C, D, \mu^{+}_{_{soft}}$ and only $2$ with the semihard contribution, $\nu_1$ and  $\nu_2$ (from $\nu_{_{SH}}(s)$ in $W_{_{SH}}(s,b)$). In addition, $4$ parameters are fixed: $m_{g} = 400$ GeV, $m_{q} = 250$ GeV, $\sqrt{s_{0}} = 5$ GeV and $\mu^{-}_{soft} = 0.5$ GeV. 

\section{\textsc{Updated PDF Sets}}
\label{secDGM19.4}

\mbox{\,\,\,\,\,\,\,\,\,}
The partons' distribution are interpreted as the probability density to find quarks and gluons within a hadron at a given momentum fraction $x$ carried by the parent hadron and in a given virtuality $Q^{2}$. There is a strong intrinsically dependence of the partonic distribution extraction, as for example, with: data sets, inclusive and exclusive processes analyses, inclusion of diffractive processes, the transition of the active flavor number in the matching-prescription scheme, the analytical structure in $x$ and $Q^{2}$ of the parametrizations, renormalization scheme, renormalization scale, interpolation and extrapolation process, uncertainty treatment, and much more. Each PDF has its own peculiarities, and for this reason, each one of them is unique. At the CM collider energy presently available, the physical explanation for the good results of some PDF set rely on the behavior of the gluon distribution at low-$x$. 

\subsection{\textsc{The Partonic Distribution CT14}}
\label{secDGM19.5}

\mbox{\,\,\,\,\,\,\,\,\,}
The CT$14$ PDF's, from a global anaysis by the CTEQ-TEA group \cite{Dulat:2015mca}, differs in several aspects from previous PDF's, more specifically CTEQ$6$L, since the former was tunned with data from the LHC experiments and also the new D$0$ charged-lepton rapidity asymmetry data. Moreover, it also uses more flexible parametrization of partonic distribution that may provide, in principle, a better fit by considering different combination sets of quark flavors. 

The functional form of the partonic parametrization are given by their $x$ dependence at low scale $Q_{0}$, is given by
\be
xf_{a}(x,Q_{0})=x^{a_{1}}(1-x)^{a_{2}}P_{a}(x),
\label{dgm19eq67}
\ee
where the behavior of $x^{a_{1}}$ at asymptotically $x\to 0$ is given by Regge theory and $(1-x)^{a_{2}}$ at $x\to 1$ is guided by quark counting rule, and $Q_{0}=1.3$ GeV stands for the initial fixed momentum scale. The $P_{a}(x)$ is written as a polynomial in $\sqrt{x}$. In particular, for the quark combination $u_{V}\equiv u - \bar{u}$, it is written as a linear combination of Bernstein polynomials \cite{Dulat:2015mca},
\be
P_{u_{V}}(y)=d_{0}p_{0}(y)+d_{1}p_{1}(y)+d_{2}p_{2}(y)+d_{3}p_{3}(y)+d_{4}p_{4}(y),
\label{dgm19eq68}
\ee
where $y$ stands for $\sqrt{x}$ and 
\bear
p_{0}(y)&=&(1-y)^{4}\nonumber\\
p_{1}(y)&=&4y(1-y)^{3}\nonumber\\
p_{2}(y)&=&6y^{2}(1-y)^{2}\\
p_{4}(y)&=&4y^{3}(1-y)\nonumber\\
p_{4}(y)&=&y^{4}.\nonumber
\label{dgm19eq69}
\eear

The same parametrizations is used for the quark distribution $d_{V}\equiv d - \bar{d}$ with the same parameter values $a_{1}$ and $a_{2}$, but with independent parameters for the coefficients of the Bernstein polynomial. Notice that the parametrization for both quark distribution shall be constrained by the counting rule normalization, see expression (\ref{chDGM.49}) and (\ref{chDGM.50}). As for the case of the gluon, it is used a similar parametrization, but with a combination of lower order polynomials, since the experimental data provide fewer constraints on the gluon distribution. Thus, the parametrization reads
\be
P_{g}(y)=g_{0}[e_{0}q_{0}(y)+e_{1}q_{1}(y)]+q_{2}(y),
\label{dgm19eq70}
\ee
where now $y=2\sqrt{x}-x$ and
\bear
q_{0}(y)&=&(1-y)^{2}\nonumber\\
q_{1}(y)&=&2y(1-y)\\
q_{0}(y)&=&y^{2}.\nonumber
\label{dgm19eq71}
\eear

The sea quark distributions $d$ and $u$, respectively, were parametrized by means of fourth-order polynomials in $y$ with the same mapping $y=2\sqrt{x}-x$ that was used for the gluon. Moreover, fewer experimental constraints apply to the strangeness asymmetry, so it was assumed $s(x)=\bar{s}(x)$. However in the light of upcoming data from the LHC, it is expected to include $s(x)\neq\bar{s}(x)$ in a forthcoming round of fits.

In the CT$14$ distribution, and more specifically by considering the CT$14$LN table file, the QCD coupling constant is written in its NLO form, see expression (\ref{chDGM.41}). As in the previous CTEQ$6$L case, the next-to-leading order coupling = is determined by the $M_{Z}$ scale where $\alpha^{NLO}_{s}(M^{2}_{Z})=0.118$ with $\Lambda_{4}=326$ MeV and $\Lambda_{5}=226$ MeV.

\subsection{\textsc{The Partonic Distribution MMHT}}
\label{secDGM19.5}

\mbox{\,\,\,\,\,\,\,\,\,}
The corresponding MMHT$(2014)$ PDF set supersede the previous MSTW$(2008)$ parton sets. However, as stressed by the authors \cite{Harland-Lang:2014zoa}, they are obtained within basically the same framework. A wide variety of new data sets was included in this newest ``MRS family'' set, \tit{e.g.} from the LHC, updated Tevatron data and the HERA combined H$1$ and ZEUS data on the total and charm structure functions. The major theoretical changes are the $u-d$ valence quark difference at low-$x$ due to an improved functional form for the partonic parametrization. 

The function form for most of the partonic parametrizations in MMHT takes a form based on Chebyshev polynomials,
\be
xf(x,Q^{2}_{0})=A(1-x)^{\eta}x^{\delta}\left(1+\sum^{n}_{i=1}a_{i}T^{Ch}_{i}(y(x))\right),
\label{dgm19eq72}
\ee
where $Q_{0}=1$ GeV is the input scale and the $T^{Ch}(y)$ stands for the Chebyshev polynomials with $y=1-2x^{k}$, where it was taken $k=0.5$ and $n=4$. The values of the corresponding set of parameters $A$, $\delta$, $\eta$ and $a_{i}$ for each partonic distribution, namely $f=u_{V}$, $d_{V}$, $S$ and $s_{+}$, is determined by a global fit. The light-quark sea contribution is $S\equiv 2(\bar{u}+\bar{d})+s+\bar{s}$ where for $s_{+}=s+\bar{s}$ it is set $\delta_{+}=\delta_{S}$.

The parametrizations for $\Delta=\bar{d}-\bar{u}$, $s_{-}=s-\bar{s}$ and for the gluon distribution are respectively given by
\be
x\Delta(x,Q^{2}_{0})=A_{\Delta}(1-x)^{\eta_{\Delta}}x^{\delta_{\Delta}}(1+\gamma_{\Delta}x+\epsilon_{\Delta}x^{2}),
\label{dgm19eq73}
\ee
\be
xs_{-}(x,Q^{2}_{0})=A_{-}(1-x)^{\eta_{-}}x^{\delta_{-}}(1-x/x_{0}),
\label{dgm19eq74}
\ee
\be
xg(x,Q^{2}_{0})=A_{g}(1-x)^{\eta_{g}}x^{\delta_{g}}\left(1+\sum^{2}_{i=1}a_{g,i}T^{Ch}_{i}(y(x))\right)+A_{g^{\prime}}(1-x)^{\eta_{g^{\prime}}}x^{\delta_{g^{\prime}}},
\label{dgm19eq75}
\ee
where $\eta_{\Delta}=\eta_{S}+2$. 

The input PDF's are subjected to the same three constraints from the counting rule, see expressions (\ref{chDGM.49}-\ref{chDGM.51}). The strong coupling is defined at the scale of the $Z$ boson mass, $M_{Z}$, \tit{i.e} $\alpha(M^{2}_{Z})$, which is considered as a free parameter when determining the best fit.

One of the basic features in QIM models are related to the fact that the high-energy dependence of the total cross sections is guided mainly by process involving at least one gluon in the initial state. Therefore, the gluon contribution is responsible for the dominant behavior at the asymptotically low-$x$ regime, as it can be seen in Figure \ref{dgm19fig4}, where it is shown the effects of the gluon distribution for three PDF's, namely CTEQ$6$L, CT$14$ and MMHT at two distinguished scales $Q^{2}=10^{2}$ and $10^{4}$ GeV$^{2}$. 
\bfg[hbtp]
\begin{center}
 \includegraphics[width=8.0cm,height=8.0cm]{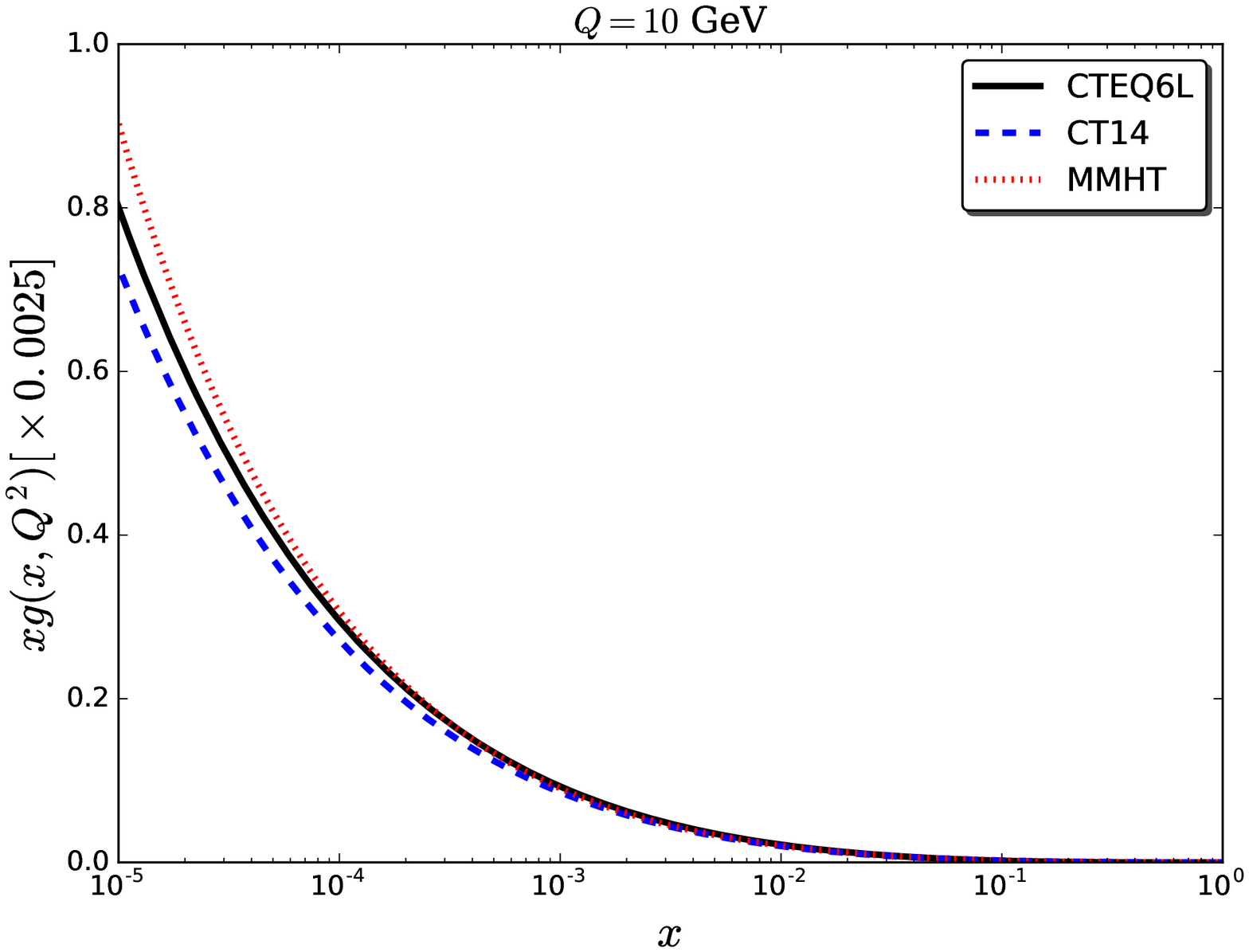}
 \includegraphics[width=8.0cm,height=8.0cm]{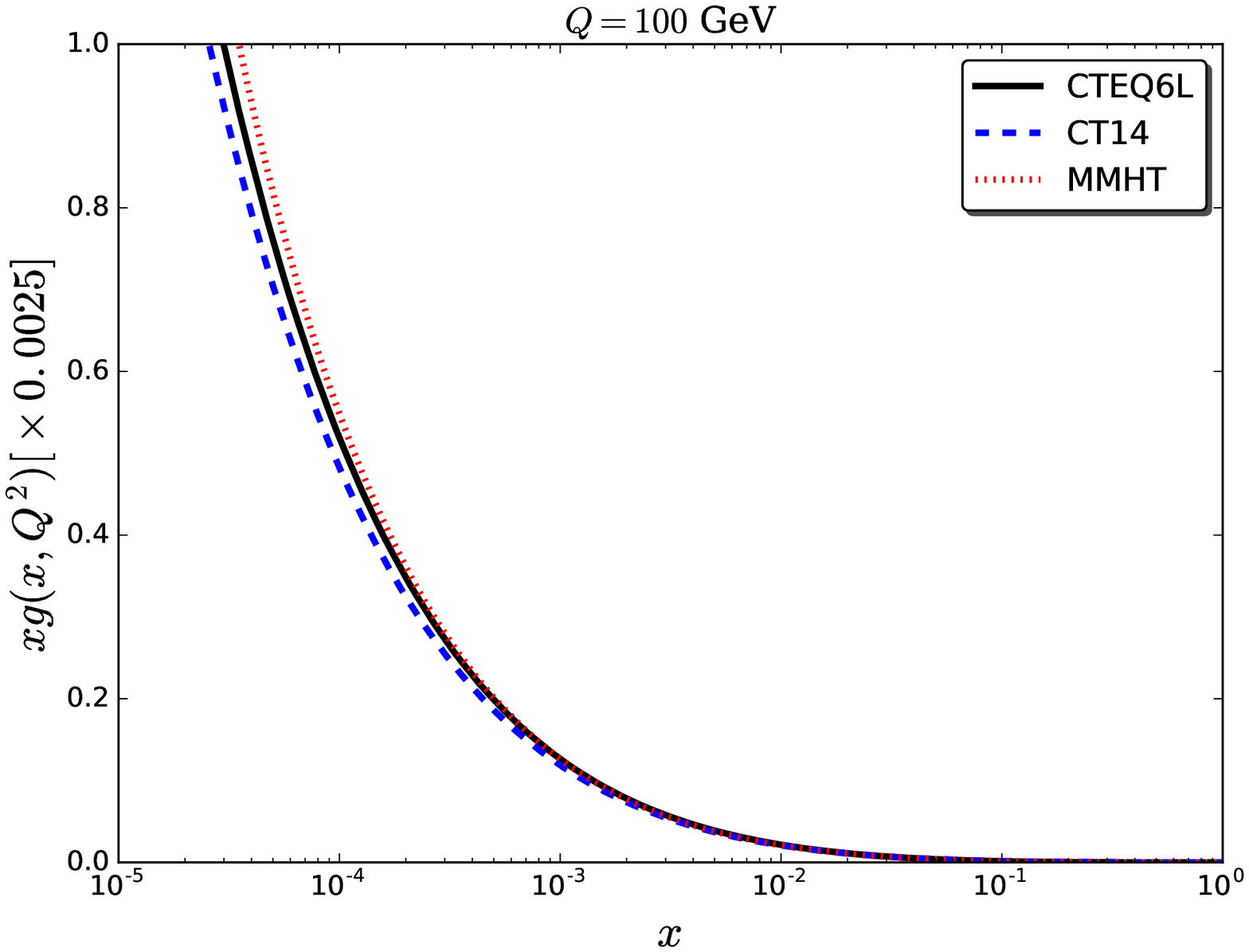}
 \caption{The contribution of the gluon distribution.}
\label{dgm19fig4}
\end{center}
\efg

\section{\textsc{Results of DGM19}}
\label{secDGM19.5}

\mbox{\,\,\,\,\,\,\,\,\,}
In order to determine the model parameters, we fix $n_{f}=4$, as before, and set the gluon and quark mass scales to $m_{g}=400$ MeV and $m_{q}=250$ MeV, as previously stressed. Our data set is compiled from a wealth of collider data on $pp$ and $\bar{p}p$ elastic scattering, available in the Particle Data Group \cite{Tanabashi:2018oca} as well as in the very recent papers of LHC Collaborations such as TOTEM \cite{Antchev:2016vpy,Antchev:2017yns,Antchev:2017dia,Antchev:2018rec} and ATLAS \cite{Aaboud:2016ijx,Aad:2014dca}, which span a large CM energy range, namely $10$ GeV $\leqslant \sqrt{s} \leqslant 13 $ TeV. For the sake of clarity and completeness see Table \ref{ch2tab1} for all the recent LHC data on $\sigma_{tot}$ ans $\rho$, still absent in the PDG$2018$ review.

In our analysis we have considered one pre-LHC PDF set, namely CTEQ$6$L, as a matter of comparison with the recent proposed updated partonic distributions. With that in mind we have investigated the effects of fine-tunned PDF's, namely CT$14$ and MMHT sets. In the specific case of CT$14$ (CT$14$LN tables), it is used the NLO formula for the $\alpha^{NLO}_{s}(M^{2}_{Z})=0.118$ consistent with the value $\Lambda^{(4 flavor)}_{CT14}=326$ MeV, and $\alpha^{NLO}_{s}(Q^{2})$ is given by expression (\ref{chDGM.53}). The MMHT set, as its previous distribution, uses an alternative definition for $\alpha_{s}$ where is considered as a free parameter. Respectively $\alpha_{s}(M^{2}_{Z})=0.135$ corresponds to $\Lambda^{(5flavor)_{MMHT}}=0.211$ GeV. Taking into account the matching-prescription scheme \cite{Owens:1992hd}, see Figure \ref{dgm19fig5}, the behavior of $\alpha_{s_{MMHT}}(Q^{2})$ can be properly reproduced from the choice $\Lambda^{(4flavor)}_{MMHT}\sim 268$ MeV.
\bfg[hbtp]
  \begin{center}
    \includegraphics[width=10cm,height=8cm]{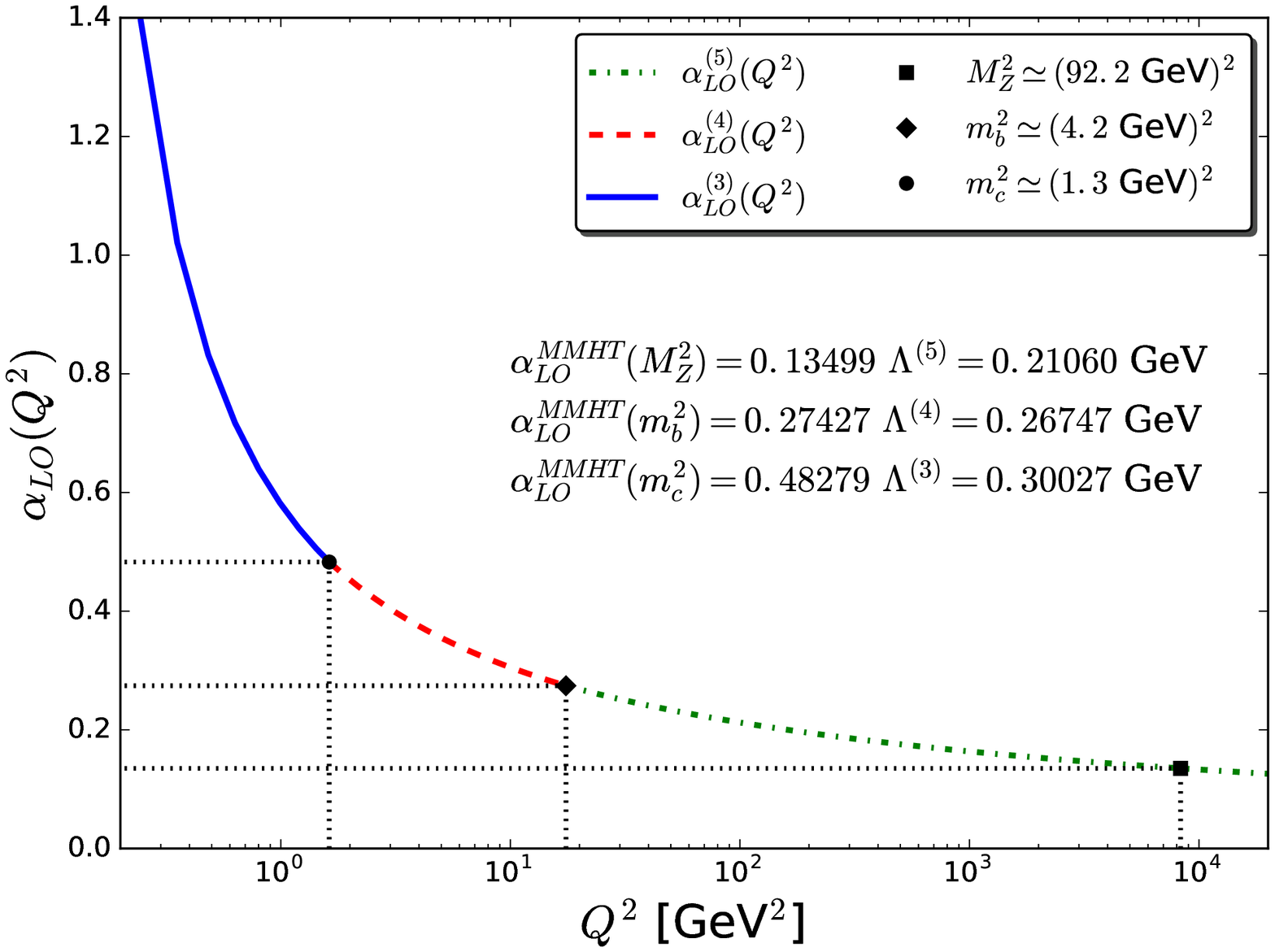}
    \caption{The matching-prescription scheme for the MMHT set.}
    \label{dgm19fig5}
  \end{center}
\efg

\subsection{\textsc{Parametrization of $\sigma_{QCD}(s)$}}
\label{secDGM19.6}

\mbox{\,\,\,\,\,\,\,\,\,}
Regarding the analytical features of $\chi_{SH}(s,b)$, we consider, as discussed above, only the steep energy rise of $\sigma_{QCD}(s)$, thus neglecting the mild energy change in $W_{SH}(b,s)$. That being said, we parametrize the QCD cross section, calculated from expression (\ref{chDGM.12}), using three distinct PDFs: CTEQ6L \cite{Pumplin:2002vw}, CT14 \cite{Dulat:2015mca} and MMHT \cite{Harland-Lang:2014zoa}. To be specific, we generate around $30$ points of $\sigma_{QCD}(S)$ for each one of these parton distributions and fit the data with less than $1\%$ error using the following analytical parametrization:
\be
\sigma_{QCD}(s)=b_{1}+b_{2}\,e^{b_{3}(X^{1.01 b_{4}})}+b_{5}\,e^{b_{6}(X^{1.05 b_{7}})}+b_{8}\,e^{b_{9}(X^{1.09 b_{10}})},
\label{dgm19eq76}
\ee
where $X=\ln\ln(-is)$ and $b_{1},...,b_{10}$ are free fit parameters, extracted for each PDF tested. In Table \ref{dgm19tab1} we display the best-fit parameters $b_{i}$ for CTEQ$6$L, CT$14$ and MMHT and in Figure \ref{dgm19fig6} we show the fits obtained for $\sigma_{QCD}(s)$, used as input to compute $\chi_{_{SH}}(s,b)$. 

\bfg[hbtp]
\begin{center}
 \includegraphics[width=10.0cm,height=8.0cm]{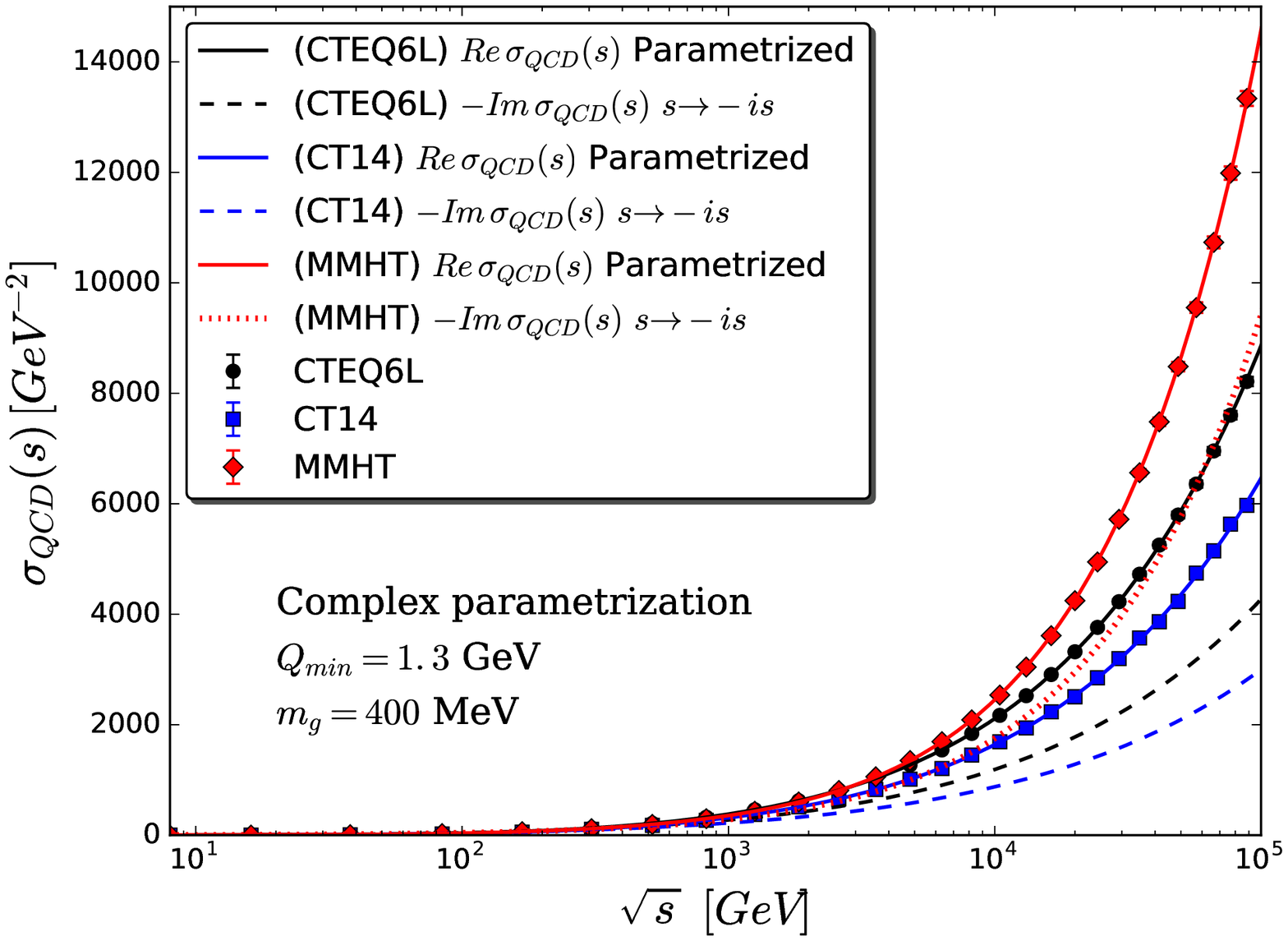}
 \caption{Real and imaginary parts of $\sigma_{QCD}(s)$ after applying the prescription $s\rightarrow -is$.}
\label{dgm19fig6}
\end{center}
\efg

\begin{table}[hbtp]
\centering
\scalebox{0.9}{
\begin{tabular}{c@{\quad}c@{\quad}c@{\quad}c@{\quad}}
\hline \hline
& & &  \\[-0.3cm]
PDF:    & CTEQ6L & CT14 & MMHT  \\[0.05ex]
\hline
& & & \\[-0.3cm]
$b_{1}$ [GeV$^{-2}$]  & 97.005                    & 100.220                  & 95.284                     \\[0.15cm]
$b_{2}$ [GeV$^{-2}$]  & 0.280 $\times$ 10$^{-1}$  & 0.434 $\times$ 10$^{-1}$ & 0.372                      \\[0.15cm]
$b_{3}$               & 1.699                     & 1.274                    & 0.600                      \\[0.15cm]
$b_{4}$               & 1.736                     & 1.919                    & 2.496                      \\[0.15cm]
$b_{5}$ [GeV$^{-2}$]  & -0.149 $\times$ 10$^{-5}$ & 0.122 $\times$ 10$^{-7}$ & -0.255 $\times$ 10$^{-5}$  \\[0.15cm]
$b_{6}$               & 14.140                    & 14.050                   & 14.281                     \\[0.15cm]
$b_{7}$               & 0.319                     & 0.504                    & 0.281                      \\[0.15cm]
$b_{8}$ [GeV$^{-2}$]  & 0.836 $\times$ 10$^{-1}$  & 3.699 $\times$ 10$^{3}$  & 0.909                      \\[0.15cm]
$b_{9}$               & 3.813                     & -80.280                  & 4.290                      \\[0.15cm]
$b_{10}$              & 0.810                     & -2.632                   & 0.673                      \\[0.15cm]
& & &  \\[-0.5cm]
\hline \hline 
\end{tabular}
}
\caption{Only central values are displayed, as our purpose is to get a very accurate description of $\textnormal{Re}\,\sigma_{QCD}(s)$, as we seek an analytical expression to interpolate/extrapole its energy dependence as well as performing the complex even prescription $s\rightarrow -is$ in order to extract $\textnormal{Im}\,\sigma_{QCD}(s)$.}
\label{dgm19tab1}
\end{table} 

\subsection{\textsc{Fit Procedures}}
\label{secDGM19.7}

\mbox{\,\,\,\,\,\,\,\,\,}
The basic idea that $\nu^{-1}$ $(\mu^{-1})$, which is the inverse of the parameters appearing in the form factors, characterizes the range that the interaction can occur resembles the work done by Durand \& Pi \cite{Durand:1988ax}. Moreover, they claim that since the odd eikonal is more sensitive to long-range Reggeon exchanges, then it would be expected that $\nu^{-1}_{-}$ turns out to be bigger than $\nu^{-1}_{+}$, \tit{i.e.} implying the constraint $\nu_{-}<\nu_{+}$. Indeed, in their first analyses \cite{Durand:1988ax}, they obtained $\nu^{2}_{-}=0.35$ GeV$^{2}$ $(\nu_{-}\simeq 0.592$ GeV$)$ and $\nu^{2}_{+}=0.77$ GeV$^{2}$ $(\nu_{+}\simeq 0.877$ GeV$)$. Therefore, it gives the ratio $\nu_{+}/\nu_{-}\simeq 1.48$. 

More or less at the same time, Margolis \tit{et al.} \cite{Margolis:1988ws} arrived at $\mu_{-}=0.58$ GeV and $\mu_{+}=0.89$ GeV, hence $\mu_{+}/\mu_{-}\simeq 1.53$. Very close to the previously ratio found by Durand \& Pi. One decade later, Block \tit{et al.} \cite{Block:1998hu} obtained $\mu_{-}=0.53$ GeV and $\mu_{qq}=0.89$ GeV, and a ratio of $\mu_{qq}/\nu_{-}\simeq 1.68$.

More recently, the first paper in which it was studied the influence of a dynamical mass in forward scattering, Luna \tit{et al.} \cite{Luna:2005nz}, it was obtained $\mu_{-}=0.58$ GeV and $\mu_{qq}=1.07$, corresponding to the ratio $\mu_{qq}/\nu_{-}\simeq 1.76$ GeV. One of the results obtained by Bahia \tit{et al.} in the DGM$15$ \cite{Bahia:2017kpi} was $\mu^{-}_{soft}=0.5$ GeV (fixed parameter) and $\mu^{+}_{soft}=0.78$ GeV by considering the CTEQ$6$L set and a monopole form factor, which gives the ratio $\mu^{+}_{soft}/\mu^{-}_{soft}\simeq 1.56$, and as for the case of a dipole $\mu^{+}_{soft}=0.82$ GeV and a ratio of $\mu^{+}_{soft}/\mu^{-}_{soft}\simeq 1.64$. The results arrived within CTEQ$6$L correspond to the closest predictions to LHC$13$. It is curious that the results obtained by considering either a monopole and a dipole form factors for the CTEQ$6$L$1$ violate the condition $\mu^{-}_{soft}<\mu^{+}_{soft}$ ($\nu_{-}<\nu_{+}$). The MSTW set violates this condition for the case of a monopole and gives a very low-ratio for the dipole, $\mu^{+}_{soft}/\mu^{-}_{soft}\simeq 1.01$.

Notice that all the ratios found, with only one exception, remains in the interval $1.4 \leq \mu^{+}_{soft}/\mu^{-}_{soft} \leq 1.8$. Since $\mu^{-}_{soft}=0.5$ GeV is fixed, we proposed to constraint $0.7 \leq \mu^{+}_{soft} \leq 0.9$. The physical argument is that, we can connect $\mu^{+}_{soft}$ with even Reggeon exchanges, as for example mesons $a_{2}(1700)$ and $f_{2}(1950)$, and $\mu^{-}_{soft}$ with odd Reggeon exchanges, for example $\rho(770)$ and $\omega(1650)$. Moreover, since the odd (even) soft eikonal now can be related to $C$-odd (even) exchanges, it gives us a physical justification for the constraint once imposed $\nu_{-}<\nu_{+}$ \cite{Durand:1988ax}, \tit{i.e.} the higher the mass of the exchanged Reggeon the shorter is its interaction range.


In the absence of \textit{ab initio} theoretical QCD arguments to determine the parameters $A, B, C, D, \mu^{+}_{_{soft}}$, $\nu_1$ and $\nu_2$, we resort to a fine-tunning fit procedure. As we are interested in the very high-energy behavior of the total cross section and of the real to imaginary ratio parameter, we shall use only $pp$ and $\bar{p}p$ elastic scattering data. Moreover, as a first approach, we only analyze forward observables, in order to test the DGM$19$ model in the $t=0$ limit, or in other words, to make an unitarized forward amplitude analysis of the elastic amplitude, $F(s,t)$ using the impact parameter representation.

In effect, we do not apply to these data set, composed of $174$ data points of $\sigma^{\bar{p}p,pp}_{tot}$ and $\rho^{\bar{p}p,pp}$, any sort of selection or sieving procedure, which might introduce bias in the analysis. In addition, as a nonlinear model, numerical data reduction is called for. Our minimizing procedure follows the standard $\chi^{2}/\nu$ hypothesis fitting test, providing statistical information on fit quality \cite{bev}. As a further test of goodness-of-fit we shall adopt the integrated probability \cite{bev}, which states that the most reliable fit is associated with the biggest $P(\chi^{2})$. Despite the limitation of treating statistical and systematical uncertainties at the same foot, we apply the $\chi^{2}/\nu$ tests to our data set with uncertainties in $\sigma^{\bar{p}p,pp}_{tot}$ and $\rho^{\bar{p}p,pp}$ summed in quadrature. In all the fits performed, it was assumed an interval $\chi^{2}-\chi^{2}_{min}$ corresponding to the projection $\chi^{2}$ hypersurface containing $68\%$, equivalent to one standard deviation. In this DGM19 analysis ($7$ free parameters) this corresponds to the interval $\chi^{2}-\chi^{2}_{min}=8.18$. As for the test-scenarios with $6$ free parameters, $\chi^{2}-\chi^{2}_{min}=7.04$.

Our fits are performed using the standard MINUIT package \cite{James:2004xla,minuit}, through the MIGRAD algorithm. While the number of calls of the MIGRAD routine may vary in the fits with PDF's CETQ$6$L, CT$14$ and MMHT, full convergence of the algorithm was always achieved. Furthermore, in all fits performed we set the low-energy cutoff, $\sqrt{s_{min}}=10$ GeV. To test the predictive power of the model we set three possible high-energy cutoffs, namely: $\sqrt{s_{max}}=7$, $8$ and $13$ TeV. Such method aims at testing possible influence of high-energy data such as those recently released by TOTEM in getting accurate description of data at and beyond LHC13.

The results for the free fit parameters, using each one of the three distinct PDF's: CTEQ$6$L \cite{Pumplin:2002vw} (pre-LHC), CT$14$ \cite{Dulat:2015mca} and MMHT \cite{Harland-Lang:2014zoa} (fine-tunned with LHC data), and for each high-energy cutoff in the data set ($\sqrt{s_{max}}=7$, $8$ and $13$ TeV), as previously discussed, are displayed in Table \ref{chDGMtab3}, together with the statistical information on the corresponding data reductions (reduced $\chi^{2}$ and corresponding integrated probability). The curves of $\sigma_{tot}$ and $\rho$ for both $pp$ and $\bar{p}p$ channels, compared with the experimental data, are depicted in Figure \ref{dgm19fig7}, respectively. The results were obtained considering a dipole form factor and using the cutoff $Q_{min}=1.3$ GeV, which corresponds to the fixed initial scale $Q_{0}$ in CTEQ$6$ and CT$14$ sets. The same discussion made for MSTW remains valid for the MMHT set.

\subsection{\textsc{Effects of the Leading Contribution in $\chi^{+}_{soft}(s,b)$}}
\label{secDGM19.8}

\mbox{\,\,\,\,\,\,\,\,\,}
As discussed throughout the text, semihard parton interactions rule the rising dynamics of the total cross section, $\sigma_{tot}$ and, as a by-product, the decrease of $\rho$. Nonetheless, our model accounts for a nonleading (Regge-inspired) soft Pomeron contribution at high energies. Specifically, we have tested the possibility that, in the absence of high-energy rising terms in $\chi^{+}_{soft}(s,b)$, namely by imposing the constraint $C=0$ in expression (\ref{dgm19eq64}), only semihard parton scatterings enables to describe the energy dependence of both, $\sigma_{tot}$ and $\rho$, from $10$ GeV onwards. 

The fit results shown in Figure \ref{dgm19fig8} and Table \ref{chDGMtab4} reveal just the opposite, that a soft high-energy term is needed.  Moreover, from both visual and statistical grounds, one sees that the fit quality is compromised, possibly indicating that logarithm-terms might be important at intermediate energies and in defining the right cross section rising curvature, which eventually yields the appropriate decrease of $\rho$ at LHC13.

\subsection{\textsc{Effects of Static Semihard Form Factor}}
\label{secDGM19.9}

\mbox{\,\,\,\,\,\,\,\,\,}
An energy dependent form factor, although not being formally established in the context of QCD, it is truly supported by the wealth of accelerator data available and seems to us more realistic than taking a static partonic configuration in $b$-space. In addition, many other phenomenological models in which the energy dependence in form factors play a crucial role in $pp$ and $\bar{p}p$ elastic scattering dynamics and, therefore, in accurate descriptions of the data beyond $\sqrt{s}\sim $10 GeV have been proposed in literature, see \cite{Fagundes:2013aja,Fagundes:2015vba} for recent studies.  

However, in order to test the possibly broadening of the gluons spatial distribution at the limit of high energies, we fixed $\nu_{2}=0$, which is the parameter responsible for the energy dependence of the semihard form factor. The fit results are depicted in Figure \ref{dgm19fig8} and the corresponding fitted parameters are shown in Table \ref{chDGMtab4}. The prediction of the total cross section at the LHC typical energy region, more specifically at $\sqrt{s}=13$ TeV, shows that a static semihard form factor is not enough to build the semihard eikonal.  

\section{\textsc{Conclusions on DGM19}}
\label{secDGM19.10}

\mbox{\,\,\,\,\,\,\,\,\,}
Some of the results obtained in the early DGM$15$ model, as for example the cutoff scale $Q_{min}=1.3$ GeV and the energy-dependent dipole form factor, served as initial inputs to investigate the asymptotic behavior of forward scattering amplitudes at high energies in the DGM$19$ model. In our analysis we have included the most recent LHC data from the TOTEM Collaboration at $\sqrt{s}=13$ TeV. Presently, we understand that the correct forward data set must include all the data available \cite{Broilo:2018els,Broilo:2018qqs}. For this reason, we also included the experimental $pp$ total cross section data points obtained by ATLAS Collaboration, respectively at $\sqrt{s}=7$ and $8$ TeV, see Table \ref{ch2tab1}. 

We selected the best result of DGM$15$, namely CTEQ$6$L (dipole), and compared with fine-tunned set of leading order parton distribution, more specifically CT$14$ and MMHT sets. In order to connect the real and imaginary parts of eikonal, we have used the complex prescription $s\to-is$ which is equivalent to an even inverse derivative dispersion relation. 

First, let us focus on the case with the complete data set, namely $\sqrt{s_{max}}=13$ TeV. From Figure \ref{dgm19fig7}, the results are in plenty agreement with all the $\sigma_{tot}$ data, independently of the PDF employed. For $\rho$ the results with CTEQ$6$L and CT$14$ also describe quite well the TOTEM data at $13$ TeV (and data at lower energies), but that is not the case with MMHT. Indeed, from Table \ref{chDGMtab3}, in this case the integrated probability is the smallest one among the three PDF's. Notice that the result with CT$14$ (fine-tunned with LHC data) gives exactly $\rho = 0.1$ at $13$ TeV. Despite a barely underestimation of the $\rho$ datum from $\bar{p}p$ at $546$ GeV, we conclude that our QCD-based model with CTEQ$6$L and CT$14$ provides a consistent description of the forward data in the interval $10$ GeV - $13$ TeV, especially a simultaneous agreement with the $\sigma_{tot}$ and $\rho$ data at $13$ TeV.

Second, and most importantly, this consistent scenario does not change if we exclude from the data set the experimental information at $13$ TeV ($\sqrt{s_{max}}=8$ TeV) and even also the data at $8$ TeV ($\sqrt{s_{max}}=7$ TeV), as shown in Figure \ref{dgm19fig7}. From Table \ref{chDGMtab3}, the integrated probability with $\sqrt{s_{max}}=7$ TeV is the highest one among the three cutoffs and the corresponding predictions at higher energies indicate the decreasing in $\rho(s)$.

These results show the powerful predictive character of the DGM$19$ model, since the $\sigma_{tot}$ and $\rho$ data at $13$ TeV are simultaneously described in all cases, even with $\sqrt{s_{max}}=7$ TeV (PDF's CT14 and CTEQ6L) and without Odderon contribution. It is most noteworthy to notice that CTEQ$6$L is pre-LHC, and for this reason does not include the most recent bulk of experimental information. However, it is curious that the low-$x$ parton distribution effects was already present in this older set.

To understand the importance and that, indeed, the high-energy dependence of the cross sections is driven mainly by semihard gluon interactions, the $\sigma_{QCD}(s)$ is depicted and compared with $\sigma_{soft}(s)$ in Figure \ref{dgm19fig9}, where the latter was calculated using the fitted parameters showed in Table \ref{chDGMtab3} with $\sqrt{s_{max}}=13$ TeV.

In addition, looking for some insights into the formalism, it may be important to notice the effects of two phenomenological inputs, one related to the soft even eikonal and the other to the semi-hard form factor. In the first case, $\chi_{soft}^{+}(s,b)$ as given by expression (\ref{dgm19eq64}), has a component which increases with the energy, namely the term with coefficient $C$. In the second case, the dipole form factor  $G^{(d)}_{_{SH}}(s,k_{\perp};\nu_{_{SH}})$, expression (\ref{chDGM.22}), also depends on the energy through the logarithmic. The effect of these terms can be investigated by assuming $C=0$ or $\nu_2=0$ and re-fitting the data set. 

Summarizing, our results show that the very good description of $\sigma^{pp,\bar{pp}}$ and $\rho^{pp,\bar{pp}}$ are strongly related to the energy dynamics present in the semihard form factor and also that the Pomeron dominance present in the even soft eikonal is of extremely importance in the mid and high-energy region. Notwithstanding the MMHT set is an updated PDF tunned with new experimental data, the reason why we obtained poorly results with this set is still obscure to us. However, notice that we obtained excellent good statistical results with two PDF's, even though they originated from the same family set, namely CTEQ$6$L and CT$14$. Both of them, indeed, gives an accurate prediction to $\rho$-parameter at $\sqrt{s}=13$ TeV. 

\begin{table}[hbtp]
\centering
\scalebox{0.9}{
\begin{tabular}{c|c@{\quad}c@{\quad}c@{\quad}c@{\quad}}
\hline \hline
& & & & \\[-0.3cm]
Energy & PDF:    & CTEQ6L & CT14 & MMHT \\[-0.2cm]
& & & & \\[-0.2cm]
\hline\hline
& & & & \\[-0.2cm]
\multirow{12}{*}{\begin{turn}{90}$\sqrt{s_{max}}=7$ TeV\end{turn}}  
  & $\mu^{+}_{soft}$ [GeV]& 0.90\,$\pm$\,0.20  & 0.90\,$\pm$\,0.16  & 0.90\,$\pm$\,0.20    \\[1.2ex]
  & $A$ [GeV$^{-2}$]      & 124.8\,$\pm$\,2.4  & 123.6\,$\pm$\,2.3  & 123.3\,$\pm$\,2.3    \\[1.2ex]
  & $B$ [GeV$^{-2}$]      & 38.6\,$\pm$\,8.2   & 42.2\,$\pm$\,7.9   & 42.1\,$\pm$\,8.0     \\[1.2ex]
  & $C$ [GeV$^{-2}$]      & 0.62\,$\pm$\,0.14  & 0.73\,$\pm$\,0.14  & 0.60\,$\pm$\,0.16    \\[1.2ex]
  & $\mu^{-}_{soft}$ [GeV]& {\bf 0.5 [fixed]}  & {\bf 0.5 [fixed]}  & {\bf 0.5 [fixed]}    \\[1.2ex]
  & $D$ [GeV$^{-2}$]      & 24.2\,$\pm$\,1.4   & 24.2\,$\pm$\,1.4   & 24.2\,$\pm$\,1.4     \\[1.2ex]
  & $\nu_{1}$ [GeV]       & 2.34\,$\pm$\,0.52  & 2.38\,$\pm$\,0.60  & 2.05\,$\pm$\,0.52    \\[1.2ex]
  & $\nu_{2}$ [GeV]       &0.052\,$\pm$\,0.036 & 0.058\,$\pm$\,0.041& 0.029\,$\pm$\,0.036  \\[1.2ex]

  & $\chi^2/156$          & 1.125               & 1.114               & 1.103              \\[1.2ex]
  & $P(\chi^2)$ & 1.4 $\times$ 10$^{-1}$ & 1.6 $\times$ 10$^{-1}$ & 1.8 $\times$ 10$^{-1}$ \\[1.2ex]
\hline \hline 
& & & & \\[-0.2cm]
\multirow{12}{*}{\begin{turn}{90}$\sqrt{s_{max}}=8$ TeV\end{turn}}  
  & $\mu^{+}_{soft}$ [GeV]& 0.90\,$\pm$\,0.20  & 0.90\,$\pm$\,0.19  & 0.90\,$\pm$\,0.19    \\[1.2ex]
  & $A$ [GeV$^{-2}$]      & 124.8\,$\pm$\,2.4  & 123.6\,$\pm$\,2.3  & 123.3\,$\pm$\,2.3    \\[1.2ex]
  & $B$ [GeV$^{-2}$]      & 38.5\,$\pm$\,8.1   & 42.1\,$\pm$\,7.9   & 42.1\,$\pm$\,7.9     \\[1.2ex]
  & $C$ [GeV$^{-2}$]      & 0.61\,$\pm$\,0.14  & 0.73\,$\pm$\,0.14  & 0.60\,$\pm$\,0.15    \\[1.2ex]
  & $\mu^{-}_{soft}$ [GeV]& {\bf 0.5 [fixed]}  & {\bf 0.5 [fixed]}  & {\bf 0.5 [fixed]}    \\[1.2ex]
  & $D$ [GeV$^{-2}$]      & 24.2\,$\pm$\,1.4   & 24.2\,$\pm$\,1.4   & 24.2\,$\pm$\,1.4     \\[1.2ex]
  & $\nu_{1}$ [GeV]       & 2.32\,$\pm$\,0.49  & 2.36\,$\pm$\,0.54  & 2.04\,$\pm$\,0.48    \\[1.2ex]
  & $\nu_{2}$ [GeV]       &0.051\,$\pm$\,0.032 & 0.057\,$\pm$\,0.035& 0.027\,$\pm$\,0.031  \\[1.2ex]
  
  & $\chi^2/163$          & 1.202               & 1.192               & 1.179              \\[1.2ex]
  & $P(\chi^2)$ & 4.0 $\times$ 10$^{-2}$ & 4.7 $\times$ 10$^{-2}$ & 5.8 $\times$ 10$^{-2}$ \\[1.2ex]
\hline \hline
& & & & \\[-0.2cm]
\multirow{12}{*}{\begin{turn}{90}$\sqrt{s_{max}}=13$ TeV\end{turn}}  
  & $\mu^{+}_{soft}$ [GeV]& 0.90\,$\pm$\,0.20  & 0.90\,$\pm$\,0.19    & 0.71\,$\pm$\,0.11  \\[1.2ex]
  & $A$ [GeV$^{-2}$]      & 121.8\,$\pm$\,4.6  & 123.5\,$\pm$\,6.4    & 107\,$\pm$\,30     \\[1.2ex]
  & $B$ [GeV$^{-2}$]      & 43.1\,$\pm$\,9.0   & 42\,$\pm$\,10        & 40.0\,$\pm$\,6.2   \\[1.2ex]
  & $C$ [GeV$^{-2}$]      & 0.51\,$\pm$\,0.21  & 0.67 \,$\pm$\,0.22   & 0.29\,$\pm$\,0.14  \\[1.2ex]
  & $\mu^{-}_{soft}$ [GeV]& {\bf 0.5 [fixed]}  & {\bf 0.5 [fixed]}    & {\bf 0.5 [fixed]}  \\[1.2ex]
  & $D$ [GeV$^{-2}$]      & 24.2\,$\pm$\,1.4   & 24.2\,$\pm$\,1.5     & 23.3\,$\pm$\,1.3   \\[1.2ex]
  & $\nu_{1}$ [GeV]       & 2.10\,$\pm$\,0.46  & 2.32\,$\pm$\,0.52    & 2.11\,$\pm$\,0.44  \\[1.2ex]
  & $\nu_{2}$ [GeV]       &0.039\,$\pm$\,0.029 & 0.055\,$\pm$\,0.034 & 0.030\,$\pm$\,0.027 \\[1.2ex]
  
  & $\chi^2/167$          & 1.188               & 1.176               & 1.210              \\[1.2ex]
  & $P(\chi^2)$ & 4.9 $\times$ 10$^{-2}$ & 5.9 $\times$ 10$^{-2}$ & 3.3 $\times$ 10$^{-2}$ \\[1.2ex]
\hline \hline  
\end{tabular}
}
\caption{Values of the fitted parameters for the DGM$19$ model by considering one standard deviation.}
\label{chDGMtab3}
\end{table}

\bfg[hbtp]
\begin{center}
 \includegraphics[width=8.0cm,height=8.0cm]{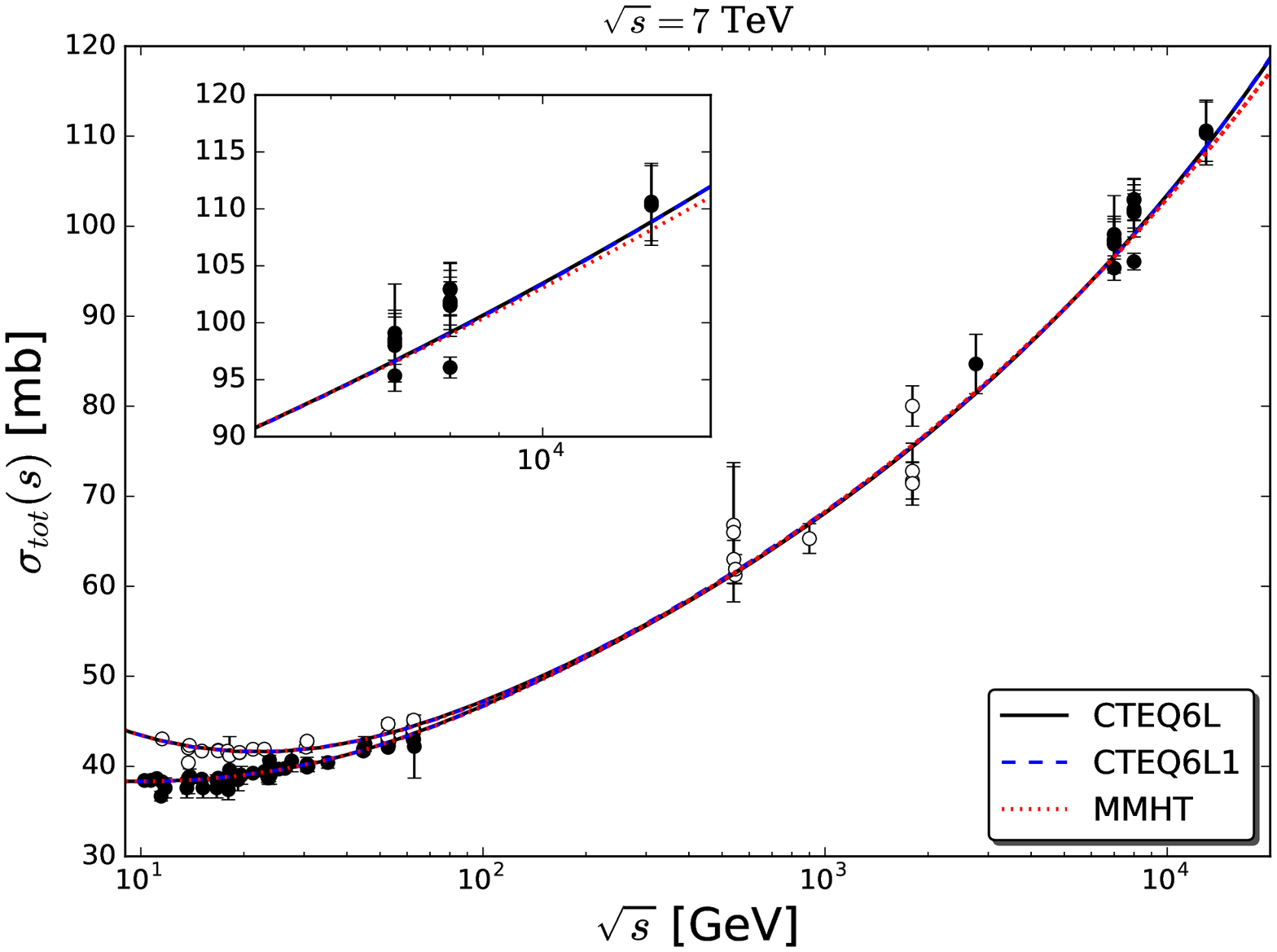}
 \includegraphics[width=8.0cm,height=8.0cm]{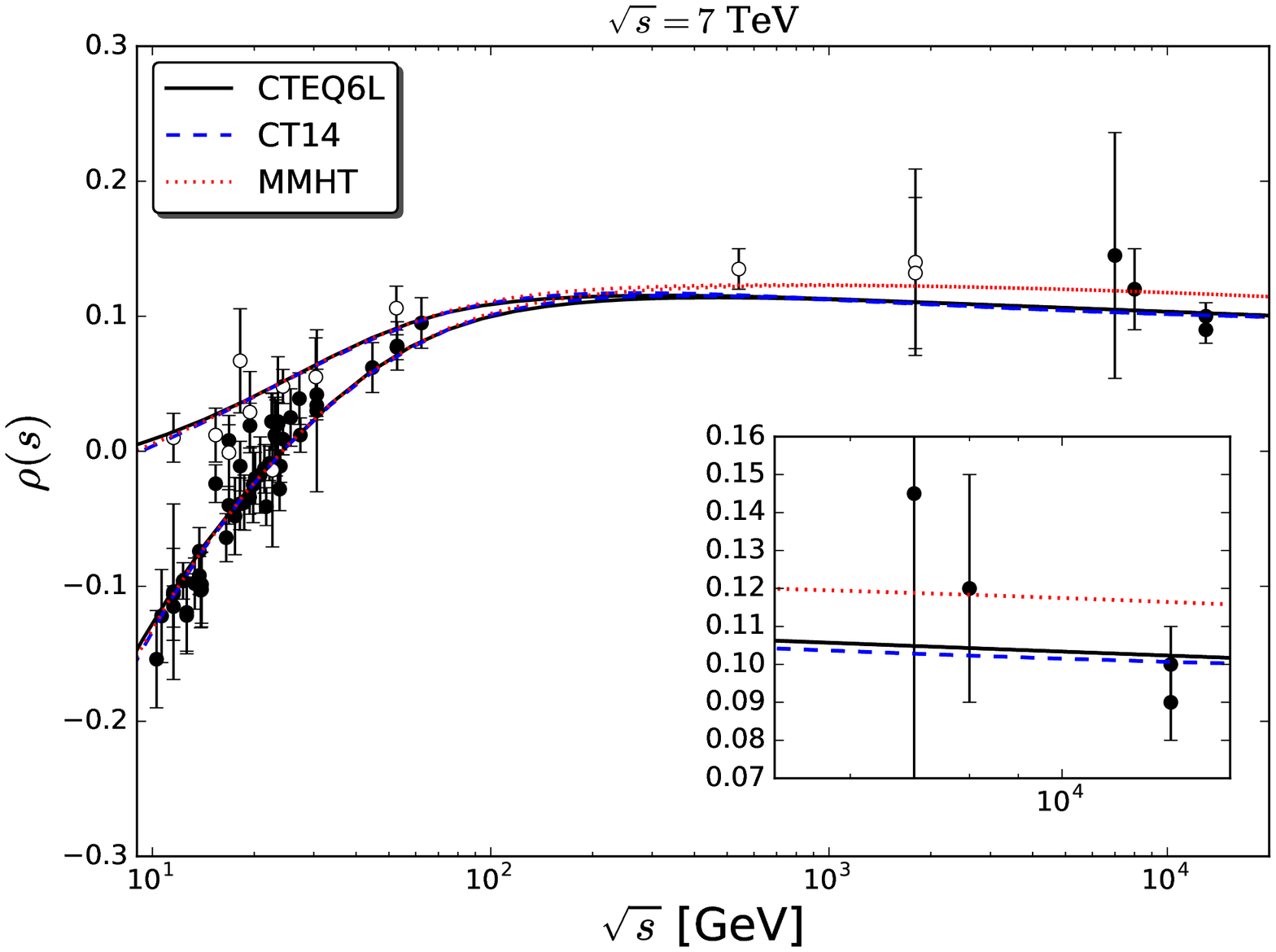}
 \includegraphics[width=8.0cm,height=8.0cm]{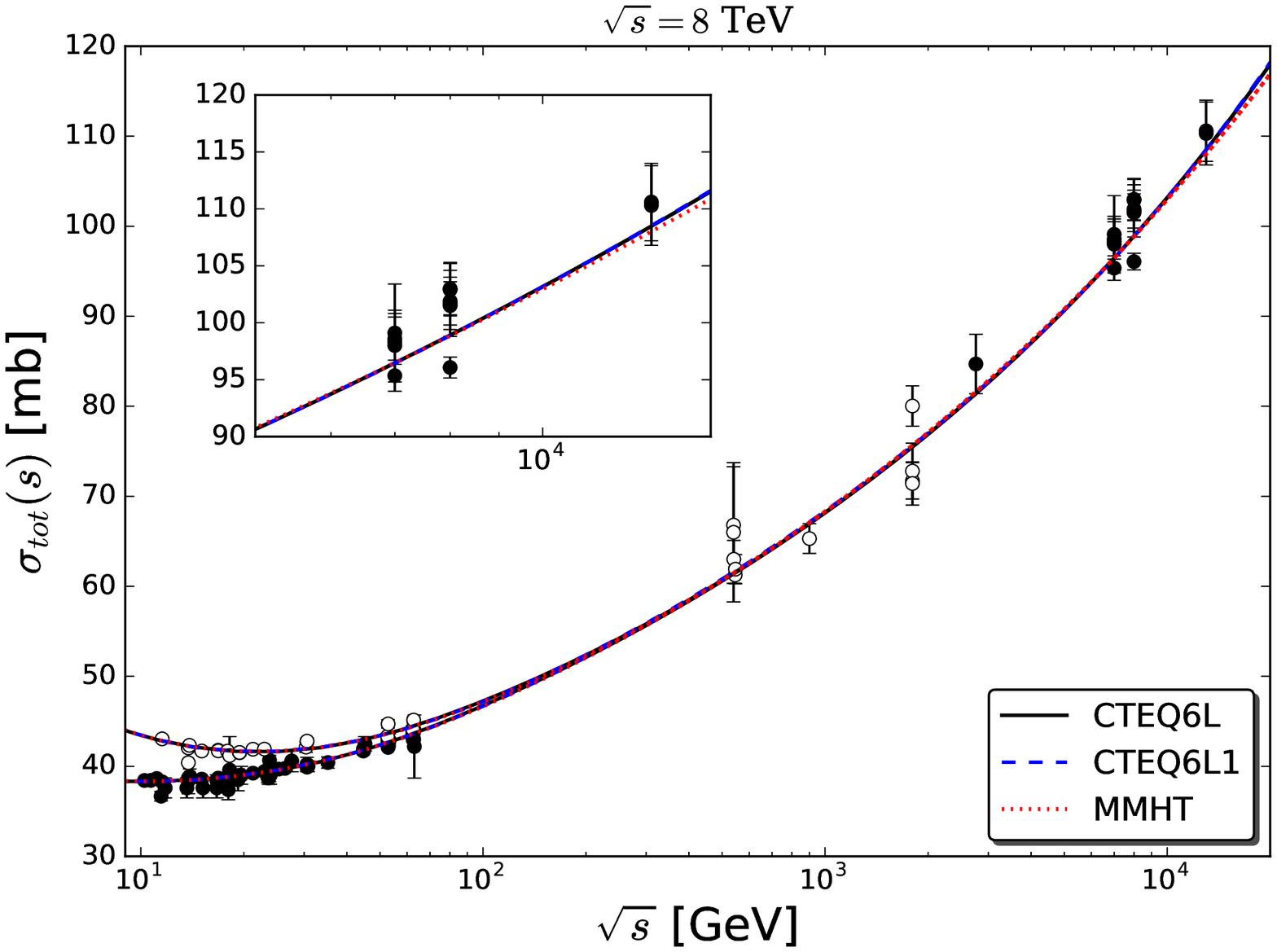}
 \includegraphics[width=8.0cm,height=8.0cm]{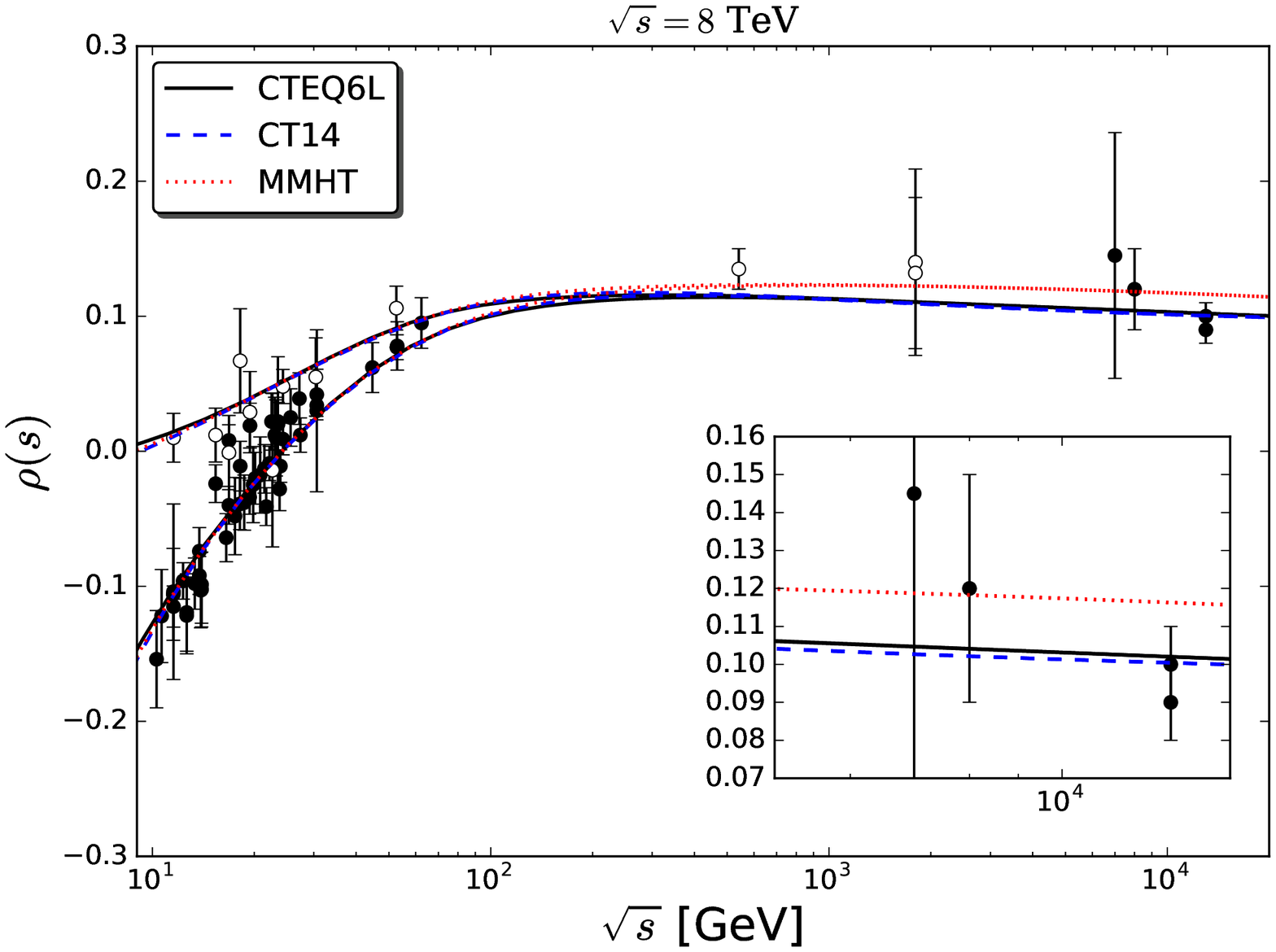}
 \includegraphics[width=8.0cm,height=8.0cm]{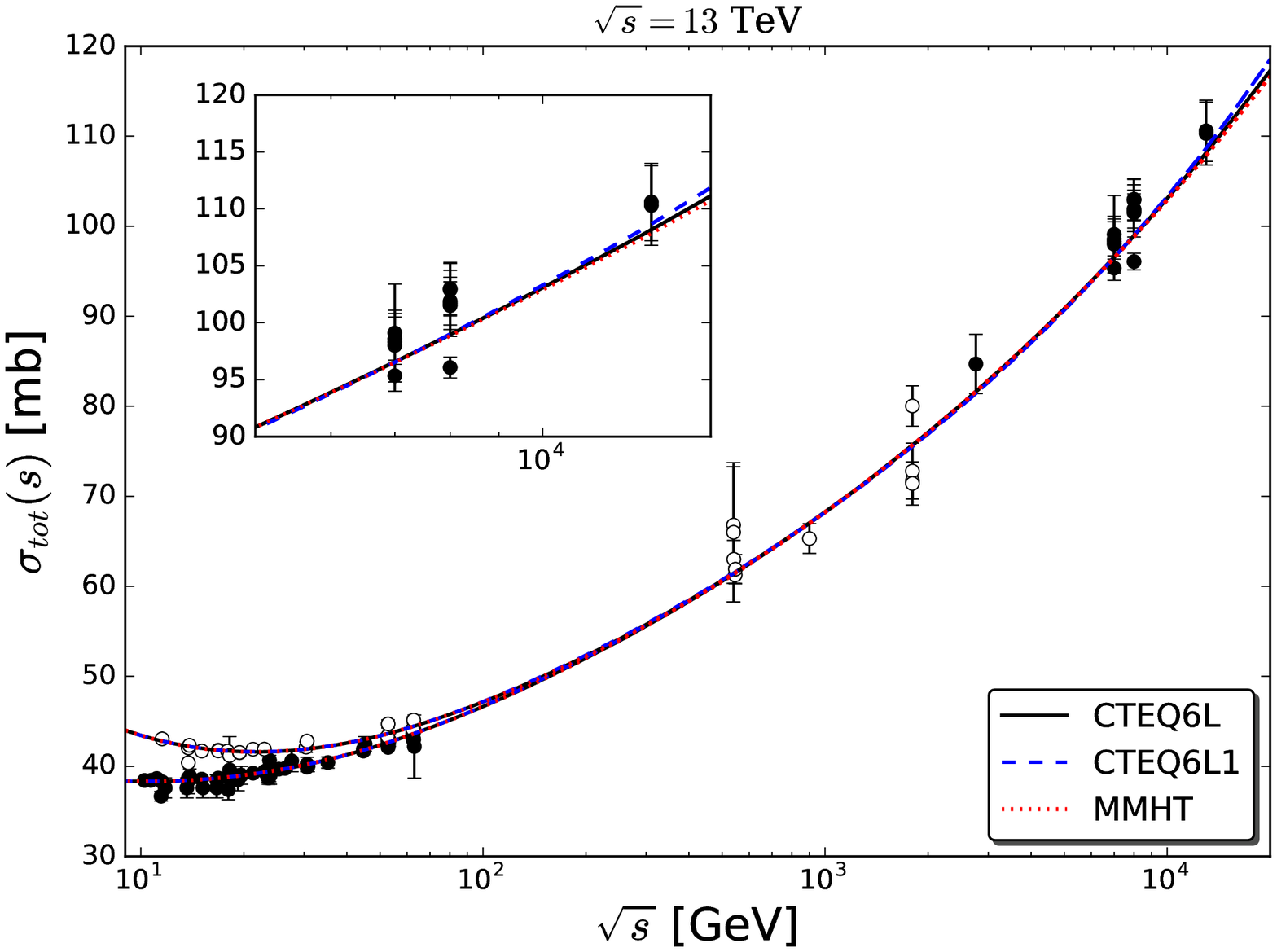}
 \includegraphics[width=8.0cm,height=8.0cm]{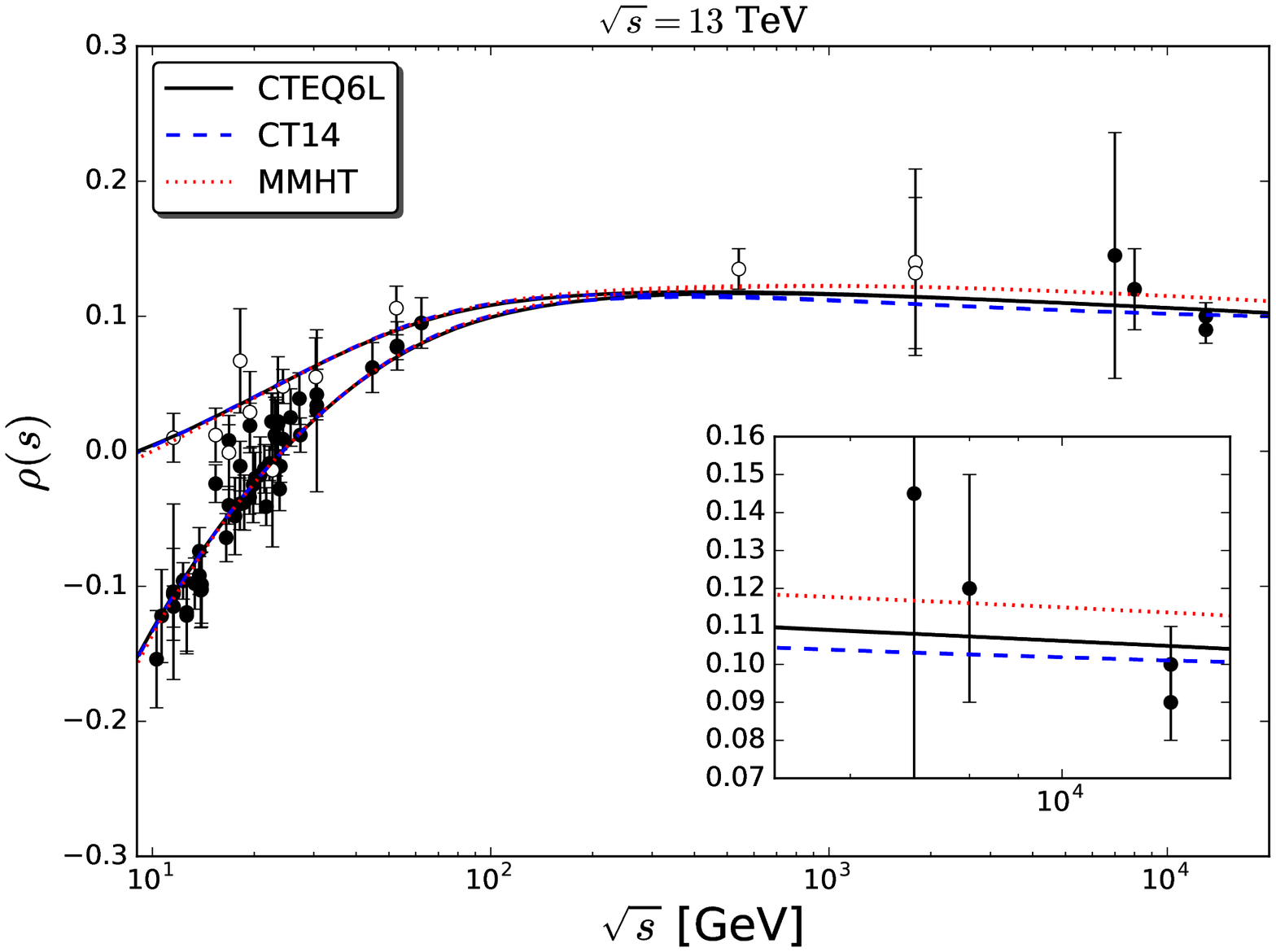}
 \caption{Total cross section and $\rho$-parameter predictions at each energy cut.}
\label{dgm19fig7}
\end{center}
\efg

\begin{table}[hbtp]
\centering
\scalebox{0.9}{
\begin{tabular}{c@{\quad}c@{\quad}c@{\quad}c@{\quad}}
\multicolumn{4}{c}{{\bf $C=0$ [fixed]}}\\
\hline \hline
& & & \\[-0.2cm]
PDF:    & CTEQ6L & CT14 & MMHT \\[0.7ex] 
\hline
& & & \\[-0.2cm]
$\mu^{+}_{soft}$ [GeV]& 0.700\,$\pm$\,0.028  & 0.700\,$\pm$\,0.011   & 0.700\,$\pm$\,0.020  \\[0.7ex]
$A$ [GeV$^{-2}$]      & 112.48\,$\pm$\,0.72  & 118.00\,$\pm$\,0.61   & 110.26\,$\pm$\,0.80  \\[0.7ex]
$B$ [GeV$^{-2}$]      & 24.2\,$\pm$\,3.1     & 11.5\,$\pm$\,2.8      & 26.9\,$\pm$\,3.3     \\[0.7ex]
$\mu^{-}_{soft}$ [GeV]& {\bf 0.5 [fixed]}    & {\bf 0.5 [fixed]}     & {\bf 0.5 [fixed]}    \\[0.7ex]
$D$ [GeV$^{-2}$]      & 23.4\,$\pm$\,1.3     & 23.6\,$\pm$\,1.3      & 23.5\,$\pm$\,1.3     \\[0.7ex]
$\nu_{1}$ [GeV]       & 1.82\,$\pm$\,0.17    & 1.74\,$\pm$\,0.20     & 1.50\,$\pm$\,0.17    \\[0.7ex]
$\nu_{2}$ [GeV]   & 0.023\,$\pm$\,0.012      & 0.022\,$\pm$\,0.014   & -0.004\,$\pm$\,0.012 \\[0.7ex]
\hline
& & & \\[-0.3cm]
$\chi^2/168$   & 1.284               & 1.757               & 1.438                        \\[0.7ex]
$P(\chi^2)$    & 7.6 $\times$ 10$^{-3}$ & 5.0 $\times$ 10$^{-9}$ & 1.7 $\times$ 10$^{-4}$ \\[0.7ex]
\hline \hline\\[0.4cm]  

\multicolumn{4}{c}{{\bf $\nu_{2}=0$ [fixed]}}\\
\hline \hline
& & & \\[-0.2cm]
PDF:    & CTEQ6L & CT14 & MMHT \\[0.7ex] 
\hline
& & & \\[-0.2cm]
$\mu^{+}_{soft}$ [GeV]& 0.90\,$\pm$\,0.14    & 0.90\,$\pm$\,0.13     & 0.80\,$\pm$\,0.10    \\[0.7ex]
$A$ [GeV$^{-2}$]      & 123.3\,$\pm$\,1.9    & 125.6\,$\pm$\,1.9     & 119.6\,$\pm$\,2.4    \\[0.7ex]
$B$ [GeV$^{-2}$]      & 38.0\,$\pm$\,6.7     & 36.0\,$\pm$\,6.7      & 46.5\,$\pm$\,6.8     \\[0.7ex]
$C$ [GeV$^{-2}$]      & 0.31\,$\pm$\,0.10    & 0.490\,$\pm$\,0.096   & 0.47\,$\pm$\,0.11    \\[0.7ex]
$\mu^{-}_{soft}$ [GeV]& {\bf 0.5 [fixed]}    & {\bf 0.5 [fixed]}     & {\bf 0.5 [fixed]}    \\[0.7ex]
$D$ [GeV$^{-2}$]      & 24.3\,$\pm$\,1.3     & 24.3\,$\pm$\,1.3      & 24.3\,$\pm$\,1.3     \\[0.7ex]
$\nu_{1}$ [GeV]       & 1.494\,$\pm$\,0.032  & 1.470\,$\pm$\,0.036   & 1.579\,$\pm$\,0.036  \\[0.7ex]
\hline
& & & \\[-0.3cm]
$\chi^2/168$   & 1.330               & 1.407               & 1.260                        \\[0.7ex]
$P(\chi^2)$    & 2.7 $\times$ 10$^{-3}$ & 4.0 $\times$ 10$^{-4}$ & 1.3 $\times$ 10$^{-2}$ \\[0.7ex]
\hline \hline 
\end{tabular}
}
\caption{Values of the fitted parameters for the DGM$19$ model fixing $C=0$ and $\nu_{2}=0$ by considering one standard deviation.}
\label{chDGMtab4}
\end{table}

\bfg[hbtp]
\begin{center}
 \includegraphics[width=8.0cm,height=8.0cm]{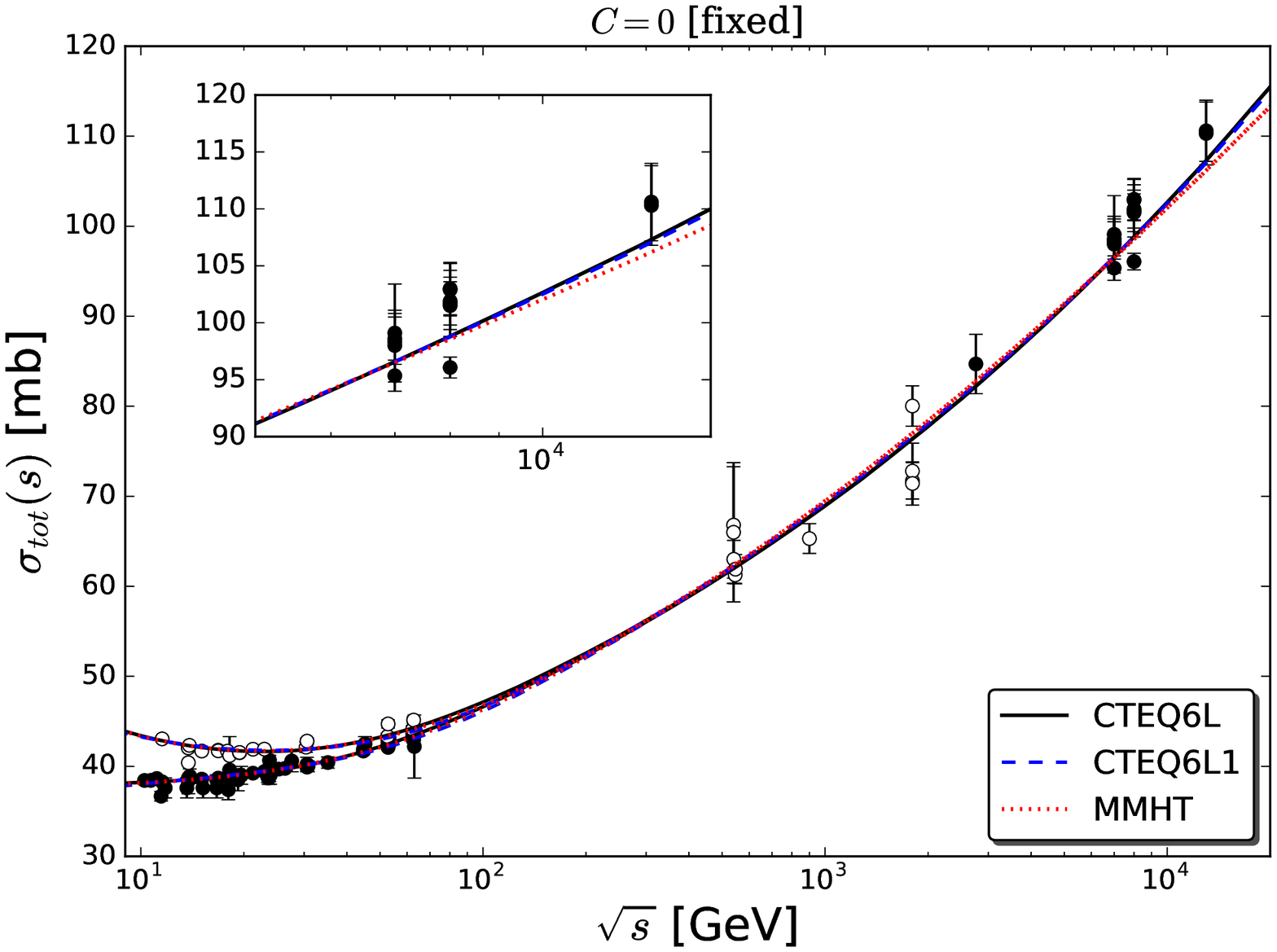}
 \includegraphics[width=8.0cm,height=8.0cm]{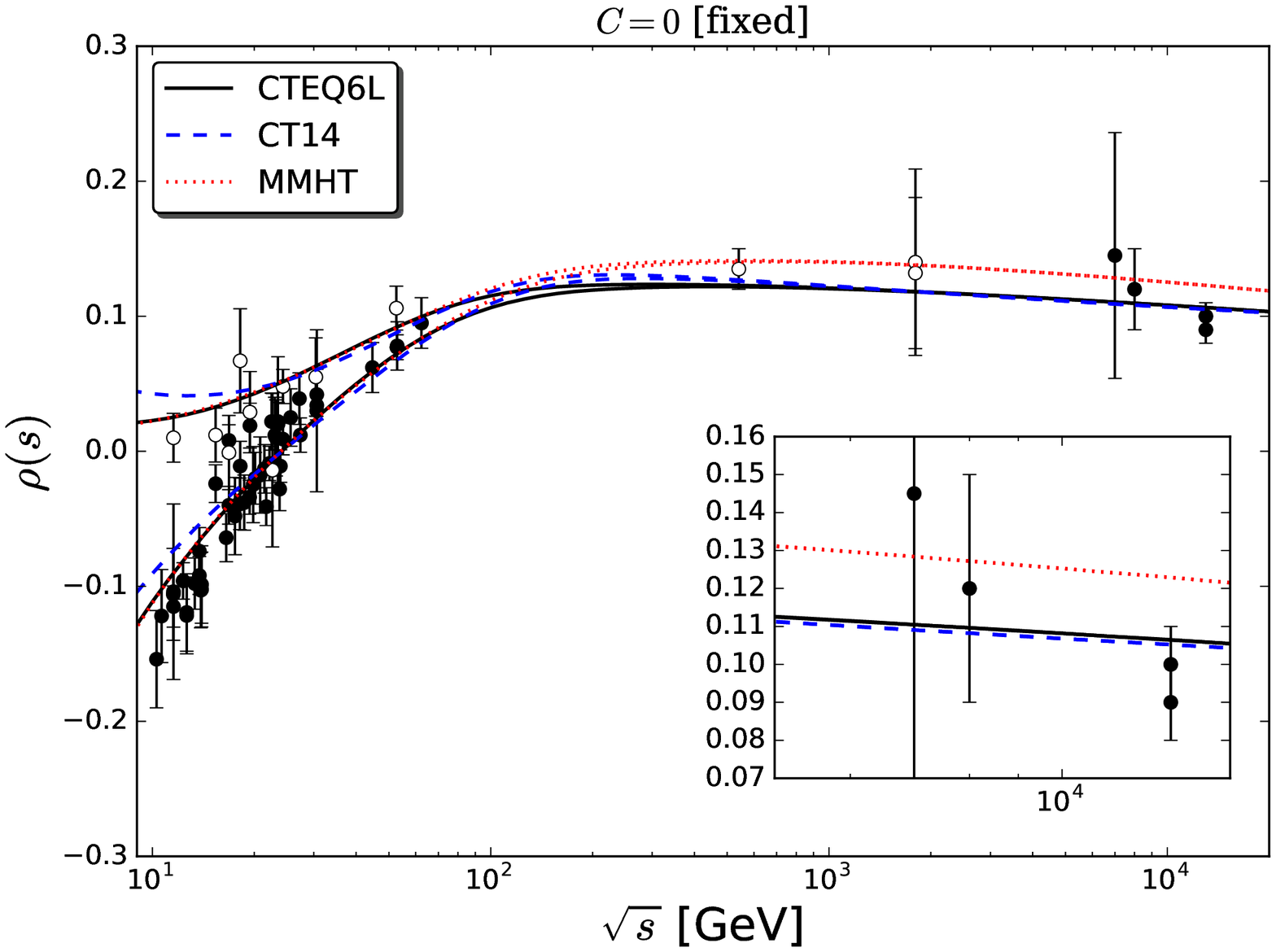}
 \includegraphics[width=8.0cm,height=8.0cm]{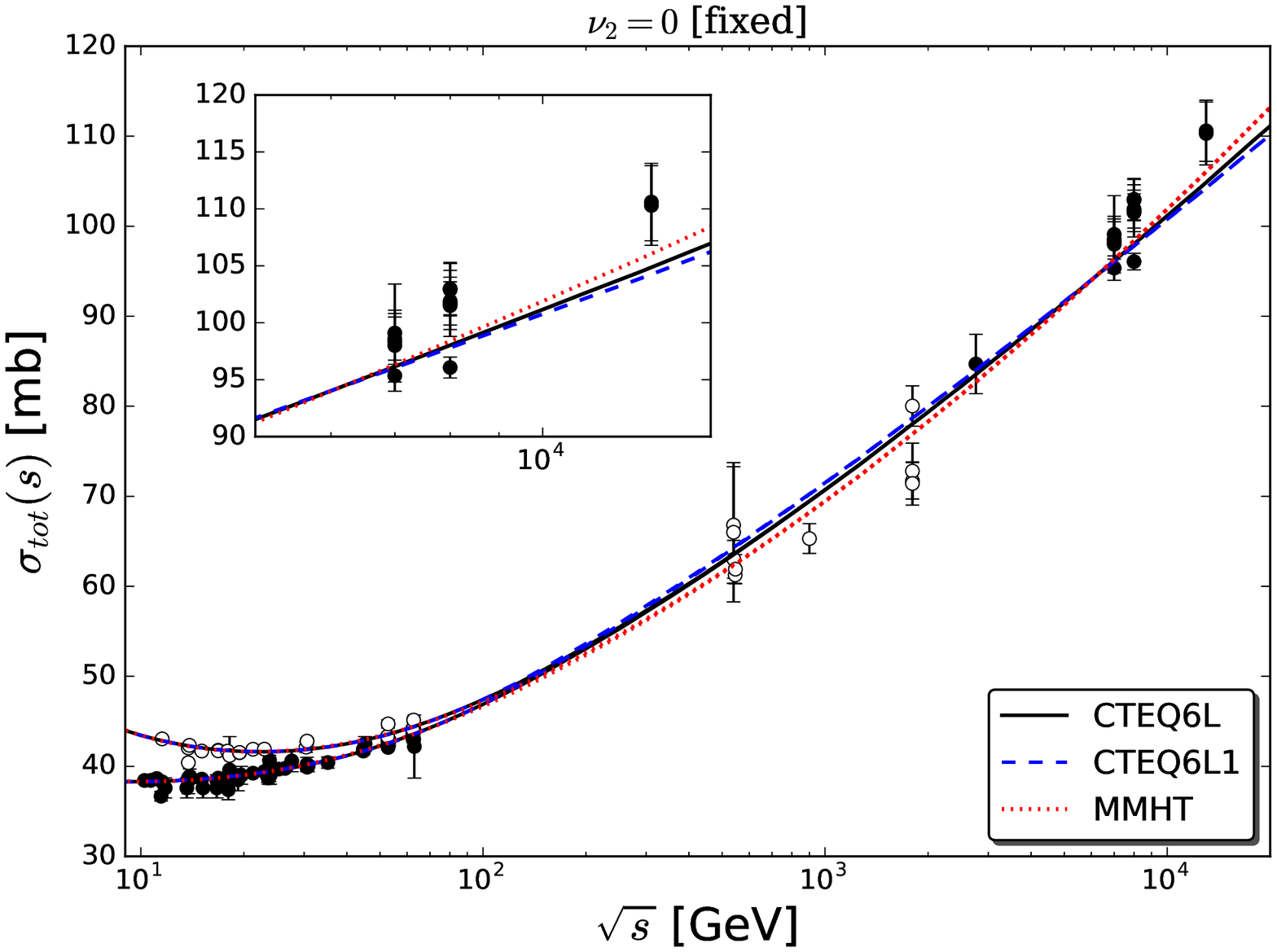}
 \includegraphics[width=8.0cm,height=8.0cm]{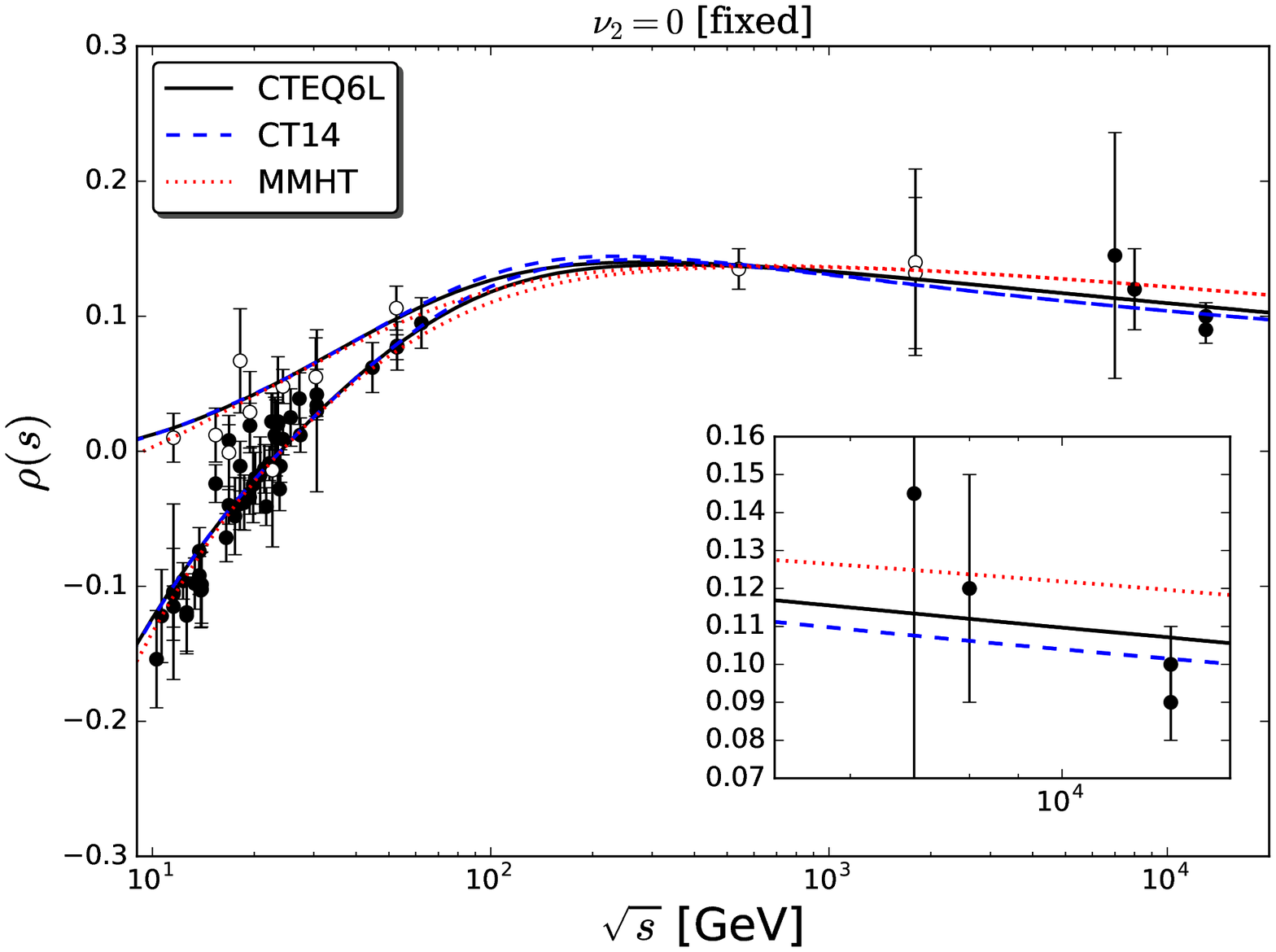}
 \caption{Total cross section and $\rho$-parameter predictions fixing $C=0$ e $\nu_{2}=0$.}
\label{dgm19fig8}
\end{center}
\efg

\bfg[hbtp]
\begin{center}
 \includegraphics[width=8.0cm,height=8.0cm]{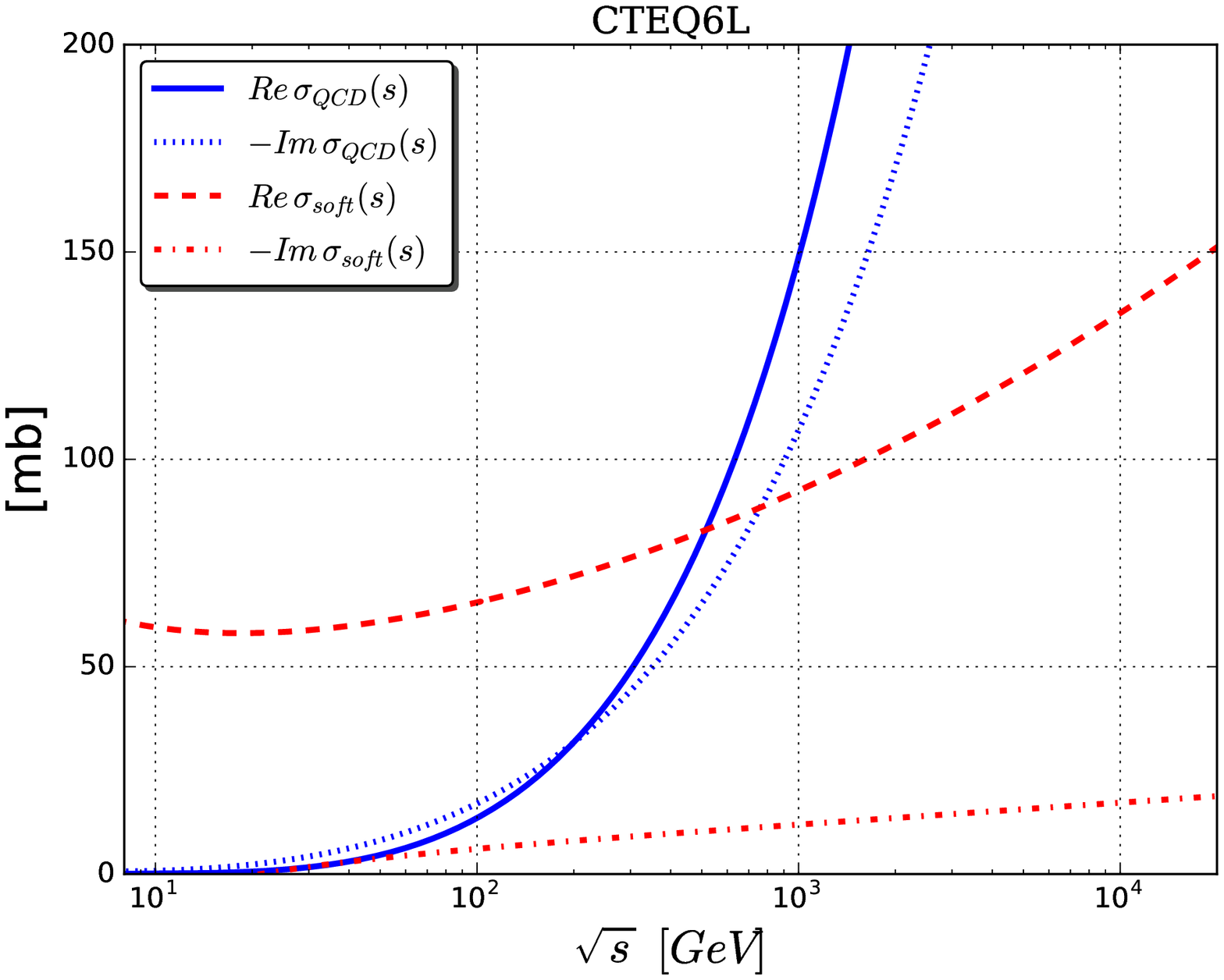}
 \includegraphics[width=8.0cm,height=8.0cm]{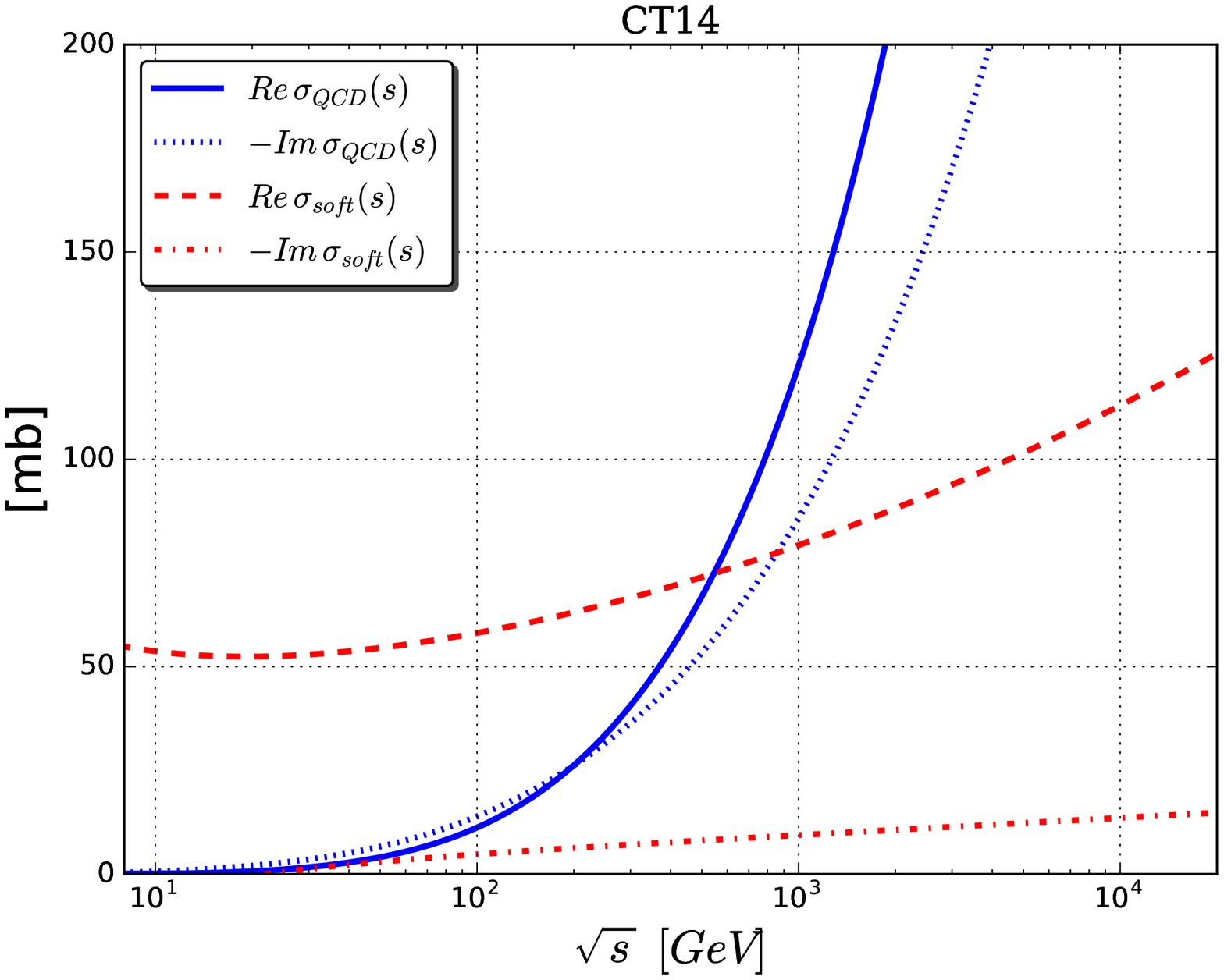}
 \includegraphics[width=8.0cm,height=8.0cm]{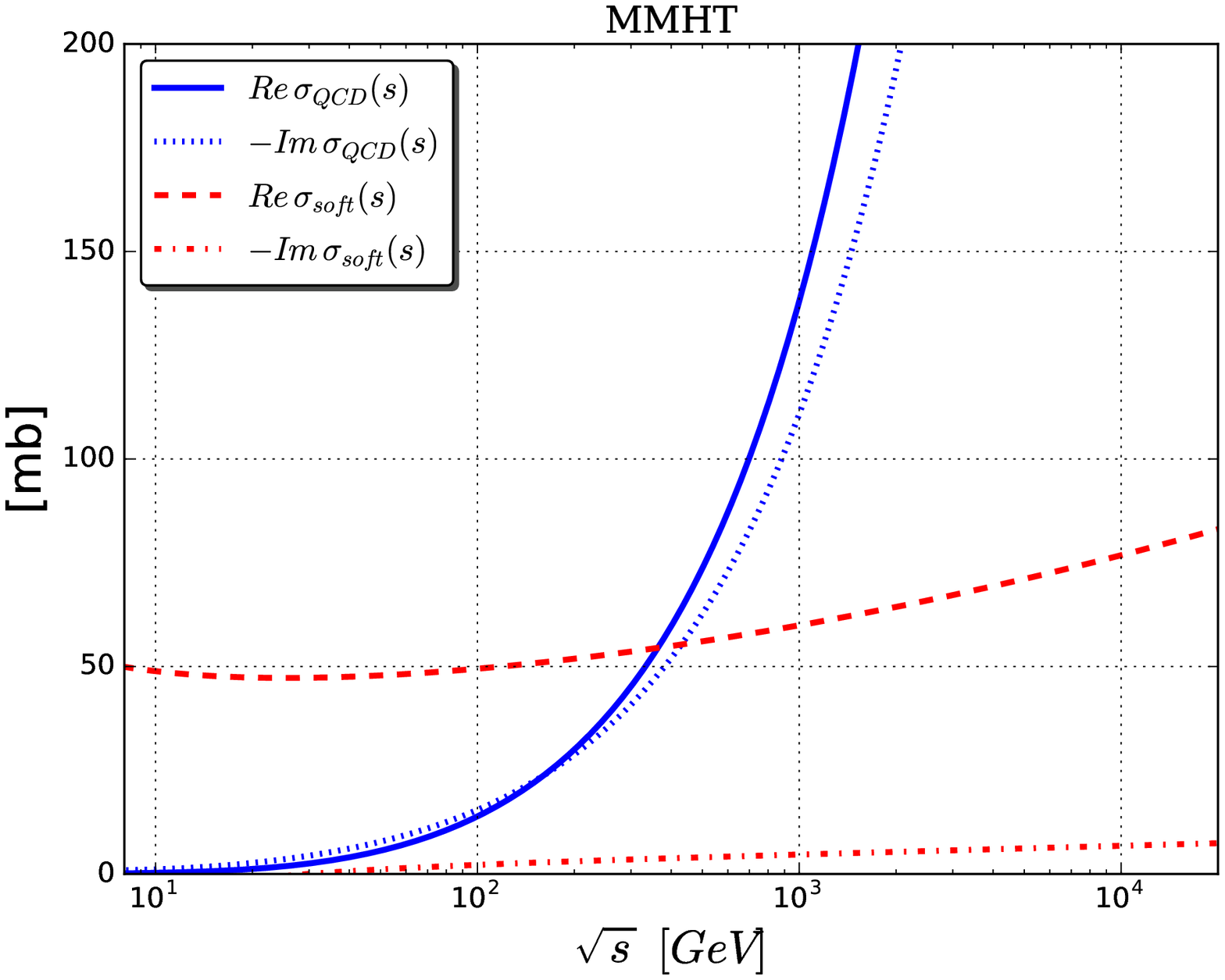}
 \caption{The comparison between $\sigma_{QCD}(s)$ and $\sigma_{soft}(s)$.}
\label{dgm19fig9}
\end{center}
\efg

\newpage

\clearpage
\thispagestyle{plain}

\chapter{\textsc{Conclusions and Final Remarks}}
\label{chap9}

\mbox{\,\,\,\,\,\,\,\,\,}
The work presented in this Thesis was devoted to study the nonperturbative dynamic aspects in elastic hadronic process. More specifically, the contribution in the high-energy behavior of proton-proton and antiproton-proton collisions. The description of elastic hadronic collision can be made by means of two different approaches, which in essence describe the same thing, both compatible with the properties of analyticity and unitarity of the scattering $S$-matrix. Therefore, this Thesis was divided in two parts: one based upon Regge theory and another inspired by Quantum Chromodynamics.

Motivated by the recent measurements of $pp$ total cross section and the respective ratio of the real to imaginary part of the forward scattering amplitude at $\sqrt{s}=13$ TeV from LHC obtained by the TOTEM Collaboration \cite{Antchev:2017dia,Antchev:2017yns}, we begin by studying the possible effects that parametrizations only with Pomeron leading contribution may cause in the asymptotic high-energy description of the forward quantities compared with the most recent exeperimental data. For the first time, as far as we known, it was considered as the most complete set of forward data from $pp$ and $\bar{p}p$ scattering two ensemble by adding either only the TOTEM data or TOTEM and ATLAS data at the LHC energy region. Another novel aspect of our analysis is that it was taked into account in the data reduction to each ensemble, the evaluation of the theoretical uncertainty through error propagation from the fit parameters with both one and two standard deviation ($\sim$ 68\% and $\sim$ 95\% CL, respectively). 

Notwithstanding the expected rise of the $\sigma^{pp}_{tot}$ \cite{Antchev:2017dia} and the corresponding unexpected decrease in the $\rho^{pp}$ value \cite{Antchev:2017yns}, indicating disagreement with all the Pomeron model predictions by the COMPETE Collaboration (2002), we have tested different analytic parametrizations, all of them characterized by Pomeron dominance. Among all the models and variants considered, one specifically, with only seven free fit parameters, led to the best results, namely Modell III characterized by two simple poles Reggeons, one double pole and one triple pole Pomerons, respectively. Although the uncertainty predictions of the fit result to $\rho$-parameter at $\sqrt{s}=13$ TeV does not cross its central value, 
\bear
\rho(\sqrt{s}=13\,\text{TeV})&=&0.1185\pm 0.0049 \,\,\,\text{(ensemble T)},\nonumber\\
\rho(\sqrt{s}=13\,\text{TeV})&=&0.1158\pm 0.0042 \,\,\,\text{(ensemble T+A)},\nonumber
\eear
where the error corresponds to $1\sigma$, the same thing also happens for the ATLAS datum on $\sigma_{tot}$ at $\sqrt{s}=8$ TeV. Therefore based on the agreement between the phenomenological model and the experimental data, compatible with the uncertainty region, we understand that the Model III cannot be excluded by the bulk of experimental data presently available on forward $pp$ and $\bar{p}p$ elastic collision. However, further insights on the discrepancies between the TOTEM and ATLAS data surely will be very helpful.

As it was mentioned in the text, at high energies the Pomeron plays a crucial part in describing the soft interactions. In fact the measurements at LHC$13$ provide an unique constraint on the Pomeron parameters, since the energy dependence of total and diffractive cross sections is driven by $\epsilon$ whilst $\alpha^{\prime}_{\IP}$ determines the relation of the forward slopes with energy. Therefore, the recent TOTEM data at $\sqrt{s}=13$ TeV give us the opportunity to tackle more effectively the functional form of the Pomeron trajectory considering the nearest $t$-channel singularity. By means of different combinations of proton-Pomeron vertices, namely exponential and a dipole (power-like) vertices, we performed analyses using Born-level and also eikonalized amplitudes. In the former case, beyond the study of the high-energy contribution of single-Pomeron exchange, we estimated the effects of double-Pomeron exchange. As for the latter one, we investigated the role of eikonalized amplitudes in both one- and two-channel models. The Born-level analysis revealed that, among the models considered, practically there is no preferred results. However, the dynamical differences found are mostly related to the choice of the functional form of the proton-Pomeron vertex considered. Inasmuch we performed the eikonalization process, either to restore $s$-unitarity and to obtain better results to the ratio of the real to imaginary part of the forward scattering amplitude, indeed we obtained better fit curves, at least for the forward quantities. However, our results, for one- and two-channel eikonalized amplitudes, showed that in the kinematical $\vert t \vert$ range considered a linear trajectory for the Pomeron still remains as a good approximation.

The Odderon is well-understood in perturbative QCD \cite{Braun:1998fs,Ewerz:2003xi,Ewerz:2005rg}, but its nonperturbative counterpart still lacks a founded formulation. In both analysis made based upon Regge-type asymptotic high-energy amplitudes, we did not consider the possibility of Odderon effects. Our attempt was to surround the Pomeron dominance in all (or at least many) possible forms to check if there is, indeed, strong arguments to exclude models based only on Pomeron dominance. On the other hand, the decrease in the $\rho^{pp}$ value has been well described with Maximal Odderon contribution in the recent analyses by Martynov and Nicolescu \cite{Martynov:2018nyb,Martynov:2017zjz}. Independently of the current scenario, \tit{``Did TOTEM experiment discover the Odderon?''} is a difficult question that requires a careful answer.
 
In the second part of the Thesis, the description of the elastic hadronic scattering is given in the context of the improved parton model. We studied infrared contributions to semihard parton-parton interactions by considering an effective charge whose infrared behavior is constrained by a dynamical mass scale, which is a purely nonperturbative dynamical effect. We have investigated $pp$ and $\bar{p}p$ scattering in the LHC energy region with the assumption that the observed increase of hadron-hadron total cross sections arises exclusively from these semihard interactions. In the calculation of $\sigma_{tot}^{pp,\bar{p}p}$ and $\rho^{pp,\bar{p}p}$, we have investigated the behavior of the forward amplitude for a range of different cutoffs and parton distribution functions, firstly the pre-LHC sets CTEQ$6$L, CTEQ$6$L$1$ and MSTW, secondly the fine-tunned sets CT14 and MMHT, and considered the phenomenological implications of a class of energy-dependent form factors for semihard partons. In what concerns the DGM$15$, we introduced integral dispersion relations specially tailored to connect the real and imaginary parts of semihard eikonals with energy-dependent form factors. 

The reanalyzed DGM$19$ model has the same footprint as DGM$15$, with some exceptions, as for example the connection between the real and imaginary parts of the semihard eikonal is now constructed by using the complex prescription $s\to -is$, which simulates an even inverse dispersion relation. Moreover, the best results obtained in the previous version of the model was used to tune the newest version. Respectively, we selected the results of CTEQ$6$L within $Q^{2}_{min}=1.3$ GeV and an energy-dependent dipole form factor, and compared with the results of post-LHC PDF's. In the context of a QCD-based model with even-under-crossing amplitude dominance at high energies, it was shown that the $pp$ and $\bar{p}p$ elastic scattering data on  $\sigma_{tot}$ and $\rho$ above $10$ GeV are quite well described, especially the recent TOTEM data at $13$ TeV. Unexpected features of the data, such as the decrease of the $\rho^{pp}$ parameter value, recently reported by the TOTEM Collaboration at LHC13 \cite{Antchev:2017yns}, was addressed using an eikonalized elastic amplitude, where unitarity and analicity properties are readily build in. The essential inputs of this model, namely the low-$x$ behavior of PDF's and the dynamical gluon mass scale were found to be crucial in the phenomenological description of present available data at energies spanning from $10$ GeV to $13$ TeV. In effect, the model provided an accurate global description of $\sigma_{tot}$ and $\rho$ with post-LHC fine-tunned PDF's such as CT14, even if the $8$ and $13$ TeV data are not included in the data set analyzed. These findings suggest that low-$x$ parton dynamics play a major role in the driving mechanism behind the pre-asymptoptic $\rho^{pp}$ decrease at LHC energies. Moreover, despite the recent claims of an Odderon discovery by TOTEM at LHC$13$, properly account of $s$-channel absorptions in the amplitude indicates that only Pomeron-like solution may be needed in describing the LHC forward elastic data.

\subsection*{\textsc{Perspectives}}
\mbox{\,\,\,\,\,\,\,\,\,}
The work so far serves as basis to study many other things and to explore in much more details other aspects of the nonperturbative character of QCD. Currently, the curious case of the Odderon is one of the ``hot spots'' in particle physics. In order to quantify the contribution of an odd-under-crossing term, we are studying some possible forms to include it in the soft odd eikonal in our QCD based model. Such analysis would be very helpful to compare Pomeron-like solutions and Pomeron plus Odderon-like solutions using a model based on the Quantum-Chromodynamics-improved parton model

We have mainly focused on forward, or nearforward, analysis. The corresponding elastic differential cross section data sets at $\sqrt{s}=7$, $8$ and $13$ TeV obtained at the LHC by the TOTEM Collaboration were not included in our DGM$19$ analysis. Presently we understand that our energy-dependent semihard form factor is not ``strong enough'' to treat the dip-shoulder region, as well as the behavior in the region of higher values of $\vert t \vert$. We are still studying some different possibilities to introduce a more convenient $\vert t\vert$-dependence in the semihard sector. 

The strong disagreement between the TOTEM and ATLAS measurements clearly indicates the possibility of different scenarios for the rise of the total cross section and consequently for the parameters of the soft Pomeron. Thus, in order to investigate this tension between both experiments results in a quantitative way, we wish to carry out global fits to $pp$ and $\bar{p}p$ data considering two distinct ensembles of data with either the TOTEM or the ATLAS $pp$ results for $\sigma_{tot}$ and $d\sigma^{pp}/d\vert t \vert$ at $7$ and $8$ TeV. The discrepancies between the TOTEM and ATLAS data certainly will result in distinct values for the Pomeron parameters.

In the first part of the Thesis, we studied and concluded that there is, at least, one combination with double and triple-pole Pomerons that cannot be excluded by the bulk of present analytical Regge-type models. However, we did not consider the $t$-dependence in those contributions and this is an important step for the eikonalization process. Perhaps the very next step is to study the theoretical error propagation in eikonalized amplitudes, firstly in Regge based models and secondly to translate this technology in the language of QCD-based models. As far as we know, this type of error evaluation in eikonalized amplitudes have never been accomplished.

We studied the nonperturbative dynamics of hadronic collisions inspired by QCD parton model and exploited its failures through out Regge-based analysis. The theory behind Regge phenomenology was built under the analytical properties of the scattering $S$-matrix and it was at that time, and still is, the correct way to access the $t=0$ physical region. However, one cannot disregard the fact that QCD is not a hypothesis, but the correct theory describing the strong interaction. With that in mind, we must keep walking forward and towards a description of hadronic diffraction fully based on QCD, even within baby steps.

As it was mentioned, there are many things to study henceforth. 

{\begin{flushright}\mbox{} \tit{``My universe is my eyes and my ears. Anything else is hearsay.''} \\-\,{\textsc{\large The Hitchhiker's Guide to the Galaxy}}, \textsc{Douglas Adams}.\end{flushright}}

\clearpage
\thispagestyle{plain}

\appendix
\pagestyle{headings}
\chapter{\textsc{The Classical Description of Diffraction}}
\label{APX1}

\mbox{\,\,\,\,\,\,\,\,\,}
In the next few sections it will be outlined in a very simple way a brief discussion on some aspects concerning the diffractive scattering via the classical approach and also, as a matter of fact, to clarify the optical and the geometrical comprehension of high-energy scattering processes. Moreover, it will be shown that the same two-particles scattering system at high energies studied in Chapter \ref{ch2} using the quantum mechanical formalism can be understood by means of this classical description of diffraction\footnote{As pointed out by V. Barone \& E. Predazzi and I quote here: \tit{``At the phenomenological level, however, the optical analogy should not be pushed too far: for instance, the shrinkage of the forward peak with increasing energy in hadronic diffraction has no optical analogue''} \cite{Barone:2002cv}.}.

\section{\textsc{Fraunhofer Regime}}
\label{chAPX1.1}

\mbox{\,\,\,\,\,\,\,\,\,}
There are three different kinds of regimes in classical optical theory: geometrical optics limit, where $kR^{2}/D\gg 1$; Fresnel regime, where $kR^{2}/D\sim 1$ and the Fraunhofer regime, where $kR^{2}/D\ll 1$. Each of these regimes has its own peculiarities and \tit{``when'', ``where''} and \tit{``how''} to use them, but despite the first two choices only the latter one is important in view of the application of optical concepts to elastic hadronic scatterings at high energies. If one considers the typical distance between the detector and a target in $pp$ and $\bar{p}p$ scattering experiments, which is $D\sim 1m$, the usual proton radius $R\sim 1fm$ and also the momenta $k\sim 5fm^{-1}$, one finds that
\be
kR^{2}\sim 5fm\to\frac{kR^{2}}{D}\ll 1.
\ee

Therefore, for this reason at high energies, the hadronic scattering process is typically diffractive, \tit{i.e.}, the optical analogy is such that a hadronic interaction is depicted in the Fraunhofer regime.

\section{\textsc{Single Slit Diffraction}}
\label{chAPX1.2}

\mbox{\,\,\,\,\,\,\,\,\,}
According to the Huygens-Fresnel principle, each point of the slit becomes the centre of a spherical wave. The initial incident beam will be scattered by each of these points in the slit such that the envelope of these waves results in the deflected wave. The diffraction pattern occurs because of the differences in the amplitudes and phases of the wavelets collected at each point $P$ of the detector plane.  

\bfg[hbtp]
  \begin{center}
    \includegraphics[width=12cm,clip=true]{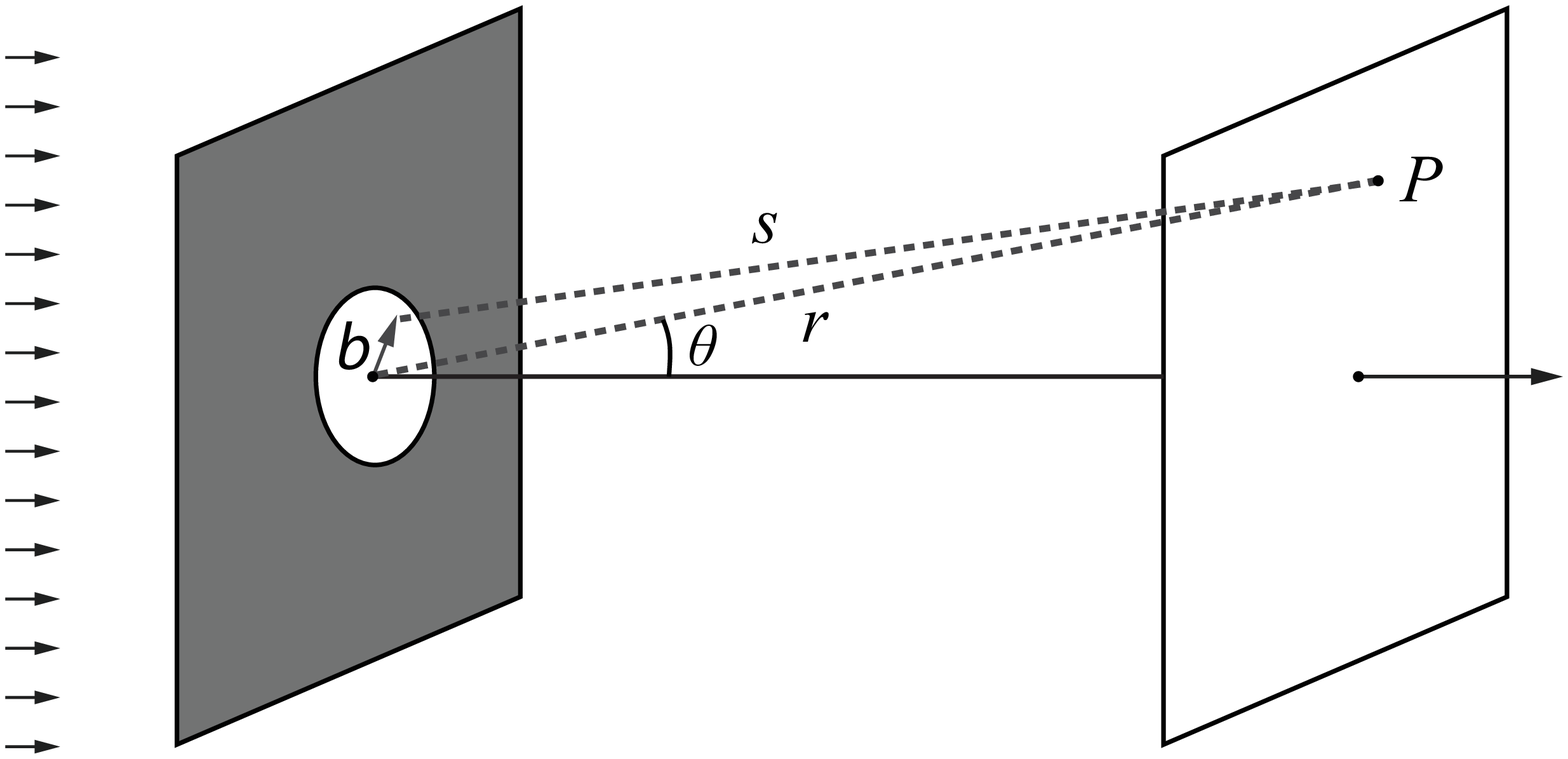}
    \caption{The diffraction of a plane-wave by a single slit.}
    \label{apx1fig1}
  \end{center}
\efg

In the context of classical optics the scattered wave function can be written using Kirchhoff theory. Within the method of separation of variables it is possible to write the wave function as a product of a spatial and a temporal function, \tit{e.g.} $\Psi(\rr,t)=U(\rr)e^{-i\omega t}$, where $U(\rr)$ satisfies the Helmholtz equation. With suitable and well defined boundary conditions, Green's theorem allow us to write what is called as the Fresnel-Kirchhoff integral,
\be
U(\rr)=-\frac{ik}{2\pi}\,U_{0}\int_{S_{0}}d^{2}\bb\,\frac{e^{iks}}{s},
\label{apx1.1}
\ee
with $U_{0}$ constant, $s$ is the distance between the slit and the point $P$, $S_{0}$ represents the slit and $\bb$ is the impact parameter, see Figure \ref{apx1fig1}.

The description of the wave function can be written in the Fraunhofer regime, which corresponds to the limit where the distance between the target and the detector $D\to\infty$, as a function of the transverse momentum in the scattering processes. Expanding $ks$ in the expression (\ref{apx1.1}) as $ks\simeq kr-\qq\cdot\bb$, the scattered wave function collected at each point $P$ is given by
\be
U(\rr)=-\frac{ik}{2\pi}\,U_{0}\,\frac{e^{ikr}}{r}\int_{S_{0}}d^{2}{\bb}\,e^{-i{\qq}\cdot{\bb}},
\label{apx1.2}
\ee
where $\vert \qq \vert=k\sin\theta\simeq k\theta$ is valid for small angles. In this limit $\qq$ is the momentum transfer $\qq\simeq \kk^{\prime}-\kk$, where $\kk^{\prime}$ is the outgoing wave vector and $\vert \kk^{\prime}\vert=\vert \kk \vert=k$.

\section{\textsc{Diffraction by an Opaque Obstacle}}
\label{chAPX1.3}

\mbox{\,\,\,\,\,\,\,\,\,}
There is a principle in optical Physics which states that a hole and an obstacle of identical form and dimension produces the same diffraction pattern, this is what is called as the Babinet's principle. Therefore, the problem related in a high-energy particle scattering experiment can be viewed by its optical analogue, the scattering by an opaque obstacle. Within this analogy and as a consequence of the Huygens-Fresnel principle, it turns out to be possible to reconstruct the incident wave function by means of summing up the diffracted waves by the slit and the obstacle, 
\be
U_{\text{slit}}(P)+U_{\text{obstacle}}(P)=U_{\text{incident}}(P).
\label{apx1.3}
\ee
\mbox{\,\,\,\,\,\,\,\,\,}
Taking into account this last result, and also considering that in particle collisions the incident plane wave is diffracted by an obstacle, \tit{i.e.} a disk, the scattered wave function takes the form
\be
U_{\text{disk}}=U_{0}\,e^{ikz}-U_{\text{slit}}.
\label{apx1.4}
\ee

\section{\textsc{The Profile Function}}
\label{chAPX1.4}

\mbox{\,\,\,\,\,\,\,\,\,}
The aforementioned expression (\ref{apx1.2}) represents, as it was already explained, the scattered wave function collected at each point $P$. But now considering the existence of a function $S(\bb)$ such that $\vert S(\bb) \vert^{2}$ gives the probability of transmission of the wave scattered by the disk, \tit{i.e.}, $S(\bb)$ defines two well unique limits of transmission: 
\begin{itemize}
	\item[i.] $S(\bb)=0$ which means no transmission, \tit{viz.} it corresponds to a completely opaque disk, 

	\item[ii.] $S(\bb)=1$ which means full transmission, \tit{viz.} it corresponds a completely transparent screen.
\end{itemize}

Therefore, the wave function beyond the obstacle is
\be
U(\rr)=-\frac{ik}{2\pi}\,U_{0}\,\frac{e^{ikr}}{r}\int d^{2}{\bb}\,S(\bb)\,e^{-{\qq}\cdot{\bb}}.
\label{apx1.5}
\ee
\mbox{\,\,\,\,\,\,\,\,\,}
It is not difficult to see that if one uses conservation of energy, this last expression could be separated into its incident and scattered components,
\bear
U(\rr)&=&U_{incident}+U_{scattered}\nonumber\\
&=&U_{0}\left(e^{ikz}+f(\qq)\,\frac{e^{ikr}}{r}\right),
\label{apx1.6}
\eear
where the factor $f(\qq)$ is physically known as the scattering amplitude,
\be
f(\qq)=\frac{ik}{2\pi}\int d^{2}\bb\,[1-S(\bb)]\,e^{-i\qq\cdot\bb}.
\label{apx1.7}
\ee
\mbox{\,\,\,\,\,\,\,\,\,}
Here, the quantity in square brackets in the \tit{rhs} of expression (\ref{apx1.7}) is defined as the profile function $\Gamma(\bb)$ which determines the diffracted wave. Moreover, it is easy to see that the scattering amplitude and the profile function are related to each other by means of the Fourier transform,
\be
f(\qq)=\frac{1}{2\pi ik}\int d^{2}\qq\,f(\qq)\,e^{i\qq\cdot\bb},
\label{apx1.8}
\ee
or either by the Hankel Transform\footnote{In general we are interested in problems with azimuthal symmetry, \tit{i.e.}, $\Gamma(\bb,\phi)\equiv\Gamma(\bb)$.},
\be
f(\qq)=ik\int^{\infty}_{0}db\,b\,J_{0}(qb)\,\Gamma(\bb),
\label{apx1.9}
\ee
where $J_{0}(x)$ is the Bessel function of the first kind and zeroth-order. It is straighforward to demonstrate this last line. Notice, however, that there are a bunch of ways of writing Bessel functions\cite{california1954tables}, but one particular form is
\be
J_{n}(x)=\frac{1}{2\pi}\int^{\pi}_{-\pi} d\phi\,e^{i(n\phi-x\sin\phi)}.
\label{apx1.10}
\ee
If one performs a phase shift of $\phi^{\prime}=\pi$,
\bear
J_{n}(x) & =&\frac{1}{2\pi}\int^{2\pi}_{0} d(\phi+\pi)\,e^{i\left[n(\phi+\pi)-x\sin(\phi+\pi)\right]}\nonumber\\
& = &\frac{1}{2\pi}\int^{2\pi}_{0} d\phi\,e^{i(n(\phi+\pi)+x\cos\phi)},
\label{apx1.11}
\eear
and finally for $n=0$,
\be
J_{0}(x)=\frac{1}{2\pi}\int^{2\pi}_{0} d\phi\,e^{ix\cos\phi}.
\label{apx1.12}
\ee

At this point, it is of extremely importance to make very clear that the scattering amplitude is the most relevant physical quantity in the processes. It is by means of expression (\ref{apx1.9}) that one may obtain all the information necessary to fully describe the scattering.

\section{\textsc{Cross Sections}}
\label{chAPX1.5}

\mbox{\,\,\,\,\,\,\,\,\,}
By definition the total cross section is written as the summation of the elastic and the absorption cross sections, \tit{i.e.}, it is linked to the partial amount of energy of the incident flux which was scattered and the amount absorbed,
\be
\sigma_{tot}=\sigma_{el}+\sigma_{abs}.
\label{apx1.13}
\ee
\mbox{\,\,\,\,\,\,\,\,\,}
In order to reveal this two missing pieces and once and for all find the ultimate expression for the total cross section, because this is kind of the meaning of this whole discussion so far, which is, using just classical optical theory, to demonstrate something called as the optical theorem, but first off one needs to look at the elastic differential cross section. It is defined as the ratio of the outgoing energy in an element of solid angle $d\Omega$ to the incident energy flux, so it is equivalent of saying that the differential cross section is simply the modulus squared of the scattering amplitude,
\be
\frac{d\sigma}{d\Omega}=\vert f(\qq)\vert^{2}.
\label{apx1.14}
\ee
\mbox{\,\,\,\,\,\,\,\,\,}
It does not seem so silly to integrate this last expression (\ref{apx1.14}). Actually this is not silly at all because when one does such thing ends up finding the scattered or elastic cross section,
\be
\sigma_{el}\equiv\int d\Omega\,\frac{d\sigma}{d\Omega}=\frac{1}{k^{2}}\,\int d^{2}\qq\,\vert f(\qq)\vert^{2}.
\label{apx1.15}
\ee
If the obstacle is considered a disk,
\be
\sigma_{el}=\int d^{2}\bb\,\vert\Gamma(\bb)\vert^{2}=\int d^{2}\bb\,\vert1-S(\bb)\vert^{2}.
\label{apx1.16}
\ee
\mbox{\,\,\,\,\,\,\,\,\,}
One thing that worths to be mentioned is that there is also the possibility that some kind of an absorption event happens during the scattering process. Hence, there will be a cross section associated with the probability of absorption occurrence such that, by means of conservation of probability, $P_{abs}=1-P_{trans}$ where $P_{trans}=\vert S(\bb)\vert^{2}$ is the probability of transmission, and then the absorption cross section is given by
\be
\begin{split}
\sigma_{abs}& =\int d^{2}\bb\,\left(1-\vert S(\bb)\vert^{2}\right)\\
& = \int d^{2}\bb\,\left[2\,\text{Re}\,\Gamma(\bb)-\vert \Gamma(\bb)\vert^{2}\right].
\end{split}
\label{apx1.17}
\ee
\mbox{\,\,\,\,\,\,\,\,\,}
The last piece is finally revealed and then the cross section can be written by summing up expressions (\ref{apx1.16}) and (\ref{apx1.17}),
\be
\begin{split}
\sigma_{tot}& =2\int d^{2}\bb\,(1-\text{Re}\,S(\bb))\\
& = 2\int d^{2}\bb\,\text{Re}\,\Gamma(\bb),
\end{split}
\label{1.18}
\ee
and then, the form which is typically known as the optical theorem \cite{Barone:2002cv} is given by
\be
\sigma_{tot}=\frac{4\pi}{k}\,\text{Im}\,f(\theta=0),
\label{apx1.19}
\ee
because from expression (\ref{apx1.9}),
\be
\text{Re}\,f(\qq=0)+i\,\text{Im}\,f(\qq=0)=\frac{ik}{2\pi}\int d^{2}\bb\,\left[\text{Re}\,\Gamma(\bb)+i\,\text{Im}\,\Gamma(\bb)\right].
\label{1.20}
\ee
\mbox{\,\,\,\,\,\,\,\,\,}
The optical theorem, which is in fact a consequence of the conservation of energy, states that the total cross section is connected to the scattering amplitude in the forward direction, \tit{i.e.}, for $\theta=0$ corresponds to $\qq=0$.

Summarizing, the cross sections expressions found here are exactly the same as those ones that were found in the Chapter \ref{ch2} via the quantum mechanical formalism in the high-energy limit.

\clearpage
\thispagestyle{plain}

\chapter{\textsc{General Properties of the Scattering Amplitude}} 
\label{APX3}

\mbox{\,\,\,\,\,\,\,\,\,}
In the early $60$'s the scattering $S$-matrix program was an attempt to pursue a theoretical approach towards a complete description of strong interactions and as motivation the belief that scattering should become simpler as the energy increases. At high-energy, it is known that a huge number of partial waves is involved, and therefore an accurate knowledge of each one of them becomes unpractical and a small number of parameters should be enough to satisfactorily describe high-energy collisions. At present, the low-momentum transfer processes which contribute to most of the total cross section and to diffractive reactions cannot be treated perturbatively and calculated in a reliable way only within QCD \cite{Matthiae:1994uw}. The original program of $S$-matrix failed as it was first conceived, but its fundamental principles remain even today in the background of hadronic physics.  

In principle it would be possible to reconstruct the entire dynamics of the interaction processes if, and only if, one has full knowledge of the $S$-matrix. Basically, it is an unitary operator, $S^{\dagger}S=SS^{\dagger}=\mathds{1}$, which transforms an initial state $\vert i \rangle$ into the final state $\vert f \rangle$ of a given scattering processes,
\be
S\vert i \rangle=\vert f \rangle,
\label{apx3.1}
\ee
where these states are defined at the time $t_{i}\to-\infty$ and $t_{f}\to+\infty$, respectively. Therefore they represent free particles before and after the interaction, and form a complete set of states. 

The $S$-matrix can be related to the time evolution unitary operator $U(t_{i},t_{f})$, which connects a state at a given time $t_{i}$ to time $t_{f}$,
\be
S=U(-\infty,+\infty),
\label{apx3.2}
\ee 
and in quantum field theory is given by the Dyson series,
\be
S=\mathds{1}+\sum^{\infty}_{n=1}\frac{i^{n}}{n!}\int d^{4}x_{1}...d^{4}x_{n}\tau(H^{\prime}_{int}(x_{1})...H^{\prime}_{int}(x_{n})),
\label{apx3.3}
\ee
where $\tau$ stands for the time-ordered product and $H^{\prime}_{int}$ is the interaction Hamiltonian.

In addition to relativistic invariance, which means that the $S$-matrix elements must depend on Lorentz-invariant combinations of the kinematic variables, the other  fundamental principles which are usually assumed in this context are given by the following:
\bi

\item[i.] Unitarity of the $S$-matrix, which is a consequence of the principle of conservation of probability. The probability that after the interaction process the system, which is initially in the in-state $\vert i \rangle$, passes to the out-state $\vert f \rangle$ is given by
\be
P_{i\to f}=\vert\langle i \vert S \vert f \rangle \vert^{2}.
\label{apx3.4}
\ee
Moreover, if one takes an orthonormal basis of vectors $\vert k \rangle$, the probability to go from an arbitrarily initial state to any of the states $\vert k \rangle$ must sum up to unity,
\be
\sum_{k}P_{i\to k}=\sum_{k}\vert\langle k \vert S \vert i \rangle \vert^{2}=\sum_{k}\langle i \vert S^{\dagger} \vert k \rangle \langle k \vert S \vert i \rangle = \langle i \vert S^{\dagger}S \vert i \rangle = 1.
\label{apx3.5}
\ee

\item[ii.] Analyticity, which originates from the principle of causality. It states that the $S$-matrix elements when expressed as functions of the kinematic variables can be analytically continued to the complex domain and the resulting analytic function has a simples singularity structure or at least the simplest one which is consistent with unitarity.

\item[iii.] Crossing symmetry, which states that the same invariant amplitudes describing elastic scattering in the $s$, $u$ and $t$ channels are actually embodied in a single analytic function expressed as $F(s,t,u)$.

\ei


\section{\textsc{Dispersion Relations}}
\label{secAPX3.1}
\mbox{\,\,\,\,\,\,\,\,\,}
A direct and very important consequence of analyticity are the dispersion relations, which relate the real and the imaginary part of the scattering amplitude. The first appearance of dispersion relations in Physics was in the $20$'s through out the work from H.A. Kramers and R. Kronig about the scattering of light by a dispersive medium. That is the reason why it is known as dispersion relations. Over the years these type of relations have been used in many different areas of modern physics and have found its way into particle physics. 

\subsection{\textsc{Integral Dispersion Relations}}
\label{secAPX3.1.1}
\mbox{\,\,\,\,\,\,\,\,\,}
Complex functions of real variables are usually found in physical systems. Although sometimes one can obtain informations on the general properties of a system if the argument of the function is complex, in physics, however, the experimental data are represented by real arguments. Therefore, it is important to verify if it is possible to connect quantities with actual physical meaning. This connection can be properly done by means of the Hilbert transform, or either known as Sokhotski-Plemelj theorem \cite{byron1992mathematics}.

Let $f(z)$ be an analytic function in the upper-half plane of a complex plane and over the $x$-axis such that $\vert f(z)\vert\to0$ at $\vert z\vert\to\infty$, and also let $C$ be a smooth closed curve,
\be
i\pi f(\alpha)=\lim\limits_{\substack{R\to \infty \\r\to 0}}\left\{\int^{\alpha-r}_{-R} dx\, \frac{f(x)}{x-\alpha}+\int^{+R}_{\alpha+r} dx\,\frac{f(x)}{x-\alpha}\right\},
\label{apx3.6}
\ee
where $R$ defines the range of the contour over the complex plane and $r$ defines the ratio of the contour over the pole. The sum of the integrals represents the principal value of Cauchy,
\be
i\pi f(\alpha)={\cal P}\int^{+\infty}_{-\infty} dx\,\frac{f(x)}{x-\alpha},
\label{apx3.7}
\ee
where $\alpha$ represents a simple pole. Since $f(x)$ is a complex function of a real value, there is no reason why one cannot write $f(x)\equiv\textnormal{Re}\,f(x)+i\,\textnormal{Im}\,f(x)$. Thus equating one finds a relation between the real and imaginary parts of the complex function,
\be
\textnormal{Re}\,f(\alpha)=\frac{1}{\pi}\,{\cal P}\int^{+\infty}_{-\infty} dx\,\frac{\textnormal{Im}\,f(x)}{x-\alpha},
\label{apx3.8}
\ee
and the inverse relation,
\be
\textnormal{Im}\,f(\alpha)=-\frac{1}{\pi}\,{\cal P}\int^{+\infty}_{-\infty} dx\,\frac{\textnormal{Re}\,f(x)}{x-\alpha},
\label{apx3.9}
\ee
\mbox{\,\,\,\,\,\,\,\,\,}
However, in some situations physical systems demand a change in the limits of integration like $(-\infty,+\infty)\to[0,+\infty)$. Therefore, $f(x)$ must obey the symmetrical relations imposed by the Schwartz reflection principle which states that if $f(z)$ is an analytic complex function, then $f^{\ast}(z)=f(z^{\ast})$. More specifically, scattering amplitudes in the case of $pp$ and $\bar{p}p$ scatterings are written in terms of even and odd functions connected by crossing symmetry where $f^{\bar{p}p}_{pp}=f^{+}\pm f^{-}$. Thus, the complex function $f(z)$ can be written as
\be
\begin{split}
f(z)&=\textnormal{Re}\,f(z)+i\,\textnormal{Im}\,f(z)\\
&=\left[\textnormal{Re}\,f_{+}(z)+\textnormal{Re}\,f_{-}(z)\right]+i\,\left[\textnormal{Im}\,f_{+}(z)+\textnormal{Im}\,f_{-}(z)\right]\\
&=\left[\textnormal{Re}\,f_{+}(z)+i\,\textnormal{Im}\,f_{+}(z)\right]+\left[\textnormal{Re}\,f_{-}(z)+i\,\textnormal{Im}\,f_{-}(z)\right]\\
&=f_{+}(z)+f_{-}(z).
\end{split}
\label{apx3.10}
\ee
\mbox{\,\,\,\,\,\,\,\,\,}
In the case of the even part of $f(z)$, \tit{i.e} when the even complex function behaves as $f_{+}(z)=f_{+}(-z)$, the Schwartz reflection principle implies that
\be
f^{\ast}_{+}(z)=f_{+}(-z^{\ast}),
\label{apx3.11}
\ee
and by applying it to the real $x$-axis,
\be
f^{\ast}_{+}(x)=f_{+}(-x^{\ast})=f_{+}(-x),
\label{apx3.12}
\ee
which gives
\be
f^{\ast}_{+}(x)=\textnormal{Re}\,f_{+}(x)-i\,\textnormal{Im}\,f_{+}(x),
\label{apx3.13}
\ee
\be
f_{+}(-x)=\textnormal{Re}\,f_{+}(-x)+i\,\textnormal{Im}\,f_{+}(-x),
\label{apx3.14}
\ee
and by means of equation (\ref{apx3.12}), one arrives at the symmetrical relations for the even term of the complex function $f(z)$,
\be
\textnormal{Re}\,f_{+}(x)=\textnormal{Re}\,f_{+}(-x),
\label{apx3.15}
\ee
\be
\textnormal{Im}\,f_{+}(x)=-\textnormal{Im}\,f_{+}(-x).
\label{apx3.16}
\ee
\mbox{\,\,\,\,\,\,\,\,\,}
Similarly, in the case of the odd part of $f(z)$, \tit{i.e.} $f_{-}(z)=-f_{-}(-z)$, then
\be
f^{\ast}_{-}(z)=-f_{-}(-z^{\ast}),
\label{apx3.17}
\ee
\be
f^{\ast}_{-}(x)=-f_{-}(-x^{\ast})=-f_{-}(-x).
\label{apx3.18}
\ee
which gives
\be
f^{\ast}_{-}(x)=\textnormal{Re}\,f_{-}(x)-i\,\textnormal{Im}\,f_{-}(x),
\label{apx3.19}
\ee
\be
f_{-}(-x)=\textnormal{Re}\,f_{-}(-x)+i\,\textnormal{Im}\,f_{-}(-x),
\label{apx3.20}
\ee
and by means of equation (\ref{apx3.12}), similarly one arrives at the odd symmetrical relations,
\be
\textnormal{Re}\,f_{-}(x)=-\textnormal{Re}\,f_{-}(-x),
\label{apx3.21}
\ee
\be
\textnormal{Im}\,f_{-}(x)=\textnormal{Im}\,f_{-}(-x).
\label{apx3.22}
\ee
\mbox{\,\,\,\,\,\,\,\,\,}
By returning to expressions (\ref{apx3.8}) and (\ref{apx3.9}), separating the limits of integration into two intervals and considering the even function,
\be
\textnormal{Re}\,f_{+}(\alpha)=\frac{1}{\pi}\,{\cal P}\int^{0}_{-\infty} dx\,\frac{\textnormal{Im}\,f_{+}(x)}{x-\alpha}+\frac{1}{\pi}\,{\cal P}\int^{+\infty}_{0} dx\,\frac{\textnormal{Im}\,f_{+}(x)}{x-\alpha},
\label{apx3.23}
\ee
\be
\textnormal{Im}\,f_{+}(\alpha)=-\frac{1}{\pi}\,{\cal P}\int^{0}_{-\infty} dx\,\frac{\textnormal{Re}\,f_{+}(x)}{x-\alpha}-\frac{1}{\pi}\,{\cal P}\int^{+\infty}_{0} dx\,\frac{\textnormal{Re}\,f_{+}(x)}{x-\alpha},
\label{apx3.24}
\ee
thus by switching $x\to-x$ in the first integral on the \tit{rhs} and within the even symmetrical relations, a straightforward calculation leads to the following integral dispersion relations:
\be
\textnormal{Re}\,f_{+}(\alpha)=\frac{2}{\pi}\,{\cal P}\int^{\infty}_{0} dx\,\frac{x}{x^{2}-\alpha^{2}}\,\textnormal{Im}\,f_{+}(x),
\label{apx3.25}
\ee
\be
\textnormal{Im}\,f_{+}(\alpha)=-\frac{2\alpha}{\pi}\,{\cal P}\int^{\infty}_{0} dx\,\frac{1}{x^{2}-\alpha^{2}}\,\textnormal{Re}\,f_{+}(x).
\label{apx3.26}
\ee
\mbox{\,\,\,\,\,\,\,\,\,}
These expressions represent even integral dispersion relations with a single pole-like singularity in the real axis. The necessary condition adopted to obtain dispersion relations is the request that $f(z)$ be a well-behaved function, \tit{i.e.} $f(z)$ must be analytic at $\textnormal{Im}\,z\geq 0$ and $\vert f(z)\vert\to 0$ at $\vert z \vert\to\infty$. A similar calculation for the odd function would therefore be written by the following relations:
\be
\textnormal{Re}\,f_{-}(\alpha)=\frac{2\alpha}{\pi}\,{\cal P}\int^{\infty}_{0} dx\,\frac{1}{x^{2}-\alpha^{2}}\,\textnormal{Im}\,f_{-}(x),
\label{apx3.27}
\ee
\be
\textnormal{Im}\,f_{-}(\alpha)=-\frac{2}{\pi}\,{\cal P}\int^{\infty}_{0} dx\,\frac{x}{x^{2}-\alpha^{2}}\,\textnormal{Re}\,f_{-}(x),
\label{apx3.28}
\ee
which represent odd integral dispersion relations with a single pole-like singularity in the real axis.

However, is common to work with complex functions $f(z)$ which are not well-behaved, \tit{i.e.} they do not converge at $\vert z \vert\to\infty$, and for this reason the approach must be slightly modified. As for example, this is the case of a limited function where $\vert f(z)\vert= \textnormal{constant}$ asymptotically for $\vert z \vert\to\infty$. To begin with, it is necessary to find a function which is asymptotically well-behaved, \tit{i.e.} analytic in the upper-half complex plane and vanishes at large values of $\vert z\vert$. Thus, defining a function $\varphi(z)$ such that
\be
\varphi(z)\equiv\frac{f(z)-f(\alpha_{0})}{z-\alpha_{0}},
\label{apx3.29}
\ee
not singular at $z=\alpha_{0}$ once it is stated that
\be
\lim_{z\to \alpha_{0}}\frac{f(z)-f(\alpha_{0})}{z-\alpha_{0}}=f^{\prime}(\alpha_{0}).
\label{apx3.30}
\ee
Then, implying that $\varphi(z)$ is, indeed, analytic in the upper-half complex plane. By taking the limit for large values of $\vert z \vert$,
\be
\lim_{\vert z\vert\to \infty}\varphi(z)=\lim_{\vert z\vert\to \infty}\frac{\vert f(z)-f(\alpha_{0})\vert }{\vert z-\alpha_{0}\vert },
\label{apx3.31}
\ee
and using the Cauchy-Schwartz inequality,
\be
\begin{split}
&\vert A-B\vert=\vert A\vert -\vert B\vert,\,\,\, \textnormal{if}\,\,\,B\geq0,\\
&\vert A-B\vert>\vert A\vert -\vert B\vert,\,\,\, \textnormal{if}\,\,\,B<0,\\
&\vert A-B\vert\geq\vert A\vert -\vert B\vert.
\end{split}
\label{apx3.32}
\ee
then one is able to prove that
\be
\lim_{\vert z\vert\to \infty}\varphi(z)\leq\lim_{\vert z\vert\to \infty}\frac{\vert f(z)-f(\alpha_{0})\vert }{\vert z\vert-\vert\alpha_{0}\vert }=0.
\label{apx3.33}
\ee
Hence, expressions (\ref{apx3.20}) and (\ref{apx3.23}) show that $\varphi(z)$ has the required properties and can be written by means of a dispersion relation,
\be
i\pi\varphi(\alpha)={\cal P}\int^{+\infty}_{-\infty}dx\,\frac{\varphi(x)}{(x-\alpha)},
\label{apx3.34}
\ee
which is exactly as
\be
\begin{split}
\!\!\!\!\!\!\!\!\!\!f(\alpha)-f(\alpha_{0})&=\frac{\alpha-\alpha_{0}}{i\pi}\,{\cal P}\int^{+\infty}_{-\infty}dx\,\frac{f(x)-f(\alpha_{0})}{(x-\alpha)(x-\alpha_{0})}\\
&=\frac{\alpha-\alpha_{0}}{i\pi}\,{\cal P}\int^{+\infty}_{-\infty}dx\,\frac{f(x)}{(x-\alpha)(x-\alpha_{0})}+\frac{f(\alpha_{0})}{i\pi}\,{\cal P}\int^{+\infty}_{-\infty}dx\,\left[\frac{1}{(x-\alpha)}-\frac{1}{(x-\alpha_{0})}\right].
\end{split}
\label{apx3.35}
\ee
It is easy to see that the second integral in the \tit{rhs} vanishes because of
\be
\int^{+\infty}_{-\infty}dx\,\frac{1}{x-y}=\lim_{\e\to\infty}\int^{+\e}_{-\e}dx\,\frac{1}{x-y}=\lim_{\e\to+\infty}\log\vert\e-y\vert-\lim_{\e\to+\infty}\log\vert\e+y\vert=0.
\label{apx3.36}
\ee
By means of the above result, the dispersion relation takes the form
\be
f(\alpha)=f(\alpha_{0})+\frac{\alpha-\alpha_{0}}{i\pi}\,{\cal P}\int^{+\infty}_{-\infty}dx\,\frac{f(x)}{(x-\alpha)(x-\alpha_{0})},
\label{apx3.37}
\ee
and by following the same procedure as before, one arrives at the relation between the real and the imaginary parts of the complex function,
\be
\textnormal{Re}\,f(\alpha)=\textnormal{Re}\,f(\alpha_{0})+\frac{\alpha-\alpha_{0}}{\pi}\,{\cal P}\int^{+\infty}_{-\infty}dx\,\frac{\textnormal{Im}\,f(x)}{(x-\alpha)(x-\alpha_{0})},
\label{apx3.38}
\ee
and the inverse relation,
\be
\textnormal{Im}\,f(\alpha)=\textnormal{Re}\,f(\alpha_{0})-\frac{\alpha-\alpha_{0}}{\pi}\,{\cal P}\int^{+\infty}_{-\infty}dx\,\frac{\textnormal{Re}\,f(x)}{(x-\alpha)(x-\alpha_{0})}.
\label{apx3.39}
\ee
\mbox{\,\,\,\,\,\,\,\,\,}
These expressions define the so-called class of singly subtracted integral dispersion relations. In general, for those cases where the complex function behaves as $\vert f(z)\vert\sim z^{\lambda}$ at $\vert z\vert\to\infty$, then in principle it is necessary to make $N$ subtractions so that $f(z)$ converges asymptotically for large $\vert z \vert$. As before, by changing the limits of integration to physical limits, \tit{i.e.}, to $(-\infty,+\infty)\to[0,+\infty)$, and using the even and odd symmetrical relation imposed by Schwartz reflection principle, a straightforward calculation for the even function would easily give
\be
\textnormal{Re}\,f_{+}(\alpha)=\textnormal{Re}\,f_{+}(\alpha_{0})+\frac{2}{\pi}\,{\cal P}\int^{+\infty}_{0}dx\,\frac{x(\alpha^{2}-\alpha^{2}_{0})}{(x^{2}-\alpha^{2})(x^{2}-\alpha^{2}_{0})}\,\textnormal{Im}\,f_{+}(x),
\label{apx3.40}
\ee
and also
\be
\textnormal{Im}\,f_{+}(\alpha)=\textnormal{Im}\,f_{+}(\alpha_{0})-\frac{2}{\pi}\,{\cal P}\int^{+\infty}_{0}dx\,\frac{(\alpha-\alpha_{0})(x^{2}-\alpha\alpha_{0})}{(x^{2}-\alpha^{2})(x^{2}-\alpha^{2}_{0})}\,\textnormal{Re}\,f_{+}(x).
\label{apx3.41}
\ee
\mbox{\,\,\,\,\,\,\,\,\,}
Just for curiosity, by switching $\alpha\to s$, $x\to s^{\prime}$ and $\alpha_{0}\to 0$, one arrives at the energy integral dispersion relations for the scattering amplitude with one subtraction constant, respectively,
\be
\textnormal{Re}\,f_{+}(s)=\textnormal{Re}\,f_{+}(0)+\frac{2s^{2}}{\pi}\,{\cal P}\int^{+\infty}_{0}ds^{\prime}\,\frac{\textnormal{Im}\,f_{+}(s^{\prime})}{s^{\prime}(s^{\prime 2}-s^{2})},
\label{apx3.42}
\ee
and also
\be
\textnormal{Im}\,f_{+}(s)=\textnormal{Im}\,f_{+}(0)-\frac{2s}{\pi}\,{\cal P}\int^{+\infty}_{0}ds^{\prime}\,\frac{\textnormal{Re}\,f_{+}(s^{\prime})}{s^{\prime 2}-s^{2}}.
\label{apx3.43}
\ee
\mbox{\,\,\,\,\,\,\,\,\,}
As for the case of the odd function, one arrives at
\be
\textnormal{Re}\,f_{-}(\alpha)=\textnormal{Re}\,f_{-}(\alpha_{0})+\frac{2}{\pi}\,{\cal P}\int^{+\infty}_{0}dx\,\frac{(\alpha-\alpha_{0})(x^{2}-\alpha\alpha_{0})}{(x^{2}-\alpha^{2})(x^{2}-\alpha^{2}_{0})}\,\textnormal{Im}\,f_{-}(x),
\label{apx3.44}
\ee
and
\be
\textnormal{Im}\,f_{-}(\alpha)=\textnormal{Im}\,f_{-}(\alpha_{0})-\frac{2}{\pi}\,{\cal P}\int^{+\infty}_{0}dx\,\frac{x(\alpha^{2}-\alpha^{2}_{0})}{(x^{2}-\alpha^{2})(x^{2}-\alpha^{2}_{0})}\,\textnormal{Re}\,f_{-}(x),
\label{apx3.45}
\ee
and finally applying the same switching prescription,
\be
\textnormal{Re}\,f_{-}(s)=\textnormal{Re}\,f_{-}(0)+\frac{2s}{\pi}\,{\cal P}\int^{+\infty}_{0}ds^{\prime}\,\frac{\textnormal{Im}\,f_{-}(s^{\prime})}{s^{\prime 2}-s^{2}},
\label{apx3.46}
\ee
and also
\be
\textnormal{Im}\,f_{-}(s)=\textnormal{Im}\,f_{-}(0)-\frac{2s^{2}}{\pi}\,{\cal P}\int^{+\infty}_{0}ds^{\prime}\,\frac{\textnormal{Re}\,f_{-}(s^{\prime})}{s^{\prime}(s^{\prime 2}-s^{2})}.
\label{apx3.47}
\ee
\mbox{\,\,\,\,\,\,\,\,\,}
Before pressing on, if ${\cal F}(E)$ is the analytic continuation of the forward elastic scattering amplitude, $f(E,t=0)$, the $pp$ and $\bar{p}p$ forward amplitudes are the limits of the analytic function ${\cal F}$ according to
\be
f^{\bar{p}p}_{pp}(E,t=0)=\lim_{\e\to0}{\cal F}(\mp E\mp i\e,t=0).
\label{apx3.48}
\ee
The Cauchy theorem implies that
\be
{\cal F}(E)=\frac{1}{2\pi i}\oint dE^{\prime}\,\frac{{\cal F}(E^{\prime})}{E^{\prime}-E},
\label{apx3.49}
\ee
and after choosing an appropriate contour, see Figure \ref{apx3fig1}, the above expression can be written as
\be
{\cal F}(E)=\frac{1}{2\pi i}\,\left[\int^{R}_{m}dE^{\prime}\,\frac{{\cal F}(E^{\prime}+i\e)-{\cal F}(E^{\prime}-i\e)}{E^{\prime}-E}+\int^{-m}_{-R}dE^{\prime}\,\frac{{\cal F}(E^{\prime}+i\e)-{\cal F}(E^{\prime}-i\e)}{E^{\prime}-E}\right],
\label{apx3.50}
\ee
where $E=-m$ and $E=m$ are cuts on the real axis and $R\to \infty$. For an even amplitude, ${\cal F}={\cal F}^{+}$, \tit{i.e.} ${\cal F}(E^{\prime}+i\e)={\cal F}(-E^{\prime}-i\e)$, thus \cite{Block:1984ru}
\be
{\cal F}^{+}(E)=\frac{1}{\pi}\int^{\infty}_{m}dE^{\prime}\,\left[\frac{1}{E^{\prime}-E}+\frac{1}{E^{\prime}+E}\right]\textnormal{Im}\,{\cal F}^{+}(E^{\prime}+i\e),
\label{apx3.51}
\ee
and for an odd amplitude, ${\cal F}={\cal F}^{-}$, \tit{i.e.} ${\cal F}(E^{\prime}+i\e)=-{\cal F}(-E^{\prime}-i\e)$,
\be
{\cal F}^{-}(E)=\frac{1}{\pi}\int^{\infty}_{m}dE^{\prime}\,\left[\frac{1}{E^{\prime}-E}-\frac{1}{E^{\prime}+E}\right]\textnormal{Im}\,{\cal F}^{-}(E^{\prime}+i\e),
\label{apx3.52}
\ee
where the Schwartz reflection principle implies that 
\be
{\cal F}(E^{\prime}+i\e)-{\cal F}(E^{\prime}-i\e)={\cal F}(E^{\prime}+i\e)-{\cal F}^{\ast}(E^{\prime}+i\e)=2i\,\textnormal{Im}\,{\cal F}(E^{\prime}-i\e).
\label{apx3.53}
\ee
The real and imaginary parts of $f(E)$ are connected by the dispersion relations,
\be
\textnormal{Re}\,f^{+}(E)=\frac{2}{\pi}\,{\cal P}\int^{\infty}_{m}dE^{\prime}\,\frac{E^{\prime}}{E^{\prime 2}-E^{2}}\,\textnormal{Im}\,f^{+}(E^{\prime}),
\label{apx3.54}
\ee
and
\be
\textnormal{Re}\,f^{-}(E)=\frac{2}{\pi}\,{\cal P}\int^{\infty}_{m}dE^{\prime}\,\frac{E}{E^{\prime 2}-E^{2}}\,\textnormal{Im}\,f^{-}(E^{\prime}),
\label{apx3.55}
\ee
which is exactly expression (\ref{apx3.25}) and (\ref{apx3.27}), except for the lower limit of integration. Hence, taking the limit $E\gg m$ and changing the variable from $E\to s$, 
\be
\textnormal{Re}\,f^{+}(s)=\frac{2}{\pi}\,{\cal P}\int^{\infty}_{s_{0}}ds^{\prime}\,\frac{s^{\prime}}{s^{\prime 2}-s^{2}}\,\textnormal{Im}\,f^{+}(s^{\prime}),
\label{apx3.56}
\ee
and
\be
\textnormal{Re}\,f^{-}(s)=\frac{2}{\pi}\,{\cal P}\int^{\infty}_{s_{0}}ds^{\prime}\,\frac{s}{s^{\prime 2}-s^{2}}\,\textnormal{Im}\,f^{-}(s^{\prime}).
\label{apx3.57}
\ee
And by means of the same procedure, the singly subtracted integral dispersion relations are given by
\be
\textnormal{Re}\,f^{+}(s)=\textnormal{Re}\,f^{+}(0)+\frac{2s^{2}}{\pi}\,{\cal P}\int^{+\infty}_{s_{0}}ds^{\prime}\,\frac{\textnormal{Im}\,f^{+}(s^{\prime})}{s^{\prime}(s^{\prime 2}-s^{2})},
\label{apx3.58}
\ee
\be
\textnormal{Re}\,f^{-}(s)=\textnormal{Re}\,f^{-}(0)+\frac{2s}{\pi}\,{\cal P}\int^{+\infty}_{s_{0}}ds^{\prime}\,\frac{\textnormal{Im}\,f^{-}(s^{\prime})}{s^{\prime 2}-s^{2}}.
\label{apx3.59}
\ee
\mbox{\,\,\,\,\,\,\,\,\,}
The same thing goes to obtain the reverse dispersion relations, but applying the Cauchy theorem into an even function such as $\phi^{+}(E)=(m^{2}=E^{2})^{-1/2}\,{\cal F}^{+}(E)$ \cite{Block:1984ru}.
\bfg[hbtp]
  \begin{center}
    \includegraphics[width=10cm,clip=true]{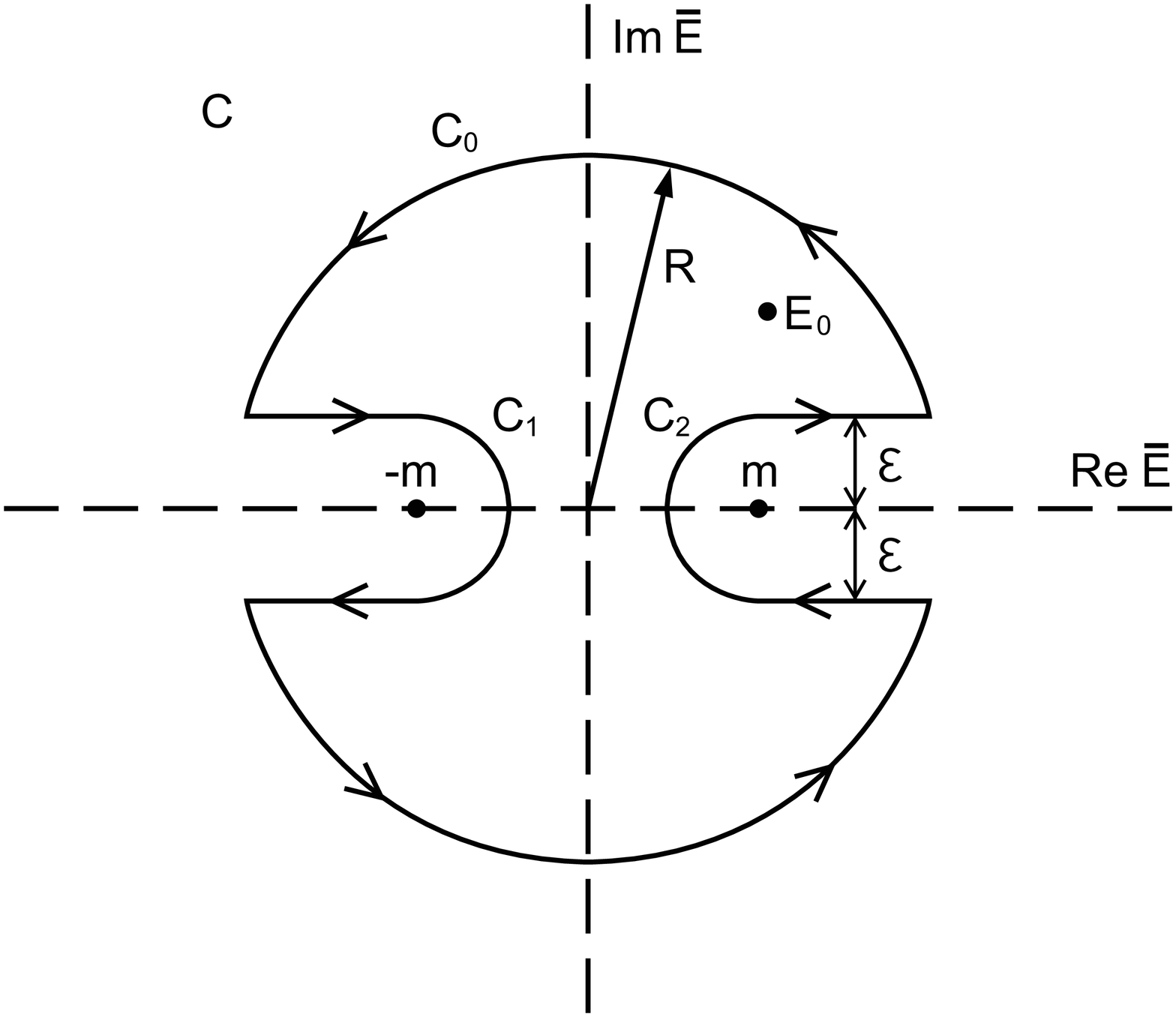}
    \caption{Complex $E$-plane and the contour over the pole singularities in the real axis.}
    \label{apx3fig1}
  \end{center}
\efg

\subsection{\textsc{Derivative Dispersion Relations}}
\label{secAPX3.1.2}
\mbox{\,\,\,\,\,\,\,\,\,}
Although the smooth increase of the total cross section experimental data at energies above $\sqrt{s}\sim20$ GeV, roughly as $\sim\log^{2}s$ and that $\sigma^{\bar{p}p}_{tot}-\sigma^{pp}_{tot}\sim0$, suggest the use of integral dispersion relations with only one subtraction constant, its nonlocal character limits the class of functions which allows analytical integration \cite{Avila:2003cu}. Under some conditions, the integral dispersion forms may be replaced by quasi-local ones, expressed in a derivative form. This is named derivative dispersion relation or either sometimes called by analyticity relations.

By considering the even amplitude in expression (\ref{apx3.58}) and rewriting $s^{\prime}=e^{\xi^{\prime}}$, $s=e^{\xi}$ and $g(\xi^{\prime})=\textnormal{Im}\,f^{+}(e^{\xi^{\prime}})/e^{\xi^{\prime}}$,
\bear
\textnormal{Re}\,f^{+}(e^{\xi})-K&=&\frac{2e^{\xi^{\prime}}}{\pi}\,{\cal P}\int^{\infty}_{\log s_{0}}d\left(e^{\xi^{\prime}}\right)\frac{\textnormal{Im}\,f^{+}(e^{\xi^{\prime}})}{e^{\xi^{\prime}}\left(e^{\xi^{\prime 2}}-e^{2\xi}\right)},\nonumber\\
&=&\frac{e^{\xi}}{\pi}\,{\cal P}\int^{\infty}_{\log s_{0}}d\xi^{\prime}\,\frac{g(\xi^{\prime})}{\sinh\left(\xi^{\prime}-\xi\right)},
\label{apx3.60}
\eear
where $K$ stands for the subtraction constant. In the case that $g(x)$ is an analytic function of its argument, then it can be expanded as
\be
g(x^{\prime})=\sum^{\infty}_{n=0}\,\frac{d^{n}}{dx^{\prime n}}\,g(x^{\prime})\bigg\vert_{x^{\prime}=x}\frac{\left(x^{\prime}-x\right)^{n}}{n!}.
\label{apx3.61}
\ee
Within the above expansion and also by considering the high-energy limit where the mass of the particles can be neglected, $s_{0}\to 0$, thus expression (\ref{apx3.60}) reads
\be
\textnormal{Re}\,f^{+}(e^{\xi})-K=\frac{e^{\xi}}{\pi}\sum^{\infty}_{n=0}\frac{g^{(n)}(\xi)}{n!}\,{\cal P}\int^{\infty}_{-\infty}d\xi^{\prime}\,\frac{\left(\xi^{\prime}-\xi\right)^{n}}{\sinh\left(\xi^{\prime}-\xi\right)}.
\label{apx3.62}
\ee
The above expression can be put into the following form if one defines $y=\left(\xi^{\prime}-\xi\right)$,
\be
\textnormal{Re}\,f^{+}(e^{\xi})-K=e^{\xi}\sum^{\infty}_{n=0}\frac{g^{(n)}(\xi)}{n!}\,I_{n}(y),
\label{apx3.63}
\ee
where
\be
I_{n}(y)=\frac{{\cal P}}{\pi}\int^{\infty}_{-\infty}dy\,\frac{y^{n}}{\sinh y}.
\label{apx3.64}
\ee
\mbox{\,\,\,\,\,\,\,\,\,}
For even values of $n$, the integrand will always be an odd function, therefore $I_{n}=0$. However, consider the following integral for odd values of $n$, 
\be
J(a)=\frac{{\cal P}}{\pi}\int^{\infty}_{-\infty}dy\,\frac{e^{ay}}{\sinh y}=\tan\left(\frac{a\pi}{2}\right),
\label{apx3.65}
\ee
in such a way that
\be
I_{n}(y)=\frac{d^{n}}{da^{n}}J(a)\bigg\vert_{a=0}=\frac{{\cal P}}{\pi}\int^{\infty}_{-\infty}dy\,\frac{y^{n}e^{ay}}{\sinh y}\bigg\vert_{a=0},
\label{apx3.66}
\ee
and then, one arrives at
\be
\textnormal{Re}\,f^{+}(e^{\xi})-K=e^{\xi}\sum^{\infty}_{n=0}\frac{1}{n!}\,\frac{d^{n}}{da^{n}}\,\tan\left(\frac{a\pi}{2}\right)\bigg\vert_{a=0}\frac{d^{n}}{d\xi^{\prime n}}\,g(\xi^{\prime})\bigg\vert_{\xi^{\prime}=\xi}.
\label{apx3.67}
\ee
Hence, by expanding the $\tan(x)$ function, differentiating it $n$-times with respect to $a$ and collecting together the survival terms when performed $a=0$ with the $n$-derivatives of $g(\xi^{\prime})$ at $\xi^{\prime}=\xi$, one is led to
\be
\textnormal{Re}\,f^{+}(e^{\xi})-K=e^{\xi}\tan\left[\frac{\pi}{2}\,\frac{d}{d\xi}\right]g(\xi),
\label{apx3.68}
\ee
notice that the tangent operator acts as a differential operator and covers the series expansion. Therefore, by means of the prescription defined before, it leads to
\be
\frac{\textnormal{Re}\,f^{+}(s)}{s}=\frac{K}{s}+\tan\left[\frac{\pi}{2}\,\frac{d}{d\log s}\right]\frac{\textnormal{Im}\,f^{+}(s)}{s}.
\label{apx3.69}
\ee
This expression represents the derivative dispersion relation for an even scattering amplitude. By means of an analogous procedure, the odd relation is written as
\be
\textnormal{Re}\,f^{-}(s)=\tan\left[\frac{\pi}{2}\,\frac{d}{d\log s}\right]\textnormal{Im}\,f^{-}(s),
\label{apx3.70}
\ee
where it was neglected the contribution of the subtraction constant, because the odd amplitude accounts for the differences between the $pp$ and $\bar{p}p$ data at low energies.

It is possible to show that by defining $x=\log s$ and $e^{x}=s$, then
\be
\textnormal{Im}\,f^{-}(s)=\left(\frac{\textnormal{Im}\,f^{-}(e^{x})}{e^{x}}\right)e^{x}\equiv G(x)\,e^{x},
\label{apx3.71}
\ee
by expanding the tangent operator,
\be
\frac{d}{dx}\left[G(x)\,e^{x}\right]=G^{\prime}(x)\,e^{x}+G(x)\,e^{x}=e^{x}\left[G^{\prime}(x)+G(x)\right],
\label{apx3.72}
\ee
\be
\begin{split}
\frac{d^{2}}{dx^{2}}\left[G(x)\,e^{x}\right] & =G^{\prime\prime}(x)\,e^{x}+G^{\prime}(x)\,e^{x}+G^{\prime}(x)\,e^{x}+G(x)\,e^{x}\\
& = e^{x}\left[G^{\prime\prime}(x)+2G^{\prime}(x)+G(x)\right],
\end{split}
\label{apx3.73}
\ee
\be
\frac{d^{3}}{dx^{3}}\left[G(x)\,e^{x}\right]=e^{x}\left[G^{\prime\prime\prime}(x)+3G^{\prime\prime}(x)+3G^{\prime}(x)+G(x)\right],
\label{apx3.74}
\ee
and collecting terms,
\bear
\tan\left[\frac{\pi}{2}\,\frac{d}{dx}\right]\left[G(x)\,e^{x}\right]&=& e^{x}\left\{\frac{\pi}{2}\left[G^{\prime}(x)+G(x)\right]+\frac{1}{3}\left(\frac{\pi}{2}\right)^{3}\left[G^{\prime\prime}(x)+2G^{\prime}(x)+G(x)\right]+...\right\}\nonumber\\
&=& e^{x}\left\{\frac{\pi}{2}\left[1+\frac{d}{dx}\right]G(x)+\frac{1}{3}\left(\frac{\pi}{2}\right)^{3}\left[1+\frac{d}{dx}\right]^{3}G(x)+...\right\}\nonumber\\
&=& e^{x}\tan\left[\frac{\pi}{2}\left(1+\frac{d}{dx}\right)\right]G(x).
\label{apx3.75}
\eear
Therefore, this result implies that expression (\ref{apx3.70}) can be rewritten as
\be
\frac{\textnormal{Re}\,f^{-}(s)}{s}=\tan\left[\frac{\pi}{2}\left(1+\frac{d}{d\log s}\right)\right]\frac{\textnormal{Im}\,f^{-}(s)}{s}.
\label{apx3.76}
\ee

By following the previous exact procedure, a similar result can be properly found for inverse derivative dispersion relations, \tit{i.e.} when the imaginary part is obtained by means of its real component,
\bear
\textnormal{Im}\,f^{+}(s)&=&K-\tan\left[\frac{\pi}{2}\,\frac{d}{d\ln s}\right]\textnormal{Re}\,f^{+}(s) \label{eq7} \\
&=& K - s\tan\left[\frac{\pi}{2}\left(1+\frac{d}{d\ln s}\right)\right]\frac{\textnormal{Re}\,f^{+}(s)}{s} 
\label{apxeq8}, 
\eear
and
\bear
\textnormal{Im}\,f^{-}(s) = K - s\tan\left[\frac{\pi}{2}\,\frac{d}{d\ln s}\right]\frac{\textnormal{Re}\,f^{-}(s)}{s} \label{eq11}
\label{apxeq12}.
\eear

\section{\textsc{Relation Between the Eikonal Function and the Scattering Amplitude}}
\label{secAPX3.2}
\mbox{\,\,\,\,\,\,\,\,\,}
As it was already mentioned before, the physical amplitude for a $pp$ and $\bar{p}p$ scattering is written by means of even and odd terms connected by crossing symmetry,
\be
F^{\bar{p}p}_{pp}(s,t)=F^{+}(s,t)\pm F^{-}(s,t),
\label{apx3.86}
\ee
then the even and odd amplitudes can be written as
\be
F^{\pm}(s,t)=\frac{F_{\bar{p}p}(s,t)\pm F_{pp}(s,t)}{2}.
\label{apx3.87}
\ee
\mbox{\,\,\,\,\,\,\,\,\,}
By means of the Durand \& Pi prescription, $\Gamma(s,b)=1-e^{-\chi(s,b)}$, and expression (\ref{ch2.101}), the forward physical scattering amplitude is written as
\be
F(s,t=0)=i\I db\,b\,\left[1-e^{-\chi_{pp}^{\bar{p}p}(s,b)}\right],
\label{apx3.88}
\ee
where $\chi_{pp}^{\bar{p}p}(s,b)=\chi^{+}(s,b)\pm\chi^{-}(s,b)$. In the case of even amplitude, one finds
\be
\begin{split}
2F^{+}(s) & = i\I db\,b\,\left[1-e^{-\chi_{\bar{p}p}(s,b)}\right]+i\I db\,b\,\left[1-e^{-\chi_{pp}(s,b)}\right]\\
& = i\I db\,b\,\left[1-e^{-\left(\chi^{+}+\chi^{-}\right)}\right]+i\I db\,b\,\left[1-e^{-\left(\chi^{+}-\chi^{-}\right)}\right],
\end{split}
\label{apx3.89}
\ee
by expanding the exponential,
\be
\begin{split}
2F^{+}(s) & = i\I db\,b\,\left\{1-\left[1+(\chi^{+}+\chi^{-})+\frac{(\chi^{+}+\chi^{-})^{2}}{2!}+\frac{(\chi^{+}+\chi^{-})^{3}}{3!}+...\right]\right\}+\\
& + i\I db\,b\,\left\{1-\left[1-(\chi^{+}-\chi^{-})+\frac{(\chi^{+}-\chi^{-})^{2}}{2!}+\frac{(\chi^{+}-\chi^{-})^{3}}{3!}+...\right]\right\},
\end{split}
\label{apx3.90}
\ee
and neglecting higher order terms ${\cal O}(\chi^{2})$, then
\bear
F^{+}(s) &=& \frac{i}{2}\I db\,b\,\left\{(\chi^{+}+\chi^{-})+(\chi^{+}-\chi^{-})\right\}\nonumber\\
&=& i\I db\,b\,\left\{2\chi^{+}\right\}\nonumber\\
&=& i\I db\,b\,\left\{\chi_{_{R}}^{+}+i\chi_{_{I}}^{+}\right\}\nonumber\\
&=& \underbrace{\I db\,b\,\left[-\chi_{_{I}}^{+}\right]}_{\textnormal{Re}\,F^{+}(s)}+\underbrace{i\I db\,b\,\left[\chi_{_{R}}^{+}\right]}_{\textnormal{Im}\,F^{+}(s)},
\label{apx3.91}
\eear
Bearing in mind that $F^{+}(s)=\textnormal{Re}\,F^{+}(s)+i\,\textnormal{Im}\,F^{+}(s)$, then at first-order of the eikonal expansion,
\begin{equation*}
\textnormal{Re}\,F^{+}(s)\longleftrightarrow -\textnormal{Im}\,\chi^{+}(s,b),
\end{equation*}
\begin{equation*}
\textnormal{Im}\,F^{+}(s)\longleftrightarrow \textnormal{Re}\,\chi^{+}(s,b).
\end{equation*}
This means that only in first-order of approximation the eikonal function has the same cut structure as the scattering amplitude. Thus justifying the inverse integral dispersion relation in expression (\ref{chDGM.26}). Following the same procedure, in the case of an odd scattering amplitude,
\be
\begin{split}
2F^{-}(s) & = i\I db\,b\,\left\{(\chi^{+}+\chi^{-})-(\chi^{+}-\chi^{-})\right\}\\
& = i\I db\,b\,\left\{2\chi^{-}\right\},
\end{split}
\label{apx3.92}
\ee
then,
\be
\begin{split}
F^{-}(s) & = i\I db\,b\,\chi^{-} \\
& = i\I db\,b\,\left\{\chi_{_{R}}^{-}+i\chi_{_{I}}^{-}\right\}\\
& = \underbrace{\I db\,b\,\left[-\chi_{_{I}}^{-}\right]}_{\textnormal{Re}\,F^{-}(s)}+\underbrace{i\I db\,b\,\left[\chi_{_{R}}^{-}\right]}_{\textnormal{Im}\,F^{-}(s)},
\end{split}
\label{apx3.93}
\ee
therefore leading to a similar result,
\begin{equation*}
\textnormal{Re}\,F^{-}(s)\longleftrightarrow -\textnormal{Im}\,\chi^{-}(s,b),
\end{equation*}
\begin{equation*}
\textnormal{Im}\,F^{-}(s)\longleftrightarrow \textnormal{Re}\,\chi^{-}(s,b).
\end{equation*}
\mbox{\,\,\,\,\,\,\,\,\,}
If, however, the profile function is written according to the usual DGM prescription, which is exactly the one in expression (\ref{ch2.102}), the relation between the eikonal function and the scattering amplitude will slightly change. Therefore, it is not difficult to show that, under this prescrition condition, the relation between the even parts of the eikonal function and the scattering amplitude reads
\begin{equation*}
\textnormal{Re}\,F^{+}(s)\longleftrightarrow \textnormal{Re}\,\chi^{+}(s,b),
\end{equation*}
\begin{equation*}
\textnormal{Im}\,F^{+}(s)\longleftrightarrow \textnormal{Im}\,\chi^{+}(s,b),
\end{equation*}
as for the case of its odd parts,
\begin{equation*}
\textnormal{Re}\,F^{-}(s)\longleftrightarrow \textnormal{Re}\,\chi^{-}(s,b),
\end{equation*}
\begin{equation*}
\textnormal{Im}\,F^{-}(s)\longleftrightarrow \textnormal{Im}\,\chi^{-}(s,b).
\end{equation*}

\clearpage
\thispagestyle{plain}

\chapter{\textsc{Some Useful Calculations}}
\label{APX2}

\mbox{\,\,\,\,\,\,\,\,\,}
Perhaps in Chapter \ref{chQIM}, which presents the newest version of the original dynamical gluon mass model, one probably felt a lack of mathematical rigorous demonstrations. And this is exactly the reason why this Appendix was written, to tie up these loose ends. 

\section{\textsc{The Overlap Density Function}}
\label{secAPX2.1}

\mbox{\,\,\,\,\,\,\,\,\,}
The hadron overlap density function at impact parameter $b$  is written as
\be
A(b)=\int d^{2}\bb^{\prime} \, \rho_{A}(\vert\bb-\bb^{\prime}\vert)\rho_{B}(b^{\prime}),
\label{apx2.1}
\ee
and normalization\footnote{In reference  \cite{LHeureux:1985qwr} the eikonal function takes its form from the contribution from each of the nine pairs of valence (anti)quarks that corresponds to each of the two (anti)protons, and the contribution of all reactions initiated by a pair of gluons even though this number cannot be determined. Then, the overlap density function for valence quarks is normalized to $9$, and for the gluons it is fixed as an effective number which is obtained by a fitting parameter.}
\be
N\int d^{2}\bb\, A(b) = 1.
\label{apx2.2}
\ee
\mbox{\,\,\,\,\,\,\,\,\,}
However, if one writes the parton density function $\rho_{B}(b^{\prime})$ as the Fourier transform of the form factor $G(q^{2})$,
\be
\rho_{B}(b^{\prime})=\frac{1}{2\pi}\int d^{2}\qq\,G(q^{2})\,\text{e}^{i\,\qq\cdot \bb^{\prime}}=\frac{1}{2\pi}\int^{2\pi}_{0}d\phi \int^{\infty}_{0}dq \, q \,G(q^{2})\,\text{e}^{i\,q\,b^{\prime}\cos{\phi}},
\label{apx2.3}
\ee
by means of the integral form of the Bessel function one can easily see that
\be
\rho_{B}(b^{\prime})=\int^{\infty}_{0} dq \, q \, J_{0}(qb)\,G(q^{2}).
\label{apx.2.4}
\ee
\mbox{\,\,\,\,\,\,\,\,\,}
Moreover, the inverse Fourier transform also defines $G(q^{2})$,
\be
G(q^2)=\frac{1}{2\pi}\int d^{2}\bb^{\prime}\, \rho_{B}(b^{\prime})\text{e}^{-i\,\qq\cdot\bb^{\prime}}.
\label{apx2.5}
\ee
\mbox{\,\,\,\,\,\,\,\,\,}
Now considering that $b^{\prime\prime}=\vert\bb-\bb^{\prime}\vert$, and following the exactly same procedure,
\bear
\rho_{A}(\vert\bb-\bb^{\prime}\vert) & =&\frac{1}{2\pi}\int d^{2}\qq\,G(q^{2})\,\text{e}^{i\,\qq\cdot (\bb-\bb^{\prime})}\nonumber\\
& =&\int^{\infty}_{0} dq \, q \, J_{0}(qb)\,G(q^{2})\,\text{e}^{-i\,\qq\cdot \bb^{\prime}},
\label{apx2.6}
\eear
and hence, $A(b)$ can be rewritten as
\be
A(b)=\int^{\infty}_{0} dq \, q \, J_{0}(qb)\,G(q^{2})\int d^{2}\bb^{\prime}\,\rho_{B}(b^{\prime})\,\text{e}^{-i\,\qq\cdot \bb^{\prime}}.
\label{apx2.7}
\ee
\mbox{\,\,\,\,\,\,\,\,\,}
The second integral in the \tit{rhs} of (\ref{apx2.7}) is simply the form factor $G(b^{\prime})$,
\bear
A(b)&=&2\pi\int^{\infty}_{0}dq\,q\,J_{0}(qb)\left[G(q^{2})\right]^{2}\nonumber\\
&=& \int d^{2}\qq\,\,J_{0}(qb)\left[G(q^{2})\right]^{2}.
\label{apx2.8}
\eear

\section{\textsc{The Distribution Function}}
\label{secAPX2.2}

\mbox{\,\,\,\,\,\,\,\,\,}
At a given impact parameter $b$ the distribution function $W(b;\mu)$ is defined as the overlap density function multiplied by the normalization factor $N$,
\be
W(b;\mu)=N A(b;\mu).
\label{apx2.9}
\ee
\mbox{\,\,\,\,\,\,\,\,\,}
Different form factors implies different distribution functions, since in essence it is simply written as the Hankel transform of $G(q^{2})$, sometimes named Fourrier-Bessel transform.

\subsection{\textsc{Monopole-like Form Factor}}
\label{secAPX2.2.1}

\mbox{\,\,\,\,\,\,\,\,\,}
A monopole distribution is simply defined as
\be
G(q^{2})=\left(1+\frac{q^{2}}{\mu^{2}}\right)^{-1}.
\label{apx2.10}
\ee
For the moment in the attempt to not overload the notation let us define the following simplified form $\left<(...)\right>\equiv \int^{\infty}_{0} dq \, q J_{0}(qb)(...)$. Therefore, the overlap density function will be given by
\be
\frac{A(b)}{2\pi}=\left<\left(\frac{\mu^{2}}{q^{2}+\mu^{2}}\right)^{2}\right>=\mu^{4}\,\left<\left(\frac{1}{q^{2}+\mu^{2}}\right)^{2}\right>.
\label{apx2.11}
\ee
\mbox{\,\,\,\,\,\,\,\,\,}
It is well known that the Hankel transform of an Yukawa-type distribution is a simple modified Bessel function\footnote{Or occasionally called as the hyperbolic Bessel function.} of the second kind and zeroth order,
\be
\left<\frac{1}{q^{2}+\mu^{2}}\right>=K_{0}(\mu b).
\label{apx2.12}
\ee
Moreover, in the literature there are many different ways that higher orders of $K_{\alpha}$ functions can be generated \cite{california1954tables}, \tit{e.g.},
\be
\left(\frac{d}{xdx}\right)^{m}\left[x^{n}K_{n}(x)\right]=(-1)^{m}x^{n-m}K_{n-m}(x),
\label{chDGM.76}
\ee
and also
\be
\left(\frac{d}{xdx}\right)^{m}\left[x^{-n}K_{n}(x)\right]=(-1)^{m}x^{-n-m}K_{n+m}(x),
\label{chDGM.77}
\ee
thus for $m=1$ and $x=\nu b$, then
\be
\partial_{\mu}\left[\nu^{n}K_{n}(\mu b)\right]=-b\mu^{n}K_{n-1}(\mu b),
\label{chDGM.78}
\ee
and
\be
\partial_{\mu}\left[\nu^{-n}K_{n}(\mu b)\right]=-b\mu^{-n}K_{n+1}(\mu b).
\label{chDGM.79}
\ee
where $\partial_{\mu}=\partial/\partial\mu$. Now, it is easy to find out that for the monopole-like form factor the overlap distribution function can be written as
\be
A(b)=2\pi\,\frac{\mu^{2}}{2}\,(\mu b)\,K_{1}(\mu b),
\label{apx2.15}
\ee
whilst that
\bear
\left<\left(\frac{1}{q^{2}+\mu^{2}}\right)^{2}\right>&=&-\frac{1}{2\mu}\,\partial_{\mu}\left<\frac{1}{q^{2}+\mu^{2}}\right>\nonumber\\
&=&-\frac{1}{2\mu}\,\partial_{\mu}\,K_{0}(\mu b)\nonumber\\
&=&\frac{b}{2\mu}\,K_{1}(\mu b).
\label{apx2.16}
\eear

The normalization of the distribution function as it was mentioned in expressions (\ref{apx2.2}) and (\ref{apx2.9}), follows as $\int d^{2}b\, W(b;\mu)=N\int d^{2}b\,A(b)=1$. For this reason one finds that for the case of a monopole-like distribution the normalization factor will be obtained by
\bear
N\int d^{2}b\,A(b)&=&N\int^{2\pi}_{0}d\phi\int^{\infty}_{0}db\,b\,A(b)=N\,2\pi\,2\pi\,\frac{\mu^{3}}{2}\int^{\infty}_{0}db\,b^{2}\,K_{1}(\nu b)\nonumber\\
&=&N\,(2\pi)^{2}\,\Gamma(2)\,\Gamma(1)=1,
\label{apx.2.17}
\eear
and hence implying the value $N=1/4\pi^{2}$. To calculate the integral of the $K_{\alpha}$ function, it was used the following relation \cite{california1954tables}:
\be 
\begin{split}
\int^{\infty}_{0}dx\, x^{\mu}\,K_{\nu}(ax)&\,=2^{\mu-1}\,a^{-\mu-1}\,\Gamma\left(\frac{1+\mu+\nu}{2}\right)\,\Gamma\left(\frac{1+\mu-\nu}{2}\right),\\
&\, \text{Re}\{\mu+1\pm\nu\}>0 \,\, \text{e} \,\, \text{Re}>0,\\
\end{split} 
\label{apx.2.18}
\ee
where the Euler gamma function is $\Gamma(n+1)=n!$. Finally, the normalized distribution function for the monopole model is written as
\be
W(b;\mu)=\frac{\mu^{2}}{4\pi}(\mu b)K_{1}(\mu b).
\label{apx2.19}
\ee

\subsection{\textsc{Dipole-like Form Factor}}
\label{secAPX2.2.2}

\mbox{\,\,\,\,\,\,\,\,\,}
A dipole distribution is defined as
\be
G(q^{2})=\left(1+\frac{q^{2}}{\nu^{2}}\right)^{-2},
\label{apx2.20}
\ee
and within the same procedure as before,
\bear
\frac{A(b)}{2\pi}&=&\left<\left(\frac{\nu^{2}}{q^{2}+\nu^{2}}\right)^{4}\right>\nonumber\\
&=& \nu^{8}\left<\left(\frac{1}{q^{2}+\nu^{2}}\right)^{2}\left(\frac{1}{q^{2}+\nu^{2}}\right)^{2}\right>\nonumber\\
&=& \frac{\nu^{6}}{4}\left\{\frac{1}{2}\,\partial_{\nu}\partial_{\nu}\left<\left(\frac{1}{q^{2}+\nu^{2}}\right)^{2}\right> - \left<\left(\frac{1}{q^{2}+\nu^{2}}\right)\partial^{2}_{\nu}\left(\frac{1}{q^{2}+\nu^{2}}\right)\right>\right\}.
\label{apx2.21}
\eear
\mbox{\,\,\,\,\,\,\,\,\,}
The math here is basically the same, but regardless it will be shown step by step. The first term in curly brackets in the \tit{rhs} of expression (\ref{apx2.21}) can be written as a linear combination of $K_{\alpha}$ functions,
\bear
\frac{1}{2}\,\partial_{\nu}\partial_{\nu}\left<\left(\frac{1}{q^{2}+\nu^{2}}\right)^{2}\right>&=&\frac{1}{2}\,\partial_{\nu}\partial_{\nu}\left(\frac{b}{2\nu}\,K_{1}(\nu b)\right)\nonumber\\
&=&\frac{b^4}{4}\left[\frac{K_{3}(\nu b)}{\nu b}-\frac{K_{2}(\nu b)}{(\nu b)^2}\right],
\label{apx2.22}
\eear
and similarly for the second term in the \tit{rhs},
\bear
\left<\left(\frac{1}{q^{2}+\nu^{2}}\right)\partial^{2}_{\nu}\left(\frac{1}{q^{2}+\nu^{2}}\right)\right>&=&\left<\frac{8\nu^{2}}{(q^{2}+\nu^{2})^{4}}\right>-\left<\frac{2}{(q^{2}+\nu^{2})^{3}}\right>\nonumber\\
&=&\left<\frac{8\nu^{2}}{(q^{2}+\nu^{2})^{4}}\right>-\frac{b^{2}}{4\nu^{2}}\,K_{2}(\nu b).
\label{apx2.23}
\eear
\mbox{\,\,\,\,\,\,\,\,\,}
Finally, substituting these last two expression into (\ref{apx2.21}),
\be
\nu^{8}\left<\left(\frac{1}{q^{2}+\nu^{2}}\right)^{4}\right>=\frac{\nu^{5}b^{3}}{16}\,K_{3}(\nu b)-2\nu^{8}\left<\left(\frac{1}{q^{2}+\nu^{2}}\right)^{4}\right>,
\label{apx2.24}
\ee 
and then the non-normalized overlap density function is found,
\be
A(b)=\left<\left(\frac{\nu^{2}}{q^{2}+\nu^{2}}\right)^{4}\right>=2\pi\,\frac{\nu^{2}}{48}(\nu b)^{3}K_{3}(\nu b).
\label{apx2.25}
\ee
\mbox{\,\,\,\,\,\,\,\,\,}
All that is left, is to determine the normalization factor for the dipole-like model. Then following the same procedure as before, one finds that the normalization $N$ is given  respectively by
\bear
N\int d^{2}b\,A(b)&=&N\int^{2\pi}_{0}d\phi\int^{\infty}_{0}db\,b\,A(b)=2\pi\,2\pi\,\frac{\nu^{5}}{48}\int^{\infty}_{0}db\,b^{4}\,K_{3}(\nu b)\nonumber\\
&=&(2\pi)^{2}\,\frac{1}{6}\,\Gamma(4)\,\Gamma(1)=1.
\label{apx2.26}
\eear
Hence, again, the normalization is given by the value $N=1/4\pi^{2}$. Thus, the normalized distribution function for the dipole model is written as
\be
W(b;\nu)=\frac{\nu^{2}}{96\pi}(\nu b)^{3}K_{3}(\nu b).
\label{apx2.27}
\ee
\mbox{\,\,\,\,\,\,\,\,\,}
Perhaps one might see that there is a quicker way to calculate the Hankel transform within the infinite integral \cite{california1954tables}
\be
\int^{\infty}_{0}dq\,q\,J_{0}(qb)\left(\frac{1}{q^{2}+\mu^{2}}\right)^{\alpha+1}=\left(\frac{b}{2\mu}\right)^{\alpha}\frac{K_{\alpha}(\mu b)}{\Gamma(\alpha+1)},
\label{apx2.28}
\ee
but the procedure shown here is very easy and simple to apply for different kinds of form factors, even in more complicated ones when there is no closed form for the Hankel transform. Such an example is the the Durand-Pi form factor where the gluons in a proton are assumed to be distributed around the valence quarks in a Yukawa-type (monopole) distribution, so that the gluon distribution in a proton is the convolution of a monopole and a dipole \cite{Durand:1988ax}.

Just for completeness, it is interesting to show the following asymptotic limits for the $x^{\alpha}K_{\alpha}$ functions:
\be
\lim_{x\rightarrow \infty} xK_{1}(x) = 0,
\label{apx2.29}
\ee
\be
\lim_{x\rightarrow 0} xK_{1}(x) = 1,
\label{apx2.30}
\ee
\be
\lim_{x\rightarrow \infty} x^{3}K_{3}(x) = 0,
\label{apx2.31}
\ee
\be
\lim_{x\rightarrow 0} x^{3}K_{3}(x) = 8.
\label{apx2.32}
\ee
\mbox{\,\,\,\,\,\,\,\,\,}
These asymptotic behavior limits can be seen in Figure \ref{apx2fig1}, where, for a better pictorical purpose, the typical monopole distribution function was multiplied by a factor of $8$ whilst the dipole remains unaltered.

\bfg[hbtp]
  \begin{center}
    \includegraphics[width=15cm,clip=true]{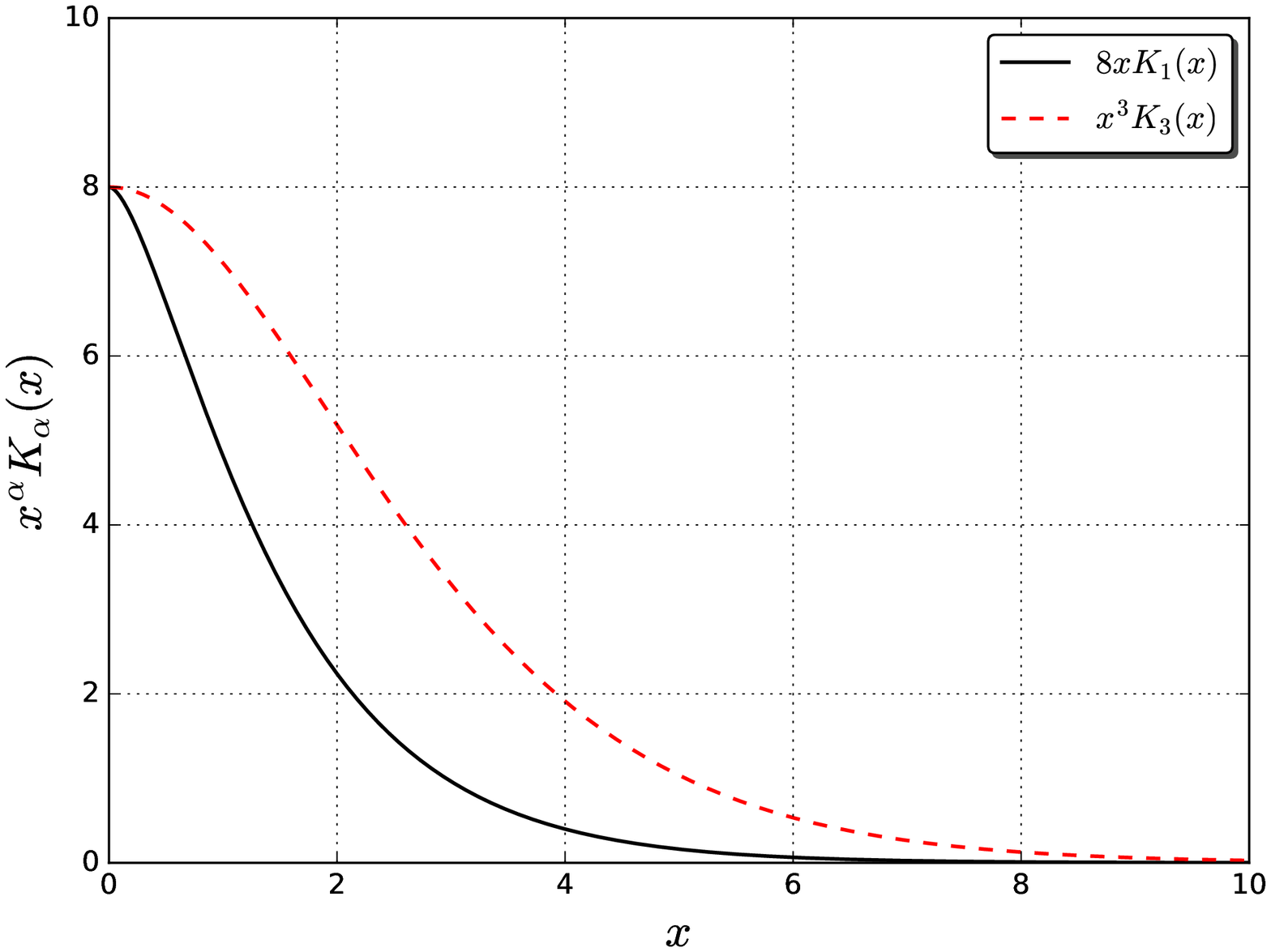}
    \caption{Comparison between the asymptotic behavior limits of typical monopole and dipole distribution functions.}
    \label{apx2fig1}
  \end{center}
\efg

\clearpage
\thispagestyle{plain}

\chapter*{\textsc{Acknowledgments}}
\addtocontents{toc}{\protect\vspace{2ex}} 
\addcontentsline{toc}{chapter}{Acknowledgments} 

\mbox{\,\,\,\,\,\,\,\,\,}
A few years ago Prof. Márcio Menon told me that to work with phenomenology Physics is like taking the risk to make mistakes everyday. The work presented here are our ``little mistakes'' \tit{per se}, and for the time being these are my biggest, and only, contributions to Physics, more specifically in the field of particle physics phenomenology. 

I would like to thank all the colleagues and friends, not just from the University, but specially from the room M$207$. I am grateful for all the quimpirinhas.

Furthermore, I am indebted to many people bonded to me, mostly to Prof. Emerson Luna who had the patience to deal with me.

Once I heard a story where the well-known latvian physicist Prof. Yuri Dokshitzer on one occasion said: \tit{``soft physics is hard and hard physics is soft''}. Well, those were hard times for me.

\clearpage
\thispagestyle{plain}

\end{document}